\documentclass[a4paper,11pt]{article}
\usepackage[T1]{fontenc}
\usepackage[utf8]{inputenc}
\usepackage{lmodern}
\usepackage{textcomp}
\usepackage[a4paper,left=15mm,right=15mm,top=16mm,bottom=18mm,footskip=10mm]{geometry}
\usepackage{amsmath,amssymb,mathtools}
\usepackage{graphicx,array,booktabs,multirow,multicol}
\providecommand{\textsubscript}[1]{\ensuremath{_{\textnormal{#1}}}}
\usepackage{microtype}
\ifdefined\pdfminorversion\pdfminorversion=7\relax\fi
\usepackage{xcolor}
\definecolor{linkblue}{RGB}{27,65,104}
\usepackage[unicode,colorlinks=true,linkcolor=linkblue,citecolor=linkblue,urlcolor=linkblue,bookmarksopen=false]{hyperref}
\hypersetup{pdftitle={Superconductivity from a Cooper surface instability in a nodal Bose metal},pdfauthor={Jian-Jian Miao and Cui Ding}}
\allowdisplaybreaks

\title{Superconductivity from a Cooper surface instability in a nodal Bose metal}
\author{Jian-Jian Miao and Cui Ding}
\newcommand{\ManuscriptAffiliation}{Quantum Science Center of Guangdong-Hong Kong-Macao Greater Bay Area, Shenzhen 518045, China}
\newcommand{\ManuscriptContact}{Corresponding author: \href{mailto:miaojianjian@quantumsc.cn}{miaojianjian@quantumsc.cn}}
\makeatletter
\renewcommand{\maketitle}{%
  \begin{center}
    {\LARGE\bfseries \@title\par}\medskip
    {\large \@author\par}\smallskip
    \ManuscriptAffiliation\par
    \ManuscriptContact\par
  \end{center}}
\makeatother
\renewenvironment{abstract}{\section*{\abstractname}}{\par}

\newsavebox{\formulaBox}
\newsavebox{\formulaNumber}
\newlength{\formulaWidth}
\newcommand{\FittedEquation}[2]{%
  \begingroup
  \setlength{\formulaWidth}{\linewidth}%
  \if\relax\detokenize{#1}\relax\else
    \sbox{\formulaNumber}{\normalfont (#1)}%
    \addtolength{\formulaWidth}{-\wd\formulaNumber}%
    \addtolength{\formulaWidth}{-\wd\formulaNumber}%
    \addtolength{\formulaWidth}{-1em}%
  \fi
  \sbox{\formulaBox}{$\displaystyle #2$}%
  \begin{equation*}
    \ifdim\wd\formulaBox>\formulaWidth
      \resizebox{\formulaWidth}{!}{\usebox{\formulaBox}}%
    \else
      \usebox{\formulaBox}%
    \fi
    \if\relax\detokenize{#1}\relax\else\tag{#1}\fi
  \end{equation*}%
  \endgroup
}

\newsavebox{\tableBox}
\newcommand{\FittedTable}[1]{%
  \par\begingroup
  \setlength{\tabcolsep}{5pt}\renewcommand{\arraystretch}{1.18}%
  \sbox{\tableBox}{#1}%
  \noindent\makebox[\linewidth][c]{%
    \ifdim\wd\tableBox>\linewidth
      \resizebox{\linewidth}{!}{\usebox{\tableBox}}%
    \else
      \usebox{\tableBox}%
    \fi
  }\par\endgroup
}
\newcommand{\ArtworkPanel}[1]{\par\noindent\includegraphics[width=\linewidth]{figures/#1}\par}
\newcommand{\PrintAsset}[1]{\csname artwork@#1\endcsname}
\expandafter\def\csname artwork@fig-1-model\endcsname{\ArtworkPanel{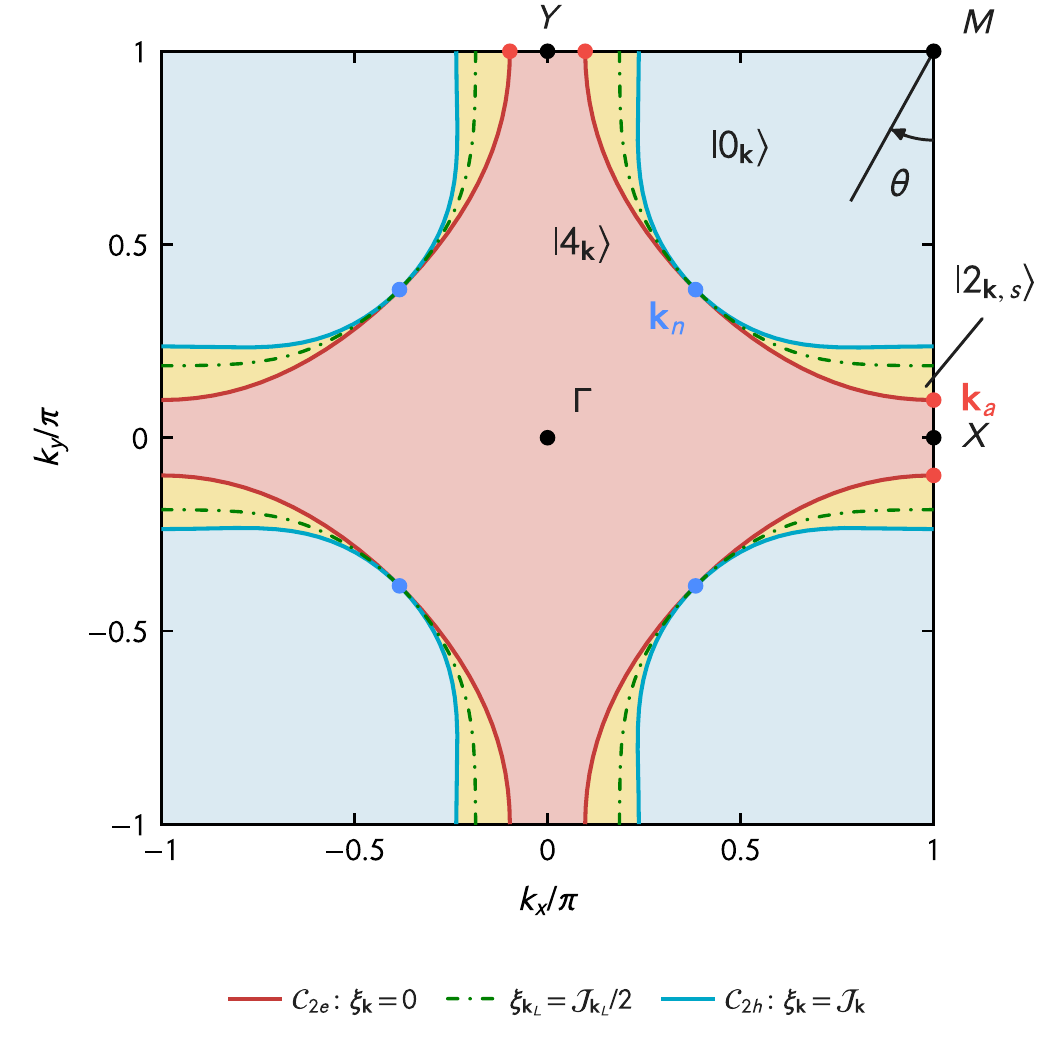}}
\expandafter\def\csname artwork@fig-2-nodal-spectra\endcsname{\ArtworkPanel{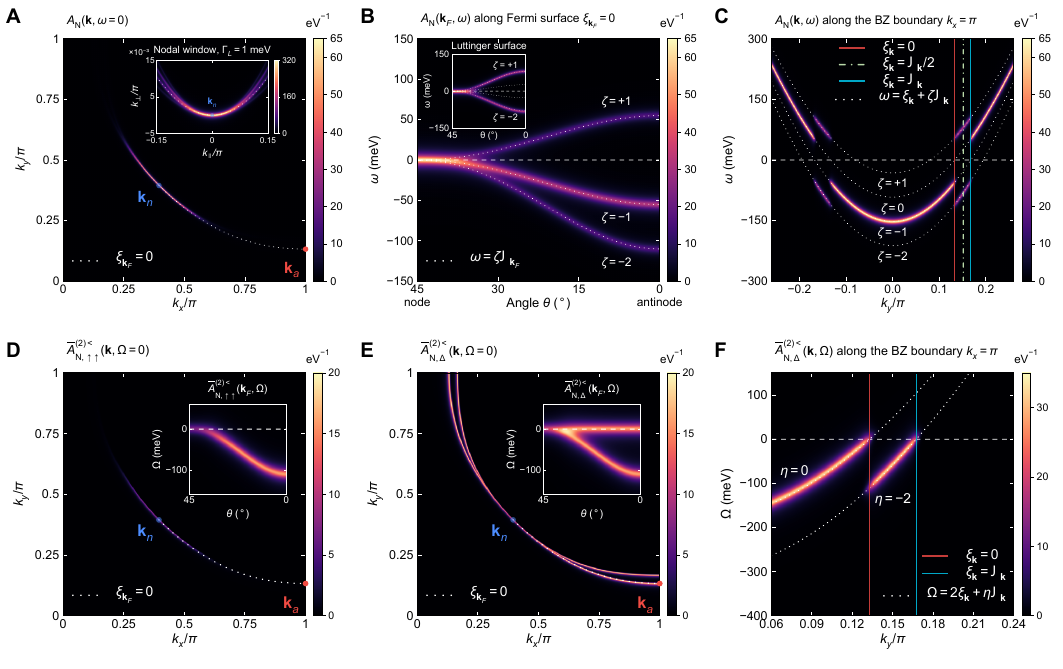}}
\expandafter\def\csname artwork@fig-3-instability\endcsname{\ArtworkPanel{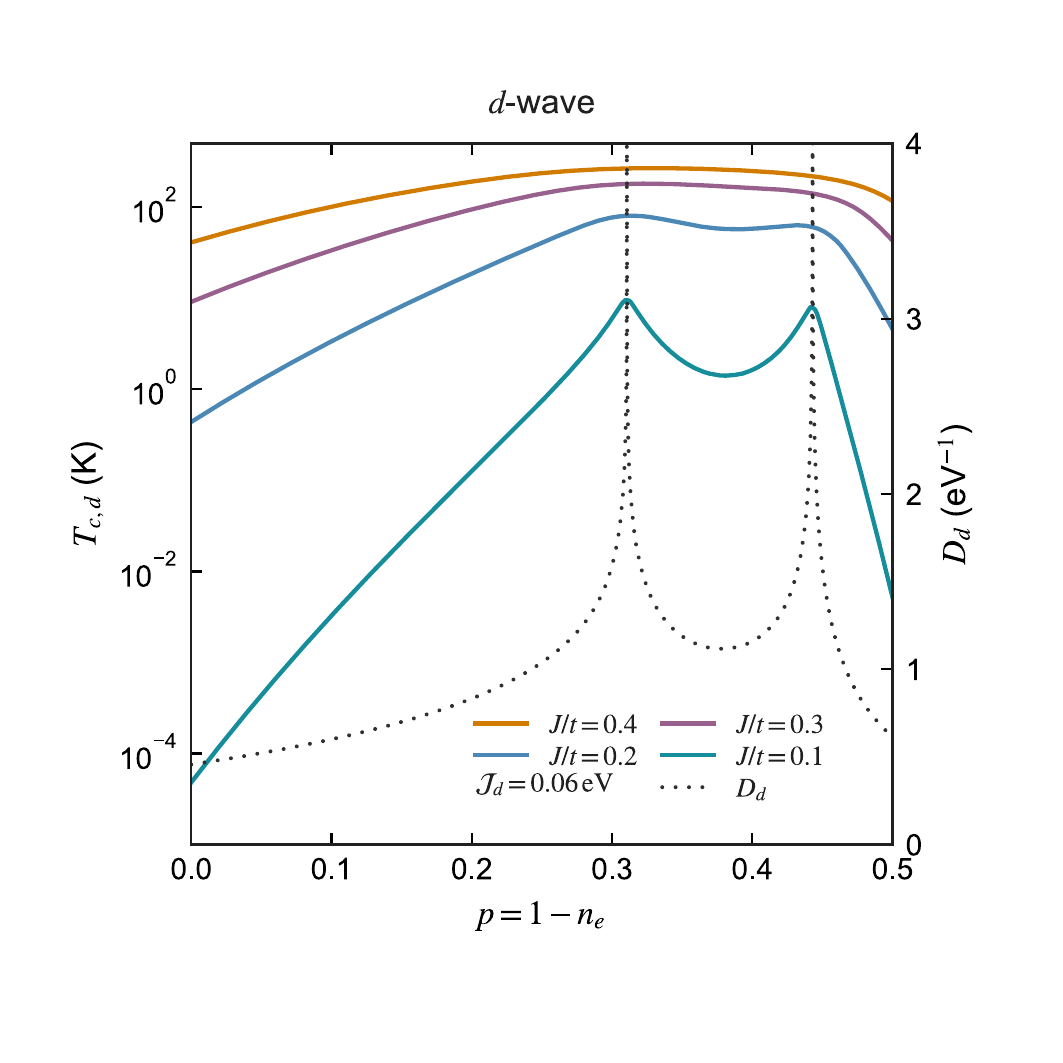}}
\expandafter\def\csname artwork@fig4-correlated-sc\endcsname{\ArtworkPanel{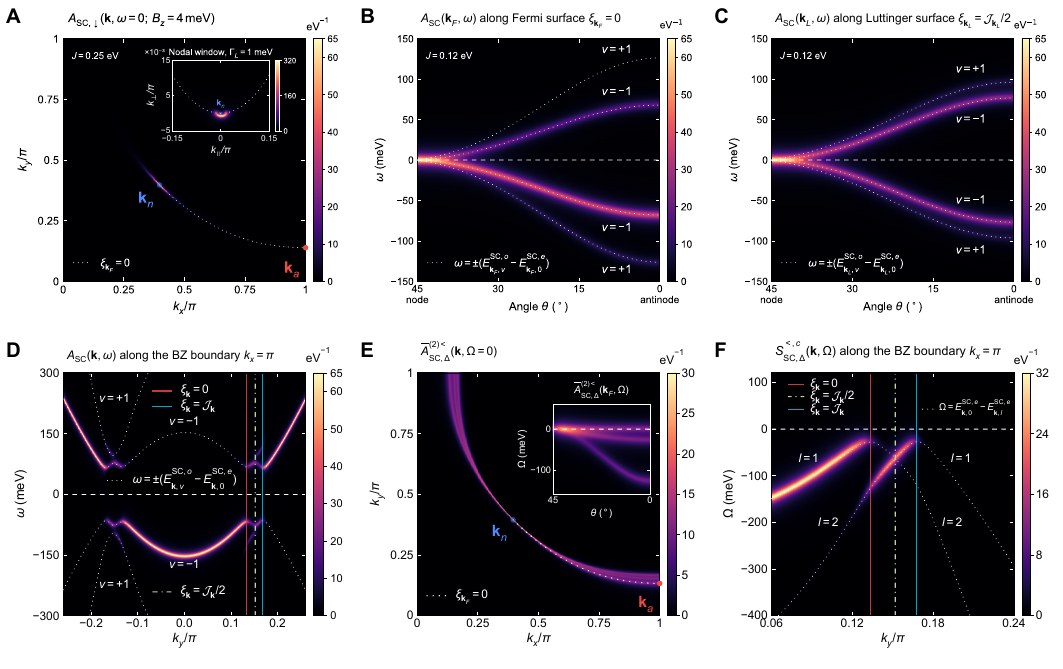}}
\expandafter\def\csname artwork@model-full-bz-Jd-0.3\endcsname{\ArtworkPanel{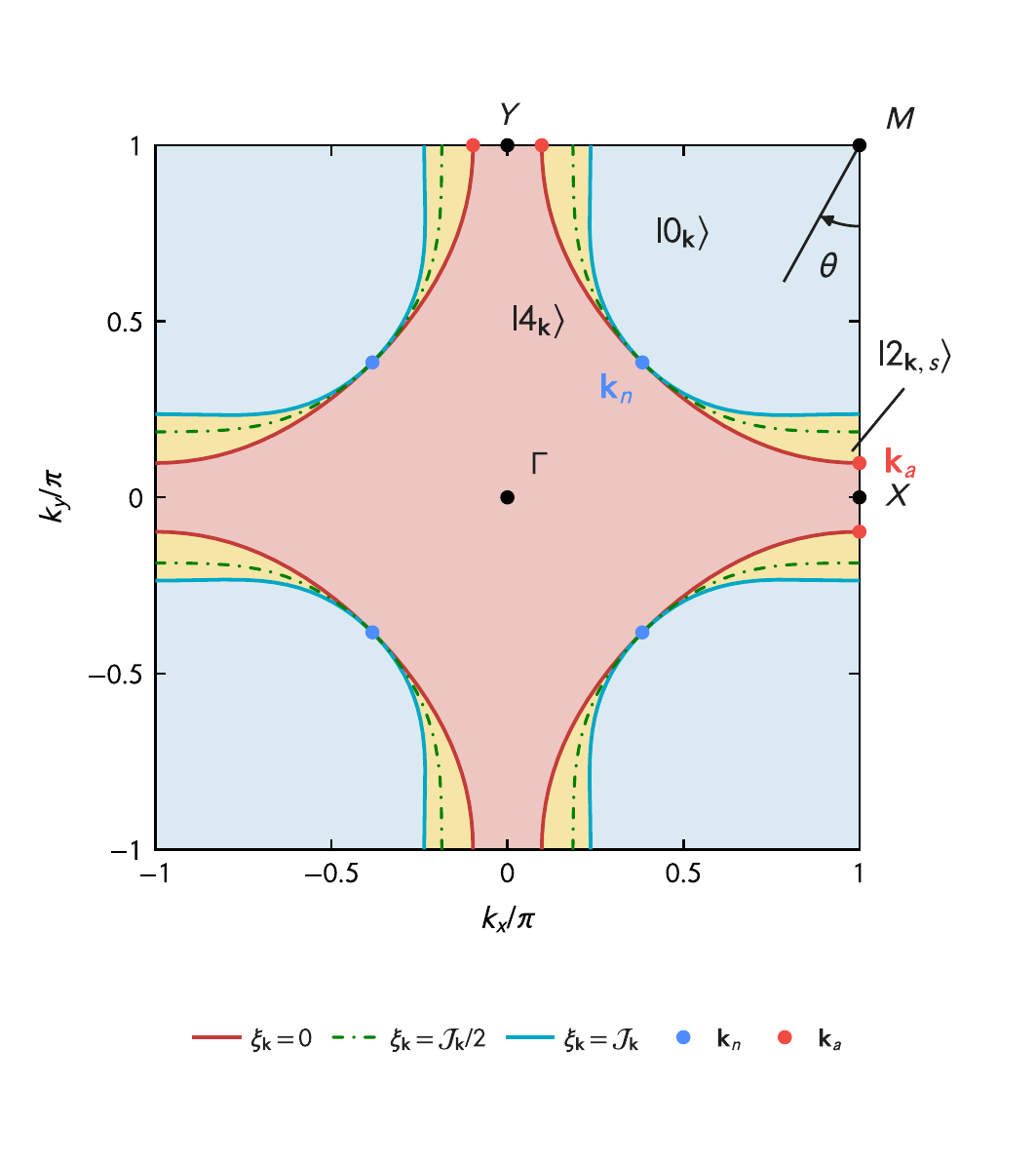}}
\expandafter\def\csname artwork@nL-over-ne-p-Jd\endcsname{\ArtworkPanel{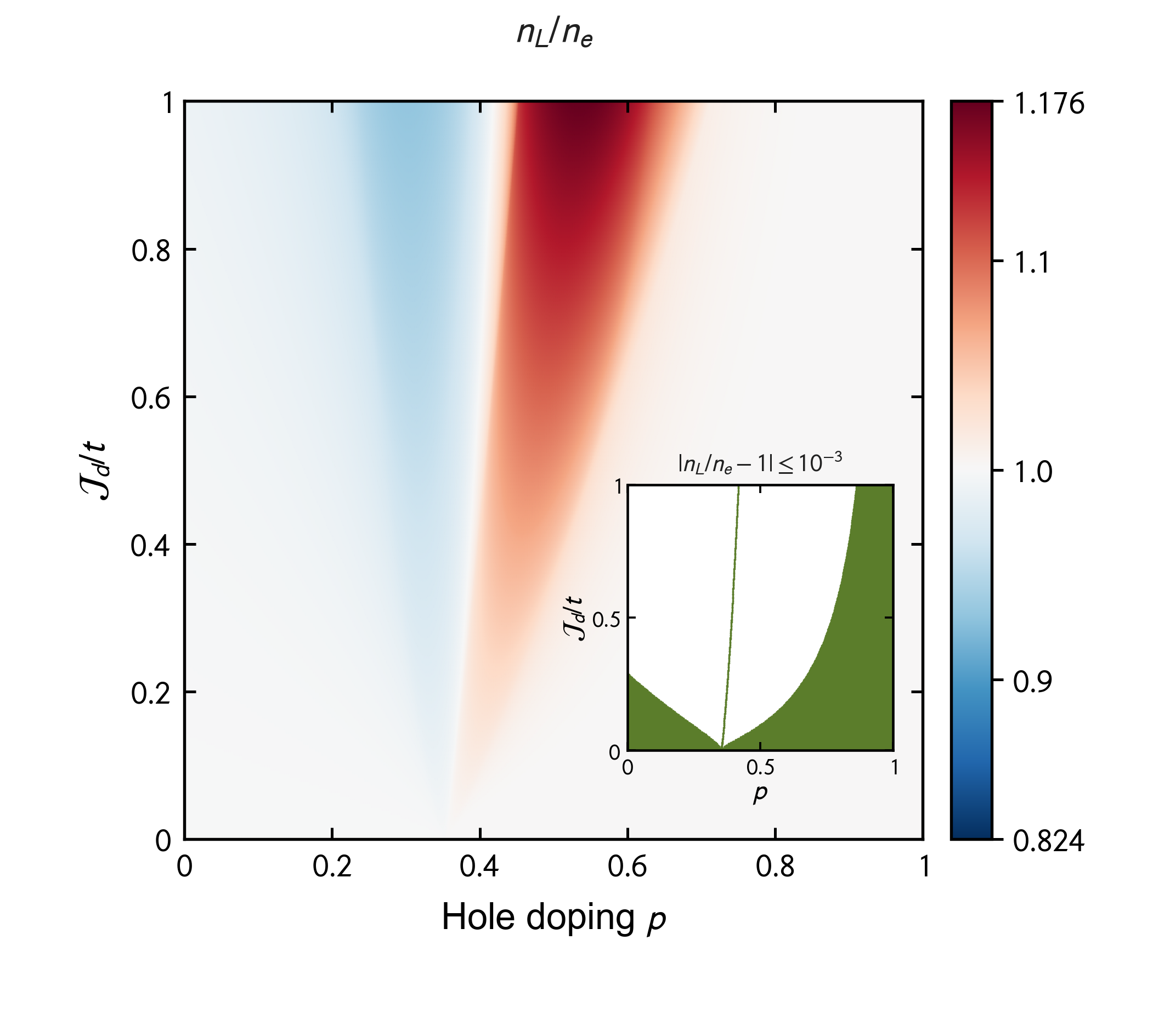}}
\expandafter\def\csname artwork@figure1-full-spectral-function-60k\endcsname{\ArtworkPanel{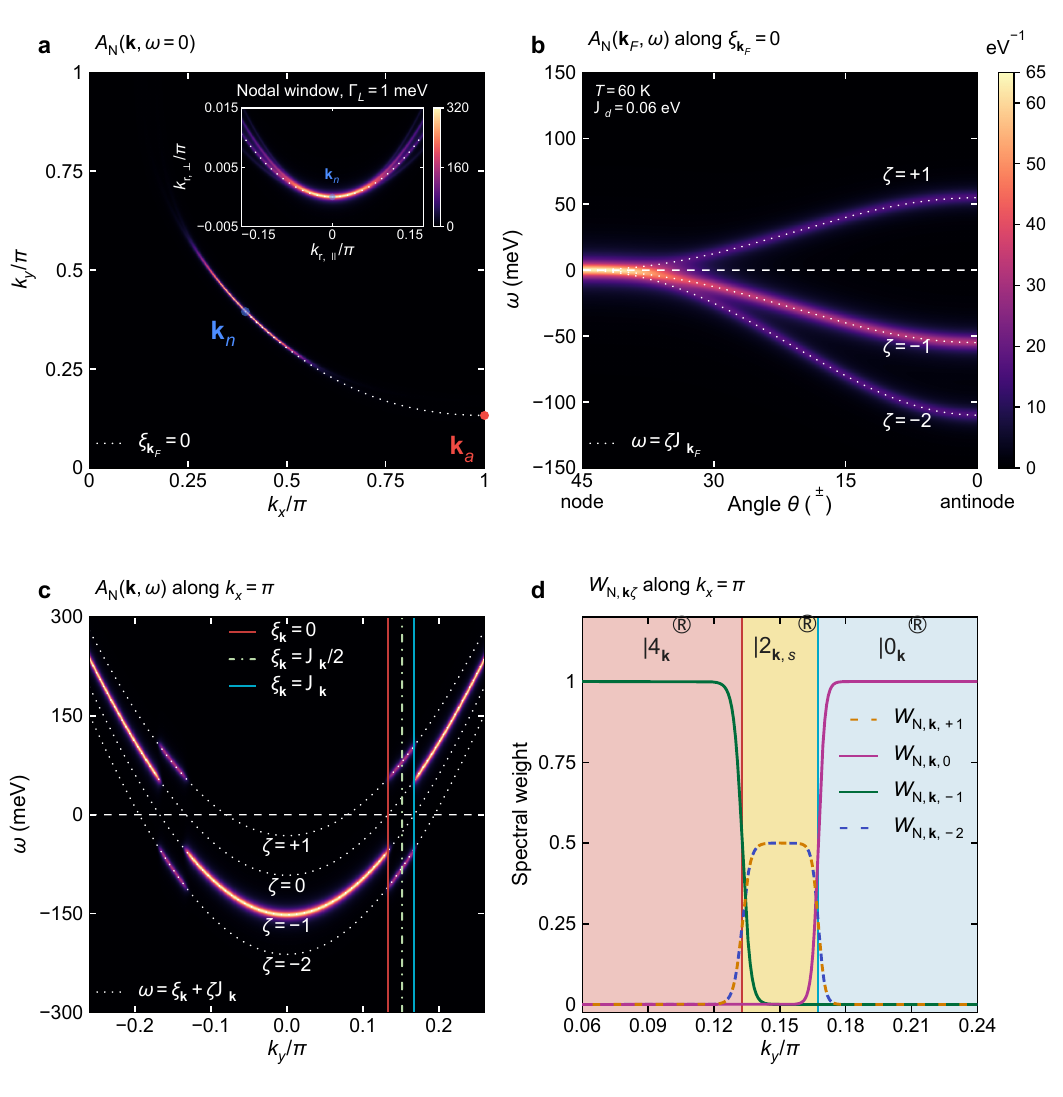}\allowbreak \ArtworkPanel{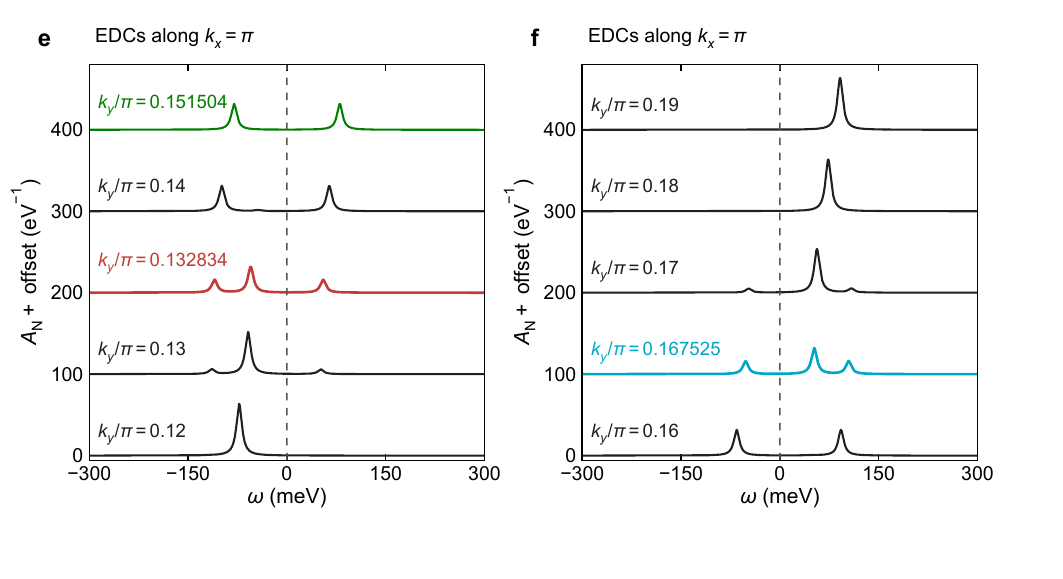}}
\expandafter\def\csname artwork@figure2-spin-resolved-carpes-60k\endcsname{\ArtworkPanel{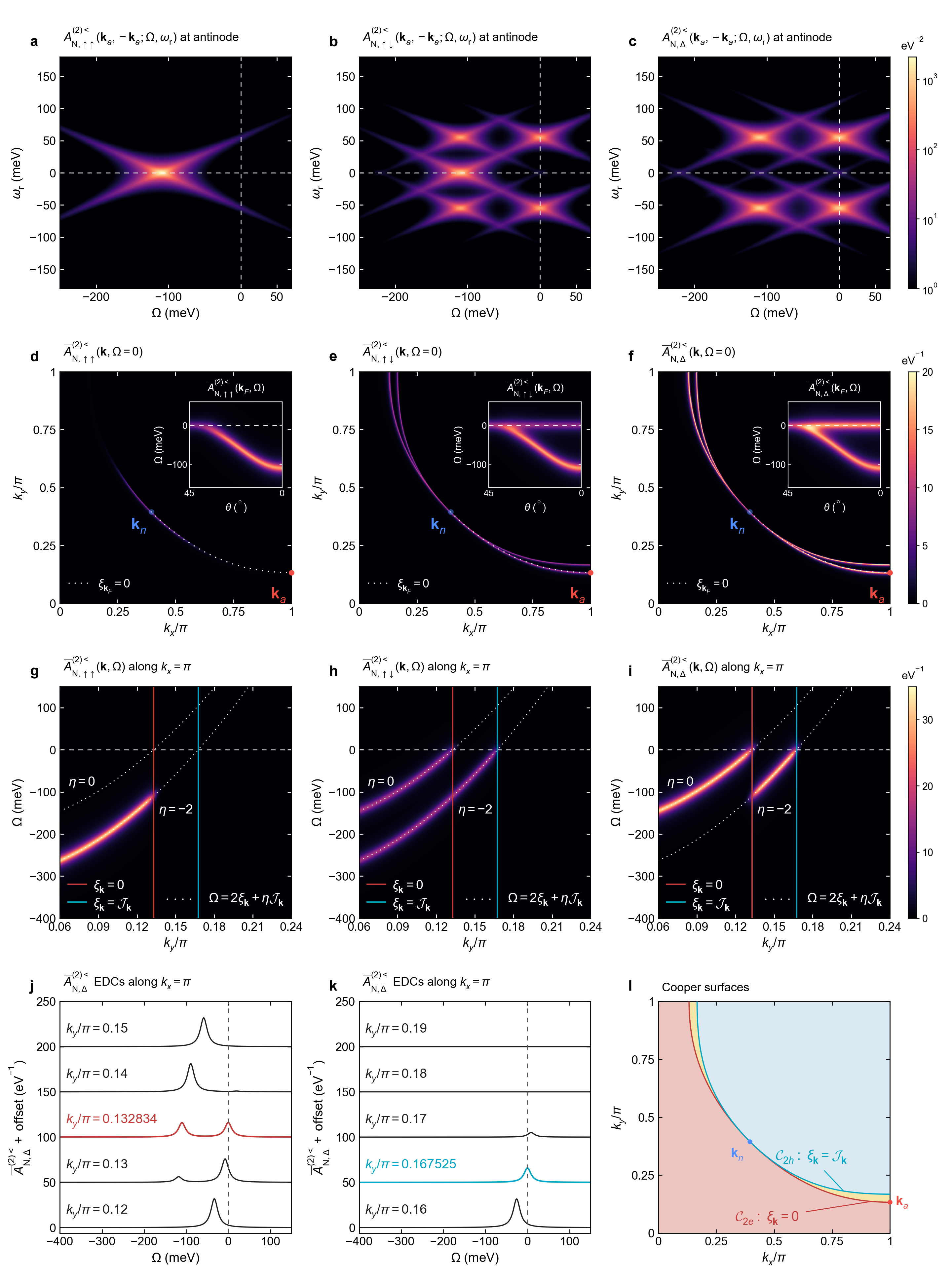}}
\expandafter\def\csname artwork@mstj-normal-observables\endcsname{\ArtworkPanel{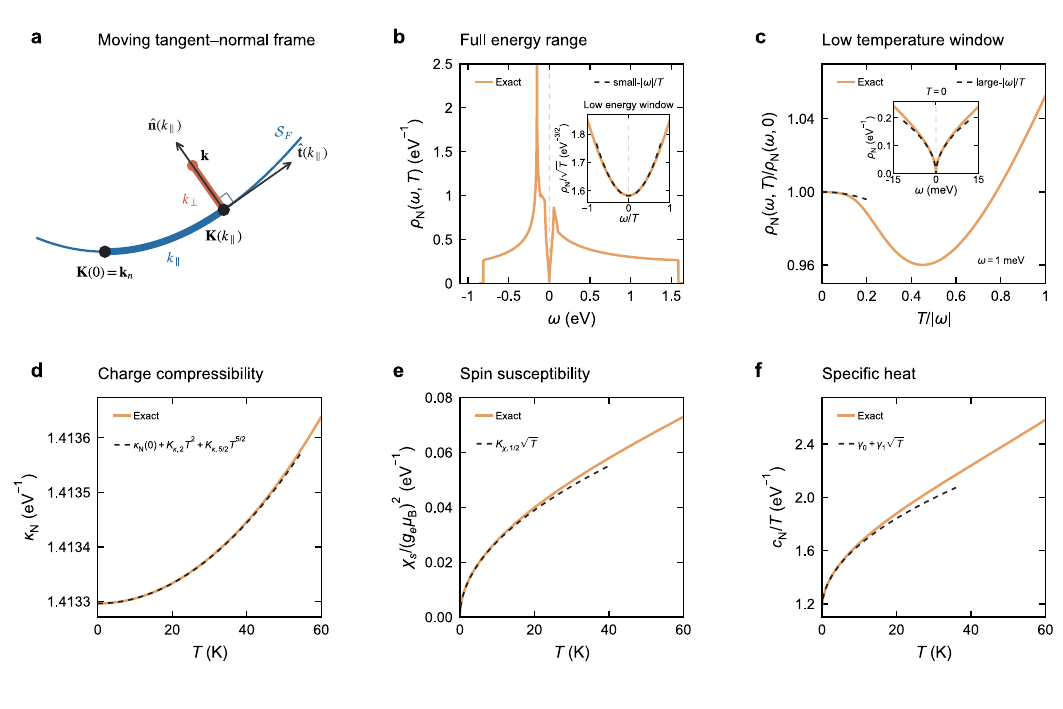}}
\expandafter\def\csname artwork@sc-instability-doping\endcsname{\ArtworkPanel{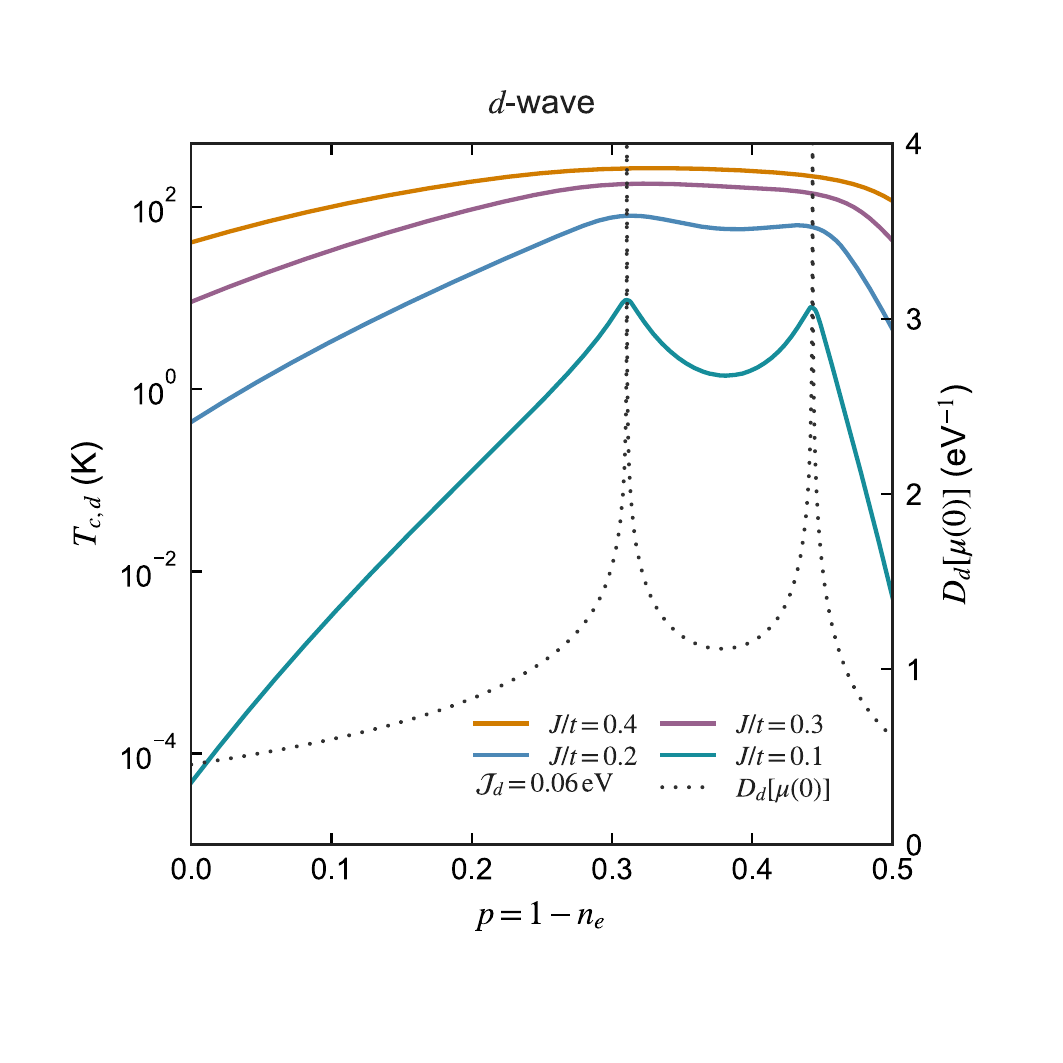}}
\expandafter\def\csname artwork@figure3-doping-dependence\endcsname{\ArtworkPanel{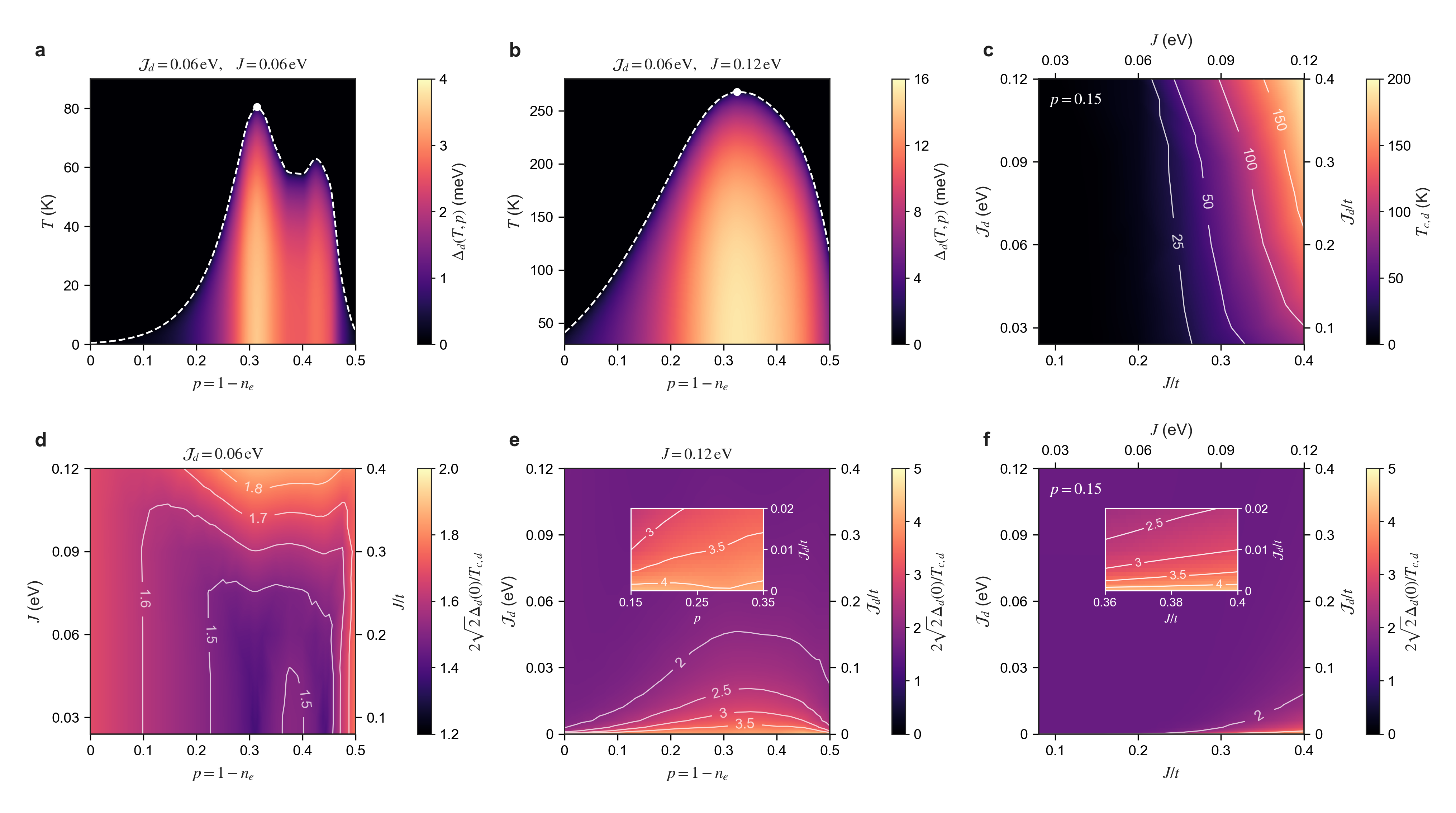}}
\expandafter\def\csname artwork@figure-sc-excitation-gap\endcsname{\ArtworkPanel{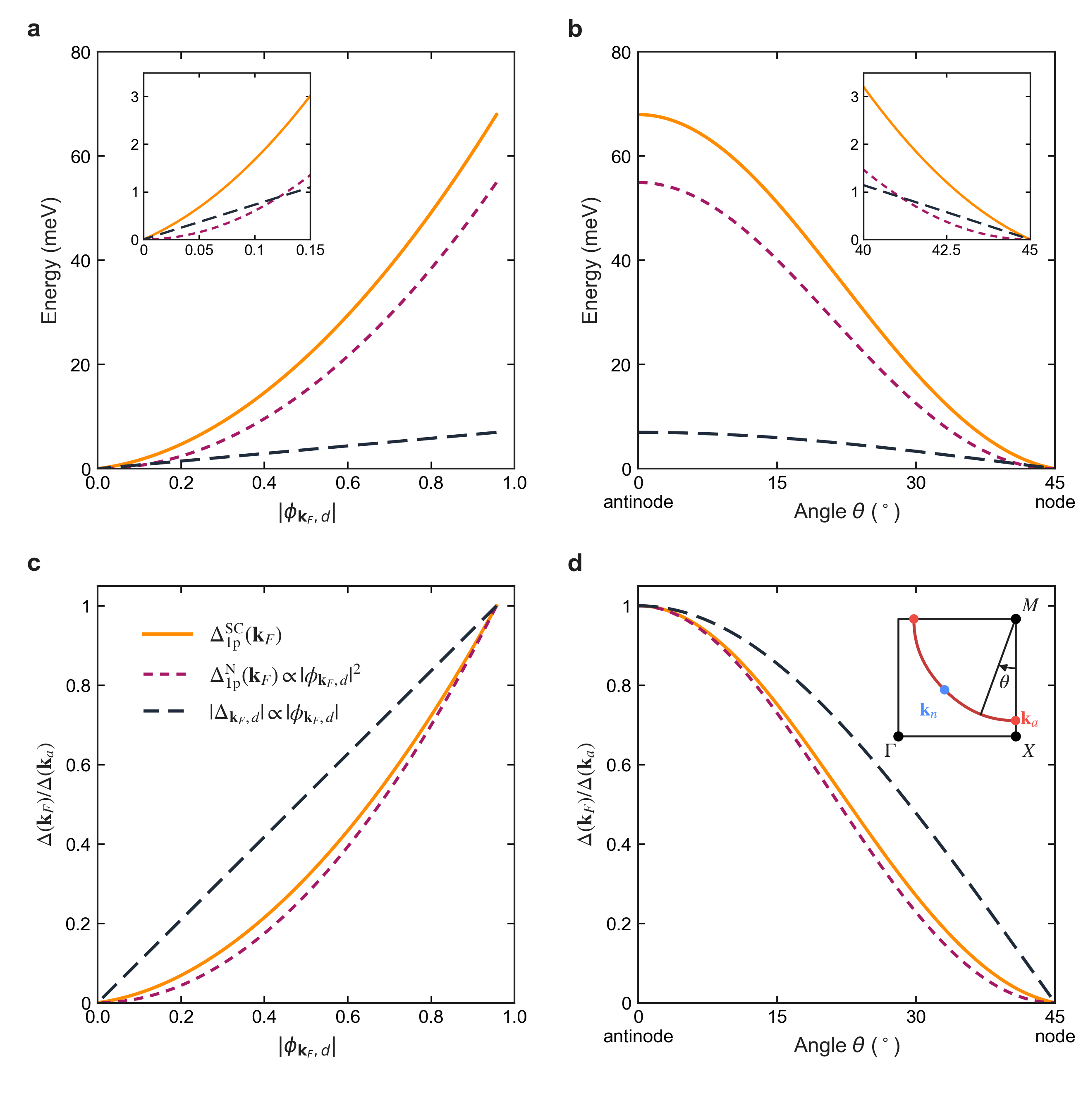}}
\expandafter\def\csname artwork@figure-constant-energy-four-states-row1\endcsname{\ArtworkPanel{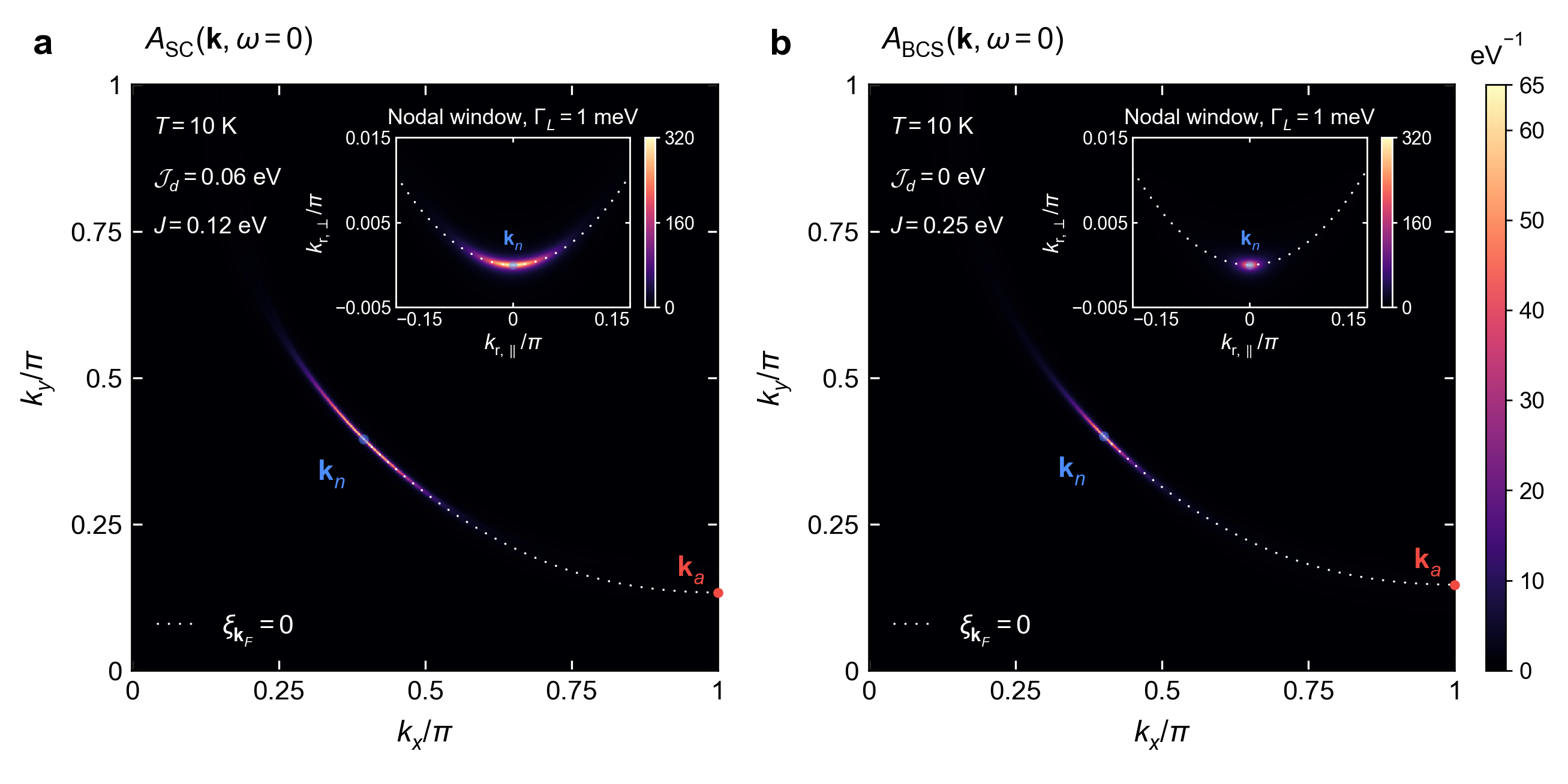}}
\expandafter\def\csname artwork@figure-constant-energy-four-states-row2\endcsname{\ArtworkPanel{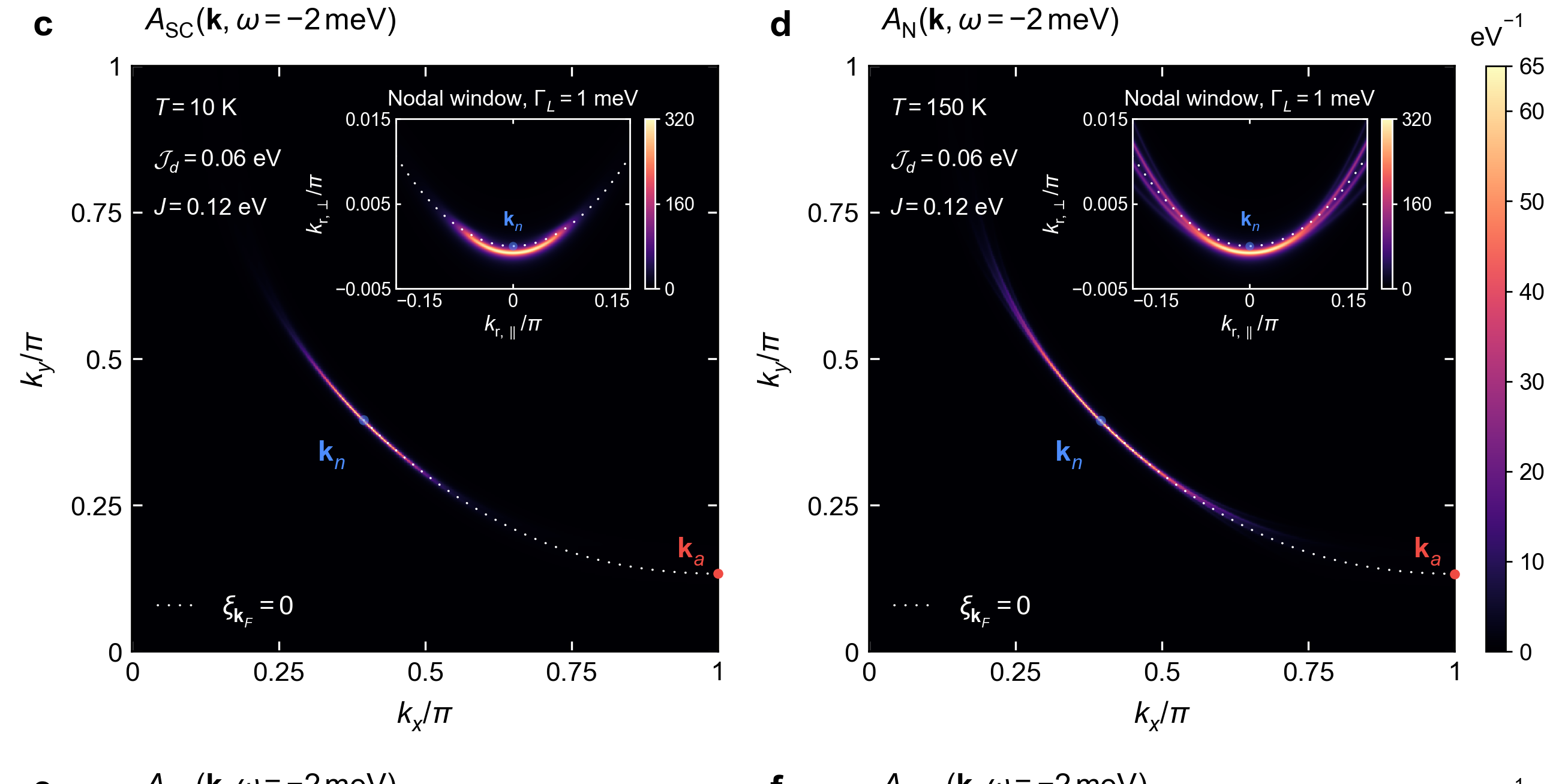}}
\expandafter\def\csname artwork@figure-constant-energy-four-states-row3\endcsname{\ArtworkPanel{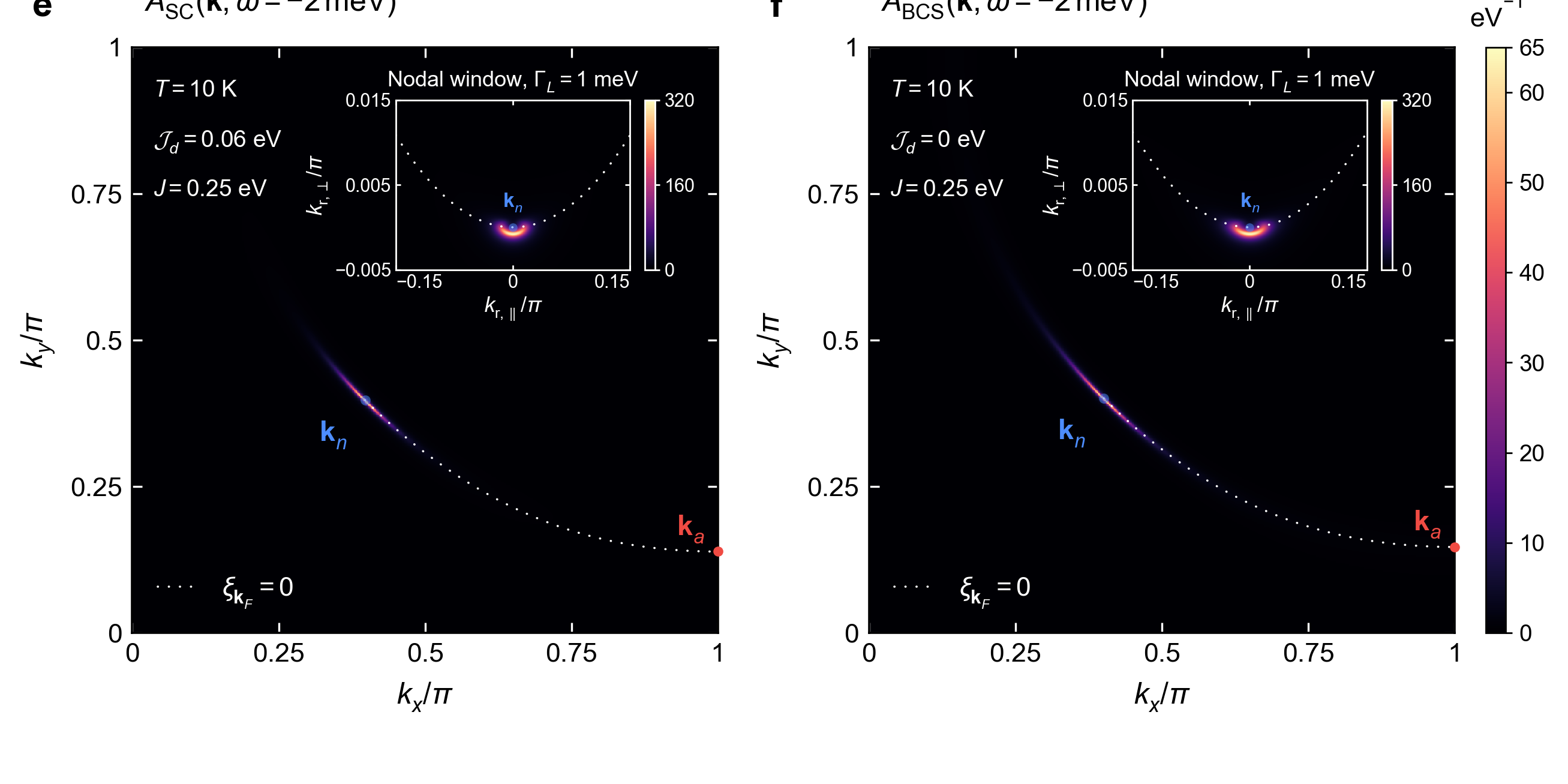}}
\expandafter\def\csname artwork@spin-resolved-zero-Bplus4meV-J025\endcsname{\ArtworkPanel{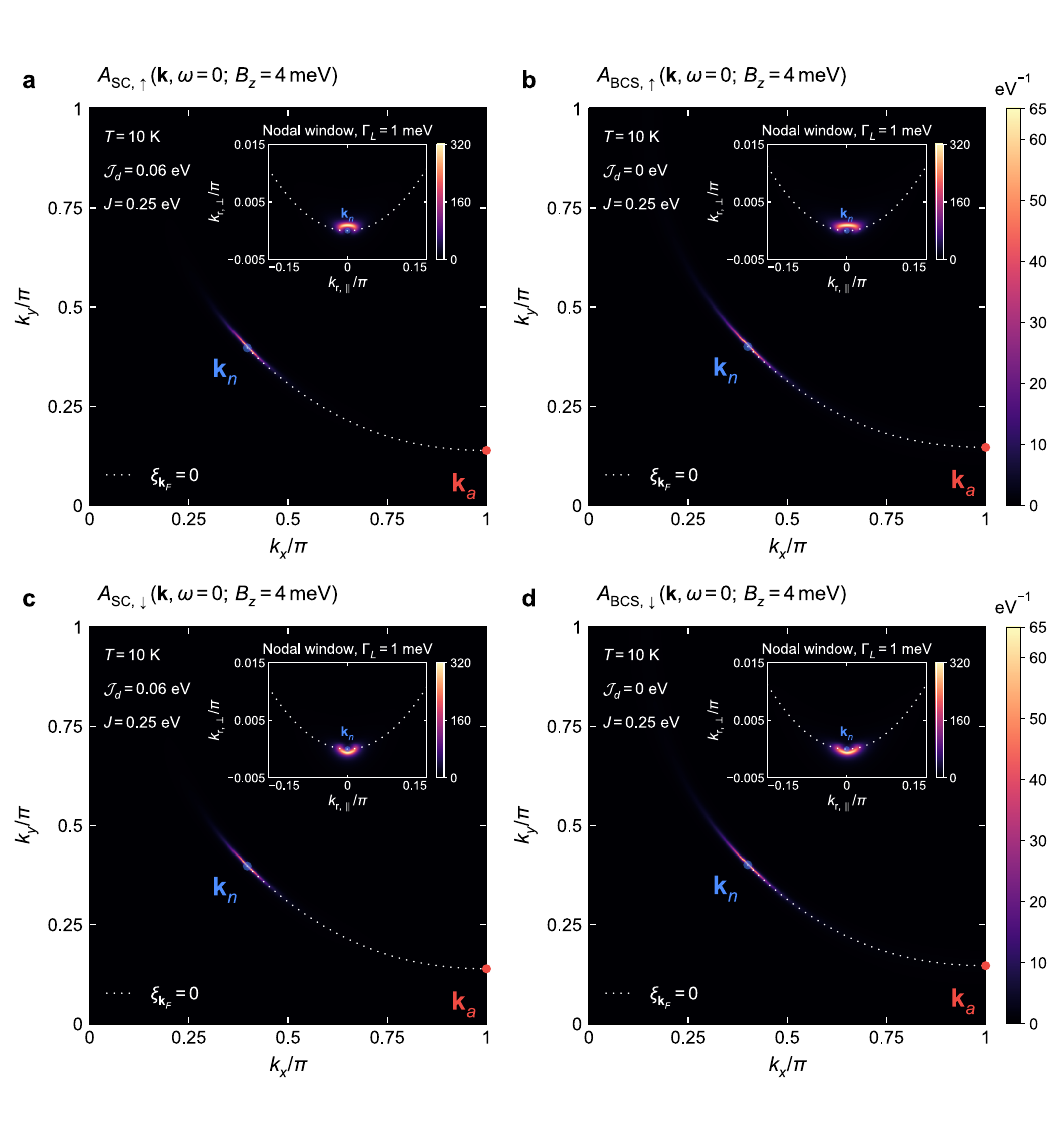}}
\expandafter\def\csname artwork@spectral-ph-symmetry-row1\endcsname{\ArtworkPanel{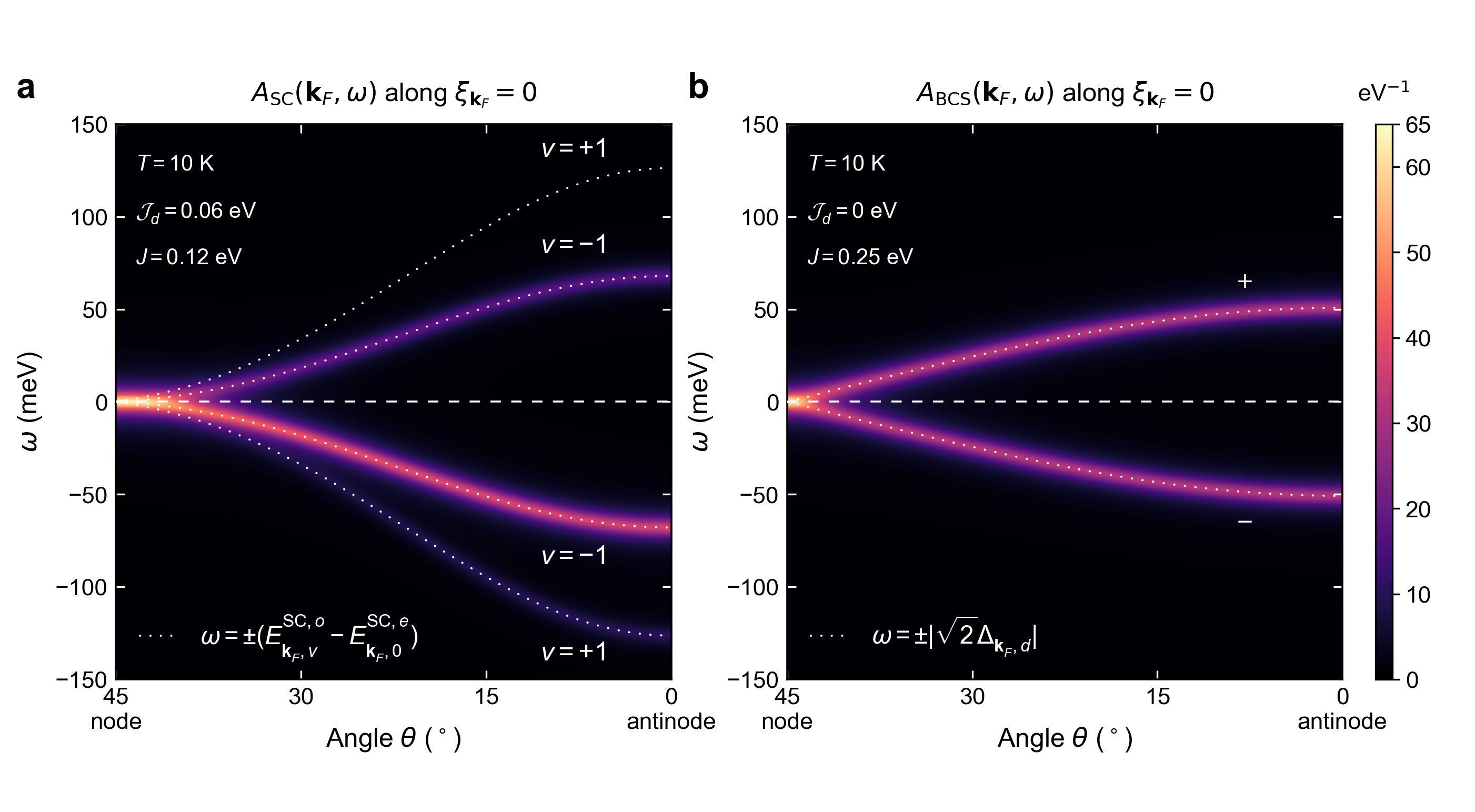}}
\expandafter\def\csname artwork@spectral-ph-symmetry-row2\endcsname{\ArtworkPanel{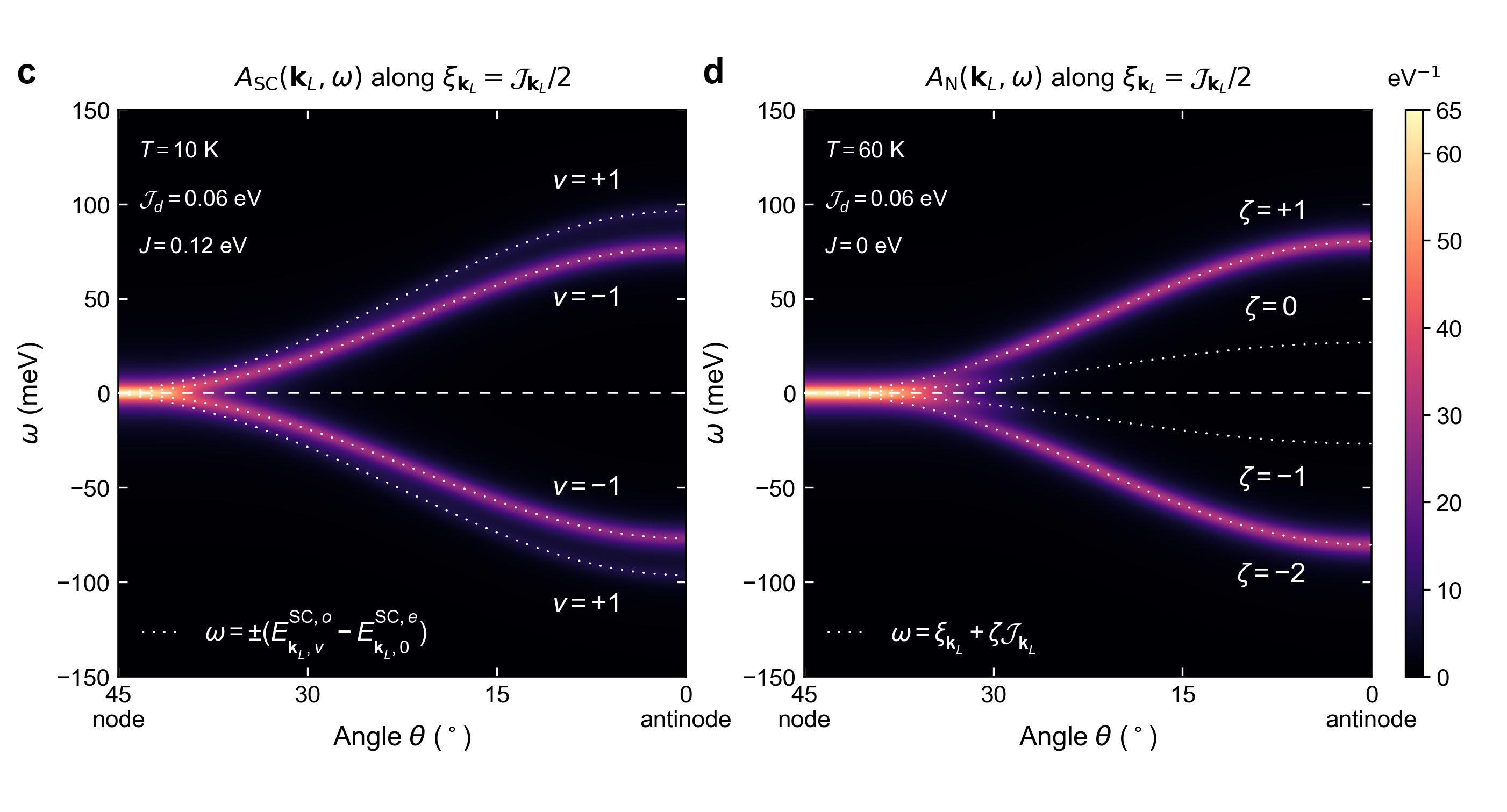}}
\expandafter\def\csname artwork@figure-sc-spectra-comparison-row1\endcsname{\ArtworkPanel{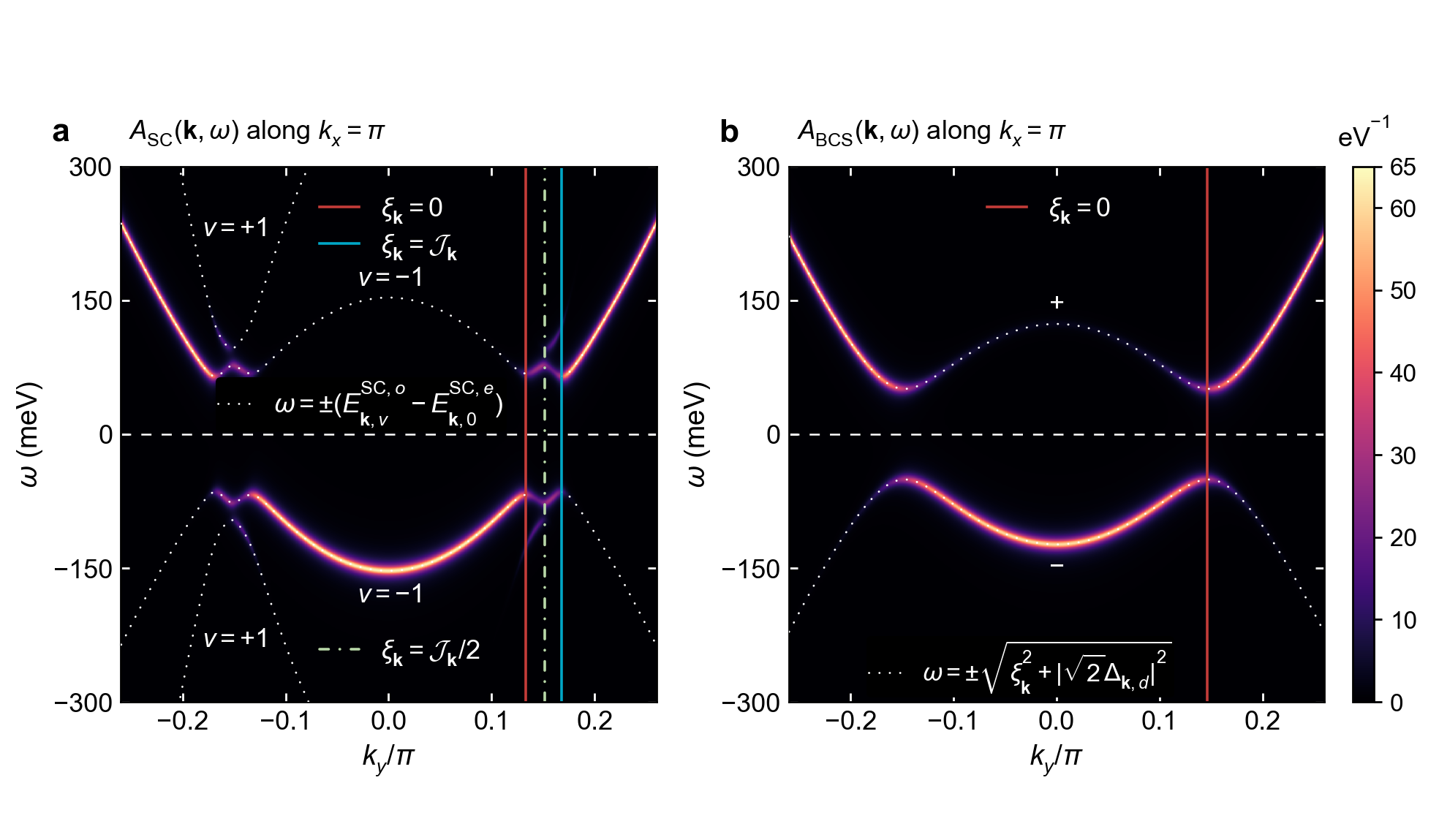}}
\expandafter\def\csname artwork@figure-sc-spectra-comparison-row2\endcsname{\ArtworkPanel{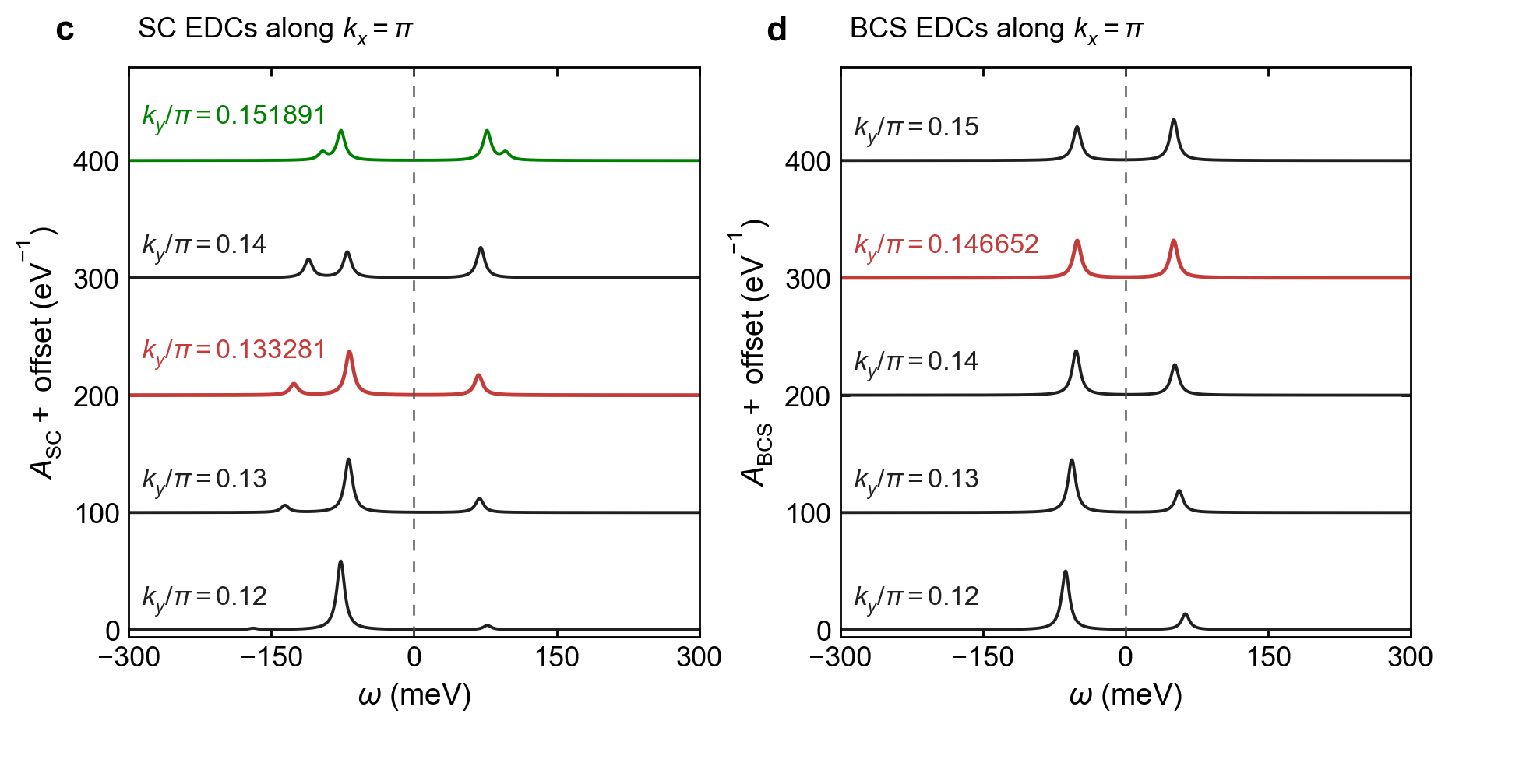}}
\expandafter\def\csname artwork@figure-sc-spectra-comparison-row3\endcsname{\ArtworkPanel{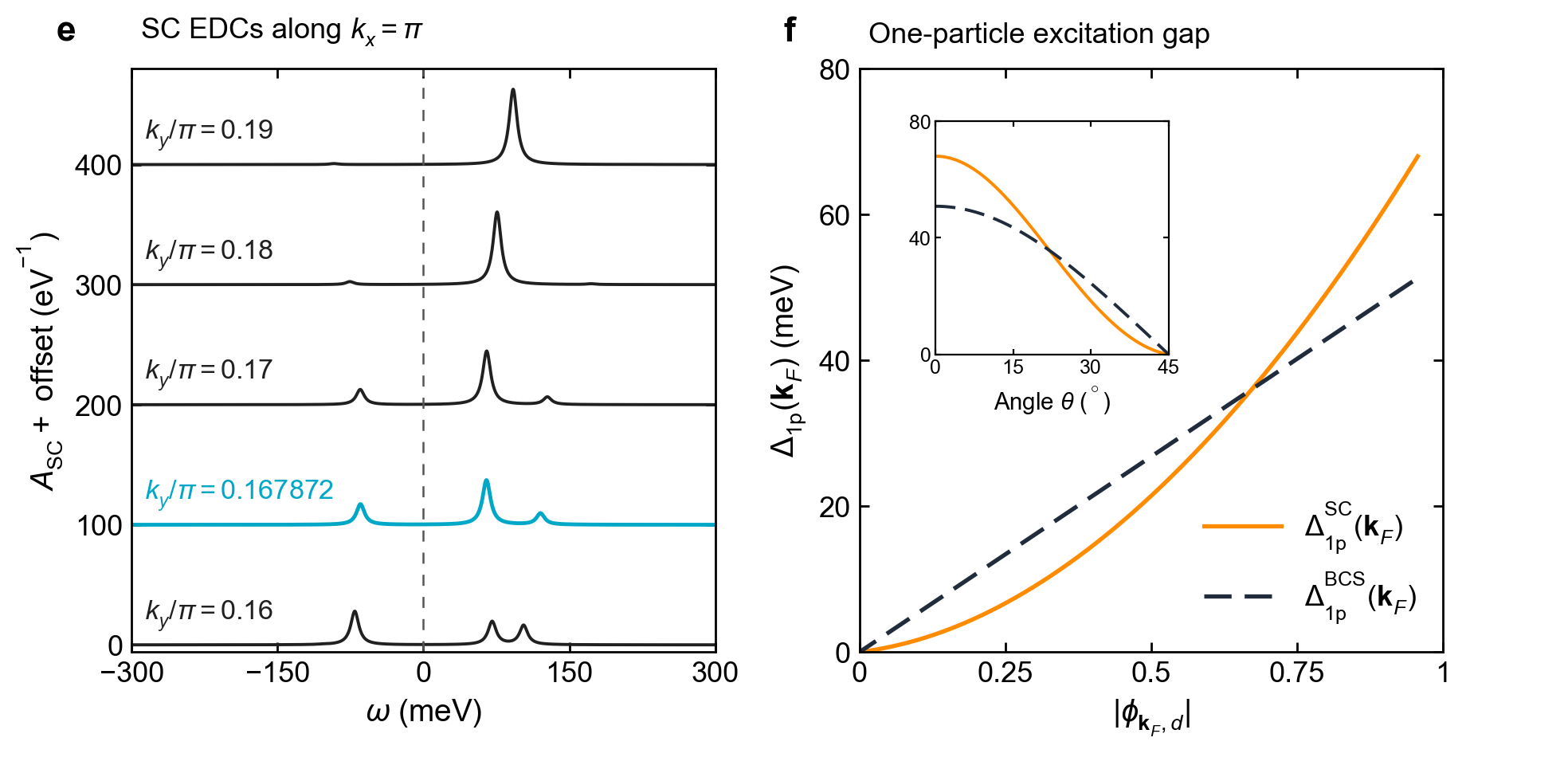}}
\expandafter\def\csname artwork@figure5-sc-spectral-redistribution\endcsname{\ArtworkPanel{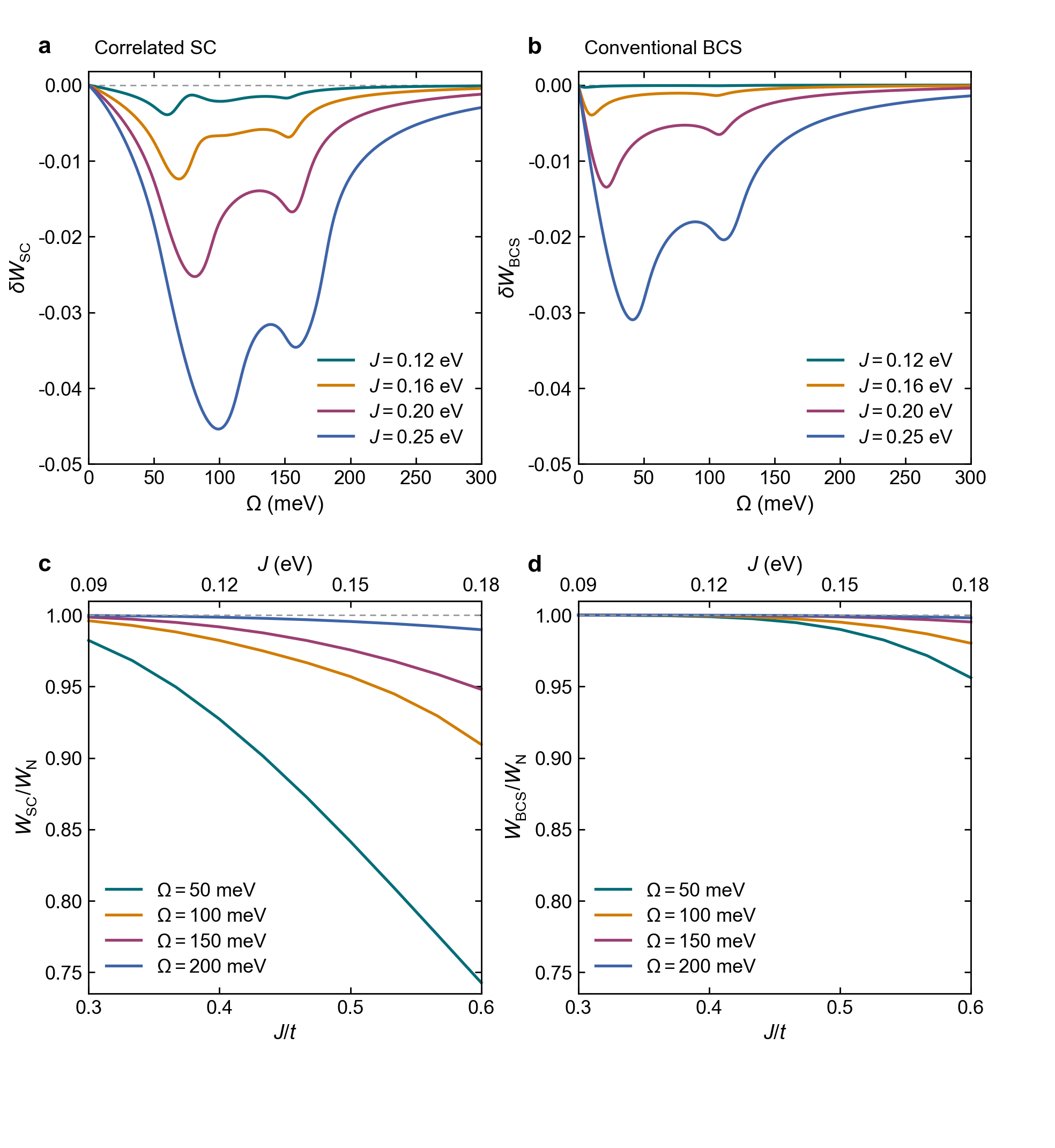}}
\expandafter\def\csname artwork@fig14-antinodal-and-maps\endcsname{\ArtworkPanel{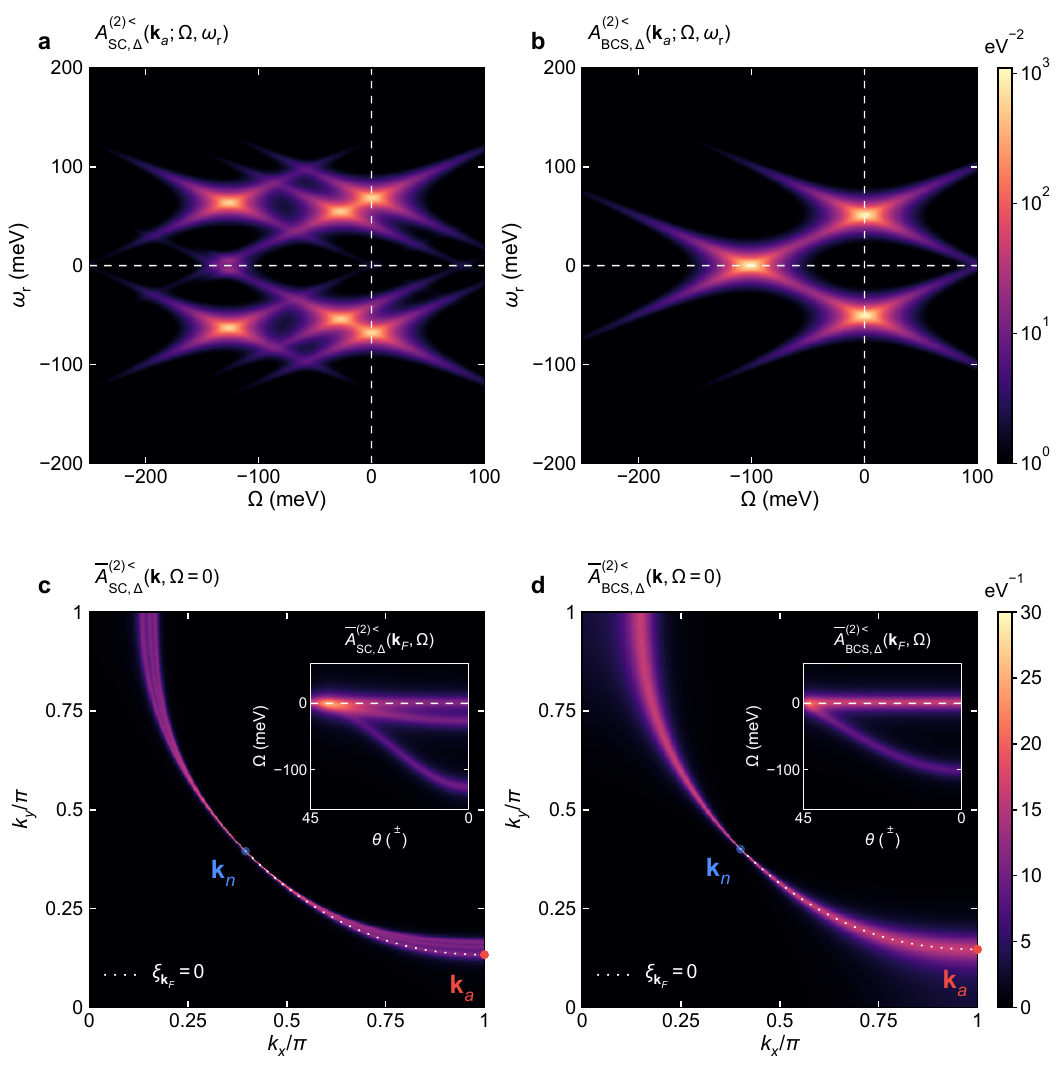}\allowbreak \ArtworkPanel{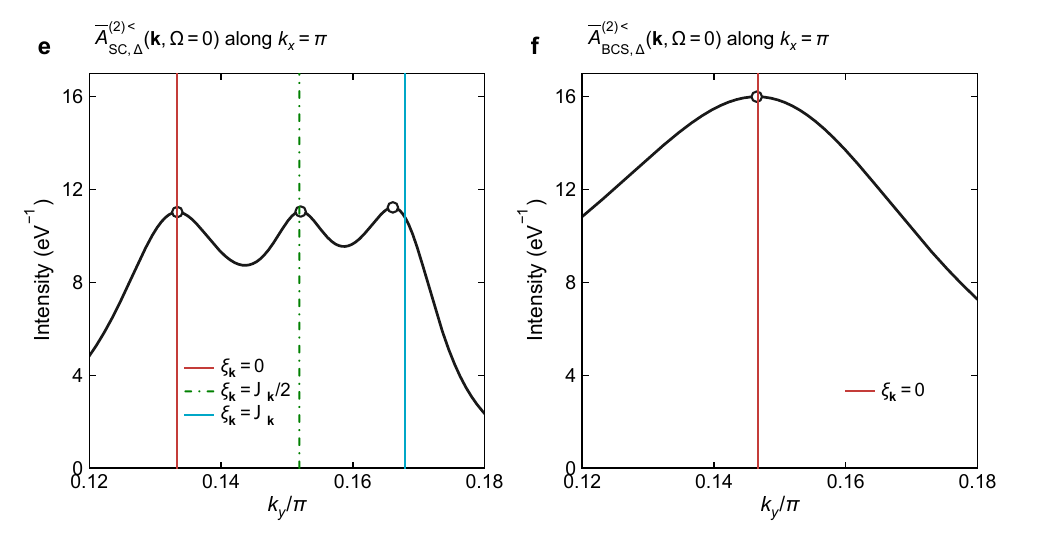}}
\expandafter\def\csname artwork@fig15-pair-cuts\endcsname{\ArtworkPanel{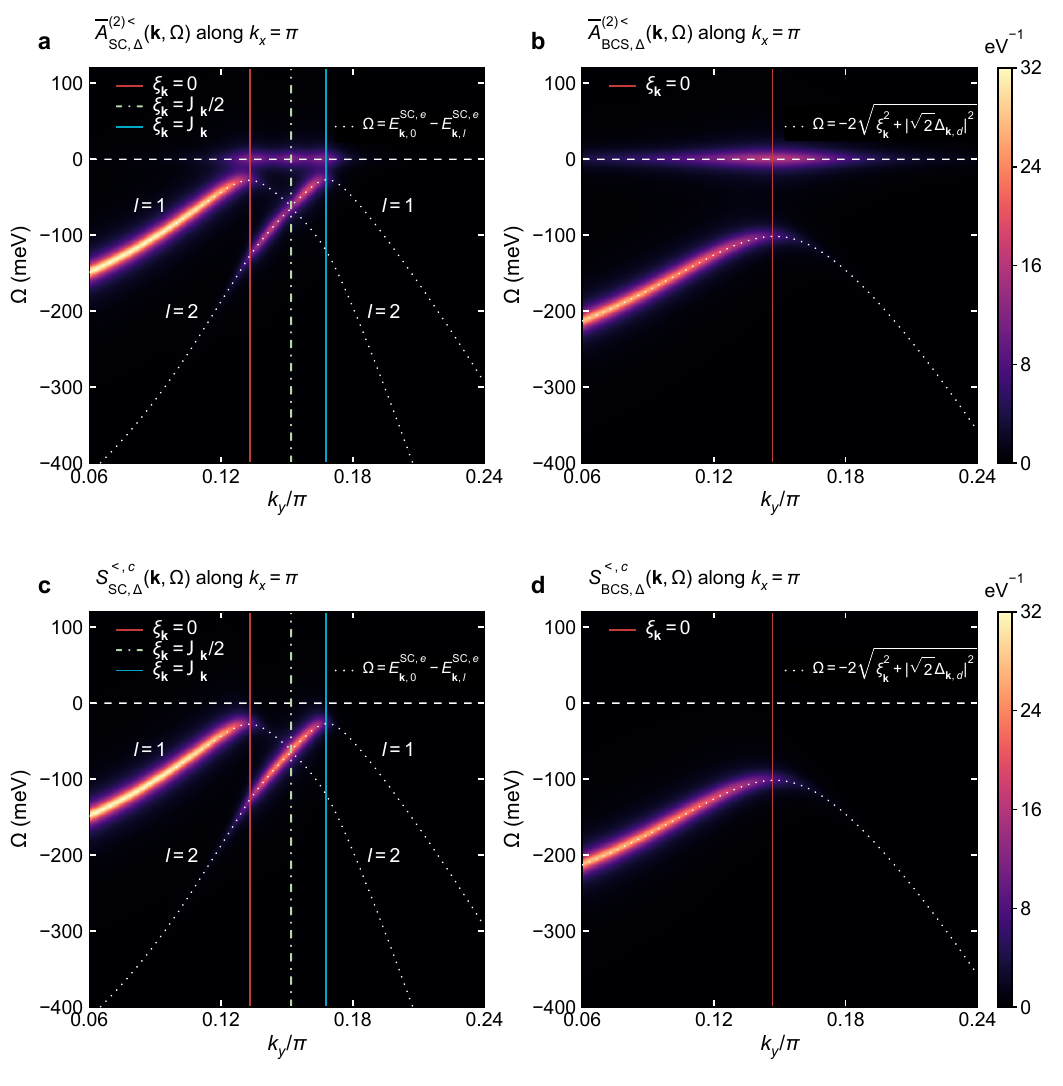}\allowbreak \ArtworkPanel{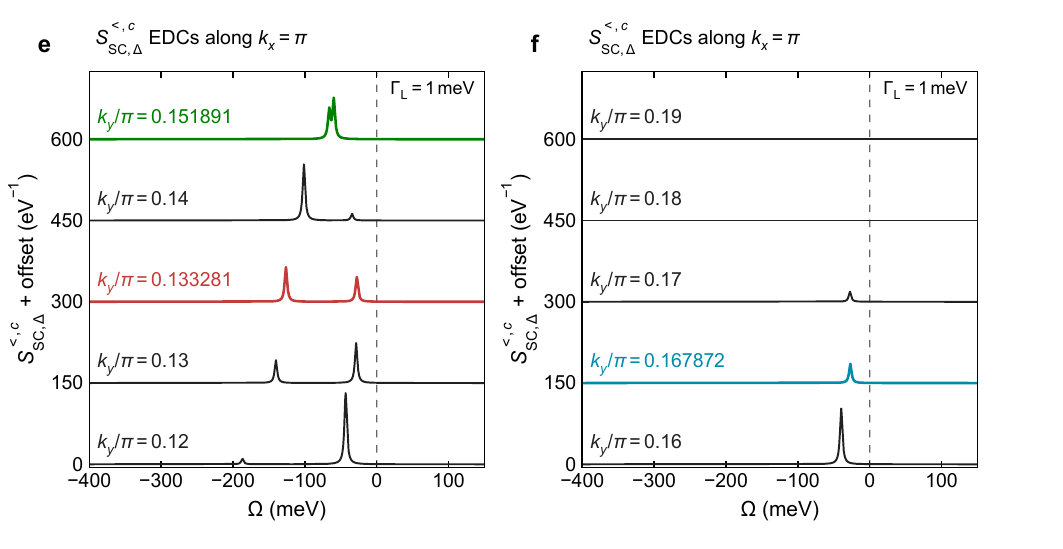}}
\expandafter\def\csname artwork@fig16-coherent-weights-and-edc\endcsname{\ArtworkPanel{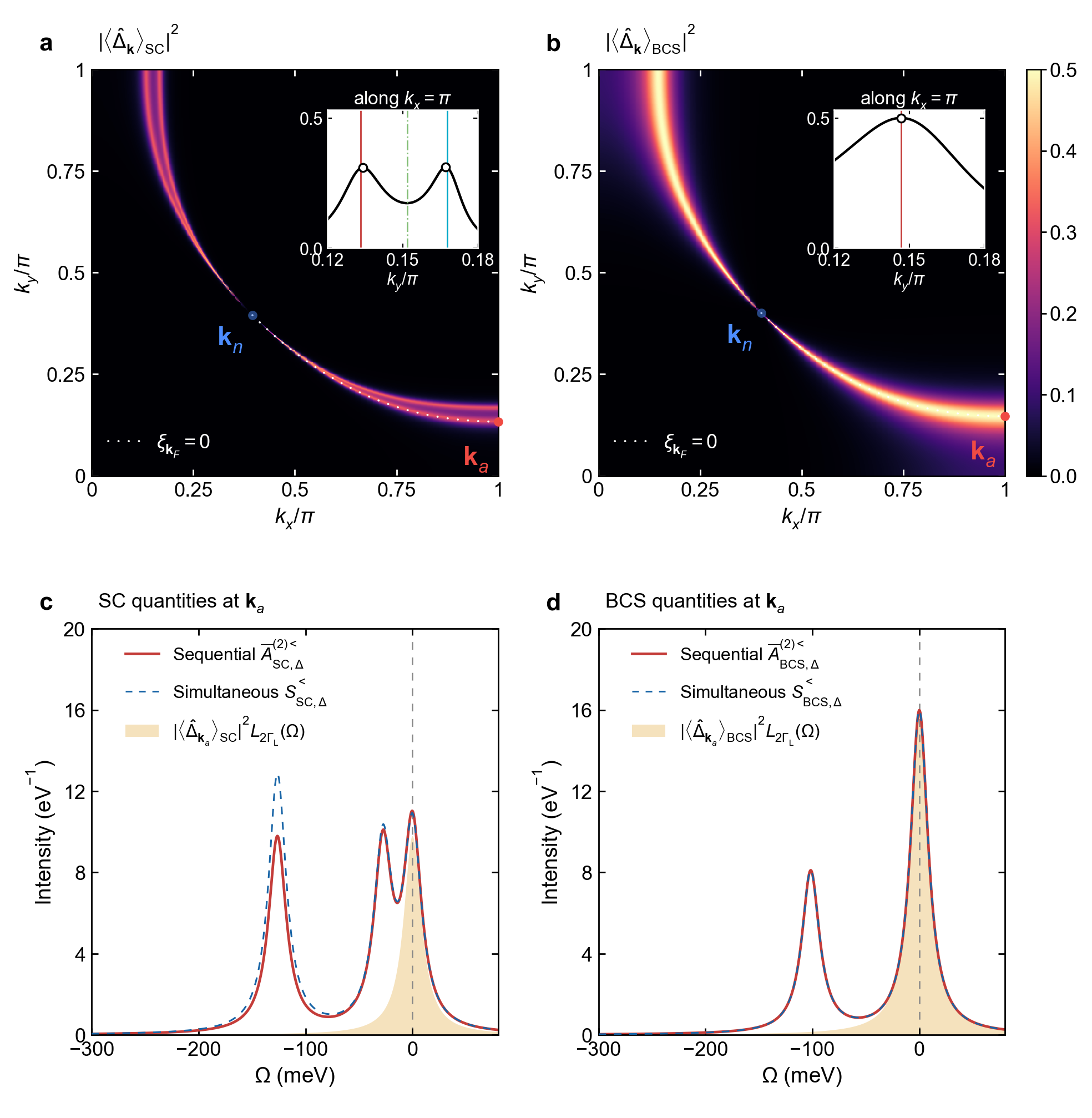}}
\expandafter\def\csname artwork@nodal-dos\endcsname{\ArtworkPanel{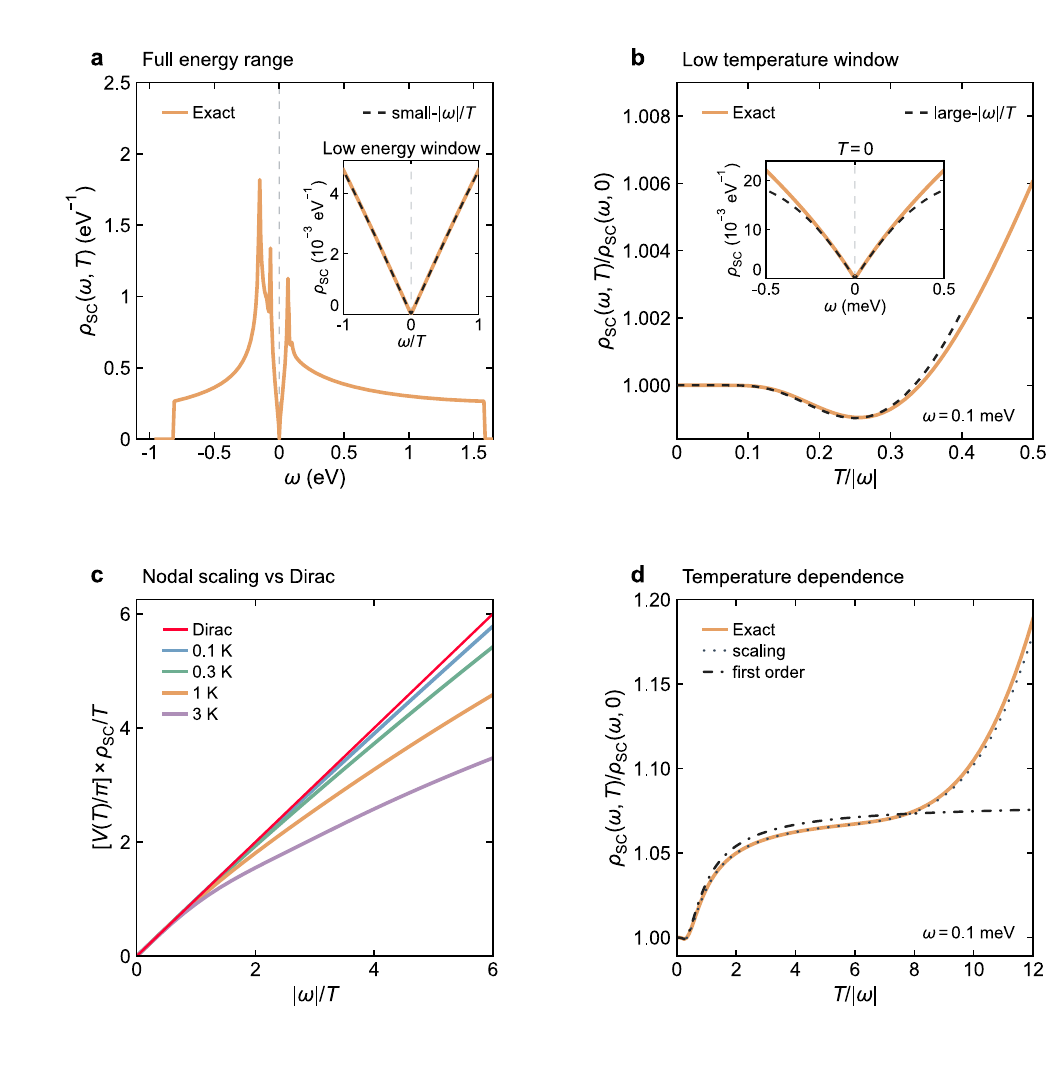}}
\expandafter\def\csname artwork@thermodynamics\endcsname{\ArtworkPanel{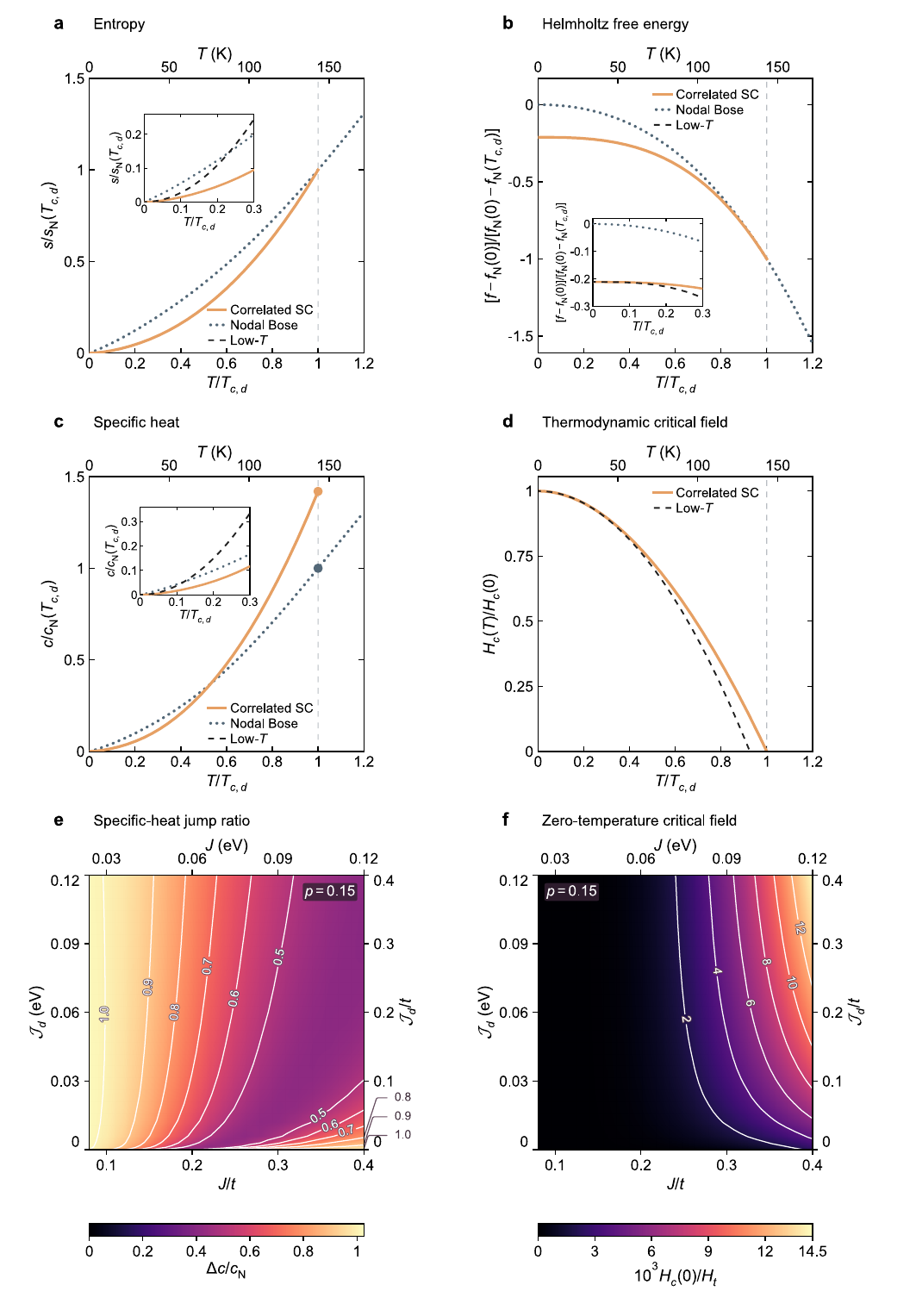}}
\expandafter\def\csname artwork@rho0-rho4-vs-doping\endcsname{\ArtworkPanel{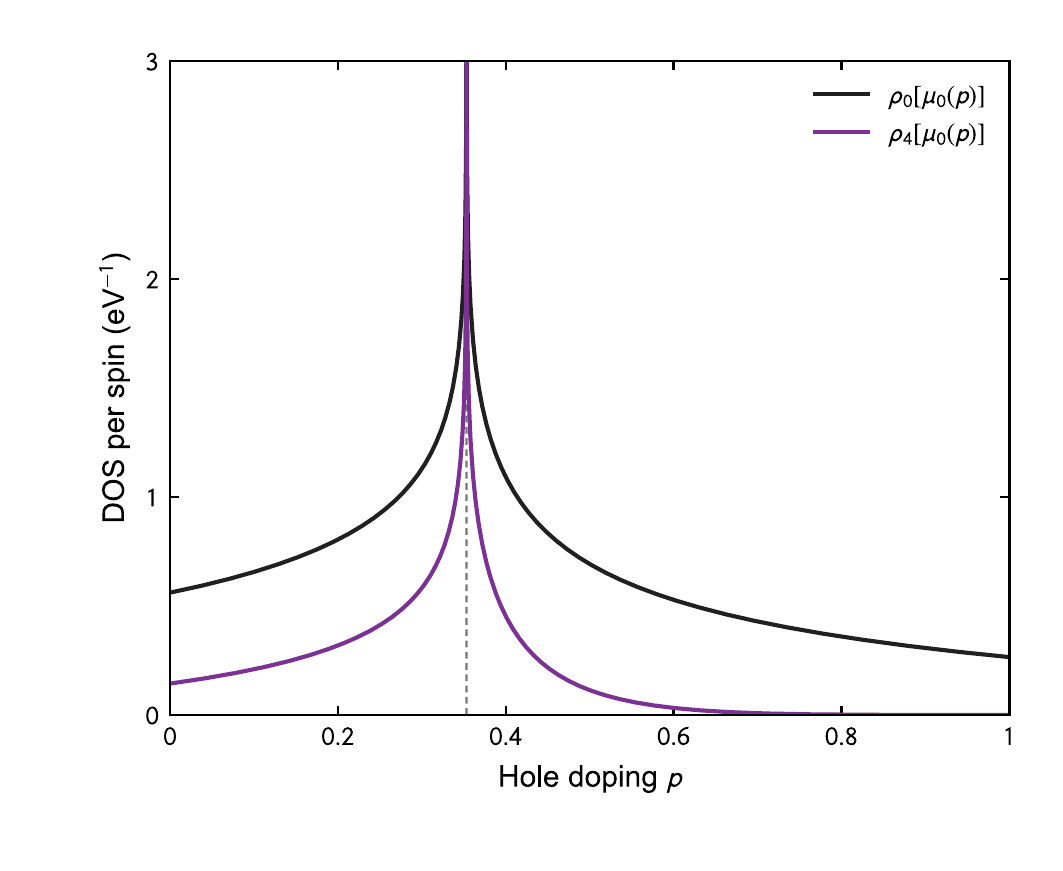}}
\expandafter\def\csname artwork@nodal-normal-coordinates\endcsname{\ArtworkPanel{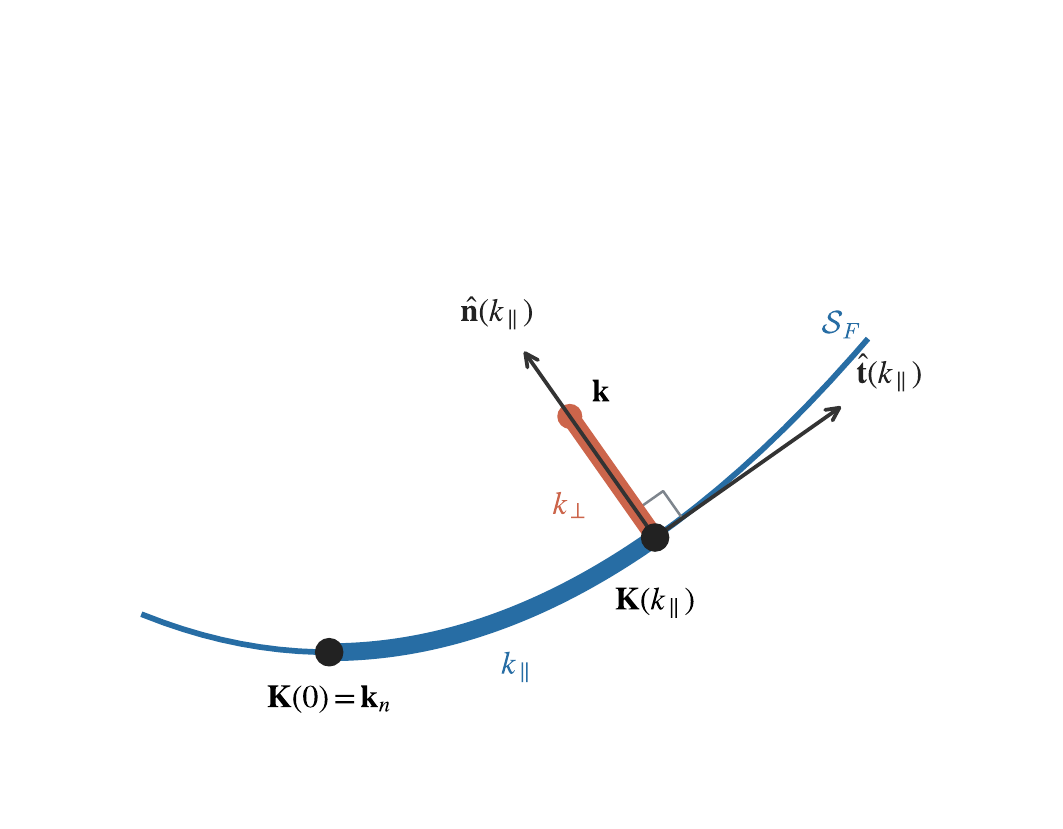}}
\expandafter\def\csname artwork@nodal-and-cooper-surface-regions\endcsname{\ArtworkPanel{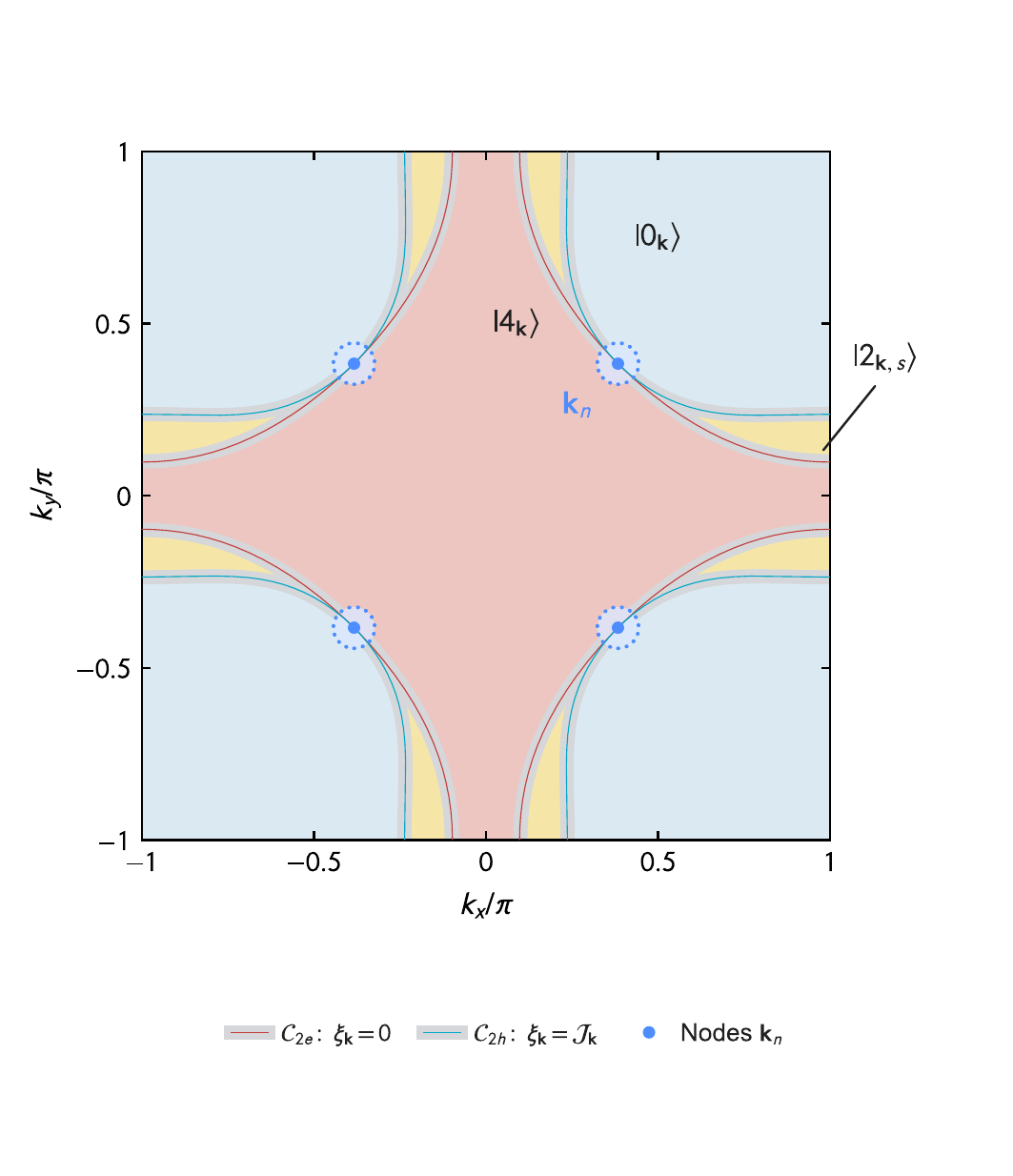}}
\expandafter\def\csname artwork@cooper-binding-np\endcsname{\ArtworkPanel{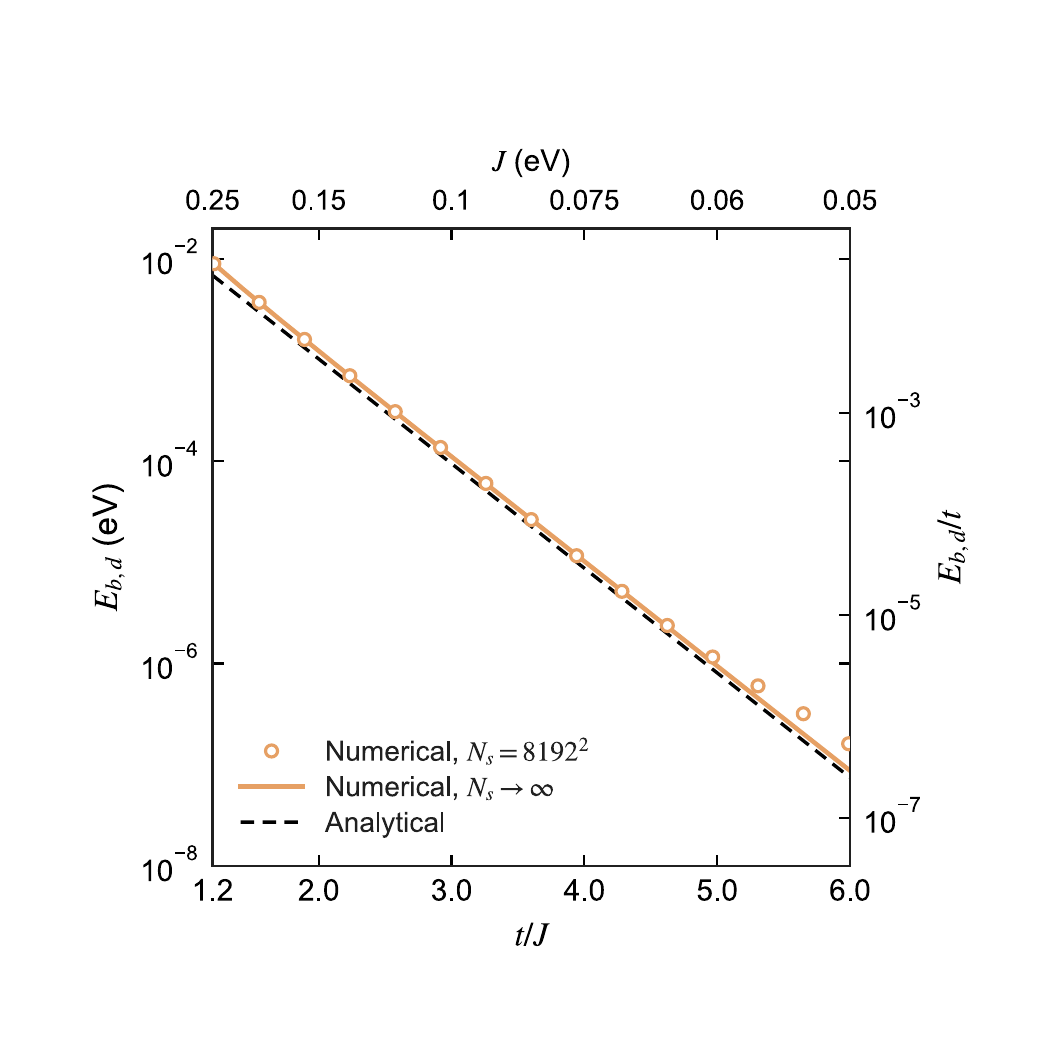}}
\expandafter\def\csname artwork@sc-instability-coupling\endcsname{\ArtworkPanel{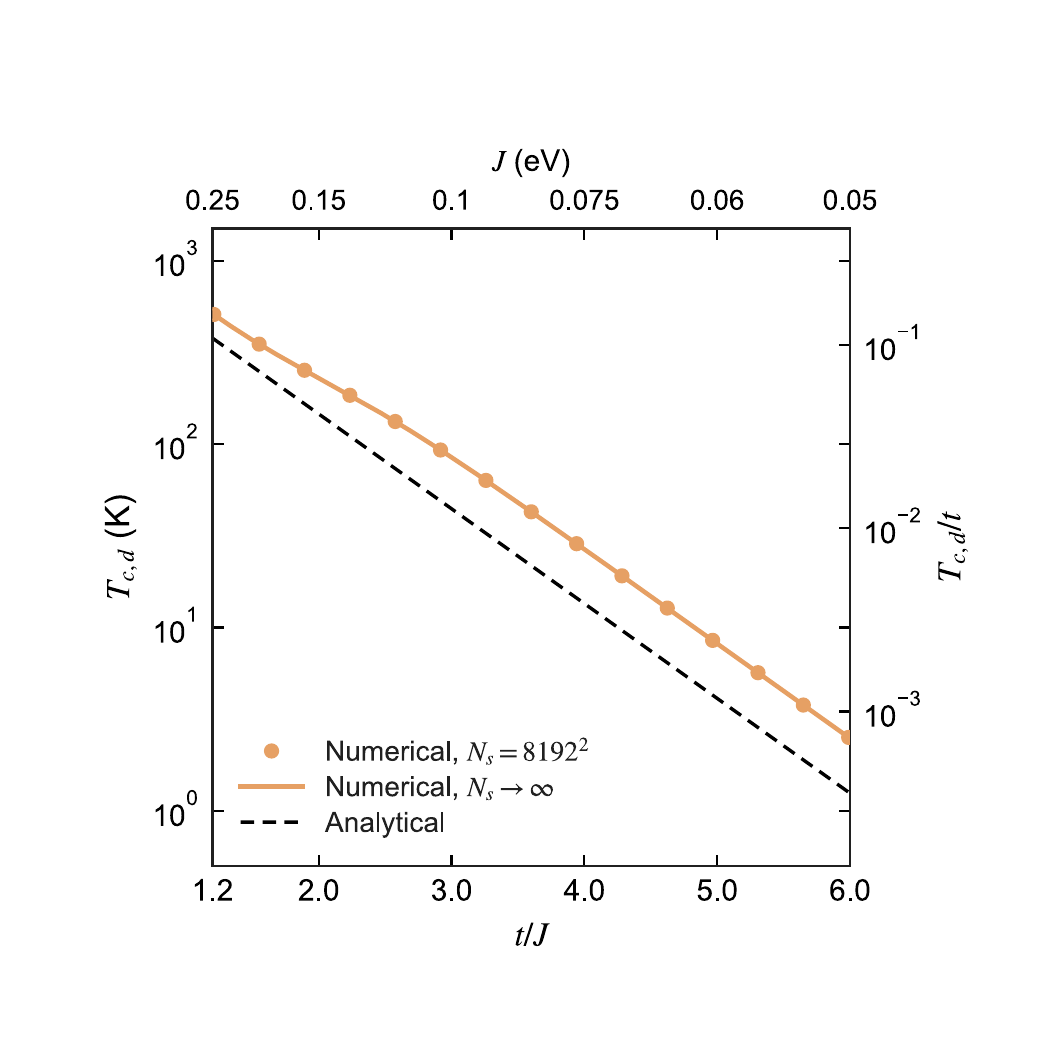}}
\expandafter\def\csname artwork@self-consistency\endcsname{\ArtworkPanel{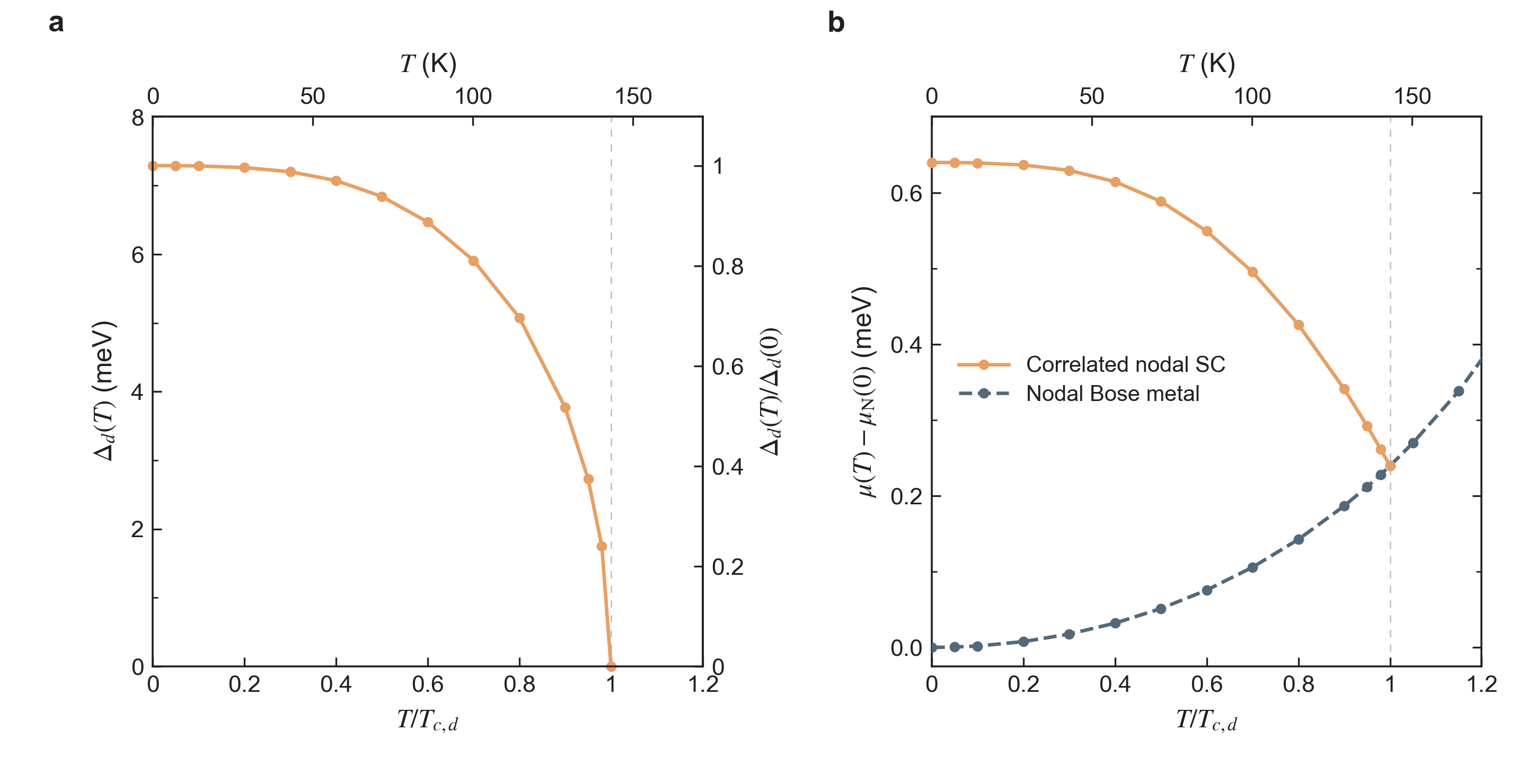}}
\expandafter\def\csname artwork@order-parameter-ratio\endcsname{\ArtworkPanel{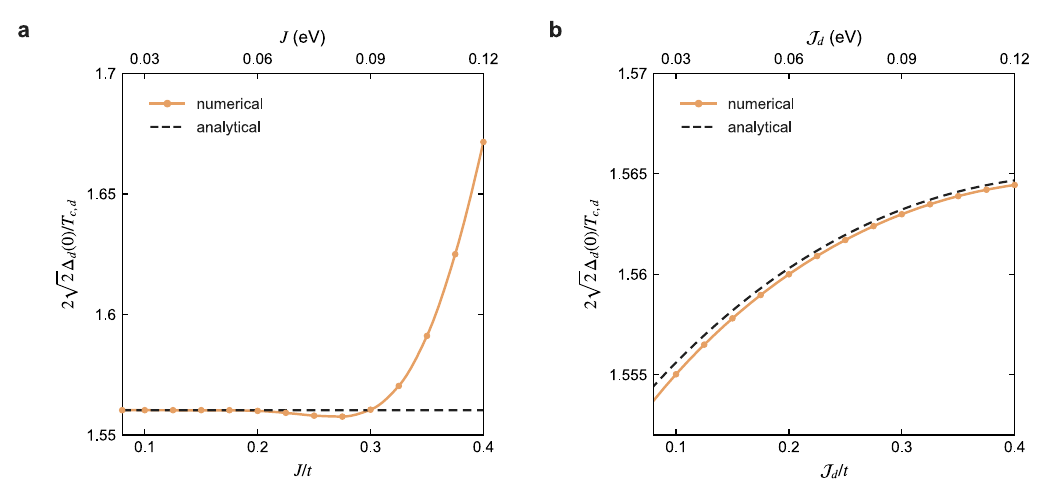}}
\expandafter\def\csname artwork@bi2201-reviewed-gap-triptych\endcsname{\ArtworkPanel{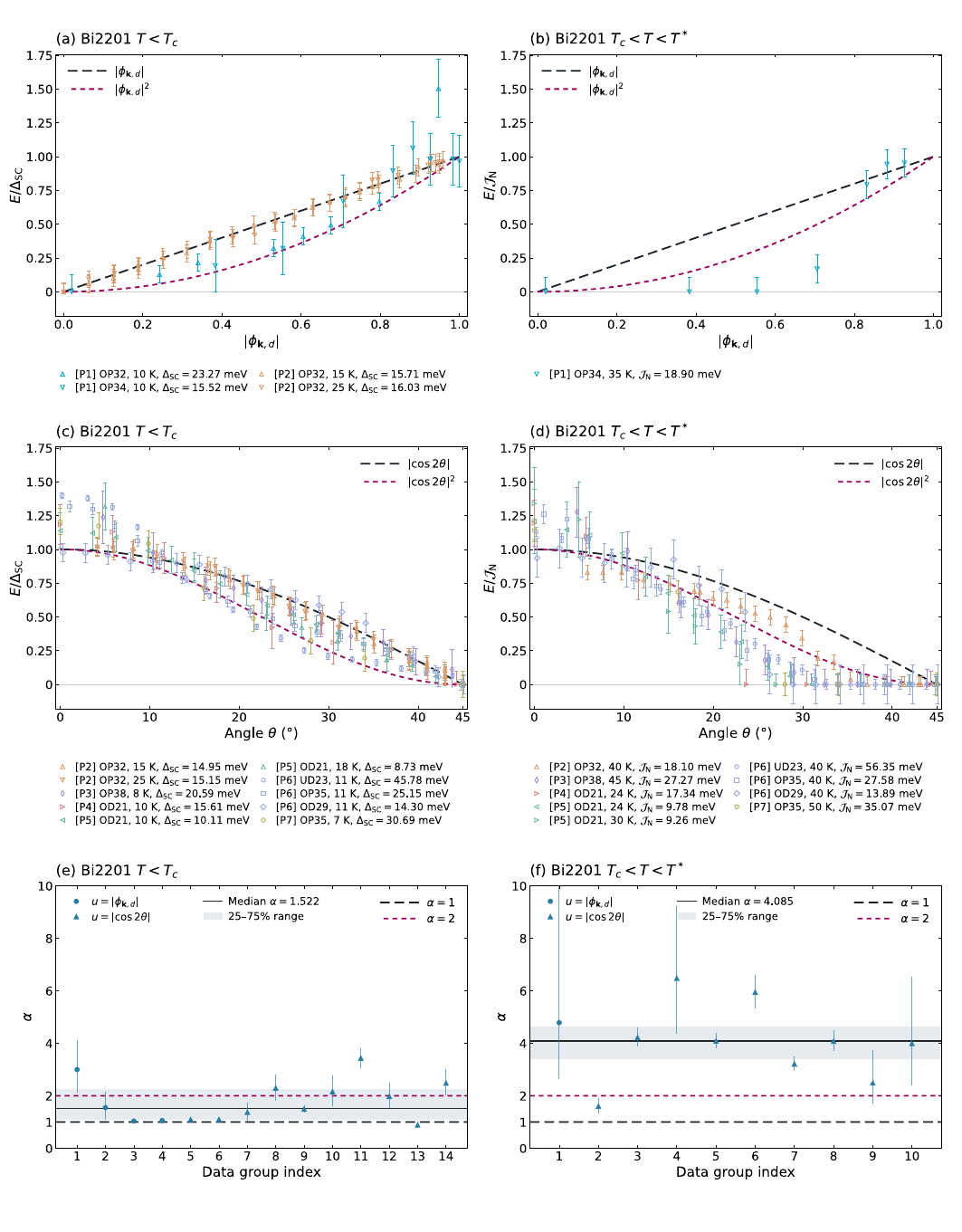}}
\expandafter\def\csname artwork@bi2212-reviewed-gap-quad\endcsname{\ArtworkPanel{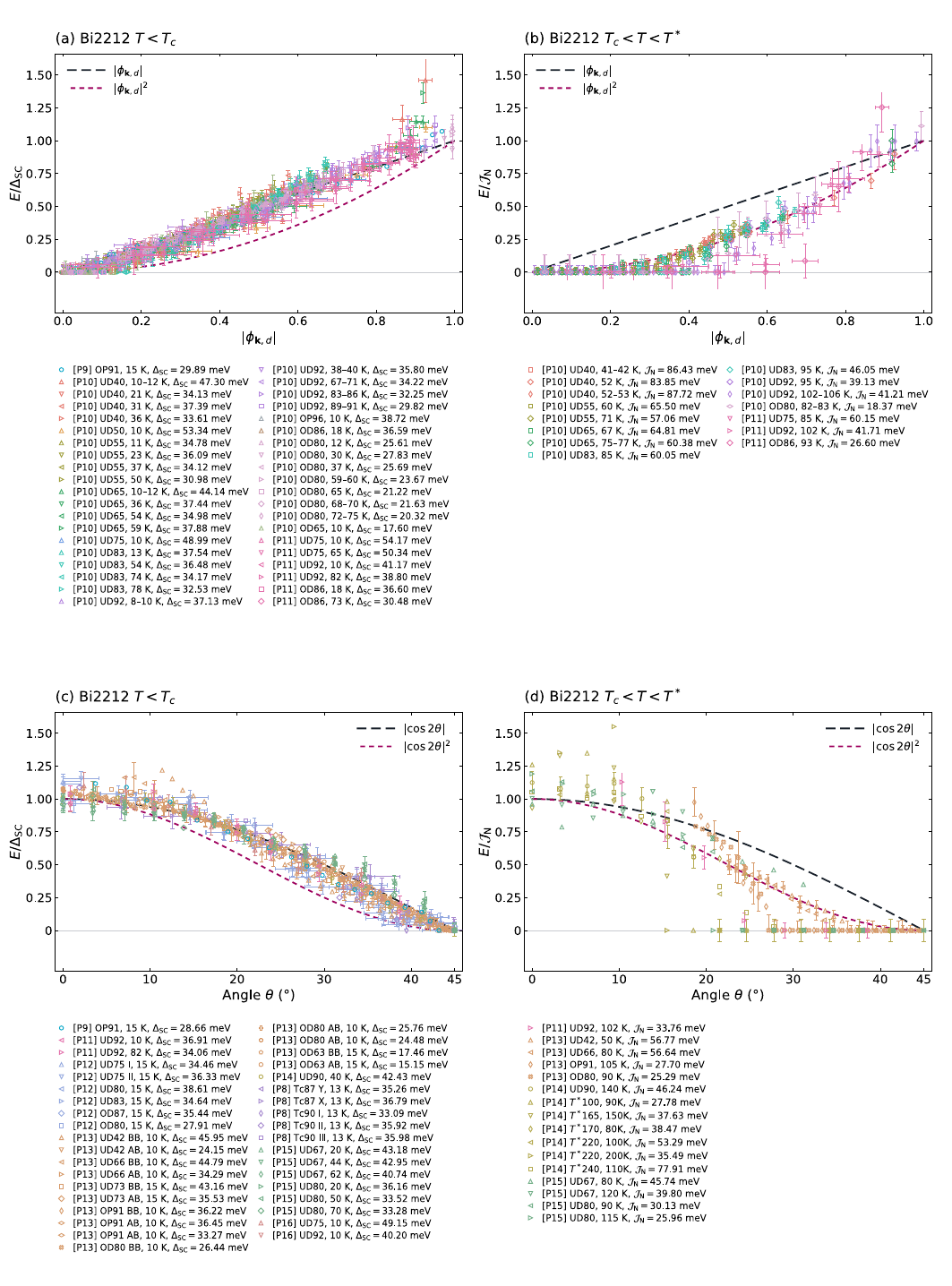}\allowbreak \ArtworkPanel{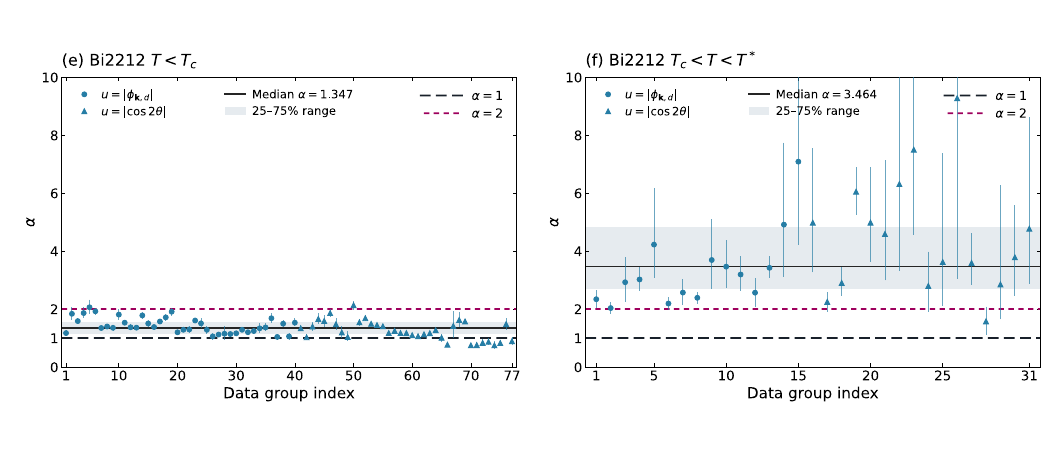}}
\expandafter\def\csname artwork@bi2223-reviewed-gap-triptych\endcsname{\ArtworkPanel{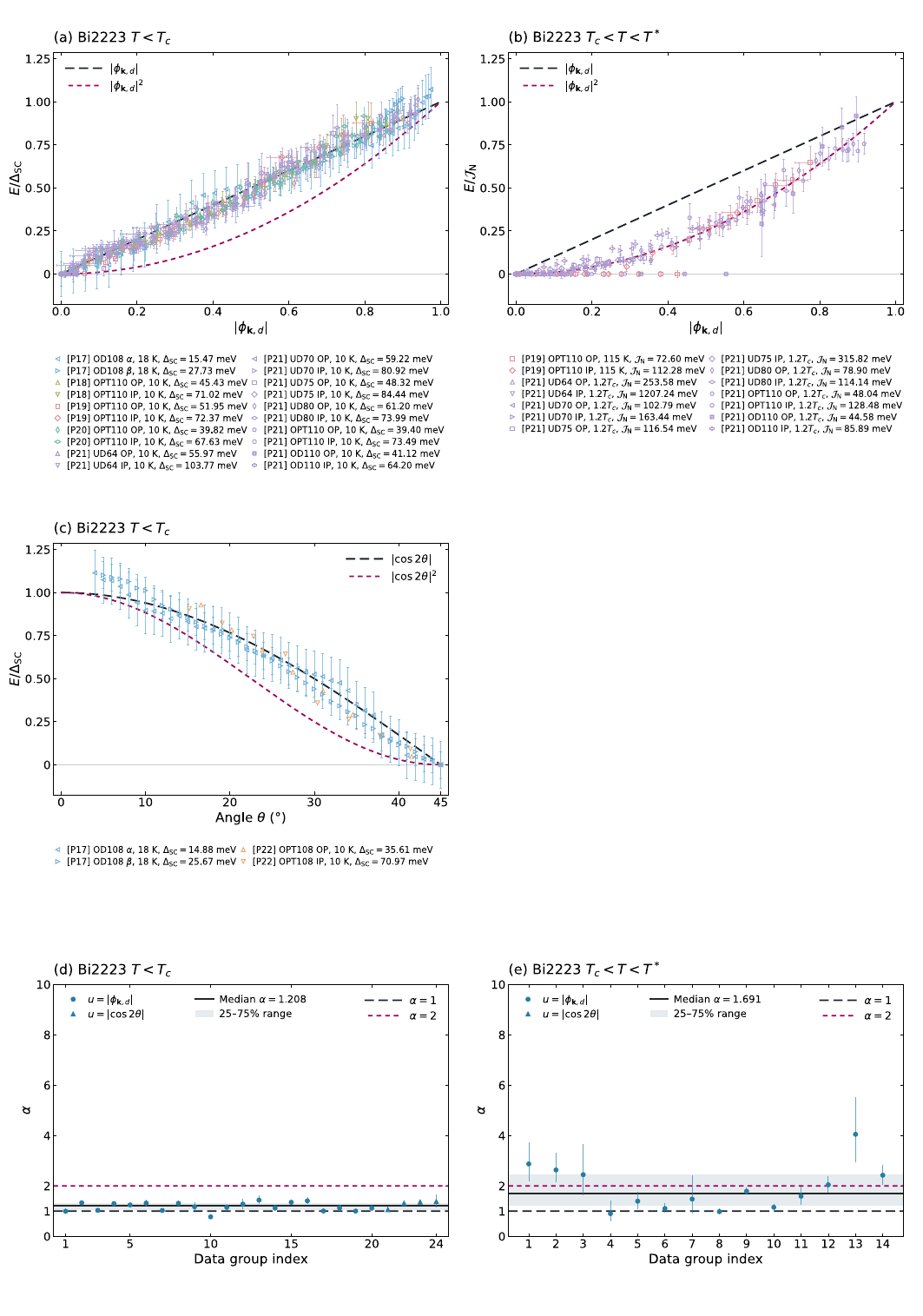}}

\begin{document}
\maketitle
 \hypertarget{ms-abstract}{}\begin{abstract}
In underdoped cuprates, the superconducting gap is largest where low-energy electron spectral weight is already depleted above the transition. How can a weak attraction drive superconductivity when the one-electron density of states vanishes? In a momentum-space \(t\)--\(J\) model, the exactly solvable normal state is a nodal Bose metal with four one-electron nodes. Along two contours called Cooper surfaces, adding or removing a spin-singlet pair of electrons with opposite momenta costs no energy. The resulting logarithmic divergence of the pair susceptibility allows an arbitrarily weak pair-transfer attraction to stabilize a correlated nodal \(d\)-wave superconductor. Spin-resolved coincidence photoemission could test this pairing mechanism by resolving two zero-energy pair-removal ridges above the transition and a third along the underlying Luttinger surface below it.
\end{abstract}

\hypertarget{ms-introduction}{}\section*{Introduction}

How does a doped Mott insulator become a high-temperature superconductor when one-electron spectral weight is already depleted near the antinodes, where the gap below the superconducting (SC) transition is largest (\hyperlink{ms-r1}{1}, \hyperlink{ms-r2}{2})? In underdoped cuprates, angle-resolved photoemission spectroscopy (ARPES) finds an antinodal pseudogap above the transition temperature \(T_c\), leaving Fermi arcs of low-energy spectral weight near the nodes (\hyperlink{ms-r3}{3}, \hyperlink{ms-r4}{4}, \hyperlink{ms-r5}{5}, \hyperlink{ms-r6}{6}, \hyperlink{ms-r7}{7}). Scanning tunneling microscopy (STM) indicates that the pseudogap evolves continuously into the SC gap on cooling through \(T_c\) (\hyperlink{ms-r8}{8}). Local shot-noise measurements deep in the SC state find a pairing scale that tracks the pseudogap energy (\hyperlink{ms-r9}{9}). Below \(T_c\), ARPES and electronic Raman scattering distinguish a near-nodal pairing scale from an antinodal gap that can grow as \(T_c\) falls (\hyperlink{ms-r10}{10}, \hyperlink{ms-r11}{11}, \hyperlink{ms-r12}{12}, \hyperlink{ms-r13}{13}). The SC coherence peak is weakest near the antinodes, where the pseudogap depletes the most spectral weight (\hyperlink{ms-r14}{14}). Specific heat measurements reveal a loss of electronic entropy on entering the pseudogap regime (\hyperlink{ms-r15}{15}). Can electrons with opposite momenta form a low-energy Cooper pair where exciting either electron costs finite energy?

Which degrees of freedom survive the loss of Landau quasiparticles in a correlated metal, and which can develop SC order? In weak-coupling Bardeen--Cooper--Schrieffer (BCS) theory, gapless electrons at the Fermi surface can form Cooper pairs under arbitrarily weak attraction (\hyperlink{ms-r16}{16}, \hyperlink{ms-r17}{17}). Almost innumerable theories of unconventional superconductivity in strongly correlated systems exist, including Anderson's resonating-valence-bond (RVB) theory (\hyperlink{ms-r18}{18}), Emery and Kivelson's phase-fluctuation theory (\hyperlink{ms-r19}{19}), and quantum-critical theories of the spin-fermion model (\hyperlink{ms-r20}{20}), etc. The enigmatic pseudogap and strange-metal regimes of the cuprates have motivated a myriad of theories, including Varma's marginal Fermi-liquid theory (\hyperlink{ms-r21}{21}), the Yang--Rice--Zhang (YRZ) theory of the pseudogap (\hyperlink{ms-r22}{22}), and studies of Sachdev--Ye--Kitaev (SYK) models (\hyperlink{ms-r23}{23}), etc. Can a non-Fermi liquid with vanishing one-electron density of states (DOS) have a divergent pair susceptibility for arbitrarily weak attraction?

The present study builds on two lines of work. The first concerns Bose metals, including states in which uncondensed electron pairs have gapless excitations along a Bose surface (\hyperlink{ms-r24}{24}, \hyperlink{ms-r25}{25}, \hyperlink{ms-r26}{26}, \hyperlink{ms-r27}{27}). In nanopatterned two-dimensional cuprate films, a low-temperature resistance plateau coexists with magnetoconductance oscillations of period \(h/(2e)\) in the flux per array cell, indicating mobile charge-\(2e\) carriers in this metal (\hyperlink{ms-r28}{28}). The second concerns exactly solvable models with spatially nonlocal interactions. The Hatsugai--Kohmoto (HK) model (\hyperlink{ms-r29}{29}, \hyperlink{ms-r30}{30}) exhibits Mottness and breaks a discrete symmetry (\hyperlink{ms-r31}{31}), and SYK models realize non-Fermi liquids without quasiparticles (\hyperlink{ms-r23}{23}).

Here, an exactly solvable momentum-space \(t\)--\(J\) (MStJ) model incorporating a relative-momentum-local (RML) attraction is proposed. This interaction breaks the local-in-momentum-space \(\mathbb Z_{2}^{\mathrm{ph}}\) particle--hole symmetry on the Fermi surface, producing a nodal Bose metal with two gapless Cooper surfaces. An arbitrarily weak pair-transfer attraction drives a pairing instability toward a correlated nodal \(d\)-wave superconductor. Two-photon coincidence ARPES (cARPES) (\hyperlink{ms-r32}{32}, \hyperlink{ms-r33}{33}, \hyperlink{ms-r34}{34}) could detect the Cooper surfaces and the Luttinger surface.
 
\hypertarget{ms-model}{}\section*{Momentum-space \emph{t}--\emph{J} model}

Within the doped Mott insulator framework, the single-band \(t\)--\(J\) model provides a minimal description of charge motion in the copper-oxide planes of cuprates (\hyperlink{ms-r2}{2}, \hyperlink{ms-r35}{35}). Antiferromagnetic (AFM) superexchange is a key ingredient of the pairing mechanism in cuprates. The nearest-neighbor exchange satisfies the exact identity \(J(\hat{\mathbf S}_i\cdot\hat{\mathbf S}_j-\hat n_i\hat n_j/4)=-J\hat\Delta_{ij}^{\dagger}\hat\Delta_{ij}\), where \(\hat\Delta_{ij}=(\hat c_{i\uparrow}\hat c_{j\downarrow}-\hat c_{i\downarrow}\hat c_{j\uparrow})/\sqrt{2}\) annihilates a singlet on the bond between sites \(i\) and \(j\). Forming a bond singlet thus lowers the local exchange energy. On the square lattice, this interaction scatters singlets in momentum space between distinct relative momenta with conserved total momentum in the extended \(s\)- and \(d\)-wave channels (\hyperlink{ms-r36}{36}).

The Fourier transform of \(\hat\Delta_{ij}\) gives \(\hat\Delta_{\mathbf k}=(\hat c_{\mathbf k\uparrow}\hat c_{-\mathbf k\downarrow}-\hat c_{\mathbf k\downarrow}\hat c_{-\mathbf k\uparrow})/\sqrt{2}\), which annihilates a spin-singlet pair with zero total momentum and relative momentum \(\mathbf k\). In the \(d\)-wave channel, the scattering term \(\hat\Delta_{\mathbf k}^{\dagger}\hat\Delta_{\mathbf k'}\) has coefficient \(-(2J/N_s)\phi_{\mathbf k,d}\phi_{\mathbf k',d}\), where \(N_s\) counts lattice sites. Diagonal terms favor singlets within each momentum block \(\{\mathbf k,-\mathbf k\}\). Off-diagonal terms transfer pairs between blocks and couple their phases. The normal-state Hamiltonian \(\hat H_{\mathrm N}\) incorporates the spin-singlet pair correlations within each block.

\hypertarget{ms-eq1}{}%
\FittedEquation{1}{%
\hat H_{\mathrm N}=\sum_{\mathbf k}\left(\xi_{\mathbf k}\hat n_{\mathbf k}-\mathcal J_{\mathbf k}\hat\Delta_{\mathbf k}^{\dagger}\hat\Delta_{\mathbf k}\right).
}
 
Here \(\hat n_{\mathbf k}=\sum_{\sigma}\hat c_{\mathbf k\sigma}^{\dagger}\hat c_{\mathbf k\sigma}\) counts electrons at momentum \(\mathbf k\), \(\xi_{\mathbf k}=\varepsilon_{\mathbf k}-\mu\) is the band dispersion, and \(\mathcal J_{\mathbf k}=\mathcal J_d\phi_{\mathbf k,d}^2\) is the RML attraction strength with \(d\)-wave form factor \(\phi_{\mathbf k,d}=(\cos k_x-\cos k_y)/2\). The hopping parameters are \(t=0.3\) eV, \(t'=-0.2t\), and \(t''=0.1t\). The electron density \(n_e=1-p\) fixes the chemical potential \(\mu\) at hole concentration \(p\). \hypertarget{ms-rml-exchange-identity}{}The diagonal term \(-\hat\Delta_{\mathbf k}^{\dagger}\hat\Delta_{\mathbf k}\) is exactly equal to \(\hat{\mathbf S}_{\mathbf k}\cdot\hat{\mathbf S}_{-\mathbf k}-\hat n_{\mathbf k}\hat n_{-\mathbf k}/4-\delta_{\mathbf k,-\mathbf k}\hat n_{\mathbf k}/2\), where \(\hat{\mathbf S}_{\mathbf k}=\tfrac12\sum_{\sigma,\sigma'}\hat c_{\mathbf k\sigma}^{\dagger}\boldsymbol{\tau}_{\sigma\sigma'}\hat c_{\mathbf k\sigma'}\) and \(\boldsymbol{\tau}\) denotes the vector of Pauli matrices. This superexchange form motivates the name momentum-space \(t\)--\(J\) model.

\hypertarget{ms-rml-real-space}{}In real space, the RML interaction transfers singlets between bonds through \(-N_s^{-1}\mathcal J_{ij,i'j'}\hat\Delta_{ij}^{\dagger}\hat\Delta_{i'j'}\), where \(\mathcal J_{ij,i'j'}\) is the Fourier transform of \(\mathcal J_{\mathbf k}\). The squared \(d\)-wave form factor restricts only the difference \((\mathbf r_i-\mathbf r_j)-(\mathbf r_{i'}-\mathbf r_{j'})\) between the two bond vectors. A singlet of any size \(|\mathbf r_i-\mathbf r_j|\) can therefore move between bond centers separated by any distance \(|\mathbf r_i+\mathbf r_j-\mathbf r_{i'}-\mathbf r_{j'}|/2\), giving an all-to-all coupling. Infinite-range couplings occur in the exactly solvable Baskaran and HK models (\hyperlink{ms-r37}{37}, \hyperlink{ms-r29}{29}) and in SYK models (\hyperlink{ms-r23}{23}).

\hypertarget{ms-normal-block-solution}{}The MStJ model is exactly solvable because \(\hat H_{\mathrm N}\) is a sum of commuting block Hamiltonians \(\hat h_{\mathbf k}^{\mathrm N}=\xi_{\mathbf k}\hat N_{\mathbf k}-2\mathcal J_{\mathbf k}\hat\Delta_{\mathbf k}^{\dagger}\hat\Delta_{\mathbf k}\), where \(\hat N_{\mathbf k}=\hat n_{\mathbf k}+\hat n_{-\mathbf k}\) counts electrons in each block. Except at time-reversal-invariant momenta, \(\hat h_{\mathbf k}^{\mathrm N}\) acts on a 16-dimensional Fock space. Away from level crossings, the block ground state (GS) is

\hypertarget{ms-eq2}{}%
\FittedEquation{2}{%
|\mathrm{GS}_{\mathbf k}\rangle=
      \begin{cases}
      |0_{\mathbf k}\rangle, & \xi_{\mathbf k}>\mathcal J_{\mathbf k},\\
      |2_{\mathbf k,s}\rangle=\hat\Delta_{\mathbf k}^{\dagger}|0_{\mathbf k}\rangle, & 0<\xi_{\mathbf k}<\mathcal J_{\mathbf k},\\
      |4_{\mathbf k}\rangle=\hat\Delta_{\mathbf k}^{\dagger}|2_{\mathbf k,s}\rangle, & \xi_{\mathbf k}<0.
      \end{cases}
}
 
The normal-state many-body GS is \(|\mathrm{GS}_{\mathrm N}\rangle=\bigl[\prod_{\{\mathbf k,-\mathbf k\}}^{\xi_{\mathbf k}<0}(\hat\Delta_{\mathbf k}^{\dagger})^2\bigr]\bigl[\prod_{\{\mathbf k',-\mathbf k'\}}^{0<\xi_{\mathbf k'}<\mathcal J_{\mathbf k'}}\hat\Delta_{\mathbf k'}^{\dagger}\bigr]|0\rangle\), where \(|0\rangle\) is the vacuum and each product counts the indicated blocks once.

\par\medskip
\hypertarget{ms-fig1}{}
\PrintAsset{fig-1-model}
{\small\noindent \textbf{Fig. 1. Block ground states, the Luttinger surface, and Cooper surfaces of the MStJ model.} Light blue, yellow, and red denote regions with block GSs \(|0_{\mathbf k}\rangle\), \(|2_{\mathbf k,s}\rangle\), and \(|4_{\mathbf k}\rangle\), respectively. Blue and red points mark the nodes \(\mathbf k_n\) and antinodes \(\mathbf k_a\), respectively. The green dash-dotted contour traces the Luttinger surface \(\xi_{\mathbf k_L}=\mathcal J_{\mathbf k_L}/2>0\), with punctures at the four nodes. The red and cyan contours trace the Cooper surfaces \(\mathcal C_{2e}:\xi_{\mathbf k}=0\) and \(\mathcal C_{2h}:\xi_{\mathbf k}=\mathcal J_{\mathbf k}\), respectively. The parameters are \(p=0.15\) and \(\mathcal J_d=0.3~\mathrm{eV}\).\par}
\medskip
 
\hypertarget{ms-nodal-bose-metal}{}\section*{Cooper surfaces in a nodal Bose metal}

\hypertarget{ms-normal-pair-order}{}Pair correlations within momentum blocks do not by themselves establish phase coherence between distinct blocks. Since \([\hat H_{\mathrm N},\hat n_{\mathbf k}]=0\), \(\hat n_{\mathbf k}\) is conserved at each momentum, implying \(\langle\hat\Delta_{\mathbf k}\rangle_{\mathrm N}=0\) for every \(\mathbf k\) and \(C_{\mathbf k\mathbf k'}=\langle\hat\Delta_{\mathbf k}^{\dagger}\hat\Delta_{\mathbf k'}\rangle_{\mathrm N}=0\) for \(\mathbf k'\ne\pm\mathbf k\). The largest eigenvalue of the pair correlation matrix \(C_{\mathbf k\mathbf k'}\) therefore remains bounded as \(N_s\to\infty\), precluding SC off-diagonal long-range order (ODLRO) in this correlated normal state.

\hypertarget{ms-normal-electron-propagator}{}Electron addition and removal energies follow from the poles of the retarded one-particle Green's function \(G^R_{\sigma}(\mathbf k,\omega)\equiv-i\int_{-\infty}^{+\infty}dt\,e^{i\omega t}\Theta(t)\langle\{\hat c_{\mathbf k\sigma}(t),\hat c_{\mathbf k\sigma}^{\dagger}(0)\}\rangle\), where \(\Theta(t)\) is the step function. In the correlated normal state at zero temperature, the exact result is 
\hypertarget{ms-eq3}{}%
\FittedEquation{3}{%
G^R_{\mathrm N,\sigma}(\mathbf k,\omega)=\begin{cases}
        \displaystyle\frac{1}{\omega+i0^+-\xi_{\mathbf k}},&\xi_{\mathbf k}>\mathcal J_{\mathbf k},\\[6pt]
        \displaystyle\frac12\left(\frac{1}{\omega+i0^+-\xi_{\mathbf k}+2\mathcal J_{\mathbf k}}+\frac{1}{\omega+i0^+-\xi_{\mathbf k}-\mathcal J_{\mathbf k}}\right),&0<\xi_{\mathbf k}<\mathcal J_{\mathbf k},\\[6pt]
        \displaystyle\frac{1}{\omega+i0^+-\xi_{\mathbf k}+\mathcal J_{\mathbf k}},&\xi_{\mathbf k}<0.
        \end{cases}
}
 The block GS \(|2_{\mathbf k,s}\rangle\) allows both electron addition and removal, each with pole weight \(1/2\). The minimum energy to add or remove an electron at \(\mathbf k\) is \(\Delta_{\mathrm{1p}}^{\mathrm N}(\mathbf k)=\mathcal J_{\mathbf k}+\min\{|\xi_{\mathbf k}|,|\xi_{\mathbf k}-\mathcal J_{\mathbf k}|\}\). On the Fermi surface (\(\xi_{\mathbf k_F}=0\)), the RML interaction opens a gap of \(\mathcal J_{\mathbf k_F}\). Zero-energy poles survive only at the four nodes (\(\xi_{\mathbf k_n}=\mathcal J_{\mathbf k_n}=0\)), giving a correlated nodal metal.

\hypertarget{ms-normal-luttinger-count}{}The two-pole structure (\hyperlink{ms-r30}{30}) in \hyperlink{ms-eq3}{Eq. 3} gives \(G^R_{\mathrm N,\sigma}=1/(\omega+i0^+-\xi_{\mathbf k}-\Sigma^R_{\mathrm N,\sigma})\) for \(0<\xi_{\mathbf k}<\mathcal J_{\mathbf k}\), with self-energy \(\Sigma^R_{\mathrm N,\sigma}=-\mathcal J_{\mathbf k}/2+(3\mathcal J_{\mathbf k}/2)^2/(\omega+i0^+-\xi_{\mathbf k}+\mathcal J_{\mathbf k}/2)\), of the same form as the self-energy in the coherent YRZ Green's function (\hyperlink{ms-r22}{22}). At zero frequency, addition and removal contributions cancel at momenta \(\mathbf k_L\) satisfying \(\xi_{\mathbf k_L}=\mathcal J_{\mathbf k_L}/2>0\), where \(\Sigma^R_{\mathrm N,\sigma}(\mathbf k_L,0)\) diverges and \(G^R_{\mathrm N,\sigma}(\mathbf k_L,0)=0\). These zeros define the Luttinger surface (\hyperlink{ms-r38}{38}). A divergent self-energy can invalidate the derivation of Luttinger's theorem from the Luttinger--Ward functional (\hyperlink{ms-r39}{39}). In the MStJ model, the violation is quantified by the Luttinger integral 
\hypertarget{ms-eq4}{}%
\FittedEquation{4}{%
I_L=\int_{\mathrm{BZ}}\frac{d^2k}{(2\pi)^2}
      \Bigl\{\bigl[\Theta(\mathcal J_{\mathbf k}/2-\xi_{\mathbf k})-\Theta(-\xi_{\mathbf k})\bigr]
      -\bigl[\Theta(\mathcal J_{\mathbf k}-\xi_{\mathbf k})-\Theta(\mathcal J_{\mathbf k}/2-\xi_{\mathbf k})\bigr]\Bigr\}.
}
 Geometrically, \(I_L\) is the signed area difference between regions \(0<\xi_{\mathbf k}<\mathcal J_{\mathbf k}/2\) and \(\mathcal J_{\mathbf k}/2<\xi_{\mathbf k}<\mathcal J_{\mathbf k}\) in units of the Brillouin-zone (BZ) area. For hole dopings \(p<0.35\) and couplings \(0<\mathcal J_d/t<1\), \(I_L>0\) signals a violation of Luttinger's theorem, revealing the non-Fermi-liquid character of the correlated nodal metal.

\hypertarget{ms-normal-electron-spectra}{}ARPES resolves the occupied one-particle spectrum in energy and momentum (\hyperlink{ms-r4}{4}). At finite temperature, the exact spectral function \(A_{\mathrm N}=-\pi^{-1}\operatorname{Im}G^R_{\mathrm N,\sigma}\) has four pole branches, \(A_{\mathrm N}(\mathbf k,\omega)=\sum_{\zeta=-2}^{1}W_{\mathrm N,\mathbf k\zeta}(T)\,\delta(\omega-\xi_{\mathbf k}-\zeta\mathcal J_{\mathbf k})\), with weights \(W_{\mathrm N,\mathbf k\zeta}\). At 60 K, the zero-energy spectral weight exhibits a Fermi-arc-like distribution near each node (\hyperlink{ms-r36}{36}) for a Lorentzian half-width at half-maximum (HWHM) of 5 meV (\hyperlink{ms-fig2}{Fig. 2A}). A 1-meV HWHM resolves nearby branches within the arc (inset). Along the Fermi surface, addition and removal branches disperse away from zero energy, and the spectrum is particle--hole asymmetric (\hyperlink{ms-r40}{40}, \hyperlink{ms-r41}{41}), with \(A_{\mathrm N}(\mathbf k_F,\omega)\ne A_{\mathrm N}(\mathbf k_F,-\omega)\) (\hyperlink{ms-fig2}{Fig. 2B}), in contrast to Landau Fermi-liquid theory. The RML interaction breaks the local-in-momentum-space \(\mathbb Z_{2}^{\mathrm{ph}}\) particle--hole symmetry identified by Anderson and Haldane (\hyperlink{ms-r42}{42}). The \(\mathbb Z_{2}^{\mathrm{ph}}\)-symmetric contour shifts from the Fermi surface to the Luttinger surface, where \(A_{\mathrm N}(\mathbf k_L,\omega)=A_{\mathrm N}(\mathbf k_L,-\omega)\) (\hyperlink{ms-fig2}{inset}). The one-particle gap \(\mathcal J_d\phi_{\mathbf k_F,d}^2\) closes quadratically at the node and peaks at about 55 meV at the antinode (\hyperlink{ms-fig2}{Fig. 2C}), producing the nodal--antinodal dichotomy characteristic of the cuprate pseudogap (\hyperlink{ms-r10}{10}). The \(\mathbb Z_{2}^{\mathrm{ph}}\)-symmetry-breaking RML interaction suppresses low-temperature entropy for arbitrarily weak \(\mathcal J_d>0\) and produces a specific-heat coefficient jump \(\gamma_0(0^+)=\gamma_0(0)/4\), where \(\gamma_0(0^+)=\lim_{\mathcal J_d\to0^+}\lim_{T\to0}c_n(T,\mathcal J_d)/T\) and \(\gamma_0(0)=\lim_{T\to0}\lim_{\mathcal J_d\to0^+}c_n(T,\mathcal J_d)/T\). Along the BZ boundary \(k_x=\pi\) (\hyperlink{ms-fig2}{Fig. 2C}), spectral weight shifts between branches at the block GS crossings \(|4_{\mathbf k}\rangle\leftrightarrow|2_{\mathbf k,s}\rangle\) at \(\xi_{\mathbf k}=0\) (red) and \(|2_{\mathbf k,s}\rangle\leftrightarrow|0_{\mathbf k}\rangle\) at \(\xi_{\mathbf k}=\mathcal J_{\mathbf k}\) (cyan). The energy distribution curves (EDCs) evolve from one dominant peak to three near the crossings. These multipeak spectra arise from spin-singlet pair correlations in the single-band MStJ model and qualitatively resemble the peak--dip--hump line shapes in cuprate ARPES (\hyperlink{ms-r43}{43}, \hyperlink{ms-r44}{44}).

\hypertarget{ms-normal-pair-spectra}{}The one-particle gap vanishes only at the nodes, while the zero-total-momentum spin-singlet pair gap \(\Delta_{\mathrm{2p}}^{\mathrm N}(\mathbf k)=2\min\!\left\{|\xi_{\mathbf k}|,\,|\xi_{\mathbf k}-\mathcal J_{\mathbf k}|\right\}\) vanishes on two contours, \(\mathcal C_{2e}:\xi_{\mathbf k}=0\) and \(\mathcal C_{2h}:\xi_{\mathbf k}=\mathcal J_{\mathbf k}\). At zero temperature, the block electron number jumps by two across each contour (\hyperlink{ms-fig1}{Fig. 1}). The pair operators satisfy \([\hat\Delta_{\mathbf k}^{\dagger},\hat\Delta_{\mathbf k}]=\hat N_{\mathbf k}/2-1\) and commute between distinct momentum blocks. The Pauli exclusion principle allows at most two Cooper pairs per block. These gapless Cooper-pair Bose surfaces, termed Cooper surfaces, are the pair analogue of a Fermi surface. Their coexistence with isolated one-electron nodes defines the nodal Bose metal. The exact spin-singlet pair-removal spectral function \(S_{\Delta}^{<}(\mathbf k,\Omega)\equiv(2\pi)^{-1}\int_{-\infty}^{+\infty}dt\,e^{i\Omega t}\langle\hat\Delta_{\mathbf k}^{\dagger}(0)\hat\Delta_{\mathbf k}(t)\rangle\) has three pole branches at finite temperature, \(S_{\mathrm N,\Delta}^{<}(\mathbf k,\Omega)=\sum_{\eta=-2}^{0}W^{\Delta}_{\mathrm N,\mathbf k\eta}\,\delta(\Omega-2\xi_{\mathbf k}-\eta\mathcal J_{\mathbf k})\), with weights \(W^{\Delta}_{\mathrm N,\mathbf k\eta}\). cARPES probes two-particle correlations and can resolve the Cooper surfaces through the coincidence probability, which is proportional to the two-particle-removal spectral function \(A^{(2)<}_{\mathrm N,\sigma_1\sigma_2}(\mathbf k_1,\mathbf k_2;\omega_1,\omega_2)\). Spin rotation symmetry gives the singlet component \(A^{(2)<}_{\mathrm N,\Delta}=2A^{(2)<}_{\mathrm N,\uparrow\downarrow}-A^{(2)<}_{\mathrm N,\uparrow\uparrow}\). For \(\mathcal J_{\mathbf k}>0\), integrating \(A^{(2)<}_{\mathrm N,\Delta}\) over relative energy \(\omega_{\mathrm r}=(\omega_1-\omega_2)/2\) at fixed total energy \(\Omega=\omega_1+\omega_2\) gives 
\hypertarget{ms-eq5}{}%
\FittedEquation{5}{%
\overline A^{(2)<}_{\mathrm N,\Delta}(\mathbf k,\Omega)=S_{\mathrm N,\Delta}^{<}(\mathbf k,\Omega)
         +\tfrac32 p^{\mathrm N}_{\mathbf k,3}\,
          \delta(\Omega-2\xi_{\mathbf k}+\mathcal J_{\mathbf k}).
}
 The spectra approach each other when the thermal probability \(p^{\mathrm N}_{\mathbf k,3}\) of each state \(|3_{\pm\mathbf k\sigma}\rangle\) is negligible. At \(\Omega=0\), equal-spin-removal weight concentrates near the node (\hyperlink{ms-fig2}{Fig. 2D}), while singlet-removal weight traces both Cooper surfaces touching there (\hyperlink{ms-fig2}{Fig. 2E}). Along the BZ boundary \(k_x=\pi\) (\hyperlink{ms-fig2}{Fig. 2F}), weight shifts from the \(\eta=0\) branch to the \(\eta=-2\) branch at \(\xi_{\mathbf k}=0\) (red). The thermally activated \(\eta=-1\) branch is barely visible because \(p^{\mathrm N}_{\mathbf k,3}\) is extremely small there. Joint ARPES and cARPES could identify both Cooper surfaces by resolving zero-total-energy spin-singlet-removal peaks where the one-particle spectrum remains gapped.

\hypertarget{ms-normal-nodal-thermodynamics}{}The nodal DOS scales as \(\rho_{\mathrm N}(\omega,0)\propto\sqrt{|\omega|}\) and \(\rho_{\mathrm N}(0,T)\propto\sqrt{T}\) because \(\xi_{\mathbf k}\propto k_\perp\) and \(\mathcal J_{\mathbf k}\propto k_\parallel^2\) near each node. For free electrons with this DOS, the charge compressibility, spin susceptibility, and specific heat would scale as \(\sqrt{T}\), \(\sqrt{T}\), and \(T^{3/2}\), respectively. In the nodal Bose metal at low temperature, \(\kappa_{\mathrm N}(T)\simeq D+K_{\kappa}T^2\), \(\chi_s(T)\propto\sqrt{T}\), and \(c_{\mathrm N}(T)/T\simeq\pi^2D/12+\gamma_1\sqrt{T}\), where \(D=\int_{\mathrm{BZ}}\frac{d^2\mathbf k}{(2\pi)^2}\left[\delta(\xi_{\mathbf k})+\delta(\xi_{\mathbf k}-\mathcal J_{\mathbf k})\right]\) is the Cooper-surface DOS. Low-energy pair excitations near the Cooper surfaces contribute to \(\kappa_{\mathrm N}\) and \(c_{\mathrm N}\) even though \(\rho_{\mathrm N}(0,0)=0\).

\par\medskip
\hypertarget{ms-fig2}{}
\PrintAsset{fig-2-nodal-spectra}
{\small\noindent \textbf{Fig. 2. One- and two-particle spectra of the nodal Bose metal at 60 K.} \textbf{(A)} Zero-energy one-particle spectrum \(A_{\mathrm N}(\mathbf k,0)\) with an enlarged nodal region (inset). \textbf{(B, C)} One-particle spectra along the Fermi surface (B), the Luttinger surface (B, inset), and the BZ boundary \(k_x=\pi\) (C). \textbf{(D--F)} Two-particle removal spectra integrated over relative energy. Equal-spin (D) and spin-singlet (E) maps at \(\Omega=0\) have insets showing the corresponding spectra along the Fermi surface. F shows the spin-singlet spectrum along \(k_x=\pi\). Dotted curves in B, C, and F mark pole energies. Red and cyan lines mark the Cooper surfaces \(\mathcal C_{2e}\) and \(\mathcal C_{2h}\), respectively. Green marks the Luttinger surface in C. Blue and red dots mark nodes and antinodes. Parameters are \(p=0.15\) and \(\mathcal J_d=0.06\) eV. Lorentzian HWHMs are 5 meV per electron (1 meV in the inset of A) and 10 meV in total energy.\par}
\medskip
 
\hypertarget{ms-superconducting-instability}{}\section*{Cooper surface instability}

Adding a single Cooper pair to the fixed GS of the nodal Bose metal tests Cooper-surface stability, as in the Cooper problem for a Fermi surface (\hyperlink{ms-r16}{16}). The pairing interaction is the leading zero-total-momentum component of AFM superexchange, \(\hat H_{J,\mathbf 0}=-2JN_s\sum_{\alpha=s,d}\hat\Lambda_\alpha^\dagger\hat\Lambda_\alpha\), where \(\hat\Lambda_\alpha=N_s^{-1}\sum_{\mathbf k}\phi_{\mathbf k,\alpha}\hat\Delta_{\mathbf k}\) and \(\phi_{\mathbf k,s}=(\cos k_x+\cos k_y)/2\) is the extended \(s\)-wave form factor. Square-lattice symmetry decouples the extended \(s\)- and \(d\)-wave pairing channels in linear response. For the single-pair addition state \(|\Psi\rangle=\sum_{\{\mathbf k,-\mathbf k\}}^{\xi_{\mathbf k}>0}a_{\mathbf k}\hat\Delta_{\mathbf k}^{\dagger}|\mathrm{GS}_{\mathrm N}\rangle\), the binding energy \(E_{b,\alpha}\) relative to the fixed GS satisfies

\hypertarget{ms-eq6}{}%
\FittedEquation{6}{%
1=4J\int_{\xi_{\mathbf k}>0}\frac{d^2\mathbf k}{(2\pi)^2}
\frac{\phi_{\mathbf k,\alpha}^2}{2\xi_{\mathbf k}-2\mathcal J_{\mathbf k}\Theta(\xi_{\mathbf k}-\mathcal J_{\mathbf k})+E_{b,\alpha}}.
}
 
At weak attraction, the \(\phi^2\)-weighted Cooper-surface DOS \(D_\alpha=\int_{\mathrm{BZ}}\frac{d^2\mathbf k}{(2\pi)^2}\phi_{\mathbf k,\alpha}^2[\delta(\xi_{\mathbf k})+\delta(\xi_{\mathbf k}-\mathcal J_{\mathbf k})]\) controls the integral. For finite \(D_\alpha\), the integral diverges logarithmically as \(E_{b,\alpha}\to0^+\), binding a pair for every positive \(J\). To leading logarithmic accuracy, the binding energy is \(E_{b,\alpha}\simeq 2E_{\mathrm{cut}}\,e^{-1/(2JD_\alpha)}\), with fixed low-energy cutoff \(E_{\mathrm{cut}}\). Despite \(\rho_{\mathrm N}(0,0)=0\), the nodal Bose metal thus exhibits a Cooper instability, with the two Cooper surfaces playing the role of the conventional Fermi surface.

The many-body instability of \(\hat H=\hat H_{\mathrm N}+\hat H_{J,\mathbf 0}\) toward macroscopic SC order is determined by the pair susceptibility \(\chi_{\alpha\alpha'}(i\Omega_m,T)\equiv N_s\int_0^\beta d\tau\,e^{i\Omega_m\tau}\langle T_\tau\hat\Lambda_\alpha(\tau)\hat\Lambda_{\alpha'}^{\dagger}(0)\rangle\), where \(\Omega_m=2\pi m/\beta\) is a bosonic Matsubara frequency and \(\beta=T^{-1}\). The susceptibility matrix satisfies the Dyson equation \(\boldsymbol\chi(i\Omega_m,T)=[\mathbf I-2J\boldsymbol\chi_{\mathrm N}(i\Omega_m,T)]^{-1}\boldsymbol\chi_{\mathrm N}(i\Omega_m,T)\). When \(\det[\mathbf I-2J\boldsymbol\chi_{\mathrm N}(0,T_c)]=0\), a collective pairing mode softens at \(T_c\), establishing phase coherence between momentum blocks with preexisting pair correlations. Evaluating the normal-state susceptibility exactly without invoking Wick's theorem (\hyperlink{ms-r30}{30}) gives \(\chi_{\mathrm N,\alpha\alpha'}(i\Omega_m,T)=(2/N_s)\delta_{\alpha\alpha'}\sum_{\mathbf k}\phi_{\mathbf k,\alpha}^2\Pi_{\mathrm N,\mathbf k}(i\Omega_m,T)\), where \(\Pi_{\mathrm N,\mathbf k}\) is the pair Matsubara Green's function. The spectral identity \(\Pi_{\mathrm N,\mathbf k}(0,T)=\int_{-\infty}^{\infty}d\Omega\,\frac{e^{\beta\Omega}-1}{\Omega}\,S^{<}_{\mathrm N,\Delta}(\mathbf k,\Omega)\) relates the static pair susceptibility to gapless pair excitations on both Cooper surfaces.

At low temperature, the singular part of the static pair susceptibility is \(\chi_{\mathrm N,\alpha\alpha}^{\mathrm{sing}}(0,T)\simeq 2D_\alpha\ln\!\left(\frac{4e^{\gamma_{\mathrm E}}E_{\mathrm{cut}}}{\pi T}\right)\), where \(\gamma_{\mathrm E}\) is Euler's constant. To leading logarithmic accuracy, the \(\alpha\)-wave transition temperature is \(T_{c,\alpha}\simeq\frac{4e^{\gamma_{\mathrm E}}}{\pi}E_{\mathrm{cut}}\,e^{-1/(4JD_\alpha)}\). For finite \(D_\alpha\), the logarithmic divergence drives pairing at arbitrarily weak attraction. For the parameters in \hyperlink{ms-fig3}{Fig. 3}, the leading pairing instability is in the \(d\)-wave channel. Each Cooper surface undergoes a doping-driven Lifshitz transition upon crossing the antinodal saddle points (\hyperlink{ms-r36}{36}), producing a van Hove singularity in \(D_d\).

\par\medskip
\hypertarget{ms-fig3}{}
\PrintAsset{fig-3-instability}
{\small\noindent \textbf{Fig. 3. Doping dependence of the \(d\)-wave pairing transition temperature \(T_{c,d}\) and the \(\phi^2\)-weighted Cooper-surface DOS \(D_d\).} Solid curves show \(T_{c,d}\) (left logarithmic axis), determined from the instability condition \(2J\chi_{\mathrm N,dd}(0,T_{c,d})=1\) at fixed electron density \(n_e\). The dotted curve shows \(D_d\) (right axis). The van Hove singularities at \(p\simeq0.311\) and \(p\simeq0.443\) correspond to \(\mathcal C_{2e}\) and \(\mathcal C_{2h}\), respectively.\par}
\medskip
 
\hypertarget{ms-correlated-superconductor}{}\section*{Correlated nodal \(d\)-wave superconductor}

\hypertarget{ms-table1}{}\par\medskip

\noindent{\small \textbf{Table 1. Block eigenstates and eigenenergies of the SC block Hamiltonian \(\hat h_{\mathbf k}^{\mathrm{SC}}\).}\par}\nopagebreak[4]
\begingroup\small\setlength{\tabcolsep}{3.5pt}\renewcommand{\arraystretch}{1.30}
\noindent\FittedTable{\begin{tabular}{@{}cccc@{}}\toprule
{\shortstack[c]{Fermion parity\\  \(\pi_{\mathbf k,\lambda}\)}} & {Eigenstate \(|\pi_{\mathbf k,\lambda}\rangle\)} & {Eigenenergy \(E^{\mathrm{SC}}_{\mathbf k,\lambda}\)} & {Degeneracy} \\
\midrule
{\(+1\)} & {\(\begin{aligned}
              &|e_{\mathbf k,l}\rangle=x_{\mathbf k,l}|0_{\mathbf k}\rangle+y_{\mathbf k,l}|2_{\mathbf k,s}\rangle+z_{\mathbf k,l}|4_{\mathbf k}\rangle\\
              &|x_{\mathbf k,l}|^2+|y_{\mathbf k,l}|^2+|z_{\mathbf k,l}|^2=1,\quad l=0,1,2
            \end{aligned}\)} & {\(E^{\mathrm{SC},e}_{\mathbf k,0}\le E^{\mathrm{SC},e}_{\mathbf k,1}\le E^{\mathrm{SC},e}_{\mathbf k,2}\)} & {1 each} \\
{\(+1\)} & {\(|2_{\pm\mathbf k}\rangle,\ |2_{\mathbf k,t_1}\rangle,\ |2_{\mathbf k,t_0}\rangle,\ |2_{\mathbf k,t_{-1}}\rangle\)} & {\(2\xi_{\mathbf k}\)} & {5} \\
{\(-1\)} & {\(\begin{aligned}
              &|o_{\pm\mathbf k\sigma,\nu}\rangle=u_{\mathbf k,\nu}|1_{\pm\mathbf k\sigma}\rangle+v_{\mathbf k,\nu}|3_{\mp\mathbf k\sigma}\rangle\\
              &|u_{\mathbf k,\nu}|^2+|v_{\mathbf k,\nu}|^2=1,\quad\nu=\pm1
            \end{aligned}\)} & {\(\begin{gathered}
              E^{\mathrm{SC},o}_{\mathbf k,\nu}=2\xi_{\mathbf k}-\mathcal J_{\mathbf k}/2\\
              {}+\nu\sqrt{(\xi_{\mathbf k}-\mathcal J_{\mathbf k}/2)^2+2|\Delta_{\mathbf k,d}|^2}
            \end{gathered}\)} & {4 each} \\
\bottomrule\end{tabular}}\par\endgroup
\smallskip
 
{\small \(|2_{\pm\mathbf k}\rangle=\hat c^{\dagger}_{\pm\mathbf k\uparrow}\hat c^{\dagger}_{\pm\mathbf k\downarrow}|0_{\mathbf k}\rangle\) denotes a state with two electrons at the same momentum. \(|2_{\mathbf k,t_m}\rangle\) denotes a spin triplet with spin projection \(m\). \(|1_{\pm\mathbf k\sigma}\rangle=\hat c^{\dagger}_{\pm\mathbf k\sigma}|0_{\mathbf k}\rangle\) and \(|3_{\pm\mathbf k\sigma}\rangle=\sqrt{2}\,\hat\Delta_{\mathbf k}^{\dagger}|1_{\mp\mathbf k\sigma}\rangle\) denote one- and three-electron states, respectively.\par}
 
Pair transfer between blocks below \(T_{c,d}\) establishes the \(d\)-wave order parameter \(\Delta_d=2J\langle\hat\Lambda_d\rangle_{\mathrm{SC}}\). Decoupling \(\hat H_{J,\mathbf 0}\) in the leading \(d\)-wave channel gives the pairing field \(\Delta_{\mathbf k,d}=\Delta_d\phi_{\mathbf k,d}\) and mean-field SC Hamiltonian \(\hat H_{\mathrm{SC}}=\sum_{\{\mathbf k,-\mathbf k\}}\hat h_{\mathbf k}^{\mathrm{SC}}+N_s|\Delta_d|^2/(2J)\). In a gauge with real \(\Delta_d\), the SC block Hamiltonian is \(\hat h_{\mathbf k}^{\mathrm{SC}}=\hat h_{\mathbf k}^{\mathrm N}-2\Delta_{\mathbf k,d}(\hat\Delta_{\mathbf k}^{\dagger}+\hat\Delta_{\mathbf k})\), which preserves block fermion parity \(\hat\pi_{\mathbf k}=(-1)^{\hat N_{\mathbf k}}\). The parameters \(\Delta_d\) and \(\mu\) are determined self-consistently at fixed \(n_e\). The many-body wavefunction \(|\mathrm{GS}_{\mathrm{SC}}\rangle=\prod_{\{\mathbf k,-\mathbf k\}}|e_{\mathbf k,0}\rangle\) reduces to the BCS wavefunction \(|\mathrm{BCS}\rangle=\prod_{\mathbf k}(u_{\mathbf k}+v_{\mathbf k}\hat c_{\mathbf k\uparrow}^{\dagger}\hat c_{-\mathbf k\downarrow}^{\dagger})|0\rangle\) when the amplitudes in \(|e_{\mathbf k,0}\rangle\) satisfy \(y_{\mathbf k,0}^{\,2}=2x_{\mathbf k,0}z_{\mathbf k,0}\) for every block. Interblock phase coherence gives \(C_{\mathbf k\mathbf k'}\) a macroscopic eigenvalue and establishes SC ODLRO.

The Hamiltonian \(\hat h_{\mathbf k}^{\mathrm{SC}}\) retains the RML interaction exactly, and the \(d\)-wave pairing field preserves all four nodes, yielding a correlated nodal SC state. The one-particle gap \(\Delta_{\mathrm{1p}}^{\mathrm{SC}}(\mathbf k)=E_{\mathbf k,-1}^{\mathrm{SC},o}-E_{\mathbf k,0}^{\mathrm{SC},e}\) reduces to the Dirac dispersion \(E_{\mathbf k}\simeq\sqrt{\xi_{\mathbf k}^2+2|\Delta_{\mathbf k,d}|^2}\) near the nodes. Zeeman splitting \(B_z\) shifts the spin-resolved excitation energies by \(\pm B_z/2\), producing a zero-energy pocket along \(E_{\mathbf k}=B_z/2\) (\hyperlink{ms-fig4}{Fig. 4A}). Near the antinodes on the Fermi surface, \(\mathcal J_{\mathbf k_F}\gg|\Delta_{\mathbf k_F,d}|\), so the gap \(\Delta_{\mathrm{1p}}^{\mathrm{SC}}(\mathbf k_F)\simeq\mathcal J_{\mathbf k_F}+2|\Delta_{\mathbf k_F,d}|\) retains the RML interaction scale. For the parameters in \hyperlink{ms-fig4}{Fig. 4B}, a small near-nodal pairing scale (\(\Delta_d\simeq7.29\) meV) coexists with a large antinodal excitation gap (\(\Delta_{\mathrm{1p}}^{\mathrm{SC}}(\mathbf k_a)\simeq68.0\) meV), resembling the nodal--antinodal dichotomy observed in underdoped cuprates below \(T_c\) (\hyperlink{ms-r11}{11}, \hyperlink{ms-r12}{12}, \hyperlink{ms-r45}{45}, \hyperlink{ms-r46}{46}).

One-electron addition and removal connect the SC block GS \(|e_{\mathbf k,0}\rangle\) to odd-fermion-parity states with \(\nu=\pm1\), giving four dominant poles at symmetric energies \(\omega=\pm(E_{\mathbf k,\nu}^{\mathrm{SC},o}-E_{\mathbf k,0}^{\mathrm{SC},e})\) (\hyperlink{ms-fig4}{Fig. 4D}). However, addition and removal weights differ on the Fermi surface (\hyperlink{ms-fig4}{Fig. 4B}) through the same \(\mathbb Z_{2}^{\mathrm{ph}}\) symmetry breaking as in the nodal Bose metal. The RML interaction shifts the \(\mathbb Z_{2}^{\mathrm{ph}}\)-symmetric contour to the underlying Luttinger surface (\hyperlink{ms-fig4}{Fig. 4C}). The pairing field opens avoided crossings on \(\mathcal C_{2e}\) and \(\mathcal C_{2h}\). Along \(k_x=\pi\), the dominant \(\nu=-1\) branches bend back near \(\mathcal C_{2e}\) and \(\mathcal C_{2h}\), while the \(\nu=+1\) branches carry little spectral weight in between (\hyperlink{ms-fig4}{Fig. 4D}). The weak intensity below the \(\nu=-1\) branches qualitatively resembles the cuprate ARPES waterfall (\hyperlink{ms-r47}{47}). In the single-band BCS limit \(\mathcal J_d=0\), the \(\nu=+1\) branches lose all spectral weight (\hyperlink{ms-r17}{17}).

The relative-energy-integrated spin-singlet-removal spectrum \(\overline A^{(2)<}_{\mathrm{SC},\Delta}(\mathbf k,\Omega)\) exhibits three ridges at zero total energy (\hyperlink{ms-fig4}{Fig. 4E}). For the dominant sequential two-electron-removal processes \(|e_{\mathbf k,0}\rangle\to|o_{\mathbf k\sigma,\nu}\rangle\to|e_{\mathbf k,0}\rangle\), the spectral weight summed over both \(\nu=\pm1\) intermediate states is 
\hypertarget{ms-eq7}{}%
\FittedEquation{7}{%
2p^{\mathrm{SC},e}_{\mathbf k,0}\sum_{\nu=\pm1}
        \left|u_{\mathbf k,\nu}^{*}x_{\mathbf k,0}+\frac{v_{\mathbf k,\nu}^{*}y_{\mathbf k,0}}{\sqrt2}\right|^2
        \left|\frac{u_{\mathbf k,\nu}^{*}y_{\mathbf k,0}}{\sqrt2}+v_{\mathbf k,\nu}^{*}z_{\mathbf k,0}\right|^2,
}
 where \(p^{\mathrm{SC},e}_{\mathbf k,0}\) is the thermal probability of \(|e_{\mathbf k,0}\rangle\). Near \(\mathcal C_{2e}\), the pairing field mixes the nearly degenerate states \(|4_{\mathbf k}\rangle\) and \(|2_{\mathbf k,s}\rangle\), giving appreciable amplitudes \(z_{\mathbf k,0}\) and \(y_{\mathbf k,0}\) in \(|e_{\mathbf k,0}\rangle\). Near \(\mathcal C_{2h}\), it mixes \(|2_{\mathbf k,s}\rangle\) and \(|0_{\mathbf k}\rangle\), giving appreciable amplitudes \(y_{\mathbf k,0}\) and \(x_{\mathbf k,0}\). The resulting enhancement of the return spectral weight in \hyperlink{ms-eq7}{Eq. 7} spans a finite momentum range around each avoided crossing, broadening the two outer ridges (\hyperlink{ms-fig4}{Fig. 4E}) relative to their normal-state counterparts (\hyperlink{ms-fig2}{Fig. 2E}). On the Luttinger surface, \(|1_{\pm\mathbf k\sigma}\rangle\) and \(|3_{\mp\mathbf k\sigma}\rangle\) mix equally, with \(|u_{\mathbf k,\nu}|^2=|v_{\mathbf k,\nu}|^2=1/2\). The dominant amplitude \(y_{\mathbf k,0}\) contributes through the same intermediate state to both squared factors in \hyperlink{ms-eq7}{Eq. 7}. This enhanced return spectral weight produces the middle ridge along the underlying Luttinger surface without requiring a third gapless Cooper surface. The middle ridge, unresolved in the nodal Bose metal, is a hallmark of the correlated nodal SC state. This Luttinger-surface signal arises primarily from the intermediate-state interference correction \(\mathcal I_{\mathrm{SC},\Delta}=\overline A^{(2)<}_{\mathrm{SC},\Delta}-S^{<}_{\mathrm{SC},\Delta}\) and is unresolved in the simultaneous pair-removal spectrum. In the single-band BCS limit, the three ridges merge into one along the Fermi surface.

The pairing field opens a spin-singlet Cooper-pair gap \(\Delta_{\mathrm{2p}}^{\mathrm{SC}}(\mathbf k)=E_{\mathbf k,1}^{\mathrm{SC},e}-E_{\mathbf k,0}^{\mathrm{SC},e}\) on both Cooper surfaces, visible in the connected pair-removal spectrum \(S^{<,c}_{\mathrm{SC},\Delta}=S^{<}_{\mathrm{SC},\Delta}-|\langle\hat\Delta_{\mathbf k}\rangle_{\mathrm{SC}}|^2\delta(\Omega)\). Along \(k_x=\pi\), the pairing-induced splitting is approximately \(4|\Delta_{\mathbf k,d}|\), giving \(4|\Delta_{\mathbf k_a,d}|\simeq27.89\) meV on \(\mathcal C_{2e}\) and \(27.17\) meV on \(\mathcal C_{2h}\), close to the respective lowest pair gaps of \(27.72\) and \(27.00\) meV (\hyperlink{ms-fig4}{Fig. 4F}). The \(l=1\) branch of \(S^{<,c}_{\mathrm{SC},\Delta}\) bends back near \(\mathcal C_{2e}\) and \(\mathcal C_{2h}\), as does the \(\nu=-1\) branch in \hyperlink{ms-fig4}{Fig. 4D}. On the Luttinger surface, the \(l=1\) and \(l=2\) pair-removal branches exhibit an avoided crossing with a minigap of \(6.33\) meV (\hyperlink{ms-fig4}{Fig. 4F}). The remaining Dirac nodes, with \(\rho_{\mathrm{SC}}(\omega,0)\propto|\omega|\) at low energies, dominate the low-temperature behavior of this state. Specific heat scales as \(c_{\mathrm{SC}}\propto T^2\) at low temperatures and jumps at \(T_{c,d}\) with a nonuniversal ratio \(\Delta c/c_{\mathrm N}(T_{c,d}^{+})\).

\par\medskip
\hypertarget{ms-fig4}{}
\PrintAsset{fig4-correlated-sc}
{\small\noindent \textbf{Fig. 4. One- and two-particle spectra of the correlated nodal superconductor at 10 K.} \textbf{(A)} Zero-energy spin-down spectrum at Zeeman splitting \(B_z=4\) meV, with an enlarged nodal region (inset). \textbf{(B--D)} One-particle spectra along the Fermi surface (B), the Luttinger surface (C), and the BZ boundary \(k_x=\pi\) (D). \textbf{(E)} Sequential spin-singlet-removal spectrum at \(\Omega=0\), integrated over relative energy. The inset shows the spectrum along the Fermi surface. \textbf{(F)} Connected simultaneous pair-removal spectrum along \(k_x=\pi\). Dotted curves mark the Fermi surface (A, E), the one-particle pole energies (B--D), and the two-particle pole energies (F). Red, green, and cyan lines in D and F mark \(\mathcal C_{2e}\), the Luttinger surface, and \(\mathcal C_{2h}\), respectively. Blue and red dots mark the node and antinode in A and E. Parameters match \hyperlink{ms-fig2}{Fig. 2}.\par}
\medskip
 
\hypertarget{ms-discussion}{}\section*{Discussion}

The nodal Bose metal is a non-Fermi liquid with an antinodal one-electron gap \(\mathcal J_{\mathbf k_a}\) and vanishing one-electron DOS \(\rho_{\mathrm N}(0,0)=0\). RML attraction induces spin-singlet pair correlations within each block. Gapless pair excitations near the two Cooper surfaces produce a logarithmically divergent pair susceptibility, so an arbitrarily weak attraction that transfers pairs between blocks can drive a pairing instability. The correlated nodal \(d\)-wave superconductor combines a small near-nodal pairing scale with a large antinodal excitation gap. The nodal Bose metal, the correlated nodal SC state, and the Cooper surface instability are strongly correlated counterparts of the Landau Fermi liquid, the BCS state (\hyperlink{ms-r17}{17}), and the Cooper instability of the Fermi surface (\hyperlink{ms-r16}{16}), respectively.

A Cooper surface traces the relative momentum of a pair with zero total momentum. Earlier Bose-liquid models track the momentum of a boson or collective mode and exhibit lines of gapless modes or singularities in the boson momentum distribution (\hyperlink{ms-r25}{25}, \hyperlink{ms-r26}{26}). A systematic renormalization-group treatment of the particle--particle and particle--hole channels on equal footing (\hyperlink{ms-r48}{48}), which could determine whether charge-density-wave, spin-density-wave, or pair-density-wave (\hyperlink{ms-r49}{49}) instabilities compete with zero-total-momentum pairing instability, is still lacking. The microscopic origin of the RML attraction remains an open question. Progressively restoring momentum mixing connects the HK and Hubbard models (\hyperlink{ms-r50}{50}), suggesting a route from the MStJ to the \(t\)--\(J\) model. However, the no-double-occupancy constraint couples different momentum blocks, substantially complicating such a construction.

The compiled ARPES gap data for Bi2201, Bi2212, and Bi2223 show stronger near-nodal suppression above \(T_c\) than below it (\hyperlink{sm-appendix-experimental-arpes-gap-data}{Appendix P}). The gaps in Bi2223, which has the highest onset \(T_c\), come closest to a quadratic dependence on \(|\phi_{\mathbf k,d}|\) above \(T_c\) and a linear dependence below, lending phenomenological support to the form \(\mathcal J_{\mathbf k}=\mathcal J_d\phi_{\mathbf k,d}^2\).

ARPES and spin-resolved cARPES could test the pairing mechanism at the same momenta across \(T_c\). In the nodal Bose metal, ARPES should show arc-like near-nodal spectral weight and an antinodal one-electron gap. The spectrum is particle--hole asymmetric on the Fermi surface but recovers particle--hole symmetry on the Luttinger surface. EDCs can have multiple peaks even in a single-layer cuprate (\hyperlink{ms-r44}{44}). cARPES should resolve both gapless Cooper surfaces as distinct peaks in the zero-total-energy spin-singlet removal spectrum at opposite electron momenta. In the correlated nodal \(d\)-wave superconductor, ARPES should resolve linear nodal dispersion and a large antinodal gap. The sequential spin-singlet-removal spectrum at zero total energy should show two broadened outer ridges near the Cooper surfaces and a middle ridge along the underlying Luttinger surface. On cooling in the nodal Bose metal, \(\kappa_{\mathrm N}\) and \(c_{\mathrm N}/T\) approach finite values, while \(\chi_s^{\mathrm N}\propto\sqrt{T}\). As in a \(d\)-wave BCS state, \(\kappa_{\mathrm{SC}}\) remains finite, while low-temperature \(c_{\mathrm{SC}}/T\) and \(\chi_s^{\mathrm{SC}}\) scale linearly with \(T\) in the correlated SC state. The specific-heat jump and gap-to-\(T_c\) ratios are nonuniversal.
 \hypertarget{ms-references}{}
\begingroup
\renewcommand{\refname}{References}

\endgroup

\hypertarget{ms-acknowledgments}{}\section*{Acknowledgments}

J.-J. M. thanks Yi Zhou, Fu-Chun Zhang, and Tao Li for insightful comments and suggestions and Yuehua Su for discussions on cARPES. Both authors thank Guangdi Zhou, Heng Wang, and Yueying Li for discussions of the experimental results.

\textbf{Funding:} This work was supported by the National Natural Science Foundation of China (grant nos. 12404171 and 12404157), the Guangdong Project (grant no. 2024QN11X176), the Guangdong Major Project of Basic Research (grant no. 2025B0303000004), and the Quantum Science Strategic Initiative of Guangdong Province (grant nos. GDZX2501001 and GDZX2401004).

\clearpage
\hypertarget{sm-supplementary-materials}{}
\begin{center}
{\Large\bfseries Supplementary Materials:\par}\smallskip
{\LARGE\bfseries Exactly solvable momentum-space \(t\)--\(J\) model I. ARPES and thermodynamics\par}\medskip
{\large Jian-Jian Miao and Cui Ding\par}\smallskip
Quantum Science Center of Guangdong-Hong Kong-Macao Greater Bay Area, Shenzhen 518045, China\par
Electronic address: \href{mailto:miaojianjian@quantumsc.cn}{miaojianjian@quantumsc.cn}\par
\end{center}
 \section*{Contents}
\begin{multicols}{2}\raggedcolumns\small
\par\noindent\textbf{\hyperlink{sm-model-hamiltonian}{I. Exactly solvable momentum-space \(t\)--\(J\) model}}\par\nopagebreak[3]
{\leftskip=0.9em\noindent \hyperlink{sm-model-exact-solution-title}{A. Exact solution}\par}
\smallskip
\par\noindent\textbf{\hyperlink{sm-model-block-spectrum}{II. Nodal Bose metal}}\par\nopagebreak[3]
{\leftskip=0.9em\noindent \hyperlink{sm-model-one-particle-spectra-title}{A. One-particle Green's function}\par}
{\leftskip=0.9em\noindent \hyperlink{sm-model-luttinger-surface-title}{B. Luttinger surface and Luttinger's theorem violation}\par}
{\leftskip=0.9em\noindent \hyperlink{sm-model-finite-temperature-spectra-title}{C. One-particle spectra: Fermi arc and nodal--antinodal dichotomy}\par}
{\leftskip=0.9em\noindent \hyperlink{sm-model-green-function-zeros-title}{D. Gapless Cooper surfaces}\par}
{\leftskip=0.9em\noindent \hyperlink{sm-model-cooper-surface-spectroscopy-title}{E. Cooper surface spectroscopy}\par}
{\leftskip=0.9em\noindent \hyperlink{sm-model-thermodynamic-quantities-title}{F. Thermodynamics beyond nodal density of states}\par}
\smallskip
\par\noindent\textbf{\hyperlink{sm-sc-instability-cnn}{III. SC instability of the nodal Bose metal}}\par\nopagebreak[3]
{\leftskip=0.9em\noindent \hyperlink{sm-sc-cooper-instability}{A. Cooper problem and Cooper instability}\par}
{\leftskip=0.9em\noindent \hyperlink{sm-sc-pair-susceptibility}{B. SC instability and pair susceptibility}\par}
\smallskip
\par\noindent\textbf{\hyperlink{sm-correlated-nodal-sc}{IV. Correlated nodal SC state}}\par\nopagebreak[3]
{\leftskip=0.9em\noindent \hyperlink{sm-sc-mean-field}{A. Mean-field theory}\par}
{\leftskip=0.9em\noindent \hyperlink{sm-sc-one-particle-gap}{B. One-particle Green's function}\par}
{\leftskip=0.9em\noindent \hyperlink{sm-sc-one-particle}{C. One-particle spectra: SC pockets, spectral particle-hole asymmetry, and spectral weight transfer}\par}
{\leftskip=0.9em\noindent \hyperlink{sm-sc-two-particle}{D. Spin-singlet removal spectra: middle ridge near Luttinger surface, and SC gap opening on Cooper surfaces}\par}
{\leftskip=0.9em\noindent \hyperlink{sm-sc-thermodynamics-main}{E. Thermodynamics with nodal density of states}\par}
\smallskip
\par\noindent\textbf{\hyperlink{sm-references}{References}}\par\nopagebreak[3]
\smallskip
\par\noindent\textbf{\hyperlink{sm-appendix-a}{Appendix A. Exact operator identities}}\par\nopagebreak[3]
{\leftskip=0.9em\noindent \hyperlink{sm-superexchange-square-lattice-title}{A.1. Superexchange interaction on the square lattice}\par}
{\leftskip=0.9em\noindent \hyperlink{sm-hubbard-pairing-title}{A.2. Hubbard interaction}\par}
\smallskip
\par\noindent\textbf{\hyperlink{sm-appendix-b}{Appendix B. Luttinger integral}}\par\nopagebreak[3]
{\leftskip=0.9em\noindent \hyperlink{sm-luttinger-exact-evaluation}{B.1. Exact evaluation}\par}
{\leftskip=0.9em\noindent \hyperlink{sm-appendix-c-title}{B.2. Weak-coupling expansion}\par}
\smallskip
\par\noindent\textbf{\hyperlink{sm-appendix-d}{Appendix C. ARPES of the conventional \(d\)-wave BCS state}}\par\nopagebreak[3]
{\leftskip=0.9em\noindent \hyperlink{sm-b-single-electron}{C.1. One-electron ARPES}\par}
{\leftskip=0.9em\noindent \hyperlink{sm-b-two-photon}{C.2. Two-photon cARPES}\par}
\smallskip
\par\noindent\textbf{\hyperlink{sm-appendix-e}{Appendix D. ARPES of the nodal Bose metal with Cooper surfaces}}\par\nopagebreak[3]
{\leftskip=0.9em\noindent \hyperlink{sm-c-single-electron}{D.1. One-electron ARPES}\par}
{\leftskip=0.9em\noindent \hyperlink{sm-c-two-photon}{D.2. Two-photon cARPES}\par}
{\leftskip=0.9em\noindent \hyperlink{sm-c-cooper-surface}{D.3. Cooper surface spectroscopy}\par}
\smallskip
\par\noindent\textbf{\hyperlink{sm-appendix-i}{Appendix E. Nodal density of states of the nodal Bose metal}}\par\nopagebreak[3]
{\leftskip=0.9em\noindent \hyperlink{sm-dos-i-exact}{E.1. Exact one-particle density of states}\par}
{\leftskip=0.9em\noindent \hyperlink{sm-dos-i-coordinates}{E.2. Moving tangent--normal frame}\par}
{\leftskip=0.9em\noindent \hyperlink{sm-dos-i-local-expansion}{E.3. Local expansion near the nodes}\par}
{\leftskip=0.9em\noindent \hyperlink{sm-dos-i-scaling}{E.4. Scaling function}\par}
{\leftskip=0.9em\noindent \hyperlink{sm-dos-i-even-small}{E.5. Small-\(|\omega|/T\) expansion}\par}
{\leftskip=0.9em\noindent \hyperlink{sm-dos-i-zero-temperature}{E.6. Large-\(|\omega|/T\) asymptotics}\par}
\smallskip
\par\noindent\textbf{\hyperlink{sm-appendix-j}{Appendix F. Thermodynamics of the nodal Bose metal}}\par\nopagebreak[3]
{\leftskip=0.9em\noindent \hyperlink{sm-thermo-f-charge}{F.1. Charge compressibility}\par}
{\leftskip=0.9em\noindent \hyperlink{sm-thermo-f-spin}{F.2. Spin susceptibility}\par}
{\leftskip=0.9em\noindent \hyperlink{sm-thermo-f-heat}{F.3. Specific heat}\par}
{\leftskip=0.9em\noindent \hyperlink{sm-thermo-f-heat-coefficient-jump}{F.4. Specific heat coefficient jump}\par}
\smallskip
\par\noindent\textbf{\hyperlink{sm-appendix-cooper-problem}{Appendix G. Cooper problem and Cooper instability}}\par\nopagebreak[3]
\smallskip
\par\noindent\textbf{\hyperlink{sm-appendix-pair-susceptibility}{Appendix H. Pair susceptibility}}\par\nopagebreak[3]
{\leftskip=0.9em\noindent \hyperlink{sm-appendix-h-dyson-equation}{H.1. Dyson equation}\par}
{\leftskip=0.9em\noindent \hyperlink{sm-appendix-h-nodal-pair-susceptibility}{H.2. SC instability temperature}\par}
\smallskip
\par\noindent\textbf{\hyperlink{sm-appendix-g}{Appendix I. SC instability at finite total momentum \(\mathbf Q\ne0\)}}\par\nopagebreak[3]
\smallskip
\par\noindent\textbf{\hyperlink{sm-appendix-sc-mean-field}{Appendix J. Mean-field theory of the correlated SC state}}\par\nopagebreak[3]
{\leftskip=0.9em\noindent \hyperlink{sm-sc-block-diagonalization}{J.1. Self-consistent equations}\par}
{\leftskip=0.9em\noindent \hyperlink{sm-sc-weak-gap-ratio}{J.2. Order-parameter-to-\(T_c\) ratio}\par}
\smallskip
\par\noindent\textbf{\hyperlink{sm-appendix-sc-excitation-gap}{Appendix K. One-particle Green's function of the correlated SC state}}\par\nopagebreak[3]
{\leftskip=0.9em\noindent \hyperlink{sm-sc-gap-explicit-resolvent}{K.1. Explicit form of the Green's function}\par}
{\leftskip=0.9em\noindent \hyperlink{sm-sc-gap-antinodal-expansion}{K.2. Antinodal expansion of the one-particle excitation gap}\par}
\smallskip
\par\noindent\textbf{\hyperlink{sm-appendix-sc-one-particle}{Appendix L. ARPES of the correlated SC state}}\par\nopagebreak[3]
{\leftskip=0.9em\noindent \hyperlink{sm-sc-spectral-lehmann}{L.1. One-electron ARPES}\par}
{\leftskip=0.9em\noindent \hyperlink{sm-sc-spectral-zeeman}{L.2. One-electron ARPES under magnetic field}\par}
\smallskip
\par\noindent\textbf{\hyperlink{sm-appendix-sc-two-particle}{Appendix M. cARPES of the correlated SC state}}\par\nopagebreak[3]
{\leftskip=0.9em\noindent \hyperlink{sm-sc-sequential-removal}{M.1. Two-photon cARPES}\par}
{\leftskip=0.9em\noindent \hyperlink{sm-sc-singlet-carpes}{M.2. Spin-singlet two-particle and pair-removal spectral functions}\par}
\smallskip
\par\noindent\textbf{\hyperlink{sm-appendix-sc-nodal-dos}{Appendix N. Nodal density of states of the correlated SC state}}\par\nopagebreak[3]
{\leftskip=0.9em\noindent \hyperlink{sm-sc-dos-exact}{N.1. Exact one-particle density of states}\par}
{\leftskip=0.9em\noindent \hyperlink{sm-sc-dos-local-scaling}{N.2. Local expansion near the nodes}\par}
{\leftskip=0.9em\noindent \hyperlink{sm-sc-dos-scaling-function}{N.3. SC scaling function}\par}
{\leftskip=0.9em\noindent \hyperlink{sm-sc-dos-frequency-limits}{N.4. Small- and large-\(|\omega|/T\) expansions}\par}
\smallskip
\par\noindent\textbf{\hyperlink{sm-appendix-sc-thermodynamics}{Appendix O. Thermodynamics of the correlated SC state}}\par\nopagebreak[3]
{\leftskip=0.9em\noindent \hyperlink{sm-sc-thermo-entropy}{O.1. Entropy, free energy, internal energy, and thermodynamic critical field}\par}
{\leftskip=0.9em\noindent \hyperlink{sm-sc-thermo-specific-heat}{O.2. Specific heat jump}\par}
{\leftskip=0.9em\noindent \hyperlink{sm-sc-thermo-nodal-limit}{O.3. Low-temperature behavior}\par}
\smallskip
\par\noindent\textbf{\hyperlink{sm-appendix-experimental-arpes-gap-data}{Appendix P. ARPES gap data}}\par\nopagebreak[3]
{\leftskip=0.9em\noindent \hyperlink{sm-appendix-p-references}{References for Appendix P}\par}
\smallskip
\end{multicols}

\hypertarget{sm-model-hamiltonian}{}\hypertarget{sm-exactly-solvable-model-title}{}\hypertarget{sm-model-title}{}
\section*{I. Exactly solvable momentum-space \(t\)--\(J\) model}

\hypertarget{sm-exactly-solvable-model}{}\hypertarget{sm-momentum-space-model-title}{} 
The single-band \(t\)--\(J\) model provides an effective description of the low-energy electronic degrees of freedom in the CuO\textsubscript{2} planes of cuprate superconductors within the framework of a doped Mott insulator \hyperlink{sm-ref-zhang-rice-1988}{[1]}, \hyperlink{sm-ref-lee-2006}{[2]}, \hyperlink{sm-ref-keimer-2015}{[3]}, \hyperlink{sm-ref-anderson-1987}{[4]}. The antiferromagnetic (AFM) superexchange interaction in the \(t\)--\(J\) model can be written exactly in terms of spin-singlet Cooper-pair operators \hyperlink{sm-ref-miao-2026}{[5]}:

\hypertarget{sm-eq-model-superexchange}{}%
\FittedEquation{1}{%
\hat H_J
        =J\sum_{\langle ij\rangle}\left(\hat{\mathbf S}_i\cdot\hat{\mathbf S}_j-\frac14\hat n_i\hat n_j\right)
        =-\frac{2J}{N_s}\sum_{\mathbf Q,\alpha=s,d}\sum_{\mathbf q,\mathbf q'}
        \phi_{\mathbf q,\alpha}\phi_{\mathbf q',\alpha}
        \hat\Delta^\dagger_{\mathbf Q,\mathbf q}\hat\Delta_{\mathbf Q,\mathbf q'}.
}
 
Here \(\hat{\mathbf S}_j=\frac12\sum_{\sigma\sigma'}\hat c^\dagger_{j\sigma}\boldsymbol\tau_{\sigma\sigma'}\hat c_{j\sigma'}\) and \(\hat n_j=\sum_\sigma\hat n_{j\sigma}=\sum_\sigma\hat c^\dagger_{j\sigma}\hat c_{j\sigma}\) are the electron spin and number operators at site \(j\), respectively. The operator \(\hat c_{j\sigma}\) annihilates an electron at site \(j\) with spin \(\sigma\), and \(\boldsymbol\tau\) denotes the vector of Pauli matrices. \(J>0\) is the AFM coupling, and the sum over \(\langle ij\rangle\) runs over nearest-neighbour bonds. \(\phi_{\mathbf q,\alpha}=(\cos q_x+\alpha\cos q_y)/2\) is the pairing form factor on the square lattice, with \(\alpha=+1\) and \(\alpha=-1\) corresponding to the extended-\(s\)- and \(d\)-wave channels, respectively. \(N_s\) is the total number of lattice sites. The spin-singlet pair-annihilation operator with total momentum \(\mathbf Q\) and relative momentum \(\mathbf q\) is defined as

\hypertarget{sm-eq-model-pair-operator}{}%
\FittedEquation{2}{%
\hat\Delta_{\mathbf Q,\mathbf q}
        \equiv\frac1{\sqrt2}\left(
        \hat c_{\mathbf Q/2+\mathbf q,\uparrow}\hat c_{\mathbf Q/2-\mathbf q,\downarrow}
        -\hat c_{\mathbf Q/2+\mathbf q,\downarrow}\hat c_{\mathbf Q/2-\mathbf q,\uparrow}
        \right).
}
 
Motivated by the exact operator identity in \hyperlink{sm-eq-model-superexchange}{Eq. (1)}, a phenomenological momentum-space model is proposed:

\hypertarget{sm-eq-model-mstj}{}%
\FittedEquation{3}{%
\hat H_{\mathrm N}
        =\sum_{\mathbf k\sigma}\xi_{\mathbf k}\hat c^\dagger_{\mathbf k\sigma}\hat c_{\mathbf k\sigma}
        -\sum_{\mathbf k}\mathcal J_{\mathbf k}\hat\Delta^\dagger_{\mathbf k}\hat\Delta_{\mathbf k},
}
 
where \(\hat c_{\mathbf k\sigma}=N_s^{-1/2}\sum_j e^{i\mathbf k\cdot\mathbf r_j}\hat c_{j\sigma}\) annihilates an electron with crystal momentum \(\mathbf k\) and spin \(\sigma\), and \(\hat\Delta_{\mathbf k}\equiv\hat\Delta_{\mathbf0,\mathbf k}=(\hat c_{\mathbf k\uparrow}\hat c_{-\mathbf k\downarrow}-\hat c_{\mathbf k\downarrow}\hat c_{-\mathbf k\uparrow})/\sqrt2\) annihilates a spin-singlet Cooper pair with zero total momentum. The no-double-occupancy constraint of the real-space \(t\)--\(J\) model, i.e. Gutzwiller projection, is not imposed on the phenomenological model. The band dispersion is \(\xi_{\mathbf k}=\varepsilon_{\mathbf k}-\mu\), with

\hypertarget{sm-eq-model-dispersion}{}%
\FittedEquation{4}{%
\varepsilon_{\mathbf k}=-2t(\cos k_x+\cos k_y)-4t'\cos k_x\cos k_y-2t''(\cos 2k_x+\cos 2k_y),
}
 
where the nearest-, second-nearest-, and third-nearest-neighbour hopping parameters are set to \(t=0.3~\mathrm{eV}\), \(t'/t=-0.2\), and \(t''/t=0.1\), respectively. Throughout, the lattice constant is set to unity, and natural units \(\hbar=k_{\mathrm B}=1\) are used. The chemical potential \(\mu(T)\) is adjusted to fix the electron density \(n_e=1-p\), where \(p\) is the hole doping concentration. The Fermi surface is the contour \(\mathcal S_F:\xi_{\mathbf k_F}=0\).

The Hamiltonian \(\hat H_{\mathrm N}\) includes a \emph{relative-momentum-local} (RML) interaction acting only within each momentum block \(\{\mathbf k,-\mathbf k\}\), with interaction strength \(\mathcal J_{\mathbf k}=\mathcal J_d\phi_{\mathbf k,d}^{2}\ge0\). Thus, the RML interaction \(\hat H_{\mathcal J}=-\sum_{\mathbf k}\mathcal J_{\mathbf k}\hat\Delta^\dagger_{\mathbf k}\hat\Delta_{\mathbf k}\) is attractive and favours spin-singlet pairing in each momentum block. \(\hat H_{\mathcal J}\) has the same operator structure as the contribution \(-2JN_s^{-1}\sum_{\mathbf k}\phi_{\mathbf k,d}^{2}\hat\Delta^\dagger_{\mathbf k}\hat\Delta_{\mathbf k}\) to \(\hat H_J\) in \hyperlink{sm-eq-model-superexchange}{Eq. (1)}, obtained by selecting \(\mathbf Q=\mathbf0\), \(\alpha=d\), and \(\mathbf q=\mathbf q'=\mathbf k\) and summing over \(\mathbf k\), which is \(O(1)\) at fixed \(J\). For fixed \(\mathcal J_d\), \(\hat H_{\mathcal J}\) is instead extensive, scaling as \(O(N_s)\). Thus, \(\hat H_{\mathcal J}\) defines a distinct interaction rather than merely a component of the superexchange interaction.

Thermal phase fluctuations \hyperlink{sm-ref-emery-kivelson-1995}{[6]} can suppress pair coherence between distinct momentum blocks while allowing RML pair correlations to persist, since the phase factors cancel in \(\hat\Delta^\dagger_{\mathbf Q,\mathbf q}\hat\Delta_{\mathbf Q,\mathbf q}\). The phenomenological model retains phase-insensitive spin-singlet pair correlations within each momentum block while omitting pair scattering between distinct blocks. Establishing whether the zero-total-momentum \(d\)-wave channel is more relevant than competing finite-total-momentum or extended-\(s\)-wave channels requires a systematic renormalization-group (RG) analysis \hyperlink{sm-ref-shankar-1994}{[7]}, \hyperlink{sm-ref-metlitski-2015}{[8]}, \hyperlink{sm-ref-metzner-2012}{[9]}. The restriction to zero total momentum and the interaction strength \(\mathcal J_d\) are phenomenological assumptions.

\hypertarget{sm-model-rml-real-space}{}The real-space representation of the RML interaction is instructive. Introducing the Fourier transform \(\hat\Delta_{\mathbf k}=N_s^{-1}\sum_{ij}e^{i\mathbf k\cdot(\mathbf r_i-\mathbf r_j)}\hat\Delta_{ij}\), where \(\hat\Delta_{ij}=(\hat c_{i\uparrow}\hat c_{j\downarrow}-\hat c_{i\downarrow}\hat c_{j\uparrow})/\sqrt2\) is the real-space spin-singlet pair operator defined on sites \(i\) and \(j\), gives

\hypertarget{sm-eq-model-rml-real-space}{}%
\FittedEquation{5}{%
\hat H_{\mathcal J}=-\frac1{N_s}\sum_{ij,i'j'}
        \mathcal J_{ij,i'j'}\hat\Delta^\dagger_{ij}\hat\Delta_{i'j'},
}
 
where the sum runs over all directed bonds \((i,j)\) and \((i',j')\) on the lattice, including \(i=j\) and \(i'=j'\), and \(\mathcal J_{ij,i'j'}=\mathcal J_{(\mathbf r_i-\mathbf r_j)-(\mathbf r_{i'}-\mathbf r_{j'})}\), with \(\mathcal J_{\mathbf R}=N_s^{-1}\sum_{\mathbf k}e^{i\mathbf k\cdot\mathbf R}\mathcal J_{\mathbf k}\). For \(\mathcal J_{\mathbf k}=\mathcal J_d\phi_{\mathbf k,d}^2\) on the square lattice,

\hypertarget{sm-eq-model-rml-real-space-kernel}{}%
\FittedEquation{6}{%
\mathcal J_{\mathbf R}=\frac{\mathcal J_d}{4}\left(
        \delta_{\mathbf R,\mathbf0}
        -\frac12\sum_{\mathbf e\in\{\pm\mathbf e_x\pm\mathbf e_y\}}\delta_{\mathbf R,\mathbf e}
        +\frac14\sum_{\mathbf e'\in\{\pm2\mathbf e_x,\pm2\mathbf e_y\}}\delta_{\mathbf R,\mathbf e'}
        \right),
}
 
where \(\mathbf e_x\) and \(\mathbf e_y\) denote the nearest-neighbour displacements along the positive \(x\) and \(y\) directions, respectively. \(\mathcal J_{\mathbf R}\) constrains only the difference \(\mathbf R=(\mathbf r_i-\mathbf r_j)-(\mathbf r_{i'}-\mathbf r_{j'})\) between the relative coordinates of the two pairs, but imposes no restriction on the size \(|\mathbf r_i-\mathbf r_j|\) of an individual pair or the separation \(\left|(\mathbf r_i+\mathbf r_j)/2-(\mathbf r_{i'}+\mathbf r_{j'})/2\right|\) of their centres. Therefore, the RML interaction is nonlocal in real space, allowing an \emph{all-to-all interaction} of spin-singlet Cooper pairs subject to the constraint imposed by \(\mathcal J_{\mathbf R}\), and such a nonlocal interaction structure is analogous to that of the Baskaran, Hatsugai--Kohmoto (HK), and Sachdev--Ye--Kitaev (SYK) models \hyperlink{sm-ref-baskaran-1991}{[10]}, \hyperlink{sm-ref-hatsugai-kohmoto-1992}{[11]}, \hyperlink{sm-ref-sachdev-ye-1993}{[12]}, \hyperlink{sm-ref-kitaev-2015}{[13]}, \hyperlink{sm-ref-chowdhury-2022}{[14]}.

Introducing the operators \(\hat{\mathbf S}_{\mathbf k}=\frac12\sum_{\sigma\sigma'}\hat c^\dagger_{\mathbf k\sigma}\boldsymbol\tau_{\sigma\sigma'}\hat c_{\mathbf k\sigma'}\) and \(\hat n_{\mathbf k}=\sum_\sigma\hat n_{\mathbf k\sigma}=\sum_\sigma\hat c^\dagger_{\mathbf k\sigma}\hat c_{\mathbf k\sigma}\), the RML interaction can be rewritten as

\hypertarget{sm-eq-model-rml-spin-exchange}{}%
\FittedEquation{7}{%
\hat H_{\mathcal J}
        =\sum_{\mathbf k}\mathcal J_{\mathbf k}
        \left(\hat{\mathbf S}_{\mathbf k}\cdot\hat{\mathbf S}_{-\mathbf k}
        -\frac14\hat n_{\mathbf k}\hat n_{-\mathbf k}
        -\frac12\delta_{\mathbf k,-\mathbf k}\hat n_{\mathbf k}\right),
}
 
Hence, the exactly solvable model \(\hat H_{\mathrm N}\), a generalized Baskaran model with a density interaction, is termed the \emph{momentum-space \(t\)--\(J\)} (MStJ) model.

\hypertarget{sm-model-exact-solution-title}{}
\subsection*{A. Exact solution}

\hypertarget{sm-c-eigenstates}{}\par\medskip

\noindent{\small Table 1 \textbar{} Block eigenstates and eigenenergies of the block Hamiltonian \(\hat h_{\mathbf k}^{\mathrm N}\) in the MStJ model.\par}\nopagebreak[4]
\begingroup\small\setlength{\tabcolsep}{3.5pt}\renewcommand{\arraystretch}{1.30}
\noindent\FittedTable{\begin{tabular}{@{}cccc@{}}\toprule
{\shortstack[c]{Block electron\\ number \(N_{\mathbf k}\)}} & {Block eigenstate \(|N_{\mathbf k,\lambda}\rangle\)} & {Block eigenenergy \(E^{\mathrm N}_{\mathbf k,\lambda}\)} & {Degeneracy} \\
\midrule
{0} & {\(|0_{\mathbf k}\rangle\)} & {\(E^{\mathrm N}_{\mathbf k,0}=0\)} & {1} \\
{1} & {\(|1_{\pm\mathbf k\sigma}\rangle=\hat c^\dagger_{\pm\mathbf k\sigma}|0_{\mathbf k}\rangle\)} & {\(E^{\mathrm N}_{\mathbf k,1}=\xi_{\mathbf k}\)} & {4} \\
{\multirow{3}{*}{2}} & {\(|2_{\mathbf k,s}\rangle=\hat\Delta^\dagger_{\mathbf k}|0_{\mathbf k}\rangle\)} & {\(E^{\mathrm N}_{\mathbf k,s}=2\xi_{\mathbf k}-2\mathcal J_{\mathbf k}\)} & {1} \\
 & {\shortstack[c]{\(|2_{\mathbf k,t_{+1}}\rangle=\hat c^\dagger_{\mathbf k\uparrow}\hat c^\dagger_{-\mathbf k\uparrow}|0_{\mathbf k}\rangle\)\\  \(|2_{\mathbf k,t_0}\rangle=\frac1{\sqrt2}\bigl(\hat c^\dagger_{\mathbf k\uparrow}\hat c^\dagger_{-\mathbf k\downarrow}+\hat c^\dagger_{\mathbf k\downarrow}\hat c^\dagger_{-\mathbf k\uparrow}\bigr)|0_{\mathbf k}\rangle\)\\  \(|2_{\mathbf k,t_{-1}}\rangle=\hat c^\dagger_{\mathbf k\downarrow}\hat c^\dagger_{-\mathbf k\downarrow}|0_{\mathbf k}\rangle\)}} & {\multirow{2}{*}{\(E^{\mathrm N}_{\mathbf k,2}=2\xi_{\mathbf k}\)}} & {\multirow{2}{*}{5}} \\
 & {\(|2_{\pm\mathbf k}\rangle=\hat c^\dagger_{\pm\mathbf k\uparrow}\hat c^\dagger_{\pm\mathbf k\downarrow}|0_{\mathbf k}\rangle\)} &  &  \\
{3} & {\(|3_{\pm\mathbf k\sigma}\rangle=\hat c^\dagger_{\pm\mathbf k\sigma}\hat c^\dagger_{\mp\mathbf k\uparrow}\hat c^\dagger_{\mp\mathbf k\downarrow}|0_{\mathbf k}\rangle\)} & {\(E^{\mathrm N}_{\mathbf k,3}=3\xi_{\mathbf k}-\mathcal J_{\mathbf k}\)} & {4} \\
{4} & {\(|4_{\mathbf k}\rangle=(\hat\Delta^\dagger_{\mathbf k})^2|0_{\mathbf k}\rangle\)} & {\(E^{\mathrm N}_{\mathbf k,4}=4\xi_{\mathbf k}-2\mathcal J_{\mathbf k}\)} & {1} \\
\bottomrule\end{tabular}}\par\endgroup
\smallskip
 \hypertarget{sm-solvable-mstj-title}{} 
The MStJ model is exactly solvable because the Hamiltonian \(\hat H_{\mathrm N}\) can be decomposed into a sum of mutually commuting independent block Hamiltonians,

\hypertarget{sm-eq2}{}%
\FittedEquation{8}{%
\hat H_{\mathrm N}=\sum_{\{\mathbf k,-\mathbf k\}}^{\mathbf k\ne-\mathbf k}\hat h_{\mathbf k}^{\mathrm N}
        +\sum_{\mathbf k_{\mathrm T}}\hat h_{\mathbf k_{\mathrm T}}^{\mathrm N},
}
 
\hypertarget{sm-eq-model-generic-block}{}%
\FittedEquation{9}{%
\hat h_{\mathbf k}^{\mathrm N}=\xi_{\mathbf k}\hat N_{\mathbf k}
        -2\mathcal J_{\mathbf k}\hat\Delta^\dagger_{\mathbf k}\hat\Delta_{\mathbf k},
}
 
\hypertarget{sm-eq-model-trim-pair-block}{}%
\FittedEquation{10}{%
\hat h_{\mathbf k_{\mathrm T}}^{\mathrm N}=\xi_{\mathbf k_{\mathrm T}}\hat n_{\mathbf k_{\mathrm T}}
        -\mathcal J_{\mathbf k_{\mathrm T}}\hat\Delta^\dagger_{\mathbf k_{\mathrm T}}\hat\Delta_{\mathbf k_{\mathrm T}},
}
 
where the first sum counts each generic momentum block \(\{\mathbf k,-\mathbf k\}\) with \(\mathbf k\ne-\mathbf k\) once, and the second sum runs over the time-reversal-invariant momenta \(\mathbf k_{\mathrm T}=-\mathbf k_{\mathrm T}\). Throughout, the relations between crystal momenta are understood modulo a reciprocal-lattice vector. \(\hat N_{\mathbf k}=\hat n_{\mathbf k}+\hat n_{-\mathbf k}\) is the block electron number and is conserved in each momentum block, \([\hat H_{\mathrm N},\hat N_{\mathbf k}]=0\). Thus each block Hamiltonian \(\hat h_{\mathbf k}^{\mathrm N}\) has an independent block \(U_{\mathbf k}(1)\) symmetry. For \(\mathbf k\ne-\mathbf k\), \(\hat h_{\mathbf k}^{\mathrm N}\) acts on a 16-dimensional Fock space and can be diagonalized exactly. \hyperlink{sm-c-eigenstates}{Table 1} lists the 16 block eigenstates, grouped by block electron number, together with their energies and degeneracies. For \(\mathbf k_{\mathrm T}=-\mathbf k_{\mathrm T}\), the block Hamiltonian \(\hat h^{\mathrm N}_{\mathbf k_{\mathrm T}}\) acting on a 4-dimensional Fock space reduces to

\hypertarget{sm-eq-model-trim-block}{}%
\FittedEquation{11}{%
\hat h_{\mathbf k_{\mathrm T}}^{\mathrm N}=\xi_{\mathbf k_{\mathrm T}}\hat n_{\mathbf k_{\mathrm T}}-2\mathcal J_{\mathbf k_{\mathrm T}}\hat n_{\mathbf k_{\mathrm T}\uparrow}\hat n_{\mathbf k_{\mathrm T}\downarrow},
}
 
which has the same operator structure as one term in the HK model \hyperlink{sm-ref-hatsugai-kohmoto-1992}{[11]}, \hyperlink{sm-ref-zhao-2025}{[15]}. On the square lattice, the four time-reversal-invariant momenta are \(\Gamma=(0,0)\), \(X=(\pi,0)\), \(Y=(0,\pi)\), and \(M=(\pi,\pi)\). These finitely many momenta contribute negligibly to bulk quantities per site and therefore require no separate treatment in calculations performed in the thermodynamic limit.

For a generic momentum block with \(\mathbf k\ne-\mathbf k\), the block ground state is

\hypertarget{sm-eq-model-block-ground-state}{}%
\FittedEquation{12}{%
|\mathrm{GS}_{\mathbf k}\rangle=
        \begin{cases}
        |0_{\mathbf k}\rangle, & \xi_{\mathbf k}>\mathcal J_{\mathbf k},\\
        |2_{\mathbf k,s}\rangle, & 0<\xi_{\mathbf k}<\mathcal J_{\mathbf k},\\
        |4_{\mathbf k}\rangle, & \xi_{\mathbf k}<0.
        \end{cases}
}
 
The two contours \(\xi_{\mathbf k}=\mathcal J_{\mathbf k}\) and \(\xi_{\mathbf k}=0\) are level crossings between block ground states whose block electron numbers \(N_{\mathbf k}\) differ by two (\hyperlink{sm-fig-model-full-bz}{Fig. 1}). The ground state of the Hamiltonian \(\hat H_{\mathrm N}\) is

\hypertarget{sm-eq-cooper-g-ground-state}{}%
\FittedEquation{13}{%
\begin{aligned}
|\mathrm{GS}_{\mathrm N}\rangle
&=\prod_{\{\mathbf k,-\mathbf k\}}^{\xi_{\mathbf k}<0}|4_{\mathbf k}\rangle
\prod_{\{\mathbf k',-\mathbf k'\}}^{0<\xi_{\mathbf k'}<\mathcal J_{\mathbf k'}}|2_{\mathbf k',s}\rangle
\prod_{\{\mathbf k'',-\mathbf k''\}}^{\xi_{\mathbf k''}>\mathcal J_{\mathbf k''}}|0_{\mathbf k''}\rangle\\
&=\left[\prod_{\{\mathbf k,-\mathbf k\}}^{\xi_{\mathbf k}<0}(\hat\Delta_{\mathbf k}^\dagger)^2\right]
\left[\prod_{\{\mathbf k',-\mathbf k'\}}^{0<\xi_{\mathbf k'}<\mathcal J_{\mathbf k'}}\hat\Delta_{\mathbf k'}^\dagger\right]
|0\rangle,
\end{aligned}
}
 
where \(|0\rangle=\prod_{\{\mathbf k,-\mathbf k\}}|0_{\mathbf k}\rangle\) is the vacuum, each product counts each indicated momentum block once, and the finitely many time-reversal-invariant momenta are omitted.

\par\medskip
\hypertarget{sm-fig-model-full-bz}{}
\PrintAsset{model-full-bz-Jd-0.3}
{\small\noindent \textbf{Figure 1 \textbar{} Block ground states and Luttinger surface of the MStJ model.} The light red, yellow, and blue regions indicate that the block ground states are \(|4_{\mathbf k}\rangle\), \(|2_{\mathbf k,s}\rangle\), and \(|0_{\mathbf k}\rangle\), respectively. The solid red and cyan contours mark \(\xi_{\mathbf k}=0\) and \(\xi_{\mathbf k}=\mathcal J_{\mathbf k}\), respectively. The green dash-dotted contours mark \(\xi_{\mathbf k}=\mathcal J_{\mathbf k}/2\). Blue and red dots indicate nodes and antinodes, with representative momenta \(\mathbf k_n/\pi=(0.38335,0.38335)\) and \(\mathbf k_a/\pi=(1,0.09749)\), respectively. Black dots mark the time-reversal-invariant momenta \(\Gamma\), \(X\), \(Y\), and \(M\). The chemical potential \(\mu(0~\mathrm K)=-0.309999~\mathrm{eV}\) is determined from the zero-temperature density equation in the thermodynamic limit \(1-p=(2\pi)^{-2}\int_{\mathrm{BZ}}d^2k\,[\Theta(-\xi_{\mathbf k})+\Theta(\mathcal J_{\mathbf k}-\xi_{\mathbf k})]\), where \(\Theta\) is the Heaviside step function. The parameters are \(\mathcal J_d=0.3~\mathrm{eV}\) and \(p=0.15\).\par}
\medskip
 
\hypertarget{sm-model-block-spectrum}{}\hypertarget{sm-model-block-spectrum-title}{}
\section*{II. Nodal Bose metal}

The MStJ model retains spin-singlet pair correlations within individual momentum blocks without establishing superconducting (SC) phase coherence between distinct blocks. Since \([\hat H_{\mathrm N},\hat n_{\mathbf k}]=0\), the Hamiltonian \(\hat H_{\mathrm N}\) conserves the electron number \(\hat n_{\mathbf k}\) at each momentum, which implies \(\langle\hat\Delta_{\mathbf k}\rangle_{\mathrm N}=0\) and \(\langle\hat\Delta^\dagger_{\mathbf k}\hat\Delta_{\mathbf k'}\rangle_{\mathrm N}=0\) for \(\mathbf k'\ne\pm\mathbf k\). Consequently, the eigenvalues of the pair correlation matrix \(C_{\mathbf k\mathbf k'}=\langle\hat\Delta^\dagger_{\mathbf k}\hat\Delta_{\mathbf k'}\rangle_{\mathrm N}\) remain \(O(1)\) as \(N_s\to\infty\), precluding the macroscopic eigenvalue required for off-diagonal long-range order \hyperlink{sm-ref-yang-1962}{[16]}. Thus, the MStJ model realizes a correlated normal state.

\hypertarget{sm-model-one-particle-spectra-title}{}
\subsection*{A. One-particle Green's function}

The one-particle Green's function can capture the electron excitations and correlation effects. Define the one-particle retarded Green's function

\hypertarget{sm-eq-one-particle-retarded-definition}{}%
\FittedEquation{14}{%
G^R_\sigma(\mathbf k,t)\equiv-i\Theta(t)
        \left\langle\left\{\hat c_{\mathbf k\sigma}(t),\hat c^\dagger_{\mathbf k\sigma}(0)\right\}\right\rangle,
}
 
where \(\Theta(t)\) is the Heaviside step function. The exact zero-temperature Green's function for the correlated normal state is

\hypertarget{sm-eq-mstj-green-zero-temperature}{}%
\FittedEquation{15}{%
\begin{aligned}
        G^R_{\mathrm N,\sigma}(\mathbf k,\omega)
        &=\int_{-\infty}^{+\infty}dt\,e^{i\omega t}G^R_{\mathrm N,\sigma}(\mathbf k,t)\\
        &=\begin{cases}
        \displaystyle\frac{1}{\omega+i0^+-\xi_{\mathbf k}},&\xi_{\mathbf k}>\mathcal J_{\mathbf k},\\[6pt]
        \displaystyle\frac12\left(\frac{1}{\omega+i0^+-\xi_{\mathbf k}+2\mathcal J_{\mathbf k}}+\frac{1}{\omega+i0^+-\xi_{\mathbf k}-\mathcal J_{\mathbf k}}\right),&0<\xi_{\mathbf k}<\mathcal J_{\mathbf k},\\[6pt]
        \displaystyle\frac{1}{\omega+i0^+-\xi_{\mathbf k}+\mathcal J_{\mathbf k}},&\xi_{\mathbf k}<0.
        \end{cases}
        \end{aligned}
}
 
For \(\xi_{\mathbf k}>\mathcal J_{\mathbf k}\), only electron addition is allowed, with an energy cost of \(\xi_{\mathbf k}\), corresponding to the pole at \(\omega=\xi_{\mathbf k}\). For \(0<\xi_{\mathbf k}<\mathcal J_{\mathbf k}\), both electron removal and addition are allowed, and each pole carries a spectral weight of \(1/2\). However, there is no zero-energy electron pole, except at the nodes \(\mathbf k_n\), where \(\xi_{\mathbf k_n}=\mathcal J_{\mathbf k_n}=0\). Correspondingly, the RML interaction opens a gap on the Fermi surface \(\xi_{\mathbf k_F}=0\) excluding the nodes, and the one-particle excitation gap, the minimum energy cost of adding or removing one electron at momentum \(\mathbf k\), is

\hypertarget{sm-eq-mstj-one-particle-gap}{}%
\FittedEquation{16}{%
\Delta_{\mathrm{1p}}^{\mathrm N}(\mathbf k)=\mathcal{J}_{\mathbf k}+\min\{|\xi_{\mathbf k}|,|\xi_{\mathbf k}-\mathcal{J}_{\mathbf k}|\},
}
 
which vanishes only at the nodes. The correlated normal state therefore is a correlated nodal metal.

\hypertarget{sm-model-luttinger-surface-title}{}
\subsection*{B. Luttinger surface and Luttinger's theorem violation}

The Green's function for \(0<\xi_{\mathbf k}<\mathcal J_{\mathbf k}\) in \hyperlink{sm-eq-mstj-green-zero-temperature}{Eq. (15)} has a two-pole structure \hyperlink{sm-ref-phillips-2020}{[17]} and can be rewritten as

\hypertarget{sm-eq-mstj-self-energy}{}%
\FittedEquation{17}{%
G^R_{\mathrm N,\sigma}(\mathbf k,\omega)
        =\frac{1}{\omega+i0^+-\xi_{\mathbf k}-\Sigma^R_{\mathrm N,\sigma}(\mathbf k,\omega)},
}
 
\hypertarget{sm-eq-mstj-self-energy-explicit}{}%
\FittedEquation{18}{%
\Sigma^R_{\mathrm N,\sigma}(\mathbf k,\omega)
        =-\mathcal J_{\mathbf k}/2
        +\frac{(3\mathcal J_{\mathbf k}/2)^2}
        {\omega+i0^+-\xi_{\mathbf k}+\mathcal J_{\mathbf k}/2}.
}
 
which has a form analogous to the coherent part of the Green's function in the phenomenological Yang--Rice--Zhang (YRZ) theory of the pseudogap state \hyperlink{sm-ref-yang-rice-zhang-2006}{[18]}, \hyperlink{sm-ref-rice-yang-zhang-2012}{[19]}. The destructive interference between the addition and removal contributions produces a zero of \(G^R_{\mathrm N,\sigma}(\mathbf k_L,0)\) on \(\xi_{\mathbf k_L}=\mathcal J_{\mathbf k_L}/2\), where the zero-frequency self-energy diverges. The surface of zeros of the zero-frequency one-particle Green's function is known as a \emph{Luttinger surface} \hyperlink{sm-ref-dzyaloshinskii-2003}{[20]}. For the MStJ model with \(\mathcal J_{\mathbf k}=\mathcal J_d\phi_{\mathbf k,d}^{2}\), the Luttinger surface is punctured at the nodes, where the self-energy vanishes. A divergence of the self-energy can invalidate the Luttinger--Ward functional derivation of Luttinger's theorem \hyperlink{sm-ref-dave-phillips-kane-2013}{[21]}.

A direct test of Luttinger's theorem is to explicitly evaluate the Luttinger count \(n_L\) and the Luttinger integral \(I_L\). The exact decomposition of the electron density is

\hypertarget{sm-model-luttinger-discussion}{}\hypertarget{sm-eq-mstj-luttinger-integral}{}%
\FittedEquation{19}{%
n_e=n_L+I_L,
}
 
\hypertarget{sm-eq-mstj-luttinger-sign-count}{}%
\FittedEquation{20}{%
\begin{aligned}
        n_L&=\int_{\mathrm{BZ}}\frac{d^2k}{(2\pi)^2}\,
        \sum_\sigma\Theta\!\left[\operatorname{Re}G^R_{\mathrm N,\sigma}(\mathbf k,0)\right]\\[6pt]
        &=2\int_{\mathrm{BZ}}\frac{d^2k}{(2\pi)^2}
        \left[
        \Theta(-\xi_{\mathbf k})+\Theta(\mathcal J_{\mathbf k}-\xi_{\mathbf k})
        -\Theta\!\left(\frac{\mathcal J_{\mathbf k}}2-\xi_{\mathbf k}\right)
        \right],
        \end{aligned}
}
 
\hypertarget{sm-eq-mstj-luttinger-integral-definition}{}%
\FittedEquation{21}{%
\begin{aligned}
        I_L&=-\frac1\pi\operatorname{Im}\int_{\mathrm{BZ}}\frac{d^2k}{(2\pi)^2}
        \int_{-\infty}^{0}d\omega\,\sum_\sigma
        G^R_{\mathrm N,\sigma}(\mathbf k,\omega)
        \frac{\partial\Sigma^R_{\mathrm N,\sigma}(\mathbf k,\omega)}{\partial\omega}\\[6pt]
        &=\int_{\mathrm{BZ}}\frac{d^2k}{(2\pi)^2}
        \left[
        2\Theta\!\left(\frac{\mathcal J_{\mathbf k}}2-\xi_{\mathbf k}\right)
        -\Theta(-\xi_{\mathbf k})-\Theta(\mathcal J_{\mathbf k}-\xi_{\mathbf k})
        \right].
        \end{aligned}
}
 
In the MStJ model, the Luttinger integral has a simple geometric interpretation: \(I_L\) is the area of the region \(0<\xi_{\mathbf k}<\mathcal J_{\mathbf k}/2\) minus that of \(\mathcal J_{\mathbf k}/2<\xi_{\mathbf k}<\mathcal J_{\mathbf k}\), normalized by the Brillouin-zone (BZ) area. The weak-coupling expansion of the Luttinger integral gives

\hypertarget{sm-eq-mstj-luttinger-weak-coupling}{}%
\FittedEquation{22}{%
I_L=\frac{\mathcal J_d^2}{2}\rho_0[\mu_0(p)]
        \frac{d\,\rho_4[\mu_0(p)]}{dp}
        +O(\mathcal J_d^3),
}
 
where \(\rho_0\) and \(\rho_4\) are the noninteracting and \(\phi_d^{4}\)-weighted densities of states per spin, respectively, defined in \hyperlink{sm-appendix-c}{Appendix B.2}. \(\mu_0(p)\) is the noninteracting chemical potential. When the quadratic term controls the weak-coupling expansion, \(I_L\) has the same sign as the slope of \(\rho_4[\mu_0(p)]\) with respect to doping. \hyperlink{sm-eq-mstj-luttinger-weak-coupling}{Eq. (22)} relates the imbalance between the areas on either side of the Luttinger surface within the \(|2_{\mathbf k,s}\rangle\) region to the band structure: doping shifts the chemical potential, whereas the interaction strength changes the extent of the two regions. Cancellation of \(I_L\) alone does not establish Fermi-liquid behavior. The weak-coupling expansion breaks down sufficiently close to the van Hove singularity, where the relevant derivatives diverge. As \(\mathcal J_d\to0^+\), the nontrivial \(I_L=0\) boundary in \hyperlink{sm-fig-model-luttinger-integral}{Fig. 2} approaches \(p_{\mathrm{vH}}\simeq0.35325\), the hole doping corresponding to the noninteracting van Hove singularity. For hole dopings \(p<0.35\) and \(0<\mathcal J_d/t\le1\), \(I_L>0\) signals a violation of Luttinger's theorem in the MStJ model, although the relative deviation is small, with a maximum ratio of \(I_L/n_e\simeq6.86\%\) (\hyperlink{sm-fig-model-luttinger-integral}{Fig. 2}). Therefore the correlated nodal metal is a non-Fermi liquid that lies beyond conventional Landau Fermi-liquid theory.

\par\medskip
\hypertarget{sm-fig-model-luttinger-integral}{}
\PrintAsset{nL-over-ne-p-Jd}
{\small\noindent \textbf{Figure 2 \textbar{} Ratio of the Luttinger count to the electron density in the MStJ model.} \(n_L/n_e\) reaches a minimum of \(0.9314\) at \((p,\mathcal J_d/t)=(0.303,1)\) and a maximum of \(1.1760\) at \((p,\mathcal J_d/t)=(0.536,1)\). The nontrivial \(n_L=n_e\) contour in the inset, across which \(I_L\) changes sign, approaches the noninteracting van Hove singularity at \(p_{\mathrm{vH}}\simeq0.35325\) as \(\mathcal J_d\to0^+\).\par}
\medskip

\hypertarget{sm-model-finite-temperature-spectra-title}{}
\subsection*{C. One-particle spectra: Fermi arc and nodal--antinodal dichotomy}

Momentum- and energy-resolved electron excitations can be probed directly by angle-resolved photoemission spectroscopy (ARPES) \hyperlink{sm-ref-damascelli-2003}{[22]}, \hyperlink{sm-ref-sobota-2021}{[23]}. ARPES detects the occupied one-particle excitation spectrum, and the photoelectron counting probability can be expressed in terms of the one-particle lesser Green's function as

\hypertarget{sm-eq3}{}%
\FittedEquation{23}{%
\Gamma^{(1)}
        =\frac{|g|^2}{i}\iint_{-\frac{\mathcal T}{2}}^{\frac{\mathcal T}{2}}
        dt\,dt'\,e^{i\omega(t-t')}
        G^<_\sigma(\mathbf k;t,t'),
}
 
where \(G^<_\sigma(\mathbf k;t,t')\equiv i\langle c^\dagger_{\mathbf k\sigma}(t')c_{\mathbf k\sigma}(t)\rangle\) describes one-electron removal processes. \(g\) is the photoemission matrix element, and \(\mathcal T\) is the measurement duration. In the long-duration limit, it follows that

\hypertarget{sm-eq4}{}%
\FittedEquation{24}{%
\lim_{\mathcal T\to\infty}\frac{\Gamma^{(1)}}{2\pi\mathcal T|g|^2}
        =f_{\mathrm F}(\omega)A_\sigma(\mathbf k,\omega),
}
 
where \(f_{\mathrm F}(\omega)=(e^{\beta\omega}+1)^{-1}\) is the Fermi function with \(\beta=1/T\), and \(A_\sigma(\mathbf k,\omega)\) is the nonnegative one-particle spectral function defined as \(A_\sigma(\mathbf k,\omega)\equiv-\pi^{-1}\operatorname{Im}G^R_\sigma(\mathbf k,\omega)\). The exact finite-temperature spectral function of the correlated nodal metal contains four branches:

 \hypertarget{sm-eq-mstj-spectral-branches}{} 
\hypertarget{sm-eq5}{}%
\FittedEquation{25}{%
A_{\mathrm N,\sigma}(\mathbf k,\omega)
        =\sum_{\zeta=-2,-1,0,1}W_{\mathrm N,\mathbf k\zeta}\,
        \delta(\omega-\omega_{\mathrm N,\mathbf k\zeta}),
}
 
where \(W_{\mathrm N,\mathbf k\zeta}\) and \(\omega_{\mathrm N,\mathbf k\zeta}=\xi_{\mathbf k}+\zeta \mathcal{J}_{\mathbf k}\) are the one-particle spectral weights and excitation energies, respectively. The correlated nodal metal preserves spin rotation symmetry, and the spectral function satisfies \(A_{\mathrm N}=A_{\mathrm N,\uparrow}=A_{\mathrm N,\downarrow}\) and the sum rule \(\int d\omega\,A_{\mathrm N}(\mathbf k,\omega)=\sum_{\zeta}W_{\mathrm N,\mathbf k\zeta}=1\). The detailed calculations are provided in \hyperlink{sm-c-single-electron}{Appendix D}.
 
\par\medskip
\hypertarget{sm-fig-c11-lorentzian}{}
\PrintAsset{figure1-full-spectral-function-60k}
{\small\noindent \textbf{Figure 3 \textbar{} One-particle excitation spectra of the correlated nodal metal at \(60~\mathrm K\).} (a) One-particle spectral function \(A_{\mathrm N}(\mathbf k,\omega=0)\) at zero energy; blue and red dots mark \(\mathbf k_n\) and \(\mathbf k_a\). The inset uses the coordinates \(k_{\mathrm r,\parallel}=(k_x-k_y)/\sqrt2\) and \(k_{\mathrm r,\perp}=(k_x+k_y-2k_n)/\sqrt2\) in a fixed rotated frame centered at \(\mathbf k_n=(k_n,k_n)\). (b) \(A_{\mathrm N}(\mathbf k_F,\omega)\) along \(\xi_{\mathbf k_F}=0\), with \(\theta=\tan^{-1}[(\pi-k_x)/(\pi-k_y)]\). (c,d) \(A_{\mathrm N}(\mathbf k,\omega)\) and spectral weights \(W_{\mathrm N,\mathbf k\zeta}\) along \(k_x=\pi\). White dotted curves denote \(\xi_{\mathbf k_F}=0\) in (a), \(\omega=\zeta \mathcal{J}_{\mathbf k_F}\) in (b), and \(\omega=\xi_{\mathbf k}+\zeta \mathcal{J}_{\mathbf k}\) in (c). Solid red and cyan vertical lines mark \(\xi_{\mathbf k}=0\) and \(\xi_{\mathbf k}=\mathcal{J}_{\mathbf k}\), and the dash-dotted green vertical line marks \(\xi_{\mathbf k}=\mathcal{J}_{\mathbf k}/2\). (e,f) Ten energy distribution curves (EDCs) along \(k_x=\pi\) over \(0.12\le k_y/\pi\le0.19\), offset by \(100~\mathrm{eV}^{-1}\). Dashed white horizontal lines in (b,c) and black vertical lines in (e,f) denote \(\omega=0\). Spectral poles are broadened by a Lorentzian \(L_{\Gamma_L}(\omega)=\pi^{-1}\Gamma_L/(\omega^2+\Gamma_L^2)\) with half-width at half-maximum (HWHM) \(\Gamma_L=5~\mathrm{meV}\), except in the inset, where \(\Gamma_L=1~\mathrm{meV}\). Main panels (a--c) share one linear colorbar. The parameters are \(\mathcal J_d=0.06~\mathrm{eV}\), \(p=0.15\), and \(\mu(60~\mathrm K)=-0.268201~\mathrm{eV}\).\par}
\medskip

\hypertarget{sm-nodal-normal-arpes}{}\hypertarget{sm-nodal-normal-arpes-title}{} 
\hypertarget{sm-figure1-analysis}{}The zero-energy spectral function in \hyperlink{sm-fig-c11-lorentzian}{Fig. 3(a)} exhibits a \emph{Fermi-arc}-like distribution of spectral weight near the node, with reduced intensity toward the antinode \hyperlink{sm-ref-kanigel-2006}{[24]}, \hyperlink{sm-ref-norman-1998}{[25]}. The arc arises from both the broadening of finite-energy poles and thermally activated spectral weight \hyperlink{sm-ref-miao-2026}{[5]}. Along the Fermi surface \(\xi_{\mathbf k_F}=0\), the one-particle excitation gap is \(\Delta_{\mathrm{1p}}^{\mathrm N}(\mathbf k_F)=\mathcal J_{\mathbf k_F}=\mathcal J_d\phi_{\mathbf k_F,d}^{2}\), which closes quadratically approaching the node and reaches a maximum at the antinodes, reflecting a \emph{nodal--antinodal dichotomy} \hyperlink{sm-ref-hashimoto-2014}{[26]}, \hyperlink{sm-ref-ding-1996}{[27]}, \hyperlink{sm-ref-loeser-1996}{[28]}. In particular, the zero-energy \(\zeta=0\) branch carries appreciable spectral weight only near the node, where \(\mathcal J_{\mathbf k_F}\lesssim T\), which defines a characteristic visible arc range \(|\phi_{\mathbf k_F,d}|\lesssim\sqrt{T/\mathcal J_d}\), whereas the weight is exponentially suppressed away from the node when \(\mathcal J_{\mathbf k_F}\gg T\) (\hyperlink{sm-fig-c11-lorentzian}{Fig. 3(b)}). The smaller broadening \(\Gamma_L\) used in the inset of \hyperlink{sm-fig-c11-lorentzian}{Fig. 3(a)} resolves nearby ridges associated with different spectral branches, whose visibility depends on the broadening and temperature-dependent spectral weights and which do not imply additional zero-temperature Fermi surfaces. As \(T,\Gamma_L\to0\), gapless one-particle excitations remain confined to the nodes.

Along \(\xi_{\mathbf k_F}=0\), where \(|2_{\mathbf k,s}\rangle\) and \(|4_{\mathbf k}\rangle\) are degenerate, their equal thermal occupation probabilities give peaks at \(-2\mathcal J_{\mathbf k_F}\), \(-\mathcal J_{\mathbf k_F}\), and \(+\mathcal J_{\mathbf k_F}\) for \(\mathcal J_{\mathbf k_F}\gg T\) (\hyperlink{sm-fig-c11-lorentzian}{Fig. 3(b)}), with weights approaching \(1/4\), \(1/2\), and \(1/4\), respectively. Along \(k_x=\pi\), \hyperlink{sm-fig-c11-lorentzian}{Figs. 3(c,d)} show how the pole dispersions and spectral weights evolve as the block ground state changes, showing transitions from \(|4_{\mathbf k}\rangle\) to \(|2_{\mathbf k,s}\rangle\) at \(\xi_{\mathbf k}=0\) with \(k_y/\pi=0.132834\) (red), and from \(|2_{\mathbf k,s}\rangle\) to \(|0_{\mathbf k}\rangle\) at \(\xi_{\mathbf k}=\mathcal J_{\mathbf k}\) with \(k_y/\pi=0.167525\) (cyan). The EDCs in \hyperlink{sm-fig-c11-lorentzian}{Figs. 3(e,f)} resolve the corresponding weight transfers at finite temperature. The one-electron addition peak from \(|0_{\mathbf k}\rangle\) and the two equal-weight peaks from \(|2_{\mathbf k,s}\rangle\) combine at the cyan boundary curve, producing a three-peak structure with low-temperature peak energies \(-\mathcal J_{\mathbf k}\), \(+\mathcal J_{\mathbf k}\), and \(+2\mathcal J_{\mathbf k}\) and corresponding weights approaching \(1/4\), \(1/2\), and \(1/4\), respectively. These multipeak spectra (qualitatively reminiscent of the \emph{peak-dip-hump} structure observed in ARPES \hyperlink{sm-ref-dessau-1991}{[29]}, \hyperlink{sm-ref-campuzano-1999}{[30]}, \hyperlink{sm-ref-wei-2008}{[31]}) are a consequence of spin-singlet pair correlations induced by the RML interaction within each momentum block, despite the \emph{single-band} character of the MStJ model.
 
\hypertarget{sm-model-green-function-zeros}{}\hypertarget{sm-model-green-function-zeros-title}{}
\subsection*{D. Gapless Cooper surfaces}

Although the RML interaction opens a one-particle excitation gap except at the nodes in the correlated nodal metal, the two-particle excitation gap for spin-singlet Cooper pairs with zero total momentum is

\hypertarget{sm-eq-two-particle-excitation-gap}{}%
\FittedEquation{26}{%
\Delta_{\mathrm{2p}}^{\mathrm N}(\mathbf k)=2\min\bigl\{|\xi_{\mathbf k}|,|\xi_{\mathbf k}-\mathcal J_{\mathbf k}|\bigr\},
}
 
which vanishes on the two contours \(\mathcal C_{2e}:\xi_{\mathbf k}=0\) and \(\mathcal C_{2h}:\xi_{\mathbf k}=\mathcal J_{\mathbf k}\). For the MStJ model with \(\mathcal J_{\mathbf k}=\mathcal J_d\phi_{\mathbf k,d}^{2}\), the two contours touch quadratically at the nodes and \(\mathcal C_{2e}\) coincides with the Fermi surface \(\mathcal S_F\). At zero temperature, \(\mathcal C_{2e}\) and \(\mathcal C_{2h}\) are precisely the contours of level crossings between block ground states, across which the block electron number changes discontinuously by two. The Cooper-pair operators obey the noncanonical bosonic commutation relation \([\hat\Delta_{\mathbf k}^{\dagger},\hat\Delta_{\mathbf k}]=\hat N_{\mathbf k}/2-1\) and \([\hat\Delta_{\mathbf k}^{\dagger},\hat\Delta_{\mathbf k'}]=[\hat\Delta_{\mathbf k},\hat\Delta_{\mathbf k'}]=0\) for \(\mathbf k'\ne\pm\mathbf k\). Such bilinear fermionic constituents remain subject to the Pauli exclusion principle, which underlies the formation of a Fermi surface, so at most two Cooper pairs can occupy a momentum block. By analogy with the gapless Fermi surface, such a contour is termed a \emph{Cooper surface}, a gapless Cooper-pair Bose surface. The correlated nodal metal therefore is a \emph{nodal Bose metal}.

Define the spin-singlet Cooper pair retarded Green's function

\hypertarget{sm-eq-pair-retarded-definition}{}%
\FittedEquation{27}{%
G^R_\Delta(\mathbf k,t)\equiv-i\Theta(t)
        \bigl\langle\bigl[\hat\Delta_{\mathbf k}(t),\hat\Delta^\dagger_{\mathbf k}(0)\bigr]\bigr\rangle.
}
 
The exact zero-temperature pair Green's function of the nodal Bose metal is

\hypertarget{sm-eq-pair-retarded-zero-temperature}{}%
\FittedEquation{28}{%
\begin{aligned}
        G^R_{\mathrm N,\Delta}(\mathbf k,\Omega)
        &=\int_{-\infty}^{+\infty}dt\,e^{i\Omega t}G^R_{\mathrm N,\Delta}(\mathbf k,t)\\
        &=\begin{cases}
        \displaystyle\frac{1}{\Omega+i0^+-2\xi_{\mathbf k}+2\mathcal J_{\mathbf k}},
        &\xi_{\mathbf k}>\mathcal J_{\mathbf k},\\[8pt]
        \displaystyle\frac{1}{\Omega+i0^+-2\xi_{\mathbf k}}
        -\frac{1}{\Omega+i0^+-2\xi_{\mathbf k}+2\mathcal J_{\mathbf k}},
        &0<\xi_{\mathbf k}<\mathcal J_{\mathbf k},\\[8pt]
        \displaystyle-\frac{1}{\Omega+i0^+-2\xi_{\mathbf k}},
        &\xi_{\mathbf k}<0.
        \end{cases}
        \end{aligned}
}
 
For \(\xi_{\mathbf k}>\mathcal J_{\mathbf k}\), only pair addition is allowed, with an energy cost of \(2\xi_{\mathbf k}-2\mathcal J_{\mathbf k}\), corresponding to the pole at \(\Omega=2\xi_{\mathbf k}-2\mathcal J_{\mathbf k}\) with residue \(+1\). For \(\xi_{\mathbf k}<0\), only pair removal is allowed, with an energy cost of \(-2\xi_{\mathbf k}\), corresponding to the pole at \(\Omega=2\xi_{\mathbf k}\) with residue \(-1\). For \(0<\xi_{\mathbf k}<\mathcal J_{\mathbf k}\), both pair addition and removal are allowed, and the corresponding poles have residues \(+1\) and \(-1\), respectively; cf. the one-particle Green's function in \hyperlink{sm-eq-mstj-green-zero-temperature}{Eq. (15)}. The two Cooper surfaces are thus the momentum-space contours where the pair-addition or pair-removal poles of \(G^R_{\mathrm N,\Delta}(\mathbf k,\Omega)\) soften to zero energy.

\hypertarget{sm-pair-correlation-definition}{}In contrast to the one-particle case, the pair spectral function \(A_\Delta(\mathbf k,\Omega)=-\pi^{-1}\operatorname{Im}G^R_\Delta(\mathbf k,\Omega)\) can have either sign due to the commutator in \hyperlink{sm-eq-pair-retarded-definition}{Eq. (27)}. To characterize the Cooper surfaces, the nonnegative spin-singlet pair-removal spectral function is introduced:

\hypertarget{sm-eq6}{}%
\FittedEquation{29}{%
S_\Delta^<(\mathbf k,\Omega)
        \equiv\int_{-\infty}^{+\infty}\frac{dt}{2\pi}\,e^{i\Omega t}
        \langle\hat\Delta^\dagger_{\mathbf k}(0)\hat\Delta_{\mathbf k}(t)\rangle.
}
 
The Kubo--Martin--Schwinger (KMS) relation \(A_\Delta(\mathbf k,\Omega)=(e^{\beta\Omega}-1)S_\Delta^<(\mathbf k,\Omega)\) implies that \(A_\Delta(\mathbf k,\Omega=0)=0\) wherever \(S_\Delta^<(\mathbf k,\Omega)\) is regular at zero energy. The exact finite-temperature pair-removal spectral function of the nodal Bose metal contains three pole branches:

\hypertarget{sm-eq-pair-removal-finite-temperature}{}%
\FittedEquation{30}{%
S_{\mathrm N,\Delta}^<(\mathbf k,\Omega)
        =\sum_{\eta=0,-1,-2}W^{\Delta}_{\mathrm N,\mathbf k\eta}\,
        \delta(\Omega-\Omega^{\Delta}_{\mathrm N,\mathbf k\eta}),
}
 
\hypertarget{sm-pair-spectrum-definitions}{}where \(W^{\Delta}_{\mathrm N,\mathbf k\eta}\) and \(\Omega^{\Delta}_{\mathrm N,\mathbf k\eta}=2\xi_{\mathbf k}+\eta \mathcal{J}_{\mathbf k}\) are the pair-removal spectral weights and energies, respectively, and the corresponding sum rule is \(\int d\Omega\,S_{\mathrm N,\Delta}^<(\mathbf k,\Omega)=\sum_\eta W^{\Delta}_{\mathrm N,\mathbf k\eta}=\langle\hat\Delta^\dagger_{\mathbf k}\hat\Delta_{\mathbf k}\rangle\). \hypertarget{sm-pair-removal-branches}{}The \(\eta=0\) and \(\eta=-2\) branches correspond to the pair-removal transitions \(|4_{\mathbf k}\rangle\to|2_{\mathbf k,s}\rangle\) and \(|2_{\mathbf k,s}\rangle\to|0_{\mathbf k}\rangle\), and reach zero energy on the Cooper surfaces \(\mathcal C_{2e}\) and \(\mathcal C_{2h}\), respectively. The middle \(\eta=-1\) branch corresponds to the thermally activated pair-removal transitions \(|3_{\pm\mathbf k\sigma}\rangle\to|1_{\mp\mathbf k\sigma}\rangle\), and \(W^{\Delta}_{\mathrm N,\mathbf k,-1}=2p^{\mathrm N}_{\mathbf k,3}=2e^{-\beta E^{\mathrm N}_{\mathbf k,3}}/Z^{\mathrm N}_{\mathbf k}\) vanishes exponentially away from the nodes as \(T\to0\), where \(Z^{\mathrm N}_{\mathbf k}=\operatorname{Tr}e^{-\beta\hat h_{\mathbf k}^{\mathrm N}}\) is the block partition function and \(p^{\mathrm N}_{\mathbf k,3}\) is the thermal occupation probability of each three-electron block eigenstate.

\hypertarget{sm-model-cooper-surface-spectroscopy-title}{}
\subsection*{E. Cooper surface spectroscopy}

Cooper surfaces reflect the nontrivial two-particle correlations induced by the RML interaction and are experimentally detectable by two-photon coincidence ARPES (cARPES) \hyperlink{sm-ref-su-2020}{[32]}, \hyperlink{sm-ref-devereaux-2023}{[33]}. In cARPES measurements, two photoelectrons, each generated by absorption of one photon, are detected in coincidence, and the coincidence counting probability can be expressed in terms of the two-particle lesser Green's function as

\hypertarget{sm-eq7}{}%
\FittedEquation{31}{%
\Gamma^{(2)}=\frac{|g_1g_2|^2}{i^2} \iiiint_{-\frac{\mathcal T}{2}}^{\frac{\mathcal T}{2}}dt_1\,dt_2\,dt'_1\,dt'_2 e^{i\sum_{j=1}^{2}\omega_j(t_j-t'_j)} G^{(2)<}_{\sigma_1\sigma_2}(\mathbf k_1,\mathbf k_2;t_1,t_2,t'_1,t'_2),
}
 
where \(G^{(2)<}_{\sigma_1\sigma_2}\equiv i^2\langle T_{t'}[c_{\mathbf k_1\sigma_1}^\dagger(t'_1)c_{\mathbf k_2\sigma_2}^\dagger(t'_2)]T_t[c_{\mathbf k_2\sigma_2}(t_2)c_{\mathbf k_1\sigma_1}(t_1)]\rangle\) describes two-electron removal processes including two removal orderings. \(T_t\) and \(T_{t'}\) denote fermionic time and anti-time orderings, respectively. In the long-duration limit, we obtain

\hypertarget{sm-eq8}{}%
\FittedEquation{32}{%
\lim_{\mathcal T\to\infty}
        \frac{\Gamma^{(2)}}{(2\pi\mathcal T)^2|g_1g_2|^2}
        =A^{(2)<}_{\sigma_1\sigma_2}(\mathbf k_1,\mathbf k_2;\omega_1,\omega_2),
}
 
where \(A^{(2)<}_{\sigma_1\sigma_2}\) is the nonnegative two-particle removal spectral function and does not generally admit a factorization with independent Fermi functions analogous to \hyperlink{sm-eq4}{Eq. (24)}.

The exact finite-temperature removal spectral function of the nodal Bose metal is

\hypertarget{sm-eq9}{}%
\FittedEquation{33}{%
A^{(2)<}_{\mathrm N,\sigma_1\sigma_2}(\mathbf k_1,\mathbf k_2;\Omega,\omega_{\mathrm r}) =\sum_\gamma W^{(2),\gamma}_{\mathrm N,\mathbf k_1\sigma_1,\mathbf k_2\sigma_2} \delta(\Omega-\Omega^{\gamma}_{\mathrm N,\mathbf k_1\sigma_1,\mathbf k_2\sigma_2}) \delta(\omega_{\mathrm r}-\omega^{\gamma}_{\mathrm N,\mathbf k_1\sigma_1,\mathbf k_2\sigma_2}),
}
 
\hypertarget{sm-carpes-process-definitions}{}where \(\Omega=\omega_1+\omega_2\) and \(\omega_{\mathrm r}=(\omega_1-\omega_2)/2\) are the total and relative energies, respectively. \(\gamma\) labels an ordered two-electron removal process from an initial state to a final state with two fewer electrons via a degenerate intermediate-state subspace, with energies \(E^{\mathrm N}_{i}\), \(E^{\mathrm N}_{f}\), and \(E^{\mathrm N}_{m}\), respectively. \(\Omega^{\gamma}_{\mathrm N,\mathbf k_1\sigma_1,\mathbf k_2\sigma_2}=E^{\mathrm N}_{i}-E^{\mathrm N}_{f}\) is the two-electron removal energy, and \(\omega^{\gamma}_{\mathrm N,\mathbf k_1\sigma_1,\mathbf k_2\sigma_2}=\pm[(E^{\mathrm N}_{i}+E^{\mathrm N}_{f})/2-E^{\mathrm N}_{m}]\) is the intermediate-state energy offset, with the choice of \(\pm\) determined by the removal ordering. \(W^{(2),\gamma}_{\mathrm N,\mathbf k_1\sigma_1,\mathbf k_2\sigma_2}\) is the coincidence spectral weight, and the corresponding sum rule is

\hypertarget{sm-eq10}{}%
\FittedEquation{34}{%
\sum_{\sigma_1,\sigma_2}\int d\Omega\,d\omega_{\mathrm r}\, A^{(2)<}_{\mathrm N,\sigma_1\sigma_2}(\mathbf k_1,\mathbf k_2;\Omega,\omega_{\mathrm r}) =\sum_{\sigma_1,\sigma_2,\gamma}W^{(2),\gamma}_{\mathrm N,\mathbf k_1\sigma_1,\mathbf k_2\sigma_2} =\bigl\langle\hat n_{\mathbf k_1}(\hat n_{\mathbf k_2}-\delta_{\mathbf k_1,\mathbf k_2})\bigr\rangle.
}
 
\hypertarget{sm-carpes-block-factorization}{}When \(\mathbf k_1\) and \(\mathbf k_2\) belong to different momentum blocks, the removal spectral function factorizes as \(A^{(2)<}_{\mathrm N,\sigma_1\sigma_2}=f_{\mathrm F}(\omega_1)A_{\mathrm N}(\mathbf k_1,\omega_1)f_{\mathrm F}(\omega_2)A_{\mathrm N}(\mathbf k_2,\omega_2)\). In the following, the focus is on generic back-to-back momenta, \(\mathbf k_1=-\mathbf k_2=\mathbf k\), which probe nontrivial two-particle correlations within the same block.

\hypertarget{sm-singlet-projection-introduction}{}Since the nodal Bose metal preserves spin rotation symmetry, the spin-singlet component can be extracted from \(A^{(2)<}_{\mathrm N,\sigma_1\sigma_2}\) using the spin-singlet projection \(A^{(2)<}_{\mathrm N,\Delta}=2A^{(2)<}_{\mathrm N,\uparrow\downarrow}-A^{(2)<}_{\mathrm N,\uparrow\uparrow}\). Integration over the relative energy gives

\hypertarget{sm-eq11}{}%
\FittedEquation{35}{%
\begin{aligned}
\overline A^{(2)<}_{\mathrm N,\Delta}(\mathbf k,\Omega)
        &=\int d\omega_{\mathrm r}\,
        A^{(2)<}_{\mathrm N,\Delta}(\mathbf k,-\mathbf k;\Omega,\omega_{\mathrm r})
        \\
        &=S_{\mathrm N,\Delta}^<(\mathbf k,\Omega)
        +\frac32p^{\mathrm N}_{\mathbf k,3}\mathbf1_{\mathcal J_{\mathbf k}>0}\,
        \delta(\Omega-\Omega^{\Delta}_{\mathrm N,\mathbf k,-1}),
\end{aligned}
}
 
\hypertarget{sm-singlet-projection-interpretation}{}where the indicator \(\mathbf1_{\mathcal J_{\mathbf k}>0}\) equals one for \(\mathcal J_{\mathbf k}>0\) and zero at \(\mathcal J_{\mathbf k}=0\). The additional term arises from the enhanced middle-branch weight associated with coherent summation over degenerate intermediate states, but does not shift the spectral poles. Thus, \(\overline A^{(2)<}_{\mathrm N,\Delta}\) for time-ordered two-electron removal approaches \(S_{\mathrm N,\Delta}^<\) for simultaneous pair removal when \(p^{\mathrm N}_{\mathbf k,3}\) is negligible. Detailed calculations are provided in \hyperlink{sm-c-cooper-surface}{Appendix D}.

\par\medskip
\hypertarget{sm-fig-two-particle-overview}{}
\PrintAsset{figure2-spin-resolved-carpes-60k}
{\small\noindent \textbf{Figure 4 \textbar{} Spin-resolved two-particle removal spectra of the nodal Bose metal at \(60~\mathrm K\).} Columns in (a--i) show the equal-spin \(A_{\mathrm N,\uparrow\uparrow}^{(2)<}\), opposite-spin \(A_{\mathrm N,\uparrow\downarrow}^{(2)<}\), and spin-singlet \(A^{(2)<}_{\mathrm N,\Delta}\) channels of the two-particle removal spectral functions for back-to-back momenta \((\mathbf k,-\mathbf k)\). An overbar denotes integration over the relative energy \(\omega_{\mathrm r}=(\omega_1-\omega_2)/2\). (a--c) Resolved spectra at \(\mathbf k_a\); white dashed horizontal and vertical guides mark \(\omega_{\mathrm r}=0\) and \(\Omega=0\), respectively. (d--f) Integrated spectra at \(\Omega=0\); white dotted curves show the Fermi-surface contour \(\xi_{\mathbf k_F}=0\), and blue and red dots mark \(\mathbf k_n\) and \(\mathbf k_a\), respectively. Insets show the corresponding spectra along \(\xi_{\mathbf k_F}=0\) versus \(\theta=\tan^{-1}[(\pi-k_x)/(\pi-k_y)]\), from the node at \(45^\circ\) to the antinode at \(0^\circ\). (g--i) Integrated spectra along \(k_x=\pi\); dotted curves show \(\Omega=2\xi_{\mathbf k}+\eta\mathcal J_{\mathbf k}\). (j,k) \(\overline A^{(2)<}_{\mathrm N,\Delta}\) EDCs along \(k_x=\pi\), offset by \(50~\mathrm{eV}^{-1}\). (l) Cooper surfaces in the Brillouin zone; solid red and cyan lines denote the Cooper surfaces \(\mathcal C_{2e}:\xi_{\mathbf k}=0\) and \(\mathcal C_{2h}:\xi_{\mathbf k}=\mathcal J_{\mathbf k}\), respectively. Dashed white horizontal lines in (d--f) insets and (g--i), and dashed black vertical lines in (j,k) denote \(\Omega=0\). Each of the three rows shares a colorbar: logarithmic in (a--c), spanning \(1\)--\(2100~\mathrm{eV}^{-2}\), and linear in (d--i). Independent broadening in the one-electron energies \(\omega_1\) and \(\omega_2\) is modeled by Lorentzians with HWHM \(\Gamma_L=5~\mathrm{meV}\), giving the joint kernel \(K(x,y)=L_{\Gamma_L}(x/2+y)L_{\Gamma_L}(x/2-y)\), where \(x=\Omega-\Omega^{\gamma}_{\mathrm N}\) and \(y=\omega_{\mathrm r}-\omega^{\gamma}_{\mathrm N}\) are the energy offsets. The tails of the joint kernel produce the diagonal streaks in (a--c). Parameters match \hyperlink{sm-fig-c11-lorentzian}{Fig. 3}: \(\mathcal J_d=0.06~\mathrm{eV}\), \(p=0.15\), and \(\mu(60~\mathrm K)=-0.268201~\mathrm{eV}\).\par}
\medskip

\hypertarget{sm-cooper-surface-carpes}{}\hypertarget{sm-cooper-surface-carpes-title}{} 
\hypertarget{sm-figure2-analysis}{}The central peak of \(A^{(2)<}_{\mathrm N,\uparrow\uparrow}\) in \hyperlink{sm-fig-two-particle-overview}{Fig. 4(a)} arises from the removal of two equal-spin electrons from \(|4_{\mathbf k_a}\rangle\). Since \(\xi_{\mathbf k_a}=0\) at the antinode, the two block eigenstates \(|4_{\mathbf k_a}\rangle\) and \(|2_{\mathbf k_a,s}\rangle\) are degenerate, and both have the same thermal occupation probability, \(p^{\mathrm N}_{\mathbf k_a,4}=p^{\mathrm N}_{\mathbf k_a,s}\simeq0.49998\), at \(T=60~\mathrm K\). The state \(|2_{\mathbf k_a,s}\rangle\) contains two electrons with opposite spins and cannot contribute to the equal-spin channel. The dominant equal-spin two-electron-removal process is \(|4_{\mathbf k_a}\rangle\to|3_{\pm\mathbf k_a\downarrow}\rangle\to|2_{\mathbf k_a,t_{-1}}\rangle\), with the two-electron-removal energy \(\Omega=E^{\mathrm N}_{\mathbf k_a,4}-E^{\mathrm N}_{\mathbf k_a,2}=-2\mathcal J_{\mathbf k_a}\), where \(\mathcal J_{\mathbf k_a}\simeq55~\mathrm{meV}\), leading to the central peak near \(\Omega=-110~\mathrm{meV}\) at \(\omega_{\mathrm r}=0\) in \hyperlink{sm-fig-two-particle-overview}{Fig. 4(a)}.

The five dominant peaks of \(A^{(2)<}_{\mathrm N,\uparrow\downarrow}\) in \hyperlink{sm-fig-two-particle-overview}{Fig. 4(b)} arise from three classes of opposite-spin two-electron-removal processes. The processes \(|2_{\mathbf k_a,s}\rangle\to\bigl\{|1_{\mathbf k_a\uparrow}\rangle,\ |1_{-\mathbf k_a\downarrow}\rangle\bigr\}\to|0_{\mathbf k_a}\rangle\) give the two side peaks near \((\Omega,\omega_{\mathrm r})=(-2\mathcal J_{\mathbf k_a},\pm\mathcal J_{\mathbf k_a})\simeq(-110,\pm55)~\mathrm{meV}\). The symmetry of \(A^{(2)<}_{\mathrm N,\uparrow\downarrow}\) under \(\omega_1\leftrightarrow\omega_2\), or equivalently \(\omega_{\mathrm r}\leftrightarrow-\omega_{\mathrm r}\), ensures that peaks away from \(\omega_{\mathrm r}=0\) always occur in pairs. The processes \(|4_{\mathbf k_a}\rangle\to\bigl\{|3_{\mathbf k_a\downarrow}\rangle,\ |3_{-\mathbf k_a\uparrow}\rangle\bigr\}\to|2_{\mathbf k_a,s}\rangle\) give the other two side peaks near \((\Omega,\omega_{\mathrm r})=(0,\pm\mathcal J_{\mathbf k_a})\simeq(0,\pm55)~\mathrm{meV}\), which resolve a gapless pair-removal transition while the one-particle excitation gap remains finite. The processes \(|4_{\mathbf k_a}\rangle\to\bigl\{|3_{\mathbf k_a\downarrow}\rangle,\ |3_{-\mathbf k_a\uparrow}\rangle\bigr\}\to|2_{\mathbf k_a,t_0}\rangle\) give the central peak near \((\Omega,\omega_{\mathrm r})=(-2\mathcal J_{\mathbf k_a},0)\simeq(-110,0)~\mathrm{meV}\). At the antinode, the integrated spectral weights are \(p^{\mathrm N}_{\mathbf k_a,4}/2\simeq0.24999\) for the central peak and \(p^{\mathrm N}_{\mathbf k_a,4}/4=p^{\mathrm N}_{\mathbf k_a,s}/4\simeq0.12499\) for each of the four side peaks at \(T=60~\mathrm K\).

The spin-singlet projection eliminates the central peak contributed by the spin-triplet final state and yields the four dominant peaks of \(A^{(2)<}_{\mathrm N,\Delta}\) in \hyperlink{sm-fig-two-particle-overview}{Fig. 4(c)} near \((\Omega,\omega_{\mathrm r})=(0,\pm\mathcal J_{\mathbf k_a})\simeq(0,\pm55)~\mathrm{meV}\) and \((-2\mathcal J_{\mathbf k_a},\pm\mathcal J_{\mathbf k_a})\simeq(-110,\pm55)~\mathrm{meV}\), with equal integrated spectral weights of \(p^{\mathrm N}_{\mathbf k_a,4}/2=p^{\mathrm N}_{\mathbf k_a,s}/2\simeq0.24999\).

The equal-spin intensity \(\overline A^{(2)<}_{\mathrm N,\uparrow\uparrow}\) at zero total energy in \hyperlink{sm-fig-two-particle-overview}{Fig. 4(d)} is concentrated near the nodes mainly because equal-spin two-electron excitations from the block ground states are gapless only at the nodes, whereas the opposite-spin and spin-singlet maps in \hyperlink{sm-fig-two-particle-overview}{Fig. 4(e,f)} resolve both \(\mathcal C_{2e}\) and \(\mathcal C_{2h}\) (\hyperlink{sm-fig-two-particle-overview}{Fig. 4(l)}). The two Cooper surfaces merge at the nodes as the RML interaction strength \(\mathcal J_{\mathbf k}\) decreases continuously to zero. The thermally activated middle branch is barely discernible near the antinode because \(p^{\mathrm N}_{\mathbf k,3}\) is extremely small there.

The dominant spectral peak of \(\overline A^{(2)<}_{\mathrm N,\Delta}\) in \hyperlink{sm-fig-two-particle-overview}{Fig. 4(i)} switches abruptly between branches across \(\mathcal C_{2e}\) because the spin-singlet spectral weight transfers from the \(\eta=0\) branch to the \(\eta=-2\) branch as the block ground state changes from \(|4_{\mathbf k}\rangle\) to \(|2_{\mathbf k,s}\rangle\). At the antinode \(\mathbf k_a\in\mathcal C_{2e}\), the equal thermal occupation probabilities of \(|4_{\mathbf k_a}\rangle\) and \(|2_{\mathbf k_a,s}\rangle\) produce two peaks near \(\Omega=0\) and \(\Omega=-2\mathcal J_{\mathbf k_a}\simeq-110~\mathrm{meV}\) in \hyperlink{sm-fig-two-particle-overview}{Fig. 4(j)}. On approaching \(\mathcal C_{2h}\), the dominant spectral peak of \(\overline A^{(2)<}_{\mathrm N,\Delta}\) in \hyperlink{sm-fig-two-particle-overview}{Fig. 4(i)} softens to zero total energy and then loses weight after the block ground state changes from \(|2_{\mathbf k,s}\rangle\) to \(|0_{\mathbf k}\rangle\). The continuation to small positive total energy is a finite-temperature effect.

Therefore, joint ARPES and cARPES measurements can identify the two Cooper surfaces \(\mathcal C_{2e}\) and \(\mathcal C_{2h}\) by tracing the two dominant spin-singlet branches of \(\overline A^{(2)<}_{\mathrm N,\Delta}\) to zero total energy while confirming that one-particle excitations remain gapped away from the nodes.
 
\hypertarget{sm-model-thermodynamic-quantities-title}{}
\subsection*{F. Thermodynamics beyond nodal density of states}

The one-particle density of states (DOS) per spin of the nodal Bose metal is defined as

\hypertarget{sm-eq-mstj-density-of-states}{}%
\FittedEquation{36}{%
\rho_{\mathrm N}(\omega,T)\equiv\int_{\mathrm{BZ}}\frac{d^2\mathbf k}{(2\pi)^2}\,A_{\mathrm N}(\mathbf k,\omega;T).
}
 
Spin rotation symmetry ensures \(\rho_{\mathrm N}=\rho_{\mathrm N,\uparrow}=\rho_{\mathrm N,\downarrow}\), and the sum rule gives the normalization \(\int_{-\infty}^{\infty}d\omega\,\rho_{\mathrm N}(\omega,T)=1\). Substituting the exact finite-temperature spectral function in \hyperlink{sm-eq5}{Eq. (25)} into \hyperlink{sm-eq-mstj-density-of-states}{Eq. (36)} gives

\hypertarget{sm-eq-mstj-dos-spectral-contours}{}%
\FittedEquation{37}{%
\rho_{\mathrm N}(\omega,T)=\int_{\mathrm{BZ}}\frac{d^2\mathbf k}{(2\pi)^2}\,\sum_\zeta W_{\mathrm N,\mathbf k\zeta}(\xi_{\mathbf k},\mathcal J_{\mathbf k};T)\,\delta(\omega-\omega_{\mathrm N,\mathbf k\zeta}).
}
 
Away from the four gapless nodes, low-energy spectral weight is exponentially suppressed at low temperatures. Thus, the dominant contributions to the low-energy DOS come from the regions close to each node. A moving tangent--normal frame is introduced near each node along the smooth Fermi surface \(\mathcal S_F:\xi_{\mathbf k_F}=0\) [\hyperlink{sm-fig-model-normal-observables}{Fig. 5(a)}]. The local coordinates \(k_\perp\) and \(k_\parallel\) denote the signed normal displacement from \(\mathcal S_F\) and the signed arc length along \(\mathcal S_F\) measured from the node, respectively. Performing the local expansion gives \(\xi_{\mathbf k}\simeq v_n k_\perp\) and \(\mathcal J_{\mathbf k}\simeq u_n k_\parallel^2\), where \(v_n=|\boldsymbol\nabla_{\mathbf k}\xi_{\mathbf k}|_{\mathbf k_n}\) and \(u_n=\mathcal J_d\sin^2k_n/2\), with \(\mathbf k_n=(k_n,k_n)\). Introduce the dimensionless variables \(x=\xi_{\mathbf k}/T\simeq v_n k_\perp/T\), \(y=\mathcal J_{\mathbf k}/T\simeq u_n k_\parallel^2/T\) and \(z=\omega/T\). Using the leading-order nodal approximation \(d^2\mathbf k\simeq dk_\parallel\,dk_\perp\), integration over \(k_\perp\) and rescaling of \(k_\parallel\), including contributions from all four nodes, give

\hypertarget{sm-eq-mstj-dos-scaling}{}%
\FittedEquation{38}{%
\rho_{\mathrm N}(\omega,T)=\frac{\sqrt{T}}{\pi^2v_n\sqrt{u_n}}\,\mathcal R_{\mathrm N}\!\left(\frac{\omega}{T}\right)+o(\sqrt{T}),
}
 
where \(\mathcal R_{\mathrm N}(z)=\int_0^\infty dy\,y^{-1/2}\sum_\zeta w_\zeta(z-\zeta y,y)\) is the dimensionless scaling function, \(w_\zeta(x,y)\equiv W_{\mathrm N,\zeta}(\xi_{\mathbf k},\mathcal J_{\mathbf k};T)\) is the dimensionless spectral weight, and the remainder \(o(\sqrt T)\) is defined as \(T\to0\) at fixed \(\omega/T\).

For \(|\omega|\ll T\) in the low-temperature regime, with \(\mathcal R_{\mathrm N}(z)=\mathcal R_{\mathrm N}(-z)\), the small-\(|z|\) expansion \(\mathcal R_{\mathrm N}(z)=\mathcal R_{\mathrm N}(0)+\mathcal R_{\mathrm N}''(0)z^2/2+O(z^4)\) gives

\hypertarget{sm-eq-mstj-dos-small-frequency}{}%
\FittedEquation{39}{%
\rho_{\mathrm N}(\omega,T)=\frac{\mathcal R_{\mathrm N}(0)\sqrt T}{\pi^2v_n\sqrt{u_n}}
        \left[1+\frac{\mathcal R_{\mathrm N}''(0)}{2\mathcal R_{\mathrm N}(0)}
        \left(\frac{\omega}{T}\right)^2+O\!\left(\frac{\omega^4}{T^4}\right)\right]+o(\sqrt T).
}
 
For \(T\ll|\omega|\) in the low-energy regime, the large-\(|\omega|/T\) asymptotic expansion gives, within the leading-order nodal approximation,

\hypertarget{sm-eq-mstj-dos-large-frequency}{}%
\FittedEquation{40}{%
\rho_{\mathrm N}(\omega,T)\simeq\frac{3-1/\sqrt2}{\pi^2v_n\sqrt{u_n}}\sqrt{|\omega|}
        \left[1-\frac{\pi^2}{96}\left(\frac{T}{|\omega|}\right)^2+O\!\left(\frac{T^4}{|\omega|^4}\right)\right].
}
 
Conventional Dirac nodes in two dimensions at half filling have a linear low-energy dispersion and a DOS \(\rho_{\mathrm N}(\omega,0)\propto|\omega|\). In contrast, the zero-energy behavior \(\rho_{\mathrm N}(0,T)\propto\sqrt T\) and the zero-temperature behavior \(\rho_{\mathrm N}(\omega,0)\propto\sqrt{|\omega|}\) in the MStJ model arise from the linear momentum scaling \(\xi_{\mathbf k}\simeq v_nk_\perp\) in the normal direction and the quadratic scaling \(\mathcal J_{\mathbf k}\simeq u_nk_\parallel^2\) in the tangential direction near the nodes, together with the spectral weights of the interacting momentum blocks. The DOS exhibits emergent \emph{particle--hole symmetry} at leading order in the low-energy, low-temperature limit, although the exact DOS over the full energy range is generally asymmetric about the chemical potential (\hyperlink{sm-fig-model-normal-observables}{Fig. 5(b)}). Comparisons between the exact DOS in \hyperlink{sm-eq-mstj-dos-spectral-contours}{Eq. (37)} and the asymptotic expansions in Eqs. \hyperlink{sm-eq-mstj-dos-small-frequency}{(39)} and \hyperlink{sm-eq-mstj-dos-large-frequency}{(40)} are shown in \hyperlink{sm-fig-model-normal-observables}{Fig. 5(b,c)}. Detailed calculations are provided in \hyperlink{sm-appendix-i}{Appendix E}.

\par\medskip
\hypertarget{sm-fig-model-normal-observables}{}
\PrintAsset{mstj-normal-observables}
{\small\noindent \textbf{Figure 5 \textbar{} Nodal density of states and thermodynamics of the nodal Bose metal.} \textbf{a,} Moving tangent--normal frame along Fermi surface \(\mathcal S_F\); the blue arc and red segment define the signed coordinates \(k_\parallel\) and \(k_\perp\), respectively. \textbf{b,} Per-spin DOS \(\rho_{\mathrm N}(\omega,T)\) over the full energy range at \(T=1~\mathrm K\). The inset shows the low-energy window as \(\rho_{\mathrm N}/\sqrt T\) versus \(\omega/T\), compared with the small-\(|\omega|/T\) expansion in \hyperlink{sm-eq-mstj-dos-small-frequency}{Eq. (39)}. The van Hove singular peak is truncated by the plotting window. \textbf{c,} \(\rho_{\mathrm N}(\omega,T)/\rho_{\mathrm N}(\omega,0)\) versus \(T/|\omega|\) at fixed \(\omega=1~\mathrm{meV}\), with \(\rho_{\mathrm N}(\omega,0)\) evaluated from the exact zero-temperature DOS, compared with the large-\(z\) asymptotic expansion in \hyperlink{sm-eq-mstj-dos-large-frequency}{Eq. (40)}. The inset compares the exact \(T=0\) DOS with its leading \(\sqrt{|\omega|}\) asymptote. \textbf{d,} Isothermal charge compressibility \(\kappa_{\mathrm N}(T)\), compared with the fixed-density expansion through \(T^{5/2}\) in \hyperlink{sm-eq-mstj-compressibility-low-temperature}{Eq. (42)}. \textbf{e,} Uniform spin susceptibility \(\chi_s(T)/(g_e\mu_{\mathrm B})^2\), compared with its leading \(\sqrt T\) behavior in \hyperlink{sm-eq-mstj-spin-susceptibility-low-temperature}{Eq. (44)}. \textbf{f,} Specific heat divided by temperature, \(c_{\mathrm N}(T)/T\), compared with \(\gamma_0+\gamma_1\sqrt T\) obtained from \hyperlink{sm-eq-mstj-heat-capacity-low-temperature}{Eq. (45)}. Light-orange solid curves in (b--f) show the exact results; black dashed curves show the analytical behaviors. The parameters are \(\mathcal J_d=0.06~\mathrm{eV}\) and \(p=0.15\).\par}
\medskip

\hypertarget{sm-model-compressibility-analysis}{}The low-temperature thermodynamic responses exhibit behavior beyond that implied by the one-particle nodal DOS. The isothermal charge compressibility per site is the electron density response to chemical potential at fixed \(T\):

\hypertarget{sm-eq-mstj-compressibility}{}%
\FittedEquation{41}{%
\kappa_{\mathrm N}(T,\mu)\equiv\left.\frac{\partial n_e(T,\mu)}{\partial\mu}\right|_T
=\frac1{2T}\int_{\mathrm{BZ}}\frac{d^2\mathbf k}{(2\pi)^2}
\left[\langle\hat N_{\mathbf k}^{\,2}\rangle-\langle\hat N_{\mathbf k}\rangle^2\right].
}
 
The low-temperature expansion at fixed electron density is

\hypertarget{sm-eq-mstj-compressibility-low-temperature}{}%
\FittedEquation{42}{%
\kappa_{\mathrm N}(T)=\kappa_{\mathrm N}(0)+K_{\kappa,2}T^2+K_{\kappa,5/2}T^{5/2}+o(T^{5/2}).
}
 
\hypertarget{sm-eq-mstj-compressibility-zero-temperature}{}In particular, the zero-temperature charge compressibility is \(\kappa_{\mathrm N}(0)=D[\mu(0)]\), where \(D(\mu)=\int_{\mathrm{BZ}}\frac{d^2\mathbf k}{(2\pi)^2}\left[\delta(\xi_{\mathbf k})+\delta(\xi_{\mathbf k}-\mathcal J_{\mathbf k})\right]\) is the DOS associated with the two Cooper surfaces \(\mathcal C_{2e}:\xi_{\mathbf k}=0\) and \(\mathcal C_{2h}:\xi_{\mathbf k}=\mathcal J_{\mathbf k}\). At zero temperature, the block electron number jumps by two across each Cooper surface, which allows the electron density to respond to changes in the chemical potential, yielding a finite charge compressibility even though the zero-energy one-particle DOS vanishes, \(\rho_{\mathrm N}(0,0)=0\). The \(T^2\) and \(T^{5/2}\) terms in \hyperlink{sm-eq-mstj-compressibility-low-temperature}{Eq. (42)} are low-temperature corrections from the regions near the two Cooper surfaces and the four nodes, respectively.
 
\hypertarget{sm-model-spin-susceptibility-analysis}{}The uniform spin susceptibility per site is the magnetization response to the uniform magnetic field \(B\):

\hypertarget{sm-eq-mstj-spin-susceptibility}{}%
\FittedEquation{43}{%
\chi_s(T,\mu)\equiv\left.\frac{\partial m(T,\mu,B)}{\partial B}\right|_{T,\mu;B=0}
=\frac{(g_e\mu_{\mathrm B})^2}{2T}\int_{\mathrm{BZ}}\frac{d^2\mathbf k}{(2\pi)^2}\left.\left[\langle(\hat{\mathbb S}^{z}_{\mathbf k})^2\rangle-\langle\hat{\mathbb S}^{z}_{\mathbf k}\rangle^2\right]\right|_{B=0},
}
 
where \(m\) is the magnetization per site and \(\hat{\mathbb S}^{z}_{\mathbf k}=\hat S^z_{\mathbf k}+\hat S^z_{-\mathbf k}\) is the block spin operator. The leading low-temperature behavior of the spin susceptibility is

\hypertarget{sm-eq-mstj-spin-susceptibility-low-temperature}{}%
\FittedEquation{44}{%
\frac{\chi_s(T)}{(g_e\mu_{\mathrm B})^2}=K_{\chi,1/2}\sqrt T+o(\sqrt T),
}
 
which vanishes at zero temperature. Physically, the low-energy states near the two Cooper surfaces have zero block spin, and spinful excitations remain gapped away from the nodes. The leading \(\sqrt T\) behavior comes only from thermally active spinful states near the nodes.
 
\hypertarget{sm-model-heat-capacity-analysis}{}The specific heat per site at fixed electron density is \(c_{\mathrm N}(T)\equiv\left.d u[T,\mu(T)]/dT\right|_{n_e}\), where \(u(T,\mu)=\frac12\int_{\mathrm{BZ}}\frac{d^2\mathbf k}{(2\pi)^2}\langle\hat h^{\mathrm N}_{\mathbf k}+\mu\hat N_{\mathbf k}\rangle\) is the internal energy per site. The low-temperature behavior of the specific heat is

 \hypertarget{sm-eq-mstj-low-temperature-thermodynamics}{} 
\hypertarget{sm-eq-mstj-heat-capacity-low-temperature}{}%
\FittedEquation{45}{%
c_{\mathrm N}(T)=\gamma_0T+\gamma_1T^{3/2}+o(T^{3/2}),
}
 
where the \(T\) and \(T^{3/2}\) terms are low-temperature Cooper-surface and nodal corrections, respectively. The gapless Cooper surfaces give rise to a finite \(\gamma_0\). Thus, at low temperature, the Cooper surfaces contribute significantly to both the charge compressibility and the specific heat, whereas the spin-singlet Cooper-pair excitations make a negligible contribution to the spin susceptibility. Detailed calculations are provided in \hyperlink{sm-appendix-j}{Appendix F}.
 
A qualitative estimate for a noninteracting system with \(\rho_0(\omega,0)\propto\sqrt{|\omega|}\) would give \(\kappa_{\mathrm N}\propto\sqrt T\) and \(c_{\mathrm N}\propto T^{3/2}\). However, in the nodal Bose metal, the simultaneous zero-temperature limits \(\kappa_{\mathrm N}(0)=D[\mu(0)]\), \(\lim_{T\to0}\chi_s(T)=0\), and \(\lim_{T\to0}c_{\mathrm N}(T)/T\propto D[\mu(0)]\) reveal the thermodynamic contributions of the Cooper surfaces beyond the nodal one-particle DOS and are incompatible with a conventional spin-degenerate Fermi surface.
 
\hypertarget{sm-sc-instability-cnn}{}\hypertarget{sm-sc-instability-cnn-title}{}
\section*{III. SC instability of the nodal Bose metal}

 \hypertarget{sm-sc-instability-zero-q}{} 
To study the SC instability of the nodal Bose metal, the pairing interaction for zero total momentum is incorporated:

\hypertarget{sm-eq12}{}%
\FittedEquation{46}{%
\hat H_{J,\mathbf0}=\sum_{\alpha=s,d}\hat H_{J,\alpha}=-\frac{2J}{N_s}\sum_{\alpha=s,d}\sum_{\mathbf k,\mathbf k'}\phi_{\mathbf k,\alpha}\phi_{\mathbf k',\alpha}\hat\Delta_{\mathbf k}^{\dagger}\hat\Delta_{\mathbf k'},
}
 
\hypertarget{sm-eq-main-channel-pairing-hamiltonian}{}%
\FittedEquation{47}{%
\hat H_{J,\alpha}=-2JN_s\hat\Lambda_\alpha^{\dagger}\hat\Lambda_\alpha,
}
 
\hypertarget{sm-eq13}{}which is the \(\mathbf Q=0\) component of the AFM superexchange interaction in \hyperlink{sm-eq-model-superexchange}{Eq. (1)}, and \(\hat\Lambda_\alpha=N_s^{-1}\sum_{\mathbf k}\phi_{\mathbf k,\alpha}\hat\Delta_{\mathbf k}\) is the collective spin-singlet pair-annihilation operator in the \(\alpha\)-wave channel. The full Hamiltonian becomes

\hypertarget{sm-eq-main-full-hamiltonian}{}%
\FittedEquation{48}{%
\hat H=\hat H_{\mathrm N}+\hat H_{J,\mathbf0},
}
 
where \(\hat H_{J,\mathbf0}\) couples distinct momentum blocks through Cooper-pair scattering. The essential issue is whether the attractive pairing interaction destabilizes the nodal Bose metal towards superconductivity.

\hypertarget{sm-sc-cooper-instability}{}
\subsection*{A. Cooper problem and Cooper instability}

The classic Cooper problem studies the stability of a Fermi surface by adding a single Cooper pair on top of a fixed Fermi sea \hyperlink{sm-ref-cooper-1956}{[34]}. Here Cooper's construction is generalized to investigate the stability of the Cooper surfaces by adding a single Cooper pair on top of the fixed nodal Bose metal background. Consider the single-pair addition state \(|\Psi\rangle=\sum_{\{\mathbf k,-\mathbf k\}}^{\xi_{\mathbf k}>0}a_{\mathbf k}\hat\Delta_{\mathbf k}^{\dagger}|\mathrm{GS}_{\mathrm N}\rangle\), where \(\sum|a_{\mathbf k}|^2=1\) and each momentum block is counted once. Here \(|\mathrm{GS}_{\mathrm N}\rangle\) is the exact nodal Bose metal ground state, and the constraint \(\xi_{\mathbf k}>0\) restricts pair addition to momentum blocks with ground states \(|0_{\mathbf k}\rangle\) or \(|2_{\mathbf k,s}\rangle\), with pair-addition excitation energy \(\epsilon_{\mathbf k}=2\xi_{\mathbf k}-2\mathcal J_{\mathbf k}\Theta(\xi_{\mathbf k}-\mathcal J_{\mathbf k})\). Consider the \(\alpha\)-wave Cooper problem, and restrict the Hamiltonian \(\hat H_{\mathrm N}+\hat H_{J,\alpha}\) to the single-pair addition subspace \(\mathcal H_{N_e+2}^{\alpha}\) with fixed electron number \(N_e+2\), where \(N_e=n_eN_s\). In Cooper's approximation, where the background \(|\mathrm{GS}_{\mathrm N}\rangle\) is held fixed, the stationary Schr\"{o}dinger equation for the added pair is derived exactly in \hyperlink{sm-appendix-cooper-problem}{Appendix G}. In the thermodynamic limit, the condition for a bound state with eigenenergy \(E=-E_{b,\alpha}<0\) is

\hypertarget{sm-eq-main-cooper-binding-condition}{}%
\FittedEquation{49}{%
1=4J\int_{\xi_{\mathbf k}>0}\frac{d^2\mathbf k}{(2\pi)^2}
\frac{\phi_{\mathbf k,\alpha}^{2}}{\epsilon_{\mathbf k}+E_{b,\alpha}},
}
 
where the energy is measured relative to \(E_{\mathrm{GS},\alpha}=\langle\mathrm{GS}_{\mathrm N}|\hat H_{\mathrm N}+\hat H_{J,\alpha}|\mathrm{GS}_{\mathrm N}\rangle\). The pair-addition excitation energy vanishes on \(\mathcal C_{2e}\) for the transition \(|2_{\mathbf k,s}\rangle\to|4_{\mathbf k}\rangle\), and on \(\mathcal C_{2h}\) for \(|0_{\mathbf k}\rangle\to|2_{\mathbf k,s}\rangle\). At weak attraction, the integral in \hyperlink{sm-eq-main-cooper-binding-condition}{Eq. (49)} is controlled by the \(\phi^2\)-weighted DOS associated with the two Cooper surfaces, \hypertarget{sm-eq-main-cooper-weighted-dos}{}\(D_\alpha(\mu)=\int_{\mathrm{BZ}}\frac{d^2\mathbf k}{(2\pi)^2}\phi_{\mathbf k,\alpha}^{2}\left[\delta(\xi_{\mathbf k})+\delta(\xi_{\mathbf k}-\mathcal J_{\mathbf k})\right]\), and diverges logarithmically as \(E_{b,\alpha}\to0^+\) for \(D_\alpha(\mu)>0\), so a bound state exists for every \(J>0\). To leading logarithmic accuracy, the binding energy is

\hypertarget{sm-eq-main-cooper-binding-energy}{}%
\FittedEquation{50}{%
E_{b,\alpha}\simeq2\Lambda_{\mathrm{cut}}
\exp\!\left[-\frac{1}{2JD_\alpha(\mu)}\right],
}
 
where \(\Lambda_{\mathrm{cut}}\) is a fixed low-energy cutoff. The asymptotic analytical solution in \hyperlink{sm-eq-main-cooper-binding-energy}{Eq. (50)} is compared with the exact numerical solutions of \hyperlink{sm-eq-main-cooper-binding-condition}{Eq. (49)} in \hyperlink{sm-appendix-cooper-problem}{Appendix G}. Thus the nodal Bose metal exhibits a Cooper instability even though the one-particle DOS vanishes at zero energy. The Cooper-surface \(\phi^2\)-weighted DOS \(D_\alpha(\mu)\) takes the role played by the Fermi-surface DOS in the conventional Cooper problem. For the parameters of \hyperlink{sm-fig-model-normal-observables}{Fig. 5}, \(D_s[\mu(0)]=0.043236~\mathrm{eV}^{-1}\) and \(D_d[\mu(0)]=0.699875~\mathrm{eV}^{-1}\), indicating a leading \(d\)-wave Cooper instability for equal, sufficiently weak attractive interactions. However, the existence of a bound state establishes an instability only within the fixed-background approximation; determining the SC instability towards macroscopic SC order requires a many-body treatment of the full interacting Hamiltonian \(\hat H\) \hyperlink{sm-ref-bcs-theory-1957}{[35]}.

\hypertarget{sm-sc-pair-susceptibility}{}
\subsection*{B. SC instability and pair susceptibility}

The SC instability is determined by the full pair susceptibility of \(\hat H\),

\hypertarget{sm-eq-main-full-pair-susceptibility}{}%
\FittedEquation{51}{%
\chi_{\alpha\alpha'}(i\Omega_m)\equiv N_s\int_0^\beta d\tau\,e^{i\Omega_m\tau}
\langle T_\tau\hat\Lambda_\alpha(\tau)\hat\Lambda_{\alpha'}^{\dagger}(0)\rangle_c,
}
 
where \(\Omega_m=2\pi m/\beta\) is a bosonic Matsubara frequency and the subscript \(c\) denotes connected correlations. On the stable normal branch, the pair susceptibility matrix satisfies the Dyson equation

\hypertarget{sm-eq-main-pair-dyson}{}%
\FittedEquation{52}{%
\boldsymbol\chi(i\Omega_m)=\boldsymbol\chi_{\mathrm N}(i\Omega_m)
+2J\boldsymbol\chi_{\mathrm N}(i\Omega_m)\boldsymbol\chi(i\Omega_m)
=\left[\mathbf I-2J\boldsymbol\chi_{\mathrm N}(i\Omega_m)\right]^{-1}\boldsymbol\chi_{\mathrm N}(i\Omega_m),
}
 
where \(\boldsymbol\chi_{\mathrm N}\) is the exact pair susceptibility of \(\hat H_{\mathrm N}\) retaining the RML interaction. The block independence of \(\hat H_{\mathrm N}\) makes the derivation within the Gaussian approximation exact in the thermodynamic limit, while higher-order connected cumulants give only \(O(N_s^{-1})\) corrections. The exact generating-functional derivation of the Dyson equation is given in \hyperlink{sm-appendix-h-dyson-equation}{Appendix H.1}. The SC instability occurs when a static eigenvalue of \(\mathbf I-2J\boldsymbol\chi_{\mathrm N}(i\Omega_m=0,T)\) vanishes:

\hypertarget{sm-eq19}{}%
\FittedEquation{53}{%
\det\!\left[\mathbf I-2J\boldsymbol\chi_{\mathrm N}(0,T_c)\right]=0,
}
 
which signals a soft collective pairing mode at \(T_c\). The soft mode concerns collective coherence between distinct momentum blocks that already possess spin-singlet pair correlations within each block.

The pair susceptibility of the nodal Bose metal can be evaluated exactly without invoking Wick's theorem (\hyperlink{sm-appendix-h-nodal-pair-susceptibility}{Appendix H.2}):

\hypertarget{sm-eq14}{}%
\FittedEquation{54}{%
\chi_{\mathrm N,\alpha\alpha'}(i\Omega_m)
=\frac2{N_s}\sum_{\mathbf k}
\phi_{\mathbf k,\alpha}\phi_{\mathbf k,\alpha'}\Pi_{\mathrm N,\mathbf k}(i\Omega_m),
}
 
where \(\Pi_{\mathrm N,\mathbf k}(i\Omega_m)\) is the spin-singlet Cooper-pair Matsubara Green's function:

\hypertarget{sm-eq15}{}%
\FittedEquation{55}{%
\begin{aligned}
\Pi_{\mathrm N,\mathbf k}(i\Omega_m)
&=\int_0^\beta d\tau\,e^{i\Omega_m\tau}
\langle T_\tau\hat\Delta_{\mathbf k}(\tau)\hat\Delta_{\mathbf k}^{\dagger}(0)\rangle_{\mathrm N}\\
&=\frac{p^{\mathrm N}_{\mathbf k,s}-p^{\mathrm N}_{\mathbf k,4}}{2\xi_{\mathbf k}-i\Omega_m}
+\frac{2(p^{\mathrm N}_{\mathbf k,1}-p^{\mathrm N}_{\mathbf k,3})}{2\xi_{\mathbf k}-\mathcal J_{\mathbf k}-i\Omega_m}
+\frac{p^{\mathrm N}_{\mathbf k,0}-p^{\mathrm N}_{\mathbf k,s}}{2\xi_{\mathbf k}-2\mathcal J_{\mathbf k}-i\Omega_m}.
\end{aligned}
}
 
Here the subscript \(\mathrm N\) denotes thermal averaging with respect to \(\hat H_{\mathrm N}\), and the three terms correspond to the transitions \(|4_{\mathbf k}\rangle\leftrightarrow|2_{\mathbf k,s}\rangle\), \(|3_{\pm\mathbf k\sigma}\rangle\leftrightarrow|1_{\mp\mathbf k\sigma}\rangle\), and \(|2_{\mathbf k,s}\rangle\leftrightarrow|0_{\mathbf k}\rangle\), respectively. \hypertarget{sm-eq-cooper-g-zero-temperature-kernel}{}The KMS relation allows the static pair Matsubara Green's function to be expressed in terms of the pair-removal spectral function in \hyperlink{sm-eq6}{Eq. (29)} as 
\hypertarget{sm-eq17}{}%
\FittedEquation{56}{%
\Pi_{\mathrm N,\mathbf k}(0)=\int_{-\infty}^{\infty}d\Omega\,
\frac{e^{\beta\Omega}-1}{\Omega}\,S_{\mathrm N,\Delta}^<(\mathbf k,\Omega),
}
 which relates the pairing susceptibility to the gapless pair excitations on the Cooper surfaces.

For the zero-total-momentum pairing interaction, the pairing susceptibility matrix is diagonal, \(\chi_{\alpha\alpha'}=\delta_{\alpha\alpha'}\chi_{\alpha\alpha}\), even though both extended-\(s\)- and \(d\)-wave channels are retained in \(\hat H_{J,\mathbf0}\). On the square lattice, exchanging \(k_x\) and \(k_y\) leaves \(\Pi_{\mathrm N,\mathbf k}\) invariant but reverses the sign of \(\phi_{\mathbf k,s}\phi_{\mathbf k,d}\), yielding \(\chi_{\mathrm N,sd}=\chi_{\mathrm N,ds}=0\) and \(\chi_{sd}=\chi_{ds}=0\), so the linear pairing responses decouple. The SC instability temperature in the \(\alpha\)-wave channel satisfies \(2J\chi_{\mathrm N,\alpha\alpha}(0,T_{c,\alpha})=1\), and the higher \(T_{c,\alpha}\) selects the leading \(\alpha\)-wave instability.

At low temperature, the singular behavior of the static pair Matsubara Green's function near a regular segment of each Cooper surface is

\hypertarget{sm-eq20}{}%
\FittedEquation{57}{%
\Pi_{\mathrm N,\mathbf k}^{\mathrm{singular}}(0,T)
\simeq\frac{\tanh(S_{\mathbf k,r}/T)}{2S_{\mathbf k,r}}
\longrightarrow
\begin{cases}
\dfrac{1}{2T},&|S_{\mathbf k,r}|\ll T\ll\mathcal J_{\mathbf k},\\[6pt]
\dfrac{1}{2|S_{\mathbf k,r}|},&T\ll|S_{\mathbf k,r}|\ll\mathcal J_{\mathbf k},
\end{cases}
}
 
where \(S_{\mathbf k,2e}=\xi_{\mathbf k}\), \(S_{\mathbf k,2h}=\xi_{\mathbf k}-\mathcal J_{\mathbf k}\), and \(r=2e,2h\). The singular contribution to the static pair susceptibility coming from tubular neighborhoods of the two Cooper surfaces is given by (\hyperlink{sm-appendix-h-nodal-pair-susceptibility}{Appendix H.2})

\hypertarget{sm-eq-cooper-log-contour-integral}{}%
\FittedEquation{58}{%
\chi_{\mathrm N,\alpha\alpha}^{\mathrm{singular}}(0,T)
=2D_\alpha(\mu)\ln\!\left(\frac{4e^{\gamma_{\mathrm E}}\Lambda_{\mathrm{cut}}}{\pi T}\right)+o(1),
}
 
\hypertarget{sm-eq-cooper-log-sum}{}where \(\gamma_{\mathrm E}\) is Euler's constant and \(\Lambda_{\mathrm{cut}}\) is a fixed low-energy cutoff. Note that the logarithmic coefficient is determined by the same Cooper-surface \(\phi^2\)-weighted DOS \(D_\alpha(\mu)\) introduced in the single-pair-addition Cooper problem. To leading logarithmic accuracy, the SC instability temperature is

\hypertarget{sm-eq22}{}%
\FittedEquation{59}{%
T_{c,\alpha}\simeq\frac{4e^{\gamma_{\mathrm E}}}{\pi}\Lambda_{\mathrm{cut}}
\exp\!\left[-\frac{1}{4JD_\alpha(\mu)}\right].
}
 
\hypertarget{sm-eq-cooper-g-thermal-logarithm}{}\hypertarget{sm-eq-cooper-g-transition-scale}{}The logarithmic divergence guarantees an SC instability of the nodal Bose metal at arbitrarily weak attraction in the \(\alpha\)-wave channel with \(D_\alpha(\mu)>0\), despite the vanishing one-particle DOS at zero energy. For the parameters of \hyperlink{sm-fig-model-normal-observables}{Fig. 5}, \(D_s[\mu(0)]=0.043236~\mathrm{eV}^{-1}\) and \(D_d[\mu(0)]=0.699875~\mathrm{eV}^{-1}\), so the leading SC instability is \(d\)-wave for equal, sufficiently weak attraction.

At finite attraction, \(T_{c,\alpha}\) is obtained by numerically solving \(2J\chi_{\mathrm N,\alpha\alpha}(0,T_{c,\alpha})=1\) under the fixed-density constraint. The doping dependence of \(T_{c,d}\), together with \(D_d[\mu(0)]\), is shown in \hyperlink{sm-fig-h-sc-doping}{Fig. 6}. The two van Hove singularities occur when the Cooper surfaces \(\mathcal C_{2e}\) and \(\mathcal C_{2h}\) cross the van Hove saddle points at \(p_{\mathrm{vH},2e}\simeq0.311\) and \(p_{\mathrm{vH},2h}\simeq0.443\), respectively, and both critical dopings approach the noninteracting van Hove value \(p_{\mathrm{vH}}\simeq0.35325\) as \(\mathcal J_d\to0^+\). The enhancement of the \(\phi^2\)-weighted DOS near van Hove singularities is accompanied by higher SC instability temperatures, most clearly resolved at weaker attraction. The Cooper surfaces thus play a role in the SC instability of the nodal Bose metal analogous to that of the Fermi surface in the pairing instability of a Landau Fermi liquid in conventional BCS theory. Moreover, Cooper surfaces can similarly undergo doping-driven Lifshitz transitions upon crossing van Hove saddle points \hyperlink{sm-ref-miao-2026}{[5]}.

\par\medskip
\hypertarget{sm-fig-h-sc-doping}{}
\PrintAsset{sc-instability-doping}
{\small\noindent \textbf{Figure 6 \textbar{} Doping dependence of the \(d\)-wave SC instability temperature and Cooper-surface \(\phi^2\)-weighted DOS.} Colored solid curves show the \(d\)-wave SC instability temperature \(T_{c,d}\) (left logarithmic axis), obtained numerically from \hyperlink{sm-eq19}{Eq. (53)} under the fixed-density constraint. The black dotted curve shows \(D_d[\mu(0)]\) (right axis), the zero-temperature Cooper-surface \(\phi^2\)-weighted DOS. The logarithmic van Hove divergences at \(p_{\mathrm{vH},2e}\simeq0.311\) and \(p_{\mathrm{vH},2h}\simeq0.443\) are truncated by the plotting window.\par}
\medskip
 
\hypertarget{sm-correlated-nodal-sc}{}\hypertarget{sm-correlated-nodal-sc-title}{}
\section*{IV. Correlated nodal SC state}

\hypertarget{sm-sc-mean-field}{}
\subsection*{A. Mean-field theory}

\hypertarget{sm-sc-eigenstates-summary}{}\hypertarget{sm-sc-block-eigenstates}{}\par\medskip

\noindent{\small Table 2 \textbar{} Block eigenstates and eigenenergies of the SC block Hamiltonian \(\hat h_{\mathbf k}^{\mathrm{SC}}\).\par}\nopagebreak[4]
\begingroup\small\setlength{\tabcolsep}{3.5pt}\renewcommand{\arraystretch}{1.30}
\noindent\FittedTable{\begin{tabular}{@{}cccc@{}}\toprule
{\shortstack[c]{Block fermion\\ parity \(\pi_{\mathbf k,\lambda}\)}} & {Block eigenstate \(|\pi_{\mathbf k,\lambda}\rangle\)} & {Block eigenenergy \(E^{\mathrm{SC}}_{\mathbf k,\lambda}\)} & {Degeneracy} \\
\midrule
{\(+1\)} & {\(\begin{aligned}
              &|e_{\mathbf k,l}\rangle=x_{\mathbf k,l}|0_{\mathbf k}\rangle+y_{\mathbf k,l}|2_{\mathbf k,s}\rangle+z_{\mathbf k,l}|4_{\mathbf k}\rangle\\
              &|x_{\mathbf k,l}|^2+|y_{\mathbf k,l}|^2+|z_{\mathbf k,l}|^2=1,\quad l=0,1,2
            \end{aligned}\)} & {\(E^{\mathrm{SC},e}_{\mathbf k,0}\le E^{\mathrm{SC},e}_{\mathbf k,1}\le E^{\mathrm{SC},e}_{\mathbf k,2}\)} & {1 each} \\
{\(+1\)} & {\(|2_{\mathbf k,t_1}\rangle,\ |2_{\mathbf k,t_0}\rangle,\ |2_{\mathbf k,t_{-1}}\rangle,\ |2_{\pm\mathbf k}\rangle\)} & {\(2\xi_{\mathbf k}\)} & {5} \\
{\(-1\)} & {\(\begin{aligned}
              &|o_{\pm\mathbf k\sigma,\nu}\rangle=u_{\mathbf k,\nu}|1_{\pm\mathbf k\sigma}\rangle+v_{\mathbf k,\nu}|3_{\mp\mathbf k\sigma}\rangle\\
              &|u_{\mathbf k,\nu}|^2+|v_{\mathbf k,\nu}|^2=1,\quad\nu=\pm1
            \end{aligned}\)} & {\(\begin{gathered}
              E^{\mathrm{SC},o}_{\mathbf k,\nu}=2\xi_{\mathbf k}-\mathcal J_{\mathbf k}/2\\
              {}+\nu\sqrt{(\xi_{\mathbf k}-\mathcal J_{\mathbf k}/2)^2+2|\Delta_{\mathbf k,d}|^2}
            \end{gathered}\)} & {4 each} \\
\bottomrule\end{tabular}}\par\endgroup
\smallskip

{\small The explicit forms of the normalized coefficients and eigenenergies are given in \hyperlink{sm-sc-block-diagonalization}{Appendix J.1}.\par}
 
Mean-field theory is employed to treat the pairing interaction in \hyperlink{sm-eq12}{Eq. (46)}. In the SC state, introduce the order parameter \(\Delta_d=2J\langle\hat\Lambda_d\rangle\) and decouple the pairing interaction in the leading \(d\)-wave channel:

\hypertarget{sm-eq-sc-collective-decoupling}{}%
\FittedEquation{60}{%
\hat H_{J,\mathbf0}\simeq-N_s\bigl(\Delta_d\hat\Lambda_d^{\dagger}+\Delta_d^{*}\hat\Lambda_d\bigr)+\frac{N_s|\Delta_d|^2}{2J}.
}
 
The mean-field SC Hamiltonian is given by

\hypertarget{sm-eq-sc-hamiltonian}{}%
\FittedEquation{61}{%
\hat H_{\mathrm{SC}}=\sum_{\{\mathbf k,-\mathbf k\}}\hat h_{\mathbf k}^{\mathrm{SC}}+\frac{N_s|\Delta_d|^2}{2J},
}
 
\hypertarget{sm-eq-sc-block-hamiltonian}{}%
\FittedEquation{62}{%
\hat h_{\mathbf k}^{\mathrm{SC}}=\xi_{\mathbf k}\hat N_{\mathbf k}-2\mathcal J_{\mathbf k}\hat\Delta_{\mathbf k}^{\dagger}\hat\Delta_{\mathbf k}-2(\Delta_{\mathbf k,d}\hat\Delta_{\mathbf k}^{\dagger}+\Delta_{\mathbf k,d}^{*}\hat\Delta_{\mathbf k}),
}
 
where \(\Delta_{\mathbf k,d}=\Delta_d\phi_{\mathbf k,d}\) is the SC pairing field. Note that only the pairing interaction is decoupled, whereas the RML interaction is retained exactly. The SC pairing field breaks the block \(U_{\mathbf k}(1)\) symmetry and mixes block states whose block electron numbers differ by two, while preserving the block fermion parity \(\hat\pi_{\mathbf k}=(-1)^{\hat N_{\mathbf k}}\), with \([\hat H_{\mathrm{SC}},\hat\pi_{\mathbf k}]=0\). Thus each SC block Hamiltonian \(\hat h_{\mathbf k}^{\mathrm{SC}}\) has an independent block \(\mathbb Z^{\mathrm f}_{2,\mathbf k}\) symmetry. \(\hat h_{\mathbf k}^{\mathrm{SC}}\) still acts on the same 16-dimensional Fock space and can be diagonalized exactly. \hyperlink{sm-sc-block-eigenstates}{Table 2} lists the 16 SC block eigenstates \(|\pi_{\mathbf k,\lambda}\rangle\), together with their eigenenergies \(E^{\mathrm{SC}}_{\mathbf k,\lambda}\), fermion parities \(\pi_{\mathbf k,\lambda}\) and degeneracies. The ground state of \(\hat H_{\mathrm{SC}}\) is

\hypertarget{sm-eq-sc-ground-state}{}%
\FittedEquation{63}{%
\begin{aligned}
|\mathrm{GS}_{\mathrm{SC}}\rangle=\prod_{\{\mathbf k,-\mathbf k\}}|e_{\mathbf k,0}\rangle&=\prod_{\{\mathbf k,-\mathbf k\}}\bigl[x_{\mathbf k,0}|0_{\mathbf k}\rangle+y_{\mathbf k,0}|2_{\mathbf k,s}\rangle+z_{\mathbf k,0}|4_{\mathbf k}\rangle\bigr]\\
&=\prod_{\{\mathbf k,-\mathbf k\}}\bigl[x_{\mathbf k,0}+y_{\mathbf k,0}\hat\Delta_{\mathbf k}^{\dagger}+z_{\mathbf k,0}(\hat\Delta_{\mathbf k}^{\dagger})^2\bigr]|0\rangle,
\end{aligned}
}
 
which has the same form as the variational wavefunction used by Phillips \emph{et al.} \hyperlink{sm-ref-phillips-2020}{[17]} to characterize the ground state of the HK model with a pairing interaction. For the conventional BCS variational wavefunction \(|\mathrm{BCS}\rangle=\prod_{\mathbf k}(u_{\mathbf k}+v_{\mathbf k}c_{\mathbf k\uparrow}^{\dagger}c_{-\mathbf k\downarrow}^{\dagger})|0\rangle\), the corresponding normalized coefficients satisfy \(y_{\mathbf k,0}^{2}=2x_{\mathbf k,0}z_{\mathbf k,0}\). The order parameter is determined by the gap equation under the fixed-density constraint

\hypertarget{sm-eq-sc-self-consistency}{}%
\FittedEquation{64}{%
\Delta_d=\frac{4J}{N_s}\sum_{\{\mathbf k,-\mathbf k\}}\phi_{\mathbf k,d}\langle\hat\Delta_{\mathbf k}\rangle_{\mathrm{SC}},
}
 
\hypertarget{sm-eq-sc-fixed-density}{}%
\FittedEquation{65}{%
n_e=\frac1{N_s}\sum_{\{\mathbf k,-\mathbf k\}}\langle\hat N_{\mathbf k}\rangle_{\mathrm{SC}},
}
 
which form the coupled self-consistent equations. The expectation values are given in \hyperlink{sm-eq-sc-block-averages}{Eqs. (J20)--(J21)}. The solution with a finite order parameter corresponds to a correlated \(d\)-wave SC state. With the global phase gauge chosen so that \(\Delta_d\) is real and nonnegative, \hyperlink{sm-fig-sc-doping}{Fig. 7} shows the temperature, doping and interaction dependence of the order parameter \(\Delta_d(T,p)\), the SC transition temperature \(T_{c,d}\) and the order-parameter-to-\(T_c\) ratio \(2\sqrt2\Delta_d(0)/T_{c,d}\), where \(\sqrt2\) originates from the normalization factor of the pair operator \(\hat\Delta_{\mathbf k}\).

\par\medskip
\hypertarget{sm-fig-sc-doping}{}
\PrintAsset{figure3-doping-dependence}
{\small\noindent \textbf{Figure 7 \textbar{} Temperature, doping and interaction dependence of the correlated \(d\)-wave SC state.} \textbf{a,b,} Order parameter \(\Delta_d(T,p)\) at \(\mathcal J_d=0.06~\mathrm{eV}\), with \(J=0.06~\mathrm{eV}\) in \textbf{a} and \(J=0.12~\mathrm{eV}\) in \textbf{b}. White dashed curves mark \(T_{c,d}(p)\), and white dots mark their maxima. \textbf{c,} SC transition temperature \(T_{c,d}\) as a function of \(J\) and \(\mathcal J_d\) at \(p=0.15\). \textbf{d--f,} Order-parameter-to-\(T_c\) ratio \(2\sqrt2\Delta_d(0)/T_{c,d}\): dependence on \(p\) and \(J\) at fixed \(\mathcal J_d=0.06~\mathrm{eV}\) in \textbf{d}; on \(p\) and \(\mathcal J_d\) at fixed \(J=0.12~\mathrm{eV}\) in \textbf{e}; and on both interaction strengths at \(p=0.15\) in \textbf{f}. The insets enlarge \(0.15\le p\le0.35\) in \textbf{e} and \(0.36\le J/t\le0.4\) in \textbf{f}, both over \(0\le\mathcal J_d/t\le0.02\). White contours give \(T_{c,d}\) in \textbf{c} and the ratio in \textbf{d--f}.\par}
\medskip

As in conventional BCS theory, the order parameter vanishes continuously at the SC transition temperature \(T_{c,d}\) shown in \hyperlink{sm-fig-sc-doping}{Fig. 7(a,b)}. Below \(T_{c,d}\), the pairing interaction establishes global phase coherence among momentum blocks that already possess spin-singlet pair correlations within each block. The weak-attraction estimate gives \(T_{c,d}\propto e^{-1/[4JD_d(\mu)]}\) (see \hyperlink{sm-sc-weak-gap-ratio}{Appendix J.2}), where \(D_d(\mu)\) is the Cooper-surface \(\phi^2\)-weighted DOS rather than the Fermi-surface DOS in conventional BCS theory. Thus, the two peaks at smaller \(J\) [\hyperlink{sm-fig-sc-doping}{Fig. 7(a)}] reflect the enhancement of \(D_d(\mu)\) by the two van Hove singularities near \(p_{\mathrm{vH},2e}\simeq0.311\) and \(p_{\mathrm{vH},2h}\simeq0.443\) (cf. the \(d\)-wave SC instability temperature in \hyperlink{sm-fig-h-sc-doping}{Fig. 6}). Increasing \(J\) from \(0.06\) to \(0.12~\mathrm{eV}\) raises the maximum \(T_{c,d}\) from about \(80\) to \(268~\mathrm K\), consistent with the weak-attraction exponential dependence on \(JD_d(\mu)\), and merges the two peaks into a broad maximum [\hyperlink{sm-fig-sc-doping}{Fig. 7(b)}]. The doping dependence in \hyperlink{sm-fig-sc-doping}{Fig. 7(a,b)} essentially reflects the geometry and form-factor weights of the two Cooper surfaces.

The two interactions play distinct roles in establishing SC order: the pairing attraction \(J\) directly drives phase coherence among momentum blocks, whereas the RML interaction \(\mathcal J_d\) alters the Cooper-surface geometry and thereby modifies the \(\phi^2\)-weighted DOS \(D_\alpha(\mu)\). For the parameter range considered here, \hyperlink{sm-fig-sc-doping}{Fig. 7(c)} shows that the SC transition temperature \(T_{c,d}\) rises steeply with increasing \(J\), whereas larger \(\mathcal J_d\) generally leads to a relatively modest increase.

For an isotropic \(s\)-wave superconductor, weak-coupling BCS theory gives the universal SC-gap-to-\(T_c\) ratio \(2\Delta(0)/T_c=2\pi/e^{\gamma_{\mathrm E}}\simeq3.53\). The same weak-attraction estimate gives \(\Delta_d(0)\propto e^{-1/[4JD_d(\mu)]-L_d}\) (see \hyperlink{sm-sc-weak-gap-ratio}{Appendix J.2}). The common factor \(e^{-1/[4JD_d(\mu)]}\) cancels in the order-parameter-to-\(T_c\) ratio, giving \(2\sqrt2\Delta_d(0)/T_{c,d}\simeq\frac{\pi}{\sqrt2e^{\gamma_{\mathrm E}}}e^{-L_d}\simeq1.24725e^{-L_d}\), where

\hypertarget{sm-eq-sc-order-parameter-log-average}{}%
\FittedEquation{66}{%
\begin{aligned}
        L_d&=\frac{\displaystyle\sum_r\oint_{\mathcal C_r}\frac{d\ell}{(2\pi)^2}\frac{\phi_{\mathbf k,d}^2\ln|\phi_{\mathbf k,d}|}{|\nabla_{\mathbf k}S_{\mathbf k,r}|}}{\displaystyle\sum_r\oint_{\mathcal C_r}\frac{d\ell}{(2\pi)^2}\frac{\phi_{\mathbf k,d}^2}{|\nabla_{\mathbf k}S_{\mathbf k,r}|}}\\[6pt]
        &=\frac{\displaystyle\int_{\mathrm{BZ}}\frac{d^2\mathbf k}{(2\pi)^2}\phi_{\mathbf k,d}^2\ln|\phi_{\mathbf k,d}|\left[\delta(\xi_{\mathbf k})+\delta(\xi_{\mathbf k}-\mathcal J_{\mathbf k})\right]}{D_d(\mu)}.
        \end{aligned}
}
 
The integrand \(\phi_{\mathbf k,d}^2\ln|\phi_{\mathbf k,d}|\) tends to zero at a node. Thus the non-universal order-parameter-to-\(T_c\) ratio mainly depends on the geometry and form-factor weights of the two Cooper surfaces, and is independent of \(J\) in the weak-attraction limit. \hyperlink{sm-fig-sc-doping}{Fig. 7(d)} exhibits nearly vertical contours at small \(J\), except near the two van Hove dopings. Near the van Hove saddle points, where \(\ln|\phi_{\mathbf k,d}|\to0^{-}\), the enhanced contribution of these regions to \(D_d(\mu)\) makes \(L_d\) less negative and reduces the ratio, producing minima near \(p_{\mathrm{vH},2e}\simeq0.311\) and \(p_{\mathrm{vH},2h}\simeq0.443\) at small \(J\).

The analytical weak-attraction estimate of the order-parameter-to-\(T_c\) ratio agrees well with the numerical results for finite \(\mathcal J_d\), as shown in \hyperlink{sm-fig-sc-order-parameter-ratio}{Fig. J2}. However, at fixed \(J\), the ratio varies substantially with \(\mathcal J_d\) in the regime of very weak RML interaction [\hyperlink{sm-fig-sc-doping}{Fig. 7(e,f)}]. When the SC pairing field or temperature becomes comparable to the RML interaction strength, \(\Delta_d\sim\mathcal J_d\) or \(T\sim\mathcal J_d\), higher-energy block eigenstates become relevant even away from the nodes. The two-state approximation used to estimate the ratio, which retains only the two low-energy states near each Cooper surface, is then no longer valid. At \(\mathcal J_d=0\), the equations reduce to conventional \(d\)-wave BCS theory, whose weak-coupling ratio is governed by the geometry and form-factor weights of the Fermi surface. The ratio in the narrow crossover regime cannot be obtained by taking \(\mathcal J_d\to0\) in the weak-attraction estimate. In general, the order-parameter-to-\(T_c\) ratio of the correlated \(d\)-wave SC state can lie either above or below the universal BCS ratio of approximately \(3.53\).

\hypertarget{sm-sc-one-particle-gap}{}
\subsection*{B. One-particle Green's function}

The zero-temperature one-particle retarded Green's function defined in \hyperlink{sm-eq-one-particle-retarded-definition}{Eq. (14)}, evaluated exactly for the correlated SC state in \hyperlink{sm-appendix-sc-excitation-gap}{Appendix K}, is

\hypertarget{sm-eq-sc-zero-temperature-green}{}%
\FittedEquation{67}{%
G^R_{\mathrm{SC},\sigma}(\mathbf k,\omega)
=\sum_{\nu=\pm1}\left[
\frac{Z^{\mathrm{SC},+}_{\mathbf k,\nu 0}}{\omega+i0^+-E^{\mathrm{SC},o}_{\mathbf k,\nu}+E^{\mathrm{SC},e}_{\mathbf k,0}}
+\frac{Z^{\mathrm{SC},-}_{\mathbf k,\nu 0}}{\omega+i0^++E^{\mathrm{SC},o}_{\mathbf k,\nu}-E^{\mathrm{SC},e}_{\mathbf k,0}}\right],
}
 
which has four poles rather than two poles in conventional BCS theory, cf. \hyperlink{sm-eq-sc-zero-temperature-green-bcs}{Eq. (74)}. Spin-rotation symmetry makes the Green's function independent of \(\sigma\). The one-electron addition and removal poles correspond to transitions from the SC block ground state \(|e_{\mathbf k,0}\rangle\) to odd-fermion-parity excited states \(|o_{\pm\mathbf k\sigma,\nu}\rangle\), with one-particle excitation energies

\hypertarget{sm-eq-sc-zero-temperature-pole-energies}{}%
\FittedEquation{68}{%
E^{\mathrm{SC},o}_{\mathbf k,\nu}-E^{\mathrm{SC},e}_{\mathbf k,0}
=2\xi_{\mathbf k}-\frac{\mathcal J_{\mathbf k}}2
+\nu\sqrt{\left(\xi_{\mathbf k}-\frac{\mathcal J_{\mathbf k}}2\right)^2+2|\Delta_{\mathbf k,d}|^2}
-E^{\mathrm{SC},e}_{\mathbf k,0}.
}
 
Denote

 \hypertarget{sm-eq-sc-zero-temperature-pole-weights}{} 
\hypertarget{sm-eq-sc-spectral-transition-weights}{}%
\FittedEquation{69}{%
\begin{aligned}
Z^{\mathrm{SC},+}_{\mathbf k,\nu l}
&=|\langle o_{\mathbf k\sigma,\nu}|\hat c_{\mathbf k\sigma}^{\dagger}|e_{\mathbf k,l}\rangle|^2=\left|u_{\mathbf k,\nu}^{*}x_{\mathbf k,l}+\frac{v_{\mathbf k,\nu}^{*}y_{\mathbf k,l}}{\sqrt2}\right|^2,
\end{aligned}
}
 
\hypertarget{sm-eq-sc-spectral-transition-weights-removal}{}%
\FittedEquation{70}{%
\begin{aligned}
Z^{\mathrm{SC},-}_{\mathbf k,\nu l}
&=|\langle o_{-\mathbf k\bar\sigma,\nu}|\hat c_{\mathbf k\sigma}|e_{\mathbf k,l}\rangle|^2=\left|\frac{u_{\mathbf k,\nu}^{*}y_{\mathbf k,l}}{\sqrt2}+v_{\mathbf k,\nu}^{*}z_{\mathbf k,l}\right|^2,
\end{aligned}
}
 
where the two terms within each amplitude interfere because \(|e_{\mathbf k,l}\rangle\) is a linear superposition of \(|0_{\mathbf k}\rangle\), \(|2_{\mathbf k,s}\rangle\) and \(|4_{\mathbf k}\rangle\). The zero-temperature spectral weights of the one-electron addition and removal poles at \(\omega=\pm(E^{\mathrm{SC},o}_{\mathbf k,\nu}-E^{\mathrm{SC},e}_{\mathbf k,0})\) are \(Z^{\mathrm{SC},\pm}_{\mathbf k,\nu 0}\), which satisfy the zero-temperature sum rule \(\sum_{\nu}(Z^{\mathrm{SC},+}_{\mathbf k,\nu 0}+Z^{\mathrm{SC},-}_{\mathbf k,\nu 0})=1\).

Performing the sum over \(\nu\) gives an explicit and equivalent form of the one-particle retarded Green's function (\hyperlink{sm-sc-gap-explicit-resolvent}{Appendix K.1}):

\hypertarget{sm-eq-sc-zero-temperature-green-rational}{}%
\FittedEquation{71}{%
\begin{aligned}
G^R_{\mathrm{SC},\sigma}(\mathbf k,\omega)&=\frac{N^+_{\mathbf k}(\omega+i0^+)}
{\begin{gathered}
(\omega+i0^++E^{\mathrm{SC},e}_{\mathbf k,0}-\xi_{\mathbf k})\\
{}\times(\omega+i0^++E^{\mathrm{SC},e}_{\mathbf k,0}-3\xi_{\mathbf k}+\mathcal J_{\mathbf k})
-2|\Delta_{\mathbf k,d}|^2
\end{gathered}}\\[6pt]
&\quad+\frac{N^-_{\mathbf k}(\omega+i0^+)}
{\begin{gathered}
(\omega+i0^+-E^{\mathrm{SC},e}_{\mathbf k,0}+\xi_{\mathbf k})\\
{}\times(\omega+i0^+-E^{\mathrm{SC},e}_{\mathbf k,0}+3\xi_{\mathbf k}-\mathcal J_{\mathbf k})
-2|\Delta_{\mathbf k,d}|^2
\end{gathered}},
\end{aligned}
}
 
where the two terms describe one-electron addition and removal, respectively, and

\hypertarget{sm-eq-sc-zero-temperature-green-numerators}{}%
\FittedEquation{72}{%
\begin{aligned}
N^+_{\mathbf k}(\omega+i0^+)
&=|x_{\mathbf k,0}|^2(\omega+i0^++E^{\mathrm{SC},e}_{\mathbf k,0}-3\xi_{\mathbf k}+\mathcal J_{\mathbf k})\\
&\quad+\frac{|y_{\mathbf k,0}|^2}{2}(\omega+i0^++E^{\mathrm{SC},e}_{\mathbf k,0}-\xi_{\mathbf k})
-2\operatorname{Re}(\Delta_{\mathbf k,d}^{*}x_{\mathbf k,0}^{*}y_{\mathbf k,0}),
\end{aligned}
}
 
\hypertarget{sm-eq-sc-zero-temperature-green-numerator-removal}{}%
\FittedEquation{73}{%
\begin{aligned}
N^-_{\mathbf k}(\omega+i0^+)
&=\frac{|y_{\mathbf k,0}|^2}{2}(\omega+i0^+-E^{\mathrm{SC},e}_{\mathbf k,0}+3\xi_{\mathbf k}-\mathcal J_{\mathbf k})\\
&\quad+|z_{\mathbf k,0}|^2(\omega+i0^+-E^{\mathrm{SC},e}_{\mathbf k,0}+\xi_{\mathbf k})
+2\operatorname{Re}(\Delta_{\mathbf k,d}^{*}y_{\mathbf k,0}^{*}z_{\mathbf k,0}),
\end{aligned}
}
 
where the real cross terms reflect interference induced by the SC pairing field, which establishes global phase coherence among momentum blocks. When \(\mathcal J_d=0\), the Green's function reduces to the conventional BCS form

\hypertarget{sm-eq-sc-zero-temperature-green-bcs}{}%
\FittedEquation{74}{%
G^R_{\mathrm{BCS},\sigma}(\mathbf k,\omega)
=\frac{\omega+i0^++\xi_{\mathbf k}}{(\omega+i0^+)^2-\xi_{\mathbf k}^2-|\sqrt2\Delta_{\mathbf k,d}|^2}.
}
 
The one-particle excitation gap of the correlated SC state is \(\Delta_{\mathrm{1p}}^{\mathrm{SC}}(\mathbf k)=E^{\mathrm{SC},o}_{\mathbf k,-}-E^{\mathrm{SC},e}_{\mathbf k,0}\), which also vanishes only at the nodes, where \(\xi_{\mathbf k_n}=\phi_{\mathbf k_n,d}=0\), and reduces to the conventional \(d\)-wave Bogoliubov quasiparticle dispersion \(E_{\mathbf k}^{\mathrm{BCS}}=\sqrt{\xi_{\mathbf k}^2+|\sqrt2\Delta_{\mathbf k,d}|^2}\) at \(\mathcal J_d=0\). The correlated \(d\)-wave SC state is therefore a correlated nodal SC state.

\par\medskip
\hypertarget{sm-fig-sc-excitation-gap}{}
\PrintAsset{figure-sc-excitation-gap}
{\small\noindent \textbf{Figure 8 \textbar{} Form factor and angle dependence of the one-particle excitation gap and SC pairing field.} \textbf{a,b,} The one-particle excitation gaps \(\Delta_{\mathrm{1p}}^{\mathrm{SC}}(\mathbf k_F)\) (bright orange solid) and \(\Delta_{\mathrm{1p}}^{\mathrm N}(\mathbf k_F)\) (deep magenta short-dashed), and the magnitude of the SC pairing field \(|\Delta_{\mathbf k_F,d}|\) (dark navy long-dashed), along the Fermi surface \(\xi_{\mathbf k_F}=0\), plotted against the magnitude of the \(d\)-wave form factor \(|\phi_{\mathbf k_F,d}|\) and the angle \(\theta=\tan^{-1}[(\pi-k_x)/(\pi-k_y)]\), respectively. The angle runs from the antinode \(\mathbf k_a\) at \(0^\circ\) to the node \(\mathbf k_n\) at \(45^\circ\), as sketched in \textbf{d}. Insets resolve the nodal regions \(0\le|\phi_{\mathbf k_F,d}|\le0.15\) in \textbf{a} and \(40^\circ\le\theta\le45^\circ\) in \textbf{b}. \textbf{c,d,} The same three quantities normalized by their respective maxima (approximately \(68\), \(55\) and \(7~\mathrm{meV}\)) at the antinode, \(\Delta(\mathbf k_F)/\Delta(\mathbf k_a)\), versus \(|\phi_{\mathbf k_F,d}|\) and \(\theta\), respectively. The zero-temperature \(d\)-wave solution gives \(\Delta_d(0)=7.2891~\mathrm{meV}\) and \(\mu_{\mathrm{SC}}(0)=-0.2675963~\mathrm{eV}\). The parameters are \(\mathcal J_d=0.06~\mathrm{eV}\), \(J=0.12~\mathrm{eV}\), and \(p=0.15\).\par}
\medskip

\hypertarget{sm-sc-gap-fermi-surface-limits}{}\hyperlink{sm-fig-sc-excitation-gap}{Fig. 8} compares the one-particle excitation gap of the correlated SC state, \(\Delta_{\mathrm{1p}}^{\mathrm{SC}}(\mathbf k_F)\), with that of the nodal Bose metal, \(\Delta_{\mathrm{1p}}^{\mathrm N}(\mathbf k_F)=\mathcal J_{\mathbf k_F}=\mathcal J_d\phi_{\mathbf k_F,d}^{2}\), and the magnitude of the SC pairing field, \(|\Delta_{\mathbf k_F,d}|=|\Delta_d\phi_{\mathbf k_F,d}|\), along the Fermi surface \(\xi_{\mathbf k_F}=0\). Note that \(\Delta_{\mathrm{1p}}^{\mathrm{SC}}(\mathbf k_F)\ge|\sqrt2\Delta_{\mathbf k_F,d}|\), in contrast to the conventional \(d\)-wave BCS state, for which \(\Delta_{\mathrm{1p}}^{\mathrm{BCS}}(\mathbf k_F)=|\sqrt2\Delta_{\mathbf k_F,d}|\). At the antinode, \(\Delta_{\mathrm{1p}}^{\mathrm{SC}}(\mathbf k_a)\simeq68~\mathrm{meV}\gg|\sqrt2\Delta_{\mathbf k_a,d}|\simeq7\sqrt2~\mathrm{meV}\). Near the antinodes, where \(\mathcal J_{\mathbf k_F}\gg|\Delta_{\mathbf k_F,d}|\), expanding the one-particle excitation gap gives (\hyperlink{sm-sc-gap-antinodal-expansion}{Appendix K.2}) 
\hypertarget{sm-eq-sc-antinodal-gap-expansion}{}%
\FittedEquation{75}{%
\Delta_{\mathrm{1p}}^{\mathrm{SC}}(\mathbf k_F)
=\mathcal J_{\mathbf k_F}+2|\Delta_{\mathbf k_F,d}|
+O\!\left(\frac{|\Delta_{\mathbf k_F,d}|^2}{\mathcal J_{\mathbf k_F}}\right).
}
 Thus \(\Delta_{\mathrm{1p}}^{\mathrm{SC}}(\mathbf k_F)\) retains the substantial RML interaction scale near the antinodes. Close to the nodes, by contrast, \(\mathcal J_{\mathbf k_F}\propto\phi_{\mathbf k_F,d}^2\) vanishes faster than the SC pairing field \(|\Delta_{\mathbf k_F,d}|\propto|\phi_{\mathbf k_F,d}|\), as shown in the insets of \hyperlink{sm-fig-sc-excitation-gap}{Fig. 8(a,b)}. Consequently, \(\Delta_{\mathrm{1p}}^{\mathrm{SC}}(\mathbf k_F)\simeq\sqrt2\,\Delta_d|\phi_{\mathbf k_F,d}|\) asymptotically near the nodes, and the nodal excitation slope with respect to \(|\phi_{\mathbf k_F,d}|\) probes the \(d\)-wave SC pairing scale. The momentum dependence of \(\Delta_{\mathrm{1p}}^{\mathrm{SC}}(\mathbf k_F)\) thus reveals the separation between the nodal SC pairing scale and the antinodal RML interaction scale, reflecting the nodal--antinodal dichotomy of the correlated \(d\)-wave SC state \hyperlink{sm-ref-tanaka-2006}{[36]}, \hyperlink{sm-ref-lee-2007}{[37]}, \hyperlink{sm-ref-mesot-1999}{[38]}, \hyperlink{sm-ref-vishik-2012}{[39]}, \hyperlink{sm-ref-anzai-2013}{[40]}, \hyperlink{sm-ref-le-tacon-2006}{[41]}.
 
\hypertarget{sm-sc-one-particle}{}
\subsection*{C. One-particle spectra: SC pockets, spectral particle-hole asymmetry, and spectral weight transfer}

The SC one-particle spectral function at finite temperature, evaluated exactly for the correlated SC state in \hyperlink{sm-sc-spectral-lehmann}{Appendix L.1}, is

\hypertarget{sm-eq-sc-spectral-one}{}%
\FittedEquation{76}{%
\begin{aligned}
&A_{\mathrm{SC},\sigma}(\mathbf k,\omega)\\
&=\sum_{\lambda,\lambda'}^{\pi_{\mathbf k,\lambda'}=-\pi_{\mathbf k,\lambda}}
(p^{\mathrm{SC}}_{\mathbf k,\lambda}+p^{\mathrm{SC}}_{\mathbf k,\lambda'})
|\langle\pi_{\mathbf k,\lambda'}|\hat c_{\mathbf k\sigma}|\pi_{\mathbf k,\lambda}\rangle|^2\,\delta\!\left(\omega-E^{\mathrm{SC}}_{\mathbf k,\lambda}+E^{\mathrm{SC}}_{\mathbf k,\lambda'}\right)\\
&=\sum_{l=0}^{2}\sum_{\nu=\pm1}
(p^{\mathrm{SC},e}_{\mathbf k,l}+p^{\mathrm{SC},o}_{\mathbf k,\nu})\,\Bigl[
Z^{\mathrm{SC},+}_{\mathbf k,\nu l}\delta(\omega-E^{\mathrm{SC},o}_{\mathbf k,\nu}+E^{\mathrm{SC},e}_{\mathbf k,l})
+Z^{\mathrm{SC},-}_{\mathbf k,\nu l}\delta(\omega+E^{\mathrm{SC},o}_{\mathbf k,\nu}-E^{\mathrm{SC},e}_{\mathbf k,l})
\Bigr]\\
&\quad+\frac{5}{2}\sum_{\nu=\pm1}\!
(p^{\mathrm{SC}}_{\mathbf k,2}+p^{\mathrm{SC},o}_{\mathbf k,\nu})\,\Bigl[
|v_{\mathbf k,\nu}|^2\delta(\omega{-}E^{\mathrm{SC},o}_{\mathbf k,\nu}{+}2\xi_{\mathbf k})
+|u_{\mathbf k,\nu}|^2\delta(\omega{+}E^{\mathrm{SC},o}_{\mathbf k,\nu}{-}2\xi_{\mathbf k})
\Bigr],
\end{aligned}
}
 
where \(|\pi_{\mathbf k,\lambda}\rangle\) denotes a SC block eigenstate with block fermion parity \(\pi_{\mathbf k,\lambda}=\pm1\) and eigenenergy \(E^{\mathrm{SC}}_{\mathbf k,\lambda}\), respectively, and \(p^{\mathrm{SC}}_{\mathbf k,\lambda}=e^{-\beta E^{\mathrm{SC}}_{\mathbf k,\lambda}}/Z^{\mathrm{SC}}_{\mathbf k}\) is the corresponding thermal occupation probability. Spin rotation symmetry of the correlated SC state ensures \(A_{\mathrm{SC}}=A_{\mathrm{SC},\uparrow}=A_{\mathrm{SC},\downarrow}\). Away from the gapless nodes, thermally activated transitions between excited SC block eigenstates (\hyperlink{sm-table-k1-sc-transitions}{Table L1}) produce additional one-electron addition and removal poles at \(\omega=\pm(E^{\mathrm{SC},o}_{\mathbf k,\nu}-E^{\mathrm{SC},e}_{\mathbf k,l})\), with \(l=1,2\), and \(\omega=\pm(E^{\mathrm{SC},o}_{\mathbf k,\nu}-2\xi_{\mathbf k})\), whose spectral weights vanish at zero temperature.

\par\medskip
\hypertarget{sm-fig-constant-energy-four-states}{}
\PrintAsset{figure-constant-energy-four-states-row1}
\PrintAsset{figure-constant-energy-four-states-row2}
\PrintAsset{figure-constant-energy-four-states-row3}
{\small\noindent \textbf{Figure 9 \textbar{} One-particle excitation spectra of the correlated SC state, conventional BCS state, and nodal Bose metal.} \textbf{a,b,} Zero-energy spectral functions at \(T=10~\mathrm K\) of the correlated SC state with \(\mathcal J_d=0.06~\mathrm{eV}\) and \(J=0.12~\mathrm{eV}\), and the conventional \(d\)-wave BCS state with \(\mathcal J_d=0\) and \(J=0.25~\mathrm{eV}\), respectively. \textbf{c,d,} Spectral functions at \(\omega=-2~\mathrm{meV}\) of the correlated SC state at \(T=10~\mathrm K\) and the nodal Bose metal at \(T=150~\mathrm K\), above the SC transition temperature \(T_{c,d}=143.13~\mathrm K\), respectively. Both panels use \(\mathcal J_d=0.06~\mathrm{eV}\) and \(J=0.12~\mathrm{eV}\). \textbf{e,f,} Spectral functions at \(\omega=-2~\mathrm{meV}\) and \(T=10~\mathrm K\) of the correlated SC state and conventional \(d\)-wave BCS state, with \(\mathcal J_d=0.06~\mathrm{eV}\) and \(0\), respectively. Both panels use \(J=0.25~\mathrm{eV}\). White dotted curves mark each state's \(\xi_{\mathbf k_F}=0\) contour; blue and red dots mark the node \(\mathbf k_n\) and antinode \(\mathbf k_a\). Insets resolve the nodal region in the coordinates of \hyperlink{sm-fig-c11-lorentzian}{Fig. 3(a)}. The Lorentzian HWHM is \(\Gamma_L=5~\mathrm{meV}\) in the main maps and \(1~\mathrm{meV}\) in the insets, and \(p=0.15\) for each panel.\par}
\medskip

\hypertarget{sm-sc-fig10-physical-discussion}{}\hyperlink{sm-fig-constant-energy-four-states}{Fig. 9(a,b)} show the zero-energy spectral functions \(A(\mathbf k,\omega=0)\) of the correlated SC state and the conventional BCS state, respectively, at \(10~\mathrm K\). Although both states are nodal, finite temperature and Lorentzian broadening produce a Fermi-arc-like distribution of spectral intensity near the nodes (cf. \hyperlink{sm-fig-c11-lorentzian}{Fig. 3(a)}), whose extent along the Fermi surface is mainly controlled by the SC pairing scale. For the parameters used in \hyperlink{sm-fig-constant-energy-four-states}{Fig. 9(a,b)}, the order parameters are \(\Delta_d^{\mathrm{SC}}\simeq7.29~\mathrm{meV}\) and \(\Delta_d^{\mathrm{BCS}}\simeq37.89~\mathrm{meV}\). The smaller order parameter gives the correlated SC state a smaller nodal excitation slope and hence a more extended bright arc than the conventional BCS state.

Slightly below zero energy, at \(\omega=-2~\mathrm{meV}\) (the Lorentzian FWHM is \(2~\mathrm{meV}\) in the nodal window), the small but nonzero SC order parameter produces a slight back-bending at the end of the arc [\hyperlink{sm-fig-constant-energy-four-states}{Fig. 9(c)}]. Increasing \(J\) from \(0.12~\mathrm{eV}\) [\hyperlink{sm-fig-constant-energy-four-states}{Fig. 9(c)}] to \(0.25~\mathrm{eV}\) [\hyperlink{sm-fig-constant-energy-four-states}{Fig. 9(e)}], with all other parameters fixed, raises the order parameter from \(\Delta_d\simeq7.29~\mathrm{meV}\) to \(44.87~\mathrm{meV}\). The increased SC pairing scale produces a \emph{pocket}-like distribution of spectral intensity around the nodes. However, increasing the temperature from \(10~\mathrm K\) [\hyperlink{sm-fig-constant-energy-four-states}{Fig. 9(c)}] through \(T_{c,d}=143.13~\mathrm K\) to \(150~\mathrm K\) [\hyperlink{sm-fig-constant-energy-four-states}{Fig. 9(d)}] only lengthens the arc, with no apparent back-bending in the nodal Bose metal. The nodal pockets are not unique to the correlated SC state: the conventional BCS state, with an order parameter \(\Delta_d^{\mathrm{BCS}}\simeq37.89~\mathrm{meV}\), also exhibits similar but slightly more extended pockets [\hyperlink{sm-fig-constant-energy-four-states}{Fig. 9(f)}]. The pockets in \hyperlink{sm-fig-constant-energy-four-states}{Fig. 9(e,f)} both arise from constant-energy contours of the nodal Dirac dispersion at negative energies. For \(\omega<0\), the nodal spectral intensity is governed by the one-electron-removal weight, which approaches \(v_{\mathbf k}^{2}=(1-\xi_{\mathbf k}/E_{\mathrm D,\mathbf k})/2\), where \(E_{\mathrm D,\mathbf k}\) is the nodal Dirac excitation energy of the corresponding state. The one-electron-removal weight is larger on the lower side of the pocket (\(\xi_{\mathbf k}<0\)) and smaller on the upper side (\(\xi_{\mathbf k}>0\)), producing a bright lower branch and a faint upper branch despite the closed constant-energy contour.

\par\medskip
\hypertarget{sm-fig-sc-zeeman-zero}{}
\PrintAsset{spin-resolved-zero-Bplus4meV-J025}
{\small\noindent \textbf{Figure 10 \textbar{} Spin-resolved zero-energy excitation spectra of the correlated SC and conventional BCS states under a magnetic field at \(10~\mathrm K\).} \textbf{a,c,} \(A_{\mathrm{SC},\sigma}(\mathbf k,0;B_z)\) for the correlated SC state with \(\mathcal J_d=0.06~\mathrm{eV}\). \textbf{b,d,} \(A_{\mathrm{BCS},\sigma}(\mathbf k,0;B_z)\) for the conventional \(d\)-wave BCS state with \(\mathcal J_d=0\). White dotted curves mark each state's \(\xi_{\mathbf k_F}=0\) contour; blue and red dots mark the node \(\mathbf k_n\) and antinode \(\mathbf k_a\). Insets resolve the nodal region in the coordinates of \hyperlink{sm-fig-c11-lorentzian}{Fig. 3(a)}. The Lorentzian HWHM is \(\Gamma_L=5~\mathrm{meV}\) in the main maps and \(1~\mathrm{meV}\) in the insets. The parameters are \(J=0.25~\mathrm{eV}\), \(p=0.15\) and \(B_z=g_e\mu_{\mathrm B}B=4~\mathrm{meV}\), with \(B\simeq34.5~\mathrm T\).\par}
\medskip

\hypertarget{sm-sc-zeeman-nodal-pockets}{}The pocket-like distributions are also observed in the zero-energy spectral functions under a magnetic field [\hyperlink{sm-fig-sc-zeeman-zero}{Fig. 10}], which is treated by adding the Zeeman term \(-B_z\hat{\mathbb S}^{z}_{\mathbf k}\) (\(\hat{\mathbb S}^{z}_{\mathbf k}=\hat S^{z}_{\mathbf k}+\hat S^{z}_{-\mathbf k}\)) to the SC block Hamiltonian \(\hat h_{\mathbf k}^{\mathrm{SC}}\) (\hyperlink{sm-sc-spectral-zeeman}{Appendix L.2}). Compared with the zero-field cases illustrated in \hyperlink{sm-fig-constant-energy-four-states}{Fig. 9(e,f)}, a magnetic field shifts the spin-resolved one-electron-removal energies by \(-\sigma B_z/2\) at fixed chemical potential and SC pairing field, allowing the zero-energy cut to probe constant-energy contours of the underlying nodal Dirac dispersion at energies \(\sigma B_z/2\). The field-dependent spectral functions satisfy \(A_{\sigma}(\mathbf k,\omega;B_z)=A_{\bar\sigma}(\mathbf k,\omega;-B_z)\). The Zeeman term produces opposite energy shifts for the two spin components, with brighter upper branches in \hyperlink{sm-fig-sc-zeeman-zero}{Fig. 10(a,b)} and brighter lower branches in \hyperlink{sm-fig-sc-zeeman-zero}{Fig. 10(c,d)}.

The competition between the linear nodal SC pairing field and the quadratically vanishing RML interaction sets a local crossover scale, \(E_{\mathrm{cross}}\sim|\Delta_d|^2/\mathcal J_d\), below which RML corrections to the leading Dirac dispersion become small. For the correlated SC states in \hyperlink{sm-fig-constant-energy-four-states}{Fig. 9(e)} and \hyperlink{sm-fig-sc-zeeman-zero}{Fig. 10(a,c)}, \(E_{\mathrm{cross}}\simeq33.6~\mathrm{meV}\). The probe energy \(|\omega|=2~\mathrm{meV}\) in \hyperlink{sm-fig-constant-energy-four-states}{Fig. 9(e)}, the thermal energy \(k_{\mathrm B}T\simeq0.86~\mathrm{meV}\) at \(10~\mathrm K\), and the Zeeman energy \(B_z=4~\mathrm{meV}\) in \hyperlink{sm-fig-sc-zeeman-zero}{Fig. 10(a,c)} all lie well below the scale \(E_{\mathrm{cross}}\). By contrast, \(E_{\mathrm{cross}}\simeq0.89~\mathrm{meV}\) in \hyperlink{sm-fig-constant-energy-four-states}{Fig. 9(c)}. Such a small crossover scale can make the pocket difficult to resolve at the probe energy despite a finite SC order parameter.
 
\par\medskip
\hypertarget{sm-fig-spectral-ph-symmetry}{}
\PrintAsset{spectral-ph-symmetry-row1}
\PrintAsset{spectral-ph-symmetry-row2}
{\small\noindent \textbf{Figure 11 \textbar{} One-particle excitation spectra along Fermi-surface and Luttinger-surface contours.} \textbf{a,b,} One-particle spectral functions of the correlated SC state and conventional \(d\)-wave BCS state along their respective Fermi-surface contours \(\xi_{\mathbf k_F}=0\) at \(T=10~\mathrm K\). \textbf{c,d,} One-particle spectral functions of the correlated SC state at \(10~\mathrm K\) and the nodal Bose metal at \(60~\mathrm K\), respectively, along their respective Luttinger-surface contours \(\xi_{\mathbf k_L}=\mathcal J_{\mathbf k_L}/2\). The correlated SC state in \textbf{a,c} has \(\mathcal J_d=0.06~\mathrm{eV}\) and \(J=0.12~\mathrm{eV}\); the conventional BCS state in \textbf{b} has \(J=0.25~\mathrm{eV}\); the nodal Bose metal in \textbf{d} has \(\mathcal J_d=0.06~\mathrm{eV}\). White dotted curves mark \(\omega=\pm(E^{\mathrm{SC},o}_{\mathbf k,\nu}-E^{\mathrm{SC},e}_{\mathbf k,0})\), with \(\nu=\pm1\), in \textbf{a,c}, \(\omega=\pm|\sqrt2\Delta_{\mathbf k_F,d}|\) in \textbf{b}, and \(\omega=\xi_{\mathbf k_L}+\zeta\mathcal J_{\mathbf k_L}\), with \(\zeta=+1,0,-1,-2\), in \textbf{d}. Dashed horizontal lines mark \(\omega=0\). The Lorentzian HWHM is \(\Gamma_L=5~\mathrm{meV}\) and \(p=0.15\) for each panel.\par}
\medskip

\hypertarget{sm-sc-fig8-fig9-physical-discussion}{}At \(10~\mathrm K\), the four zero-temperature poles of the correlated SC state at \(\omega=\pm(E^{\mathrm{SC},o}_{\mathbf k_F,\nu}-E^{\mathrm{SC},e}_{\mathbf k_F,0})\), with \(\nu=\pm1\), dominate the low-temperature excitation spectra away from the nodes, while other thermally activated poles produce only weak satellite features that are barely discernible on the chosen color scale in \hyperlink{sm-fig-spectral-ph-symmetry}{Fig. 11(a)}. By contrast, the conventional BCS state has only two poles at \(\omega=\pm|\sqrt2\Delta_{\mathbf k_F,d}|\), as shown in \hyperlink{sm-fig-spectral-ph-symmetry}{Fig. 11(b)}, with no additional thermally activated poles. The BCS Hamiltonian \(\hat H_{\mathrm{BCS}}\) in \hyperlink{sm-eq-b3}{Eq. (C3)} has a block \(U_{\mathbf k}(1)\) symmetry associated with Bogoliubov quasiparticle-number conservation. Since the electron operators are linear in Bogoliubov quasiparticle operators, one-electron addition or removal from the BCS ground state creates only one Bogoliubov quasiparticle, producing the two Bogoliubov quasiparticle poles. However, \(\hat H_{\mathrm{SC}}\) retains only the block fermion-parity \(\mathbb Z^{\mathrm f}_{2,\mathbf k}\) symmetry. The odd-fermion-parity SC block eigenstates are superpositions of one- and three-electron states, and a single electron operator acting on \(|\mathrm{GS}_{\mathrm{SC}}\rangle\) can access both \(|o_{\pm\mathbf k\sigma,\nu}\rangle\), with \(\nu=\pm1\). From the wavefunction perspective, \(|\mathrm{GS}_{\mathrm{SC}}\rangle\) in \hyperlink{sm-eq-sc-ground-state}{Eq. (63)} encodes the same selection rule. At \(\mathcal J_d=0\), the normalized coefficients satisfy the BCS constraint \(y_{\mathbf k,0}^{2}=2x_{\mathbf k,0}z_{\mathbf k,0}\), giving \(Z^{\mathrm{SC},\pm}_{\mathbf k,\nu=+1,l=0}=0\). Only the addition and removal poles associated with \(\nu=-1\) retain nonzero spectral weight in the conventional BCS state.

The spectral weights along \(\xi_{\mathbf k_F}=0\) further distinguish the two states: the one-electron addition and removal weights of the conventional BCS state are both \(1/2\), giving a symmetric spectral function \(A_{\mathrm{BCS}}(\mathbf k_F,\omega)=A_{\mathrm{BCS}}(\mathbf k_F,-\omega)\) in \hyperlink{sm-fig-spectral-ph-symmetry}{Fig. 11(b)}, whereas the correlated SC state has unequal weights \(Z^{\mathrm{SC},+}_{\mathbf k_F,\nu 0}\ne Z^{\mathrm{SC},-}_{\mathbf k_F,\nu 0}\), producing an asymmetric spectral function \(A_{\mathrm{SC}}(\mathbf k_F,\omega)\ne A_{\mathrm{SC}}(\mathbf k_F,-\omega)\) in \hyperlink{sm-fig-spectral-ph-symmetry}{Fig. 11(a)}. The \emph{spectral asymmetry} arises because the RML interaction makes one-electron addition and removal inequivalent along \(\xi_{\mathbf k_F}=0\). At zero temperature, the imbalance of the total one-electron addition and removal spectral weights, \(\sum_\nu(Z^{\mathrm{SC},+}_{\mathbf k_F,\nu0}-Z^{\mathrm{SC},-}_{\mathbf k_F,\nu0})=|x_{\mathbf k_F,0}|^2-|z_{\mathbf k_F,0}|^2=1-2\langle\hat n_{\mathbf k_F\sigma}\rangle_{\mathrm{SC}}\) [\hyperlink{sm-eq-gap-weight-normalization}{Eqs. (K11)--(K12)}], directly measures the difference between the probabilities of \(|0_{\mathbf k_F}\rangle\) and \(|4_{\mathbf k_F}\rangle\) in the ground state \(|\mathrm{GS}_{\mathrm{SC}}\rangle\). In particular, near the antinodes, where \(|\Delta_{\mathbf k_F,d}|\ll\mathcal J_{\mathbf k_F}\), the normalized ground-state coefficients satisfy \(|x_{\mathbf k_F,0}|^2\to0\) and \(|y_{\mathbf k_F,0}|^2,|z_{\mathbf k_F,0}|^2\to1/2\). Consequently, \(\sum_\nu Z^{\mathrm{SC},+}_{\mathbf k_F,\nu0}\to1/4\), whereas \(\sum_\nu Z^{\mathrm{SC},-}_{\mathbf k_F,\nu0}\to3/4\), so the negative-energy electron-removal spectrum can carry substantially more weight than the positive-energy addition spectrum [\hyperlink{sm-fig-spectral-ph-symmetry}{Fig. 11(a)}].

\hypertarget{sm-sc-block-particle-hole-transformation}{}The spectral asymmetry is a direct consequence of the broken RML \emph{particle--hole symmetry}, a \(\mathbb Z_{2}^{\mathrm{ph}}\) symmetry local in momentum space \hyperlink{sm-ref-anderson-haldane-2001}{[42]}, \hyperlink{sm-ref-huang-la-nave-phillips-2022}{[43]}. Introduce the unitary charge conjugation operator 
\hypertarget{sm-eq-sc-particle-hole-unitary}{}%
\FittedEquation{77}{%
\hat{\mathcal C}_{\mathbf k}=\hat{\mathcal C}_{\mathbf k}^{\dagger}
=-\bigl(\hat c_{\mathbf k\uparrow}-\hat c_{\mathbf k\uparrow}^{\dagger}\bigr)
\bigl(\hat c_{\mathbf k\downarrow}+\hat c_{\mathbf k\downarrow}^{\dagger}\bigr)
\bigl(\hat c_{-\mathbf k\uparrow}-\hat c_{-\mathbf k\uparrow}^{\dagger}\bigr)
\bigl(\hat c_{-\mathbf k\downarrow}+\hat c_{-\mathbf k\downarrow}^{\dagger}\bigr),
}
 and define the RML particle--hole transformation by 
\hypertarget{sm-eq-sc-particle-hole-transformation}{}%
\FittedEquation{78}{%
\hat{\mathcal C}_{\mathbf k}\hat c_{\pm\mathbf k\sigma}\hat{\mathcal C}_{\mathbf k}^{-1}
=\sigma\hat c_{\pm\mathbf k\sigma}^{\dagger},
}
 which is a canonical transformation preserving the fermionic anticommutation relations and satisfying \(\hat{\mathcal C}_{\mathbf k}^{2}=1\). The charge conjugation operator \(\hat{\mathcal C}_{\mathbf k}\) transforms the block states as 
\hypertarget{sm-eq-sc-particle-hole-block-states}{}\hypertarget{sm-eq-sc-particle-hole-singlet-state}{}\hypertarget{sm-eq-sc-particle-hole-full-state}{}%
\FittedEquation{79}{%
\hat{\mathcal C}_{\mathbf k}|0_{\mathbf k}\rangle=|4_{\mathbf k}\rangle,\quad \hat{\mathcal C}_{\mathbf k}|2_{\mathbf k,s}\rangle=|2_{\mathbf k,s}\rangle,\quad \hat{\mathcal C}_{\mathbf k}|4_{\mathbf k}\rangle=|0_{\mathbf k}\rangle.
}
 For systems with time-reversal symmetry, a global phase gauge is chosen so that \(\Delta_d\) is real. The block operators and SC block Hamiltonian transform as 
\hypertarget{sm-eq-sc-particle-hole-block-operators}{}%
\FittedEquation{80}{%
\begin{aligned}
\hat{\mathcal C}_{\mathbf k}\hat N_{\mathbf k}\hat{\mathcal C}_{\mathbf k}^{-1}
&=4-\hat N_{\mathbf k},
\end{aligned}
}
 
\hypertarget{sm-eq-sc-particle-hole-pair-operator}{}%
\FittedEquation{81}{%
\begin{aligned}
\hat{\mathcal C}_{\mathbf k}\hat\Delta_{\mathbf k}\hat{\mathcal C}_{\mathbf k}^{-1}
&=\hat\Delta_{\mathbf k}^{\dagger},
\end{aligned}
}
 
\hypertarget{sm-eq-sc-particle-hole-hamiltonian}{}%
\FittedEquation{82}{%
\begin{aligned}
\hat{\mathcal C}_{\mathbf k}\hat h_{\mathbf k}^{\mathrm{SC}}\hat{\mathcal C}_{\mathbf k}^{-1}
&=\hat h_{\mathbf k}^{\mathrm{SC}}+(\mathcal J_{\mathbf k}-2\xi_{\mathbf k})(\hat N_{\mathbf k}-2).
\end{aligned}
}
 Thus the SC block Hamiltonian has the RML particle--hole symmetry along \(\xi_{\mathbf k_L}=\mathcal J_{\mathbf k_L}/2\), which defines the Luttinger surface in the nodal Bose metal. Consequently, one-electron addition and removal transitions at the same momentum and spin occur with equal weights and opposite energies, producing \(A_{\mathrm{SC}}(\mathbf k_L,\omega)=A_{\mathrm{SC}}(\mathbf k_L,-\omega)\) at any temperature [\hyperlink{sm-fig-spectral-ph-symmetry}{Fig. 11(c)}]. From the wavefunction perspective, the RML particle--hole symmetry enforces \(x_{\mathbf k_L,l}=\pm z_{\mathbf k_L,l}\) for the normalized coefficients of the even-fermion-parity block eigenstates, giving \(\sum_\nu Z^{\mathrm{SC},+}_{\mathbf k_L,\nu0}=\sum_\nu Z^{\mathrm{SC},-}_{\mathbf k_L,\nu0}=1/2\) at zero temperature. Moreover, the spectral-function sum rule, together with this symmetry, gives \(\langle\hat n_{\mathbf k_L\sigma}\rangle_{\mathrm{SC}}=\int_{-\infty}^{\infty}d\omega\,f_{\mathrm F}(\omega)A_{\mathrm{SC},\sigma}(\mathbf k_L,\omega)=1/2\) at any temperature. In the limit \(\mathcal J_d=0\), the Luttinger-surface contour \(\xi_{\mathbf k_L}=\mathcal J_{\mathbf k_L}/2\) coincides with the Fermi surface \(\xi_{\mathbf k_F}=0\) of the conventional BCS state [\hyperlink{sm-fig-spectral-ph-symmetry}{Fig. 11(b)}]. The nodal Bose metal also exhibits spectral particle--hole symmetry along the Luttinger surface [\hyperlink{sm-fig-spectral-ph-symmetry}{Fig. 11(d)}; cf. \hyperlink{sm-fig-c11-lorentzian}{Fig. 3(b)}], and its spectral function satisfies \(A_{\mathrm N}(\mathbf k_L,\omega)=A_{\mathrm N}(\mathbf k_L,-\omega)\) at any temperature.
 
\hypertarget{sm-sc-luttinger-spectral-symmetry-discussion}{}
 
\par\medskip
\hypertarget{sm-fig-sc-one-particle}{}
\PrintAsset{figure-sc-spectra-comparison-row1}
\PrintAsset{figure-sc-spectra-comparison-row2}
\PrintAsset{figure-sc-spectra-comparison-row3}
{\small\noindent \textbf{Figure 12 \textbar{} One-particle excitation spectra of the correlated SC state and conventional \(d\)-wave BCS state at \(10~\mathrm K\).} \textbf{a,b,} One-particle spectral functions \(A_{\mathrm{SC}}(\mathbf k,\omega)\) and \(A_{\mathrm{BCS}}(\mathbf k,\omega)\) along \(k_x=\pi\), respectively. White dotted curves denote \(\omega=\pm(E^{\mathrm{SC},o}_{\mathbf k,\nu}-E^{\mathrm{SC},e}_{\mathbf k,0})\) in \textbf{a} and \(\omega=\pm\sqrt{\xi_{\mathbf k}^2+|\sqrt2\Delta_{\mathbf k,d}|^2}\) in \textbf{b}. Solid red vertical lines mark \(\xi_{\mathbf k}=0\) in \textbf{a,b}; the solid cyan and dash-dotted green vertical lines in \textbf{a} mark \(\xi_{\mathbf k}=\mathcal J_{\mathbf k}\) and \(\xi_{\mathbf k}=\mathcal J_{\mathbf k}/2\), respectively. \textbf{c--e,} EDCs along \(k_x=\pi\), offset by \(100~\mathrm{eV}^{-1}\): ten correlated SC EDCs over \(0.12\le k_y/\pi\le0.19\) in \textbf{c,e} and five BCS EDCs over \(0.12\le k_y/\pi\le0.15\) in \textbf{d}. Red, cyan, and green EDCs correspond to the contours marked by the vertical lines in \textbf{a,b}. \textbf{f,} One-particle excitation gaps \(\Delta_{\mathrm{1p}}^{\mathrm{SC}}(\mathbf k_F)\) (orange solid) and \(\Delta_{\mathrm{1p}}^{\mathrm{BCS}}(\mathbf k_F)\) (navy dashed) versus \(|\phi_{\mathbf k_F,d}|\) along the respective Fermi-surface contours \(\xi_{\mathbf k_F}=0\). The inset shows the same gaps versus \(\theta\). Dashed white horizontal lines in \textbf{a,b} and gray vertical lines in \textbf{c--e} denote \(\omega=0\). Spectral poles are broadened by a Lorentzian with HWHM \(\Gamma_L=5~\mathrm{meV}\). The parameters are \(p=0.15\), with \(\mathcal J_d=0.06~\mathrm{eV}\) and \(J=0.12~\mathrm{eV}\) for the correlated SC state, and \(\mathcal J_d=0\) and \(J=0.25~\mathrm{eV}\) for the conventional BCS state.\par}
\medskip

\hypertarget{sm-sc-fig11-physical-discussion}{}The SC pairing field mixes block states whose block electron numbers differ by two, converting the normal-state Cooper-surface level crossings at \(\xi_{\mathbf k}=0\) and \(\xi_{\mathbf k}=\mathcal J_{\mathbf k}\) into avoided crossings. In the nodal Bose metal, the dominant spectral weight shifts rapidly between distinct branches across the \(\xi_{\mathbf k}=0\) and \(\xi_{\mathbf k}=\mathcal J_{\mathbf k}\) contours [\hyperlink{sm-fig-c11-lorentzian}{Fig. 3(c)}]. The dominant \(\nu=-1\) branch of the correlated SC state instead evolves continuously and exhibits back-bending [\hyperlink{sm-fig-sc-one-particle}{Fig. 12(a)}], resembling the Bogoliubov quasiparticle dispersion near the Fermi surface in the conventional BCS state [\hyperlink{sm-fig-sc-one-particle}{Fig. 12(b)}]. However, the higher-energy \(\nu=+1\) branch retains weak but finite spectral weight in the region between \(\xi_{\mathbf k}=0\) and \(\xi_{\mathbf k}=\mathcal J_{\mathbf k}\) (qualitatively reminiscent of the \emph{waterfall} structure observed in ARPES \hyperlink{sm-ref-graf-2007}{[44]}, \hyperlink{sm-ref-valla-2007}{[45]}, \hyperlink{sm-ref-meevasana-2007}{[46]}), despite the \emph{single-band} character of the correlated SC state.

\hyperlink{sm-fig-sc-one-particle}{Fig. 12(a,c,e)} shows the evolution of the one-electron addition and removal weights of the correlated SC state along \(k_x=\pi\) at \(10~\mathrm K\). Although the SC block ground state \(|e_{\mathbf k,0}\rangle\) is nondegenerate, the SC EDCs also exhibit a multipeak structure, qualitatively resembling those of the nodal Bose metal [\hyperlink{sm-fig-c11-lorentzian}{Fig. 3(e,f)}]. The red \(\xi_{\mathbf k}=0\) EDC has a dominant removal peak near \(-68~\mathrm{meV}\) and two weaker peaks of unequal weight, corresponding to removal near \(-126~\mathrm{meV}\) and addition near \(+68~\mathrm{meV}\). The cyan \(\xi_{\mathbf k}=\mathcal J_{\mathbf k}\) EDC instead has a dominant addition peak near \(+65~\mathrm{meV}\) and two weaker peaks of unequal weight, corresponding to addition near \(+120~\mathrm{meV}\) and removal near \(-65~\mathrm{meV}\). Thus, the relative spectral weights of the one-electron removal and addition poles evolve with momentum along the \(k_x=\pi\) cut. When the cut crosses the Luttinger-surface contour \(\xi_{\mathbf k_L}=\mathcal J_{\mathbf k_L}/2\), the RML particle--hole symmetry enforces \(A_{\mathrm{SC}}(\mathbf k_L,\omega)=A_{\mathrm{SC}}(\mathbf k_L,-\omega)\), producing an EDC symmetric about \(\omega=0\) in \hyperlink{sm-fig-sc-one-particle}{Fig. 12(c)} (cf. the green EDC in \hyperlink{sm-fig-c11-lorentzian}{Fig. 3(e)}). In the conventional BCS comparison [\hyperlink{sm-fig-sc-one-particle}{Fig. 12(d)}], the spectral weight is redistributed only between the two Bogoliubov poles as \(\xi_{\mathbf k}\) changes sign, with equal weights at the Fermi surface \(\xi_{\mathbf k_F}=0\).

\hyperlink{sm-fig-sc-one-particle}{Fig. 12(f)} compares the one-particle excitation gaps of the correlated SC state, with \(\Delta_d^{\mathrm{SC}}\simeq7.29~\mathrm{meV}\) and \(\Delta_{\mathrm{1p}}^{\mathrm{SC}}(\mathbf k_a)\simeq68.0~\mathrm{meV}\), and the conventional BCS state, with \(\Delta_d^{\mathrm{BCS}}\simeq37.89~\mathrm{meV}\) and \(\Delta_{\mathrm{1p}}^{\mathrm{BCS}}(\mathbf k_a)\simeq50.8~\mathrm{meV}\), along their respective Fermi-surface contours. Therefore, a small nodal SC pairing field and a large antinodal excitation gap can coexist in the correlated SC state.
 
\par\medskip
\hypertarget{sm-fig-sc-spectral-redistribution}{}
\PrintAsset{figure5-sc-spectral-redistribution}
{\small\noindent \textbf{Figure 13 \textbar{} Integrated spectral weight of the correlated SC state and the conventional BCS state at \(0~\mathrm K\).} \textbf{a,b,} Changes in the integrated spectral weight, \(\delta W=W-W_{\mathrm N}\), relative to the respective normal states within a fixed energy window \([-\Omega,\Omega]\) for the correlated SC state in \textbf{a} and the conventional BCS state in \textbf{b}. \textbf{c,d,} Corresponding ratios \(W/W_{\mathrm N}\) versus \(J\). Dashed lines mark \(\delta W=0\) in \textbf{a,b} and \(W/W_{\mathrm N}=1\) in \textbf{c,d}. The parameters are \(p=0.15\), Lorentzian HWHM \(\Gamma_L=5~\mathrm{meV}\), and \(\mathcal J_d=0.06~\mathrm{eV}\) in \textbf{a,c} and \(0\) in \textbf{b,d}. For \(J=(0.12,0.16,0.20,0.25)~\mathrm{eV}\) in \textbf{a,b}, the order parameters are \(\Delta_d^{\mathrm{SC}}=(7.289,15.969,27.459,44.871)~\mathrm{meV}\) and \(\Delta_d^{\mathrm{BCS}}=(2.071,8.567,19.734,37.895)~\mathrm{meV}\), respectively.\par}
\medskip

\hypertarget{sm-sc-spectral-weight}{}The integrated spectral weight within a fixed energy window \([-\Omega,\Omega]\) (distinct from the integrated weight of the lower Hubbard band in the HK model \hyperlink{sm-ref-phillips-2020}{[17]}) is defined by 
\hypertarget{sm-eq-integrated-spectral-weight-definition}{}%
\FittedEquation{83}{%
W(\Omega)\equiv\int_{-\Omega}^{\Omega}d\omega\,\frac{1}{N_s}\sum_{\mathbf k}A(\mathbf k,\omega).
}
 The spectral function sum rule ensures \(W(\infty)=1\), so a loss from a finite window must be compensated outside that window. \hyperlink{sm-fig-sc-spectral-redistribution}{Fig. 13} displays the integrated spectral weight transfer in the correlated SC state and the conventional BCS state relative to the nodal Bose metal and the noninteracting normal state, respectively. Both pairing states exhibit negative changes \(\delta W(\Omega)=W(\Omega)-W_{\mathrm N}(\Omega)<0\), and their magnitudes increase with the pairing attraction \(J\) [\hyperlink{sm-fig-sc-spectral-redistribution}{Fig. 13(a,b)}]. Correspondingly, both ratios \(W/W_{\mathrm N}\) decrease with \(J\) at each fixed window and approach unity as the window is enlarged [\hyperlink{sm-fig-sc-spectral-redistribution}{Fig. 13(c,d)}]. The reduction in the integrated spectral weight is a direct consequence of SC gap opening (except at the nodes), which lowers the many-body ground-state energy while increasing the energy cost of adding or removing a single electron. The correlated SC state exhibits a greater relative reduction because the RML interaction and the pairing attraction together produce a larger SC pairing field.

\hypertarget{sm-sc-two-particle}{}
\subsection*{D. Spin-singlet removal spectra: middle ridge near Luttinger surface, and SC gap opening on Cooper surfaces}

\par\medskip
\hypertarget{sm-fig-sc-pair}{}
\PrintAsset{fig14-antinodal-and-maps}
{\small\noindent \textbf{Figure 14 \textbar{} Spin-singlet two-particle removal spectral functions of the correlated SC and conventional BCS states at \(10~\mathrm K\).} The left and right columns show the correlated SC and conventional BCS states, respectively. \textbf{a,b,} Two-particle removal spectra \(A^{(2)<}_{\Delta}(\mathbf k_a,-\mathbf k_a;\Omega,\omega_{\mathrm r})\) at the respective antinodes. \textbf{c,d,} Relative-energy-integrated two-particle removal spectra \(\overline A^{(2)<}_{\Delta}(\mathbf k,\Omega)\) at zero total energy \(\Omega=0\). White dotted curves denote the respective Fermi-surface contours \(\xi_{\mathbf k_F}=0\). Blue and red dots mark the node \(\mathbf k_n\) and antinode \(\mathbf k_a\), respectively. Insets show the corresponding spectra along their respective \(\xi_{\mathbf k_F}=0\) contours versus \(\theta=\tan^{-1}[(\pi-k_x)/(\pi-k_y)]\), from the node at \(45^\circ\) to the antinode at \(0^\circ\). \textbf{e,f,} Cuts through the maps in \textbf{c,d} along \(k_x=\pi\). Open circles mark local maxima of the broadened spectra. Solid red vertical lines mark \(\xi_{\mathbf k}=0\) in \textbf{e,f}; the dash-dotted green and solid cyan vertical lines in \textbf{e} mark \(\xi_{\mathbf k}=\mathcal J_{\mathbf k}/2\) and \(\xi_{\mathbf k}=\mathcal J_{\mathbf k}\), respectively. White dashed lines in \textbf{a,b} mark \(\Omega=0\) and \(\omega_{\mathrm r}=0\); those in the \textbf{c,d} insets mark \(\Omega=0\). Independent one-electron Lorentzian broadening has HWHM \(\Gamma_{\mathrm L}=5~\mathrm{meV}\), giving total-energy HWHM \(2\Gamma_{\mathrm L}=10~\mathrm{meV}\) in \textbf{c--f}. The parameters are \(p=0.15\), with \(J=0.12~\mathrm{eV}\) and \(\mathcal J_d=0.06~\mathrm{eV}\) for the correlated SC state, and \(J=0.25~\mathrm{eV}\) and \(\mathcal J_d=0\) for the conventional BCS state.\par}
\medskip

\hyperlink{sm-fig-sc-pair}{Fig. 14(a,b)} compares the spin-singlet two-particle removal spectral functions of the correlated SC and conventional BCS states at their respective antinodes. The dominant peaks in \hyperlink{sm-fig-sc-pair}{Fig. 14(a,b)} resolve not only the final states but also the intermediate states. In the correlated SC state, the dominant transitions \(|e_{\mathbf k_a,0}\rangle\to|e_{\mathbf k_a,l}\rangle\), with \(l=0,1,2\), produce peaks at total energies \(\Omega=0\), \(-27.72~\mathrm{meV}\), and \(-126.37~\mathrm{meV}\), respectively. The relative-energy structure resolves the two odd-fermion-parity intermediate-state energies [\hyperlink{sm-eq-sc-carpes-singlet-explicit}{Eq. (M20)}]. The symmetry of \(A^{(2)<}_{\Delta}\) under \(\omega_{\mathrm r}\to-\omega_{\mathrm r}\) also ensures that peaks away from \(\omega_{\mathrm r}=0\) always occur in pairs. In the conventional BCS state, the dominant peaks instead occur at \((\Omega,\omega_{\mathrm r})=(0,\pm E^{\mathrm{BCS}}_{\mathbf k_a})\) and \((-2E^{\mathrm{BCS}}_{\mathbf k_a},0)\), with \(E^{\mathrm{BCS}}_{\mathbf k_a}=50.80~\mathrm{meV}\), corresponding to final states with zero and two Bogoliubov quasiparticles, respectively.

Note that both pairing states exhibit a pair of side peaks pinned to the total energy \(\Omega=0\), associated with SC anomalous pair correlations; see the contributions with \(\delta(\Omega)\) in \hyperlink{sm-eq-b25}{Eqs. (C26)} and \hyperlink{sm-eq-sc-carpes-singlet-explicit}{(M20)}. By contrast, the two symmetric side peaks accidentally centered at \(\Omega=0\) in the nodal Bose metal [\hyperlink{sm-fig-two-particle-overview}{Fig. 4(c)}] arise from the contribution with \(\delta(\Omega-2\xi_{\mathbf k_a})=\delta(\Omega)\) [\hyperlink{sm-eq-c37}{Eq. (D42)}]. The SC pairing field lifts the degeneracy between \(|4_{\mathbf k}\rangle\) and \(|2_{\mathbf k,s}\rangle\) at the antinodes and shifts the corresponding side peaks induced by the RML interaction to finite total energy \(\Omega=E^{\mathrm{SC},e}_{\mathbf k_a,0}-E^{\mathrm{SC},e}_{\mathbf k_a,1}\simeq-27.72~\mathrm{meV}\) in \hyperlink{sm-fig-sc-pair}{Fig. 14(a)}.

Integrating over \(\omega_{\mathrm r}\) gives the sequential two-particle removal spectral function \(\overline A^{(2)<}_{\mathrm{SC},\Delta}(\mathbf k,\Omega)\) [\hyperlink{sm-eq-sc-carpes-singlet-integrated}{Eq. (M21)}]. \hyperlink{sm-fig-sc-pair}{Fig. 14(c)} shows that the zero-total-energy spectral function \(\overline A^{(2)<}_{\mathrm{SC},\Delta}(\mathbf k,0)\) has three ridges. Near each Cooper surface, the SC pairing field strongly mixes two nearly degenerate even-fermion-parity block states, such as \(|2_{\mathbf k,s}\rangle\) and \(|0_{\mathbf k}\rangle\) near \(\mathcal C_{2h}\). Such mixing enhances the spectral weight for two-electron removal back to the SC block ground state \(|e_{\mathbf k,0}\rangle\), with \(\langle e_{\mathbf k,0}|\hat\Delta_{\mathbf k}|e_{\mathbf k,0}\rangle\ne0\), producing the two outer ridges in \hyperlink{sm-fig-sc-pair}{Fig. 14(c)}.

\hypertarget{sm-sc-zero-energy-ridges-interpretation}{}Near the Luttinger surface, the SC pairing field strongly mixes the two nearly degenerate odd-fermion-parity block states \(|1_{\pm\mathbf k\sigma}\rangle\) and \(|3_{\mp\mathbf k\sigma}\rangle\). For example, in the return process \(|e_{\mathbf k,0}\rangle\to|o_{\mathbf k\sigma,\nu}\rangle\to|e_{\mathbf k,0}\rangle\), the first electron removal can reach the \(|3_{-\mathbf k\sigma}\rangle\) component of the intermediate state, and the second can act on its \(|1_{\mathbf k\sigma}\rangle\) component. Such mixing enhances the zero-total-energy return spectral weight and produces the middle ridge in \hyperlink{sm-fig-sc-pair}{Fig. 14(c)} without requiring a third gapless Cooper surface. No such ridge is resolved in the nodal Bose metal [\hyperlink{sm-fig-two-particle-overview}{Fig. 4(f)}]. The \emph{middle ridge} is therefore a hallmark of the correlated SC state in the MStJ model.

In the conventional BCS state, the zero-total-energy spectral function \(\overline A^{(2)<}_{\mathrm{BCS},\Delta}(\mathbf k,0)\) shows a single bright ridge near the Fermi surface [\hyperlink{sm-fig-sc-pair}{Fig. 14(d)}], where the SC pairing field strongly mixes the nearly degenerate empty and doubly occupied states. Correspondingly, the cut along \(k_x=\pi\) shows only a single maximum near the antinode [\hyperlink{sm-fig-sc-pair}{Fig. 14(f)}]. By contrast, the three ridges of the correlated SC state [\hyperlink{sm-fig-sc-pair}{Fig. 14(c)}] give three maxima along the same cut [\hyperlink{sm-fig-sc-pair}{Fig. 14(e)}].

\par\medskip
\hypertarget{sm-fig-sc-pair-cuts}{}
\PrintAsset{fig15-pair-cuts}
{\small\noindent \textbf{Figure 15 \textbar{} Spin-singlet two-particle and pair-removal spectral functions along \(k_x=\pi\) at \(10~\mathrm K\).} Panels \textbf{a,c} show the correlated SC state and \textbf{b,d} the conventional BCS state. \textbf{a,b,} Sequential spin-singlet two-particle removal spectra \(\overline A^{(2)<}_{\Delta}(\mathbf k,\Omega)\), integrated over relative energy. \textbf{c,d,} Connected simultaneous pair-removal spectra \(S^{<,c}_{\Delta}(\mathbf k,\Omega)\). In \textbf{a,c}, the solid red and cyan lines mark the two Cooper surfaces \(\mathcal C_{2e}:\xi_{\mathbf k}=0\) and \(\mathcal C_{2h}:\xi_{\mathbf k}=\mathcal J_{\mathbf k}\). The dash-dotted green line marks the Luttinger surface \(\xi_{\mathbf k}=\mathcal J_{\mathbf k}/2\). The solid red line in \textbf{b,d} marks the Fermi surface \(\xi_{\mathbf k}=0\). White dotted curves denote \(\Omega=E^{\mathrm{SC},e}_{\mathbf k,0}-E^{\mathrm{SC},e}_{\mathbf k,l}\) for \(l=1,2\) in \textbf{a,c} and \(\Omega=-2\sqrt{\xi_{\mathbf k}^2+|\sqrt{2}\Delta_{\mathbf k,d}|^2}\) in \textbf{b,d}. \textbf{e,f,} EDCs of \(S^{<,c}_{\mathrm{SC},\Delta}(\mathbf k,\Omega)\) along \(k_x=\pi\), with adjacent curves offset by \(150~\mathrm{eV}^{-1}\). Red, green and cyan EDCs mark \(\xi_{\mathbf k}=0\), \(\xi_{\mathbf k}=\mathcal J_{\mathbf k}/2\) and \(\xi_{\mathbf k}=\mathcal J_{\mathbf k}\), respectively. Horizontal white dashed lines in \textbf{a--d} and vertical gray dashed lines in \textbf{e,f} mark \(\Omega=0\). The one-electron Lorentzian HWHM is \(\Gamma_{\mathrm L}=5~\mathrm{meV}\) in \textbf{a--d} and \(1~\mathrm{meV}\) in \textbf{e,f}. The remaining parameters match \hyperlink{sm-fig-sc-pair}{Fig. 14}: \(p=0.15\), with \(J=0.12~\mathrm{eV}\) and \(\mathcal J_d=0.06~\mathrm{eV}\) for the correlated SC state, and \(J=0.25~\mathrm{eV}\) and \(\mathcal J_d=0\) for the conventional BCS state.\par}
\medskip

The gapless spin-singlet Cooper-pair excitations of the nodal Bose metal arise from level crossings between even-fermion-parity block ground states. The SC pairing field mixes these states, converts both level crossings into avoided crossings, and opens a pair-excitation gap \(\Delta_{\mathrm{2p}}^{\mathrm{SC}}\) on the Cooper surfaces except at the nodes, where the spin-singlet two-particle excitation gap is

\hypertarget{sm-eq-sc-two-particle-gap}{}%
\FittedEquation{84}{%
\Delta_{\mathrm{2p}}^{\mathrm{SC}}(\mathbf k)=E^{\mathrm{SC},e}_{\mathbf k,1}-E^{\mathrm{SC},e}_{\mathbf k,0}.
}
 
\hyperlink{sm-fig-sc-pair-cuts}{Fig. 15(a,b)} shows the sequential two-particle removal spectral function \(\overline A^{(2)<}_{\Delta}(\mathbf k,\Omega)\) along \(k_x=\pi\). For the correlated SC state, \(\overline A^{(2)<}_{\mathrm{SC},\Delta}=S^<_{\mathrm{SC},\Delta}+\mathcal I_{\mathrm{SC},\Delta}\) [\hyperlink{sm-eq-sc-singlet-pair-relation}{Eq. (M22)}], where \(S^<_{\mathrm{SC},\Delta}\) is the simultaneous pair-removal spectral function [\hyperlink{sm-eq6}{Eq. (29)}]. The interference correction \(\mathcal I_{\mathrm{SC},\Delta}\) arises because sequential removal adds probabilities for paths through distinct intermediate-state subspaces, whereas simultaneous pair removal adds their amplitudes before squaring (see \hyperlink{sm-sc-singlet-carpes}{Appendix M.2}). Both removal spectra carry spectral weight at \(\Omega=0\) because two-electron removal can return the SC block ground state \(|e_{\mathbf k,0}\rangle\) to itself. The conventional BCS state likewise exhibits a zero-total-energy signal in \hyperlink{sm-fig-sc-pair-cuts}{Fig. 15(b)}.

To isolate the pair excitations, the connected pair-removal spectral function is defined by

 \hypertarget{sm-eq-sc-spectral-pair}{} 
\hypertarget{sm-eq-sc-pair-connected-definition}{}%
\FittedEquation{85}{%
\begin{aligned}
S^{<,c}_{\Delta}(\mathbf k,\Omega)
&\equiv\int_{-\infty}^{\infty}\frac{dt}{2\pi}\,e^{i\Omega t}
\Big\langle\big(\hat\Delta_{\mathbf k}^{\dagger}(0)-\langle\hat\Delta_{\mathbf k}^{\dagger}(0)\rangle\big)
\big(\hat\Delta_{\mathbf k}(t)-\langle\hat\Delta_{\mathbf k}(t)\rangle\big)\Big\rangle\\
&=S^<_{\Delta}(\mathbf k,\Omega)-|\langle\hat\Delta_{\mathbf k}\rangle|^2\delta(\Omega).
\end{aligned}
}
 
For the correlated SC state, subtracting the zero-total-energy contribution from \hyperlink{sm-eq-sc-pair-lehmann}{Eq. (M17)} gives

\hypertarget{sm-eq-sc-pair-connected-explicit}{}%
\FittedEquation{86}{%
\begin{aligned}
S^{<,c}_{\mathrm{SC},\Delta}(\mathbf k,\Omega)
={}&\sum_{l,l'=0}^{2}p^{\mathrm{SC},e}_{\mathbf k,l}
|x_{\mathbf k,l'}^*y_{\mathbf k,l}+y_{\mathbf k,l'}^*z_{\mathbf k,l}|^2
\delta(\Omega-E^{\mathrm{SC},e}_{\mathbf k,l}+E^{\mathrm{SC},e}_{\mathbf k,l'})\\
&+2\sum_{\nu,\nu'=\pm1}p^{\mathrm{SC},o}_{\mathbf k,\nu}
|u_{\mathbf k,\nu'}^*v_{\mathbf k,\nu}|^2
\delta(\Omega-E^{\mathrm{SC},o}_{\mathbf k,\nu}+E^{\mathrm{SC},o}_{\mathbf k,\nu'})
-\big|\langle\hat\Delta_{\mathbf k}\rangle_{\mathrm{SC}}\big|^2\delta(\Omega).
\end{aligned}
}
 
In the zero-temperature limit, the nondegeneracy of the SC block ground state ensures the cancellation of the zero-total-energy contributions, leaving

\hypertarget{sm-eq-sc-pair-connected-ground}{}%
\FittedEquation{87}{%
\left.S^{<,c}_{\mathrm{SC},\Delta}(\mathbf k,\Omega)\right|_{T=0}
=\sum_{l=1}^{2}
\left|x_{\mathbf k,l}^*y_{\mathbf k,0}+y_{\mathbf k,l}^*z_{\mathbf k,0}\right|^2
\delta\!\left(\Omega+E^{\mathrm{SC},e}_{\mathbf k,l}-E^{\mathrm{SC},e}_{\mathbf k,0}\right).
}
 
At low temperatures, the connected pair-removal spectral function thus probes transitions from the block ground state to excited states. The two-state approximation gives a splitting of \(2\sqrt{\xi_{\mathbf k}^2+4|\Delta_{\mathbf k,d}|^2}\) near \(\mathcal C_{2e}\) and \(2\sqrt{(\xi_{\mathbf k}-\mathcal J_{\mathbf k})^2+4|\Delta_{\mathbf k,d}|^2}\) near \(\mathcal C_{2h}\), which reduces to \(4|\Delta_{\mathbf k,d}|\) on either contour. Along \(k_x=\pi\), the two-state splitting gives \(27.89~\mathrm{meV}\) on \(\mathcal C_{2e}\) and \(27.17~\mathrm{meV}\) on \(\mathcal C_{2h}\). For comparison, \hyperlink{sm-fig-sc-pair-cuts}{Fig. 15(c)} shows exact pair-excitation gaps [\hyperlink{sm-eq-sc-two-particle-gap}{Eq. (84)}] of \(27.72~\mathrm{meV}\) and \(27.00~\mathrm{meV}\) on the respective Cooper surfaces.

The two avoided crossings cause the \(l=1\) pair-removal branch to bend back near each Cooper surface, much as the \(\nu=-1\) one-electron removal branch does in \hyperlink{sm-fig-sc-one-particle}{Fig. 12(a)}. The EDCs in \hyperlink{sm-fig-sc-pair-cuts}{Fig. 15(e,f)} resolve how the peak energies and weights vary along the \(l=1\) and \(l=2\) branches. On the Luttinger surface, the SC pairing field lifts the degeneracy between \(|0_{\mathbf k_L}\rangle\) and \(|4_{\mathbf k_L}\rangle\) by coupling both to \(|2_{\mathbf k_L,s}\rangle\), producing a small splitting of \(6.33~\mathrm{meV}\) between the two pair-removal branches [\hyperlink{sm-fig-sc-pair-cuts}{Fig. 15(e)}]. In the conventional BCS state, there is only one pair-removal branch [\hyperlink{sm-fig-sc-pair-cuts}{Fig. 15(b,d)}], with excitation gap

\hypertarget{sm-eq-bcs-two-particle-gap}{}%
\FittedEquation{88}{%
\Delta_{\mathrm{2p}}^{\mathrm{BCS}}(\mathbf k)=2E_{\mathbf k}^{\mathrm{BCS}}
=2\sqrt{\xi_{\mathbf k}^2+|\sqrt{2}\Delta_{\mathbf k,d}|^2}.
}
 
\par\medskip
\hypertarget{sm-fig-sc-coherent-pair}{}
\PrintAsset{fig16-coherent-weights-and-edc}
{\small\noindent \textbf{Figure 16 \textbar{} Coherent pair-removal weights and sequential and simultaneous pair-removal spectral functions at \(\mathbf k_a\) for \(T=10~\mathrm K\).} The left and right columns show the correlated SC and conventional BCS states, respectively. \textbf{a,b,} Coherent pair-removal weights \(|\langle\hat\Delta_{\mathbf k}\rangle_{\mathrm{SC}}|^2\) and \(|\langle\hat\Delta_{\mathbf k}\rangle_{\mathrm{BCS}}|^2\). White dotted curves denote the respective Fermi-surface contours \(\xi_{\mathbf k_F}=0\). Blue and red dots mark the node \(\mathbf k_n\) and antinode \(\mathbf k_a\), respectively. The insets show the coherent pair-removal weights along \(k_x=\pi\). Open circles mark local maxima. Vertical lines mark the same contours as in \hyperlink{sm-fig-sc-pair}{Fig. 14(e,f)}. \textbf{c,d,} Pair-removal spectra at the respective antinodes. Solid red curves show the sequential spectra \(\overline A^{(2)<}_{\Delta}\); dashed blue curves show the simultaneous spectra \(S^<_{\Delta}\). Shading shows the coherent zero-total-energy contribution \(|\langle\hat\Delta_{\mathbf k_a}\rangle|^2L_{2\Gamma_{\mathrm L}}(\Omega)\). Dashed vertical lines mark \(\Omega=0\). The parameters match \hyperlink{sm-fig-sc-pair}{Fig. 14}: one-electron Lorentzian HWHM \(\Gamma_{\mathrm L}=5~\mathrm{meV}\) and \(p=0.15\), with \(J=0.12~\mathrm{eV}\) and \(\mathcal J_d=0.06~\mathrm{eV}\) for the correlated SC state, and \(J=0.25~\mathrm{eV}\) and \(\mathcal J_d=0\) for the conventional BCS state.\par}
\medskip

\hyperlink{sm-fig-sc-coherent-pair}{Fig. 16(a,b)} shows the distributions of the coherent pair-removal weight \(\lvert\langle\hat\Delta_{\mathbf k}\rangle\rvert^2\) in the correlated SC and conventional BCS states. \(\lvert\langle\hat\Delta_{\mathbf k}\rangle\rvert^2\) is the coefficient of the coherent \(\delta(\Omega)\) term in the simultaneous pair-removal spectral function [\hyperlink{sm-eq-sc-pair-connected-definition}{Eq. (85)}]. Compared with \hyperlink{sm-fig-sc-pair}{Fig. 14(c,d)}, \hyperlink{sm-fig-sc-coherent-pair}{Fig. 16(a,b)} shows one striking difference: the middle ridge in \hyperlink{sm-fig-sc-pair}{Fig. 14(c)} is absent from \hyperlink{sm-fig-sc-coherent-pair}{Fig. 16(a)}. Such a difference has the same physical origin as the interference correction. Near the Luttinger surface, the two amplitudes for return to \(|e_{\mathbf k,0}\rangle\) through the intermediate states \(|o_{\pm\mathbf k\sigma,\nu}\rangle\) with \(\nu=\pm1\) have opposite signs for a given removal ordering. Sequential removal adds the squared moduli of the two amplitudes, whereas simultaneous removal adds the amplitudes before taking the squared modulus [\hyperlink{sm-eq-sc-singlet-interference-correction}{Eq. (M23)}]. The two amplitudes therefore partially cancel in simultaneous removal, suppressing the coherent weight near the Luttinger surface and explaining the absence of the middle ridge in \hyperlink{sm-fig-sc-coherent-pair}{Fig. 16(a)}.

\hyperlink{sm-fig-sc-coherent-pair}{Fig. 16(c)} shows that the broadened coherent zero-total-energy contribution \(\lvert\langle\hat\Delta_{\mathbf k_a}\rangle_{\mathrm{SC}}\rvert^2 L_{2\Gamma_{\mathrm L}}(\Omega)\) nearly overlaps the central peaks of both the sequential and simultaneous pair-removal spectra at \(\Omega=0\). Both spectra of the correlated SC state also have poles at \(-27.72\) and \(-126.37~\mathrm{meV}\), corresponding to the \(l=1\) and \(l=2\) excitations in \hyperlink{sm-fig-sc-pair-cuts}{Fig. 15(a,c)}, respectively. In general, the interference correction changes the spectral weight of a pole without shifting its energy.

The two spectra of the conventional BCS state overlap at the resolution shown in \hyperlink{sm-fig-sc-coherent-pair}{Fig. 16(d)} because the interference correction contributes only at \(\Omega=0\) and its weight is exponentially small at the antinode \(\mathbf k_a\) for \(T=10~\mathrm K\) [\hyperlink{sm-eq-sc-singlet-bcs-interference}{Eq. (M25)}]. In both the nodal Bose metal [\hyperlink{sm-eq11}{Eq. (35)}] and the conventional BCS state, the interference correction requires thermal occupation of excited states. In the correlated SC state, the interference correction generally remains nonzero even at zero temperature when \(\mathcal J_{\mathbf k}\ne0\) and \(\Delta_{\mathbf k,d}\ne0\), because both odd-fermion-parity intermediate states are accessible from the SC block ground state and their amplitudes interfere.

\hypertarget{sm-sc-thermodynamics-main}{}
\subsection*{E. Thermodynamics with nodal density of states}

The one-particle density of states (DOS) per spin of the correlated SC state is

\hypertarget{sm-eq-main-sc-full-dos}{}%
\FittedEquation{89}{%
\begin{aligned}
\rho_{\mathrm{SC}}(\omega,T)&\equiv\int_{\mathrm{BZ}}\frac{d^2\mathbf k}{(2\pi)^2}\,A_{\mathrm{SC}}(\mathbf k,\omega;T)\\
&=\int_{\mathrm{BZ}}\frac{d^2\mathbf k}{(2\pi)^2}\sum_{\nu=\pm1}
\Biggl\{\sum_{l=0}^{2}\bigl(p^{\mathrm{SC},e}_{\mathbf k,l}+p^{\mathrm{SC},o}_{\mathbf k,\nu}\bigr)
\Bigl[Z^{\mathrm{SC},+}_{\mathbf k,\nu l}\delta(\omega-E^{\mathrm{SC},o}_{\mathbf k,\nu}+E^{\mathrm{SC},e}_{\mathbf k,l})\\
&\qquad\qquad+Z^{\mathrm{SC},-}_{\mathbf k,\nu l}\delta(\omega+E^{\mathrm{SC},o}_{\mathbf k,\nu}-E^{\mathrm{SC},e}_{\mathbf k,l})\Bigr]\\
&\quad+\frac52\bigl(p^{\mathrm{SC}}_{\mathbf k,2}+p^{\mathrm{SC},o}_{\mathbf k,\nu}\bigr)
\Bigl[|v_{\mathbf k,\nu}|^2\delta(\omega{-}E^{\mathrm{SC},o}_{\mathbf k,\nu}{+}2\xi_{\mathbf k})+|u_{\mathbf k,\nu}|^2\delta(\omega{+}E^{\mathrm{SC},o}_{\mathbf k,\nu}{-}2\xi_{\mathbf k})\Bigr]\Biggr\},
\end{aligned}
}
 
where \(\rho_{\mathrm{SC}}(0,T\ne0)>0\) and \(\int_{-\infty}^{\infty}d\omega\,\rho_{\mathrm{SC}}(\omega,T)=1\). At low temperatures, the dominant contributions to the low-energy DOS come from regions near the nodes. A local expansion near each node in the moving tangent--normal frame [\hyperlink{sm-fig-model-normal-observables}{Fig. 5(a)}] reduces the one-particle excitation gap to the Dirac dispersion \hypertarget{sm-eq-sc-nodal-spectrum}{}\(\Delta_{\mathrm{1p}}^{\mathrm{SC}}(\mathbf k)\simeq\sqrt{v_n^2k_\perp^2+v_\Delta^2k_\parallel^2}\), where \hypertarget{sm-eq-sc-nodal-fermi-velocity}{}\(v_n=\left|\nabla_{\mathbf k}\xi_{\mathbf k}\right|_{\mathbf k_n}\) and \hypertarget{sm-eq-sc-nodal-gap-velocity}{}\(v_\Delta=\sqrt2\,\left|\partial_{k_\parallel}\Delta_{\mathbf k,d}\right|_{\mathbf k_n}\). The linear SC pairing field \(\sqrt2\Delta_{\mathbf k,d}\simeq v_\Delta k_\parallel\) controls the lowest-energy excitations, and the competition with the quadratic RML interaction \(\mathcal J_{\mathbf k}\simeq u_n k_\parallel^2\) sets the crossover energy scale \(E_*=v_\Delta^2/u_n=2|\Delta_d|^2/\mathcal J_d\). The local expansion near the nodes requires both \(T\) and \(|\omega|\) to be small compared with \(E_*\).

Introduce the dimensionless variables \(z=\omega/T\) and \(g_*=T/E_*\). The SC DOS within the leading-order nodal approximation takes the scaling form \(\rho_{\mathrm{SC}}(\omega,T)\simeq V(T)^{-1}T\,\mathcal R_{\mathrm{SC}}(z,g_*)\), where \(V(T)=\pi^2 v_n(T)v_\Delta(T)\) [\hyperlink{sm-eq-sc-dos-scaling-density}{Eq. (N13)}]. Expanding the SC scaling function \(\mathcal R_{\mathrm{SC}}(z,g_*)\) to first order in \(g_*\) at fixed nonzero \(z\) gives [\hyperlink{sm-eq-sc-dos-first-correction}{Eq. (N22)}]

\hypertarget{sm-eq-sc-nodal-clean-dos}{}%
\FittedEquation{90}{%
\rho_{\mathrm{SC}}(\omega,T)\simeq \frac{\pi|\omega|}{V(T)}
\left\{1-\frac{7g_*}{16}\left[3|z|\tanh\!\left(\frac{|z|}{2}\right)
+\frac{z^2}{2}\operatorname{sech}^2\!\left(\frac{|z|}{2}\right)\right]\right\},
}
 
where the leading term is the conventional Dirac DOS \(V(T)^{-1}\pi|\omega|\). For \(|\omega|\ll T\ll E_*\), the small-\(|\omega|/T\) expansion gives [\hyperlink{sm-eq-sc-dos-small-z}{Eq. (N23)}]

\hypertarget{sm-eq-main-sc-dos-small-frequency}{}%
\FittedEquation{91}{%
\rho_{\mathrm{SC}}(\omega,T)\simeq \frac{\pi|\omega|}{V(T)}
\left[1-\frac{7\omega^2}{8TE_*}+\frac{7\omega^4}{64T^3E_*}
+O\!\left(\frac{\omega^6}{T^5E_*}\right)\right].
}
 
For \(T\ll|\omega|\ll E_*\), the large-\(|\omega|/T\) expansion gives [\hyperlink{sm-eq-sc-dos-large-z}{Eq. (N24)}]

\hypertarget{sm-eq-main-sc-dos-large-frequency}{}%
\FittedEquation{92}{%
\rho_{\mathrm{SC}}(\omega,T)\simeq \frac{\pi|\omega|}{V(T)}\Biggl[1-\frac{21|\omega|}{16E_*}+\frac{7|\omega|}{8E_*}\left(3-\frac{|\omega|}{T}\right)e^{-|\omega|/T}+O\!\left(\frac{\omega^2}{TE_*}\,e^{-2|\omega|/T}\right)\Biggr].
}
 
The SC DOS also exhibits emergent particle--hole symmetry within the leading-order nodal approximation and approaches the conventional Dirac DOS for \(T,|\omega|\ll E_*\) [\hyperlink{sm-fig-sc-nodal-dos}{Fig. 17(a--c)}]. The departure from the Dirac limit becomes more pronounced at higher temperatures and larger values of \(|\omega|/T\) [\hyperlink{sm-fig-sc-nodal-dos}{Fig. 17(c)}]. The RML interaction produces a shallow minimum near \(T/|\omega|\simeq0.26\) in \hyperlink{sm-fig-sc-nodal-dos}{Fig. 17(b)} at fixed \(\omega=0.1~\mathrm{meV}\) [cf. the minimum in \hyperlink{sm-fig-model-normal-observables}{Fig. 5(c)}]. Since the term \((3-|z|)e^{-|z|}\) in the large-\(|z|\) expansion [\hyperlink{sm-eq-main-sc-dos-large-frequency}{Eq. (92)}] reaches its minimum at \(|z|=4\), the first-order RML interaction correction predicts a nearby value of \(T/|\omega|\simeq0.25\). \hyperlink{sm-fig-sc-nodal-dos}{Fig. 17(d)} compares the nodal scaling form [\hyperlink{sm-eq-sc-dos-scaling-density}{Eq. (N13)}] and the first-order RML interaction correction [\hyperlink{sm-eq-sc-nodal-clean-dos}{Eq. (90)}] with the SC DOS [\hyperlink{sm-eq-main-sc-full-dos}{Eq. (89)}]. Retaining the RML interaction in the nodal \(d\)-wave BCS Hamiltonian captures the temperature dependence of the normalized DOS in \hyperlink{sm-fig-sc-nodal-dos}{Fig. 17(d)} at fixed \(\omega=0.1~\mathrm{meV}\), below the crossover energy scale \(E_*(0)/|\omega|=17.71\).

\par\medskip
\hypertarget{sm-fig-sc-nodal-dos}{}
\PrintAsset{nodal-dos}
{\small\noindent \textbf{Figure 17 \textbar{} Nodal density of states of the correlated SC state.} \textbf{a,} Per-spin SC DOS \(\rho_{\mathrm{SC}}(\omega,T)\) over the full energy range at \(T=1~\mathrm K\). The inset shows \(\rho_{\mathrm{SC}}\) versus \(\omega/T\) in the low-energy window and compares it with the small-\(|\omega|/T\) expansion in \hyperlink{sm-eq-main-sc-dos-small-frequency}{Eq. (91)}. \textbf{b,} \(\rho_{\mathrm{SC}}(\omega,T)/\rho_{\mathrm{SC}}(\omega,0)\) versus \(T/|\omega|\) at fixed \(\omega=0.1~\mathrm{meV}\), with \(\rho_{\mathrm{SC}}(\omega,0)\) evaluated from the zero-temperature SC DOS, compared with the large-\(|\omega|/T\) expansion in \hyperlink{sm-eq-main-sc-dos-large-frequency}{Eq. (92)}. The inset compares the zero-temperature SC DOS with the large-\(|\omega|/T\) expansion at \(T=0\). \textbf{c,} Scaled SC DOS compared with the Dirac limit [\hyperlink{sm-eq-sc-dos-dirac-integral}{Eq. (N10)}]. \textbf{d,} Temperature dependence of the SC DOS [\hyperlink{sm-eq-main-sc-full-dos}{Eq. (89)}], the nodal scaling form [\hyperlink{sm-eq-sc-dos-scaling-density}{Eq. (N13)}], and the first-order RML interaction correction [\hyperlink{sm-eq-sc-nodal-clean-dos}{Eq. (90)}]. The parameters are: \(\mathcal J_d=0.06~\mathrm{eV}\), \(J=0.12~\mathrm{eV}\), and \(p=0.15\).\par}
\medskip

\hyperlink{sm-fig-sc-thermodynamics}{Fig. 18(a--d)} compares the thermodynamic quantities of the correlated SC state with those of the nodal Bose metal at the same electron density \(n_e\). For \(T\ll\min\{T_{c,d},E_*(0)\}\), the linear SC pairing field dominates the quadratic RML interaction near the four gapless Dirac nodes. The low-temperature behavior of the correlated SC state thus has the same leading thermodynamics as a conventional nodal \(d\)-wave BCS state. The resulting entropy and specific heat are [\hyperlink{sm-eq-sc-thermo-nodal-entropy}{Eqs. (O25)} and \hyperlink{sm-eq-sc-thermo-nodal-heat-capacity}{(O26)}]

\hypertarget{sm-eq-main-sc-nodal-thermodynamics}{}%
\FittedEquation{93}{%
s_{\mathrm{SC}}(T)\simeq \frac{9\pi\zeta(3)T^2}{V(0)},\qquad
c_{\mathrm{SC}}(T)\simeq \frac{18\pi\zeta(3)T^2}{V(0)},
}
 
with \(c_{\mathrm{SC}}/s_{\mathrm{SC}}\to2\) as \(T\to0\), whereas gapless pair excitations on the two Cooper surfaces of the nodal Bose metal instead give \(s_{\mathrm N}\simeq c_{\mathrm N}\simeq\gamma_0T\) [\hyperlink{sm-eq-mstj-heat-capacity-low-temperature}{Eq. (45)}]. The different powers also account for the free-energy curves in \hyperlink{sm-fig-sc-thermodynamics}{Fig. 18(b)}: \(f_{\mathrm{SC}}(T)-f_{\mathrm{SC}}(0)\simeq-3\pi\zeta(3)T^3/V(0)\), whereas \(f_{\mathrm N}(T)-f_{\mathrm N}(0)\simeq-\gamma_0T^2/2\). The insets of \hyperlink{sm-fig-sc-thermodynamics}{Fig. 18(a--c)} show that the low-temperature asymptotes visibly deviate from the exact results as \(T/T_{c,d}\) approaches the small crossover scale \(E_*(0)/T_{c,d}\simeq0.144\).

For \(0<T<T_{c,d}\), the SC entropy is lower than that of the nodal Bose metal, and the two entropies coincide at \(T_{c,d}=143.13~\mathrm K\) [\hyperlink{sm-fig-sc-thermodynamics}{Fig. 18(a)}]. The area between the two entropy curves up to \(T_{c,d}\) gives the condensation energy, \(\epsilon_{\mathrm{conden}}=f_{\mathrm N}(0)-f_{\mathrm{SC}}(0)=0.03793~\mathrm{meV}\) per site [\hyperlink{sm-eq-sc-thermo-condensation-energy-definition}{Eq. (O6)}], which is the free-energy gain from SC pairing at zero temperature. At the continuous SC transition, the free energies of the correlated SC state and the nodal Bose metal meet at \(T_{c,d}\) with the same slope, so there is no latent heat. Thus the entropy, free energy, and specific heat follow the pattern of a conventional BCS transition, even though the RML interaction induces additional excitation branches in the correlated SC state [\hyperlink{sm-fig-sc-pair-cuts}{Fig. 15}].

The specific-heat jump ratio at \(T_{c,d}\), defined by \(\Delta c/c_{\mathrm N}(T_{c,d}^{+})=[c_{\mathrm{SC}}(T_{c,d}^{-})-c_{\mathrm N}(T_{c,d}^{+})]/c_{\mathrm N}(T_{c,d}^{+})\), is nonuniversal in the correlated SC state. At fixed \(J/t=0.4\) and \(p=0.15\), \hyperlink{sm-fig-sc-thermodynamics}{Fig. 18(e)} shows that the ratio changes from \(1.023\) for the conventional \(d\)-wave BCS state at \(\mathcal J_d=0\) to \(0.42\) for the correlated SC state at \(\mathcal J_d=0.06~\mathrm{eV}\).

The thermodynamic critical field in \hyperlink{sm-fig-sc-thermodynamics}{Fig. 18(d)} is determined by the free-energy difference, \(H_c(T)/H_c(0)=\sqrt{[f_{\mathrm N}(T)-f_{\mathrm{SC}}(T)]/\epsilon_{\mathrm{conden}}}\) [\hyperlink{sm-eq-sc-thermo-critical-field-ratio}{Eq. (O8)}], and the leading low-temperature behavior is \(H_c(T)/H_c(0)\simeq1-\gamma_0 T^2/(4\epsilon_{\mathrm{conden}})\) [\hyperlink{sm-eq-sc-thermo-low-t-critical-field}{Eq. (O29)}]. As \(T\) approaches \(T_{c,d}\) from below, \(f_{\mathrm N}(T)-f_{\mathrm{SC}}(T)\simeq\Delta c\,(T_{c,d}-T)^2/(2T_{c,d})\), so \(H_c(T)\) vanishes linearly at the SC transition.

\hyperlink{sm-fig-sc-thermodynamics}{Fig. 18(f)} shows the zero-temperature thermodynamic critical field \(H_c(0)\) at different interaction strengths in units of \(H_t=\sqrt{8\pi t/v_0}\). Since \(H_c(0)/H_t=\sqrt{\epsilon_{\mathrm{conden}}/t}\), \hyperlink{sm-fig-sc-thermodynamics}{Fig. 18(f)} essentially measures the zero-temperature condensation energy, whereas the specific-heat jump ratio in \hyperlink{sm-fig-sc-thermodynamics}{Fig. 18(e)} measures the relative change in free-energy curvature at the SC transition.

\par\medskip
\hypertarget{sm-fig-sc-thermodynamics}{}
\PrintAsset{thermodynamics}
{\small\noindent \textbf{Figure 18 \textbar{} Thermodynamics of the correlated SC state and nodal Bose metal at fixed electron density \(n_e=1-p\).} \textbf{a,} Entropy per site normalized by \(s_{\mathrm N}(T_{c,d})\). \textbf{b,} Helmholtz free energy per site relative to \(f_{\mathrm N}(0)\), normalized by \(f_{\mathrm N}(0)-f_{\mathrm N}(T_{c,d})\). \textbf{c,} Specific heat per site normalized by \(c_{\mathrm N}(T_{c,d})\). The dots mark the one-sided limits at the transition. The insets in \textbf{a--c} show both states for \(0\le T/T_{c,d}\le0.3\) with the same normalization as the main panels. Black dashed curves in the insets show the leading low-temperature forms for the SC state. \textbf{d,} Thermodynamic critical field \(H_c(T)/H_c(0)\) from \hyperlink{sm-eq-sc-thermo-critical-field-ratio}{Eq. (O8)} and its low-temperature expansion [\hyperlink{sm-eq-sc-thermo-low-t-critical-field}{Eq. (O29)}]. \textbf{e,} Specific-heat jump ratio \(\Delta c/c_{\mathrm N}(T_{c,d}^{+})\) versus \(J\) and \(\mathcal J_d\) at \(p=0.15\). \textbf{f,} Zero-temperature thermodynamic critical field on the same grid, in units of \(H_t=\sqrt{8\pi t/v_0}\): \(H_c(0)/H_t=\sqrt{\epsilon_{\mathrm{conden}}/t}\) [\hyperlink{sm-eq-sc-thermo-critical-field}{Eq. (O7)}]. Panels \textbf{a--d} use \(\mathcal J_d=0.06~\mathrm{eV}\), \(J=0.12~\mathrm{eV}\), and \(p=0.15\).\par}
\medskip
 \hypertarget{sm-references}{}\hypertarget{sm-references-title}{}
\begingroup
\renewcommand{\refname}{References}

\endgroup

\hypertarget{sm-appendix-a}{}\hypertarget{sm-appendix-a-title}{}
\section*{Appendix A. Exact operator identities}

\hypertarget{sm-superexchange-square-lattice-title}{}
\subsection*{A.1. Superexchange interaction on the square lattice}

\hypertarget{sm-appendix-a-derivation}{}Define the spin-singlet bond operator 
\hypertarget{sm-eq-a3}{}%
\FittedEquation{A1}{%
\hat b_{ij}\equiv\frac1{\sqrt2}\bigl(
        \hat c_{i\uparrow}\hat c_{j\downarrow}
        -\hat c_{i\downarrow}\hat c_{j\uparrow}\bigr).
}
 Using \(\hat S_i^+=\hat S_i^x+i\hat S_i^y\), \(\hat S_i^-=\hat S_i^x-i\hat S_i^y\), and \(\hat S_i^z=(\hat n_{i\uparrow}-\hat n_{i\downarrow})/2\) gives 
\hypertarget{sm-eq-a4}{}%
\FittedEquation{A2}{%
\begin{aligned}
        \hat b^\dagger_{ij}\hat b_{ij}
        &=\frac12\bigl(\hat c^\dagger_{j\downarrow}\hat c^\dagger_{i\uparrow}
        -\hat c^\dagger_{j\uparrow}\hat c^\dagger_{i\downarrow}\bigr)
        \bigl(\hat c_{i\uparrow}\hat c_{j\downarrow}-\hat c_{i\downarrow}\hat c_{j\uparrow}\bigr)\\
        &=\frac12\bigl(\hat n_{i\uparrow}\hat n_{j\downarrow}
        +\hat n_{i\downarrow}\hat n_{j\uparrow}
        -\hat S_i^+\hat S_j^- -\hat S_i^-\hat S_j^+\bigr),
        \end{aligned}
}
 and 
\hypertarget{sm-eq-a5}{}%
\FittedEquation{A3}{%
\hat S_i^z\hat S_j^z-\frac14\hat n_i\hat n_j
        =-\frac12\bigl(\hat n_{i\uparrow}\hat n_{j\downarrow}
        +\hat n_{i\downarrow}\hat n_{j\uparrow}\bigr).
}
 Then the superexchange term becomes 
\hypertarget{sm-eq-a6}{}%
\FittedEquation{A4}{%
\hat{\mathbf S}_i\cdot\hat{\mathbf S}_j-\frac14\hat n_i\hat n_j =\hat S_i^z\hat S_j^z-\frac14\hat n_i\hat n_j +\frac12\bigl(\hat S_i^+\hat S_j^-+\hat S_i^-\hat S_j^+\bigr) =-\hat b^\dagger_{ij}\hat b_{ij}.
}
 On the square lattice, the nearest-neighbor superexchange interaction becomes 
\hypertarget{sm-eq-a7}{}%
\FittedEquation{A5}{%
\hat H_J=-J\sum_{\langle ij\rangle}\hat b^\dagger_{ij}\hat b_{ij}
        =-J\sum_j\sum_{\mathbf e\in\{\mathbf e_x,\mathbf e_y\}}
        \hat b^\dagger_{j,j+\mathbf e}\hat b_{j,j+\mathbf e},
}
 where \(\mathbf e_x\) and \(\mathbf e_y\) denote the nearest-neighbour displacements along the positive \(x\) and \(y\) directions, respectively. Introducing the Fourier transform of the bond operator with respect to the bond center, 
\hypertarget{sm-eq-a8}{}%
\FittedEquation{A6}{%
\hat b_{j,j+\mathbf e}
        =\frac1{\sqrt{N_s}}\sum_{\mathbf Q}
        e^{-i\mathbf Q\cdot(\mathbf r_j+\mathbf e/2)}
        \hat b_{\mathbf Q,\mathbf e},\qquad \mathbf e\in\{\mathbf e_x,\mathbf e_y\},
}
 and using \(N_s^{-1}\sum_j e^{-i(\mathbf Q'-\mathbf Q)\cdot\mathbf r_j}=\delta_{\mathbf Q,\mathbf Q'}\) gives 
\hypertarget{sm-eq-a9}{}%
\FittedEquation{A7}{%
\hat H_J=-J\sum_j\sum_{\mathbf e\in\{\mathbf e_x,\mathbf e_y\}}\frac1{N_s} \sum_{\mathbf Q,\mathbf Q'} e^{-i(\mathbf Q'-\mathbf Q)\cdot(\mathbf r_j+\mathbf e/2)} \hat b^\dagger_{\mathbf Q,\mathbf e} \hat b_{\mathbf Q',\mathbf e} =-J\sum_{\mathbf Q}\sum_{\mathbf e\in\{\mathbf e_x,\mathbf e_y\}} \hat b^\dagger_{\mathbf Q,\mathbf e} \hat b_{\mathbf Q,\mathbf e}.
}
 From the Fourier transform of the electron operators, \hypertarget{sm-eq-a10}{}\(\hat c_{j\sigma}=N_s^{-1/2}\sum_{\mathbf k}e^{-i\mathbf k\cdot\mathbf r_j}\hat c_{\mathbf k\sigma}\), the momentum-space bond operator is 
\hypertarget{sm-eq-a11}{}%
\FittedEquation{A8}{%
\begin{aligned}
        \hat b_{\mathbf Q,\mathbf e}
        &=\frac1{\sqrt{N_s}}\sum_j
        e^{i\mathbf Q\cdot(\mathbf r_j+\mathbf e/2)}
        \hat b_{j,j+\mathbf e}\\
        &=\frac1{\sqrt{2N_s}}\sum_j
        e^{i\mathbf Q\cdot(\mathbf r_j+\mathbf e/2)}
        \bigl(\hat c_{j\uparrow}\hat c_{j+\mathbf e,\downarrow}
        -\hat c_{j\downarrow}\hat c_{j+\mathbf e,\uparrow}\bigr)\\
        &=\frac1{\sqrt{2N_s}}\sum_j\frac1{N_s}
        \sum_{\mathbf k_1,\mathbf k_2}
        e^{-i(\mathbf k_1+\mathbf k_2-\mathbf Q)\cdot\mathbf r_j}
        e^{-i(\mathbf k_2-\mathbf Q/2)\cdot\mathbf e}
        \bigl(\hat c_{\mathbf k_1\uparrow}\hat c_{\mathbf k_2\downarrow}
        -\hat c_{\mathbf k_1\downarrow}\hat c_{\mathbf k_2\uparrow}\bigr)\\
        &=\frac1{\sqrt{2N_s}}\sum_{\mathbf q}
        e^{i\mathbf q\cdot\mathbf e}
        \bigl(\hat c_{\mathbf Q/2+\mathbf q,\uparrow}\hat c_{\mathbf Q/2-\mathbf q,\downarrow}
        -\hat c_{\mathbf Q/2+\mathbf q,\downarrow}\hat c_{\mathbf Q/2-\mathbf q,\uparrow}\bigr)\\
        &=\frac1{\sqrt{N_s}}\sum_{\mathbf q}
        \cos(\mathbf q\cdot\mathbf e)
        \hat\Delta_{\mathbf Q,\mathbf q},
        \end{aligned}
}
 where the site sum enforces \(\mathbf k_1+\mathbf k_2=\mathbf Q\) modulo a reciprocal-lattice vector. The summation variable has been changed to \(\mathbf q=\mathbf k_1-\mathbf Q/2\), so that \(\mathbf k_1=\mathbf Q/2+\mathbf q\) and \(\mathbf k_2=\mathbf Q/2-\mathbf q\), and the identity \(\hat\Delta_{\mathbf Q,\mathbf q}=\hat\Delta_{\mathbf Q,-\mathbf q}\) eliminates the sine contributions. Substituting \hyperlink{sm-eq-a11}{Eq. (A8)} into \hyperlink{sm-eq-a9}{Eq. (A7)} yields the exact operator identity in \hyperlink{sm-eq-model-superexchange}{Eq. (1)} 
\hypertarget{sm-eq-a12}{}%
\FittedEquation{A9}{%
\begin{aligned}
        \hat H_J&=-\frac{J}{N_s}\sum_{\mathbf Q}\sum_{\mathbf q,\mathbf q'}
        \bigl(\cos q_x\cos q'_x+\cos q_y\cos q'_y\bigr)
        \hat\Delta^\dagger_{\mathbf Q,\mathbf q}
        \hat\Delta_{\mathbf Q,\mathbf q'}\\
        &=-\frac{2J}{N_s}\sum_{\mathbf Q}\sum_{\mathbf q,\mathbf q'}
        \bigl(\phi_{\mathbf q,s}\phi_{\mathbf q',s}
        +\phi_{\mathbf q,d}\phi_{\mathbf q',d}\bigr)
        \hat\Delta^\dagger_{\mathbf Q,\mathbf q}
        \hat\Delta_{\mathbf Q,\mathbf q'}.
        \end{aligned}
}

Introducing the Fourier transforms of the electron spin and number operators gives

\hypertarget{sm-eq-superexchange-spin-fourier}{}%
\FittedEquation{A10}{%
\begin{aligned}
        \hat{\mathbf S}_j
        &=\frac1{2N_s}\sum_{\mathbf k,\mathbf k'}\sum_{\sigma\sigma'}
        e^{i(\mathbf k-\mathbf k')\cdot\mathbf r_j}
        \hat c^\dagger_{\mathbf k\sigma}\boldsymbol\tau_{\sigma\sigma'}\hat c_{\mathbf k'\sigma'}\\
        &=\frac1{N_s}\sum_{\mathbf K}e^{i\mathbf K\cdot\mathbf r_j}
        \sum_{\mathbf k}\hat{\mathbf S}_{\mathbf K,\mathbf k},
        \end{aligned}
}
 
and

\hypertarget{sm-eq-superexchange-number-fourier}{}%
\FittedEquation{A11}{%
\hat n_j=\frac1{N_s}\sum_{\mathbf K}e^{i\mathbf K\cdot\mathbf r_j}
        \sum_{\mathbf k}\hat n_{\mathbf K,\mathbf k},
}
 
where \(\hat{\mathbf S}_{\mathbf K,\mathbf k}=\frac12\sum_{\sigma\sigma'}\hat c^\dagger_{\mathbf K+\mathbf k,\sigma}\boldsymbol\tau_{\sigma\sigma'}\hat c_{\mathbf k\sigma'}\) and \(\hat n_{\mathbf K,\mathbf k}=\sum_\sigma\hat n_{\mathbf K,\mathbf k,\sigma}=\sum_\sigma\hat c^\dagger_{\mathbf K+\mathbf k,\sigma}\hat c_{\mathbf k\sigma}\), with \(\mathbf K\) denoting the momentum transfer. The nearest-neighbour superexchange interaction on the square lattice can be rewritten as

\hypertarget{sm-eq-superexchange-spin-transfer}{}%
\FittedEquation{A12}{%
\begin{aligned}
        \hat H_J
        &=\frac J2\sum_j\sum_{\mathbf e\in\{\pm\mathbf e_x,\pm\mathbf e_y\}}
        \left(\hat{\mathbf S}_j\cdot\hat{\mathbf S}_{j+\mathbf e}
        -\frac14\hat n_j\hat n_{j+\mathbf e}\right)\\
        &=\frac J{2N_s}\sum_{\mathbf K}\sum_{\mathbf k,\mathbf k'}
        \left(\sum_{\mathbf e}e^{-i\mathbf K\cdot\mathbf e}\right)
        \left(\hat{\mathbf S}_{\mathbf K,\mathbf k}\cdot\hat{\mathbf S}_{-\mathbf K,\mathbf k'}
        -\frac14\hat n_{\mathbf K,\mathbf k}\hat n_{-\mathbf K,\mathbf k'}\right)\\
        &=\frac{2J}{N_s}\sum_{\mathbf K}\sum_{\mathbf k,\mathbf k'}
        \phi_{\mathbf K,s}
        \left(\hat{\mathbf S}_{\mathbf K,\mathbf k}\cdot\hat{\mathbf S}_{-\mathbf K,\mathbf k'}
        -\frac14\hat n_{\mathbf K,\mathbf k}\hat n_{-\mathbf K,\mathbf k'}\right),
        \end{aligned}
}
 
where the two operators in each product transfer opposite momenta, conserving the total momentum.

\hypertarget{sm-hubbard-pairing-title}{}
\subsection*{A.2. Hubbard interaction}

The on-site Hubbard interaction is

\hypertarget{sm-eq-hubbard-real-space}{}%
\FittedEquation{A13}{%
\hat H_U=U\sum_j\hat n_{j\uparrow}\hat n_{j\downarrow},
}
 
where \(U\) is the on-site interaction strength. Define the site operator \(\hat a_j\equiv\hat c_{j\uparrow}\hat c_{j\downarrow}\), which gives \(\hat a_j^\dagger\hat a_j=\hat n_{j\uparrow}\hat n_{j\downarrow}\). Introducing the Fourier transform of the site operator, \(\hat a_j=N_s^{-1/2}\sum_{\mathbf Q}e^{-i\mathbf Q\cdot\mathbf r_j}\hat a_{\mathbf Q}\), the on-site Hubbard interaction becomes

\hypertarget{sm-eq-hubbard-pair-hamiltonian}{}%
\FittedEquation{A14}{%
\hat H_U=U\sum_j\hat a_j^\dagger\hat a_j
        =U\sum_{\mathbf Q}\hat a_{\mathbf Q}^\dagger\hat a_{\mathbf Q}.
}
 
The momentum-space site operator is

\hypertarget{sm-eq-hubbard-pair-fourier}{}%
\FittedEquation{A15}{%
\begin{aligned}
        \hat a_{\mathbf Q}
        &=\frac1{\sqrt{N_s}}\sum_j
        e^{i\mathbf Q\cdot\mathbf r_j}
        \hat c_{j\uparrow}\hat c_{j\downarrow}\\
        &=\frac1{\sqrt{N_s}}\sum_{\mathbf q}
        \hat c_{\mathbf Q/2+\mathbf q,\uparrow}
        \hat c_{\mathbf Q/2-\mathbf q,\downarrow}\\
        &=\frac1{2\sqrt{N_s}}\sum_{\mathbf q}\bigl(
        \hat c_{\mathbf Q/2+\mathbf q,\uparrow}\hat c_{\mathbf Q/2-\mathbf q,\downarrow}
        -\hat c_{\mathbf Q/2+\mathbf q,\downarrow}\hat c_{\mathbf Q/2-\mathbf q,\uparrow}\bigr)\\
        &=\frac1{\sqrt{2N_s}}\sum_{\mathbf q}\hat\Delta_{\mathbf Q,\mathbf q}.
        \end{aligned}
}
 
Substituting \hyperlink{sm-eq-hubbard-pair-fourier}{Eq. (A15)} into \hyperlink{sm-eq-hubbard-pair-hamiltonian}{Eq. (A14)} yields the exact operator identity

\hypertarget{sm-eq-hubbard-singlet-identity}{}%
\FittedEquation{A16}{%
\hat H_U=\frac{U}{2N_s}\sum_{\mathbf Q}\sum_{\mathbf q,\mathbf q'}
        \hat\Delta^\dagger_{\mathbf Q,\mathbf q}\hat\Delta_{\mathbf Q,\mathbf q'},
}
 
which corresponds to the on-site \(s\)-wave channel.

The on-site Hubbard interaction can also be rewritten as

\hypertarget{sm-eq-hubbard-density-transfer}{}%
\FittedEquation{A17}{%
\hat H_U=\frac{U}{N_s}\sum_{\mathbf K}\sum_{\mathbf k,\mathbf k'}
        \hat n_{\mathbf K,\mathbf k,\uparrow}\hat n_{-\mathbf K,\mathbf k',\downarrow}.
}
 
\hypertarget{sm-appendix-b}{}\hypertarget{sm-appendix-b-title}{}
\section*{Appendix B. Luttinger integral}

\hypertarget{sm-luttinger-exact-evaluation}{}
\subsection*{B.1. Exact evaluation}

At zero temperature in the thermodynamic limit, the Luttinger integral, defined in \hyperlink{sm-eq-mstj-luttinger-integral-definition}{Eq. (21)}, is

\hypertarget{sm-eq-luttinger-app-integral-definition}{}%
\FittedEquation{B1}{%
I_L=-\frac1\pi\operatorname{Im}\int_{\mathrm{BZ}}\frac{d^2k}{(2\pi)^2}
        \int_{-\infty}^{0}d\omega\,\sum_\sigma
        G^R_{\mathrm N,\sigma}(\mathbf k,\omega)
        \frac{\partial\Sigma^R_{\mathrm N,\sigma}(\mathbf k,\omega)}{\partial\omega}.
}
 
The frequency integral above can be evaluated analytically for the exactly solvable MStJ model. Recall that the exact Green's function and self-energy for \(0<\xi_{\mathbf k}<\mathcal J_{\mathbf k}\), given in Eqs. \hyperlink{sm-eq-mstj-green-zero-temperature}{(15)} and \hyperlink{sm-eq-mstj-self-energy-explicit}{(18)}, are

\hypertarget{sm-eq-luttinger-app-green-self-energy}{}%
\FittedEquation{B2}{%
\begin{aligned}
G^R_{\mathrm N,\sigma}(\mathbf k,\omega)
        &=\frac{\omega+i0^+-\xi_{\mathbf k}+\mathcal J_{\mathbf k}/2}
        {(\omega+i0^+-\xi_{\mathbf k}+\mathcal J_{\mathbf k}/2)^2-(3\mathcal J_{\mathbf k}/2)^2}
\end{aligned}
}
 
\hypertarget{sm-eq-luttinger-app-self-energy}{}%
\FittedEquation{B3}{%
\begin{aligned}
\Sigma^R_{\mathrm N,\sigma}(\mathbf k,\omega)
        &=-\mathcal J_{\mathbf k}/2
        +\frac{(3\mathcal J_{\mathbf k}/2)^2}
        {\omega+i0^+-\xi_{\mathbf k}+\mathcal J_{\mathbf k}/2}.
\end{aligned}
}
 
It follows that

\hypertarget{sm-eq-luttinger-app-integrand-identity}{}%
\FittedEquation{B4}{%
\begin{aligned}
        G^R_{\mathrm N,\sigma}\frac{\partial\Sigma^R_{\mathrm N,\sigma}}{\partial\omega}
        &=-\frac{(3\mathcal J_{\mathbf k}/2)^2}
        {(\omega+i0^+-\xi_{\mathbf k}+\mathcal J_{\mathbf k}/2)^2}\,G^R_{\mathrm N,\sigma}\\[6pt]
        &=\frac1{\omega+i0^+-\xi_{\mathbf k}+\mathcal J_{\mathbf k}/2}-G^R_{\mathrm N,\sigma}\\[6pt]
        &=\frac1{\omega+i0^+-\xi_{\mathbf k}+\mathcal J_{\mathbf k}/2}\\[6pt]
        &\quad-\frac12\left[
        \frac1{\omega+i0^+-\xi_{\mathbf k}+2\mathcal J_{\mathbf k}}
        +\frac1{\omega+i0^+-\xi_{\mathbf k}-\mathcal J_{\mathbf k}}
        \right].
        \end{aligned}
}
 
Each term in \hyperlink{sm-eq-luttinger-app-integrand-identity}{Eq. (B4)} can be integrated using

\hypertarget{sm-eq-luttinger-app-pole-integral}{}%
\FittedEquation{B5}{%
-\frac1\pi\int_{-\infty}^{0}d\omega\,
        \operatorname{Im}\frac1{\omega-E+i0^+}
        =\Theta(-E),
}
 
The contribution to \(I_L\) at each momentum is

\hypertarget{sm-eq-luttinger-app-momentum-integral}{}%
\FittedEquation{B6}{%
\begin{aligned}
        i_L(\mathbf k)
        &=-\frac1\pi\operatorname{Im}
        \int_{-\infty}^{0}d\omega\,\sum_\sigma
        G^R_{\mathrm N,\sigma}\,\partial_\omega\Sigma^R_{\mathrm N,\sigma}\\
        &=2\Theta\!\left(\frac{\mathcal J_{\mathbf k}}2-\xi_{\mathbf k}\right)
        -\Theta(2\mathcal J_{\mathbf k}-\xi_{\mathbf k})
        -\Theta(-\mathcal J_{\mathbf k}-\xi_{\mathbf k})\\
        &=2\Theta\!\left(\frac{\mathcal J_{\mathbf k}}2-\xi_{\mathbf k}\right)-1,
        \qquad 0<\xi_{\mathbf k}<\mathcal J_{\mathbf k}.
        \end{aligned}
}
 
For \(\xi_{\mathbf k}>\mathcal J_{\mathbf k}\) and \(\xi_{\mathbf k}<0\), \(\Sigma^R_{\mathrm N,\sigma}=0\) and \(\Sigma^R_{\mathrm N,\sigma}=-\mathcal J_{\mathbf k}\), respectively. Both are frequency independent and hence contribute zero to \(I_L\). Away from the level crossings of the block ground states and the Luttinger surface, the result is

\hypertarget{sm-eq-luttinger-app-piecewise}{}%
\FittedEquation{B7}{%
i_L(\mathbf k)=
        \begin{cases}
        +1,&0<\xi_{\mathbf k}<\mathcal J_{\mathbf k}/2,\\
        -1,&\mathcal J_{\mathbf k}/2<\xi_{\mathbf k}<\mathcal J_{\mathbf k},\\
        0,&\xi_{\mathbf k}<0\ \text{or}\ \xi_{\mathbf k}>\mathcal J_{\mathbf k}.
        \end{cases}
}
 
The Brillouin-zone integral can consequently be written as

\hypertarget{sm-eq-luttinger-app-area-difference}{}%
\FittedEquation{B8}{%
\begin{aligned}
        I_L
        &=\int_{\mathrm{BZ}}\frac{d^2k}{(2\pi)^2}\,i_L(\mathbf k)\\
        &=\int_{\mathrm{BZ}}\frac{d^2k}{(2\pi)^2}
        \left[
        2\Theta\!\left(\frac{\mathcal J_{\mathbf k}}2-\xi_{\mathbf k}\right)
        -\Theta(-\xi_{\mathbf k})
        -\Theta(\mathcal J_{\mathbf k}-\xi_{\mathbf k})
        \right].
        \end{aligned}
}
 
The Luttinger surface at \(\xi_{\mathbf k}=\mathcal J_{\mathbf k}/2\) and level crossings at \(\xi_{\mathbf k}=0\) and \(\xi_{\mathbf k}=\mathcal J_{\mathbf k}\) have zero area measure in the Brillouin zone and contribute negligibly to the Luttinger integral in the thermodynamic limit.

\hypertarget{sm-appendix-c}{}\hypertarget{sm-appendix-c-title}{}
\subsection*{B.2. Weak-coupling expansion}

The weak-coupling expansion relates the Luttinger integral to the doping dependence of a weighted density of states. Define the density of states per spin weighted by the \(d\)-wave form factor as

\hypertarget{sm-eq-wc-weighted-dos}{}%
\FittedEquation{B9}{%
\rho_l(\epsilon)=\int_{\mathrm{BZ}}\frac{d^2k}{(2\pi)^2}
        \phi_{\mathbf k,d}^{\,l}\delta(\epsilon-\varepsilon_{\mathbf k}).
}
 
On the Fermi surface, the magnitude of the \(d\)-wave form factor \(\lvert\phi_{\mathbf k,d}\rvert\) is maximal at the antinodes \(\mathbf k_a\). Therefore, \(\rho_4\) emphasizes the antinodal regions compared with the noninteracting density of states per spin \(\rho_0\). Let \(\mu_0(p)\) denote the noninteracting chemical potential at electron density \(1-p\). \hyperlink{sm-fig-weak-coupling-dos}{Fig. B1} compares \(\rho_0[\mu_0(p)]\) and \(\rho_4[\mu_0(p)]\) for \(t=0.3~\mathrm{eV}\), \(t'/t=-0.2\), and \(t''/t=0.1\). Both diverge logarithmically at the van Hove singularity, where

\hypertarget{sm-eq-wc-van-hove}{}%
\FittedEquation{B10}{%
\begin{aligned}
p_{\mathrm{vH}}\simeq0.35325
\end{aligned}
}
 
\hypertarget{sm-eq-wc-van-hove-chemical-potential}{}%
\FittedEquation{B11}{%
\begin{aligned}
\mu_0(p_{\mathrm{vH}})=4t'-4t''=-0.36~\mathrm{eV}.
\end{aligned}
}
 
At the van Hove saddle points \(X\) and \(Y\), \(\lvert\phi_{\mathbf k_{X/Y},d}\rvert=1\), so the logarithmic divergence is preserved in \(\rho_4\). The plotting window truncates the divergent curves; no finite peak height is assigned at \(p_{\mathrm{vH}}\).

\par\medskip
\hypertarget{sm-fig-weak-coupling-dos}{}
\PrintAsset{rho0-rho4-vs-doping}
{\small\noindent \textbf{Figure B1 \textbar{} Noninteracting and weighted density of states.} The black and purple curves show \(\rho_0[\mu_0(p)]\) and \(\rho_4[\mu_0(p)]\), respectively, in units of \(\mathrm{eV}^{-1}\) per spin. The dashed line marks \(p_{\mathrm{vH}}\simeq0.35325\); the logarithmic van Hove divergence is truncated by the plotting window.\par}
\medskip

Recall that the Luttinger integral of the MStJ model in \hyperlink{sm-eq-mstj-luttinger-integral-definition}{Eq. (21)} is

\hypertarget{sm-eq-wc-luttinger-area-integral}{}%
\FittedEquation{B12}{%
I_L=\int_{\mathrm{BZ}}\frac{d^2k}{(2\pi)^2}
        \left[
        2\Theta\!\left(\frac{\mathcal J_{\mathbf k}}2-\xi_{\mathbf k}\right)
        -\Theta(-\xi_{\mathbf k})-\Theta(\mathcal J_{\mathbf k}-\xi_{\mathbf k})
        \right].
}
 
With \(\mathcal J_{\mathbf k}=\mathcal J_d\phi_{\mathbf k,d}^{2}\), define

\hypertarget{sm-eq-wc-area-definition}{}%
\FittedEquation{B13}{%
\mathcal A_\upsilon(\mu,\mathcal J_d)=\int_{\mathrm{BZ}}\frac{d^2k}{(2\pi)^2}
        \Theta\!\left(\mu+\upsilon\mathcal J_d\phi_{\mathbf k,d}^{2}
        -\varepsilon_{\mathbf k}\right).
}
 
The electron density and Luttinger integral then satisfy

\hypertarget{sm-eq-wc-exact-density-integral}{}%
\FittedEquation{B14}{%
\begin{aligned}
n_e=\mathcal A_0+\mathcal A_1=1-p
\end{aligned}
}
 
\hypertarget{sm-eq-wc-exact-luttinger-integral}{}%
\FittedEquation{B15}{%
\begin{aligned}
I_L=2\mathcal A_{1/2}-\mathcal A_0-\mathcal A_1.
\end{aligned}
}
 
For chemical potentials away from the band edges and van Hove singularities, the weighted densities of states are smooth near \(E=\mu\). With \(\Theta'(x)=\delta(x)\) and \(\Theta''(x)=\delta'(x)\), expanding the step function in \hyperlink{sm-eq-wc-area-definition}{Eq. (B13)} in powers of \(\mathcal J_d\) gives

\hypertarget{sm-eq-wc-area-expansion}{}%
\FittedEquation{B16}{%
\mathcal A_\upsilon(\mu,\mathcal J_d)=\mathcal A_0(\mu)
        +\upsilon\mathcal J_d\rho_2(\mu)
        +\frac{\upsilon^2\mathcal J_d^2}{2}\rho_4'(\mu)
        +O(\mathcal J_d^3),
}
 
where a prime denotes differentiation with respect to energy and \(\mathcal A_0\) is independent of \(\mathcal J_d\). Substitution into \hyperlink{sm-eq-wc-exact-luttinger-integral}{Eq. (B15)} cancels the constant and linear terms, leaving

\hypertarget{sm-eq-wc-integral-fixed-mu}{}%
\FittedEquation{B17}{%
I_L=-\frac{\mathcal J_d^2}{4}\rho_4'(\mu)
        +O(\mathcal J_d^3).
}
 
At fixed doping, the chemical potential also changes with the coupling. The noninteracting chemical potential is determined by

\hypertarget{sm-eq-wc-noninteracting-density}{}%
\FittedEquation{B18}{%
2\mathcal A_0[\mu_0(p)]=1-p.
}
 
To first order in \(\mathcal J_d\), the interacting density equation is

\hypertarget{sm-eq-wc-density-expansion}{}%
\FittedEquation{B19}{%
1-p=2\mathcal A_0(\mu)+\mathcal J_d\rho_2(\mu)
        +O(\mathcal J_d^2).
}
 
Using \(\mathcal A_0'(\mu_0)=\rho_0(\mu_0)\) and expanding about \(\mu_0\) yields

\hypertarget{sm-eq-wc-chemical-potential-shift}{}%
\FittedEquation{B20}{%
\mu=\mu_0-
        \frac{\mathcal J_d\rho_2(\mu_0)}{2\rho_0(\mu_0)}
        +O(\mathcal J_d^2).
}
 
Thus \(\mu-\mu_0=O(\mathcal J_d)\). Replacing \(\mu\) by \(\mu_0\) in the quadratic term of \hyperlink{sm-eq-wc-integral-fixed-mu}{Eq. (B17)} changes only terms of cubic and higher order. The leading result at fixed doping is therefore

\hypertarget{sm-eq-wc-integral-fixed-density}{}%
\FittedEquation{B21}{%
I_L(p,\mathcal J_d)=-\frac{\mathcal J_d^2}{4}
        \rho_4'[\mu_0(p)]+O(\mathcal J_d^3).
}
 
Differentiation of \hyperlink{sm-eq-wc-noninteracting-density}{Eq. (B18)} gives

\hypertarget{sm-eq-wc-mu-doping-derivative}{}%
\FittedEquation{B22}{%
\frac{d\mu_0}{dp}=-\frac{1}{2\rho_0[\mu_0(p)]}.
}
 
The chain rule then gives the equivalent form

\hypertarget{sm-eq-wc-integral-doping-slope}{}%
\FittedEquation{B23}{%
I_L=\frac{\mathcal J_d^2}{2}\rho_0[\mu_0(p)]
        \frac{d\,\rho_4[\mu_0(p)]}{dp}
        +O(\mathcal J_d^3).
}
 
\hypertarget{sm-appendix-d}{}\hypertarget{sm-appendix-d-title}{}
\section*{Appendix C. ARPES of the conventional \(d\)-wave BCS state}

\hypertarget{sm-b-conventions}{}The mean-field BCS Hamiltonian with d-wave SC pairing is

\hypertarget{sm-eq-b1}{}%
\FittedEquation{C1}{%
\hat H_{\mathrm{BCS}}=\sum_{\mathbf k\sigma}\xi_{\mathbf k}\hat c^\dagger_{\mathbf k\sigma}\hat c_{\mathbf k\sigma}
        +\sum_{\mathbf k}\left(\Delta_{\mathbf k}
        \hat c^\dagger_{\mathbf k\uparrow}\hat c^\dagger_{-\mathbf k\downarrow}
        +\mathrm{H.c.}\right).
}
 
Here \(\xi_{\mathbf k}\) is the band dispersion, with \(\xi_{-\mathbf k}=\xi_{\mathbf k}\). The d-wave SC pairing field is \(\Delta_{\mathbf k}=\Delta_d\phi_{\mathbf k,d}=\Delta_{-\mathbf k}\), where \(\phi_{\mathbf k,d}=(\cos k_x-\cos k_y)/2\). The Bogoliubov transformation

\hypertarget{sm-eq-b2}{}%
\FittedEquation{C2}{%
\begin{pmatrix}
        \hat c_{\mathbf k\uparrow}\\
        \hat c^\dagger_{-\mathbf k\downarrow}
        \end{pmatrix}
        =
        \begin{pmatrix}
        u_{\mathbf k} & -v_{\mathbf k}\\
        v_{\mathbf k}^* & u_{\mathbf k}
        \end{pmatrix}
        \begin{pmatrix}
        \hat\gamma_{\mathbf k\uparrow}\\
        \hat\gamma^\dagger_{-\mathbf k\downarrow}
        \end{pmatrix}
}
 
diagonalizes the BCS Hamiltonian

\hypertarget{sm-eq-b3}{}%
\FittedEquation{C3}{%
\hat H_{\mathrm{BCS}}=\sum_{\mathbf k}E^{\mathrm{BCS}}_{\mathbf k}
        \left(\hat\gamma^\dagger_{\mathbf k\uparrow}\hat\gamma_{\mathbf k\uparrow}
        +\hat\gamma^\dagger_{-\mathbf k\downarrow}\hat\gamma_{-\mathbf k\downarrow}\right)
        +\sum_{\mathbf k}\left(\xi_{\mathbf k}-E^{\mathrm{BCS}}_{\mathbf k}\right),
}
 
where the Bogoliubov quasiparticle dispersion is \(E^{\mathrm{BCS}}_{\mathbf k}=\sqrt{\xi_{\mathbf k}^2+|\Delta_{\mathbf k}|^2}\), and the BCS coherence factors are

\hypertarget{sm-eq-b4}{}\hypertarget{sm-eq-bcs-coherence-v}{}%
\FittedEquation{C4}{%
u_{\mathbf k}=\sqrt{\frac12\left(1+\frac{\xi_{\mathbf k}}{E^{\mathrm{BCS}}_{\mathbf k}}\right)},\quad v_{\mathbf k}=\frac{\Delta_{\mathbf k}}{|\Delta_{\mathbf k}|} \sqrt{\frac12\left(1-\frac{\xi_{\mathbf k}}{E^{\mathrm{BCS}}_{\mathbf k}}\right)}.
}
 
The time evolution of the Bogoliubov quasiparticle operators is \(\hat\gamma_{\mathbf k\sigma}(t)=e^{-iE_{\mathbf k}t}\hat\gamma_{\mathbf k\sigma}(0)\), and the finite-temperature occupation is \(\langle\hat\gamma^\dagger_{\mathbf k\sigma}\hat\gamma_{\mathbf k\sigma}\rangle=f_{\mathrm F}(E^{\mathrm{BCS}}_{\mathbf k})\), where \(f_{\mathrm F}(E^{\mathrm{BCS}}_{\mathbf k})=(e^{\beta E^{\mathrm{BCS}}_{\mathbf k}}+1)^{-1}\) is the Fermi function.

\hypertarget{sm-b-single-electron}{}
\subsection*{C.1. One-electron ARPES}

The occupied one-particle spectrum of conventional d-wave BCS state measured by one-electron cARPES is

\hypertarget{sm-eq-b5}{}%
\FittedEquation{C5}{%
f_{\mathrm F}(\omega)A_{\mathrm{BCS},\sigma}(\mathbf k,\omega)
        \equiv\lim_{\mathcal T\to\infty}^{\mathcal D'}\frac{1}{2\pi i\mathcal T}
        \iint_{-\frac{\mathcal T}{2}}^{\frac{\mathcal T}{2}}dt\,dt'\,
        e^{i\omega(t-t')}G^<_{\mathrm{BCS},\sigma}(\mathbf k;t,t').
}
 
Here \(\mathcal T\) is the measurement duration, \(\mathcal T^{-1}\) removes the overall exposure factor, and \(\mathcal D'\) denotes distributional convergence with respect to the energy variable \(\omega\). For the spin-singlet BCS state, the one-particle lesser Green's function is

\hypertarget{sm-eq-b6}{}\hypertarget{sm-eq-b6a}{}%
\FittedEquation{C6}{%
G^<_{\mathrm{BCS}}(\mathbf k\sigma;\mathbf k'\sigma';t,t') \equiv i\langle \hat c^\dagger_{\mathbf k'\sigma'}(t')\hat c_{\mathbf k\sigma}(t)\rangle=\delta_{\mathbf k,\mathbf k'}\delta_{\sigma\sigma'}\,i\langle \hat c^\dagger_{\mathbf k\sigma}(t')\hat c_{\mathbf k\sigma}(t)\rangle,
}
 
\hypertarget{sm-eq-b6b}{}%
\FittedEquation{C7}{%
\begin{aligned}
G^<_{\mathrm{BCS},\sigma}(\mathbf k;t,t') &=i\langle \hat c^\dagger_{\mathbf k\sigma}(t')\hat c_{\mathbf k\sigma}(t)\rangle\\
&=i\Bigl[u_{\mathbf k}^2e^{-iE_{\mathbf k}(t-t')} \langle\hat\gamma^\dagger_{\mathbf k\sigma}\hat\gamma_{\mathbf k\sigma}\rangle+|v_{\mathbf k}|^2e^{iE_{\mathbf k}(t-t')} \langle\hat\gamma_{-\mathbf k\bar\sigma}\hat\gamma^\dagger_{-\mathbf k\bar\sigma}\rangle\Bigr]\\
&=i\Bigl[u_{\mathbf k}^2f_{\mathrm F}(E^{\mathrm{BCS}}_{\mathbf k})e^{-iE_{\mathbf k}(t-t')}+|v_{\mathbf k}|^2f_{\mathrm F}(-E^{\mathrm{BCS}}_{\mathbf k})e^{iE_{\mathbf k}(t-t')}\Bigr].
\end{aligned}
}
 
Here \(\bar\sigma\) denotes the spin opposite to \(\sigma\). Introducing the center and relative times

\hypertarget{sm-eq-b7}{}\hypertarget{sm-eq-b7a}{}\hypertarget{sm-eq-bcs-relative-time}{}\hypertarget{sm-eq-b7b}{}%
\FittedEquation{C8}{%
t_c=\frac{t+t'}{2},\quad t_r=t-t',\quad |t_r|\le\mathcal T,\quad |t_c|\le\frac{\mathcal T-|t_r|}{2}
}
 
with unit Jacobian \(dt\,dt'=dt_c\,dt_r\), and substituting \hyperlink{sm-eq-b6b}{Eq. (C7)} into \hyperlink{sm-eq-b5}{Eq. (C5)} yields

\hypertarget{sm-eq-b8}{}%
\FittedEquation{C9}{%
\begin{aligned}
        &f_{\mathrm F}(\omega)A_{\mathrm{BCS},\sigma}(\mathbf k,\omega)\\
        &=\lim_{\mathcal T\to\infty}^{\mathcal D'}\frac{1}{2\pi\mathcal T}
        \int_{-\mathcal T}^{\mathcal T}dt_r
        \int_{-(\mathcal T-|t_r|)/2}^{(\mathcal T-|t_r|)/2}dt_c\\
        &\qquad\times\Bigl[u_{\mathbf k}^2f_{\mathrm F}(E^{\mathrm{BCS}}_{\mathbf k})e^{i(\omega-E^{\mathrm{BCS}}_{\mathbf k})t_r}\\
        &\qquad\quad+|v_{\mathbf k}|^2f_{\mathrm F}(-E^{\mathrm{BCS}}_{\mathbf k})e^{i(\omega+E^{\mathrm{BCS}}_{\mathbf k})t_r}\Bigr]\\
        &=\lim_{\mathcal T\to\infty}^{\mathcal D'}\frac{1}{2\pi}
        \int_{-\mathcal T}^{\mathcal T}dt_r\left(1-\frac{|t_r|}{\mathcal T}\right)\\
        &\qquad\times\Bigl[u_{\mathbf k}^2f_{\mathrm F}(E^{\mathrm{BCS}}_{\mathbf k})e^{i(\omega-E^{\mathrm{BCS}}_{\mathbf k})t_r}\\
        &\qquad\quad+|v_{\mathbf k}|^2f_{\mathrm F}(-E^{\mathrm{BCS}}_{\mathbf k})e^{i(\omega+E^{\mathrm{BCS}}_{\mathbf k})t_r}\Bigr]\\
        &=f_{\mathrm F}(\omega)\Bigl[u_{\mathbf k}^2\delta(\omega-E^{\mathrm{BCS}}_{\mathbf k})
        +|v_{\mathbf k}|^2\delta(\omega+E^{\mathrm{BCS}}_{\mathbf k})\Bigr],
        \end{aligned}
}
 
Thus, the nonnegative one-particle spectral function of the conventional d-wave BCS state is

\hypertarget{sm-eq-b9}{}%
\FittedEquation{C10}{%
A_{\mathrm{BCS},\sigma}(\mathbf k,\omega)=u_{\mathbf k}^2\delta(\omega-E^{\mathrm{BCS}}_{\mathbf k})
        +|v_{\mathbf k}|^2\delta(\omega+E^{\mathrm{BCS}}_{\mathbf k}).
}
 
\hypertarget{sm-b-two-photon}{}
\subsection*{C.2. Two-photon cARPES}

The two-particle removal spectral function of conventional d-wave BCS state measured by two-photon cARPES is

\hypertarget{sm-eq-b10}{}%
\FittedEquation{C11}{%
\begin{aligned}
        &A^{(2)<}_{\mathrm{BCS},\sigma_1\sigma_2}(\mathbf k_1,\mathbf k_2;\omega_1,\omega_2)
        =\lim_{\mathcal T\to\infty}^{\mathcal D'}\frac{1}{(2\pi i)^2(\mathcal T)^2}
        \iiiint_{-\frac{\mathcal T}{2}}^{\frac{\mathcal T}{2}}dt_1\,dt_2\,dt'_1\,dt'_2\,\\
        &\qquad\times e^{i[\omega_1(t_1-t'_1)+\omega_2(t_2-t'_2)]}
        G^{(2)<}_{\mathrm{BCS},\sigma_1\sigma_2}(\mathbf k_1,\mathbf k_2;t_1,t_2,t'_1,t'_2).
        \end{aligned}
}
 
Here \((\mathcal T)^{-2}\) removes the two overall exposure factors. The two-particle lesser Green's function is defined as

\hypertarget{sm-eq-b11}{}%
\FittedEquation{C12}{%
G^{(2)<}_{\mathrm{BCS},\sigma_1\sigma_2}(\mathbf k_1,\mathbf k_2;t_1,t_2,t'_1,t'_2)
        \equiv i^2\Bigl\langle
        T_{t'}[\hat c^\dagger_{\mathbf k_1\sigma_1}(t'_1)\hat c^\dagger_{\mathbf k_2\sigma_2}(t'_2)]
        T_t[\hat c_{\mathbf k_2\sigma_2}(t_2)\hat c_{\mathbf k_1\sigma_1}(t_1)]\Bigr\rangle,
}
 
where \(T_t\) and \(T_{t'}\) denote fermionic time ordering and anti-time ordering, respectively. For the mean-field BCS Hamiltonian, Wick's theorem gives

\hypertarget{sm-eq-b12}{}%
\FittedEquation{C13}{%
\begin{aligned}
        G^{(2)<}_{\mathrm{BCS},\sigma_1\sigma_2}(\mathbf k_1,\mathbf k_2;t_1,t_2,t'_1,t'_2)
        &=G^<_{\mathrm{BCS},\sigma_1}(\mathbf k_1;t_1,t'_1)G^<_{\mathrm{BCS},\sigma_2}(\mathbf k_2;t_2,t'_2)\\
        &\quad-\delta_{\mathbf k_1,\mathbf k_2}\delta_{\sigma_1\sigma_2}
        G^<_{\mathrm{BCS},\sigma_1}(\mathbf k_1;t_1,t'_2)G^<_{\mathrm{BCS},\sigma_2}(\mathbf k_2;t_2,t'_1)\\
        &\quad-\delta_{\mathbf k_1,-\mathbf k_2}(1-\delta_{\sigma_1\sigma_2})
        \mathcal F(\mathbf k_1;t_1,t_2)[\mathcal F(\mathbf k_1;t'_1,t'_2)]^*.
        \end{aligned}
}
 
The normal contractions give the one-particle lesser Green's function in \hyperlink{sm-eq-b6b}{Eq. (C7)}. For the anomalous contractions, define the time-ordered and anti-time-ordered anomalous Green's functions for the spin-singlet BCS state as

\hypertarget{sm-eq-b13}{}\hypertarget{sm-eq-b13a}{}\hypertarget{sm-eq-b13b}{}%
\FittedEquation{C14}{%
\mathcal F^{T_t}(\mathbf k_1\sigma_1,\mathbf k_2\sigma_2;t_1,t_2)
        \equiv-i\langle T_t[\hat c_{\mathbf k_1\sigma_1}(t_1)\hat c_{\mathbf k_2\sigma_2}(t_2)]\rangle
        =\delta_{\mathbf k_1,-\mathbf k_2}\epsilon_{\sigma_1\sigma_2}\mathcal F(\mathbf k_1;t_1,t_2),
}
\FittedEquation{C15}{%
\bar{\mathcal F}^{T_{t'}}(\mathbf k_1\sigma_1,\mathbf k_2\sigma_2;t'_1,t'_2)
        \equiv-i\langle T_{t'}[\hat c^\dagger_{\mathbf k_1\sigma_1}(t'_1)\hat c^\dagger_{\mathbf k_2\sigma_2}(t'_2)]\rangle
        =\delta_{\mathbf k_1,-\mathbf k_2}\epsilon_{\sigma_1\sigma_2}[\mathcal F(\mathbf k_1;t'_1,t'_2)]^*.
}
 
Here \(\epsilon_{\uparrow\downarrow}=-\epsilon_{\downarrow\uparrow}=1\), \(\epsilon_{\uparrow\uparrow}=\epsilon_{\downarrow\downarrow}=0\), and \(\epsilon_{\sigma_1\sigma_2}^{2}=1-\delta_{\sigma_1\sigma_2}\). The anomalous contractions are given by

\hypertarget{sm-eq-b14}{}\hypertarget{sm-eq-b15}{}%
\FittedEquation{C16}{%
\begin{aligned}
        \mathcal F(\mathbf k_1;t_1,t_2)
        &=-i\epsilon_{\sigma_1\bar\sigma_1}\langle T_t[\hat c_{\mathbf k_1\sigma_1}(t_1)\hat c_{-\mathbf k_1\bar\sigma_1}(t_2)]\rangle\\
        &=-iu_{\mathbf k_1}v_{\mathbf k_1}\Bigl\{
        \Theta(t_1-t_2)\Bigl[e^{-iE_{\mathbf k_1}(t_1-t_2)}\langle\hat\gamma_{\mathbf k_1\sigma_1}\hat\gamma^\dagger_{\mathbf k_1\sigma_1}\rangle
        -e^{iE_{\mathbf k_1}(t_1-t_2)}\langle\hat\gamma^\dagger_{-\mathbf k_1\bar\sigma_1}\hat\gamma_{-\mathbf k_1\bar\sigma_1}\rangle\Bigr]\\
        &\hphantom{{}=-iu_{\mathbf k_1}v_{\mathbf k_1}\Bigl\{}\mathllap{{}-{}}\Theta(t_2-t_1)\Bigl[e^{-iE_{\mathbf k_1}(t_1-t_2)}\langle\hat\gamma^\dagger_{\mathbf k_1\sigma_1}\hat\gamma_{\mathbf k_1\sigma_1}\rangle
        -e^{iE_{\mathbf k_1}(t_1-t_2)}\langle\hat\gamma_{-\mathbf k_1\bar\sigma_1}\hat\gamma^\dagger_{-\mathbf k_1\bar\sigma_1}\rangle\Bigr]\Bigr\}\\
        &=-iu_{\mathbf k_1}v_{\mathbf k_1}\Bigl\{
        \Theta(t_1-t_2)\Bigl[f_{\mathrm F}(-E^{\mathrm{BCS}}_{\mathbf k_1})e^{-iE_{\mathbf k_1}(t_1-t_2)}-f_{\mathrm F}(E^{\mathrm{BCS}}_{\mathbf k_1})e^{iE_{\mathbf k_1}(t_1-t_2)}\Bigr]\\
        &\hphantom{{}=-iu_{\mathbf k_1}v_{\mathbf k_1}\Bigl\{}\mathllap{{}-{}}\Theta(t_2-t_1)\Bigl[f_{\mathrm F}(E^{\mathrm{BCS}}_{\mathbf k_1})e^{-iE_{\mathbf k_1}(t_1-t_2)}-f_{\mathrm F}(-E^{\mathrm{BCS}}_{\mathbf k_1})e^{iE_{\mathbf k_1}(t_1-t_2)}\Bigr]\Bigr\}\\
        &=iu_{\mathbf k_1}v_{\mathbf k_1}\Bigl[f_{\mathrm F}(E^{\mathrm{BCS}}_{\mathbf k_1})e^{iE_{\mathbf k_1}|t_1-t_2|}-f_{\mathrm F}(-E^{\mathrm{BCS}}_{\mathbf k_1})e^{-iE_{\mathbf k_1}|t_1-t_2|}\Bigr].
        \end{aligned}
}
 
Here \(\Theta\) is the Heaviside step function.

The two-particle removal spectral function in \hyperlink{sm-eq-b10}{Eq. (C11)} can be decomposed into direct, exchange, and anomalous contributions:

\hypertarget{sm-eq-b16}{}%
\FittedEquation{C17}{%
A^{(2)<}_{\mathrm{BCS},\sigma_1\sigma_2}=A^{(2)<}_{\mathrm{BCS},\sigma_1\sigma_2,\mathrm{direct}}
        +A^{(2)<}_{\mathrm{BCS},\sigma_1\sigma_2,\mathrm{exchange}}+A^{(2)<}_{\mathrm{BCS},\sigma_1\sigma_2,\mathrm{anomalous}}.
}
 
The center and relative times for each photoemission channel are introduced as

\hypertarget{sm-eq-b17}{}\hypertarget{sm-eq-b17a}{}\hypertarget{sm-eq-bcs-two-photon-relative-time}{}\hypertarget{sm-eq-b17b}{}%
\FittedEquation{C18}{%
t_{cj}=\frac{t_j+t'_j}{2},\quad t_{rj}=t_j-t'_j,\quad |t_{rj}|\le\mathcal T,\quad |t_{cj}|\le\frac{\mathcal T-|t_{rj}|}{2}
}
 
with unit Jacobian \(dt_j\,dt'_j=dt_{cj}\,dt_{rj}\). The direct contribution is

\hypertarget{sm-eq-b18}{}%
\FittedEquation{C19}{%
\begin{aligned}
        &A^{(2)<}_{\mathrm{BCS},\sigma_1\sigma_2,\mathrm{direct}}(\mathbf k_1,\mathbf k_2;\omega_1,\omega_2)\\
        &\quad=\lim_{\mathcal T\to\infty}^{\mathcal D'}\prod_{j=1}^2\Biggl[
        \frac{1}{2\pi i\mathcal T}\int_{-\mathcal T}^{\mathcal T}dt_{rj}
        \int_{-(\mathcal T-|t_{rj}|)/2}^{(\mathcal T-|t_{rj}|)/2}dt_{cj}\,
        e^{i\omega_jt_{rj}}G^<_{\mathrm{BCS},\sigma_j}(\mathbf k_j;t_{rj},0)\Biggr]\\
        &\quad=\lim_{\mathcal T\to\infty}^{\mathcal D'}\prod_{j=1}^2\Biggl[
        \frac{1}{2\pi i}\int_{-\mathcal T}^{\mathcal T}dt_{rj}
        \left(1-\frac{|t_{rj}|}{\mathcal T}\right)
        e^{i\omega_jt_{rj}}G^<_{\mathrm{BCS},\sigma_j}(\mathbf k_j;t_{rj},0)\Biggr]\\
        &\quad=\prod_{j=1}^2f_{\mathrm F}(\omega_j)A_{\mathrm{BCS},\sigma_j}(\mathbf k_j,\omega_j).
        \end{aligned}
}
 
For the exchange contribution, the center-time integral eliminates the oscillating cosine term as \(\mathcal T\to\infty\):

\hypertarget{sm-eq-b19}{}%
\FittedEquation{C20}{%
\begin{aligned}
        &\lim_{\mathcal T\to\infty}\frac{1}{(\mathcal T)^2}
        \prod_{j=1}^2\left[\int_{-(\mathcal T-|t_{rj}|)/2}^{(\mathcal T-|t_{rj}|)/2}dt_{cj}\right]
        G^<_{\mathrm{BCS},\sigma_1}\!\left(\mathbf k_1;t_{c1}+\frac{t_{r1}}{2},t_{c2}-\frac{t_{r2}}{2}\right)
        G^<_{\mathrm{BCS},\sigma_1}\!\left(\mathbf k_1;t_{c2}+\frac{t_{r2}}{2},t_{c1}-\frac{t_{r1}}{2}\right)\\
        {}={}&\lim_{\mathcal T\to\infty}\frac{1}{(\mathcal T)^2}
        \prod_{j=1}^2\left[\int_{-(\mathcal T-|t_{rj}|)/2}^{(\mathcal T-|t_{rj}|)/2}dt_{cj}\right]
        \times\Bigl[-u_{\mathbf k_1}^4f_{\mathrm F}(E^{\mathrm{BCS}}_{\mathbf k_1})^2e^{-iE_{\mathbf k_1}(t_{r1}+t_{r2})}
        -|v_{\mathbf k_1}|^4f_{\mathrm F}(-E^{\mathrm{BCS}}_{\mathbf k_1})^2e^{iE_{\mathbf k_1}(t_{r1}+t_{r2})}\\
        &\hphantom{\lim_{\mathcal T\to\infty}\frac{1}{(\mathcal T)^2}\prod_{j=1}^2\left[\int_{-(\mathcal T-|t_{rj}|)/2}^{(\mathcal T-|t_{rj}|)/2}dt_{cj}\right]\times\Bigl[}\,\mathord{-}2|u_{\mathbf k_1}v_{\mathbf k_1}|^2f_{\mathrm F}(E^{\mathrm{BCS}}_{\mathbf k_1})f_{\mathrm F}(-E^{\mathrm{BCS}}_{\mathbf k_1})
        \cos\!\left[2E^{\mathrm{BCS}}_{\mathbf k_1}(t_{c1}-t_{c2})\right]\Bigr]\\
        {}={}&-u_{\mathbf k_1}^4f_{\mathrm F}(E^{\mathrm{BCS}}_{\mathbf k_1})^2e^{-iE_{\mathbf k_1}(t_{r1}+t_{r2})}
        -|v_{\mathbf k_1}|^4f_{\mathrm F}(-E^{\mathrm{BCS}}_{\mathbf k_1})^2e^{iE_{\mathbf k_1}(t_{r1}+t_{r2})}.
        \end{aligned}
}
 
The remaining Fourier transforms over the relative times yield

\hypertarget{sm-eq-b20}{}%
\FittedEquation{C21}{%
\begin{aligned}
        &A^{(2)<}_{\mathrm{BCS},\sigma_1\sigma_2,\mathrm{exchange}}(\mathbf k_1,\mathbf k_2;\omega_1,\omega_2)\\
        &\quad=-\delta_{\mathbf k_1,\mathbf k_2}\delta_{\sigma_1\sigma_2}
        \Bigl[u_{\mathbf k_1}^4f_{\mathrm F}(E^{\mathrm{BCS}}_{\mathbf k_1})^2
        \delta(\omega_1-E^{\mathrm{BCS}}_{\mathbf k_1})\delta(\omega_2-E^{\mathrm{BCS}}_{\mathbf k_1})\\
        &\hphantom{\quad=-\delta_{\mathbf k_1,\mathbf k_2}\delta_{\sigma_1\sigma_2}\Bigl[}\mathllap{{}+{}}|v_{\mathbf k_1}|^4f_{\mathrm F}(-E^{\mathrm{BCS}}_{\mathbf k_1})^2
        \delta(\omega_1+E^{\mathrm{BCS}}_{\mathbf k_1})\delta(\omega_2+E^{\mathrm{BCS}}_{\mathbf k_1})\Bigr].
        \end{aligned}
}
 
For the anomalous contribution, the center-time integral similarly gives

\hypertarget{sm-eq-b21}{}%
\FittedEquation{C22}{%
\begin{aligned}
        &\lim_{\mathcal T\to\infty}\frac{1}{(\mathcal T)^2}
        \prod_{j=1}^2\left[\int_{-(\mathcal T-|t_{rj}|)/2}^{(\mathcal T-|t_{rj}|)/2}dt_{cj}\right]
        \mathcal F\!\left(\mathbf k_1;t_{c1}+\frac{t_{r1}}{2},t_{c2}+\frac{t_{r2}}{2}\right)
        \left[\mathcal F\!\left(\mathbf k_1;t_{c1}-\frac{t_{r1}}{2},t_{c2}-\frac{t_{r2}}{2}\right)\right]^*\\
        {}={}&|u_{\mathbf k_1}v_{\mathbf k_1}|^2
        \bigl[f_{\mathrm F}(E^{\mathrm{BCS}}_{\mathbf k_1})^2+f_{\mathrm F}(-E^{\mathrm{BCS}}_{\mathbf k_1})^2\bigr]
        \cos\!\left[E^{\mathrm{BCS}}_{\mathbf k_1}(t_{r1}-t_{r2})\right].
        \end{aligned}
}
 
The remaining Fourier transforms over the relative times yield

\hypertarget{sm-eq-b22}{}%
\FittedEquation{C23}{%
\begin{aligned}
        &A^{(2)<}_{\mathrm{BCS},\sigma_1\sigma_2,\mathrm{anomalous}}(\mathbf k_1,\mathbf k_2;\omega_1,\omega_2)\\
        &\quad=\delta_{\mathbf k_1,-\mathbf k_2}(1-\delta_{\sigma_1\sigma_2})
        \frac{|u_{\mathbf k_1}v_{\mathbf k_1}|^2}{2}
        \bigl[f_{\mathrm F}(E^{\mathrm{BCS}}_{\mathbf k_1})^2+f_{\mathrm F}(-E^{\mathrm{BCS}}_{\mathbf k_1})^2\bigr]\\
        &\qquad\times\Bigl[\delta(\omega_1-E^{\mathrm{BCS}}_{\mathbf k_1})\delta(\omega_2+E^{\mathrm{BCS}}_{\mathbf k_1})
        +\delta(\omega_1+E^{\mathrm{BCS}}_{\mathbf k_1})\delta(\omega_2-E^{\mathrm{BCS}}_{\mathbf k_1})\Bigr].
        \end{aligned}
}
 
For generic back-to-back momenta \(\mathbf k_1=-\mathbf k_2=\mathbf k\), the exchange contribution vanishes. Introducing the total and relative energies

\hypertarget{sm-eq-b23}{}\hypertarget{sm-eq-bcs-relative-energy}{}%
\FittedEquation{C24}{%
\Omega=\omega_1+\omega_2,\quad \omega_{\mathrm r}=\frac{\omega_1-\omega_2}{2}
}
 
with unit Jacobian \(d\omega_1\,d\omega_2=d\Omega\,d\omega_{\mathrm r}\), the direct contribution becomes

\hypertarget{sm-eq-b24}{}%
\FittedEquation{C25}{%
\begin{aligned}
        &A^{(2)<}_{\mathrm{BCS},\sigma_1\sigma_2,\mathrm{direct}}(\mathbf k,-\mathbf k;\Omega,\omega_{\mathrm r})\\
        &\quad=u_{\mathbf k}^4f_{\mathrm F}(E^{\mathrm{BCS}}_{\mathbf k})^2
        \delta(\Omega-2E^{\mathrm{BCS}}_{\mathbf k})\delta(\omega_{\mathrm r})
        +|v_{\mathbf k}|^4f_{\mathrm F}(-E^{\mathrm{BCS}}_{\mathbf k})^2
        \delta(\Omega+2E^{\mathrm{BCS}}_{\mathbf k})\delta(\omega_{\mathrm r})\\
        &\qquad+|u_{\mathbf k}v_{\mathbf k}|^2f_{\mathrm F}(E^{\mathrm{BCS}}_{\mathbf k})f_{\mathrm F}(-E^{\mathrm{BCS}}_{\mathbf k})
        \delta(\Omega)\bigl[\delta(\omega_{\mathrm r}-E^{\mathrm{BCS}}_{\mathbf k})+\delta(\omega_{\mathrm r}+E^{\mathrm{BCS}}_{\mathbf k})\bigr].
        \end{aligned}
}
 
The anomalous contribution becomes

\hypertarget{sm-eq-b25}{}%
\FittedEquation{C26}{%
\begin{aligned}
        &A^{(2)<}_{\mathrm{BCS},\sigma_1\sigma_2,\mathrm{anomalous}}(\mathbf k,-\mathbf k;\Omega,\omega_{\mathrm r})\\
        &\quad=
        (1-\delta_{\sigma_1\sigma_2})\frac{|u_{\mathbf k}v_{\mathbf k}|^2}{2}
        \bigl[f_{\mathrm F}(E^{\mathrm{BCS}}_{\mathbf k})^2+f_{\mathrm F}(-E^{\mathrm{BCS}}_{\mathbf k})^2\bigr]\delta(\Omega)\\
        &\qquad\times\bigl[\delta(\omega_{\mathrm r}-E^{\mathrm{BCS}}_{\mathbf k})+\delta(\omega_{\mathrm r}+E^{\mathrm{BCS}}_{\mathbf k})\bigr].
        \end{aligned}
}
 
Combining the direct and anomalous contributions, the two-particle removal spectral function for generic back-to-back momenta is

\hypertarget{sm-eq-b26}{}%
\FittedEquation{C27}{%
\begin{aligned}
        &A^{(2)<}_{\mathrm{BCS},\sigma_1\sigma_2}(\mathbf k,-\mathbf k;\Omega,\omega_{\mathrm r})=u_{\mathbf k}^4f_{\mathrm F}(E^{\mathrm{BCS}}_{\mathbf k})^2\delta(\Omega-2E^{\mathrm{BCS}}_{\mathbf k})\delta(\omega_{\mathrm r})\\
        &\qquad+|v_{\mathbf k}|^4f_{\mathrm F}(-E^{\mathrm{BCS}}_{\mathbf k})^2\delta(\Omega+2E^{\mathrm{BCS}}_{\mathbf k})\delta(\omega_{\mathrm r})\\
        &\qquad+\frac{|u_{\mathbf k}v_{\mathbf k}|^2}{2}\delta(\Omega)
        \bigl[\delta(\omega_{\mathrm r}-E^{\mathrm{BCS}}_{\mathbf k})+\delta(\omega_{\mathrm r}+E^{\mathrm{BCS}}_{\mathbf k})\bigr]\\
        &\qquad-\delta_{\sigma_1\sigma_2}\frac{|u_{\mathbf k}v_{\mathbf k}|^2}{2}
        \bigl[f_{\mathrm F}(E^{\mathrm{BCS}}_{\mathbf k})^2+f_{\mathrm F}(-E^{\mathrm{BCS}}_{\mathbf k})^2\bigr]\delta(\Omega)\\
        &\qquad\quad\times\bigl[\delta(\omega_{\mathrm r}-E^{\mathrm{BCS}}_{\mathbf k})+\delta(\omega_{\mathrm r}+E^{\mathrm{BCS}}_{\mathbf k})\bigr].
        \end{aligned}
}
 
\hypertarget{sm-appendix-e}{}\hypertarget{sm-appendix-e-title}{}
\section*{Appendix D. ARPES of the nodal Bose metal with Cooper surfaces}

\hypertarget{sm-c-single-electron}{}
\subsection*{D.1. One-electron ARPES}

The occupied one-particle spectrum of the nodal Bose metal measured by one-electron ARPES is

\hypertarget{sm-eq-c5}{}%
\FittedEquation{D1}{%
f_{\mathrm F}(\omega)A_{\mathrm N,\sigma}(\mathbf k,\omega)
=\lim_{\mathcal T\to\infty}^{\mathcal D'}\frac{1}{2\pi i\mathcal T}
\iint_{-\frac{\mathcal T}{2}}^{\frac{\mathcal T}{2}}dt\,dt'\,
e^{i\omega(t-t')}G^<_{\mathrm N,\sigma}(\mathbf k;t,t').
}
 
For the spin-rotation-invariant nodal Bose metal, the one-particle lesser Green's function has the exact Lehmann representation

\hypertarget{sm-eq-c6}{}%
\FittedEquation{D2}{%
G^<_{\mathrm N}(\mathbf k\sigma;\mathbf k'\sigma';t,t') \equiv i\langle c^\dagger_{\mathbf k'\sigma'}(t')c_{\mathbf k\sigma}(t)\rangle_{\mathrm N} =\delta_{\mathbf k\mathbf k'}\delta_{\sigma\sigma'}G^<_{\mathrm N,\sigma}(\mathbf k;t,t'),
}
 
\hypertarget{sm-eq-normal-lesser-lehmann}{}%
\FittedEquation{D3}{%
G^<_{\mathrm N,\sigma}(\mathbf k;t,t')
        =i\sum\limits_{\lambda,\lambda'}^{N_{\mathbf k,\lambda}=N_{\mathbf k,\lambda'}+1}\frac{e^{-\beta E^{\mathrm N}_{\mathbf k,\lambda}}}{Z^{\mathrm N}_{\mathbf k}}
        |\langle N_{\mathbf k,\lambda'}|c_{\mathbf k\sigma}|N_{\mathbf k,\lambda}\rangle|^2
        e^{-i(E^{\mathrm N}_{\mathbf k,\lambda}-E^{\mathrm N}_{\mathbf k,\lambda'})(t-t')}
}
 
where \(|N_{\mathbf k,\lambda}\rangle\) denotes a block eigenstate with block electron number and energy \(N_{\mathbf k,\lambda}\) and \(E^{\mathrm N}_{\mathbf k,\lambda}\), respectively. The six classes of allowed one-electron removal processes between block eigenstates in \hyperlink{sm-c-one-electron-transitions}{Table D1} yield four distinct one-electron removal energies:

\hypertarget{sm-eq-c7}{}%
\FittedEquation{D4}{%
\omega_{\mathrm N,\mathbf k\zeta}=\xi_{\mathbf k}+\zeta\mathcal J_{\mathbf k},
        \qquad \zeta=-2,-1,0,+1.
}
 
\hypertarget{sm-c-one-electron-transitions}{}\par\medskip

\noindent{\small Table D1 \textbar{} One-electron removal from the fixed mode \((\mathbf k,\uparrow)\).\par}\nopagebreak[4]
\begingroup\small\setlength{\tabcolsep}{3.5pt}\renewcommand{\arraystretch}{1.30}
\noindent\FittedTable{\begin{tabular}{@{}ccc@{}}\toprule
{\(|N_{\mathbf k,\lambda}\rangle\to|N_{\mathbf k,\lambda'}\rangle\)} & {\(\omega_{\mathrm N,\mathbf k\zeta}=E^{\mathrm N}_{\mathbf k,\lambda}-E^{\mathrm N}_{\mathbf k,\lambda'}\)} & {\(|\langle N_{\mathbf k,\lambda'}|c_{\mathbf k\uparrow}|N_{\mathbf k,\lambda}\rangle|^2\)} \\
\midrule
{\(|2_{\mathbf k,s}\rangle\to|1_{-\mathbf k\downarrow}\rangle\)} & {\(\xi_{\mathbf k}-2\mathcal J_{\mathbf k}\)} & {\(1/2\)} \\
{\(\begin{aligned}&|3_{\mathbf k\downarrow}\rangle\to|2_{-\mathbf k}\rangle,\\&|3_{-\mathbf k\uparrow}\rangle\to|2_{\mathbf k,t_{-1}}\rangle,\\&|3_{-\mathbf k\downarrow}\rangle\to|2_{\mathbf k,t_0}\rangle\end{aligned}\)} & {\(\xi_{\mathbf k}-\mathcal J_{\mathbf k}\)} & {\(\left.\begin{gathered}1\\1\\1/2\end{gathered}\right\}\,5/2\)} \\
{\(|4_{\mathbf k}\rangle\to|3_{\mathbf k\uparrow}\rangle\)} & {\(\xi_{\mathbf k}-\mathcal J_{\mathbf k}\)} & {\(1\)} \\
{\(|1_{\mathbf k\uparrow}\rangle\to|0_{\mathbf k}\rangle\)} & {\(\xi_{\mathbf k}\)} & {\(1\)} \\
{\(\begin{aligned}&|2_{\mathbf k,t_0}\rangle\to|1_{-\mathbf k\downarrow}\rangle,\\&|2_{\mathbf k,t_{+1}}\rangle\to|1_{-\mathbf k\uparrow}\rangle,\\&|2_{\mathbf k}\rangle\to|1_{\mathbf k\downarrow}\rangle\end{aligned}\)} & {\(\xi_{\mathbf k}\)} & {\(\left.\begin{gathered}1/2\\1\\1\end{gathered}\right\}\,5/2\)} \\
{\(|3_{-\mathbf k\downarrow}\rangle\to|2_{\mathbf k,s}\rangle\)} & {\(\xi_{\mathbf k}+\mathcal J_{\mathbf k}\)} & {\(1/2\)} \\
\bottomrule\end{tabular}}\par\endgroup
\smallskip

{\small Each row groups processes between initial states \(|N_{\mathbf k,\lambda}\rangle\) with degenerate energies \(E^{\mathrm N}_{\mathbf k,\lambda}\) and final states \(|N_{\mathbf k,\lambda'}\rangle\) with degenerate energies \(E^{\mathrm N}_{\mathbf k,\lambda'}\). The last column gives the individual squared matrix elements, with braces indicating their sums for grouped processes. With spin rotation symmetry, the results for \(\sigma=\downarrow\) are identical to those for \(\sigma=\uparrow\).\par}
 
The resulting one-particle lesser Green's function is

\hypertarget{sm-eq-c8}{}%
\FittedEquation{D5}{%
\begin{aligned}
&G^<_{\mathrm N,\sigma}(\mathbf k;t,t')
        =\frac{i}{Z^{\mathrm N}_{\mathbf k}}\Bigl[\frac12e^{-\beta(2\xi_{\mathbf k}-2\mathcal J_{\mathbf k})}
        e^{-i(\xi_{\mathbf k}-2\mathcal J_{\mathbf k})(t-t')}\\
&\quad +\left(\frac52e^{-\beta(3\xi_{\mathbf k}-\mathcal J_{\mathbf k})}+e^{-\beta(4\xi_{\mathbf k}-2\mathcal J_{\mathbf k})}\right)
        e^{-i(\xi_{\mathbf k}-\mathcal J_{\mathbf k})(t-t')}\\
&\quad +\left(e^{-\beta\xi_{\mathbf k}}+\frac52e^{-2\beta\xi_{\mathbf k}}\right)
        e^{-i\xi_{\mathbf k}(t-t')} +\frac12e^{-\beta(3\xi_{\mathbf k}-\mathcal J_{\mathbf k})}
        e^{-i(\xi_{\mathbf k}+\mathcal J_{\mathbf k})(t-t')}\Bigr].
\end{aligned}
}
 
Introducing the center and relative times

\hypertarget{sm-eq-c9}{}\hypertarget{sm-eq-normal-relative-time}{}%
\FittedEquation{D6}{%
t_c=\frac{t+t'}{2},\qquad t_r=t-t'
}
 
with unit Jacobian \(dt\,dt'=dt_c\,dt_r\), and substituting Eqs. \hyperlink{sm-eq-c6}{(D2)--(D3)} into Eq. \hyperlink{sm-eq-c5}{(D1)} yields

\hypertarget{sm-eq-c10}{}%
\FittedEquation{D7}{%
\begin{aligned}
f_{\mathrm F}(\omega)A_{\mathrm N,\sigma}(\mathbf k,\omega)
&=\lim_{\mathcal T\to\infty}^{\mathcal D'}\frac{1}{2\pi\mathcal T Z^{\mathrm N}_{\mathbf k}}
\sum\limits_{\lambda,\lambda'}^{N_{\mathbf k,\lambda}=N_{\mathbf k,\lambda'}+1}
e^{-\beta E^{\mathrm N}_{\mathbf k,\lambda}}
|\langle N_{\mathbf k,\lambda'}|c_{\mathbf k\sigma}|N_{\mathbf k,\lambda}\rangle|^2\\
&\qquad\times\int_{-\mathcal T}^{\mathcal T}dt_r
\int_{-(\mathcal T-|t_r|)/2}^{(\mathcal T-|t_r|)/2}dt_c\,
e^{i(\omega-E^{\mathrm N}_{\mathbf k,\lambda}+E^{\mathrm N}_{\mathbf k,\lambda'})t_r}\\
&=\lim_{\mathcal T\to\infty}^{\mathcal D'}\frac{1}{2\pi Z^{\mathrm N}_{\mathbf k}}
\sum\limits_{\lambda,\lambda'}^{N_{\mathbf k,\lambda}=N_{\mathbf k,\lambda'}+1}
e^{-\beta E^{\mathrm N}_{\mathbf k,\lambda}}
|\langle N_{\mathbf k,\lambda'}|c_{\mathbf k\sigma}|N_{\mathbf k,\lambda}\rangle|^2\\
&\qquad\times\int_{-\mathcal T}^{\mathcal T}dt_r
\left(1-\frac{|t_r|}{\mathcal T}\right)
e^{i(\omega-E^{\mathrm N}_{\mathbf k,\lambda}+E^{\mathrm N}_{\mathbf k,\lambda'})t_r}\\
&=\frac{1}{Z^{\mathrm N}_{\mathbf k}}
\sum\limits_{\lambda,\lambda'}^{N_{\mathbf k,\lambda}=N_{\mathbf k,\lambda'}+1}
e^{-\beta E^{\mathrm N}_{\mathbf k,\lambda}}
|\langle N_{\mathbf k,\lambda'}|c_{\mathbf k\sigma}|N_{\mathbf k,\lambda}\rangle|^2
\delta(\omega-E^{\mathrm N}_{\mathbf k,\lambda}+E^{\mathrm N}_{\mathbf k,\lambda'}).
\end{aligned}
}
 
Using the explicit weights in Eq. \hyperlink{sm-eq-c8}{(D5)}, the occupied one-particle spectrum at finite temperature is

\hypertarget{sm-eq-c11}{}%
\FittedEquation{D8}{%
\begin{aligned}
f_{\mathrm F}(\omega)A_{\mathrm N,\sigma}(\mathbf k,\omega)
&=\frac{f_{\mathrm F}(\omega)}{Z^{\mathrm N}_{\mathbf k}}\Bigl[
\frac12\left(e^{-\beta(2\xi_{\mathbf k}-2\mathcal J_{\mathbf k})}+e^{-\beta\xi_{\mathbf k}}\right)
\delta(\omega-\xi_{\mathbf k}+2\mathcal J_{\mathbf k})\\
&\quad+\left(\frac52e^{-2\beta\xi_{\mathbf k}}+\frac72e^{-\beta(3\xi_{\mathbf k}-\mathcal J_{\mathbf k})}
+e^{-\beta(4\xi_{\mathbf k}-2\mathcal J_{\mathbf k})}\right)
\delta(\omega-\xi_{\mathbf k}+\mathcal J_{\mathbf k})\\
&\quad+\left(1+\frac72e^{-\beta\xi_{\mathbf k}}+\frac52e^{-2\beta\xi_{\mathbf k}}\right)
\delta(\omega-\xi_{\mathbf k})\\
&\quad+\frac12\left(e^{-\beta(2\xi_{\mathbf k}-2\mathcal J_{\mathbf k})}
+e^{-\beta(3\xi_{\mathbf k}-\mathcal J_{\mathbf k})}\right)
\delta(\omega-\xi_{\mathbf k}-\mathcal J_{\mathbf k})\Bigr],
\end{aligned}
}
 
Thus, the one-particle spectral weights in \hyperlink{sm-eq5}{Eq. (25)} are

\hypertarget{sm-eq-c12}{}\hypertarget{sm-eq-normal-weight-minus-one}{}\hypertarget{sm-eq-normal-weight-zero}{}\hypertarget{sm-eq-normal-weight-plus-one}{}%
\FittedEquation{D9}{%
W_{\mathrm N,\mathbf k,-2}=\frac{e^{-\beta E^{\mathrm N}_{\mathbf k,s}}+e^{-\beta E^{\mathrm N}_{\mathbf k,1}}}{2Z^{\mathrm N}_{\mathbf k}}
}
\FittedEquation{D10}{%
W_{\mathrm N,\mathbf k,-1}=\frac{\frac52e^{-\beta E^{\mathrm N}_{\mathbf k,2}}+\frac72e^{-\beta E^{\mathrm N}_{\mathbf k,3}}+e^{-\beta E^{\mathrm N}_{\mathbf k,4}}}{Z^{\mathrm N}_{\mathbf k}}
}
\FittedEquation{D11}{%
W_{\mathrm N,\mathbf k,0}=\frac{1+\frac72e^{-\beta E^{\mathrm N}_{\mathbf k,1}}+\frac52e^{-\beta E^{\mathrm N}_{\mathbf k,2}}}{Z^{\mathrm N}_{\mathbf k}}
}
\FittedEquation{D12}{%
W_{\mathrm N,\mathbf k,1}=\frac{e^{-\beta E^{\mathrm N}_{\mathbf k,s}}+e^{-\beta E^{\mathrm N}_{\mathbf k,3}}}{2Z^{\mathrm N}_{\mathbf k}}.
}
 
\hypertarget{sm-c-two-photon}{}
\subsection*{D.2. Two-photon cARPES}

The two-particle removal spectral function of the nodal Bose metal measured by two-photon cARPES is

\hypertarget{sm-eq-c13}{}%
\FittedEquation{D13}{%
\begin{aligned}
&A^{(2)<}_{\mathrm N,\sigma_1\sigma_2}(\mathbf k_1,\mathbf k_2;\omega_1,\omega_2)
=\lim_{\mathcal T\to\infty}^{\mathcal D'}\frac{1}{(2\pi i)^2(\mathcal T)^2}
\iiiint_{-\frac{\mathcal T}{2}}^{\frac{\mathcal T}{2}}dt_1\,dt_2\,dt'_1\,dt'_2\,\\
&\quad\times e^{i[\omega_1(t_1-t'_1)+\omega_2(t_2-t'_2)]}
G^{(2)<}_{\mathrm N,\sigma_1\sigma_2}(\mathbf k_1,\mathbf k_2;t_1,t_2,t'_1,t'_2).
\end{aligned}
}
 
The two-particle lesser Green's function is defined as

\hypertarget{sm-eq-c14}{}%
\FittedEquation{D14}{%
G^{(2)<}_{\mathrm N,\sigma_1\sigma_2}(\mathbf k_1,\mathbf k_2;t_1,t_2,t'_1,t'_2) \equiv i^2\Bigl\langle
        T_{t'}[c^\dagger_{\mathbf k_1\sigma_1}(t'_1)c^\dagger_{\mathbf k_2\sigma_2}(t'_2)]
        T_t[c_{\mathbf k_2\sigma_2}(t_2)c_{\mathbf k_1\sigma_1}(t_1)]\Bigr\rangle_{\mathrm N},
}
 
where \(T_t\) and \(T_{t'}\) denote fermionic time ordering and anti-time ordering, respectively. Wick's theorem does not generally apply to the interacting Hamiltonian \(\hat H_{\mathrm N}\). For generic back-to-back momenta \(\mathbf k_1=-\mathbf k_2=\mathbf k\), the two-particle lesser Green's function is evaluated from the exact block eigenstates. Define the time-ordered two-electron removal amplitude

\hypertarget{sm-eq-c15}{}%
\FittedEquation{D15}{%
\varphi^{(2)}_{\mathrm N,\lambda\lambda'';\sigma_1\sigma_2}(\mathbf k,-\mathbf k;t_1,t_2)
        \equiv\langle N_{\mathbf k,\lambda''}|T_t[c_{-\mathbf k\sigma_2}(t_2)c_{\mathbf k\sigma_1}(t_1)]|N_{\mathbf k,\lambda}\rangle,
}
 
and express the two-particle lesser Green's function as

\hypertarget{sm-eq-c16}{}%
\FittedEquation{D16}{%
\begin{aligned}
        &G^{(2)<}_{\mathrm N,\sigma_1\sigma_2}(\mathbf k,-\mathbf k;t_1,t_2,t'_1,t'_2)\\
        &\quad=i^2
        \sum\limits_{\lambda,\lambda''}^{N_{\mathbf k,\lambda''}=N_{\mathbf k,\lambda}-2}
        \frac{e^{-\beta E^{\mathrm N}_{\mathbf k,\lambda}}}{Z^{\mathrm N}_{\mathbf k}}
        \varphi^{(2)}_{\mathrm N,\lambda\lambda'';\sigma_1\sigma_2}(\mathbf k,-\mathbf k;t_1,t_2)
        [\varphi^{(2)}_{\mathrm N,\lambda\lambda'';\sigma_1\sigma_2}(\mathbf k,-\mathbf k;t'_1,t'_2)]^*.
        \end{aligned}
}
 
Inserting the intermediate block eigenstates gives the time-ordered removal amplitude

\hypertarget{sm-eq-c17}{}%
\FittedEquation{D17}{%
\begin{aligned}
        &\varphi^{(2)}_{\mathrm N,\lambda\lambda'';\sigma_1\sigma_2}(\mathbf k,-\mathbf k;t_1,t_2)\\
        &=\Theta(t_2-t_1)
        \sum\limits_{\lambda'}^{N_{\mathbf k,\lambda'}=N_{\mathbf k,\lambda}-1}
        \langle N_{\mathbf k,\lambda''}|c_{-\mathbf k\sigma_2}|N_{\mathbf k,\lambda'}\rangle
        \langle N_{\mathbf k,\lambda'}|c_{\mathbf k\sigma_1}|N_{\mathbf k,\lambda}\rangle\\
        &\qquad\times e^{i(E^{\mathrm N}_{\mathbf k,\lambda''}-E^{\mathrm N}_{\mathbf k,\lambda'})t_2}
        e^{i(E^{\mathrm N}_{\mathbf k,\lambda'}-E^{\mathrm N}_{\mathbf k,\lambda})t_1}\\
        &\quad-\Theta(t_1-t_2)
        \sum\limits_{\lambda'}^{N_{\mathbf k,\lambda'}=N_{\mathbf k,\lambda}-1}
        \langle N_{\mathbf k,\lambda''}|c_{\mathbf k\sigma_1}|N_{\mathbf k,\lambda'}\rangle
        \langle N_{\mathbf k,\lambda'}|c_{-\mathbf k\sigma_2}|N_{\mathbf k,\lambda}\rangle\\
        &\qquad\times e^{i(E^{\mathrm N}_{\mathbf k,\lambda''}-E^{\mathrm N}_{\mathbf k,\lambda'})t_1}
        e^{i(E^{\mathrm N}_{\mathbf k,\lambda'}-E^{\mathrm N}_{\mathbf k,\lambda})t_2}.
        \end{aligned}
}
 
Intermediate states \(|N_{\mathbf k,\lambda'}\rangle\) with the same energy \(E^{\mathrm N}_{\mathbf k,\lambda'}=E^{\Xi_{\mathrm N}}_{\mathbf k}\) are grouped into a degenerate subspace \(\Xi_{\mathrm N}\) satisfying \(N_{\mathbf k,\lambda'}=N_{\mathbf k,\lambda}-1\). The coherent sums of matrix-element products for the two removal orderings are

\hypertarget{sm-eq-c18}{}\hypertarget{sm-eq-c18a}{}\hypertarget{sm-eq-c18b}{}%
\FittedEquation{D18}{%
M^{\sigma_2\sigma_1}_{\lambda,\lambda'';\Xi_{\mathrm N}}(\mathbf k)
        =\sum_{\lambda'\in\Xi_{\mathrm N}}
        \langle N_{\mathbf k,\lambda''}|c_{-\mathbf k\sigma_2}|N_{\mathbf k,\lambda'}\rangle
        \langle N_{\mathbf k,\lambda'}|c_{\mathbf k\sigma_1}|N_{\mathbf k,\lambda}\rangle,
}
\FittedEquation{D19}{%
M^{\sigma_1\sigma_2}_{\lambda,\lambda'';\Xi_{\mathrm N}}(-\mathbf k)
        =\sum_{\lambda'\in\Xi_{\mathrm N}}
        \langle N_{\mathbf k,\lambda''}|c_{\mathbf k\sigma_1}|N_{\mathbf k,\lambda'}\rangle
        \langle N_{\mathbf k,\lambda'}|c_{-\mathbf k\sigma_2}|N_{\mathbf k,\lambda}\rangle,
}
 
The intermediate-state energy offset is

\hypertarget{sm-eq-c19}{}%
\FittedEquation{D20}{%
\omega^{\Xi_{\mathrm N}}_{\lambda,\lambda''}(\mathbf k)
=\frac{E^{\mathrm N}_{\mathbf k,\lambda}+E^{\mathrm N}_{\mathbf k,\lambda''}}{2}-E^{\Xi_{\mathrm N}}_{\mathbf k}.
}
 
The allowed two-electron removal processes between block eigenstates are listed in \hyperlink{sm-c-two-electron-transitions}{Table D2} for the equal-spin channel and \hyperlink{sm-c-two-electron-transitions-opposite-spin}{Table D3} for the opposite-spin channel.

\hypertarget{sm-c-two-electron-transitions}{}\par\medskip

\noindent{\small Table D2 \textbar{} Two-electron removal from the fixed equal-spin back-to-back modes \((\mathbf k,\uparrow)\) and \((-\mathbf k,\uparrow)\).\par}\nopagebreak[4]
\begingroup\small\setlength{\tabcolsep}{3.5pt}\renewcommand{\arraystretch}{1.30}
\noindent\FittedTable{\begin{tabular}{@{}ccccc@{}}\toprule
{\(|N_{\mathbf k,\lambda}\rangle\to|N_{\mathbf k,\lambda''}\rangle\)} & {\(E^{\Xi_{\mathrm N}}_{\mathbf k}\)} & {\(\omega^{\Xi_{\mathrm N}}_{\lambda,\lambda''}(\mathbf k)\)} & {\(M^{\sigma_2\sigma_1}_{\lambda,\lambda'';\Xi_{\mathrm N}}(\mathbf k)\)} & {\(M^{\sigma_1\sigma_2}_{\lambda,\lambda'';\Xi_{\mathrm N}}(-\mathbf k)\)} \\
\midrule
{\(|2_{\mathbf k,t_{+1}}\rangle\to|0_{\mathbf k}\rangle\)} & {\(\xi_{\mathbf k}\)} & {\(0\)} & {\(1\)} & {\(-1\)} \\
{\(|3_{\mathbf k\downarrow}\rangle\to|1_{-\mathbf k\downarrow}\rangle\)} & {\(2\xi_{\mathbf k}-2\mathcal J_{\mathbf k}\)} & {\(3\mathcal J_{\mathbf k}/2\)} & {\(0\)} & {\(-1/2\)} \\
{\(|3_{\mathbf k\downarrow}\rangle\to|1_{-\mathbf k\downarrow}\rangle\)} & {\(2\xi_{\mathbf k}\)} & {\(-\mathcal J_{\mathbf k}/2\)} & {\(1\)} & {\(-1/2\)} \\
{\(|3_{-\mathbf k\downarrow}\rangle\to|1_{\mathbf k\downarrow}\rangle\)} & {\(2\xi_{\mathbf k}-2\mathcal J_{\mathbf k}\)} & {\(3\mathcal J_{\mathbf k}/2\)} & {\(-1/2\)} & {\(0\)} \\
{\(|3_{-\mathbf k\downarrow}\rangle\to|1_{\mathbf k\downarrow}\rangle\)} & {\(2\xi_{\mathbf k}\)} & {\(-\mathcal J_{\mathbf k}/2\)} & {\(-1/2\)} & {\(1\)} \\
{\(|4_{\mathbf k}\rangle\to|2_{\mathbf k,t_{-1}}\rangle\)} & {\(3\xi_{\mathbf k}-\mathcal J_{\mathbf k}\)} & {\(0\)} & {\(1\)} & {\(-1\)} \\
\bottomrule\end{tabular}}\par\endgroup
\smallskip

{\small Each row specifies an initial state \(|N_{\mathbf k,\lambda}\rangle\), a final state \(|N_{\mathbf k,\lambda''}\rangle\), and an intermediate degenerate subspace \(\Xi_{\mathrm N}\) with energy \(E^{\Xi_{\mathrm N}}_{\mathbf k}\). With spin rotation symmetry, the results for \((\downarrow,\downarrow)\) are identical to those for \((\uparrow,\uparrow)\).\par}
 
For the equal-spin channel \(\sigma_1=\sigma_2=\sigma\),

\hypertarget{sm-eq-c20}{}%
\FittedEquation{D21}{%
\begin{aligned}
        &G^{(2)<}_{\mathrm N,\sigma\sigma}(\mathbf k,-\mathbf k;t_1,t_2,t'_1,t'_2)\\
        &\quad=\frac{i^2}{Z^{\mathrm N}_{\mathbf k}}\Biggl\{
        e^{-2\beta\xi_{\mathbf k}}
        e^{-i\xi_{\mathbf k}(t_1+t_2-t'_1-t'_2)}\\
        &\qquad+e^{-\beta(3\xi_{\mathbf k}-\mathcal J_{\mathbf k})}
        e^{-i(2\xi_{\mathbf k}-\mathcal J_{\mathbf k})(t_1+t_2-t'_1-t'_2)/2}
        \Biggl[e^{-i\mathcal J_{\mathbf k}(|t_1-t_2|-|t'_1-t'_2|)/2}\\
        &\qquad\qquad+e^{i\mathcal J_{\mathbf k}(|t_1-t_2|-|t'_1-t'_2|)/2}
        \cos\!\left[\mathcal J_{\mathbf k}(|t_1-t_2|+|t'_1-t'_2|)\right]\\
        &\qquad\qquad+
        [\Theta(t_1-t_2)\Theta(t'_1-t'_2)+\Theta(t_2-t_1)\Theta(t'_2-t'_1)]
        e^{i\mathcal J_{\mathbf k}(|t_1-t_2|-|t'_1-t'_2|)/2}\\
        &\qquad\qquad\quad\times\sin(\mathcal J_{\mathbf k}|t_1-t_2|)
        \sin(\mathcal J_{\mathbf k}|t'_1-t'_2|)\Biggr]\\
        &\qquad+e^{-\beta(4\xi_{\mathbf k}-2\mathcal J_{\mathbf k})}
        e^{-i(2\xi_{\mathbf k}-2\mathcal J_{\mathbf k})(t_1+t_2-t'_1-t'_2)/2}
        \Biggr\}.
        \end{aligned}
}
 
\hypertarget{sm-c-two-electron-transitions-opposite-spin}{}\par\medskip

\noindent{\small Table D3 \textbar{} Two-electron removal from the fixed opposite-spin back-to-back modes \((\mathbf k,\uparrow)\) and \((-\mathbf k,\downarrow)\).\par}\nopagebreak[4]
\begingroup\small\setlength{\tabcolsep}{3.5pt}\renewcommand{\arraystretch}{1.30}
\noindent\FittedTable{\begin{tabular}{@{}ccccc@{}}\toprule
{\(|N_{\mathbf k,\lambda}\rangle\to|N_{\mathbf k,\lambda''}\rangle\)} & {\(E^{\Xi_{\mathrm N}}_{\mathbf k}\)} & {\(\omega^{\Xi_{\mathrm N}}_{\lambda,\lambda''}(\mathbf k)\)} & {\(M^{\sigma_2\sigma_1}_{\lambda,\lambda'';\Xi_{\mathrm N}}(\mathbf k)\)} & {\(M^{\sigma_1\sigma_2}_{\lambda,\lambda'';\Xi_{\mathrm N}}(-\mathbf k)\)} \\
\midrule
{\(|2_{\mathbf k,s}\rangle\to|0_{\mathbf k}\rangle\)} & {\(\xi_{\mathbf k}\)} & {\(-\mathcal J_{\mathbf k}\)} & {\(-1/\sqrt2\)} & {\(1/\sqrt2\)} \\
{\(|2_{\mathbf k,t_0}\rangle\to|0_{\mathbf k}\rangle\)} & {\(\xi_{\mathbf k}\)} & {\(0\)} & {\(1/\sqrt2\)} & {\(-1/\sqrt2\)} \\
{\(|3_{\mathbf k\uparrow}\rangle\to|1_{-\mathbf k\downarrow}\rangle\)} & {\(2\xi_{\mathbf k}-2\mathcal J_{\mathbf k}\)} & {\(3\mathcal J_{\mathbf k}/2\)} & {\(0\)} & {\(1/2\)} \\
{\(|3_{\mathbf k\uparrow}\rangle\to|1_{-\mathbf k\downarrow}\rangle\)} & {\(2\xi_{\mathbf k}\)} & {\(-\mathcal J_{\mathbf k}/2\)} & {\(0\)} & {\(-1/2\)} \\
{\(|3_{\mathbf k\downarrow}\rangle\to|1_{-\mathbf k\uparrow}\rangle\)} & {\(2\xi_{\mathbf k}\)} & {\(-\mathcal J_{\mathbf k}/2\)} & {\(-1\)} & {\(1\)} \\
{\(|3_{-\mathbf k\uparrow}\rangle\to|1_{\mathbf k\downarrow}\rangle\)} & {\(2\xi_{\mathbf k}\)} & {\(-\mathcal J_{\mathbf k}/2\)} & {\(1\)} & {\(-1\)} \\
{\(|3_{-\mathbf k\downarrow}\rangle\to|1_{\mathbf k\uparrow}\rangle\)} & {\(2\xi_{\mathbf k}-2\mathcal J_{\mathbf k}\)} & {\(3\mathcal J_{\mathbf k}/2\)} & {\(1/2\)} & {\(0\)} \\
{\(|3_{-\mathbf k\downarrow}\rangle\to|1_{\mathbf k\uparrow}\rangle\)} & {\(2\xi_{\mathbf k}\)} & {\(-\mathcal J_{\mathbf k}/2\)} & {\(-1/2\)} & {\(0\)} \\
{\(|4_{\mathbf k}\rangle\to|2_{\mathbf k,s}\rangle\)} & {\(3\xi_{\mathbf k}-\mathcal J_{\mathbf k}\)} & {\(-\mathcal J_{\mathbf k}\)} & {\(-1/\sqrt2\)} & {\(1/\sqrt2\)} \\
{\(|4_{\mathbf k}\rangle\to|2_{\mathbf k,t_0}\rangle\)} & {\(3\xi_{\mathbf k}-\mathcal J_{\mathbf k}\)} & {\(0\)} & {\(-1/\sqrt2\)} & {\(1/\sqrt2\)} \\
\bottomrule\end{tabular}}\par\endgroup
\smallskip

{\small Each row specifies an initial state \(|N_{\mathbf k,\lambda}\rangle\), a final state \(|N_{\mathbf k,\lambda''}\rangle\), and an intermediate degenerate subspace \(\Xi_{\mathrm N}\) with energy \(E^{\Xi_{\mathrm N}}_{\mathbf k}\). With spin rotation symmetry, the results for \((\downarrow,\uparrow)\) are identical to those for \((\uparrow,\downarrow)\).\par}
 
For the opposite-spin channel \(\sigma_1=\bar\sigma_2=\sigma\),

\hypertarget{sm-eq-c21}{}%
\FittedEquation{D22}{%
\begin{aligned}
        &G^{(2)<}_{\mathrm N,\sigma\bar\sigma}(\mathbf k,-\mathbf k;t_1,t_2,t'_1,t'_2)\\
        &\quad=\frac{i^2}{Z^{\mathrm N}_{\mathbf k}}\Biggl\{
        \frac12e^{-\beta(2\xi_{\mathbf k}-2\mathcal J_{\mathbf k})}
        e^{-i(2\xi_{\mathbf k}-2\mathcal J_{\mathbf k})(t_1+t_2-t'_1-t'_2)/2}
        e^{-i\mathcal J_{\mathbf k}(|t_1-t_2|-|t'_1-t'_2|)}\\
        &\qquad+\frac12e^{-2\beta\xi_{\mathbf k}}
        e^{-i\xi_{\mathbf k}(t_1+t_2-t'_1-t'_2)}\\
        &\qquad+e^{-\beta(3\xi_{\mathbf k}-\mathcal J_{\mathbf k})}
        e^{-i(2\xi_{\mathbf k}-\mathcal J_{\mathbf k})(t_1+t_2-t'_1-t'_2)/2}
        \Biggl[2e^{-i\mathcal J_{\mathbf k}(|t_1-t_2|-|t'_1-t'_2|)/2}\\
        &\qquad\qquad+
        [\Theta(t_1-t_2)\Theta(t'_1-t'_2)+\Theta(t_2-t_1)\Theta(t'_2-t'_1)]
        e^{i\mathcal J_{\mathbf k}(|t_1-t_2|-|t'_1-t'_2|)/2}\\
        &\qquad\qquad\quad\times
        \sin(\mathcal J_{\mathbf k}|t_1-t_2|)
        \sin(\mathcal J_{\mathbf k}|t'_1-t'_2|)\Biggr]\\
        &\qquad+\frac12e^{-\beta(4\xi_{\mathbf k}-2\mathcal J_{\mathbf k})}
        \Bigl[e^{-i\xi_{\mathbf k}(t_1+t_2-t'_1-t'_2)}
        e^{-i\mathcal J_{\mathbf k}(|t_1-t_2|-|t'_1-t'_2|)}\\
        &\qquad\qquad+e^{-i(2\xi_{\mathbf k}-2\mathcal J_{\mathbf k})(t_1+t_2-t'_1-t'_2)/2}\Bigr]\Biggr\}.
        \end{aligned}
}
 
The cosine and sine terms in Eqs. \hyperlink{sm-eq-c20}{(D21)} and \hyperlink{sm-eq-c21}{(D22)} arise from the coherent sum over \(|2_{\mathbf k,s}\rangle\) and other two-electron intermediate states, whose energies differ by \(2\mathcal J_{\mathbf k}\).

The center and relative times for each photoemission channel are introduced as

\hypertarget{sm-eq-c22}{}\hypertarget{sm-eq-c22a}{}\hypertarget{sm-eq-normal-two-photon-relative-time}{}\hypertarget{sm-eq-c22b}{}%
\FittedEquation{D23}{%
t_{cj}=\frac{t_j+t'_j}{2},\quad t_{rj}=t_j-t'_j,\quad |t_{rj}|\le\mathcal T,\quad |t_{cj}|\le\frac{\mathcal T-|t_{rj}|}{2}
}
 
with unit Jacobian \(dt_j\,dt'_j=dt_{cj}\,dt_{rj}\). At fixed relative times, the center-time integral gives

\hypertarget{sm-eq-c23}{}%
\FittedEquation{D24}{%
\begin{aligned}
&\lim_{\mathcal T\to\infty}\frac{1}{(\mathcal T)^2}
\prod_{j=1}^2\left[\int_{-(\mathcal T-|t_{rj}|)/2}^{(\mathcal T-|t_{rj}|)/2}dt_{cj}\right]
\times G^{(2)<}_{\mathrm N,\sigma_1\sigma_2}(\mathbf k_1,\mathbf k_2;t_1,t_2,t'_1,t'_2)\\
&\quad=\frac{i^2}{2Z^{\mathrm N}_{\mathbf k}}
\sum\limits_{\lambda,\lambda'',\Xi_{\mathrm N}}^{N_{\mathbf k,\lambda}=N_{\mathbf k,\lambda''}+2}
e^{-\beta E^{\mathrm N}_{\mathbf k,\lambda}}\times
\Bigl[|M^{\sigma_2\sigma_1}_{\lambda,\lambda'';\Xi_{\mathrm N}}(\mathbf k)|^2
e^{-i(E^{\mathrm N}_{\mathbf k,\lambda}-E^{\Xi_{\mathrm N}}_{\mathbf k})t_{r1}}
e^{-i(E^{\Xi_{\mathrm N}}_{\mathbf k}-E^{\mathrm N}_{\mathbf k,\lambda''})t_{r2}}\\
&\qquad\qquad+|M^{\sigma_1\sigma_2}_{\lambda,\lambda'';\Xi_{\mathrm N}}(-\mathbf k)|^2
e^{-i(E^{\Xi_{\mathrm N}}_{\mathbf k}-E^{\mathrm N}_{\mathbf k,\lambda''})t_{r1}}
e^{-i(E^{\mathrm N}_{\mathbf k,\lambda}-E^{\Xi_{\mathrm N}}_{\mathbf k})t_{r2}}\Bigr].
\end{aligned}
}
 
The center-time integral eliminates terms with phase factors \(e^{\pm i(E^{\Xi_{\mathrm N}}_{\mathbf k}-E^{\Xi'_{\mathrm N}}_{\mathbf k})(t_{c1}-t_{c2})}\) arising from distinct degenerate subspaces \(\Xi_{\mathrm N}\neq\Xi_{\mathrm N}'\) with different intermediate energies, \(E^{\Xi_{\mathrm N}}_{\mathbf k}\neq E^{\Xi'_{\mathrm N}}_{\mathbf k}\). The intermediate states within the same degenerate subspace \(\Xi_{\mathrm N}\) remain coherent. Introducing the total and relative energies

\hypertarget{sm-eq-c24}{}\hypertarget{sm-eq-normal-relative-energy}{}%
\FittedEquation{D25}{%
\Omega=\omega_1+\omega_2,\qquad \omega_{\mathrm r}=\frac{\omega_1-\omega_2}{2}
}
 
with unit Jacobian \(d\omega_1\,d\omega_2=d\Omega\,d\omega_{\mathrm r}\), the remaining Fourier transforms over the relative times yield

\hypertarget{sm-eq-c25}{}%
\FittedEquation{D26}{%
\begin{aligned}
&A^{(2)<}_{\mathrm N,\sigma_1\sigma_2}(\mathbf k_1,\mathbf k_2;\Omega,\omega_{\mathrm r})=\frac{1}{2Z^{\mathrm N}_{\mathbf k}}
\sum\limits_{\lambda,\lambda'',\Xi_{\mathrm N}}^{N_{\mathbf k,\lambda}=N_{\mathbf k,\lambda''}+2}
e^{-\beta E^{\mathrm N}_{\mathbf k,\lambda}}
\delta(\Omega-E^{\mathrm N}_{\mathbf k,\lambda}+E^{\mathrm N}_{\mathbf k,\lambda''})\\
&\quad\times\Bigl[|M^{\sigma_2\sigma_1}_{\lambda,\lambda'';\Xi_{\mathrm N}}(\mathbf k)|^2
\delta\!\left(\omega_{\mathrm r}-\omega^{\Xi_{\mathrm N}}_{\lambda,\lambda''}(\mathbf k)\right)+|M^{\sigma_1\sigma_2}_{\lambda,\lambda'';\Xi_{\mathrm N}}(-\mathbf k)|^2
\delta\!\left(\omega_{\mathrm r}+\omega^{\Xi_{\mathrm N}}_{\lambda,\lambda''}(\mathbf k)\right)\Bigr].
\end{aligned}
}
 
Eq. \hyperlink{sm-eq-c25}{(D26)} has the form of \hyperlink{sm-eq9}{Eq. (33)}: the process label \(\gamma\) specifies the initial state \(\lambda\), final state \(\lambda''\), degenerate intermediate subspace \(\Xi_{\mathrm N}\), and removal ordering \(21\) or \(12\). The weights for these two orderings are \(p^{\mathrm N}_{\mathbf k,\lambda}|M^{\sigma_2\sigma_1}_{\lambda,\lambda'';\Xi_{\mathrm N}}(\mathbf k)|^2/2\) and \(p^{\mathrm N}_{\mathbf k,\lambda}|M^{\sigma_1\sigma_2}_{\lambda,\lambda'';\Xi_{\mathrm N}}(-\mathbf k)|^2/2\), respectively. The total energy is \(\Omega^{\gamma}_{\mathrm N,\mathbf k_1\sigma_1,\mathbf k_2\sigma_2}=E^{\mathrm N}_{\mathbf k,\lambda}-E^{\mathrm N}_{\mathbf k,\lambda''}\), and the relative energy is \(\omega^{\gamma}_{\mathrm N,\mathbf k_1\sigma_1,\mathbf k_2\sigma_2}=\pm\omega^{\Xi_{\mathrm N}}_{\lambda,\lambda''}(\mathbf k)\), with the positive (negative) sign for ordering \(21\) (\(12\)). The intermediate-state sum within each degenerate subspace is taken before the modulus squared.

For generic back-to-back momenta and \(\mathcal J_{\mathbf k}>0\), the two-photon cARPES spectral function at finite temperature for the equal-spin channel \(\sigma_1=\sigma_2=\sigma\) is

\hypertarget{sm-eq-c26}{}%
\FittedEquation{D27}{%
\begin{aligned}
        &A^{(2)<}_{\mathrm N,\sigma\sigma}(\mathbf k,-\mathbf k;\Omega,\omega_{\mathrm r})\\
        &\quad=\frac{1}{Z^{\mathrm N}_{\mathbf k}}\Biggl\{
        e^{-2\beta\xi_{\mathbf k}}\delta(\Omega-2\xi_{\mathbf k})\delta(\omega_{\mathrm r})\\
        &\qquad+\frac18e^{-\beta(3\xi_{\mathbf k}-\mathcal J_{\mathbf k})}
        \delta(\Omega-2\xi_{\mathbf k}+\mathcal J_{\mathbf k})
        \Biggl[5\bigl[\delta(\omega_{\mathrm r}-\mathcal J_{\mathbf k}/2)
        +\delta(\omega_{\mathrm r}+\mathcal J_{\mathbf k}/2)\bigr]\\
        &\qquad\qquad+\delta(\omega_{\mathrm r}-3\mathcal J_{\mathbf k}/2)
        +\delta(\omega_{\mathrm r}+3\mathcal J_{\mathbf k}/2)\Biggr]\\
        &\qquad+e^{-\beta(4\xi_{\mathbf k}-2\mathcal J_{\mathbf k})}
        \delta(\Omega-2\xi_{\mathbf k}+2\mathcal J_{\mathbf k})\delta(\omega_{\mathrm r})
        \Biggr\},
        \end{aligned}
}
 
and for the opposite-spin channel \(\sigma_1=\bar\sigma_2=\sigma\) is

\hypertarget{sm-eq-c27}{}%
\FittedEquation{D28}{%
\begin{aligned}
        &A^{(2)<}_{\mathrm N,\sigma\bar\sigma}(\mathbf k,-\mathbf k;\Omega,\omega_{\mathrm r})\\
        &\quad=\frac{1}{Z^{\mathrm N}_{\mathbf k}}\Biggl\{
        \frac14e^{-\beta(2\xi_{\mathbf k}-2\mathcal J_{\mathbf k})}
        \delta(\Omega-2\xi_{\mathbf k}+2\mathcal J_{\mathbf k})
        \bigl[\delta(\omega_{\mathrm r}-\mathcal J_{\mathbf k})
        +\delta(\omega_{\mathrm r}+\mathcal J_{\mathbf k})\bigr]\\
        &\qquad+\frac12e^{-2\beta\xi_{\mathbf k}}
        \delta(\Omega-2\xi_{\mathbf k})\delta(\omega_{\mathrm r})\\
        &\qquad+\frac18e^{-\beta(3\xi_{\mathbf k}-\mathcal J_{\mathbf k})}
        \delta(\Omega-2\xi_{\mathbf k}+\mathcal J_{\mathbf k})
        \Biggl[9\bigl[\delta(\omega_{\mathrm r}-\mathcal J_{\mathbf k}/2)
        +\delta(\omega_{\mathrm r}+\mathcal J_{\mathbf k}/2)\bigr]\\
        &\qquad\qquad+\delta(\omega_{\mathrm r}-3\mathcal J_{\mathbf k}/2)
        +\delta(\omega_{\mathrm r}+3\mathcal J_{\mathbf k}/2)\Biggr]\\
        &\qquad+\frac14e^{-\beta(4\xi_{\mathbf k}-2\mathcal J_{\mathbf k})}
        \delta(\Omega-2\xi_{\mathbf k})
        \bigl[\delta(\omega_{\mathrm r}-\mathcal J_{\mathbf k})
        +\delta(\omega_{\mathrm r}+\mathcal J_{\mathbf k})\bigr]\\
        &\qquad+\frac12e^{-\beta(4\xi_{\mathbf k}-2\mathcal J_{\mathbf k})}
        \delta(\Omega-2\xi_{\mathbf k}+2\mathcal J_{\mathbf k})\delta(\omega_{\mathrm r})\Biggr\}.
        \end{aligned}
}
 
For \(\mathcal J_{\mathbf k}=0\), \(|2_{\mathbf k,s}\rangle\) and the other two-electron states are degenerate and must be summed coherently in Eqs. \hyperlink{sm-eq-c18a}{(D18)} and \hyperlink{sm-eq-c18b}{(D19)} before taking the modulus squared. Then the spectral functions reduce to the noninteracting results

\hypertarget{sm-eq-c27-free}{}%
\FittedEquation{D29}{%
A^{(2)<}_{\mathrm N,\sigma_1\sigma_2}(\mathbf k,-\mathbf k;\Omega,\omega_{\mathrm r})\big|_{\mathcal J_{\mathbf k}=0}
=f_{\mathrm F}(\xi_{\mathbf k})^2
\delta(\Omega-2\xi_{\mathbf k})\delta(\omega_{\mathrm r}).
}
 
The integrals of the two spin channels in Eqs. \hyperlink{sm-eq-c26}{(D27)} and \hyperlink{sm-eq-c27}{(D28)} give

\hypertarget{sm-eq-c28}{}%
\FittedEquation{D30}{%
\int d\Omega\,d\omega_{\mathrm r}\,
A^{(2)<}_{\mathrm N,\sigma\sigma}(\mathbf k,-\mathbf k;\Omega,\omega_{\mathrm r})
=\langle \hat n_{\mathbf k\sigma}\hat n_{-\mathbf k\sigma}\rangle
-\frac{e^{-\beta(3\xi_{\mathbf k}-\mathcal J_{\mathbf k})}}{2Z^{\mathrm N}_{\mathbf k}},
}
 
and

\hypertarget{sm-eq-c29}{}%
\FittedEquation{D31}{%
\int d\Omega\,d\omega_{\mathrm r}\,
A^{(2)<}_{\mathrm N,\sigma\bar\sigma}(\mathbf k,-\mathbf k;\Omega,\omega_{\mathrm r})
=\langle \hat n_{\mathbf k\sigma}\hat n_{-\mathbf k\bar\sigma}\rangle
+\frac{e^{-\beta(3\xi_{\mathbf k}-\mathcal J_{\mathbf k})}}{2Z^{\mathrm N}_{\mathbf k}}.
}
 
The two-particle removal spectral function thus satisfies

\hypertarget{sm-eq-c30}{}%
\FittedEquation{D32}{%
\sum_{\sigma_1,\sigma_2}
\int d\Omega\,d\omega_{\mathrm r}\,
A^{(2)<}_{\mathrm N,\sigma_1\sigma_2}(\mathbf k,-\mathbf k;\Omega,\omega_{\mathrm r})
=
\langle \hat n_{\mathbf k}\hat n_{-\mathbf k}\rangle,
}
 
where \(\hat n_{\mathbf k}\) is conserved at each momentum by the exactly solvable Hamiltonian \(\hat H_{\mathrm N}\).

\hypertarget{sm-c-cooper-surface}{}
\subsection*{D.3. Cooper surface spectroscopy}

For generic back-to-back momenta, the time-domain correlation corresponding to the pair-removal spectral function in \hyperlink{sm-eq6}{Eq. (29)} has the exact Lehmann representation

\hypertarget{sm-eq-c31}{}%
\FittedEquation{D33}{%
\begin{aligned}
S_{\mathrm N,\Delta}^<(\mathbf k;t,t')
&\equiv\langle\hat\Delta^\dagger_{\mathbf k}(t')\hat\Delta_{\mathbf k}(t)\rangle\\
&=\frac{1}{Z^{\mathrm N}_{\mathbf k}}
\sum\limits_{\lambda,\lambda''}^{N_{\mathbf k,\lambda}=N_{\mathbf k,\lambda''}+2}
e^{-\beta E^{\mathrm N}_{\mathbf k,\lambda}}
|\langle N_{\mathbf k,\lambda''}|\hat\Delta_{\mathbf k}|N_{\mathbf k,\lambda}\rangle|^2 \times e^{-i(E^{\mathrm N}_{\mathbf k,\lambda}-E^{\mathrm N}_{\mathbf k,\lambda''})(t-t')}.
\end{aligned}
}
 
The allowed pair-removal processes at finite temperature are \(|4_{\mathbf k}\rangle\to|2_{\mathbf k,s}\rangle\), \(|3_{\pm\mathbf k\sigma}\rangle\to|1_{\mp\mathbf k\sigma}\rangle\), and \(|2_{\mathbf k,s}\rangle\to|0_{\mathbf k}\rangle\), with the corresponding squared matrix elements \(1\), \(1/2\), and \(1\), respectively. Setting \(t'=0\) gives

\hypertarget{sm-eq-c32}{}%
\FittedEquation{D34}{%
S_{\mathrm N,\Delta}^<(\mathbf k;t,0)
=\frac{e^{-\beta(4\xi_{\mathbf k}-2\mathcal J_{\mathbf k})}}{Z^{\mathrm N}_{\mathbf k}}e^{-2i\xi_{\mathbf k}t}
+4\times\frac{1}{2}\,\frac{e^{-\beta(3\xi_{\mathbf k}-\mathcal J_{\mathbf k})}}{Z^{\mathrm N}_{\mathbf k}}e^{-i(2\xi_{\mathbf k}-\mathcal J_{\mathbf k})t} +\frac{e^{-\beta(2\xi_{\mathbf k}-2\mathcal J_{\mathbf k})}}{Z^{\mathrm N}_{\mathbf k}}e^{-i(2\xi_{\mathbf k}-2\mathcal J_{\mathbf k})t}.
}
 
The corresponding nonnegative spin-singlet pair-removal spectral function is

\hypertarget{sm-eq-c33}{}%
\FittedEquation{D35}{%
S_{\mathrm N,\Delta}^<(\mathbf k,\Omega)
\equiv\int_{-\infty}^{\infty}\frac{dt}{2\pi}\,
e^{i\Omega t}S_{\mathrm N,\Delta}^<(\mathbf k;t,0) =\sum_{\eta=0,-1,-2}W^{\Delta}_{\mathrm N,\mathbf k\eta}
\delta(\Omega-\Omega^{\Delta}_{\mathrm N,\mathbf k\eta}),
}
 
where \(\Omega^{\Delta}_{\mathrm N,\mathbf k\eta}=2\xi_{\mathbf k}+\eta\mathcal J_{\mathbf k}\) is the pair-removal energy. The corresponding spectral weights are

\hypertarget{sm-eq-c33-weights}{}\hypertarget{sm-eq-normal-pair-weight-minus-one}{}\hypertarget{sm-eq-normal-pair-weight-minus-two}{}%
\FittedEquation{D36}{%
W^{\Delta}_{\mathrm N,\mathbf k0}=\frac{e^{-\beta(4\xi_{\mathbf k}-2\mathcal J_{\mathbf k})}}{Z^{\mathrm N}_{\mathbf k}},\quad W^{\Delta}_{\mathrm N,\mathbf k,-1}=\frac{2e^{-\beta(3\xi_{\mathbf k}-\mathcal J_{\mathbf k})}}{Z^{\mathrm N}_{\mathbf k}},\quad W^{\Delta}_{\mathrm N,\mathbf k,-2}=\frac{e^{-\beta(2\xi_{\mathbf k}-2\mathcal J_{\mathbf k})}}{Z^{\mathrm N}_{\mathbf k}}
}
 
and the sum rule is

\hypertarget{sm-eq-c33-sum-rule}{}%
\FittedEquation{D37}{%
\int d\Omega\,S_{\mathrm N,\Delta}^<(\mathbf k,\Omega)
=\sum_{\eta=0,-1,-2}W^{\Delta}_{\mathrm N,\mathbf k\eta}
=\langle\hat\Delta^\dagger_{\mathbf k}\hat\Delta_{\mathbf k}\rangle.
}
 
In particular, for \(\mathcal J_{\mathbf k}=0\), the pair-removal spectral function in Eq. \hyperlink{sm-eq-c33}{(D35)} reduces to the noninteracting result

\hypertarget{sm-eq-c38}{}%
\FittedEquation{D38}{%
S_{\mathrm N,\Delta}^<(\mathbf k,\Omega)\big|_{\mathcal J_{\mathbf k}=0}
=f_{\mathrm F}(\xi_{\mathbf k})^2\delta(\Omega-2\xi_{\mathbf k}).
}
 
With spin rotation symmetry, the three spin-triplet components of the two-particle lesser Green's function are identical. An equal-spin channel measures one spin-triplet component, whereas an opposite-spin channel contains one half of the spin-singlet and one half of a spin-triplet component. Thus, the spin-singlet component of the two-particle lesser Green's function is

\hypertarget{sm-eq-c34}{}%
\FittedEquation{D39}{%
\begin{aligned}
G_{\mathrm N,\Delta}^{(2)<}(\mathbf k;t_1,t_2,t'_1,t'_2)
&=2G^{(2)<}_{\mathrm N,\uparrow\downarrow}(\mathbf k,-\mathbf k;t_1,t_2,t'_1,t'_2)
-G^{(2)<}_{\mathrm N,\uparrow\uparrow}(\mathbf k,-\mathbf k;t_1,t_2,t'_1,t'_2)\\
&=\frac{i^2}{2}\Bigl\langle
T_{t'}\Bigl[\hat c^\dagger_{\mathbf k\uparrow}(t'_1)\hat c^\dagger_{-\mathbf k\downarrow}(t'_2)
-\hat c^\dagger_{\mathbf k\downarrow}(t'_1)\hat c^\dagger_{-\mathbf k\uparrow}(t'_2)\Bigr]\\
&\quad\times T_t\Bigl[\hat c_{-\mathbf k\downarrow}(t_2)\hat c_{\mathbf k\uparrow}(t_1)
-\hat c_{-\mathbf k\uparrow}(t_2)\hat c_{\mathbf k\downarrow}(t_1)\Bigr]\Bigr\rangle.
\end{aligned}
}
 
The equal-time relation following from Eq. \hyperlink{sm-eq-c31}{(D33)} is

\hypertarget{sm-eq-c35}{}%
\FittedEquation{D40}{%
G_{\mathrm N,\Delta}^{(2)<}(\mathbf k;t,t,t',t')=i^2S_{\mathrm N,\Delta}^<(\mathbf k;t,t').
}
 
Applying the Fourier transform and long-duration normalization of Eq. \hyperlink{sm-eq-c13}{(D13)} to the first equality in Eq. \hyperlink{sm-eq-c34}{(D39)} gives the spin-singlet two-particle removal spectral function

\hypertarget{sm-eq-c36}{}%
\FittedEquation{D41}{%
A^{(2)<}_{\mathrm N,\Delta}(\mathbf k,-\mathbf k;\Omega,\omega_{\mathrm r})
=2A^{(2)<}_{\mathrm N,\uparrow\downarrow}(\mathbf k,-\mathbf k;\Omega,\omega_{\mathrm r}) -A^{(2)<}_{\mathrm N,\uparrow\uparrow}(\mathbf k,-\mathbf k;\Omega,\omega_{\mathrm r}),
}
 
which is a nonnegative spin-singlet projection. For \(\mathcal J_{\mathbf k}>0\), substituting Eqs. \hyperlink{sm-eq-c26}{(D27)} and \hyperlink{sm-eq-c27}{(D28)} and integrating over relative energy yields

\hypertarget{sm-eq-c37}{}%
\FittedEquation{D42}{%
\begin{aligned}
\overline A^{(2)<}_{\mathrm N,\Delta}(\mathbf k,\Omega)
&=\int_{-\infty}^{\infty}d\omega_{\mathrm r}\,
A^{(2)<}_{\mathrm N,\Delta}(\mathbf k,-\mathbf k;\Omega,\omega_{\mathrm r})\\
&=\frac{1}{Z^{\mathrm N}_{\mathbf k}}\Bigl[
e^{-\beta(2\xi_{\mathbf k}-2\mathcal J_{\mathbf k})}
\delta(\Omega-2\xi_{\mathbf k}+2\mathcal J_{\mathbf k})\\
&\qquad+\frac72e^{-\beta(3\xi_{\mathbf k}-\mathcal J_{\mathbf k})}
\delta(\Omega-2\xi_{\mathbf k}+\mathcal J_{\mathbf k}) +e^{-\beta(4\xi_{\mathbf k}-2\mathcal J_{\mathbf k})}
\delta(\Omega-2\xi_{\mathbf k})\Bigr].
\end{aligned}
}
 
The middle-branch coefficient is \(2(5/2)-3/2=7/2\).

Combining Eqs. \hyperlink{sm-eq-c27-free}{(D29)}, \hyperlink{sm-eq-c33}{(D35)}, \hyperlink{sm-eq-c38}{(D38)}, \hyperlink{sm-eq-c36}{(D41)}, and \hyperlink{sm-eq-c37}{(D42)} gives the relation between the spin-singlet two-particle removal and the pair-removal spectral functions as in \hyperlink{sm-eq11}{Eq. (35)}:

\hypertarget{sm-eq-c39}{}%
\FittedEquation{D43}{%
\begin{aligned}
\overline A^{(2)<}_{\mathrm N,\Delta}(\mathbf k,\Omega)
&=2\overline A^{(2)<}_{\mathrm N,\uparrow\downarrow}(\mathbf k,\Omega)
-\overline A^{(2)<}_{\mathrm N,\uparrow\uparrow}(\mathbf k,\Omega)\\
&=S_{\mathrm N,\Delta}^<(\mathbf k,\Omega)
+\frac32\frac{e^{-\beta(3\xi_{\mathbf k}-\mathcal J_{\mathbf k})}}{Z^{\mathrm N}_{\mathbf k}}\mathbf1_{\mathcal J_{\mathbf k}>0}
\delta(\Omega-\Omega^{\Delta}_{\mathrm N,\mathbf k,-1}),
\end{aligned}
}
 
where the overbar denotes integration over \(\omega_{\mathrm r}\). The indicator \(\mathbf1_{\mathcal J_{\mathbf k}>0}\) is one for \(\mathcal J_{\mathbf k}>0\) and zero at \(\mathcal J_{\mathbf k}=0\). Compared with simultaneous pair removal, for which each of the four three-electron initial states has a squared matrix element of \(1/2\) in \(S_{\mathrm N,\Delta}^<\), sequential spin-singlet removal in \(A^{(2)<}_{\mathrm N,\Delta}\) has a total integrated spectral-weight coefficient of \(7/2\) from these four states, yielding the coefficient \(7/2-4\times1/2=3/2\) in the additional term in Eq. \hyperlink{sm-eq-c39}{(D43)}. Physically, sequential removal involves different intermediate states and both time orderings. In the limit of a long measurement duration, interference terms between intermediate states with different energies average to zero. Simultaneous pair removal retains the coherent combination encoded in the pair operator, leading to different spectral weights. Integrating over \(\omega_{\mathrm r}\) enforces \(t_1-t'_1=t_2-t'_2\), rather than simultaneous removal at \(t_1=t_2\) and \(t'_1=t'_2\).
 
\hypertarget{sm-appendix-i}{}\hypertarget{sm-appendix-i-title}{}
\section*{Appendix E. Nodal density of states of the nodal Bose metal}

\hypertarget{sm-table-dos-i-thermal-weights}{}\par\medskip

\noindent{\small Table E1 \textbar{} Thermal occupation probabilities of the block Hamiltonian \(\hat h_{\mathbf k}^{\mathrm N}\) in the MStJ model.\par}\nopagebreak[4]
\begingroup\small\setlength{\tabcolsep}{3.5pt}\renewcommand{\arraystretch}{1.30}
\noindent\FittedTable{\begin{tabular}{@{}ccc@{}}\toprule
{Block eigenenergy} & {Degeneracy} & {Thermal occupation probability} \\
\midrule
{\(E^{\mathrm N}_{\mathbf k,0}=0\)} & {1} & {\(p^{\mathrm N}_{\mathbf k,0}=e^{-\beta E^{\mathrm N}_{\mathbf k,0}}/Z^{\mathrm N}_{\mathbf k}=1/Z^{\mathrm N}_{\mathbf k}\)} \\
{\(E^{\mathrm N}_{\mathbf k,1}=\xi_{\mathbf k}\)} & {4} & {\(p^{\mathrm N}_{\mathbf k,1}=e^{-\beta E^{\mathrm N}_{\mathbf k,1}}/Z^{\mathrm N}_{\mathbf k}=e^{-\beta\xi_{\mathbf k}}/Z^{\mathrm N}_{\mathbf k}\)} \\
{\(E^{\mathrm N}_{\mathbf k,s}=2\xi_{\mathbf k}-2\mathcal J_{\mathbf k}\)} & {1} & {\(p^{\mathrm N}_{\mathbf k,s}=e^{-\beta E^{\mathrm N}_{\mathbf k,s}}/Z^{\mathrm N}_{\mathbf k}=e^{-\beta(2\xi_{\mathbf k}-2\mathcal J_{\mathbf k})}/Z^{\mathrm N}_{\mathbf k}\)} \\
{\(E^{\mathrm N}_{\mathbf k,2}=2\xi_{\mathbf k}\)} & {5} & {\(p^{\mathrm N}_{\mathbf k,2}=e^{-\beta E^{\mathrm N}_{\mathbf k,2}}/Z^{\mathrm N}_{\mathbf k}=e^{-2\beta\xi_{\mathbf k}}/Z^{\mathrm N}_{\mathbf k}\)} \\
{\(E^{\mathrm N}_{\mathbf k,3}=3\xi_{\mathbf k}-\mathcal J_{\mathbf k}\)} & {4} & {\(p^{\mathrm N}_{\mathbf k,3}=e^{-\beta E^{\mathrm N}_{\mathbf k,3}}/Z^{\mathrm N}_{\mathbf k}=e^{-\beta(3\xi_{\mathbf k}-\mathcal J_{\mathbf k})}/Z^{\mathrm N}_{\mathbf k}\)} \\
{\(E^{\mathrm N}_{\mathbf k,4}=4\xi_{\mathbf k}-2\mathcal J_{\mathbf k}\)} & {1} & {\(p^{\mathrm N}_{\mathbf k,4}=e^{-\beta E^{\mathrm N}_{\mathbf k,4}}/Z^{\mathrm N}_{\mathbf k}=e^{-\beta(4\xi_{\mathbf k}-2\mathcal J_{\mathbf k})}/Z^{\mathrm N}_{\mathbf k}\)} \\
\bottomrule\end{tabular}}\par\endgroup
\smallskip
 
\hypertarget{sm-dos-i-exact}{}
\subsection*{E.1. Exact one-particle density of states}

The exact finite-temperature one-particle density of states (DOS) per spin in \hyperlink{sm-eq-mstj-dos-spectral-contours}{Eq. (37)} is

\hypertarget{sm-eq-dos-i-exact}{}%
\FittedEquation{E1}{%
\rho_{\mathrm N}(\omega,T)=\int_{\mathrm{BZ}}\frac{d^2\mathbf k}{(2\pi)^2}\,\sum_\zeta W_{\mathrm N,\mathbf k\zeta}(\xi_{\mathbf k},\mathcal J_{\mathbf k};T)\,\delta(\omega-\omega_{\mathrm N,\mathbf k\zeta}),
}
 
where \(W_{\mathrm N,\mathbf k\zeta}\) and \(\omega_{\mathrm N,\mathbf k\zeta}=\xi_{\mathbf k}+\zeta\mathcal J_{\mathbf k}\) are the one-particle spectral weights and excitation energies, respectively. \hyperlink{sm-table-dos-i-thermal-weights}{Table E1} lists the block eigenenergies \(E^{\mathrm N}_{\mathbf k,\iota}\), their degeneracies \(g_\iota\), and the thermal occupation probabilities \(p^{\mathrm N}_{\mathbf k,\iota}\). The block partition function is

\hypertarget{sm-eq-dos-i-partition}{}%
\FittedEquation{E2}{%
\begin{aligned}
        Z^{\mathrm N}_{\mathbf k}&=\operatorname{Tr}e^{-\beta\hat h_{\mathbf k}^{\mathrm N}}=\sum_\iota g_\iota e^{-\beta E^{\mathrm N}_{\mathbf k,\iota}}\\
        &=1+4e^{-\beta\xi_{\mathbf k}}+e^{-2\beta(\xi_{\mathbf k}-\mathcal J_{\mathbf k})}+5e^{-2\beta\xi_{\mathbf k}}+4e^{-\beta(3\xi_{\mathbf k}-\mathcal J_{\mathbf k})}+e^{-\beta(4\xi_{\mathbf k}-2\mathcal J_{\mathbf k})},
        \end{aligned}
}
 
where \(\iota\) labels the six sectors in \hyperlink{sm-table-dos-i-thermal-weights}{Table E1}, and \(\sum_\iota g_\iota p^{\mathrm N}_{\mathbf k,\iota}=1\). The four spectral weights in \hyperlink{sm-eq-dos-i-exact}{Eq. (E1)} are

\hypertarget{sm-eq-dos-i-weights}{}%
\FittedEquation{E3}{%
\begin{aligned}
W_{\mathrm N,\mathbf k,-2}&=\tfrac12(p^{\mathrm N}_{\mathbf k,s}+p^{\mathrm N}_{\mathbf k,1})
          =\frac{e^{-\beta E^{\mathrm N}_{\mathbf k,s}}+e^{-\beta E^{\mathrm N}_{\mathbf k,1}}}{2Z^{\mathrm N}_{\mathbf k}}
\end{aligned}
}
 
\hypertarget{sm-eq-dos-i-weight-minus-one}{}%
\FittedEquation{E4}{%
\begin{aligned}
W_{\mathrm N,\mathbf k,-1}&=\tfrac52p^{\mathrm N}_{\mathbf k,2}+\tfrac72p^{\mathrm N}_{\mathbf k,3}+p^{\mathrm N}_{\mathbf k,4}
          =\frac{\frac52e^{-\beta E^{\mathrm N}_{\mathbf k,2}}+\frac72e^{-\beta E^{\mathrm N}_{\mathbf k,3}}+e^{-\beta E^{\mathrm N}_{\mathbf k,4}}}{Z^{\mathrm N}_{\mathbf k}}
\end{aligned}
}
 
\hypertarget{sm-eq-dos-i-weight-zero}{}%
\FittedEquation{E5}{%
\begin{aligned}
W_{\mathrm N,\mathbf k,0}&=p^{\mathrm N}_{\mathbf k,0}+\tfrac72p^{\mathrm N}_{\mathbf k,1}+\tfrac52p^{\mathrm N}_{\mathbf k,2}
          =\frac{1+\frac72e^{-\beta E^{\mathrm N}_{\mathbf k,1}}+\frac52e^{-\beta E^{\mathrm N}_{\mathbf k,2}}}{Z^{\mathrm N}_{\mathbf k}}
\end{aligned}
}
 
\hypertarget{sm-eq-dos-i-weight-plus-one}{}%
\FittedEquation{E6}{%
\begin{aligned}
W_{\mathrm N,\mathbf k,1}&=\tfrac12(p^{\mathrm N}_{\mathbf k,s}+p^{\mathrm N}_{\mathbf k,3})
          =\frac{e^{-\beta E^{\mathrm N}_{\mathbf k,s}}+e^{-\beta E^{\mathrm N}_{\mathbf k,3}}}{2Z^{\mathrm N}_{\mathbf k}}.
\end{aligned}
}
 
\hypertarget{sm-dos-i-coordinates}{}
\subsection*{E.2. Moving tangent--normal frame}

\par\medskip
\hypertarget{sm-fig-dos-i-coordinates}{}
\PrintAsset{nodal-normal-coordinates}
{\small\noindent \textbf{Figure E1 \textbar{} Moving tangent--normal frame.} The thick blue arc along \(\mathcal S_F\) from the node \(\mathbf k_n\) to \(\mathbf K(k_\parallel)\) defines the signed tangential coordinate \(k_\parallel\). The red segment from \(\mathbf K(k_\parallel)\) to \(\mathbf k\) defines the signed normal coordinate \(k_\perp\). The unit tangent and normal vectors, \(\hat{\mathbf t}(k_\parallel)\) and \(\hat{\mathbf n}(k_\parallel)\), are shown at \(\mathbf K(k_\parallel)\) and rotate along the Fermi surface.\par}
\medskip

Let \(\mathbf K(k_\parallel)\) parameterize the smooth Fermi surface \(\mathcal S_F:\xi_{\mathbf K(k_\parallel)}=0\) by signed arc length \(k_\parallel\) along \(\mathcal S_F\), measured from the node, with \(\mathbf K(0)=\mathbf k_n\) and \(\boldsymbol\nabla_{\mathbf k}\xi_{\mathbf K(k_\parallel)}\ne0\). The unit tangent and normal vectors are

\hypertarget{sm-eq-dos-i-frame}{}\hypertarget{sm-eq-dos-i-normal-vector}{}%
\FittedEquation{E7}{%
\hat{\mathbf t}(k_\parallel)=\frac{d\mathbf K(k_\parallel)}{dk_\parallel},\qquad \hat{\mathbf n}(k_\parallel)=\frac{\boldsymbol\nabla_{\mathbf k}\xi_{\mathbf K(k_\parallel)}}{|\boldsymbol\nabla_{\mathbf k}\xi_{\mathbf K(k_\parallel)}|},
}
 
and the signed curvature \(\kappa_F\) satisfies

\hypertarget{sm-eq-dos-i-signed-curvature}{}\hypertarget{sm-eq-dos-i-normal-curvature}{}%
\FittedEquation{E8}{%
\frac{d\hat{\mathbf t}(k_\parallel)}{dk_\parallel}=\kappa_F(k_\parallel)\hat{\mathbf n}(k_\parallel),\qquad \frac{d\hat{\mathbf n}(k_\parallel)}{dk_\parallel}=-\kappa_F(k_\parallel)\hat{\mathbf t}(k_\parallel).
}
 
A generic momentum \(\mathbf k\) near the curve is specified by the footpoint \(\mathbf K(k_\parallel)\) and its signed normal displacement \(k_\perp\):

\hypertarget{sm-eq-dos-i-coordinate-map}{}%
\FittedEquation{E9}{%
\mathbf k(k_\parallel,k_\perp)=\mathbf K(k_\parallel)+k_\perp\hat{\mathbf n}(k_\parallel).
}
 
Differentiating \hyperlink{sm-eq-dos-i-coordinate-map}{Eq. (E9)} gives \(\partial_{k_\parallel}\mathbf k=(1-\kappa_F k_\perp)\hat{\mathbf t}\) and \(\partial_{k_\perp}\mathbf k=\hat{\mathbf n}\). The absolute value of the local Jacobian is

\hypertarget{sm-eq-dos-i-jacobian-determinant}{}%
\FittedEquation{E10}{%
\left|\frac{\partial(k_x,k_y)}{\partial(k_\parallel,k_\perp)}\right|=\left|\det\begin{pmatrix}(1-\kappa_F k_\perp)t_x & n_x\\(1-\kappa_F k_\perp)t_y & n_y\end{pmatrix}\right|=|1-\kappa_F k_\perp |\,|t_xn_y-t_yn_x|=|1-\kappa_F k_\perp |.
}
 
Then the area element in the local coordinates is

\hypertarget{sm-eq-dos-i-jacobian}{}%
\FittedEquation{E11}{%
d^2\mathbf k=|1-\kappa_F k_\perp |\,dk_\parallel \,dk_\perp .
}
 
The coordinate map is locally one-to-one in a sufficiently narrow neighborhood of the smooth curve. In particular, \(|\kappa_F k_\perp |<1\) avoids a local Jacobian singularity.

\hypertarget{sm-dos-i-local-expansion}{}
\subsection*{E.3. Local expansion near the nodes}

Without loss of generality, consider the node \(\mathbf k_n=(k_n,k_n)\) in the first quadrant. The unit normal and tangent vectors at the node are \(\hat{\mathbf n}_n=(1,1)/\sqrt2\) and \(\hat{\mathbf t}_n=(1,-1)/\sqrt2\), respectively.

Expanding \(\xi_{\mathbf k}\) near the node in the normal displacement \(k_\perp\) gives

\hypertarget{sm-eq-dos-i-dispersion}{}%
\FittedEquation{E12}{%
\xi_{\mathbf k(k_\parallel,k_\perp)}=v(k_\parallel)k_\perp+O(k_\perp^2)=v_nk_\perp+O(k_\perp k_\parallel^2,k_\perp^2).
}
 
The \(k_x\leftrightarrow k_y\) symmetry of \(\xi_{\mathbf k}\) ensures \(v(-k_\parallel)=v(k_\parallel)\), so \(v(k_\parallel)=|\boldsymbol\nabla_{\mathbf k}\xi_{\mathbf K(k_\parallel)}|=v_n+O(k_\parallel^2)\), with \(v_n=v(0)\). There is no term in \hyperlink{sm-eq-dos-i-dispersion}{Eq. (E12)} depending only on \(k_\parallel\), because \(\xi_{\mathbf k(k_\parallel,0)}=\xi_{\mathbf K(k_\parallel)}=0\) on the Fermi surface \(\mathcal S_F\).

The local expansion gives

\hypertarget{sm-eq-dos-i-momentum-expansion}{}%
\FittedEquation{E13}{%
\delta\mathbf k=\mathbf k-\mathbf k_n=k_\parallel\hat{\mathbf t}_n+k_\perp\hat{\mathbf n}_n+O(k_\parallel^2,k_\perp k_\parallel),
}
 
so \(\delta k_x-\delta k_y=\sqrt2k_\parallel+O(k_\perp k_\parallel,k_\parallel^3)\). Expanding the \emph{d}-wave form factor gives

\hypertarget{sm-eq-dos-i-interaction}{}%
\FittedEquation{E14}{%
\begin{aligned}
        \phi_{\mathbf k,d}&=\frac{\cos k_x-\cos k_y}{2}
        =-\frac{\sin k_n}{2}(\delta k_x-\delta k_y)+O(|\delta\mathbf k|^2)\\
        &=-\frac{\sin k_n}{\sqrt2}k_\parallel+O(k_\perp k_\parallel,k_\parallel^3),
        \end{aligned}
}
 
The \(k_x\leftrightarrow k_y\) antisymmetry of \(\phi_{\mathbf k,d}\) ensures that \(\phi_{\mathbf k,d}\) is odd in \(k_\parallel\), so there is no \(k_\parallel^2\) term in \hyperlink{sm-eq-dos-i-interaction}{Eq. (E14)}. The local expansion of \(\mathcal J_{\mathbf k}\) is

\hypertarget{sm-eq-dos-i-interaction-expansion}{}%
\FittedEquation{E15}{%
\mathcal J_{\mathbf k}=\mathcal J_d\phi_{\mathbf k,d}^2=u_nk_\parallel^2+O(k_\perp k_\parallel^2,k_\parallel^4),
}
 
where \(u_n=\mathcal J_d\sin^2k_n/2\). The \(k_x\leftrightarrow k_y\) symmetry of \(\mathcal J_{\mathbf k}\) ensures that \(\mathcal J_{\mathbf k}\) is even in \(k_\parallel\). There is no term in \hyperlink{sm-eq-dos-i-interaction-expansion}{Eq. (E15)} depending only on \(k_\perp\), because \(\mathcal J_{\mathbf k(0,k_\perp)}=0\) along \(k_x=k_y\).

As \(T\to0\) with \(\omega/T\) fixed, the dominant energy scales near the node are \(\xi_{\mathbf k},\mathcal J_{\mathbf k}=O(T)\), so

\hypertarget{sm-eq-dos-i-scales}{}\hypertarget{sm-eq-dos-i-tangential-scale}{}%
\FittedEquation{E16}{%
k_\perp=O(T/v_n),\qquad k_\parallel=O\!\left(\sqrt{T/u_n}\right).
}
 
\hypertarget{sm-dos-i-scaling}{}
\subsection*{E.4. Scaling function}

Keeping the leading terms in Eqs. \hyperlink{sm-eq-dos-i-jacobian}{(E11)}, \hyperlink{sm-eq-dos-i-dispersion}{(E12)} and \hyperlink{sm-eq-dos-i-interaction-expansion}{(E15)}, and including the contributions from all four nodes, the finite-temperature DOS in \hyperlink{sm-eq-dos-i-exact}{Eq. (E1)} becomes

\hypertarget{sm-eq-dos-i-nodal-integral}{}%
\FittedEquation{E17}{%
\begin{aligned}
\rho_{\mathrm N}(\omega,T)&\simeq\frac{4}{(2\pi)^2}
        \int_{-\infty}^{+\infty}dk_\parallel \int_{-\infty}^{+\infty}dk_\perp \,\\
        &\qquad\times\sum_\zeta W_{\mathrm N,\zeta}(v_nk_\perp,u_nk_\parallel^2;T)\,
        \delta(\omega-v_nk_\perp -\zeta u_nk_\parallel^2).
\end{aligned}
}
 
where \(W_{\mathrm N,\mathbf k\zeta}(\xi_{\mathbf k},\mathcal J_{\mathbf k};T)\simeq W_{\mathrm N,\zeta}(v_nk_\perp,u_nk_\parallel^2;T)\), and the infinite integration ranges correspond to the leading-order nodal approximation. Using \(\delta(\omega-v_nk_\perp-\zeta u_nk_\parallel^2)=v_n^{-1}\delta\!\left(k_\perp-(\omega-\zeta u_nk_\parallel^2)/v_n\right)\), performing the normal integral gives

\hypertarget{sm-eq-dos-i-normal-integrated}{}%
\FittedEquation{E18}{%
\rho_{\mathrm N}(\omega,T)\simeq\frac1{\pi^2v_n}
        \int_{-\infty}^{\infty}dk_\parallel \sum_\zeta
        W_{\mathrm N,\zeta}(\omega-\zeta u_nk_\parallel^2,u_nk_\parallel^2;T).
}
 
Introduce the dimensionless variables \(x=\xi_{\mathbf k}/T\simeq v_nk_\perp/T\), \(y=\mathcal J_{\mathbf k}/T\simeq u_nk_\parallel^2/T\) and \(z=\omega/T\), and denote the dimensionless spectral weights by \(w_\zeta(x,y)\equiv W_{\mathrm N,\zeta}(\xi_{\mathbf k},\mathcal J_{\mathbf k};T)\). The finite-temperature DOS in the leading-order nodal approximation is

\hypertarget{sm-eq-dos-i-scaling-result}{}%
\FittedEquation{E19}{%
\rho_{\mathrm N}(\omega,T)=\frac{\sqrt T}{\pi^2v_n\sqrt{u_n}}\,\mathcal R_{\mathrm N}(\omega/T)+o(\sqrt T),
}
 
where the scaling function is

\hypertarget{sm-eq-dos-i-scaling-function}{}%
\FittedEquation{E20}{%
\mathcal R_{\mathrm N}(z)=\int_0^\infty dy\,y^{-1/2}\sum_\zeta w_\zeta(z-\zeta y,y).
}
 
The remainder \(o(\sqrt T)\) is taken as \(T\to0\) at fixed \(\omega/T\).

\hypertarget{sm-dos-i-even-small}{}
\subsection*{E.5. Small-\(|\omega|/T\) expansion}

The transformation \(x\mapsto y-x\) exchanges the empty and fully occupied sectors and the one- and three-electron sectors, while leaving both two-electron sectors unchanged, which gives \(w_\zeta(x,y)=w_{-1-\zeta}(y-x,y)\). Then, with \(\zeta^{\prime}=-1-\zeta\), the scaling function satisfies

\hypertarget{sm-eq-dos-i-evenness}{}%
\FittedEquation{E21}{%
\begin{aligned}
        \mathcal R_{\mathrm N}(-z)&=\int_0^\infty dy\,y^{-1/2}\sum_\zeta w_{-1-\zeta}(z+(1+\zeta)y,y)\\
        &=\int_0^\infty dy\,y^{-1/2}\sum_{\zeta^{\prime}}w_{\zeta^{\prime}}(z-\zeta^{\prime}y,y)=\mathcal R_{\mathrm N}(z).
        \end{aligned}
}
 
At low temperatures with \(|z|=|\omega|/T\ll1\), the scaling function in the leading-order nodal approximation has the expansion

\hypertarget{sm-eq-dos-i-small-series}{}%
\FittedEquation{E22}{%
\mathcal R_{\mathrm N}(z)=\mathcal R_{\mathrm N}(0)+\tfrac12\mathcal R_{\mathrm N}^{\prime\prime}(0)z^2+O(z^4),
}
 
where \(\mathcal R_{\mathrm N}(0)\simeq2.05533\) and \(\mathcal R_{\mathrm N}^{\prime\prime}(0)\simeq0.792892\) are obtained from numerical evaluation of \hyperlink{sm-eq-dos-i-scaling-function}{Eq. (E20)}. Substitution into \hyperlink{sm-eq-dos-i-scaling-result}{Eq. (E19)} gives

\hypertarget{sm-eq-dos-i-small-frequency-result}{}%
\FittedEquation{E23}{%
\rho_{\mathrm N}(\omega,T)=\frac{\mathcal R_{\mathrm N}(0)\sqrt T}{\pi^2v_n\sqrt{u_n}}
        \left[1+\frac{\mathcal R_{\mathrm N}''(0)}{2\mathcal R_{\mathrm N}(0)}
        \left(\frac{\omega}{T}\right)^2+O\!\left(\frac{\omega^4}{T^4}\right)\right]+o(\sqrt T).
}
 
\hypertarget{sm-dos-i-zero-temperature}{}
\subsection*{E.6. Large-\(|\omega|/T\) asymptotics}

At low temperatures with \(|z|=|\omega|/T\gg1\) and the pole condition \(x=z-\zeta y\), the dimensionless spectral weights obtained from \hyperlink{sm-eq-dos-i-weights}{Eqs. (E3)--(E6)} take the following asymptotic forms for \(z>0\)

\hypertarget{sm-eq-dos-i-large-frequency-weights}{}%
\FittedEquation{E24}{%
\begin{aligned}
w_0(z,y)&\simeq\frac{1}{1+e^{2(y-z)}}
\end{aligned}
}
 
\hypertarget{sm-eq-dos-i-large-frequency-weight-one}{}%
\FittedEquation{E25}{%
\begin{aligned}
w_1(z-y,y)&\simeq\frac12\left[\frac{1}{1+e^{2(y-z)}}-\frac{1}{1+e^{4(y-z/2)}}\right].
\end{aligned}
}
 
The two leading contributions belong to distinct spectral branches, \(\zeta=0\) and \(\zeta=1\), while the omitted terms and the contributions from the \(\zeta=-2,-1\) branches are exponentially small in \(z\). Substitution into \hyperlink{sm-eq-dos-i-scaling-function}{Eq. (E20)} and application of the Sommerfeld expansion, together with \(\mathcal R_{\mathrm N}(-z)=\mathcal R_{\mathrm N}(z)\), yield

\hypertarget{sm-eq-dos-i-zero-temperature-integrals}{}%
\FittedEquation{E26}{%
\begin{aligned}
        \mathcal R_{\mathrm N}(z)&\simeq\frac{3-1/\sqrt2}{2}\int_0^\infty\frac{dy\,y^{-1/2}}{1+e^{2(y-|z|)}}\\
        &=(3-1/\sqrt2)\sqrt{|z|}\left[1-\frac{\pi^2}{96|z|^2}+O(|z|^{-4})\right],\qquad |z|\to\infty.
        \end{aligned}
}
 
Matching the above expansion to \hyperlink{sm-eq-dos-i-scaling-result}{Eq. (E19)} in the low-energy regime, within the leading-order nodal approximation, yields

\hypertarget{sm-eq-dos-i-zero-temperature-result}{}%
\FittedEquation{E27}{%
\rho_{\mathrm N}(\omega,T)\simeq\frac{3-1/\sqrt2}{\pi^2v_n\sqrt{u_n}}\sqrt{|\omega|}
        \left[1-\frac{\pi^2}{96}\left(\frac{T}{|\omega|}\right)^2+O\!\left(\frac{T^4}{|\omega|^4}\right)\right].
}
 
\hypertarget{sm-appendix-j}{}\hypertarget{sm-appendix-j-title}{}
\section*{Appendix F. Thermodynamics of the nodal Bose metal}

\hypertarget{sm-thermo-f-charge}{}
\subsection*{F.1. Charge compressibility}

The isothermal charge compressibility per site, by definition, is the electron density response to chemical potential at fixed \(T\):

\hypertarget{sm-eq-thermo-f-charge-exact}{}%
\FittedEquation{F1}{%
\kappa_{\mathrm N}(T,\mu)\equiv\left.\frac{\partial n_e(T,\mu)}{\partial\mu}\right|_T.
}
 
The grand potential is defined by

\hypertarget{sm-eq-thermo-f-grand-potential}{}%
\FittedEquation{F2}{%
\Omega_{\mathrm N,\mathrm G}(T,\mu)\equiv-T\ln Z_{\mathrm N}
=-\frac T2\sum_{\mathbf k}\ln Z^{\mathrm N}_{\mathbf k},
}
 
where \(Z_{\mathrm N}=\operatorname{Tr}e^{-\beta\hat H_{\mathrm N}}\), \(Z^{\mathrm N}_{\mathbf k}=\operatorname{Tr}e^{-\beta\hat h^{\mathrm N}_{\mathbf k}}\), and \(\hat h^{\mathrm N}_{\mathbf k}=(\varepsilon_{\mathbf k}-\mu)\hat N_{\mathbf k}-2\mathcal J_{\mathbf k}\hat\Delta^\dagger_{\mathbf k}\hat\Delta_{\mathbf k}\). The block independence of the Hamiltonian \(\hat H_{\mathrm N}\) allows the partition function to factorize over momentum blocks, and the factor \(1/2\) removes the double counting of \(\{\mathbf k,-\mathbf k\}\) blocks. The grand potential per site in the thermodynamic limit is

\hypertarget{sm-eq-thermo-f-grand-potential-per-site}{}%
\FittedEquation{F3}{%
\omega_{\mathrm N,\mathrm G}(T,\mu)=\frac{\Omega_{\mathrm N,\mathrm G}(T,\mu)}{N_s}
=-\frac T2\int_{\mathrm{BZ}}\frac{d^2\mathbf k}{(2\pi)^2}\ln Z^{\mathrm N}_{\mathbf k}.
}
 
The electron density is

\hypertarget{sm-eq-thermo-f-potential-density}{}%
\FittedEquation{F4}{%
\begin{aligned}
n_e(T,\mu)=-\left.\frac{\partial\omega_{\mathrm N,\mathrm G}(T,\mu)}{\partial\mu}\right|_T=-\frac T2&\int_{\mathrm{BZ}}\frac{d^2\mathbf k}{(2\pi)^2}\,\frac{(-\beta)\operatorname{Tr}[\hat N_{\mathbf k}e^{-\beta\hat h^{\mathrm N}_{\mathbf k}}]}{Z^{\mathrm N}_{\mathbf k}}\\
=\frac12&\int_{\mathrm{BZ}}\frac{d^2\mathbf k}{(2\pi)^2}\langle\hat N_{\mathbf k}\rangle,
\end{aligned}
}
 
where \hypertarget{sm-eq-thermo-f-number-thermal-average}{}\(\langle\hat N_{\mathbf k}\rangle=\frac{\operatorname{Tr}[\hat N_{\mathbf k}e^{-\beta\hat H_{\mathrm N}}]}{Z_{\mathrm N}}=\frac{\operatorname{Tr}[\hat N_{\mathbf k}e^{-\beta\hat h^{\mathrm N}_{\mathbf k}}]}{Z^{\mathrm N}_{\mathbf k}}\). Differentiating the electron density in \hyperlink{sm-eq-thermo-f-potential-density}{Eq. (F4)} gives the exact expression

\hypertarget{sm-eq-thermo-f-probability-derivative}{}%
\FittedEquation{F5}{%
\begin{aligned}
\kappa_{\mathrm N}(T,\mu)
&=\frac12\int_{\mathrm{BZ}}\frac{d^2\mathbf k}{(2\pi)^2}\,
\partial_\mu\!\left[
\frac{\operatorname{Tr}[\hat N_{\mathbf k}e^{-\beta\hat h^{\mathrm N}_{\mathbf k}}]}{Z^{\mathrm N}_{\mathbf k}}\right]\\
&=\frac12\int_{\mathrm{BZ}}\frac{d^2\mathbf k}{(2\pi)^2}\,
\Biggl[\frac{\beta}{Z^{\mathrm N}_{\mathbf k}}
\operatorname{Tr}[\hat N_{\mathbf k}^{\,2}e^{-\beta\hat h^{\mathrm N}_{\mathbf k}}]-\frac{\partial_\mu Z^{\mathrm N}_{\mathbf k}}{(Z^{\mathrm N}_{\mathbf k})^2}
\operatorname{Tr}[\hat N_{\mathbf k}e^{-\beta\hat h^{\mathrm N}_{\mathbf k}}]\Biggr]\\
&=\frac1{2T}\int_{\mathrm{BZ}}\frac{d^2\mathbf k}{(2\pi)^2}
\left[\langle\hat N_{\mathbf k}^{\,2}\rangle
-\langle\hat N_{\mathbf k}\rangle^2\right],
\end{aligned}
}
 
which gives the charge compressibility in \hyperlink{sm-eq-mstj-compressibility}{Eq. (41)}. The difference \(\langle\hat N_{\mathbf k}^{\,2}\rangle-\langle\hat N_{\mathbf k}\rangle^2\) measures the block electron-number fluctuations within a momentum block, and

\hypertarget{sm-eq-thermo-f-number-moments}{}%
\FittedEquation{F6}{%
\begin{aligned}
\langle\hat N_{\mathbf k}\rangle
&=4p^{\mathrm N}_{\mathbf k,1}+10p^{\mathrm N}_{\mathbf k,2}+2p^{\mathrm N}_{\mathbf k,s}
+12p^{\mathrm N}_{\mathbf k,3}+4p^{\mathrm N}_{\mathbf k,4}
\end{aligned}
}
 
\hypertarget{sm-eq-thermo-f-number-second-moment}{}%
\FittedEquation{F7}{%
\begin{aligned}
\langle\hat N_{\mathbf k}^{\,2}\rangle
&=4p^{\mathrm N}_{\mathbf k,1}+20p^{\mathrm N}_{\mathbf k,2}+4p^{\mathrm N}_{\mathbf k,s}
+36p^{\mathrm N}_{\mathbf k,3}+16p^{\mathrm N}_{\mathbf k,4}.
\end{aligned}
}
 
The plotted electron density response in \hyperlink{sm-fig-model-normal-observables}{Fig. 5(d)} is \hypertarget{sm-eq-thermo-f-charge-fixed-density-evaluation}{}\(\kappa_{\mathrm N}(T)=\left.\frac{\partial n_e(T,\mu)}{\partial\mu}\right|_{\mu=\mu(T)}\): the partial derivative defining the response is taken at fixed \(T\), before imposing the fixed-density constraint. At fixed electron density, the equation \(n_e[T,\mu(T)]=1-p\) determines \(\mu(T)\).

At zero temperature, \(\left.N_{\mathbf k}\right|_{T=0}=2\Theta(-\xi_{\mathbf k})+2\Theta(\mathcal J_{\mathbf k}-\xi_{\mathbf k})\), so the electron density is

\hypertarget{sm-eq-thermo-f-charge-zero-density}{}%
\FittedEquation{F8}{%
n_e(0,\mu)=\int_{\mathrm{BZ}}\frac{d^2\mathbf k}{(2\pi)^2}
\left[\Theta(-\xi_{\mathbf k})+\Theta(\mathcal J_{\mathbf k}-\xi_{\mathbf k})\right],
}
 
and the charge compressibility is

 \hypertarget{sm-eq-thermo-f-charge-contours}{} 
\hypertarget{sm-eq-thermo-f-charge-zero}{}%
\FittedEquation{F9}{%
\begin{aligned}
\kappa_{\mathrm N}(0,\mu)=\frac{\partial n_e(0,\mu)}{\partial\mu}
&=\int_{\mathrm{BZ}}\frac{d^2\mathbf k}{(2\pi)^2}
\left[\delta(\xi_{\mathbf k})+\delta(\xi_{\mathbf k}-\mathcal J_{\mathbf k})\right]\\
&=\frac1{(2\pi)^2}\left[
\oint_{\mathcal C_{2e}}\frac{d\ell}{|\nabla_{\mathbf k}\xi_{\mathbf k}|}
+\oint_{\mathcal C_{2h}}\frac{d\ell}{|\nabla_{\mathbf k}(\xi_{\mathbf k}-\mathcal J_{\mathbf k})|}
\right],
\end{aligned}
}
 
where the two contributions come from the Cooper surfaces \(\mathcal C_{2e}:\xi_{\mathbf k}=0\) and \(\mathcal C_{2h}:\xi_{\mathbf k}=\mathcal J_{\mathbf k}\), respectively, even though the zero-temperature one-particle DOS vanishes at zero energy, \(\rho_{\mathrm N}(0,0)=0\). At zero temperature, the block electron number jumps by two across each Cooper surface, and these block ground-state level crossings allow the electron density to respond to changes in the chemical potential, yielding a finite zero-temperature charge compressibility.

A low-energy analysis of finite-temperature block electron-number fluctuations yields the same limiting result as \(T\to0\). Away from the nodes, the two low-energy states near \(\mathcal C_{2e}\) are \(|4_{\mathbf k}\rangle\) and \(|2_{\mathbf k,s}\rangle\), with energy difference \(E^{\mathrm N}_{\mathbf k,4}-E^{\mathrm N}_{\mathbf k,s}=2\xi_{\mathbf k}\). Keeping only the two low-energy states near each Cooper surface is termed the \emph{two-state approximation}. Within the two-state subspace, the thermal occupancy probability of \(|4_{\mathbf k}\rangle\) is \(f_{\mathrm F}(2\xi_{\mathbf k})\), so the block electron-number fluctuations are

\hypertarget{sm-eq-thermo-f-charge-two-state-fluctuations}{}%
\FittedEquation{F10}{%
\langle\hat N_{\mathbf k}^{\,2}\rangle-\langle\hat N_{\mathbf k}\rangle^2
\simeq2^2f_{\mathrm F}(2\xi_{\mathbf k})-\left[2f_{\mathrm F}(2\xi_{\mathbf k})\right]^2
=\operatorname{sech}^2(\xi_{\mathbf k}/T).
}
 
In the zero-temperature limit,

\hypertarget{sm-eq-thermo-f-charge-delta-limit}{}%
\FittedEquation{F11}{%
\lim_{T\to0}\frac{\operatorname{sech}^2(\xi_{\mathbf k}/T)}{2T}
=\delta(\xi_{\mathbf k}).
}
 
Replacing \(\xi_{\mathbf k}\) by \(\xi_{\mathbf k}-\mathcal J_{\mathbf k}\) gives the contribution from the level crossing between \(|2_{\mathbf k,s}\rangle\) and \(|0_{\mathbf k}\rangle\) near \(\mathcal C_{2h}\), and the contributions from both Cooper surfaces reproduce the zero-temperature result.

\par\medskip
\hypertarget{sm-fig-thermo-f-regions}{}
\PrintAsset{nodal-and-cooper-surface-regions}
{\small\noindent \textbf{Figure F1 \textbar{} Schematic Brillouin zone division for the low-temperature expansion.} Blue dotted circles centered on the four nodes delimit the fixed nodal patches \(\mathcal R_n\). Light-gray shading marks neighborhoods of the Cooper surfaces \(\mathcal C_{2e}:\xi_{\mathbf k}=0\) (red) and \(\mathcal C_{2h}:\xi_{\mathbf k}=\mathcal J_{\mathbf k}\) (cyan) outside the nodal patches, indicating the thermally accessible regions within \(\mathcal R_c=\mathrm{BZ}\setminus\mathcal R_n\).\par}
\medskip

To obtain the low-temperature corrections, the Brillouin zone is divided into small patches \(\mathcal R_n\) centered on the nodes and their complement \(\mathcal R_c=\mathrm{BZ}\setminus\mathcal R_n\) (\hyperlink{sm-fig-thermo-f-regions}{Fig. F1}). In \(\mathcal R_c\), only regions near the two Cooper surfaces contribute low-temperature corrections, whereas contributions from regions away from both surfaces are exponentially small.

Away from the nodes, the two low-energy states near \(\mathcal C_{2h}\) are \(|2_{\mathbf k,s}\rangle\) and \(|0_{\mathbf k}\rangle\), with energy difference \(E^{\mathrm N}_{\mathbf k,s}-E^{\mathrm N}_{\mathbf k,0}=2(\xi_{\mathbf k}-\mathcal J_{\mathbf k})\). The temperature \(T\) under consideration is much smaller than the gaps separating these two states from the other block eigenstates. The block grand potential \(\Omega^{\mathrm G}_{\mathrm N,\mathbf k}(T,\mu)=-T\ln Z^{\mathrm N}_{\mathbf k}\) acquires the low-temperature correction

\hypertarget{sm-eq-thermo-f-two-state-free-energy}{}%
\FittedEquation{F12}{%
\Delta\Omega^{\mathrm G}_{\mathrm N,\mathbf k}(T,\mu)
=\Omega^{\mathrm G}_{\mathrm N,\mathbf k}(T,\mu)-\Omega^{\mathrm G}_{\mathrm N,\mathbf k}(0,\mu)
\simeq-T\ln\!\left(1+e^{-2|\xi_{\mathbf k}-\mathcal J_{\mathbf k}|/T}\right).
}
 
For \(2|\xi_{\mathbf k}-\mathcal J_{\mathbf k}|\gg T\), \(\Delta\Omega^{\mathrm G}_{\mathrm N,\mathbf k}\simeq-Te^{-2|\xi_{\mathbf k}-\mathcal J_{\mathbf k}|/T}\). Thus, the dominant contribution comes from a neighborhood of \(\mathcal C_{2h}\) with characteristic energy width \(|\xi_{\mathbf k}-\mathcal J_{\mathbf k}|\sim T\). Replacing \(\xi_{\mathbf k}-\mathcal J_{\mathbf k}\) by \(\xi_{\mathbf k}\) gives the corresponding result near \(\mathcal C_{2e}\).

Define the density of states associated with the two Cooper surfaces in a momentum space region \(\mathcal R\) by

\hypertarget{sm-eq-thermo-f-region-contour-density}{}%
\FittedEquation{F13}{%
D_{\mathcal R}(\epsilon)=\int_{\mathcal R}\frac{d^2\mathbf k}{(2\pi)^2}
\left[\delta(\epsilon-\varepsilon_{\mathbf k})
+\delta(\epsilon-\varepsilon_{\mathbf k}+\mathcal J_{\mathbf k})\right].
}
 
For the full Brillouin zone,

\hypertarget{sm-eq-thermo-f-full-contour-density}{}%
\FittedEquation{F14}{%
D(\epsilon)=\int_{\mathrm{BZ}}\frac{d^2\mathbf k}{(2\pi)^2}
\left[\delta(\epsilon-\varepsilon_{\mathbf k})
+\delta(\epsilon-\varepsilon_{\mathbf k}+\mathcal J_{\mathbf k})\right]
=D_{\mathcal R_n}(\epsilon)+D_{\mathcal R_c}(\epsilon).
}
 
Here \(D(\mu)=\kappa_{\mathrm N}(0,\mu)\). Let \(\Delta\omega_{\mathrm N,\mathrm G,\mathcal R}\) denote the low-temperature correction to the grand potential per site contributed by region \(\mathcal R\). In \(\mathcal R_c\), adding the contributions from the two Cooper surfaces gives, up to exponentially small corrections,

\hypertarget{sm-eq-thermo-f-crossing-potential}{}%
\FittedEquation{F15}{%
\begin{aligned}
\Delta\omega_{\mathrm N,\mathrm G,\mathcal R_c}(T,\mu)
&\simeq-\frac T2\int_{\mathcal R_c}\frac{d^2\mathbf k}{(2\pi)^2}
\left[\ln(1+e^{-2|\xi_{\mathbf k}|/T})
+\ln(1+e^{-2|\xi_{\mathbf k}-\mathcal J_{\mathbf k}|/T})\right]\\
&=-\frac T2\int_{-\infty}^{\infty}d\epsilon\,
D_{\mathcal R_c}(\mu+\epsilon)\ln(1+e^{-2|\epsilon|/T}).
\end{aligned}
}
 
Expand \(D_{\mathcal R_c}(\mu+\epsilon)\) around \(\epsilon=0\). The odd terms in \(\epsilon\) integrate to zero because the logarithm is even. The constant term uses

\hypertarget{sm-eq-thermo-f-crossing-integral}{}%
\FittedEquation{F16}{%
\int_{-\infty}^{\infty}d\epsilon\,\ln(1+e^{-2|\epsilon|/T})=T\int_0^\infty dz\,\ln(1+e^{-z})=\frac{\pi^2}{12}T.
}
 
The next even term is proportional to \(\epsilon^2\) and contributes at order \(T^4\); thus,

\hypertarget{sm-eq-thermo-f-sommerfeld}{}%
\FittedEquation{F17}{%
\Delta\omega_{\mathrm N,\mathrm G,\mathcal R_c}(T,\mu)
=-\frac{\pi^2}{24}D_{\mathcal R_c}(\mu)T^2+O(T^4).
}
 
Inside the nodal patches \(\mathcal R_n\), the complete set of block eigenstates must be retained. The zero-temperature grand potential per site contributed by \(\mathcal R_n\) is

\hypertarget{sm-eq-thermo-f-nodal-ground-potential}{}%
\FittedEquation{F18}{%
\omega_{\mathrm N,\mathrm G,\mathcal R_n}(0,\mu)
=\frac12\int_{\mathcal R_n}\frac{d^2\mathbf k}{(2\pi)^2}E^{\mathrm N,\min}_{\mathbf k},
}
 
where \(E^{\mathrm N,\min}_{\mathbf k}=\min\{0,2\xi_{\mathbf k}\}+\min\{0,2\xi_{\mathbf k}-2\mathcal J_{\mathbf k}\}\). The exact finite-temperature correction is

\hypertarget{sm-eq-thermo-f-nodal-thermal-potential}{}%
\FittedEquation{F19}{%
\begin{aligned}
\Delta\omega_{\mathrm N,\mathrm G,\mathcal R_n}(T,\mu)&=\omega_{\mathrm N,\mathrm G,\mathcal R_n}(T,\mu)-\omega_{\mathrm N,\mathrm G,\mathcal R_n}(0,\mu)\\
&=-\frac T2\int_{\mathcal R_n}\frac{d^2\mathbf k}{(2\pi)^2}\left[\ln Z^{\mathrm N}_{\mathbf k}+\frac{E^{\mathrm N,\min}_{\mathbf k}}T\right].
\end{aligned}
}
 
The nodal patches \(\mathcal R_n\) also contain segments of the two Cooper surfaces. Subtracting the two logarithmic terms associated with both surfaces as in \hyperlink{sm-eq-thermo-f-crossing-potential}{Eq. (F15)}, whose leading contribution to the grand potential is \(-\pi^2D_{\mathcal R_n}(\mu)T^2/24\) as in \hyperlink{sm-eq-thermo-f-sommerfeld}{Eq. (F17)}, isolates the remaining nodal contribution:

\hypertarget{sm-eq-thermo-f-nodal-subtraction}{}%
\FittedEquation{F20}{%
\begin{aligned}
&\ln Z^{\mathrm N}_{\mathbf k}+\frac{E^{\mathrm N,\min}_{\mathbf k}}T
-\ln(1+e^{-2|\xi_{\mathbf k}|/T})
-\ln(1+e^{-2|\xi_{\mathbf k}-\mathcal J_{\mathbf k}|/T})\\
={}&\ln\frac{Z^{\mathrm N}_{\mathbf k}}
{(1+e^{-2\xi_{\mathbf k}/T})
(1+e^{-2(\xi_{\mathbf k}-\mathcal J_{\mathbf k})/T})},
\end{aligned}
}
 
which is appreciable only where both \(\xi_{\mathbf k}\) and \(\mathcal J_{\mathbf k}\) are of order \(T\). Introducing the moving tangent--normal frame of \hyperlink{sm-dos-i-coordinates}{Appendix E.2} and performing the local expansion near the nodes gives

\hypertarget{sm-eq-thermo-f-nodal-variables}{}\hypertarget{sm-eq-thermo-f-nodal-interaction-variable}{}\hypertarget{sm-eq-thermo-f-nodal-measure}{}%
\FittedEquation{F21}{%
x=\frac{\xi_{\mathbf k}}T\simeq\frac{v_nk_\perp}T,\qquad y=\frac{\mathcal J_{\mathbf k}}T\simeq\frac{u_nk_\parallel^2}T,\qquad d^2\mathbf k\simeq\frac{T^{3/2}}{2v_n\sqrt{u_n}}\,\frac{dx\,dy}{\sqrt y}.
}
 
where the area element applies to either branch \(k_\parallel=\pm\sqrt{Ty/u_n}\) for \(y\ge0\). The two branches contribute equally at leading order, giving a combined measure \(T^{3/2}dx\,dy/(v_n\sqrt{u_n}\sqrt y)\) for each node. Since \(|k_\perp|\sim T/v_n\) and \(|k_\parallel|\sim\sqrt{T/u_n}\), the thermally active nodal area scales as \(T^{3/2}\). Multiplication by the factor \(T\) in the grand potential in \hyperlink{sm-eq-thermo-f-nodal-thermal-potential}{Eq. (F19)} therefore gives a \(T^{5/2}\) correction.

Extending the scaled patch boundaries to infinity, including both branches of \(k_\parallel\), and summing the four nodes yields

\hypertarget{sm-eq-thermo-f-nodal-potential}{}%
\FittedEquation{F22}{%
\Delta\omega_{\mathrm N,\mathrm G,\mathcal R_n}(T,\mu)
=-\frac{\pi^2}{24}D_{\mathcal R_n}(\mu)T^2
-\frac{4T}{2(2\pi)^2}\frac{T^{3/2}}{v_n\sqrt{u_n}}\mathcal I_\Omega
+o(T^{5/2}),
}
 
where

\hypertarget{sm-eq-thermo-f-nodal-integral}{}%
\FittedEquation{F23}{%
\mathcal I_\Omega
=\int_{-\infty}^{\infty}dx\int_0^{\infty}\frac{dy}{\sqrt y}\,
\ln\frac{\mathcal Z_{\mathrm N}(x,y)}
{(1+e^{-2x})(1+e^{-2(x-y)})}
=11.336402,
}
 
and the exact block partition function in dimensionless form is

\hypertarget{sm-eq-thermo-f-scaled-partition}{}%
\FittedEquation{F24}{%
\mathcal Z_{\mathrm N}(x,y)=1+4e^{-x}+5e^{-2x}+e^{-2x+2y}
+4e^{-3x+y}+e^{-4x+2y}.
}
 
Adding the two regions gives

\hypertarget{sm-eq-thermo-f-potential-expansion}{}%
\FittedEquation{F25}{%
\begin{aligned}
&\omega_{\mathrm N,\mathrm G}(T,\mu)-\omega_{\mathrm N,\mathrm G}(0,\mu)\\
&=\Delta\omega_{\mathrm N,\mathrm G,\mathcal R_c}+\Delta\omega_{\mathrm N,\mathrm G,\mathcal R_n}\\
&=-\frac{\pi^2}{24}
\left[D_{\mathcal R_c}(\mu)+D_{\mathcal R_n}(\mu)\right]T^2
-\frac{\mathcal I_\Omega}{2\pi^2v_n(\mu)\sqrt{u_n(\mu)}}T^{5/2}+o(T^{5/2})\\
&=-\frac{\pi^2}{24}D(\mu)T^2
-Q(\mu)T^{5/2}+o(T^{5/2}).
\end{aligned}
}
 
The \(T^2\) term inside \(\mathcal R_n\) in \hyperlink{sm-eq-thermo-f-nodal-potential}{Eq. (F22)} is essential: the entire nodal-patch contribution is not of order \(T^{5/2}\). Changing the patch boundary transfers the same contribution between \(D_{\mathcal R_n}\) and \(D_{\mathcal R_c}\), leaving their sum unchanged. The coefficient \(Q(\mu)\) of the \(T^{5/2}\) nodal contribution is independent of the patch boundary.

Differentiating the grand potential per site in \hyperlink{sm-eq-thermo-f-potential-expansion}{Eq. (F25)} at fixed temperature gives the low-temperature corrections to the electron density

\hypertarget{sm-eq-thermo-f-density-response-expansion}{}%
\FittedEquation{F26}{%
n_e(T,\mu)=n_e(0,\mu)+\frac{\pi^2}{24}D'(\mu)T^2
+Q'(\mu)T^{5/2}+o(T^{5/2}),
}
 
and the charge compressibility

\hypertarget{sm-eq-thermo-f-charge-response-expansion}{}%
\FittedEquation{F27}{%
\kappa_{\mathrm N}(T,\mu)=D(\mu)+\frac{\pi^2}{24}D''(\mu)T^2
+Q''(\mu)T^{5/2}+o(T^{5/2}).
}
 
Finally, impose the fixed-density constraint on the electron density, \(n_e[T,\mu(T)]=n_e[0,\mu(0)]\). Substituting the fixed-density constraint and \(\mu=\mu(T)\) into \hyperlink{sm-eq-thermo-f-density-response-expansion}{Eq. (F26)} gives

\hypertarget{sm-eq-thermo-f-fixed-density-substitution}{}%
\FittedEquation{F28}{%
n_e[0,\mu(0)]=n_e[0,\mu(T)]+\frac{\pi^2}{24}D'[\mu(T)]T^2+Q'[\mu(T)]T^{5/2}+o(T^{5/2}).
}
 
Denote the chemical-potential shift as \hypertarget{sm-eq-thermo-f-shift-definition}{}\(\delta\mu(T)=\mu(T)-\mu(0)\). Expanding \(n_e[0,\mu(T)]\) around \(\mu(0)\) and using \(\partial_\mu n_e(0,\mu)=D(\mu)\) gives

\hypertarget{sm-eq-thermo-f-zero-density-taylor}{}%
\FittedEquation{F29}{%
n_e[0,\mu(T)]=n_e[0,\mu(0)]+D[\mu(0)]\,\delta\mu+\frac12D'[\mu(0)](\delta\mu)^2+O((\delta\mu)^2).
}
 
Similarly, expanding the coefficients in \hyperlink{sm-eq-thermo-f-fixed-density-substitution}{Eq. (F28)} gives

\hypertarget{sm-eq-thermo-f-coefficient-taylor}{}%
\FittedEquation{F30}{%
D'[\mu(T)]=D'[\mu(0)]+D''[\mu(0)]\delta\mu+O((\delta\mu)^2),
}
 
\hypertarget{sm-eq-thermo-f-q-coefficient-taylor}{}%
\FittedEquation{F31}{%
Q'[\mu(T)]=Q'[\mu(0)]+Q''[\mu(0)]\delta\mu+O((\delta\mu)^2).
}
 
Combining \hyperlink{sm-eq-thermo-f-fixed-density-substitution}{Eqs. (F28)} and \hyperlink{sm-eq-thermo-f-zero-density-taylor}{(F29)} gives

\hypertarget{sm-eq-thermo-f-fixed-density-balance}{}%
\FittedEquation{F32}{%
\begin{aligned}
0={}&D[\mu(0)]\,\delta\mu+\frac12D'[\mu(0)](\delta\mu)^2\\
&+\frac{\pi^2}{24}D'[\mu(0)]T^2+Q'[\mu(0)]T^{5/2}\\
&+\frac{\pi^2}{24}D''[\mu(0)]\delta\mu\,T^2+Q''[\mu(0)]\delta\mu\,T^{5/2}+O((\delta\mu)^2)+o(T^{5/2}).
\end{aligned}
}
 
Since \(D[\mu(0)]\) is positive and finite, balancing the leading thermal correction requires \(\delta\mu=O(T^2)\). Consequently, \hypertarget{sm-eq-thermo-f-shift-order-counting}{}\((\delta\mu)^2=O(T^4)\), \(\delta\mu\,T^2=O(T^4)\), and \(\delta\mu\,T^{5/2}=O(T^{9/2})\), which are all of higher order than \(T^{5/2}\), so retaining terms up to order \(T^{5/2}\) gives

\hypertarget{sm-eq-thermo-f-leading-density-balance}{}%
\FittedEquation{F33}{%
0=D[\mu(0)]\,\delta\mu+\frac{\pi^2}{24}D'[\mu(0)]T^2
+Q'[\mu(0)]T^{5/2}+o(T^{5/2}).
}
 
Dividing by \(D[\mu(0)]\) yields

\hypertarget{sm-eq-thermo-f-chemical-shift}{}%
\FittedEquation{F34}{%
\mu(T)-\mu(0)=-\frac{\pi^2}{24}\frac{D'[\mu(0)]}{D[\mu(0)]}T^2
-\frac{Q'[\mu(0)]}{D[\mu(0)]}T^{5/2}+o(T^{5/2}).
}
 
Physically, the minus signs in \hyperlink{sm-eq-thermo-f-chemical-shift}{Eq. (F34)} indicate that the density change \(D[\mu(0)]\,\delta\mu\) induced by the chemical-potential shift compensates for the thermal correction to maintain a fixed electron density.

Substituting \hyperlink{sm-eq-thermo-f-chemical-shift}{Eq. (F34)} into \hyperlink{sm-eq-thermo-f-charge-response-expansion}{Eq. (F27)} gives

\hypertarget{sm-eq-thermo-f-charge-low-temperature}{}%
\FittedEquation{F35}{%
\begin{aligned}
\kappa_{\mathrm N}(T)&=D[\mu(0)]+\frac{\pi^2}{24}\left(D''[\mu(0)]-\frac{\{D'[\mu(0)]\}^2}{D[\mu(0)]}\right)T^2\\
&\quad+\left(Q''[\mu(0)]-\frac{D'[\mu(0)]Q'[\mu(0)]}{D[\mu(0)]}\right)T^{5/2}+o(T^{5/2})\\
&=D[\mu(0)]+K_{\kappa,2}T^2+K_{\kappa,5/2}T^{5/2}+o(T^{5/2}).
\end{aligned}
}
 
For \hyperlink{sm-fig-model-normal-observables}{Fig. 5(d)}, the coefficients are \(D[\mu(0)]=1.41330~\mathrm{eV}^{-1}\), \(D'[\mu(0)]=-3.96464~\mathrm{eV}^{-2}\), \(D''[\mu(0)]=41.05696~\mathrm{eV}^{-3}\), \(Q'[\mu(0)]=-2.62856~\mathrm{eV}^{-5/2}\), and \(Q''[\mu(0)]=9.93664~\mathrm{eV}^{-7/2}\), giving \(K_{\kappa,2}=12.31035~\mathrm{eV}^{-3}\) and \(K_{\kappa,5/2}=2.56290~\mathrm{eV}^{-7/2}\).

\hypertarget{sm-thermo-f-spin}{}
\subsection*{F.2. Spin susceptibility}

\hypertarget{sm-table-thermo-f-zeeman}{}\par\medskip

\noindent{\small Table F1 \textbar{} Block eigenstates and eigenenergies in the MStJ model under a magnetic field.\par}\nopagebreak[4]
\begingroup\small\setlength{\tabcolsep}{3.5pt}\renewcommand{\arraystretch}{1.30}
\noindent\FittedTable{\begin{tabular}{@{}ccccc@{}}\toprule
{\shortstack[c]{Block electron\\ number \(N_{\mathbf k}\)}} & {Block eigenstate} & {\shortstack[c]{Block spin\\ \(\mathbb S_{\mathbf k}^z\)}} & {Block eigenenergy} & {Degeneracy} \\
\midrule
{0} & {\(\lvert0_{\mathbf k}\rangle\)} & {\(0\)} & {\(0\)} & {1} \\
{\multirow{2}{*}{1}} & {\(\lvert1_{\pm\mathbf k\uparrow}\rangle\)} & {\(+1/2\)} & {\(\xi_{\mathbf k}-B_z/2\)} & {2} \\
 & {\(\lvert1_{\pm\mathbf k\downarrow}\rangle\)} & {\(-1/2\)} & {\(\xi_{\mathbf k}+B_z/2\)} & {2} \\
{\multirow{4}{*}{2}} & {\(\lvert2_{\mathbf k,s}\rangle\)} & {\(0\)} & {\(2\xi_{\mathbf k}-2\mathcal J_{\mathbf k}\)} & {1} \\
 & {\(\lvert2_{\mathbf k,t_{+1}}\rangle\)} & {\(+1\)} & {\(2\xi_{\mathbf k}-B_z\)} & {1} \\
 & {\(\lvert2_{\mathbf k,t_0}\rangle,\ \lvert2_{\pm\mathbf k}\rangle\)} & {\(0\)} & {\(2\xi_{\mathbf k}\)} & {3} \\
 & {\(\lvert2_{\mathbf k,t_{-1}}\rangle\)} & {\(-1\)} & {\(2\xi_{\mathbf k}+B_z\)} & {1} \\
{\multirow{2}{*}{3}} & {\(\lvert3_{\pm\mathbf k\uparrow}\rangle\)} & {\(+1/2\)} & {\(3\xi_{\mathbf k}-\mathcal J_{\mathbf k}-B_z/2\)} & {2} \\
 & {\(\lvert3_{\pm\mathbf k\downarrow}\rangle\)} & {\(-1/2\)} & {\(3\xi_{\mathbf k}-\mathcal J_{\mathbf k}+B_z/2\)} & {2} \\
{4} & {\(\lvert4_{\mathbf k}\rangle\)} & {\(0\)} & {\(4\xi_{\mathbf k}-2\mathcal J_{\mathbf k}\)} & {1} \\
\bottomrule\end{tabular}}\par\endgroup
\smallskip
 
The uniform static spin susceptibility per site, by definition, is the magnetization response to a uniform magnetic field at fixed \(T\) and \(\mu\), evaluated at zero field:

\hypertarget{sm-eq-thermo-f-spin-definition}{}%
\FittedEquation{F36}{%
\chi_s(T,\mu)\equiv\left.\frac{\partial m(T,\mu,B)}{\partial B}\right|_{T,\mu;\,B=0}.
}
 
Denote the Zeeman energy by \(B_z=g_e\mu_{\mathrm B}B\) and the block spin operator along the magnetic field axis by \(\hat{\mathbb S}^{z}_{\mathbf k}=\hat S^{z}_{\mathbf k}+\hat S^{z}_{-\mathbf k}\). The field-dependent block Hamiltonian is \(\hat h^{\mathrm N}_{\mathbf k}(B_z)=\hat h^{\mathrm N}_{\mathbf k}-B_z\hat{\mathbb S}^{z}_{\mathbf k}\). Summing all sixteen block eigenstates in \hyperlink{sm-table-thermo-f-zeeman}{Table F1} (cf. \hyperlink{sm-c-eigenstates}{Table 1}) gives the field-dependent block partition function

\hypertarget{sm-eq-thermo-f-field-partition}{}%
\FittedEquation{F37}{%
\begin{aligned}
Z^{\mathrm N}_{\mathbf k}(T,\mu,B_z)={}&1+4\cosh\!\left(\frac{B_z}{2T}\right)
\left[e^{-\xi_{\mathbf k}/T}+e^{-(3\xi_{\mathbf k}-\mathcal J_{\mathbf k})/T}\right]\\
&+\left[3+2\cosh\!\left(\frac{B_z}{T}\right)\right]e^{-2\xi_{\mathbf k}/T}
+e^{-(2\xi_{\mathbf k}-2\mathcal J_{\mathbf k})/T}
+e^{-(4\xi_{\mathbf k}-2\mathcal J_{\mathbf k})/T}.
\end{aligned}
}
 
The field-dependent grand potential per site is

\hypertarget{sm-eq-thermo-f-field-grand-potential}{}%
\FittedEquation{F38}{%
\omega_{\mathrm N,\mathrm G}(T,\mu,B)=-\frac T2\int_{\mathrm{BZ}}\frac{d^2\mathbf k}{(2\pi)^2}\ln Z^{\mathrm N}_{\mathbf k}(T,\mu,B_z).
}
 
Differentiating with respect to \(B\) gives the magnetization per site

\hypertarget{sm-eq-thermo-f-magnetic-derivatives}{}%
\FittedEquation{F39}{%
m(T,\mu,B)=-\left.\frac{\partial\omega_{\mathrm N,\mathrm G}(T,\mu,B)}{\partial B}\right|_{T,\mu}
=\frac{g_e\mu_{\mathrm B}}2\int_{\mathrm{BZ}}\frac{d^2\mathbf k}{(2\pi)^2}\langle\hat{\mathbb S}^{z}_{\mathbf k}\rangle,
}
 
where \(\langle\hat{\mathbb S}^{z}_{\mathbf k}\rangle=\operatorname{Tr}[\hat{\mathbb S}^{z}_{\mathbf k}e^{-\beta\hat h^{\mathrm N}_{\mathbf k}(B_z)}]/Z^{\mathrm N}_{\mathbf k}(T,\mu,B_z)\). Since \([\hat h^{\mathrm N}_{\mathbf k},\hat{\mathbb S}^{z}_{\mathbf k}]=0\), differentiating \(\langle\hat{\mathbb S}^{z}_{\mathbf k}\rangle\) at fixed \(T\) and \(\mu\) gives

\hypertarget{sm-eq-thermo-f-spin-average-derivative}{}%
\FittedEquation{F40}{%
\begin{aligned}
\left.\frac{\partial\langle\hat{\mathbb S}^{z}_{\mathbf k}\rangle}{\partial B_z}\right|_{T,\mu}
&=\frac{\beta\,\operatorname{Tr}[(\hat{\mathbb S}^{z}_{\mathbf k})^2e^{-\beta\hat h^{\mathrm N}_{\mathbf k}(B_z)}]}{Z^{\mathrm N}_{\mathbf k}}
-\frac{\partial_{B_z} Z^{\mathrm N}_{\mathbf k}}{(Z^{\mathrm N}_{\mathbf k})^2}
\operatorname{Tr}[\hat{\mathbb S}^{z}_{\mathbf k}e^{-\beta\hat h^{\mathrm N}_{\mathbf k}(B_z)}]\\
&=\frac1T\left[\langle(\hat{\mathbb S}^{z}_{\mathbf k})^2\rangle-\langle\hat{\mathbb S}^{z}_{\mathbf k}\rangle^2\right].
\end{aligned}
}
 
At zero magnetic field, spin rotation symmetry gives \(\left.\langle\hat{\mathbb S}^{z}_{\mathbf k}\rangle\right|_{B=0}=0\), and summing the squared block spin projections gives \(\left.\langle(\hat{\mathbb S}^{z}_{\mathbf k})^2\rangle\right|_{B=0}=4(1/2)^2p^{\mathrm N}_{\mathbf k,1}+4(1/2)^2p^{\mathrm N}_{\mathbf k,3}+2p^{\mathrm N}_{\mathbf k,2}\). The exact finite-temperature spin susceptibility is therefore

\hypertarget{sm-eq-thermo-f-spin-exact}{}%
\FittedEquation{F41}{%
\begin{aligned}
\frac{\chi_s(T,\mu)}{(g_e\mu_{\mathrm B})^2}
&=\frac1{2T}\int_{\mathrm{BZ}}\frac{d^2\mathbf k}{(2\pi)^2}
\left[\langle(\hat{\mathbb S}^{z}_{\mathbf k})^2\rangle-\langle\hat{\mathbb S}^{z}_{\mathbf k}\rangle^2\right]\big|_{B=0}\\
&=\frac1{2T}\int_{\mathrm{BZ}}\frac{d^2\mathbf k}{(2\pi)^2}
\left(p^{\mathrm N}_{\mathbf k,1}+p^{\mathrm N}_{\mathbf k,3}+2p^{\mathrm N}_{\mathbf k,2}\right),
\end{aligned}
}
 
which gives \hyperlink{sm-eq-mstj-spin-susceptibility}{Eq. (43)}. The difference \(\langle(\hat{\mathbb S}^{z}_{\mathbf k})^2\rangle-\langle\hat{\mathbb S}^{z}_{\mathbf k}\rangle^2\) measures the block spin fluctuations within a momentum block. The plotted spin susceptibility in \hyperlink{sm-fig-model-normal-observables}{Fig. 5(e)} is \(\chi_s(T)=\chi_s[T,\mu(T)]\), where the fixed electron density \(n_e[T,\mu(T)]=1-p\) determines the chemical potential \(\mu(T)\). The derivative defining the magnetization response is taken before imposing the fixed-density constraint.

At zero magnetic field, the linear magnetization response at fixed electron density equals that at fixed chemical potential because the chemical-potential adjustment required to keep the density fixed starts at quadratic order \(B^2\). Since reversing the magnetic field does not change the electron density, \hypertarget{sm-eq-thermo-f-density-field-symmetry}{}\(n_e(T,\mu,B)=n_e(T,\mu,-B)\), it follows that \(\left.\frac{\partial n_e(T,\mu,B)}{\partial B}\right|_{T,\mu;\,B=0}=0\). Differentiating the fixed-density constraint \(n_e[T,\mu(T,B),B]=1-p\) at fixed \(T\) gives

\hypertarget{sm-eq-thermo-f-density-field-constraint}{}%
\FittedEquation{F42}{%
0=\left.\frac{dn_e}{dB}\right|_{T,n_e}
=\left.\frac{\partial n_e}{\partial B}\right|_{T,\mu}
+\left.\frac{\partial n_e}{\partial\mu}\right|_{T,B}
\left.\frac{d\mu}{dB}\right|_{T,n_e}.
}
 
At \(B=0\), the first partial derivative vanishes, while the zero-field charge compressibility \(\kappa_{\mathrm N}(T,\mu)\) is finite. Therefore, \hypertarget{sm-eq-thermo-f-chemical-field-derivative}{}\(\left.\frac{d\mu}{dB}\right|_{T,n_e;\,B=0}=0\), which gives \(\mu(T,B)=\mu(T,0)+O(B^2)\). The magnetization response under the fixed-density constraint is

\hypertarget{sm-eq-thermo-f-magnetization-density-chain}{}%
\FittedEquation{F43}{%
\left.\frac{dm}{dB}\right|_{T,n_e}
=\left.\frac{\partial m}{\partial B}\right|_{T,\mu}
+\left.\frac{\partial m}{\partial\mu}\right|_{T,B}
\left.\frac{d\mu}{dB}\right|_{T,n_e}.
}
 
The last term vanishes at zero field, so

\hypertarget{sm-eq-thermo-f-spin-ensemble-equivalence}{}%
\FittedEquation{F44}{%
\left.\frac{dm}{dB}\right|_{T,n_e;\,B=0}
=\left.\frac{\partial m}{\partial B}\right|_{T,\mu;\,B=0},
}
 
where both magnetization responses are evaluated at \(\mu=\mu(T,0)\). The zero-field spin susceptibility can therefore be calculated at fixed chemical potential and then evaluated at \(\mu(T)=\mu(T,0)\), as determined by the fixed-density constraint.

To obtain the low-temperature behavior, use the division into \(\mathcal R_n\) and \(\mathcal R_c\) in \hyperlink{sm-fig-thermo-f-regions}{Fig. F1}. The low-energy states near the two Cooper surfaces have zero total spin. Away from the nodes, the lowest excitation with nonzero spin at either Cooper surface costs \(\mathcal J_{\mathbf k}>0\), so the contribution to the spin susceptibility from \(\mathcal R_c\) is exponentially small. Inside \(\mathcal R_n\), the zero-field block spin fluctuations in the local coordinates of \hyperlink{sm-eq-thermo-f-nodal-variables}{Eq. (F21)} are

\hypertarget{sm-eq-thermo-f-scaled-spin-fluctuations}{}%
\FittedEquation{F45}{%
\left.\langle(\hat{\mathbb S}^{z}_{\mathbf k})^2\rangle\right|_{B=0}
=\frac{e^{-x}+e^{-3x+y}+2e^{-2x}}{\mathcal Z_{\mathrm N}(x,y)}.
}
 
The thermally active nodal area scales as \(T^{3/2}\), and the factor \(1/T\) in the spin susceptibility in \hyperlink{sm-eq-thermo-f-spin-exact}{Eq. (F41)} therefore gives a \(\sqrt T\) contribution. The fixed-density chemical-potential shift in \hyperlink{sm-eq-thermo-f-chemical-shift}{Eq. (F34)} is of order \(T^2\), so \(v_n[\mu(T)]\) and \(u_n[\mu(T)]\) can be replaced by \(v_n[\mu(0)]\) and \(u_n[\mu(0)]\) in the leading term. Extending the scaled patch boundaries to infinity, including both branches of \(k_\parallel\), and summing the four nodes yields

\hypertarget{sm-eq-thermo-f-spin-scaling}{}%
\FittedEquation{F46}{%
\begin{aligned}
\frac{\chi_s(T)}{(g_e\mu_{\mathrm B})^2}
&=\frac{4}{2T(2\pi)^2}\frac{T^{3/2}}{v_n[\mu(0)]\sqrt{u_n[\mu(0)]}}
\int_{-\infty}^{\infty}dx\int_0^\infty\frac{dy}{\sqrt y}\,\left.\langle(\hat{\mathbb S}^{z}_{\mathbf k})^2\rangle\right|_{B=0}+o(\sqrt T)\\
&=\frac{\mathcal I_\chi}{2\pi^2v_n[\mu(0)]\sqrt{u_n[\mu(0)]}}\sqrt T+o(\sqrt T)\\
&=K_{\chi,1/2}\sqrt T+o(\sqrt T),
\end{aligned}
}
 
where

\hypertarget{sm-eq-thermo-f-spin-integral}{}%
\FittedEquation{F47}{%
\mathcal I_\chi=\int_{-\infty}^{\infty}dx\int_0^\infty\frac{dy}{\sqrt y}\,
\frac{e^{-x}+e^{-3x+y}+2e^{-2x}}{\mathcal Z_{\mathrm N}(x,y)}
=2.438324.
}
 
For \hyperlink{sm-fig-model-normal-observables}{Fig. 5(e)}, the coefficients are \(v_n[\mu(0)]=0.802893~\mathrm{eV}\) and \(u_n[\mu(0)]=0.026860~\mathrm{eV}\), giving \(\mathcal I_\chi/\{2\pi^2v_n[\mu(0)]\sqrt{u_n[\mu(0)]}\}=0.938755~\mathrm{eV}^{-3/2}\).

\hypertarget{sm-thermo-f-heat}{}
\subsection*{F.3. Specific heat}

The specific heat per site at fixed electron density, by definition, is the temperature derivative of the internal energy per site:

\hypertarget{sm-eq-thermo-f-heat-definition}{}%
\FittedEquation{F48}{%
c_{\mathrm N}(T)\equiv\left.\frac{d u[T,\mu(T)]}{dT}\right|_{n_e},
}
 
where the internal energy per site is

\hypertarget{sm-eq-thermo-f-internal-energy}{}%
\FittedEquation{F49}{%
u(T,\mu)=\frac12\int_{\mathrm{BZ}}\frac{d^2\mathbf k}{(2\pi)^2}\langle\hat h^{\mathrm N}_{\mathbf k}+\mu\hat N_{\mathbf k}\rangle
=\frac12\int_{\mathrm{BZ}}\frac{d^2\mathbf k}{(2\pi)^2}\langle\hat h^{\mathrm N}_{\mathbf k}\rangle+\mu n_e(T,\mu).
}
 
Differentiating the internal energy in \hyperlink{sm-eq-thermo-f-internal-energy}{Eq. (F49)} under the fixed-density constraint gives the exact finite-temperature specific heat

\hypertarget{sm-eq-thermo-f-heat-exact}{}%
\FittedEquation{F50}{%
c_{\mathrm N}(T)=\frac1{2T^2}\Biggl\{ \int_{\mathrm{BZ}}\frac{d^2\mathbf k}{(2\pi)^2} \left[\langle(\hat h^{\mathrm N}_{\mathbf k})^2\rangle -\langle\hat h^{\mathrm N}_{\mathbf k}\rangle^2\right] -\frac{\displaystyle\left\{ \int_{\mathrm{BZ}}\frac{d^2\mathbf k}{(2\pi)^2} \left[\langle\hat h^{\mathrm N}_{\mathbf k}\hat N_{\mathbf k}\rangle -\langle\hat h^{\mathrm N}_{\mathbf k}\rangle\langle\hat N_{\mathbf k}\rangle\right]\right\}^2} {\displaystyle\int_{\mathrm{BZ}}\frac{d^2\mathbf k}{(2\pi)^2} \left[\langle\hat N_{\mathbf k}^{\,2}\rangle-\langle\hat N_{\mathbf k}\rangle^2\right]} \Biggr\},
}
 
where the first term measures the block energy fluctuations, the second term accounts for the change in chemical potential required to keep the electron density fixed, and the thermal averages and \(\hat h^{\mathrm N}_{\mathbf k}\) are evaluated at \(\mu=\mu(T)\),

\hypertarget{sm-eq-thermo-f-covariances}{}%
\FittedEquation{F51}{%
\begin{aligned}
\hphantom{\langle\hat h^{\mathrm N}_{\mathbf k}\hat N_{\mathbf k}\rangle}\mathllap{\langle\hat h^{\mathrm N}_{\mathbf k}\rangle}
&=4p^{\mathrm N}_{\mathbf k,1}E^{\mathrm N}_{\mathbf k,1}+5p^{\mathrm N}_{\mathbf k,2}E^{\mathrm N}_{\mathbf k,2}
+p^{\mathrm N}_{\mathbf k,s}E^{\mathrm N}_{\mathbf k,s}\\
&\quad+4p^{\mathrm N}_{\mathbf k,3}E^{\mathrm N}_{\mathbf k,3}+p^{\mathrm N}_{\mathbf k,4}E^{\mathrm N}_{\mathbf k,4}
\end{aligned}
}
 
\hypertarget{sm-eq-thermo-f-energy-second-moment}{}%
\FittedEquation{F52}{%
\begin{aligned}
\hphantom{\langle\hat h^{\mathrm N}_{\mathbf k}\hat N_{\mathbf k}\rangle}\mathllap{\langle(\hat h^{\mathrm N}_{\mathbf k})^2\rangle}
&=4p^{\mathrm N}_{\mathbf k,1}(E^{\mathrm N}_{\mathbf k,1})^2+5p^{\mathrm N}_{\mathbf k,2}(E^{\mathrm N}_{\mathbf k,2})^2
+p^{\mathrm N}_{\mathbf k,s}(E^{\mathrm N}_{\mathbf k,s})^2\\
&\quad+4p^{\mathrm N}_{\mathbf k,3}(E^{\mathrm N}_{\mathbf k,3})^2+p^{\mathrm N}_{\mathbf k,4}(E^{\mathrm N}_{\mathbf k,4})^2
\end{aligned}
}
 
\hypertarget{sm-eq-thermo-f-energy-number-moment}{}%
\FittedEquation{F53}{%
\begin{aligned}
\hphantom{\langle\hat h^{\mathrm N}_{\mathbf k}\hat N_{\mathbf k}\rangle}\mathllap{\langle\hat h^{\mathrm N}_{\mathbf k}\hat N_{\mathbf k}\rangle}
&=4p^{\mathrm N}_{\mathbf k,1}E^{\mathrm N}_{\mathbf k,1}+10p^{\mathrm N}_{\mathbf k,2}E^{\mathrm N}_{\mathbf k,2}
+2p^{\mathrm N}_{\mathbf k,s}E^{\mathrm N}_{\mathbf k,s}\\
&\quad+12p^{\mathrm N}_{\mathbf k,3}E^{\mathrm N}_{\mathbf k,3}+4p^{\mathrm N}_{\mathbf k,4}E^{\mathrm N}_{\mathbf k,4}.
\end{aligned}
}
 
To derive \hyperlink{sm-eq-thermo-f-heat-exact}{Eq. (F50)}, first differentiate the electron density with respect to \(T\) and \(\mu\) before imposing \(\mu=\mu(T)\):

\hypertarget{sm-eq-thermo-f-density-derivatives}{}%
\FittedEquation{F54}{%
\left.\frac{\partial n_e(T,\mu)}{\partial T}\right|_\mu
=\frac1{2T^2}\int_{\mathrm{BZ}}\frac{d^2\mathbf k}{(2\pi)^2}
\left[\langle\hat N_{\mathbf k}\hat h^{\mathrm N}_{\mathbf k}\rangle
-\langle\hat N_{\mathbf k}\rangle\langle\hat h^{\mathrm N}_{\mathbf k}\rangle\right],
}
 
\hypertarget{sm-eq-thermo-f-density-mu-derivative}{}%
\FittedEquation{F55}{%
\left.\frac{\partial n_e(T,\mu)}{\partial\mu}\right|_T
=\frac1{2T}\int_{\mathrm{BZ}}\frac{d^2\mathbf k}{(2\pi)^2}
\left[\langle\hat N_{\mathbf k}^{\,2}\rangle-\langle\hat N_{\mathbf k}\rangle^2\right].
}
 
Under the fixed-density constraint, \(0=d n_e[T,\mu(T)]/dT=(\partial_T n_e)_\mu+(\partial_\mu n_e)_T\,d\mu/dT\), which gives

\hypertarget{sm-eq-thermo-f-fixed-density-mu-derivative}{}%
\FittedEquation{F56}{%
\frac{d\mu(T)}{dT}
=-\frac1T\,
\frac{\displaystyle\int_{\mathrm{BZ}}\frac{d^2\mathbf k}{(2\pi)^2}
\left[\langle\hat N_{\mathbf k}\hat h^{\mathrm N}_{\mathbf k}\rangle
-\langle\hat N_{\mathbf k}\rangle\langle\hat h^{\mathrm N}_{\mathbf k}\rangle\right]}
{\displaystyle\int_{\mathrm{BZ}}\frac{d^2\mathbf k}{(2\pi)^2}
\left[\langle\hat N_{\mathbf k}^{\,2}\rangle-\langle\hat N_{\mathbf k}\rangle^2\right]},
}
 
The temperature and chemical-potential derivatives of the block energy are

\hypertarget{sm-eq-thermo-f-energy-derivatives}{}%
\FittedEquation{F57}{%
\begin{aligned}
\hphantom{\left.\frac{\partial\langle\hat h^{\mathrm N}_{\mathbf k}\rangle}{\partial\mu}\right|_T}\mathllap{\left.\frac{\partial\langle\hat h^{\mathrm N}_{\mathbf k}\rangle}{\partial T}\right|_\mu}
&=\frac1{T^2}\left[\langle(\hat h^{\mathrm N}_{\mathbf k})^2\rangle
-\langle\hat h^{\mathrm N}_{\mathbf k}\rangle^2\right],
\end{aligned}
}
 
\hypertarget{sm-eq-thermo-f-energy-mu-derivative}{}%
\FittedEquation{F58}{%
\begin{aligned}
\hphantom{\left.\frac{\partial\langle\hat h^{\mathrm N}_{\mathbf k}\rangle}{\partial\mu}\right|_T}\mathllap{\left.\frac{\partial\langle\hat h^{\mathrm N}_{\mathbf k}\rangle}{\partial\mu}\right|_T}
&=-\langle\hat N_{\mathbf k}\rangle
+\frac1T\left[\langle\hat h^{\mathrm N}_{\mathbf k}\hat N_{\mathbf k}\rangle
-\langle\hat h^{\mathrm N}_{\mathbf k}\rangle\langle\hat N_{\mathbf k}\rangle\right].
\end{aligned}
}
 
When the internal energy in \hyperlink{sm-eq-thermo-f-internal-energy}{Eq. (F49)} is differentiated under the fixed-density constraint, the contributions from the \(-\langle\hat N_{\mathbf k}\rangle\) term in \hyperlink{sm-eq-thermo-f-energy-mu-derivative}{Eq. (F58)} and the \(\mu n_e\) term in \hyperlink{sm-eq-thermo-f-internal-energy}{Eq. (F49)} cancel, giving

\hypertarget{sm-eq-thermo-f-heat-substitution}{}%
\FittedEquation{F59}{%
\begin{aligned}
c_{\mathrm N}(T)={}&\frac1{2T^2}\int_{\mathrm{BZ}}\frac{d^2\mathbf k}{(2\pi)^2}
\left[\langle(\hat h^{\mathrm N}_{\mathbf k})^2\rangle
-\langle\hat h^{\mathrm N}_{\mathbf k}\rangle^2\right]\\
&+\frac1{2T}\frac{d\mu(T)}{dT}\int_{\mathrm{BZ}}\frac{d^2\mathbf k}{(2\pi)^2}
\left[\langle\hat h^{\mathrm N}_{\mathbf k}\hat N_{\mathbf k}\rangle
-\langle\hat h^{\mathrm N}_{\mathbf k}\rangle\langle\hat N_{\mathbf k}\rangle\right].
\end{aligned}
}
 
Substituting \hyperlink{sm-eq-thermo-f-fixed-density-mu-derivative}{Eq. (F56)} into \hyperlink{sm-eq-thermo-f-heat-substitution}{Eq. (F59)} gives the exact specific heat in \hyperlink{sm-eq-thermo-f-heat-exact}{Eq. (F50)}, where the full momentum integrals enter the ratio because the fixed-density constraint applies to the total electron number, while the electron number within each momentum block can still fluctuate.

An equivalent exact calculation uses the entropy per site,

\hypertarget{sm-eq-thermo-f-entropy}{}%
\FittedEquation{F60}{%
s(T,\mu)=\frac12\int_{\mathrm{BZ}}\frac{d^2\mathbf k}{(2\pi)^2}\left[\ln Z^{\mathrm N}_{\mathbf k}
+\frac{\langle\hat h^{\mathrm N}_{\mathbf k}\rangle}{T}\right],
}
 
with \(c_{\mathrm N}(T)=T\left.d s[T,\mu(T)]/dT\right|_{n_e}\).

To obtain the low-temperature behavior of the specific heat under the fixed-density constraint, introduce the Helmholtz free energy per site \(f_n(T)=\mu(T)n_e+\omega_{\mathrm N,\mathrm G}[T,\mu(T)]\). Substituting \hyperlink{sm-eq-thermo-f-potential-expansion}{Eq. (F25)} gives

\hypertarget{sm-eq-thermo-f-thermal-coefficient-taylor}{}%
\FittedEquation{F61}{%
f_n(T)=\mu(T)n_e+\omega_{\mathrm N,\mathrm G}[0,\mu(T)]-\frac{\pi^2}{24}D[\mu(T)]T^2-Q[\mu(T)]T^{5/2}+o(T^{5/2}).
}
 
Expanding around \(\mu(0)\) gives

\hypertarget{sm-eq-thermo-f-zero-temperature-potential-taylor}{}%
\FittedEquation{F62}{%
\begin{aligned}
f_n(T)&=\mu(0)n_e+n_e\delta\mu\\
&\quad+\omega_{\mathrm N,\mathrm G}[0,\mu(0)]-n_e\delta\mu-\frac12D[\mu(0)](\delta\mu)^2+O((\delta\mu)^3)\\
&\quad-\frac{\pi^2}{24}\left\{D[\mu(0)]+D'[\mu(0)]\,\delta\mu+O((\delta\mu)^2)\right\}T^2\\
&\quad-\left\{Q[\mu(0)]+Q'[\mu(0)]\,\delta\mu+O((\delta\mu)^2)\right\}T^{5/2}+o(T^{5/2}),
\end{aligned}
}
 
where \hypertarget{sm-eq-thermo-f-zero-temperature-potential-derivatives}{}\(\left.\partial_\mu\omega_{\mathrm N,\mathrm G}(0,\mu)\right|_{\mu=\mu(0)}=-n_e\) and \(\left.\partial_\mu^2\omega_{\mathrm N,\mathrm G}(0,\mu)\right|_{\mu=\mu(0)}=-D[\mu(0)]\). The terms \(+n_e\delta\mu\) and \(-n_e\delta\mu\) cancel. Since \(\delta\mu(T)=O(T^2)\), as in \hyperlink{sm-eq-thermo-f-chemical-shift}{Eq. (F34)}, the remaining chemical-potential shift corrections are of order \(T^4\) or higher, so retaining terms through order \(T^{5/2}\) gives

\hypertarget{sm-eq-thermo-f-fixed-density-free-energy}{}%
\FittedEquation{F63}{%
f_n(T)=f_n(0)-\frac{\pi^2}{24}D[\mu(0)]T^2-Q[\mu(0)]T^{5/2}+o(T^{5/2}).
}
 
Differentiating \(f_n(T)\) gives the low-temperature specific heat

\hypertarget{sm-eq-thermo-f-heat-low-temperature}{}%
\FittedEquation{F64}{%
\begin{aligned}
c_{\mathrm N}(T)&=-T\frac{d^2f_n(T)}{dT^2}
=\frac{\pi^2}{12}D[\mu(0)]T+\frac{15}{4}Q[\mu(0)]T^{3/2}+o(T^{3/2})\\
&=\gamma_0T+\gamma_1T^{3/2}+o(T^{3/2}),
\end{aligned}
}
 
For \hyperlink{sm-fig-model-normal-observables}{Fig. 5(f)}, the coefficients are \(D[\mu(0)]=1.413296~\mathrm{eV}^{-1}\) and \(Q[\mu(0)]=4.364518~\mathrm{eV}^{-3/2}\), giving \(\gamma_0=1.162390~\mathrm{eV}^{-1}\) and \(\gamma_1=16.366943~\mathrm{eV}^{-3/2}\). The linear term comes from contributions near the two Cooper surfaces, and the \(T^{3/2}\) term comes from contributions near the nodes.

\hypertarget{sm-thermo-f-heat-coefficient-jump}{}
\subsection*{F.4. Specific heat coefficient jump}

Consider the dependence of the normal-state specific heat per site \(c_{\mathrm N}(T,\mathcal J_d)\) on the RML interaction strength \(\mathcal J_d\) at fixed electron density \(n_e\). At any fixed \(T>0\), the block partition function and its thermal derivatives vary smoothly with \(\mathcal J_d\). Differentiating the density constraint \(n(T,\mu,\mathcal J_d)=n_e\) with respect to \(\mathcal J_d\) at fixed \(T\) and \(n_e\) gives

\hypertarget{sm-eq-thermo-f-mu-rml-derivative}{}%
\FittedEquation{F65}{%
\left.\frac{\partial\mu}{\partial\mathcal J_d}\right|_{T,n_e}
=-\frac{\left.(\partial n/\partial\mathcal J_d)\right|_{T,\mu}}
       {\left.(\partial n/\partial\mu)\right|_{T,\mathcal J_d}}
=-\frac{\left.(\partial n/\partial\mathcal J_d)\right|_{T,\mu}}{\kappa_{\mathrm N}}.
}
 
The charge compressibility \(\kappa_{\mathrm N}\) defined in \hyperlink{sm-eq-thermo-f-charge-exact}{Eq. (F1)} is positive, so the chemical potential varies smoothly with \(\mathcal J_d\). The exact specific heat in \hyperlink{sm-eq-thermo-f-heat-exact}{Eq. (F50)} therefore approaches the free-electron value continuously,

\hypertarget{sm-eq-thermo-f-heat-finite-t-continuity}{}%
\FittedEquation{F66}{%
\lim_{\mathcal J_d\to0^+}c_{\mathrm N}(T,\mathcal J_d)=c_{\mathrm N}(T,0).
}
 
However, the order of limits matters for the linear coefficient \(\gamma_0(\mathcal J_d)\equiv\lim_{T\to0}c_{\mathrm N}(T,\mathcal J_d)/T\). For each fixed \(\mathcal J_d>0\), \hyperlink{sm-eq-thermo-f-heat-low-temperature}{Eq. (F64)} gives \(\gamma_0(\mathcal J_d)=\pi^2 D[\mu(0)]/12\). As \(\mathcal J_d\to0^+\), the two contours merge into the free-electron Fermi surface and \(\mu(0)\to\mu_0(0)\). For a regular Fermi surface with a smooth, finite density of states, \hyperlink{sm-eq-thermo-f-full-contour-density}{Eq. (F14)} gives

\hypertarget{sm-eq-thermo-f-contour-density-free-limit}{}%
\FittedEquation{F67}{%
\lim_{\mathcal J_d\to0^+}D[\mu(0)]=2\rho_0[\mu_0(0)],
}
 
where \(\rho_0\) is the density of states per spin defined in \hyperlink{sm-eq-wc-weighted-dos}{Eq. (B9)}. The free-electron linear coefficient is \(\gamma_0(0)=2\pi^2\rho_0[\mu_0(0)]/3\). The two orders of limits give

\hypertarget{sm-eq-thermo-f-heat-coefficient-limits}{}%
\FittedEquation{F68}{%
\lim_{\mathcal J_d\to0^+}\lim_{T\to0}\frac{c_{\mathrm N}(T,\mathcal J_d)}{T}=\frac{\pi^2}{6}\rho_0[\mu_0(0)]=\gamma_0(0^+),
}
 
\hypertarget{sm-eq-thermo-f-heat-coefficient-free-first}{}%
\FittedEquation{F69}{%
\lim_{T\to0}\lim_{\mathcal J_d\to0^+}\frac{c_{\mathrm N}(T,\mathcal J_d)}{T}=\frac{2\pi^2}{3}\rho_0[\mu_0(0)]=\gamma_0(0).
}
 
At fixed \(\mathcal J_d>0\), cooling resolves the dominant pair excitations at the two Cooper surfaces. These excitations determine the leading linear term in the specific heat, while the isolated nodes contribute only a subleading \(T^{3/2}\) term. At fixed \(T>0\), reducing \(\mathcal J_d\) makes the RML interaction-induced energy splittings smaller than \(T\), and the specific heat approaches the free-electron value. In both cases, the specific heat vanishes at zero temperature, but the linear coefficient changes by a factor of four, \(\gamma_0(0^+)=\gamma_0(0)/4\).
 
\hypertarget{sm-appendix-cooper-problem}{}\hypertarget{sm-appendix-cooper-problem-title}{}
\section*{Appendix G. Cooper problem and Cooper instability}

Consider the pairing interaction for zero total momentum in the \(\alpha\) channel:

\hypertarget{sm-eq-cooper-g-hamiltonian}{}%
\FittedEquation{G1}{%
\hat H_{J,\alpha}=-\frac{2J}{N_s}\sum_{\mathbf k,\mathbf k'}\phi_{\mathbf k,\alpha}\phi_{\mathbf k',\alpha}\hat\Delta_{\mathbf k}^\dagger\hat\Delta_{\mathbf k'},
}
 
which is the \(\mathbf Q=0\), \(\alpha\)-component of the AFM superexchange interaction in \hyperlink{sm-eq-model-superexchange}{Eq. (1)}. The Hamiltonian considered in the Cooper problem is

\hypertarget{sm-eq-cooper-g-total-hamiltonian}{}%
\FittedEquation{G2}{%
\hat H_\alpha=\hat H_{\mathrm N}+\hat H_{J,\alpha}.
}
 
The ground state \(|\mathrm{GS}_{\mathrm N}\rangle\) of \(\hat H_{\mathrm N}\) is given in \hyperlink{sm-eq-cooper-g-ground-state}{Eq. (13)}.

\hypertarget{sm-eq-cooper-g-excitation-energies}{}Creating a spin-singlet Cooper pair on top of \(|\mathrm{GS}_{\mathrm N}\rangle\) gives the pair-addition state \(|\mathbf k,-\mathbf k\rangle=\hat\Delta_{\mathbf k}^\dagger|\mathrm{GS}_{\mathrm N}\rangle\) with implicit \(\xi_{\mathbf k}>0\), and the excitation energy is \(\epsilon_{\mathbf k}=2\xi_{\mathbf k}-2\mathcal J_{\mathbf k}\Theta(\xi_{\mathbf k}-\mathcal J_{\mathbf k})\). The set \(\{|\mathbf k,-\mathbf k\rangle\}\) forms an orthonormal basis for the single-pair addition subspace \(\mathcal H_{N_e+2}\) with fixed electron number \(N_e+2\), where \(N_e=n_eN_s\), and the background \(|\mathrm{GS}_{\mathrm N}\rangle\) is held fixed in the Cooper approximation \hyperlink{sm-ref-cooper-1956}{[34]}. The single-pair removal problem is related to the addition problem by a particle-hole transformation.

Consider a normalized single-pair trial state:

\hypertarget{sm-eq-cooper-g-ansatz}{}%
\FittedEquation{G3}{%
|\Psi\rangle=\sum_{\{\mathbf k,-\mathbf k\}}^{\xi_{\mathbf k}>0}a_{\mathbf k}|\mathbf k,-\mathbf k\rangle,
}
 
where the normalization condition is \(\sum_{\{\mathbf k,-\mathbf k\}}^{\xi_{\mathbf k}>0}|a_{\mathbf k}|^2=1\). After factoring out the phase \(e^{-iE_{\mathrm{GS},\alpha}t}\), the pair equation of motion in the Cooper approximation is

\hypertarget{sm-eq-cooper-g-general-equation-of-motion}{}%
\FittedEquation{G4}{%
i\frac{\partial a_{\mathbf k}(t)}{\partial t}=\sum_{\{\mathbf k',-\mathbf k'\}}^{\xi_{\mathbf k'}>0}M_{\mathbf k\mathbf k'}a_{\mathbf k'}(t),
}
 
where \(M_{\mathbf k\mathbf k'}=\langle\mathbf k,-\mathbf k|\left(\hat H_\alpha-E_{\mathrm{GS},\alpha}\right)|\mathbf k',-\mathbf k'\rangle\) is the matrix element in the subspace \(\mathcal H_{N_e+2}\). In the fixed background, energies are measured relative to the expectation of \(\hat H_\alpha\):

\hypertarget{sm-eq-cooper-g-reference-energy}{}%
\FittedEquation{G5}{%
E_{\mathrm{GS},\alpha}=\langle\mathrm{GS}_{\mathrm N}|\hat H_\alpha|\mathrm{GS}_{\mathrm N}\rangle
=E_{G,\mathrm N}-\frac{8J}{N_s}\sum_{\{\mathbf k,-\mathbf k\}}^{\xi_{\mathbf k}<\mathcal J_{\mathbf k}}\phi_{\mathbf k,\alpha}^{2},
}
 
where \(E_{G,\mathrm N}\) is the ground-state energy of \(\hat H_{\mathrm N}\). Using

\hypertarget{sm-eq-cooper-g-addition-action}{}%
\FittedEquation{G6}{%
\begin{aligned}
\left(\hat H_\alpha-E_{\mathrm{GS},\alpha}\right)|\mathbf k',-\mathbf k'\rangle
&=\left(\hat H_\alpha-E_{\mathrm{GS},\alpha}\right)\hat\Delta_{\mathbf k'}^\dagger|\mathrm{GS}_{\mathrm N}\rangle\\
&=\left[\hat H_\alpha,\hat\Delta_{\mathbf k'}^\dagger\right]|\mathrm{GS}_{\mathrm N}\rangle
+\hat\Delta_{\mathbf k'}^\dagger\left(\hat H_\alpha-E_{\mathrm{GS},\alpha}\right)|\mathrm{GS}_{\mathrm N}\rangle
\end{aligned}
}
 
and the commutators

\hypertarget{sm-eq-cooper-g-commutators}{}\hypertarget{sm-eq-cooper-g-pairing-commutator}{}%
\FittedEquation{G7}{%
\left[\hat H_{\mathrm N},\hat\Delta_{\mathbf k'}^\dagger\right]=\hat\Delta_{\mathbf k'}^\dagger
\left(2\xi_{\mathbf k'}-2\mathcal J_{\mathbf k'}+\mathcal J_{\mathbf k'}\hat N_{\mathbf k'}\right),
}
\FittedEquation{G8}{%
\left[\hat H_{J,\alpha},\hat\Delta_{\mathbf k'}^\dagger\right]=-\frac{8J}{N_s}\phi_{\mathbf k',\alpha}
\sum_{\{\mathbf k'',-\mathbf k''\}}\phi_{\mathbf k'',\alpha}\hat\Delta_{\mathbf k''}^\dagger
\left(1-\frac{\hat N_{\mathbf k'}}2\right)
}
 
gives

\hypertarget{sm-eq-cooper-g-normal-commutator-matrix}{}\hypertarget{sm-eq-cooper-g-addition-commutator-matrix}{}\hypertarget{sm-eq-cooper-g-addition-background-matrix}{}%
\FittedEquation{G9}{%
\langle\mathbf k,-\mathbf k|\left[\hat H_{\mathrm N},\hat\Delta_{\mathbf k'}^\dagger\right]|\mathrm{GS}_{\mathrm N}\rangle=\epsilon_{\mathbf k'}\langle\mathbf k,-\mathbf k|\mathbf k',-\mathbf k'\rangle=\epsilon_{\mathbf k'}\delta_{\mathbf k\mathbf k'},
}
\FittedEquation{G10}{%
\langle\mathbf k,-\mathbf k|\left[\hat H_{J,\alpha},\hat\Delta_{\mathbf k'}^\dagger\right]|\mathrm{GS}_{\mathrm N}\rangle=-\frac{8J}{N_s}\phi_{\mathbf k,\alpha}\phi_{\mathbf k',\alpha}\Theta(\xi_{\mathbf k'}-\mathcal J_{\mathbf k'}),
}
\FittedEquation{G11}{%
\langle\mathbf k,-\mathbf k|\hat\Delta_{\mathbf k'}^\dagger\left(\hat H_\alpha-E_{\mathrm{GS},\alpha}\right)|\mathrm{GS}_{\mathrm N}\rangle=-\left(1-\delta_{\mathbf k\mathbf k'}\right)\frac{8J}{N_s}\phi_{\mathbf k,\alpha}\phi_{\mathbf k',\alpha}\Theta(\mathcal J_{\mathbf k'}-\xi_{\mathbf k'}),
}
 
where the conditions \(\xi_{\mathbf k},\xi_{\mathbf k'}>0\) are implicit. Adding the contributions in \hyperlink{sm-eq-cooper-g-normal-commutator-matrix}{Eqs. (G9)}--\hyperlink{sm-eq-cooper-g-addition-background-matrix}{(G11)} gives the exact matrix element in the subspace \(\mathcal H_{N_e+2}\):

\hypertarget{sm-eq-cooper-g-projected-matrix}{}%
\FittedEquation{G12}{%
\begin{aligned}
M_{\mathbf k\mathbf k'}
&=\langle\mathbf k,-\mathbf k|\left(\hat H_\alpha-E_{\mathrm{GS},\alpha}\right)|\mathbf k',-\mathbf k'\rangle\\
&=\begin{cases}
-\dfrac{8J}{N_s}\phi_{\mathbf k,\alpha}\phi_{\mathbf k',\alpha},&\mathbf k\ne\mathbf k',\\[4pt]
\epsilon_{\mathbf k}-\dfrac{8J}{N_s}\phi_{\mathbf k,\alpha}^{2}\Theta(\xi_{\mathbf k}-\mathcal J_{\mathbf k}),&\mathbf k=\mathbf k'
\end{cases}\\
&=\widetilde\epsilon_{\mathbf k}\delta_{\mathbf k\mathbf k'}-\frac{8J}{N_s}\phi_{\mathbf k,\alpha}\phi_{\mathbf k',\alpha},
\end{aligned}
}
 
\hypertarget{sm-eq-cooper-g-diagonal-shifts}{}where \(\widetilde\epsilon_{\mathbf k}=\epsilon_{\mathbf k}+\frac{8J}{N_s}\phi_{\mathbf k,\alpha}^{2}\Theta(\mathcal J_{\mathbf k}-\xi_{\mathbf k})\). The off-diagonal matrix elements transfer the added pair between allowed momentum blocks.

Substituting the matrix elements in \hyperlink{sm-eq-cooper-g-projected-matrix}{Eq. (G12)} into \hyperlink{sm-eq-cooper-g-general-equation-of-motion}{Eq. (G4)} gives

\hypertarget{sm-eq-cooper-g-addition-time-evolution}{}%
\FittedEquation{G13}{%
i\frac{\partial a_{\mathbf k}(t)}{\partial t}
=\widetilde\epsilon_{\mathbf k}a_{\mathbf k}(t)
-\frac{8J}{N_s}\phi_{\mathbf k,\alpha}
\sum_{\{\mathbf k',-\mathbf k'\}}^{\xi_{\mathbf k'}>0}\phi_{\mathbf k',\alpha}a_{\mathbf k'}(t).
}
 
For a stationary state, let \(a_{\mathbf k}(t)=a_{\mathbf k}e^{-iEt}\), where \(E\) is measured relative to \(E_{\mathrm{GS},\alpha}\). Then \hyperlink{sm-eq-cooper-g-addition-time-evolution}{Eq. (G13)} becomes

\hypertarget{sm-eq-cooper-g-amplitude-equation}{}%
\FittedEquation{G14}{%
\left(\widetilde\epsilon_{\mathbf k}-E\right)a_{\mathbf k}
=\frac{8J}{N_s}\phi_{\mathbf k,\alpha}
\sum_{\{\mathbf k',-\mathbf k'\}}^{\xi_{\mathbf k'}>0}\phi_{\mathbf k',\alpha}a_{\mathbf k'}.
}
 
Multiplying both sides of \hyperlink{sm-eq-cooper-g-amplitude-equation}{Eq. (G14)} by \(\phi_{\mathbf k,\alpha}/(\widetilde\epsilon_{\mathbf k}-E)\), summing over the allowed momentum blocks, and canceling the common nonzero factor \(\sum_{\{\mathbf k,-\mathbf k\}}^{\xi_{\mathbf k}>0}\phi_{\mathbf k,\alpha}a_{\mathbf k}\) gives the condition for a nontrivial solution:

\hypertarget{sm-eq-cooper-g-finite-secular}{}%
\FittedEquation{G15}{%
1=\frac{8J}{N_s}\sum_{\{\mathbf k,-\mathbf k\}}^{\xi_{\mathbf k}>0}
\frac{\phi_{\mathbf k,\alpha}^2}{\widetilde\epsilon_{\mathbf k}-E}.
}
 
In the thermodynamic limit, the continuum of unbound pair-addition excitations has a lower threshold at \(E=0\). Consider a bound state with eigenvalue \(E=-E_{b,\alpha}<0\). Since \(N_s^{-1}\sum_{\{\mathbf k,-\mathbf k\}}\to\tfrac12\int_{\mathrm{BZ}}d^2\mathbf k/(2\pi)^2\), the condition for a bound-state solution in \hyperlink{sm-eq-cooper-g-finite-secular}{Eq. (G15)} becomes

\hypertarget{sm-eq-cooper-g-continuum-secular}{}%
\FittedEquation{G16}{%
\begin{aligned}
1&=4J\int_{\xi_{\mathbf k}>0}\frac{d^2\mathbf k}{(2\pi)^2}\frac{\phi_{\mathbf k,\alpha}^2}{\epsilon_{\mathbf k}+E_{b,\alpha}}\\
&=4J\int_{0<\xi_{\mathbf k}<\mathcal J_{\mathbf k}}\frac{d^2\mathbf k}{(2\pi)^2}\frac{\phi_{\mathbf k,\alpha}^2}{2\xi_{\mathbf k}+E_{b,\alpha}}+4J\int_{\xi_{\mathbf k}>\mathcal J_{\mathbf k}}\frac{d^2\mathbf k}{(2\pi)^2}\frac{\phi_{\mathbf k,\alpha}^2}{2(\xi_{\mathbf k}-\mathcal J_{\mathbf k})+E_{b,\alpha}},
\end{aligned}
}
 
where the \(O(J/N_s)\) correction in \(\widetilde\epsilon_{\mathbf k}\) vanishes in the thermodynamic limit.

In the weak-attraction limit \(J\to0^+\), the binding energy satisfies \(E_{b,\alpha}\to0^+\), and the dominant contributions to the integrals in \hyperlink{sm-eq-cooper-g-continuum-secular}{Eq. (G16)} come from momenta near the two Cooper surfaces, which touch at the nodes. Introduce the weighted density of states associated with the two Cooper surfaces (cf. \hyperlink{sm-eq-thermo-f-full-contour-density}{Eq. (F14)}):

\hypertarget{sm-eq-cooper-g-contour-coefficient}{}%
\FittedEquation{G17}{%
D_\alpha(\epsilon)=\int_{\mathrm{BZ}}\frac{d^2\mathbf k}{(2\pi)^2}\phi_{\mathbf k,\alpha}^{2}\left[\delta(\epsilon-\varepsilon_{\mathbf k})+\delta(\epsilon-\varepsilon_{\mathbf k}+\mathcal J_{\mathbf k})\right].
}
 
To leading logarithmic accuracy, the contributions from the regions near the two Cooper surfaces are approximated by

\hypertarget{sm-eq-cooper-g-normal-integral}{}%
\FittedEquation{G18}{%
\begin{aligned}
4J\int_0^{\Lambda_{\mathrm{cut}}}d\epsilon\,
\frac{D_\alpha(\mu+\epsilon)}{2\epsilon+E_{b,\alpha}}
&\simeq4JD_\alpha(\mu)\int_0^{\Lambda_{\mathrm{cut}}}\frac{d\epsilon}{2\epsilon+E_{b,\alpha}}\\
&=2JD_\alpha(\mu)\ln\frac{2\Lambda_{\mathrm{cut}}+E_{b,\alpha}}{E_{b,\alpha}},
\end{aligned}
}
 
where \(\Lambda_{\mathrm{cut}}>0\) is a fixed low-energy cutoff. The area of the regions around the isolated nodes with \(|\xi_{\mathbf k}|,\mathcal J_{\mathbf k}\sim E_{b,\alpha}\) scales as \(E_{b,\alpha}^{3/2}\). Since the denominator in \hyperlink{sm-eq-cooper-g-continuum-secular}{Eq. (G16)} is of order \(E_{b,\alpha}\), the contributions from these shrinking nodal regions are at most \(O(E_{b,\alpha}^{1/2})\) and vanish as \(E_{b,\alpha}\to0^+\). The binding energy of the bound state in the weak-attraction limit and to leading logarithmic accuracy is

\hypertarget{sm-eq-cooper-g-binding-scale}{}%
\FittedEquation{G19}{%
E_{b,\alpha}\simeq2\Lambda_{\mathrm{cut}}\exp\!\left[-\frac{1}{2JD_\alpha(\mu)}\right].
}
 
Therefore, for an \(\alpha\) channel with \(D_\alpha(\mu)>0\), the Cooper problem in the thermodynamic limit admits a bound-state solution for every \(J>0\), even for an infinitesimal attractive interaction. Within the Cooper approximation, the existence of a bound pair for arbitrarily weak attraction signals a Cooper instability of the nodal Bose metal in the \(\alpha\) channel. For the parameters of \hyperlink{sm-fig-model-normal-observables}{Fig. 5}, \(D_s[\mu(0)]=0.043236~\mathrm{eV}^{-1}\) and \(D_d[\mu(0)]=0.699875~\mathrm{eV}^{-1}\). Thus the \(d\)-wave binding energy is exponentially larger than that of the extended-\(s\)-wave channel for equal, sufficiently weak attractive interactions. Comparisons between the numerical solution of \hyperlink{sm-eq-cooper-g-finite-secular}{Eq. (G15)} and the analytical solution in \hyperlink{sm-eq-cooper-g-binding-scale}{Eq. (G19)} in the \(d\)-wave channel are shown in \hyperlink{sm-fig-cooper-g-binding}{Fig. G1}, whereas the extended-\(s\)-wave binding energy is exponentially smaller.

\par\medskip
\hypertarget{sm-fig-cooper-g-binding}{}
\PrintAsset{cooper-binding-np}
{\small\noindent \textbf{Figure G1 \textbar{} Cooper-pair binding energy in the \(d\)-wave channel.} The light-orange solid curve shows numerical solutions of \hyperlink{sm-eq-cooper-g-finite-secular}{Eq. (G15)} in the thermodynamic limit under the fixed-density constraint. Open circles show finite-size solutions for \(N_s=8192^2\), including the \(O(J/N_s)\) correction in \(\widetilde\epsilon_{\mathbf k}\). The black dashed curve shows the leading-logarithmic analytical solution in \hyperlink{sm-eq-cooper-g-binding-scale}{Eq. (G19)}, with the fixed cutoff \(\Lambda_{\mathrm{cut}}=0.06~\mathrm{eV}\). Parameters match \hyperlink{sm-fig-model-normal-observables}{Fig. 5}: \(\mathcal J_d=0.06~\mathrm{eV}\), \(p=0.15\), and \(\mu(0)=-0.268236~\mathrm{eV}\).\par}
\medskip
 
\hypertarget{sm-appendix-pair-susceptibility}{}\hypertarget{sm-appendix-pair-susceptibility-title}{}
\section*{Appendix H. Pair susceptibility}

\hypertarget{sm-appendix-h-dyson-equation}{}
\subsection*{H.1. Dyson equation}

Consider the pairing interaction for zero total momentum

\hypertarget{sm-eq-hs-pairing-hamiltonian}{}%
\FittedEquation{H1}{%
\hat H_{J,\mathbf0}=-\frac{2J}{N_s}\sum_{\alpha=s,d}\sum_{\mathbf k,\mathbf k'}\phi_{\mathbf k,\alpha}\phi_{\mathbf k',\alpha}\hat\Delta_{\mathbf k}^\dagger\hat\Delta_{\mathbf k'}=-2JN_s\sum_{\alpha=s,d}\hat\Lambda_\alpha^\dagger\hat\Lambda_\alpha,
}
 
\hypertarget{sm-eq-hs-pair-operator-definitions}{}which is the \(\mathbf Q=0\) component of the AFM superexchange interaction in \hyperlink{sm-eq-model-superexchange}{Eq. (1)}, and \(\hat\Lambda_\alpha=N_s^{-1}\sum_{\mathbf k}\phi_{\mathbf k,\alpha}\hat\Delta_{\mathbf k}\) is the collective spin-singlet \(\alpha\)-wave pair-annihilation operator.

The full pair susceptibility is evaluated for the Hamiltonian

\hypertarget{sm-eq-hs-full-hamiltonian}{}%
\FittedEquation{H2}{%
\hat H=\hat H_{\mathrm N}+\hat H_{J,\mathbf0},
}
 
and is defined by

\hypertarget{sm-eq-hs-full-susceptibility}{}%
\FittedEquation{H3}{%
\chi_{\alpha\alpha'}(i\Omega_m)\equiv N_s\int_0^\beta d\tau\,e^{i\Omega_m\tau}
        \langle T_\tau\hat\Lambda_\alpha(\tau)
        \hat\Lambda_{\alpha'}^\dagger(0)\rangle_c,
}
 
where the subscript \(c\) denotes connected correlations, and \(\Omega_m=2\pi m/\beta\) is a bosonic Matsubara frequency.

The generating-functional approach to linear response is employed. A complex source \(h_\alpha(\tau)\) coupled to \(\hat\Lambda_\alpha(\tau)\) is introduced through

\hypertarget{sm-eq-hs-source}{}%
\FittedEquation{H4}{%
\hat H_h(\tau)=-N_s\sum_\alpha
        [h_\alpha^*(\tau)\hat\Lambda_\alpha(\tau)
        +\hat\Lambda_\alpha^\dagger(\tau) h_\alpha(\tau)].
}
 
The source-dependent partition function \(Z[h]\) generates the full pair susceptibility via

\hypertarget{sm-eq-hs-source-response}{}%
\FittedEquation{H5}{%
\chi_{\alpha\alpha'}(i\Omega_m)=
        \left.\frac1{N_s}\frac{\partial^2\ln Z[h]}
        {\partial h_{\alpha m}^*\,\partial h_{\alpha' m}}\right|_{h=0},
}
 
where \(h_\alpha(\tau)=\beta^{-1/2}\sum_m h_{\alpha m}e^{-i\Omega_m\tau}\).

In the fermionic coherent-state path integral, the source-dependent partition function is

\hypertarget{sm-eq-hs-fermion-integral}{}%
\FittedEquation{H6}{%
\begin{aligned}
Z[h]&=\int\mathcal D[\bar c,c]\,e^{-S[\bar c,c;h]}
\end{aligned}
}
 
\hypertarget{sm-eq-hs-sourced-action}{}%
\FittedEquation{H7}{%
S[\bar c,c;h]=S_{\mathrm N}[\bar c,c]-N_s\int_0^\beta d\tau\sum_\alpha \left(2J\bar\Lambda_\alpha(\tau)\Lambda_\alpha(\tau)+h_\alpha^*(\tau)\Lambda_\alpha(\tau)+\bar\Lambda_\alpha(\tau) h_\alpha(\tau)\right)
}
 
where \(c(\tau)\) and \(\bar c(\tau)\) denote the Grassmann variables corresponding to the fermionic operators \(\hat c\) and \(\hat c^\dagger\), respectively, and \(\Lambda_\alpha(\tau)\) and \(\bar\Lambda_\alpha(\tau)\) denote the Grassmann symbols corresponding to \(\hat\Lambda_\alpha\) and \(\hat\Lambda_\alpha^\dagger\), respectively. The normal-state action retaining the RML interaction is

\hypertarget{sm-eq-hs-reference-action}{}%
\FittedEquation{H8}{%
S_{\mathrm N}[\bar c,c]=\int_0^\beta d\tau\left[
        \sum_{\mathbf k,\sigma}\bar c_{\mathbf k\sigma}(\tau)\partial_\tau c_{\mathbf k\sigma}(\tau)
        +H_{\mathrm N}[\bar c(\tau),c(\tau)]\right].
}
 
The attractive pairing interaction \(\hat H_{J,\mathbf0}\) in \hyperlink{sm-eq-hs-pairing-hamiltonian}{Eq. (H1)} is decoupled by the Hubbard--Stratonovich transformation

\hypertarget{sm-eq-hs-identity}{}%
\FittedEquation{H9}{%
\begin{aligned}
e^{2JN_s\int_0^\beta d\tau\sum_\alpha\bar\Lambda_\alpha(\tau)\Lambda_\alpha(\tau)}
        &=\int\mathcal D[\Xi^*,\Xi]e^{-S_{\mathrm{HS}}}
\end{aligned}
}
 
\hypertarget{sm-eq-hs-auxiliary-action}{}%
\FittedEquation{H10}{%
\begin{aligned}
S_{\mathrm{HS}}&=N_s\int_0^\beta d\tau\sum_\alpha
        \left[\frac{|\Xi_\alpha(\tau)|^2}{2J}
        -\Xi_\alpha^*(\tau)\Lambda_\alpha(\tau)-\bar\Lambda_\alpha(\tau)\Xi_\alpha(\tau)\right]
\end{aligned}
}
 
where \(\Xi_\alpha(\tau)\) is a complex auxiliary pairing field in channel \(\alpha\). Integrating out the electrons while treating the RML interaction exactly gives

\hypertarget{sm-eq-hs-effective-action}{}%
\FittedEquation{H11}{%
\begin{aligned}
Z[h]&=Z_{\mathrm N}\int\mathcal D[\Xi^*,\Xi]
        e^{-S_{\mathrm{eff}}[\Xi,h]}
\end{aligned}
}
 
\hypertarget{sm-eq-hs-effective-action-definition}{}%
\FittedEquation{H12}{%
\begin{aligned}
S_{\mathrm{eff}}[\Xi,h]&=\frac{N_s}{2J}\int_0^\beta d\tau\sum_\alpha
        |\Xi_\alpha(\tau)|^2-\mathcal W_{\mathrm N}[\Xi+h]
\end{aligned}
}
 
where \(Z_{\mathrm N}=\operatorname{Tr}e^{-\beta\hat H_{\mathrm N}}\) is the normal-state partition function, and the exact connected generating functional is

\hypertarget{sm-eq-hs-generator}{}%
\FittedEquation{H13}{%
\mathcal W_{\mathrm N}[h]=\ln\left\langle T_\tau
        \exp\!\left[N_s\int_0^\beta d\tau\sum_\alpha
        (h_\alpha^*(\tau)\hat\Lambda_\alpha(\tau)+\hat\Lambda_\alpha^\dagger(\tau) h_\alpha(\tau))\right]
        \right\rangle_{\mathrm N}.
}
 
The subscript \(\mathrm N\) denotes thermal averaging with respect to the Hamiltonian \(\hat H_{\mathrm N}\).

In the normal state, \(\langle\hat\Lambda_\alpha(\tau)\rangle_{\mathrm N}=\langle\hat\Lambda_\alpha^2(\tau)\rangle_{\mathrm N}=0\). The lowest-order nonvanishing connected cumulant is

\hypertarget{sm-eq-hs-cumulant}{}%
\FittedEquation{H14}{%
\begin{aligned}
        \mathcal W_{\mathrm N}^{(2)}[h]
        &=N_s^2\sum_{\alpha\alpha'}\int_0^\beta d\tau\,d\tau'\,
        h_\alpha^*(\tau)\langle T_\tau\hat\Lambda_\alpha(\tau)
        \hat\Lambda_{\alpha'}^\dagger(\tau')\rangle_{\mathrm N,c}h_{\alpha'}(\tau')\\
        &=N_s\sum_{m,\alpha\alpha'}
        h_{\alpha m}^*\chi_{\mathrm N,\alpha\alpha'}(i\Omega_m)h_{\alpha' m}.
        \end{aligned}
}
 
where the pair susceptibility for the Hamiltonian \(\hat H_{\mathrm N}\) is

\hypertarget{sm-eq-hs-reference-susceptibility}{}%
\FittedEquation{H15}{%
\chi_{\mathrm N,\alpha\alpha'}(i\Omega_m)=N_s\int_0^\beta d\tau\,e^{i\Omega_m\tau}\langle T_\tau\hat\Lambda_\alpha(\tau)\hat\Lambda_{\alpha'}^\dagger(0)\rangle_{\mathrm N,c}.
}
 
The cumulant expansion retains the RML interaction without invoking Wick's theorem. The quadratic expansion about the normal saddle \(\Xi_\alpha(\tau)=0\) gives

\hypertarget{sm-eq-hs-sourced-quadratic}{}%
\FittedEquation{H16}{%
\begin{aligned}
        S_{\mathrm{eff}}^{(2)}[\Xi,h]
        &=N_s\sum_{m,\alpha\alpha'}\Big[
        \Xi_{\alpha m}^*K_{\alpha\alpha'}(i\Omega_m)\Xi_{\alpha' m}
        -\Xi_{\alpha m}^*\chi_{\mathrm N,\alpha\alpha'}(i\Omega_m)h_{\alpha' m}\\
        &\qquad-h_{\alpha m}^*\chi_{\mathrm N,\alpha\alpha'}(i\Omega_m)\Xi_{\alpha' m}
        -h_{\alpha m}^*\chi_{\mathrm N,\alpha\alpha'}(i\Omega_m)h_{\alpha' m}\Big],
        \end{aligned}
}
 
where \(K_{\alpha\alpha'}(i\Omega_m)=\delta_{\alpha\alpha'}/(2J)-\chi_{\mathrm N,\alpha\alpha'}(i\Omega_m)\). Introduce the matrices \(\boldsymbol\chi(i\Omega_m)\), \(\boldsymbol\chi_{\mathrm N}(i\Omega_m)\), and \(\mathbf K(i\Omega_m)\), with entries \(\chi_{\alpha\alpha'}(i\Omega_m)\), \(\chi_{\mathrm N,\alpha\alpha'}(i\Omega_m)\), and \(K_{\alpha\alpha'}(i\Omega_m)\), respectively. On the stable normal branch, performing the Gaussian integral over the fields \(\Xi_\alpha(\tau)\) gives

\hypertarget{sm-eq-hs-gaussian-source}{}%
\FittedEquation{H17}{%
\ln\frac{Z[h]}{Z[0]}
        \simeq N_s\sum_{m,\alpha\alpha'}h_{\alpha m}^*
        \Big[\boldsymbol\chi_{\mathrm N}(i\Omega_m)
        +\boldsymbol\chi_{\mathrm N}(i\Omega_m)\mathbf K^{-1}(i\Omega_m)
        \boldsymbol\chi_{\mathrm N}(i\Omega_m)\Big]_{\alpha\alpha'}h_{\alpha' m}.
}
 
Taking the source derivatives in \hyperlink{sm-eq-hs-source-response}{Eq. (H5)} yields

\hypertarget{sm-eq18}{}%
\FittedEquation{H18}{%
\boldsymbol\chi(i\Omega_m)=\boldsymbol\chi_{\mathrm N}(i\Omega_m)+\boldsymbol\chi_{\mathrm N}(i\Omega_m)\mathbf K^{-1}(i\Omega_m)\boldsymbol\chi_{\mathrm N}(i\Omega_m).
}
 
Substituting \(\mathbf K^{-1}(i\Omega_m)=2J[\mathbf I-2J\boldsymbol\chi_{\mathrm N}(i\Omega_m)]^{-1}\) gives the Dyson equation

\hypertarget{sm-eq-hs-dyson}{}%
\FittedEquation{H19}{%
\boldsymbol\chi(i\Omega_m)=\boldsymbol\chi_{\mathrm N}(i\Omega_m)+2J\boldsymbol\chi_{\mathrm N}(i\Omega_m)\boldsymbol\chi(i\Omega_m).
}
 
On the stable normal branch in the thermodynamic limit, the Dyson equation in \hyperlink{sm-eq-hs-dyson}{Eq. (H19)}, derived in the Gaussian approximation, becomes exact. Note that the independent momentum blocks make the exact connected generating functional additive:

\hypertarget{sm-eq-hs-block-additivity}{}%
\FittedEquation{H20}{%
\mathcal W_{\mathrm N}[\Xi]=\sum_{\{\mathbf k,-\mathbf k\}}\mathcal W_{\mathrm N,\mathbf k}[\Xi],
}
 
where \(\mathcal W_{\mathrm N,\mathbf k}\) is the connected block generating functional. Performing the expansion at fixed finite temperature gives

\hypertarget{sm-eq-hs-connected-cumulant-expansion}{}%
\FittedEquation{H21}{%
\mathcal W_{\mathrm N}[\Xi]=\sum_{l=1}^{\infty}\mathcal W_{\mathrm N}^{(2l)}[\Xi]
=\sum_{l=1}^{\infty}\sum_{\{\mathbf k,-\mathbf k\}}\mathcal W_{\mathrm N,\mathbf k}^{(2l)}[\Xi],
}
 
where normal-state \(U(1)\) symmetry allows only terms of even orders. At fixed auxiliary pairing field amplitude, each \(\mathcal W_{\mathrm N,\mathbf k}^{(2l)}[\Xi]\) is finite and \(O(1)\), so every \(\mathcal W_{\mathrm N}^{(2l)}[\Xi]\) is extensive and \(O(N_s)\). Rescaling \(\Xi_\alpha(\tau)=\widetilde\Xi_\alpha(\tau)/\sqrt{N_s}\) gives

\hypertarget{sm-eq-hs-scaling}{}%
\FittedEquation{H22}{%
S_{\mathrm{eff}}[\widetilde\Xi/\sqrt{N_s},0]
        =\sum_{m,\alpha\alpha'}\widetilde\Xi_{\alpha m}^*K_{\alpha\alpha'}(i\Omega_m)\widetilde\Xi_{\alpha' m}
        +\sum_{l=2}^{\infty}\frac{S_{2l}[\widetilde\Xi]}{N_s^{l-1}}.
}
 
Here \(S_{2l}\) collects terms of order \(2l\) in \(\widetilde\Xi\) and \(\widetilde\Xi^*\), with coefficients finite as \(N_s\to\infty\). When \(K_{\alpha\alpha'}(i\Omega_m)\) remains nonsingular, higher-order connected cumulants give only \(O(N_s^{-1})\) corrections to the pair susceptibility.

\hypertarget{sm-appendix-h-nodal-pair-susceptibility}{}
\subsection*{H.2. SC instability temperature}

The pair susceptibility of the nodal Bose metal can be evaluated exactly without invoking Wick's theorem:

\hypertarget{sm-eq-h2-block-reduction}{}%
\FittedEquation{H23}{%
\chi_{\mathrm N,\alpha\alpha'}(i\Omega_m)
=\frac2{N_s}\sum_{\mathbf k}
\phi_{\mathbf k,\alpha}\phi_{\mathbf k,\alpha'}\Pi_{\mathrm N,\mathbf k}(i\Omega_m),
}
 
where the spin-singlet Cooper-pair Matsubara Green's function is

\hypertarget{sm-eq-h2-block-green-definition}{}%
\FittedEquation{H24}{%
\Pi_{\mathrm N,\mathbf k}(i\Omega_m)=\int_0^\beta d\tau\,e^{i\Omega_m\tau}
\langle T_\tau\hat\Delta_{\mathbf k}(\tau)\hat\Delta_{\mathbf k}^{\dagger}(0)\rangle_{\mathrm N},
}
 
with \(\hat\Delta_{\mathbf k}(\tau)=e^{\tau\hat h^{\mathrm N}_{\mathbf k}}\hat\Delta_{\mathbf k}e^{-\tau\hat h^{\mathrm N}_{\mathbf k}}\). The exact Lehmann representation of the pair Matsubara Green's function is

\hypertarget{sm-eq-h2-pair-lehmann}{}%
\FittedEquation{H25}{%
\Pi_{\mathrm N,\mathbf k}(i\Omega_m) =\frac{1}{Z^{\mathrm N}_{\mathbf k}} \sum\limits_{\lambda,\lambda'}^{N_{\mathbf k,\lambda}=N_{\mathbf k,\lambda'}+2} \frac{e^{-\beta E^{\mathrm N}_{\mathbf k,\lambda'}}-e^{-\beta E^{\mathrm N}_{\mathbf k,\lambda}}} {E^{\mathrm N}_{\mathbf k,\lambda}-E^{\mathrm N}_{\mathbf k,\lambda'}-i\Omega_m} |\langle N_{\mathbf k,\lambda'}|\hat\Delta_{\mathbf k}|N_{\mathbf k,\lambda}\rangle|^2,
}
 
where the ratio is defined continuously as \(\beta e^{-\beta E^{\mathrm N}_{\mathbf k,\lambda}}\) for \(\Omega_m=0\) and zero for \(\Omega_m\ne0\) at a degenerate transition with \(E^{\mathrm N}_{\mathbf k,\lambda}=E^{\mathrm N}_{\mathbf k,\lambda'}\). The three groups of allowed transitions are listed in \hyperlink{sm-table-h-pair-transitions}{Table H1} and yield

\hypertarget{sm-eq-h2-exact-pair-kernel}{}%
\FittedEquation{H26}{%
\Pi_{\mathrm N,\mathbf k}(i\Omega_m) =\frac{p^{\mathrm N}_{\mathbf k,s}-p^{\mathrm N}_{\mathbf k,4}}{2\xi_{\mathbf k}-i\Omega_m} +4\times\frac{1}{2}\frac{p^{\mathrm N}_{\mathbf k,1}-p^{\mathrm N}_{\mathbf k,3}}{2\xi_{\mathbf k}-\mathcal J_{\mathbf k}-i\Omega_m} +\frac{p^{\mathrm N}_{\mathbf k,0}-p^{\mathrm N}_{\mathbf k,s}}{2\xi_{\mathbf k}-2\mathcal J_{\mathbf k}-i\Omega_m},
}
 
which is the exact expression in \hyperlink{sm-eq15}{Eq. (55)}.

\hypertarget{sm-table-h-pair-transitions}{}\par\medskip

\noindent{\small Table H1 \textbar{} Spin-singlet pair-removal transitions\par}\nopagebreak[4]
\begingroup\small\setlength{\tabcolsep}{3.5pt}\renewcommand{\arraystretch}{1.30}
\noindent\FittedTable{\begin{tabular}{@{}cccc@{}}\toprule
{\(|N_{\mathbf k,\lambda}\rangle\to|N_{\mathbf k,\lambda'}\rangle\)} & {\(E^{\mathrm N}_{\mathbf k,\lambda}-E^{\mathrm N}_{\mathbf k,\lambda'}\)} & {\(|\langle N_{\mathbf k,\lambda'}|\hat\Delta_{\mathbf k}|N_{\mathbf k,\lambda}\rangle|^2\)} & {Multiplicity} \\
\midrule
{\(|4_{\mathbf k}\rangle\to|2_{\mathbf k,s}\rangle\)} & {\(2\xi_{\mathbf k}\)} & {\(1\)} & {1} \\
{\(|3_{\pm\mathbf k\sigma}\rangle\to|1_{\mp\mathbf k\sigma}\rangle\)} & {\(2\xi_{\mathbf k}-\mathcal J_{\mathbf k}\)} & {\(1/2\)} & {4} \\
{\(|2_{\mathbf k,s}\rangle\to|0_{\mathbf k}\rangle\)} & {\(2\xi_{\mathbf k}-2\mathcal J_{\mathbf k}\)} & {\(1\)} & {1} \\
\bottomrule\end{tabular}}\par\endgroup
\smallskip
 
The pair-addition spectral function is defined by

\hypertarget{sm-eq-h2-lesser-greater-spectra}{}%
\FittedEquation{H27}{%
S_\Delta^>(\mathbf k,\Omega)
\equiv\int_{-\infty}^{\infty}\frac{dt}{2\pi}e^{i\Omega t}
\langle\hat\Delta_{\mathbf k}(t)\hat\Delta_{\mathbf k}^{\dagger}(0)\rangle.
}
 
The KMS relation gives \(S_\Delta^>(\mathbf k,\Omega)=e^{\beta\Omega}S_\Delta^<(\mathbf k,\Omega)\). Analytically continuing the real time \(t\) to imaginary time \(t=-i\tau\) and performing the inverse Fourier integral over \(\Omega\) gives

\hypertarget{sm-eq-h2-kms-matsubara}{}%
\FittedEquation{H28}{%
\begin{aligned}
\Pi_{\mathrm N,\mathbf k}(i\Omega_m)
&=\int_0^\beta d\tau\,e^{i\Omega_m\tau}
\int_{-\infty}^{\infty}d\Omega\,e^{-\Omega\tau}e^{\beta\Omega}S_{\mathrm N,\Delta}^<(\mathbf k,\Omega)\\
&=\int_{-\infty}^{\infty}d\Omega\,
\frac{e^{\beta\Omega}-1}{\Omega-i\Omega_m}S_{\mathrm N,\Delta}^<(\mathbf k,\Omega).
\end{aligned}
}
 
The static limit gives \hyperlink{sm-eq17}{Eq. (56)}:

\hypertarget{sm-eq-h2-kms-static}{}%
\FittedEquation{H29}{%
\Pi_{\mathrm N,\mathbf k}(0)=\int_{-\infty}^{\infty}d\Omega\,
\frac{e^{\beta\Omega}-1}{\Omega}S_{\mathrm N,\Delta}^<(\mathbf k,\Omega),
}
 
where the ratio is defined continuously as \(\beta\) for \(\Omega=0\).

To obtain the low-temperature behavior of \(\Pi_{\mathrm N,\mathbf k}\), let \(S_{\mathbf k,2e}=\xi_{\mathbf k}\) and \(S_{\mathbf k,2h}=\xi_{\mathbf k}-\mathcal J_{\mathbf k}\). Near a regular segment of \(\mathcal C_{2e}\), the two low-energy states are \(|4_{\mathbf k}\rangle\) and \(|2_{\mathbf k,s}\rangle\), with energy difference \(2S_{\mathbf k,2e}\). Near \(\mathcal C_{2h}\), the two low-energy states are \(|2_{\mathbf k,s}\rangle\) and \(|0_{\mathbf k}\rangle\), with energy difference \(2S_{\mathbf k,2h}\). The remaining states are separated by nonzero gaps near the Cooper surfaces away from the nodes. At low temperature, the thermal occupation probabilities in the two-state approximation are

\hypertarget{sm-eq-h2-two-state-probabilities}{}%
\FittedEquation{H30}{%
\begin{aligned}
p^{\mathrm N}_{\mathbf k,s}-p^{\mathrm N}_{\mathbf k,4}
&\simeq\frac{1-e^{-2S_{\mathbf k,2e}/T}}{1+e^{-2S_{\mathbf k,2e}/T}}
=\tanh\frac{S_{\mathbf k,2e}}T
\end{aligned}
}
 
\hypertarget{sm-eq-h2-two-hole-probabilities}{}%
\FittedEquation{H31}{%
\begin{aligned}
p^{\mathrm N}_{\mathbf k,0}-p^{\mathrm N}_{\mathbf k,s}
&\simeq\frac{1-e^{-2S_{\mathbf k,2h}/T}}{1+e^{-2S_{\mathbf k,2h}/T}}
=\tanh\frac{S_{\mathbf k,2h}}T
\end{aligned}
}
 
and the singular behavior of \(\Pi_{\mathrm N,\mathbf k}\) in the static limit is

\hypertarget{sm-eq-h2-soft-kernel}{}%
\FittedEquation{H32}{%
\Pi_{\mathrm N,\mathbf k}^{\mathrm{singular}}(0,T)
\simeq\frac{\tanh(S_{\mathbf k,r}/T)}{2S_{\mathbf k,r}}
\longrightarrow
\begin{cases}
\dfrac{1}{2T},&|S_{\mathbf k,r}|\ll T\ll\mathcal J_{\mathbf k},\\[6pt]
\dfrac{1}{2|S_{\mathbf k,r}|},&T\ll|S_{\mathbf k,r}|\ll\mathcal J_{\mathbf k},
\end{cases}
}
 
where \(r=2e,2h\). Other transitions only give nonsingular contributions.

Parameterize a regular segment of each Cooper surface \(\mathcal C_r\) by arc length \(\ell\), with \(\mathbf k_r(\ell)\in\mathcal C_r\), \(S_{\mathbf k_r,r}=0\), and unit normal \(\hat{\mathbf n}_r=\left.\frac{\nabla S_{\mathbf k,r}}{|\nabla S_{\mathbf k,r}|}\right|_{\mathbf k=\mathbf k_r}\). In a tubular neighborhood of \(\mathcal C_r\), expansion in the normal displacement gives

\hypertarget{sm-eq-h2-coarea-coordinates}{}\hypertarget{sm-eq-h2-coarea-detuning}{}%
\FittedEquation{H33}{%
\mathbf k=\mathbf k_r(\ell)+q_\perp\hat{\mathbf n}_r(\ell),\qquad S_{\mathbf k,r}=\left|\nabla S_{\mathbf k,r}\right|_{\mathbf k=\mathbf k_r}q_\perp+O(q_\perp^2).
}
 
The area element becomes

\hypertarget{sm-eq-h2-coarea-measure}{}%
\FittedEquation{H34}{%
d^2\mathbf k
=\frac{d\ell\,dS}{\left|\nabla S_{\mathbf k,r}\right|_{\mathbf k=\mathbf k_r}}\bigl[1+O(q_\perp)\bigr].
}
 
The integral of the squared form factor over the two Cooper surfaces to leading order reproduces the nonnegative \(\phi^2\)-weighted DOS defined in \hyperlink{sm-eq-cooper-g-contour-coefficient}{Eq. (G17)}:

\hypertarget{sm-eq-h2-weighted-dos}{}%
\FittedEquation{H35}{%
\begin{aligned}
D_\alpha(\mu)
&=\sum_{r=2e,2h}\oint_{\mathcal C_r}\frac{d\ell}{(2\pi)^2}
\left.\frac{\phi_{\mathbf k,\alpha}^2}{|\nabla S_{\mathbf k,r}|}\right|_{\mathbf k=\mathbf k_r}\\
&=\int_{\mathrm{BZ}}\frac{d^2\mathbf k}{(2\pi)^2}\phi_{\mathbf k,\alpha}^2
\left[\delta(\xi_{\mathbf k})+\delta(\xi_{\mathbf k}-\mathcal J_{\mathbf k})\right].
\end{aligned}
}
 
At low temperature, the singular contribution to the static pair susceptibility in \hyperlink{sm-eq-h2-block-reduction}{Eq. (H23)} is

\hypertarget{sm-eq-h2-contour-integral}{}%
\FittedEquation{H36}{%
\chi_{\mathrm N,\alpha\alpha}^{\mathrm{singular}}(0,T)
=2\sum_r\oint_{\mathcal C_r}\frac{d\ell}{(2\pi)^2}
\frac{\phi_{\mathbf k,\alpha}^2}{|\nabla S_{\mathbf k,r}|}
\int_{-\Lambda_{\mathrm{cut},r}(\ell)}^{\Lambda_{\mathrm{cut},r}(\ell)}dS\,
\frac{\tanh(S/T)}{2S},
}
 
where \(\Lambda_{\mathrm{cut},r}(\ell)\) is a low-energy cutoff. The normal integral includes both sides of each Cooper surface (cf. the integral for the pair-addition Cooper problem in \hyperlink{sm-eq-cooper-g-normal-integral}{Eq. (G18)}) and is given by

\hypertarget{sm-eq-h2-logarithmic-integral}{}%
\FittedEquation{H37}{%
\int_{-\Lambda_{\mathrm{cut},r}(\ell)}^{\Lambda_{\mathrm{cut},r}(\ell)}dS\,
\frac{\tanh(S/T)}{2S}
=\ln\frac{\Lambda_{\mathrm{cut},r}(\ell)}T+\gamma_{\mathrm E}+\ln\frac4\pi+o(1),
\qquad \frac{\Lambda_{\mathrm{cut},r}(\ell)}T\to\infty,
}
 
where \(\gamma_{\mathrm E}\) is Euler's constant. Choosing a fixed cutoff \(\Lambda_{\mathrm{cut},r}(\ell)=\Lambda_{\mathrm{cut}}\) therefore gives

\hypertarget{sm-eq-h2-thermal-logarithm}{}%
\FittedEquation{H38}{%
\chi_{\mathrm N,\alpha\alpha}^{\mathrm{singular}}(0,T)
=2D_\alpha(\mu)\ln\!\left(\frac{4e^{\gamma_{\mathrm E}}\Lambda_{\mathrm{cut}}}{\pi T}\right)+o(1).
}
 
Other nonsingular contributions do not alter the logarithmic coefficient \(2D_\alpha(\mu)\). For smooth \(D_\alpha\), the shift \(\delta\mu(T)=O(T^2)\) at fixed density [\hyperlink{sm-eq-thermo-f-chemical-shift}{Eq. (F34)}] gives

\hypertarget{sm-eq-h2-fixed-density-log}{}%
\FittedEquation{H39}{%
\bigl(D_\alpha[\mu(T)]-D_\alpha[\mu(0)]\bigr)\ln\frac{\Lambda_{\mathrm{cut}}}T
=O\!\left(T^2\ln\frac{\Lambda_{\mathrm{cut}}}T\right)=o(1).
}
 
The asymptotic coefficient thus can be approximately evaluated at \(\mu(0)\).

Using the nodal scalings in \hyperlink{sm-eq-thermo-f-nodal-variables}{Eq. (F21)}, the shrinking region \(|\xi_{\mathbf k}|,\mathcal J_{\mathbf k}\lesssim T\) has momentum-space area \(O(T^{3/2})\) and \(\Pi_{\mathrm N,\mathbf k}(0,T)=O(T^{-1})\). Since \(\phi_{\mathbf k,s}^2=O(1)\) and \(\phi_{\mathbf k,d}^2=O(T)\), the nodal contributions scale as

\hypertarget{sm-eq-h2-nodal-orders}{}%
\FittedEquation{H40}{%
\begin{aligned}
\left.\chi_{\mathrm N,ss}(0,T)\right|_{|\xi_{\mathbf k}|,\mathcal J_{\mathbf k}\lesssim T}
=O(T^{1/2})
\end{aligned}
}
 
\hypertarget{sm-eq-h2-nodal-d-order}{}%
\FittedEquation{H41}{%
\begin{aligned}
\left.\chi_{\mathrm N,dd}(0,T)\right|_{|\xi_{\mathbf k}|,\mathcal J_{\mathbf k}\lesssim T}
=O(T^{3/2}).
\end{aligned}
}
 
The logarithmic contributions from Cooper-surface segments within fixed nodal patches are included in \(D_\alpha(\mu)\).

To leading logarithmic accuracy, the SC instability temperature is

\hypertarget{sm-eq-h2-leading-transition}{}%
\FittedEquation{H42}{%
T_{c,\alpha}\simeq\frac{4e^{\gamma_{\mathrm E}}}{\pi}\Lambda_{\mathrm{cut}}
\exp\!\left[-\frac1{4JD_\alpha(\mu)}\right].
}
 
The logarithmic divergence guarantees an SC instability of the nodal Bose metal at arbitrarily weak \(J>0\) in both channels with \(D_\alpha(\mu)>0\). For the parameters of \hyperlink{sm-fig-model-normal-observables}{Fig. 5}, \(D_s[\mu(0)]=0.043236~\mathrm{eV}^{-1}\) and \(D_d[\mu(0)]=0.699875~\mathrm{eV}^{-1}\), so the \(d\)-wave instability is dominant for equal, sufficiently weak attraction.

At finite attraction, the SC instability temperature is obtained by numerically solving

\hypertarget{sm-eq-h2-fixed-density-system}{}%
\FittedEquation{H43}{%
1=4J\int_{\mathrm{BZ}}\frac{d^2\mathbf k}{(2\pi)^2}
\phi_{\mathbf k,\alpha}^2\Pi_{\mathrm N,\mathbf k}\!\left(0,T_{c,\alpha};\mu(T_{c,\alpha})\right),
}
 
together with the fixed-density constraint \(n_e=1-p\), which determines \(\mu(T_{c,\alpha})\). Comparisons between the numerical solution of \hyperlink{sm-eq-h2-fixed-density-system}{Eq. (H43)} and the analytical solution in \hyperlink{sm-eq-h2-leading-transition}{Eq. (H42)} in the \(d\)-wave channel are shown in \hyperlink{sm-fig-h-sc-coupling}{Fig. H1}, whereas the extended-\(s\)-wave instability temperature is exponentially smaller.

\par\medskip
\hypertarget{sm-fig-h-sc-coupling}{}
\PrintAsset{sc-instability-coupling}
{\small\noindent \textbf{Figure H1 \textbar{} SC instability temperature in the \(d\)-wave channel.} The light-orange solid curve shows numerical solutions of \hyperlink{sm-eq-h2-fixed-density-system}{Eq. (H43)} in the thermodynamic limit under the fixed-density constraint. Circles show finite-size solutions for \(N_s=8192^2\). The black dashed curve shows the leading-logarithmic analytical solution in \hyperlink{sm-eq-h2-leading-transition}{Eq. (H42)}, with the fixed cutoff \(\Lambda_{\mathrm{cut}}=0.06~\mathrm{eV}\). Parameters match \hyperlink{sm-fig-model-normal-observables}{Fig. 5}: \(\mathcal J_d=0.06~\mathrm{eV}\), \(p=0.15\), and \(\mu(0)=-0.268236~\mathrm{eV}\).\par}
\medskip
 
\hypertarget{sm-appendix-g}{}\hypertarget{sm-appendix-g-title}{}
\section*{Appendix I. SC instability at finite total momentum \(\mathbf Q\ne0\)}

\hypertarget{sm-finite-q-response}{}Consider the full AFM superexchange interaction in \hyperlink{sm-eq-model-superexchange}{Eq. (1)} as the pairing interaction that includes all total momenta of Cooper pairs, and the full Hamiltonian is

\hypertarget{sm-eq-fq-interaction}{}%
\FittedEquation{I1}{%
\begin{aligned}
\hat H&=\hat H_{\mathrm N}+\hat H_J
\end{aligned}
}
 
\hypertarget{sm-eq-fq-pairing-interaction}{}%
\FittedEquation{I2}{%
\hat H_J=-2JN_s\sum_{\mathbf Q,\alpha=s,d}\hat\Lambda_{\mathbf Q,\alpha}^{\dagger}\hat\Lambda_{\mathbf Q,\alpha},
}
 
where \(\hat\Lambda_{\mathbf Q,\alpha}=N_s^{-1}\sum_{\mathbf k}\phi_{\mathbf k,\alpha}\hat\Delta_{\mathbf Q,\mathbf k}\), and \(\hat\Delta_{\mathbf Q,\mathbf k}\) is the spin-singlet pair operator in \hyperlink{sm-eq-model-pair-operator}{Eq. (2)}, with electron momenta \(\mathbf Q/2\pm\mathbf k\). The exact pair susceptibility of the nodal Bose metal is

\hypertarget{sm-eq-fq-susceptibility}{}%
\FittedEquation{I3}{%
\chi_{\mathrm N,\alpha\alpha'}(\mathbf Q,i\Omega_m)
=N_s\int_0^\beta d\tau\,e^{i\Omega_m\tau}
\langle T_\tau\hat\Lambda_{\mathbf Q,\alpha}(\tau)\hat\Lambda_{\mathbf Q,\alpha'}^{\dagger}(0)\rangle_{\mathrm N,c}.
}
 
The generating-functional construction of \hyperlink{sm-appendix-h-dyson-equation}{Appendix H.1} extends directly to auxiliary pairing fields \(\Xi_{\mathbf Q,\alpha}(\tau)\). The quadratic expansion about the normal saddle \(\Xi_{\mathbf Q,\alpha}^{(0)}=0\) gives

\hypertarget{sm-eq-fq-quadratic}{}%
\FittedEquation{I4}{%
S_{\mathrm{eff}}^{(2)}=N_s\sum_{\mathbf Q,m,\alpha\alpha'}\Xi_{\mathbf Q,\alpha m}^{*}K_{\alpha\alpha'}(\mathbf Q,i\Omega_m)\Xi_{\mathbf Q,\alpha' m},
}
 
where

\hypertarget{sm-eq-fq-quadratic-kernel}{}%
\FittedEquation{I5}{%
K_{\alpha\alpha'}(\mathbf Q,i\Omega_m)=\frac{\delta_{\alpha\alpha'}}{2J}-\chi_{\mathrm N,\alpha\alpha'}(\mathbf Q,i\Omega_m).
}
 
Consequently, the Gaussian pair susceptibility in the ladder approximation is

\hypertarget{sm-eq-fq-gaussian}{}%
\FittedEquation{I6}{%
\begin{aligned}
\boldsymbol\chi(\mathbf Q,i\Omega_m)\simeq
[\mathbf I-2J\boldsymbol\chi_{\mathrm N}(\mathbf Q,i\Omega_m)]^{-1}
\boldsymbol\chi_{\mathrm N}(\mathbf Q,i\Omega_m),
\end{aligned}
}
 
and the Gaussian SC instability condition is

\hypertarget{sm-eq-fq-instability-condition}{}%
\FittedEquation{I7}{%
2J\,\lambda_{\max}\!\left[\boldsymbol\chi_{\mathrm N}(\mathbf Q,0,T_c)\right]=1.
}
 
where \(\lambda_{\max}\) denotes the largest eigenvalue of the static pair susceptibility matrix \(\boldsymbol\chi_{\mathrm N}(\mathbf Q,0,T_c)\).

At any nonzero \(\mathbf Q\) modulo a reciprocal-lattice vector, the two electrons with momenta \(\mathbf Q/2\pm\mathbf k\) generally belong to different momentum blocks. Using block independence and conservation of the electron number \(\hat n_{\mathbf k}\), without invoking Wick's theorem, gives

\hypertarget{sm-eq-fq-bubble}{}%
\FittedEquation{I8}{%
\chi_{\mathrm N,\alpha\alpha'}(\mathbf Q,i\Omega_m)=\frac2{N_s}\sum_{\mathbf k}\phi_{\mathbf k,\alpha}\phi_{\mathbf k,\alpha'}\Pi_{\mathrm N,\mathbf k}(\mathbf Q,i\Omega_m),
}
 
and

\hypertarget{sm-eq-fq-pair-propagator}{}%
\FittedEquation{I9}{%
\begin{aligned}
\Pi_{\mathrm N,\mathbf k}(\mathbf Q,i\Omega_m)&=\int_0^\beta d\tau\,e^{i\Omega_m\tau}\left\langle T_\tau\hat\Delta_{\mathbf Q,\mathbf k}(\tau)\hat\Delta_{\mathbf Q,\mathbf k}^{\dagger}(0)\right\rangle_{\mathrm N}\\
&=\frac1\beta\sum_{\nu_n}G_{\mathrm N}(\mathbf Q/2+\mathbf k,i\nu_n)G_{\mathrm N}(\mathbf Q/2-\mathbf k,i\Omega_m-i\nu_n),
\end{aligned}
}
 
where \(G_{\mathrm N}(\mathbf k,i\nu_n)=-\int_0^\beta d\tau\,e^{i\nu_n\tau}\langle T_\tau\hat c_{\mathbf k\sigma}(\tau)\hat c_{\mathbf k\sigma}^{\dagger}(0)\rangle_{\mathrm N}\) is the one-particle Matsubara Green's function, and \(\nu_n=(2n+1)\pi/\beta\) is the fermionic Matsubara frequency. For finite total momentum \(\mathbf Q\ne0\), square-lattice symmetry does not generally enforce the vanishing of the off-diagonal elements of \(\boldsymbol\chi_{\mathrm N}(\mathbf Q,i\Omega_m)\).

\hypertarget{sm-finite-q-bound}{}The static pair susceptibility at finite total momentum can be bounded in the thermodynamic limit. Introduce a normalized vector \(\mathbf v=(v_s,v_d)^{\mathsf T}\), with \(|v_s|^2+|v_d|^2=1\), and the Cauchy--Schwarz inequality gives

\hypertarget{sm-eq-fq-channel-form-factor-bound}{}%
\FittedEquation{I10}{%
\left|\sum_\alpha v_\alpha\phi_{\mathbf k,\alpha}\right|^2
\le\phi_{\mathbf k,s}^2+\phi_{\mathbf k,d}^2
=\frac{\cos^2k_x+\cos^2k_y}{2}\le1.
}
 
For \(0<\tau<\beta\),

\hypertarget{sm-eq-fq-green-spectral-representation}{}%
\FittedEquation{I11}{%
G_{\mathrm N}(\mathbf k,\tau)=-\left\langle\hat c_{\mathbf k\sigma}(\tau)\hat c_{\mathbf k\sigma}^{\dagger}(0)\right\rangle_{\mathrm N}=-\int_{-\infty}^{\infty}d\omega\,A_{\mathrm N}(\mathbf k,\omega)\frac{e^{-\tau\omega}}{1+e^{-\beta\omega}}\le0.
}
 
where \(A_{\mathrm N}(\mathbf k,\omega)=-\pi^{-1}\operatorname{Im}G_{\mathrm N}(\mathbf k,i\nu_n\to\omega+i0^+)\). Then \hyperlink{sm-eq-fq-bubble}{Eqs. (I8)}--\hyperlink{sm-eq-fq-pair-propagator}{(I9)} in the static limit give

\hypertarget{sm-eq-fq-channel-quadratic-bound}{}%
\FittedEquation{I12}{%
\begin{aligned}
\mathbf v^\dagger\boldsymbol\chi_{\mathrm N}(\mathbf Q,0,T)\mathbf v&=2\int_{\mathrm{BZ}}\frac{d^2\mathbf k}{(2\pi)^2}\int_0^\beta d\tau\,\left|\sum_\alpha v_\alpha\phi_{\mathbf k,\alpha}\right|^2G_{\mathrm N}(\mathbf Q/2+\mathbf k,\tau)G_{\mathrm N}(\mathbf Q/2-\mathbf k,\tau)\\
&\le2\int_{\mathrm{BZ}}\frac{d^2\mathbf k}{(2\pi)^2}\int_0^\beta d\tau\,
G_{\mathrm N}(\mathbf Q/2+\mathbf k,\tau)G_{\mathrm N}(\mathbf Q/2-\mathbf k,\tau)\\
&\le\int_{\mathrm{BZ}}\frac{d^2\mathbf k}{(2\pi)^2}\int_0^\beta d\tau\,
\left[|G_{\mathrm N}(\mathbf Q/2+\mathbf k,\tau)|^2+|G_{\mathrm N}(\mathbf Q/2-\mathbf k,\tau)|^2\right]\\
&=2\int_{\mathrm{BZ}}\frac{d^2\mathbf k}{(2\pi)^2}\int_0^\beta d\tau\,
|G_{\mathrm N}(\mathbf k,\tau)|^2.
\end{aligned}
}
 
where the upper bound is independent of \(\mathbf Q\). The Rayleigh--Ritz variational principle gives

\hypertarget{sm-eq-fq-rayleigh-ritz}{}%
\FittedEquation{I13}{%
\max_{\mathbf v^\dagger\mathbf v=1}\mathbf v^\dagger\boldsymbol\chi_{\mathrm N}(\mathbf Q,0,T)\mathbf v=\lambda_{\max}\!\left[\boldsymbol\chi_{\mathrm N}(\mathbf Q,0,T)\right].
}
 
The Lehmann representation of \(G_{\mathrm N}(\mathbf k,\tau)\), for \(0<\tau<\beta\), gives

\hypertarget{sm-eq-fq-lehmann-green}{}%
\FittedEquation{I14}{%
|G_{\mathrm N}(\mathbf k,\tau)|=\frac1{Z^{\mathrm N}_{\mathbf k}}\sum_{\lambda,\lambda'}e^{-(\beta-\tau)E^{\mathrm N}_{\mathbf k,\lambda}-\tau E^{\mathrm N}_{\mathbf k,\lambda'}}\left|\langle N_{\mathbf k,\lambda}|\hat c_{\mathbf k\sigma}|N_{\mathbf k,\lambda'}\rangle\right|^2.
}
 
where \(|N_{\mathbf k,\lambda}\rangle\) and \(|N_{\mathbf k,\lambda'}\rangle\) have opposite block fermion parity for every nonzero matrix element. Denoting the block ground-state energy by \(E^{\mathrm N}_{\mathbf k,\mathrm{GS}}=\min_\lambda E^{\mathrm N}_{\mathbf k,\lambda}\), it follows that

\hypertarget{sm-eq-fq-lehmann-exponent-bound}{}%
\FittedEquation{I15}{%
(\beta-\tau)E^{\mathrm N}_{\mathbf k,\lambda}+\tau E^{\mathrm N}_{\mathbf k,\lambda'}
\ge\beta E^{\mathrm N}_{\mathbf k,\mathrm{GS}}+\Delta_{\mathrm{1p}}^{\mathrm N}(\mathbf k)\min(\tau,\beta-\tau).
}
 
Using the completeness of the block eigenstates gives

\hypertarget{sm-eq-fq-electron-matrix-element-trace}{}%
\FittedEquation{I16}{%
\sum_{\lambda,\lambda'}\left|\langle N_{\mathbf k,\lambda}|\hat c_{\mathbf k\sigma}|N_{\mathbf k,\lambda'}\rangle\right|^2=\operatorname{Tr}\!\left(\hat c_{\mathbf k\sigma}\hat c_{\mathbf k\sigma}^{\dagger}\right)=\operatorname{Tr}(1-\hat n_{\mathbf k\sigma})=8.
}
 
Since \(Z^{\mathrm N}_{\mathbf k}\ge e^{-\beta E^{\mathrm N}_{\mathbf k,\mathrm{GS}}}\), combining \hyperlink{sm-eq-fq-lehmann-green}{Eqs. (I14)}--\hyperlink{sm-eq-fq-electron-matrix-element-trace}{(I16)} yields

\hypertarget{sm-eq-fq-green-exponential-bound}{}%
\FittedEquation{I17}{%
|G_{\mathrm N}(\mathbf k,\tau)|\le\frac{8e^{-\beta E^{\mathrm N}_{\mathbf k,\mathrm{GS}}}}{Z^{\mathrm N}_{\mathbf k}}e^{-\Delta_{\mathrm{1p}}^{\mathrm N}(\mathbf k)\min(\tau,\beta-\tau)}\le8e^{-\Delta_{\mathrm{1p}}^{\mathrm N}(\mathbf k)\min(\tau,\beta-\tau)}.
}
 
For nonzero \(\mathbf Q\), combining \hyperlink{sm-eq-fq-channel-quadratic-bound}{Eqs. (I12)}--\hyperlink{sm-eq-fq-rayleigh-ritz}{(I13)} and \hyperlink{sm-eq-fq-green-exponential-bound}{(I17)} and performing the imaginary-time integral give

\hypertarget{sm-eq-fq-bound}{}%
\FittedEquation{I18}{%
\begin{aligned}
\lambda_{\max}\!\left[\boldsymbol\chi_{\mathrm N}(\mathbf Q,0,T)\right]
&\le2\int_{\mathrm{BZ}}\frac{d^2\mathbf k}{(2\pi)^2}\int_0^\beta d\tau\,|G_{\mathrm N}(\mathbf k,\tau)|^2\\
&\le128\int_{\mathrm{BZ}}\frac{d^2\mathbf k}{(2\pi)^2}\frac1{\Delta_{\mathrm{1p}}^{\mathrm N}(\mathbf k)}.
\end{aligned}
}
 
The only possible singularities of the momentum integral occur at the nodes. Denote the independent local coordinates near each node by \(\epsilon=\xi_{\mathbf k}\) and \(\varphi=\phi_{\mathbf k,d}\), with a finite Jacobian. For fixed positive cutoffs \(\epsilon_c\) and \(\varphi_c\), using \(\Delta_{\mathrm{1p}}^{\mathrm N}(\mathbf k)\ge\tfrac12(|\xi_{\mathbf k}|+\mathcal J_d\phi_{\mathbf k,d}^2)\) [\hyperlink{sm-eq-mstj-one-particle-gap}{Eq. (16)}], the contributions near the nodes are bounded by a finite constant times

\hypertarget{sm-eq-fq-nodal-integrability}{}%
\FittedEquation{I19}{%
\int_0^{\varphi_c}d\varphi\int_0^{\epsilon_c}\frac{d\epsilon}{\epsilon+\mathcal J_d\varphi^2}
=\int_0^{\varphi_c}d\varphi\,\ln\!\left(1+\frac{\epsilon_c}{\mathcal J_d\varphi^2}\right)<\infty.
}
 
The contributions away from the nodes are finite. \hyperlink{sm-eq-fq-bound}{Eq. (I18)} therefore gives \(\lambda_{\max}[\boldsymbol\chi_{\mathrm N}(\mathbf Q,0,T)]\le C<\infty\), uniformly in nonzero \(\mathbf Q\) as \(T\to0\). Thus the finite-total-momentum susceptibility has no logarithmic divergence of the type found at \(\mathbf Q=0\) in \hyperlink{sm-eq-cooper-log-contour-integral}{Eq. (58)}.

For sufficiently weak attraction such that \(2JC<1\), the Gaussian SC instability condition in \hyperlink{sm-eq-fq-instability-condition}{Eq. (I7)} cannot be satisfied at any nonzero \(\mathbf Q\). In contrast, the logarithmic divergence in \hyperlink{sm-eq-cooper-log-contour-integral}{Eq. (58)} guarantees a zero-total-momentum SC instability for arbitrarily weak attraction whenever \(D_\alpha(\mu)>0\). The leading pairing instability at sufficiently weak attraction therefore occurs at \(\mathbf Q=0\).
 
\hypertarget{sm-appendix-sc-mean-field}{}\hypertarget{sm-appendix-sc-mean-field-title}{}
\section*{Appendix J. Mean-field theory of the correlated SC state}

\hypertarget{sm-sc-block-diagonalization}{}
\subsection*{J.1. Self-consistent equations}

\hypertarget{sm-appendix-j-eigenstates-summary}{}\hypertarget{sm-table-j1-sc-eigenstates}{}\par\medskip

\noindent{\small Table J1 \textbar{} Block eigenstates and eigenenergies of the SC block Hamiltonian \(\hat h_{\mathbf k}^{\mathrm{SC}}\).\par}\nopagebreak[4]
\begingroup\small\setlength{\tabcolsep}{3.5pt}\renewcommand{\arraystretch}{1.30}
\noindent\FittedTable{\begin{tabular}{@{}cccc@{}}\toprule
{Block eigenstate \(|\pi_{\mathbf k,\lambda}\rangle\)} & {Block eigenenergy \(E^{\mathrm{SC}}_{\mathbf k,\lambda}\)} & {Degeneracy} & {\shortstack[c]{Thermal occupation\\ probability \(p^{\mathrm{SC}}_{\mathbf k,\lambda}\)}} \\
\midrule
{\(\begin{aligned}
              &|e_{\mathbf k,l}\rangle=x_{\mathbf k,l}|0_{\mathbf k}\rangle+y_{\mathbf k,l}|2_{\mathbf k,s}\rangle+z_{\mathbf k,l}|4_{\mathbf k}\rangle\\
              &|x_{\mathbf k,l}|^2+|y_{\mathbf k,l}|^2+|z_{\mathbf k,l}|^2=1,\quad l=0,1,2
            \end{aligned}\)} & {\(E^{\mathrm{SC},e}_{\mathbf k,0}\le E^{\mathrm{SC},e}_{\mathbf k,1}\le E^{\mathrm{SC},e}_{\mathbf k,2}\)} & {1 each} & {\(p^{\mathrm{SC},e}_{\mathbf k,l}=\dfrac{e^{-\beta E^{\mathrm{SC},e}_{\mathbf k,l}}}{Z^{\mathrm{SC}}_{\mathbf k}}\)} \\
{\(|2_{\mathbf k,t_1}\rangle,\ |2_{\mathbf k,t_0}\rangle,\ |2_{\mathbf k,t_{-1}}\rangle,\ |2_{\pm\mathbf k}\rangle\)} & {\(2\xi_{\mathbf k}\)} & {5} & {\(p^{\mathrm{SC}}_{\mathbf k,2}=\dfrac{e^{-2\beta\xi_{\mathbf k}}}{Z^{\mathrm{SC}}_{\mathbf k}}\)} \\
{\(\begin{aligned}
              &|o_{\pm\mathbf k\sigma,\nu}\rangle=u_{\mathbf k,\nu}|1_{\pm\mathbf k\sigma}\rangle+v_{\mathbf k,\nu}|3_{\mp\mathbf k\sigma}\rangle\\
              &|u_{\mathbf k,\nu}|^2+|v_{\mathbf k,\nu}|^2=1,\quad\nu=\pm1
            \end{aligned}\)} & {\(\begin{gathered}
              E^{\mathrm{SC},o}_{\mathbf k,\nu}=2\xi_{\mathbf k}-\mathcal J_{\mathbf k}/2\\
              {}+\nu\sqrt{(\xi_{\mathbf k}-\mathcal J_{\mathbf k}/2)^2+2|\Delta_{\mathbf k,d}|^2}
            \end{gathered}\)} & {4 each} & {\(p^{\mathrm{SC},o}_{\mathbf k,\nu}=\dfrac{e^{-\beta E^{\mathrm{SC},o}_{\mathbf k,\nu}}}{Z^{\mathrm{SC}}_{\mathbf k}}\)} \\
\bottomrule\end{tabular}}\par\endgroup
\smallskip
 
The mean-field SC Hamiltonian in the leading \(d\)-wave channel is

\hypertarget{sm-eq-sc-order}{}%
\FittedEquation{J1}{%
\hat H_{\mathrm{SC}}=\sum_{\{\mathbf k,-\mathbf k\}}\hat h_{\mathbf k}^{\mathrm{SC}}+\frac{N_s|\Delta_d|^2}{2J},
}
 
\hypertarget{sm-eq-sc-appendix-block-hamiltonian}{}%
\FittedEquation{J2}{%
\hat h_{\mathbf k}^{\mathrm{SC}}=\xi_{\mathbf k}\hat N_{\mathbf k}-2\mathcal J_{\mathbf k}\hat\Delta_{\mathbf k}^{\dagger}\hat\Delta_{\mathbf k}-2(\Delta_{\mathbf k,d}\hat\Delta_{\mathbf k}^{\dagger}+\Delta_{\mathbf k,d}^{*}\hat\Delta_{\mathbf k}),
}
 
where \(\Delta_{\mathbf k,d}=\Delta_d\phi_{\mathbf k,d}\) and \(\Delta_d=2J\langle\hat\Lambda_d\rangle\).

In the even-fermion-parity subspace spanned by \(\{|0_{\mathbf k}\rangle,|2_{\mathbf k,s}\rangle,|4_{\mathbf k}\rangle\}\), the SC block Hamiltonian \(\hat h_{\mathbf k}^{\mathrm{SC}}\) has the matrix representation

\hypertarget{sm-eq-sc-even-matrix}{}%
\FittedEquation{J3}{%
h_{\mathbf k}^{\mathrm{SC},e}=\begin{pmatrix}
0&-2\Delta_{\mathbf k,d}^{*}&0\\
-2\Delta_{\mathbf k,d}&2\xi_{\mathbf k}-2\mathcal J_{\mathbf k}&-2\Delta_{\mathbf k,d}^{*}\\
0&-2\Delta_{\mathbf k,d}&4\xi_{\mathbf k}-2\mathcal J_{\mathbf k}
\end{pmatrix}.
}
 
The three eigenstates are \(|e_{\mathbf k,l}\rangle=x_{\mathbf k,l}|0_{\mathbf k}\rangle+y_{\mathbf k,l}|2_{\mathbf k,s}\rangle+z_{\mathbf k,l}|4_{\mathbf k}\rangle\), \(l=0,1,2\), with normalization \(|x_{\mathbf k,l}|^2+|y_{\mathbf k,l}|^2+|z_{\mathbf k,l}|^2=1\) and eigenenergies \(E^{\mathrm{SC},e}_{\mathbf k,0}\le E^{\mathrm{SC},e}_{\mathbf k,1}\le E^{\mathrm{SC},e}_{\mathbf k,2}\). The analytical expressions for the eigenenergies are

\hypertarget{sm-eq-sc-even-energies}{}%
\FittedEquation{J4}{%
E^{\mathrm{SC},e}_{\mathbf k,l}=2\xi_{\mathbf k}-\frac{4\mathcal J_{\mathbf k}}3+2\sqrt{P_{\mathbf k}}\cos\!\left[\frac13\arccos\!\left(\frac{Q_{\mathbf k}}{P_{\mathbf k}^{3/2}}\right)+\frac{2\pi(l+1)}3\right],
}
 
where

\hypertarget{sm-eq-sc-even-cubic-parameters}{}%
\FittedEquation{J5}{%
\begin{aligned}
P_{\mathbf k}&=\frac{(2\xi_{\mathbf k}-\mathcal J_{\mathbf k})^2+8|\Delta_{\mathbf k,d}|^2}{3}+\frac{\mathcal J_{\mathbf k}^2}{9},
\end{aligned}
}
 
\hypertarget{sm-eq-sc-even-cubic-q}{}%
\FittedEquation{J6}{%
\begin{aligned}
Q_{\mathbf k}&=\frac{\mathcal J_{\mathbf k}}3\left[(2\xi_{\mathbf k}-\mathcal J_{\mathbf k})^2-4|\Delta_{\mathbf k,d}|^2-\frac{\mathcal J_{\mathbf k}^2}{9}\right].
\end{aligned}
}
 
The explicit normalized coefficients for \(\Delta_{\mathbf k,d}\ne0\) are

\hypertarget{sm-eq-sc-even-coefficients}{}%
\FittedEquation{J7}{%
\begin{aligned}
x_{\mathbf k,l}&=\frac{4(\Delta_{\mathbf k,d}^{*})^2}{\mathcal N_{\mathbf k,l}},
\end{aligned}
}
 
\hypertarget{sm-eq-sc-even-coefficient-y}{}%
\FittedEquation{J8}{%
\begin{aligned}
y_{\mathbf k,l}&=-\frac{2E^{\mathrm{SC},e}_{\mathbf k,l}\Delta_{\mathbf k,d}^{*}}{\mathcal N_{\mathbf k,l}},
\end{aligned}
}
 
\hypertarget{sm-eq-sc-even-coefficient-z}{}%
\FittedEquation{J9}{%
\begin{aligned}
z_{\mathbf k,l}&=\frac{E^{\mathrm{SC},e}_{\mathbf k,l}(E^{\mathrm{SC},e}_{\mathbf k,l}-2\xi_{\mathbf k}+2\mathcal J_{\mathbf k})-4|\Delta_{\mathbf k,d}|^2}{\mathcal N_{\mathbf k,l}},
\end{aligned}
}
 
where

\hypertarget{sm-eq-sc-even-normalization}{}%
\FittedEquation{J10}{%
\mathcal N_{\mathbf k,l}=\sqrt{16|\Delta_{\mathbf k,d}|^4+4(E^{\mathrm{SC},e}_{\mathbf k,l})^2|\Delta_{\mathbf k,d}|^2+\left[E^{\mathrm{SC},e}_{\mathbf k,l}(E^{\mathrm{SC},e}_{\mathbf k,l}-2\xi_{\mathbf k}+2\mathcal J_{\mathbf k})-4|\Delta_{\mathbf k,d}|^2\right]^2}.
}
 
In each of the four odd-fermion-parity subspaces spanned by \(\{|1_{\pm\mathbf k\sigma}\rangle,|3_{\mp\mathbf k\sigma}\rangle\}\), the SC block Hamiltonian \(\hat h_{\mathbf k}^{\mathrm{SC}}\) has the matrix representation

\hypertarget{sm-eq-sc-odd-matrix}{}%
\FittedEquation{J11}{%
h_{\mathbf k}^{\mathrm{SC},o}=\begin{pmatrix}
\xi_{\mathbf k}&-\sqrt2\Delta_{\mathbf k,d}^{*}\\
-\sqrt2\Delta_{\mathbf k,d}&3\xi_{\mathbf k}-\mathcal J_{\mathbf k}
\end{pmatrix}.
}
 
The two eigenstates in each subspace are \(|o_{\pm\mathbf k\sigma,\nu}\rangle=u_{\mathbf k,\nu}|1_{\pm\mathbf k\sigma}\rangle+v_{\mathbf k,\nu}|3_{\mp\mathbf k\sigma}\rangle\), \(\nu=\pm1\), with normalized coefficients

\hypertarget{sm-eq-sc-odd-coherence-factors}{}%
\FittedEquation{J12}{%
\begin{aligned}
u_{\mathbf k,\nu}&=\sqrt{\frac12\left(1-\nu\frac{\xi_{\mathbf k}-\mathcal J_{\mathbf k}/2}{\sqrt{(\xi_{\mathbf k}-\mathcal J_{\mathbf k}/2)^2+2|\Delta_{\mathbf k,d}|^2}}\right)},
\end{aligned}
}
 
\hypertarget{sm-eq-sc-odd-coherence-v}{}%
\FittedEquation{J13}{%
\begin{aligned}
v_{\mathbf k,\nu}&=-\nu\frac{\Delta_{\mathbf k,d}}{|\Delta_{\mathbf k,d}|}\sqrt{\frac12\left(1+\nu\frac{\xi_{\mathbf k}-\mathcal J_{\mathbf k}/2}{\sqrt{(\xi_{\mathbf k}-\mathcal J_{\mathbf k}/2)^2+2|\Delta_{\mathbf k,d}|^2}}\right)},
\end{aligned}
}
 
and eigenenergies

\hypertarget{sm-eq-sc-odd-energies}{}%
\FittedEquation{J14}{%
E^{\mathrm{SC},o}_{\mathbf k,\nu}=2\xi_{\mathbf k}-\frac{\mathcal J_{\mathbf k}}2+\nu\sqrt{\left(\xi_{\mathbf k}-\frac{\mathcal J_{\mathbf k}}2\right)^2+2|\Delta_{\mathbf k,d}|^2}.
}
 
The five remaining two-electron states are annihilated by both pair operators \(\hat\Delta_{\mathbf k}^{\dagger}\) and \(\hat\Delta_{\mathbf k}\), retaining the fivefold degenerate eigenenergy \(2\xi_{\mathbf k}\). The 16 block eigenstates of \(\hat h_{\mathbf k}^{\mathrm{SC}}\) are listed in \hyperlink{sm-table-j1-sc-eigenstates}{Table J1}. The SC partition function is

\hypertarget{sm-eq-sc-partition}{}%
\FittedEquation{J15}{%
Z_{\mathrm{SC}}=e^{-\beta N_s|\Delta_d|^2/(2J)}\prod_{\{\mathbf k,-\mathbf k\}}Z^{\mathrm{SC}}_{\mathbf k},
}
 
where the SC block partition function is

\hypertarget{sm-eq-sc-block-partition}{}%
\FittedEquation{J16}{%
Z^{\mathrm{SC}}_{\mathbf k}=\operatorname{Tr}e^{-\beta\hat h_{\mathbf k}^{\mathrm{SC}}}=\sum_{\lambda}e^{-\beta E^{\mathrm{SC}}_{\mathbf k,\lambda}}=\sum_{l=0}^{2}e^{-\beta E^{\mathrm{SC},e}_{\mathbf k,l}}+5e^{-2\beta\xi_{\mathbf k}}+4\sum_{\nu=\pm1}e^{-\beta E^{\mathrm{SC},o}_{\mathbf k,\nu}}.
}
 
The Helmholtz free-energy density at fixed electron density \(n_e=1-p\) is

\hypertarget{sm-eq-sc-free-energy}{}%
\FittedEquation{J17}{%
f_{\mathrm{SC}}(T,n_e)=-\frac{1}{N_s\beta}\ln Z_{\mathrm{SC}}+\mu n_e=\frac{|\Delta_d|^2}{2J}-\frac{T}{N_s}\sum_{\{\mathbf k,-\mathbf k\}}\ln Z^{\mathrm{SC}}_{\mathbf k}+\mu n_e.
}
 
Stationarity with respect to \(\Delta_d^{*}\) and the fixed-density constraint give the coupled self-consistent equations

\hypertarget{sm-eq-sc-appendix-gap-equation}{}%
\FittedEquation{J18}{%
\Delta_d=\frac{4J}{N_s}\sum_{\{\mathbf k,-\mathbf k\}}\phi_{\mathbf k,d}\langle\hat\Delta_{\mathbf k}\rangle_{\mathrm{SC}},
}
 
\hypertarget{sm-eq-sc-appendix-density-equation}{}%
\FittedEquation{J19}{%
n_e=\frac1{N_s}\sum_{\{\mathbf k,-\mathbf k\}}\langle\hat N_{\mathbf k}\rangle_{\mathrm{SC}},
}
 
where the expectation values are given by

\hypertarget{sm-eq-sc-block-averages}{}%
\FittedEquation{J20}{%
\begin{aligned}
\langle\hat\Delta_{\mathbf k}\rangle_{\mathrm{SC}}&=\sum_{l=0}^{2}p^{\mathrm{SC},e}_{\mathbf k,l}(x_{\mathbf k,l}^{*}y_{\mathbf k,l}+y_{\mathbf k,l}^{*}z_{\mathbf k,l})+\frac4{\sqrt2}\sum_{\nu=\pm1}p^{\mathrm{SC},o}_{\mathbf k,\nu}u_{\mathbf k,\nu}^{*}v_{\mathbf k,\nu},
\end{aligned}
}
 
\hypertarget{sm-eq-sc-block-number-average}{}%
\FittedEquation{J21}{%
\begin{aligned}
\langle\hat N_{\mathbf k}\rangle_{\mathrm{SC}}&=\sum_{l=0}^{2}p^{\mathrm{SC},e}_{\mathbf k,l}(2|y_{\mathbf k,l}|^2+4|z_{\mathbf k,l}|^2)+10p^{\mathrm{SC}}_{\mathbf k,2}+4\sum_{\nu=\pm1}p^{\mathrm{SC},o}_{\mathbf k,\nu}(|u_{\mathbf k,\nu}|^2+3|v_{\mathbf k,\nu}|^2).
\end{aligned}
}
 
Linearizing the gap equation in \hyperlink{sm-eq-sc-appendix-gap-equation}{Eq. (J18)} near \(T_{c,d}\) gives \(\Delta_d=2J\chi_{\mathrm N,dd}(0,T)\Delta_d+O(|\Delta_d|^3)\), reproducing the SC instability condition in \hyperlink{sm-eq19}{Eq. (53)}. With the global phase gauge chosen so that \(\Delta_d\) is real and nonnegative, \hyperlink{sm-fig-sc-self-consistency}{Fig. J1} shows the \(d\)-wave order parameter \(\Delta_d(T)\) and chemical potential \(\mu(T)\) obtained by numerically solving the self-consistent equations.

\par\medskip
\hypertarget{sm-fig-sc-self-consistency}{}
\PrintAsset{self-consistency}
{\small\noindent \textbf{Figure J1 \textbar{} \(d\)-wave solution of the self-consistent equations at fixed electron density.} \textbf{a,} SC order parameter \(\Delta_d(T)\). \textbf{b,} Chemical potential \(\mu(T)\) relative to \(\mu_{\mathrm N}(0)=-0.268236~\mathrm{eV}\), the zero-temperature chemical potential of the nodal Bose metal. Orange circles show the \(d\)-wave solution, with \(\Delta_d\ne0\) below \(T_{c,d}\), and blue-gray circles show the \(\Delta_d=0\) solution corresponding to the nodal Bose metal. The light-orange solid curves and blue-gray dashed curve are guides to the eye. The vertical dashed lines mark \(T/T_{c,d}=1\). \(\Delta_d(0)=7.2891~\mathrm{meV}\), \(T_{c,d}=143.13~\mathrm K\), and the order-parameter-to-\(T_c\) ratio is \(2\sqrt2\Delta_d(0)/T_{c,d}=1.6715\). Parameters are \(\mathcal J_d=0.06~\mathrm{eV}\), \(J=0.12~\mathrm{eV}\), and \(p=0.15\).\par}
\medskip

\hypertarget{sm-sc-weak-gap-ratio}{}
\subsection*{J.2. Order-parameter-to-\(T_c\) ratio}

The gap equation is used to estimate the order-parameter-to-\(T_c\) ratio \(2\sqrt2\Delta_d(0)/T_{c,d}\) in the weak-attraction limit, where \(\sqrt2\) originates from the normalization factor of the pair operator \(\hat\Delta_{\mathbf k}\). Near the Cooper surfaces away from nodes, with \(T,|\Delta_{\mathbf k,d}|,|S_{\mathbf k,r}|\ll\mathcal J_{\mathbf k}\), projecting \(\hat h_{\mathbf k}^{\mathrm{SC}}\) onto the two-state subspace in the two-state approximation gives

\hypertarget{sm-eq-sc-mft-two-state}{}%
\FittedEquation{J22}{%
\hat h^{\mathrm{SC},(2)}_{\mathbf k,r}=\overline E^{\mathrm N}_{\mathbf k,r}\mathbf I+
        \begin{pmatrix}
        -S_{\mathbf k,r}&-2\Delta_{\mathbf k,d}^{*}\\
        -2\Delta_{\mathbf k,d}&S_{\mathbf k,r}
        \end{pmatrix},
}
 
where \(\overline E^{\mathrm N}_{\mathbf k,2e}=(E^{\mathrm N}_{\mathbf k,s}+E^{\mathrm N}_{\mathbf k,4})/2=3\xi_{\mathbf k}-2\mathcal J_{\mathbf k}\) and \(\overline E^{\mathrm N}_{\mathbf k,2h}=(E^{\mathrm N}_{\mathbf k,0}+E^{\mathrm N}_{\mathbf k,s})/2=\xi_{\mathbf k}-\mathcal J_{\mathbf k}\), and recall that \(S_{\mathbf k,2e}=\xi_{\mathbf k}\), \(S_{\mathbf k,2h}=\xi_{\mathbf k}-\mathcal J_{\mathbf k}\), and \(\Delta_{\mathbf k,d}=\Delta_d\phi_{\mathbf k,d}\). The two eigenenergies of \(\hat h^{\mathrm{SC},(2)}_{\mathbf k,r}\) are \(\overline E^{\mathrm N}_{\mathbf k,r}\pm E^{\mathrm{SC},(2)}_{\mathbf k,r}\), where \(E^{\mathrm{SC},(2)}_{\mathbf k,r}=\sqrt{S_{\mathbf k,r}^2+4|\Delta_{\mathbf k,d}|^2}\). The SC block partition function in the two-state approximation is \(Z^{\mathrm{SC},(2)}_{\mathbf k,r}=2e^{-\beta\overline E^{\mathrm N}_{\mathbf k,r}}\cosh(\beta E^{\mathrm{SC},(2)}_{\mathbf k,r})\), and the expectation value is approximately

\hypertarget{sm-eq-sc-mft-two-state-pair-average}{}%
\FittedEquation{J23}{%
\langle\hat\Delta_{\mathbf k}\rangle_{\mathrm{SC}}=\frac1{2\beta}\frac{\partial\ln Z^{\mathrm{SC}}_{\mathbf k}}{\partial\Delta_{\mathbf k,d}^{*}}\simeq\Delta_{\mathbf k,d}\frac{\tanh(\beta E^{\mathrm{SC},(2)}_{\mathbf k,r})}{E^{\mathrm{SC},(2)}_{\mathbf k,r}}.
}
 
Dividing the gap equation in \hyperlink{sm-eq-sc-appendix-gap-equation}{Eq. (J18)} by \(2J\Delta_d\) gives

\hypertarget{sm-eq-sc-mft-pair-kernel}{}%
\FittedEquation{J24}{%
\begin{aligned}
\frac1{2J}
&=\frac{2}{N_s\Delta_d}\sum_{\{\mathbf k,-\mathbf k\}}
\phi_{\mathbf k,d}\langle\hat\Delta_{\mathbf k}\rangle_{\mathrm{SC}}\\
&\simeq\sum_r\int_{\mathbf k\,\text{near}\,\mathcal C_r}
\frac{d^2k}{(2\pi)^2}\phi_{\mathbf k,d}^2
\frac{\tanh(\beta E^{\mathrm{SC},(2)}_{\mathbf k,r})}
{E^{\mathrm{SC},(2)}_{\mathbf k,r}}+O(1).
\end{aligned}
}
 
The momentum integrals cover tubular neighborhoods of the two Cooper surfaces, and \(O(1)\) denotes nonsingular contributions. To logarithmic accuracy, the form factor \(\phi_{\mathbf k,d}^2\) can be evaluated on the Cooper surfaces. Following the integration in \hyperlink{sm-appendix-h-nodal-pair-susceptibility}{Appendix H.2}, the normal integrals at zero temperature and at the transition temperature are

\hypertarget{sm-eq-sc-mft-log-matching}{}%
\FittedEquation{J25}{%
\int_{-\Lambda_{\mathrm{cut}}}^{\Lambda_{\mathrm{cut}}}\!
        \frac{dS}{\sqrt{S^2+4|\Delta_d(0)|^2\phi_{\mathbf k,d}^2}}
        =2\ln\frac{\Lambda_{\mathrm{cut}}}
        {|\Delta_d(0)||\phi_{\mathbf k,d}|}+o(1),
}
 
\hypertarget{sm-eq-sc-mft-log-transition}{}%
\FittedEquation{J26}{%
\int_{-\Lambda_{\mathrm{cut}}}^{\Lambda_{\mathrm{cut}}}\!
        \frac{dS}{S}\tanh\frac{S}{T_{c,d}}
        =2\ln\frac{4e^{\gamma_{\mathrm E}}\Lambda_{\mathrm{cut}}}
        {\pi T_{c,d}}+o(1).
}
 
Performing the line integrals with weight \(\phi_{\mathbf k,d}^2/|\nabla_{\mathbf k}S_{\mathbf k,r}|\) along the Cooper surfaces gives the zero-temperature order parameter and SC transition temperature (cf. the SC instability temperature in \hyperlink{sm-eq-h2-leading-transition}{Eq. (H42)}) in the weak-attraction limit

\hypertarget{sm-eq-sc-mft-weak-scales}{}%
\FittedEquation{J27}{%
\Delta_d(0)\simeq\Lambda_{\mathrm{cut}}\exp\!\left[-\frac{1}{4JD_d(\mu)}-L_d\right],
}
 
\hypertarget{sm-eq-sc-mft-weak-temperature}{}%
\FittedEquation{J28}{%
T_{c,d}\simeq\frac{4e^{\gamma_{\mathrm E}}}{\pi}\Lambda_{\mathrm{cut}}\exp\!\left[-\frac{1}{4JD_d(\mu)}\right].
}
 
To leading logarithmic accuracy, the order-parameter-to-\(T_c\) ratio is

\hypertarget{sm-eq-sc-mft-weak-ratio}{}%
\FittedEquation{J29}{%
\frac{2\sqrt2\Delta_d(0)}{T_{c,d}}
        \simeq\frac{\pi}{\sqrt2e^{\gamma_{\mathrm E}}}e^{-L_d}\simeq1.24725e^{-L_d},
}
 
where

\hypertarget{sm-eq-sc-mft-log-average}{}%
\FittedEquation{J30}{%
\begin{aligned}
        L_d&=\frac{\displaystyle\sum_r\oint_{\mathcal C_r}\frac{d\ell}{(2\pi)^2}\frac{\phi_{\mathbf k,d}^2\ln|\phi_{\mathbf k,d}|}{|\nabla_{\mathbf k}S_{\mathbf k,r}|}}{\displaystyle\sum_r\oint_{\mathcal C_r}\frac{d\ell}{(2\pi)^2}\frac{\phi_{\mathbf k,d}^2}{|\nabla_{\mathbf k}S_{\mathbf k,r}|}}\\[6pt]
        &=\frac{\displaystyle\int_{\mathrm{BZ}}\frac{d^2\mathbf k}{(2\pi)^2}\phi_{\mathbf k,d}^2\ln|\phi_{\mathbf k,d}|\left[\delta(\xi_{\mathbf k})+\delta(\xi_{\mathbf k}-\mathcal J_{\mathbf k})\right]}{D_d(\mu)}.
        \end{aligned}
}
 
The integrand \(\phi_{\mathbf k,d}^2\ln|\phi_{\mathbf k,d}|\) tends to zero at a node. Thus the order-parameter-to-\(T_c\) ratio mainly depends on the geometry and form-factor weights of the two Cooper surfaces. \hyperlink{sm-fig-sc-order-parameter-ratio}{Fig. J2} compares the numerical order-parameter-to-\(T_c\) ratio with the analytical result in \hyperlink{sm-eq-sc-mft-weak-ratio}{Eq. (J29)} as \(J\) and \(\mathcal J_d\) are varied.

\par\medskip
\hypertarget{sm-fig-sc-order-parameter-ratio}{}
\PrintAsset{order-parameter-ratio}
{\small\noindent \textbf{Figure J2 \textbar{} Order-parameter-to-\(T_c\) ratio in the correlated SC state.} \textbf{a,} Dependence on the pairing attraction strength \(J\) at fixed \(\mathcal J_d=0.06~\mathrm{eV}\). \textbf{b,} Dependence on the RML interaction strength \(\mathcal J_d\) at fixed \(J=0.06~\mathrm{eV}\). Light-orange circles show \(2\sqrt2\Delta_d(0)/T_{c,d}\) obtained numerically from the self-consistent equations under the fixed-density constraint; the solid curves are guides to the eye. The black dashed curves show the weak-attraction analytical result in \hyperlink{sm-eq-sc-mft-weak-ratio}{Eq. (J29)}. The electron density is \(p=0.15\).\par}
\medskip
 
\hypertarget{sm-appendix-sc-excitation-gap}{}\hypertarget{sm-appendix-sc-excitation-gap-title}{}
\section*{Appendix K. One-particle Green's function of the correlated SC state}

The zero-temperature one-particle retarded Green's function of the correlated SC state is

\hypertarget{sm-eq-gap-retarded-time}{}%
\FittedEquation{K1}{%
\begin{aligned}
G^R_{\mathrm{SC},\sigma}(\mathbf k,t)
&=-i\Theta(t)\langle\mathrm{GS}_{\mathrm{SC}}|
\{\hat c_{\mathbf k\sigma}(t),\hat c_{\mathbf k\sigma}^{\dagger}(0)\}
|\mathrm{GS}_{\mathrm{SC}}\rangle\\
&=-i\Theta(t)\langle e_{\mathbf k,0}|
\{\hat c_{\mathbf k\sigma}(t),\hat c_{\mathbf k\sigma}^{\dagger}(0)\}
|e_{\mathbf k,0}\rangle,
\end{aligned}
}
 
where the product structure of the ground state \(|\mathrm{GS}_{\mathrm{SC}}\rangle=\prod_{\{\mathbf k,-\mathbf k\}}|e_{\mathbf k,0}\rangle\) reduces the expectation value to a single momentum block. The time evolution \(\hat c_{\mathbf k\sigma}(t)=e^{i\hat h_{\mathbf k}^{\mathrm{SC}}t}\hat c_{\mathbf k\sigma}e^{-i\hat h_{\mathbf k}^{\mathrm{SC}}t}\) and the relations

\hypertarget{sm-eq-gap-electron-actions}{}%
\FittedEquation{K2}{%
\begin{aligned}
\hat c_{\mathbf k\sigma}^{\dagger}|e_{\mathbf k,0}\rangle
&=x_{\mathbf k,0}|1_{\mathbf k\sigma}\rangle+\frac{y_{\mathbf k,0}}{\sqrt2}|3_{-\mathbf k\sigma}\rangle,
\end{aligned}
}
 
\hypertarget{sm-eq-gap-electron-removal-action}{}%
\FittedEquation{K3}{%
\begin{aligned}
\hat c_{\mathbf k\sigma}|e_{\mathbf k,0}\rangle
&=-\sigma\left[\frac{y_{\mathbf k,0}}{\sqrt2}|1_{-\mathbf k\bar\sigma}\rangle+z_{\mathbf k,0}|3_{\mathbf k\bar\sigma}\rangle\right],
\end{aligned}
}
 
yield

\hypertarget{sm-eq-gap-time-correlators}{}%
\FittedEquation{K4}{%
\begin{aligned}
\langle e_{\mathbf k,0}|\hat c_{\mathbf k\sigma}(t)\hat c_{\mathbf k\sigma}^{\dagger}|e_{\mathbf k,0}\rangle
&=\sum_{\nu=\pm1}Z^{\mathrm{SC},+}_{\mathbf k,\nu 0}e^{-i(E^{\mathrm{SC},o}_{\mathbf k,\nu}-E^{\mathrm{SC},e}_{\mathbf k,0})t},
\end{aligned}
}
 
\hypertarget{sm-eq-gap-removal-time-correlator}{}%
\FittedEquation{K5}{%
\begin{aligned}
\langle e_{\mathbf k,0}|\hat c_{\mathbf k\sigma}^{\dagger}\hat c_{\mathbf k\sigma}(t)|e_{\mathbf k,0}\rangle
&=\sum_{\nu=\pm1}Z^{\mathrm{SC},-}_{\mathbf k,\nu 0}e^{+i(E^{\mathrm{SC},o}_{\mathbf k,\nu}-E^{\mathrm{SC},e}_{\mathbf k,0})t},
\end{aligned}
}
 
and

\hypertarget{sm-eq-gap-transition-weights}{}%
\FittedEquation{K6}{%
\begin{aligned}
Z^{\mathrm{SC},+}_{\mathbf k,\nu 0}
&=|\langle o_{\mathbf k\sigma,\nu}|\hat c_{\mathbf k\sigma}^{\dagger}|e_{\mathbf k,0}\rangle|^2
=\left|u_{\mathbf k,\nu}^{*}x_{\mathbf k,0}+\frac{v_{\mathbf k,\nu}^{*}y_{\mathbf k,0}}{\sqrt2}\right|^2,
\end{aligned}
}
 
\hypertarget{sm-eq-gap-removal-transition-weight}{}%
\FittedEquation{K7}{%
\begin{aligned}
Z^{\mathrm{SC},-}_{\mathbf k,\nu 0}
&=|\langle o_{-\mathbf k\bar\sigma,\nu}|\hat c_{\mathbf k\sigma}|e_{\mathbf k,0}\rangle|^2
=\left|\frac{u_{\mathbf k,\nu}^{*}y_{\mathbf k,0}}{\sqrt2}+v_{\mathbf k,\nu}^{*}z_{\mathbf k,0}\right|^2.
\end{aligned}
}
 
Fourier transformation uses the retarded convergence factor,

\hypertarget{sm-eq-gap-retarded-transform}{}%
\FittedEquation{K8}{%
-i\lim_{\eta\to0^+}\int_0^{\infty}dt\,e^{i(\omega\mp E^{\mathrm{SC},o}_{\mathbf k,\nu}\pm E^{\mathrm{SC},e}_{\mathbf k,0})t-\eta t}
=\frac1{\omega+i0^+\mp E^{\mathrm{SC},o}_{\mathbf k,\nu}\pm E^{\mathrm{SC},e}_{\mathbf k,0}},
}
 
and therefore gives \hyperlink{sm-eq-sc-zero-temperature-green}{Eq. (67)}:

\hypertarget{sm-eq-gap-four-pole-green}{}%
\FittedEquation{K9}{%
G^R_{\mathrm{SC},\sigma}(\mathbf k,\omega)
=\sum_{\nu=\pm1}\left[
\frac{Z^{\mathrm{SC},+}_{\mathbf k,\nu 0}}{\omega+i0^+-E^{\mathrm{SC},o}_{\mathbf k,\nu}+E^{\mathrm{SC},e}_{\mathbf k,0}}
+\frac{Z^{\mathrm{SC},-}_{\mathbf k,\nu 0}}{\omega+i0^++E^{\mathrm{SC},o}_{\mathbf k,\nu}-E^{\mathrm{SC},e}_{\mathbf k,0}}\right].
}
 
The completeness relation \(\sum_{\nu=\pm1}|o_{\pm\mathbf k\sigma,\nu}\rangle\langle o_{\pm\mathbf k\sigma,\nu}|=\mathbf I_2\) implies

\hypertarget{sm-eq-gap-odd-completeness}{}\hypertarget{sm-eq-gap-odd-orthogonality}{}%
\FittedEquation{K10}{%
\sum_{\nu}|u_{\mathbf k,\nu}|^2=\sum_{\nu}|v_{\mathbf k,\nu}|^2=1,\qquad \sum_{\nu}u_{\mathbf k,\nu}v_{\mathbf k,\nu}^{*}=0.
}
 
Expanding the squared moduli in \hyperlink{sm-eq-gap-transition-weights}{Eqs. (K6)--(K7)} and summing over \(\nu\) gives

\hypertarget{sm-eq-gap-weight-normalization}{}%
\FittedEquation{K11}{%
\begin{aligned}
\sum_{\nu}Z^{\mathrm{SC},+}_{\mathbf k,\nu 0}
&=|x_{\mathbf k,0}|^2+\frac{|y_{\mathbf k,0}|^2}{2}
=1-\langle\hat n_{\mathbf k\sigma}\rangle_{\mathrm{SC}},
\end{aligned}
}
 
\hypertarget{sm-eq-gap-removal-weight-normalization}{}%
\FittedEquation{K12}{%
\begin{aligned}
\sum_{\nu}Z^{\mathrm{SC},-}_{\mathbf k,\nu 0}
&=\frac{|y_{\mathbf k,0}|^2}{2}+|z_{\mathbf k,0}|^2
=\langle\hat n_{\mathbf k\sigma}\rangle_{\mathrm{SC}},
\end{aligned}
}
 
and the sum rule \(\sum_{\nu}(Z^{\mathrm{SC},+}_{\mathbf k,\nu 0}+Z^{\mathrm{SC},-}_{\mathbf k,\nu 0})=1\).

\hypertarget{sm-sc-gap-explicit-resolvent}{}
\subsection*{K.1. Explicit form of the Green's function}

Write the one-electron-addition and one-electron-removal states in \hyperlink{sm-eq-gap-electron-actions}{Eqs. (K2)--(K3)} as two-component columns in their respective odd-fermion-parity subspaces

\hypertarget{sm-eq-gap-source-vectors}{}\hypertarget{sm-eq-gap-removal-source-vector}{}%
\FittedEquation{K13}{%
\boldsymbol a_{\mathbf k}^{+}=\begin{pmatrix}x_{\mathbf k,0}\\y_{\mathbf k,0}/\sqrt2\end{pmatrix},\qquad \boldsymbol a_{\mathbf k}^{-}=\begin{pmatrix}y_{\mathbf k,0}/\sqrt2\\z_{\mathbf k,0}\end{pmatrix},
}
 
and recall

\hypertarget{sm-eq-gap-odd-matrix}{}%
\FittedEquation{K14}{%
h^{\mathrm{SC},o}_{\mathbf k}=\begin{pmatrix}
\xi_{\mathbf k}&-\sqrt2\Delta_{\mathbf k,d}^{*}\\
-\sqrt2\Delta_{\mathbf k,d}&3\xi_{\mathbf k}-\mathcal J_{\mathbf k}
\end{pmatrix}.
}
 
The Green's function in \hyperlink{sm-eq-gap-four-pole-green}{Eq. (K9)} can be rewritten as

\hypertarget{sm-eq-gap-resolvent-green}{}%
\FittedEquation{K15}{%
\begin{aligned}
G^R_{\mathrm{SC},\sigma}(\mathbf k,\omega)
&=(\boldsymbol a_{\mathbf k}^{+})^{\dagger}
\bigl[(\omega+i0^++E^{\mathrm{SC},e}_{\mathbf k,0})\mathbf I_2-h^{\mathrm{SC},o}_{\mathbf k}\bigr]^{-1}\boldsymbol a_{\mathbf k}^{+}\\
&\quad+(\boldsymbol a_{\mathbf k}^{-})^{\dagger}
\bigl[(\omega+i0^+-E^{\mathrm{SC},e}_{\mathbf k,0})\mathbf I_2+h^{\mathrm{SC},o}_{\mathbf k}\bigr]^{-1}\boldsymbol a_{\mathbf k}^{-}.
\end{aligned}
}
 
Inverting the two-by-two matrices

\hypertarget{sm-eq-gap-explicit-inverses}{}%
\FittedEquation{K16}{%
\begin{aligned}
\bigl[(\omega+i0^++E^{\mathrm{SC},e}_{\mathbf k,0})\mathbf I_2-h^{\mathrm{SC},o}_{\mathbf k}\bigr]^{-1}
&=\frac1{\prod_{\nu=\pm1}(\omega+i0^+-E^{\mathrm{SC},o}_{\mathbf k,\nu}+E^{\mathrm{SC},e}_{\mathbf k,0})}\\[-3pt]
&\hspace{-31mm}\times\begin{pmatrix}
\omega+i0^++E^{\mathrm{SC},e}_{\mathbf k,0}-3\xi_{\mathbf k}+\mathcal J_{\mathbf k}&-\sqrt2\Delta_{\mathbf k,d}^{*}\\
-\sqrt2\Delta_{\mathbf k,d}&\omega+i0^++E^{\mathrm{SC},e}_{\mathbf k,0}-\xi_{\mathbf k}
\end{pmatrix},
\end{aligned}
}
 
\hypertarget{sm-eq-gap-removal-explicit-inverse}{}%
\FittedEquation{K17}{%
\begin{aligned}
\bigl[(\omega+i0^+-E^{\mathrm{SC},e}_{\mathbf k,0})\mathbf I_2+h^{\mathrm{SC},o}_{\mathbf k}\bigr]^{-1}
&=\frac1{\prod_{\nu=\pm1}(\omega+i0^++E^{\mathrm{SC},o}_{\mathbf k,\nu}-E^{\mathrm{SC},e}_{\mathbf k,0})}\\[-3pt]
&\hspace{-31mm}\times\begin{pmatrix}
\omega+i0^+-E^{\mathrm{SC},e}_{\mathbf k,0}+3\xi_{\mathbf k}-\mathcal J_{\mathbf k}&\sqrt2\Delta_{\mathbf k,d}^{*}\\
\sqrt2\Delta_{\mathbf k,d}&\omega+i0^+-E^{\mathrm{SC},e}_{\mathbf k,0}+\xi_{\mathbf k}
\end{pmatrix},
\end{aligned}
}
 
yields \hyperlink{sm-eq-sc-zero-temperature-green-rational}{Eq. (71)},

\hypertarget{sm-eq-gap-rational-green}{}%
\FittedEquation{K18}{%
\begin{aligned}
G^R_{\mathrm{SC},\sigma}(\mathbf k,\omega)
&=\frac{N_{\mathbf k}^{+}(\omega+i0^+)}{\prod_{\nu=\pm1}(\omega+i0^+-E^{\mathrm{SC},o}_{\mathbf k,\nu}+E^{\mathrm{SC},e}_{\mathbf k,0})}\\
&\quad+\frac{N_{\mathbf k}^{-}(\omega+i0^+)}{\prod_{\nu=\pm1}(\omega+i0^++E^{\mathrm{SC},o}_{\mathbf k,\nu}-E^{\mathrm{SC},e}_{\mathbf k,0})},
\end{aligned}
}
 
and \hyperlink{sm-eq-sc-zero-temperature-green-numerators}{Eqs. (72)--(73)},

\hypertarget{sm-eq-gap-rational-numerators}{}%
\FittedEquation{K19}{%
\begin{aligned}
N^+_{\mathbf k}(\omega+i0^+)
&=|x_{\mathbf k,0}|^2(\omega+i0^++E^{\mathrm{SC},e}_{\mathbf k,0}-3\xi_{\mathbf k}+\mathcal J_{\mathbf k})\\
&\quad+\frac{|y_{\mathbf k,0}|^2}{2}(\omega+i0^++E^{\mathrm{SC},e}_{\mathbf k,0}-\xi_{\mathbf k})
-2\operatorname{Re}(\Delta_{\mathbf k,d}^{*}x_{\mathbf k,0}^{*}y_{\mathbf k,0}),
\end{aligned}
}
 
\hypertarget{sm-eq-gap-removal-rational-numerator}{}%
\FittedEquation{K20}{%
\begin{aligned}
N^-_{\mathbf k}(\omega+i0^+)
&=\frac{|y_{\mathbf k,0}|^2}{2}(\omega+i0^+-E^{\mathrm{SC},e}_{\mathbf k,0}+3\xi_{\mathbf k}-\mathcal J_{\mathbf k})\\
&\quad+|z_{\mathbf k,0}|^2(\omega+i0^+-E^{\mathrm{SC},e}_{\mathbf k,0}+\xi_{\mathbf k})
+2\operatorname{Re}(\Delta_{\mathbf k,d}^{*}y_{\mathbf k,0}^{*}z_{\mathbf k,0}),
\end{aligned}
}
 
where the real cross terms arise from the off-diagonal matrix elements in \hyperlink{sm-eq-gap-odd-matrix}{Eq. (K14)}.

\hypertarget{sm-sc-gap-antinodal-expansion}{}
\subsection*{K.2. Antinodal expansion of the one-particle excitation gap}

Along the Fermi surface \(\xi_{\mathbf k_F}=0\) near the antinodes, where \(\mathcal J_{\mathbf k_F}\gg|\Delta_{\mathbf k_F,d}|\), the matrix \(h^{\mathrm{SC},e}_{\mathbf k}\) in \hyperlink{sm-eq-sc-even-matrix}{Eq. (J3)}, restricted to the low-energy even-fermion-parity subspace spanned by \(\{|0_{\mathbf k}\rangle,|2_{\mathbf k,s}\rangle,|4_{\mathbf k}\rangle\}\), reduces to

\hypertarget{sm-eq-gap-antinodal-even-matrix}{}%
\FittedEquation{K21}{%
h^{\mathrm{SC},e}_{\mathbf k_F}=
\begin{pmatrix}
0&-2\Delta_{\mathbf k_F,d}^{*}&0\\
-2\Delta_{\mathbf k_F,d}&-2\mathcal J_{\mathbf k_F}&-2\Delta_{\mathbf k_F,d}^{*}\\
0&-2\Delta_{\mathbf k_F,d}&-2\mathcal J_{\mathbf k_F}
\end{pmatrix}.
}
 
At zero pairing field, \(|2_{\mathbf k_F,s}\rangle\) and \(|4_{\mathbf k_F}\rangle\) are degenerate ground states with energy \(-2\mathcal J_{\mathbf k_F}\), while \(|0_{\mathbf k_F}\rangle\) has energy zero. The leading correction can be obtained by diagonalizing the perturbation within the two-dimensional ground-state subspace, in which \(h^{\mathrm{SC},e}_{\mathbf k_F}\) further reduces to

\hypertarget{sm-eq-gap-antinodal-degenerate-matrix}{}%
\FittedEquation{K22}{%
\begin{pmatrix}
-2\mathcal J_{\mathbf k_F}&-2\Delta_{\mathbf k_F,d}^{*}\\
-2\Delta_{\mathbf k_F,d}&-2\mathcal J_{\mathbf k_F}
\end{pmatrix},
}
 
with eigenvalues \(-2\mathcal J_{\mathbf k_F}\pm2|\Delta_{\mathbf k_F,d}|\). The leading perturbed ground state is 
\hypertarget{sm-eq-gap-antinodal-even-leading}{}%
\FittedEquation{K23}{%
|e_{\mathbf k_F,0}^{(0)}\rangle
=\frac1{\sqrt2}\left[|2_{\mathbf k_F,s}\rangle
+\frac{\Delta_{\mathbf k_F,d}}{|\Delta_{\mathbf k_F,d}|}|4_{\mathbf k_F}\rangle\right],
}
 and the first-order energy correction caused by the SC pairing field is \(\delta E^{\mathrm{SC},e,(1)}_{\mathbf k_F,0}=-2|\Delta_{\mathbf k_F,d}|\). \hypertarget{sm-eq-gap-antinodal-empty-coupling}{}Since \(\langle0_{\mathbf k_F}|h^{\mathrm{SC},e}_{\mathbf k_F}|e_{\mathbf k_F,0}^{(0)}\rangle=-\sqrt2\Delta_{\mathbf k_F,d}^{*}\), the second-order energy correction is 
\hypertarget{sm-eq-gap-antinodal-even-second-correction}{}%
\FittedEquation{K24}{%
\delta E^{\mathrm{SC},e,(2)}_{\mathbf k_F,0}
=\frac{|\langle0_{\mathbf k_F}|h^{\mathrm{SC},e}_{\mathbf k_F}|e_{\mathbf k_F,0}^{(0)}\rangle|^2}
{(-2\mathcal J_{\mathbf k_F})-0}
=-\frac{|\Delta_{\mathbf k_F,d}|^2}{\mathcal J_{\mathbf k_F}}.
}
 Combining the first- and second-order energy corrections gives 
\hypertarget{sm-eq-gap-antinodal-even-energy}{}%
\FittedEquation{K25}{%
E^{\mathrm{SC},e}_{\mathbf k_F,0}
=-2\mathcal J_{\mathbf k_F}-2|\Delta_{\mathbf k_F,d}|
-\frac{|\Delta_{\mathbf k_F,d}|^2}{\mathcal J_{\mathbf k_F}}
+O\!\left(\frac{|\Delta_{\mathbf k_F,d}|^3}{\mathcal J_{\mathbf k_F}^{2}}\right).
}

The matrix \(h^{\mathrm{SC},o}_{\mathbf k}\) in \hyperlink{sm-eq-sc-odd-matrix}{Eq. (J11)}, restricted to the odd-fermion-parity subspace spanned by \(\{|1_{\pm\mathbf k\sigma}\rangle,|3_{\mp\mathbf k\sigma}\rangle\}\), becomes

\hypertarget{sm-eq-gap-antinodal-odd-matrix}{}%
\FittedEquation{K26}{%
h^{\mathrm{SC},o}_{\mathbf k_F}=
\begin{pmatrix}
0&-\sqrt2\Delta_{\mathbf k_F,d}^{*}\\
-\sqrt2\Delta_{\mathbf k_F,d}&-\mathcal J_{\mathbf k_F}
\end{pmatrix},
}
 
with eigenvalues \(E^{\mathrm{SC},o}_{\mathbf k_F,\pm}=-\mathcal J_{\mathbf k_F}/2\pm\sqrt{\mathcal J_{\mathbf k_F}^{2}/4+2|\Delta_{\mathbf k_F,d}|^2}\). Expanding the lower eigenenergy gives

\hypertarget{sm-eq-gap-antinodal-odd-energy}{}%
\FittedEquation{K27}{%
\begin{aligned}
E^{\mathrm{SC},o}_{\mathbf k_F,-}
&=-\frac{\mathcal J_{\mathbf k_F}}2
-\sqrt{\frac{\mathcal J_{\mathbf k_F}^{2}}4+2|\Delta_{\mathbf k_F,d}|^2}\\
&=-\mathcal J_{\mathbf k_F}
-\frac{2|\Delta_{\mathbf k_F,d}|^2}{\mathcal J_{\mathbf k_F}}
+O\!\left(\frac{|\Delta_{\mathbf k_F,d}|^4}{\mathcal J_{\mathbf k_F}^{3}}\right),
\end{aligned}
}
 
where the first-order energy correction to each of the two nondegenerate states \(|1_{\pm\mathbf k_F\sigma}\rangle\) and \(|3_{\mp\mathbf k_F\sigma}\rangle\) vanishes because the pairing perturbation is off-diagonal. Subtracting \hyperlink{sm-eq-gap-antinodal-even-energy}{Eq. (K25)} from \hyperlink{sm-eq-gap-antinodal-odd-energy}{Eq. (K27)} yields the antinodal expansion of the one-particle excitation gap in \hyperlink{sm-eq-sc-antinodal-gap-expansion}{Eq. (75)}

\hypertarget{sm-eq-gap-antinodal-gap-series}{}%
\FittedEquation{K28}{%
\begin{aligned}
\Delta_{\mathrm{1p}}^{\mathrm{SC}}(\mathbf k_F)
&=E^{\mathrm{SC},o}_{\mathbf k_F,-}-E^{\mathrm{SC},e}_{\mathbf k_F,0}\\
&=\mathcal J_{\mathbf k_F}+2|\Delta_{\mathbf k_F,d}|
-\frac{|\Delta_{\mathbf k_F,d}|^2}{\mathcal J_{\mathbf k_F}}
+O\!\left(\frac{|\Delta_{\mathbf k_F,d}|^3}{\mathcal J_{\mathbf k_F}^{2}}\right)\\
&=\mathcal J_{\mathbf k_F}+2|\Delta_{\mathbf k_F,d}|
+O\!\left(\frac{|\Delta_{\mathbf k_F,d}|^2}{\mathcal J_{\mathbf k_F}}\right).
\end{aligned}
}
 
\hypertarget{sm-appendix-sc-one-particle}{}\hypertarget{sm-appendix-sc-one-particle-title}{}
\section*{Appendix L. ARPES of the correlated SC state}

\hypertarget{sm-sc-spectral-lehmann}{}
\subsection*{L.1. One-electron ARPES}

The occupied one-particle spectrum of the correlated SC state measured by one-electron ARPES is

\hypertarget{sm-eq-sc-arpes-occupied-definition}{}%
\FittedEquation{L1}{%
f_{\mathrm F}(\omega)A_{\mathrm{SC},\sigma}(\mathbf k,\omega)
=\lim_{\mathcal T\to\infty}^{\mathcal D'}\frac{1}{2\pi i\mathcal T}
\iint_{-\frac{\mathcal T}{2}}^{\frac{\mathcal T}{2}}dt\,dt'\,
e^{i\omega(t-t')}G^<_{\mathrm{SC},\sigma}(\mathbf k;t,t'),
}
 
where \(A_{\mathrm{SC},\sigma}(\mathbf k,\omega)=-\pi^{-1}\operatorname{Im}G^R_{\mathrm{SC},\sigma}(\mathbf k,\omega)\) is the SC one-particle spectral function. The one-particle lesser Green's function of the correlated SC state with spin rotation symmetry is

\hypertarget{sm-eq-sc-lesser-lehmann}{}%
\FittedEquation{L2}{%
\begin{aligned}
G^<_{\mathrm{SC}}(\mathbf k\sigma,\mathbf k'\sigma';t,t')
&\equiv i\langle\hat c_{\mathbf k'\sigma'}^{\dagger}(t')\hat c_{\mathbf k\sigma}(t)\rangle_{\mathrm{SC}}
=\delta_{\mathbf k,\mathbf k'}\delta_{\sigma,\sigma'}G^<_{\mathrm{SC},\sigma}(\mathbf k;t,t'),
\end{aligned}
}
 
\hypertarget{sm-eq-sc-lesser-block-lehmann}{}%
\FittedEquation{L3}{%
\begin{aligned}
G^<_{\mathrm{SC},\sigma}(\mathbf k;t,t')
&=i\sum_{\lambda,\lambda'}^{\pi_{\mathbf k,\lambda'}=-\pi_{\mathbf k,\lambda}}
p^{\mathrm{SC}}_{\mathbf k,\lambda}
|\langle\pi_{\mathbf k,\lambda'}|\hat c_{\mathbf k\sigma}|\pi_{\mathbf k,\lambda}\rangle|^2
e^{-i(E^{\mathrm{SC}}_{\mathbf k,\lambda}-E^{\mathrm{SC}}_{\mathbf k,\lambda'})(t-t')},
\end{aligned}
}
 
where \(|\pi_{\mathbf k,\lambda}\rangle\) denotes a SC block eigenstate with block fermion parity \(\pi_{\mathbf k,\lambda}=\pm1\) and eigenenergy \(E^{\mathrm{SC}}_{\mathbf k,\lambda}\), respectively, and \(p^{\mathrm{SC}}_{\mathbf k,\lambda}=e^{-\beta E^{\mathrm{SC}}_{\mathbf k,\lambda}}/Z^{\mathrm{SC}}_{\mathbf k}\) is the corresponding thermal occupation probability. The long-duration limit in \hyperlink{sm-eq-sc-arpes-occupied-definition}{Eq. (L1)} is equivalent to a Fourier transformation

\hypertarget{sm-eq-sc-occupied-spectrum}{}%
\FittedEquation{L4}{%
\begin{aligned}
&f_{\mathrm F}(\omega)A_{\mathrm{SC},\sigma}(\mathbf k,\omega)
=\frac1{2\pi i}\int_{-\infty}^{\infty}dt_r\,e^{i\omega t_r}G^<_{\mathrm{SC},\sigma}(\mathbf k;t_r,0)\\
&=\sum_{\lambda,\lambda'}^{\pi_{\mathbf k,\lambda'}=-\pi_{\mathbf k,\lambda}}
f_{\mathrm F}(E^{\mathrm{SC}}_{\mathbf k,\lambda}-E^{\mathrm{SC}}_{\mathbf k,\lambda'})
(p^{\mathrm{SC}}_{\mathbf k,\lambda}+p^{\mathrm{SC}}_{\mathbf k,\lambda'})
|\langle\pi_{\mathbf k,\lambda'}|\hat c_{\mathbf k\sigma}|\pi_{\mathbf k,\lambda}\rangle|^2
\delta\!\left(\omega-E^{\mathrm{SC}}_{\mathbf k,\lambda}+E^{\mathrm{SC}}_{\mathbf k,\lambda'}\right)\\
&=f_{\mathrm F}(\omega)\sum_{\lambda,\lambda'}^{\pi_{\mathbf k,\lambda'}=-\pi_{\mathbf k,\lambda}}
(p^{\mathrm{SC}}_{\mathbf k,\lambda}+p^{\mathrm{SC}}_{\mathbf k,\lambda'})
|\langle\pi_{\mathbf k,\lambda'}|\hat c_{\mathbf k\sigma}|\pi_{\mathbf k,\lambda}\rangle|^2
\delta\!\left(\omega-E^{\mathrm{SC}}_{\mathbf k,\lambda}+E^{\mathrm{SC}}_{\mathbf k,\lambda'}\right).
\end{aligned}
}
 
which yields the SC one-particle spectral function in \hyperlink{sm-eq-sc-spectral-one}{Eq. (76)},

\hypertarget{sm-eq-sc-spectral-lehmann}{}%
\FittedEquation{L5}{%
\begin{aligned}
&A_{\mathrm{SC},\sigma}(\mathbf k,\omega)\\
&=\sum_{\lambda,\lambda'}^{\pi_{\mathbf k,\lambda'}=-\pi_{\mathbf k,\lambda}}
(p^{\mathrm{SC}}_{\mathbf k,\lambda}+p^{\mathrm{SC}}_{\mathbf k,\lambda'})
|\langle\pi_{\mathbf k,\lambda'}|\hat c_{\mathbf k\sigma}|\pi_{\mathbf k,\lambda}\rangle|^2\\
&\quad\times\delta\!\left(\omega-E^{\mathrm{SC}}_{\mathbf k,\lambda}+E^{\mathrm{SC}}_{\mathbf k,\lambda'}\right)\\
&=\sum_{l=0}^{2}\sum_{\nu=\pm1}
(p^{\mathrm{SC},e}_{\mathbf k,l}+p^{\mathrm{SC},o}_{\mathbf k,\nu})\\
&\quad\times\Bigl[
Z^{\mathrm{SC},+}_{\mathbf k,\nu l}\delta(\omega-E^{\mathrm{SC},o}_{\mathbf k,\nu}+E^{\mathrm{SC},e}_{\mathbf k,l})
+Z^{\mathrm{SC},-}_{\mathbf k,\nu l}\delta(\omega+E^{\mathrm{SC},o}_{\mathbf k,\nu}-E^{\mathrm{SC},e}_{\mathbf k,l})
\Bigr]\\
&\quad+(1{+}1{+}1/2)\sum_{\nu=\pm1}\!
(p^{\mathrm{SC}}_{\mathbf k,2}+p^{\mathrm{SC},o}_{\mathbf k,\nu})\\
&\qquad\times\Bigl[
|v_{\mathbf k,\nu}|^2\delta(\omega{-}E^{\mathrm{SC},o}_{\mathbf k,\nu}{+}2\xi_{\mathbf k})
+|u_{\mathbf k,\nu}|^2\delta(\omega{+}E^{\mathrm{SC},o}_{\mathbf k,\nu}{-}2\xi_{\mathbf k})
\Bigr].
\end{aligned}
}
 
The four classes of allowed one-electron removal processes between SC block eigenstates are listed in \hyperlink{sm-table-k1-sc-transitions}{Table L1}.

\hypertarget{sm-table-k1-sc-transitions}{}\par\medskip

\noindent{\small \textbf{Table L1 \textbar{} One-electron removal processes between SC block eigenstates.}\par}\nopagebreak[4]
\begingroup\small\setlength{\tabcolsep}{3.5pt}\renewcommand{\arraystretch}{1.30}
\noindent\FittedTable{\begin{tabular}{@{}cccc@{}}\toprule
{\shortstack[c]{Initial state\\ \(|\pi_{\mathbf k,\lambda}\rangle\)}} & {\shortstack[c]{Final state\\ \(|\pi_{\mathbf k,\lambda'}\rangle\)}} & {\shortstack[c]{Removal energy\\ \(E^{\mathrm{SC}}_{\mathbf k,\lambda}-E^{\mathrm{SC}}_{\mathbf k,\lambda'}\)}} & {\shortstack[c]{Squared matrix elements\\ \(|\langle\pi_{\mathbf k,\lambda'}|\hat c_{\mathbf k\sigma}|\pi_{\mathbf k,\lambda}\rangle|^2\)}} \\
\midrule
{\(|o_{\pm\mathbf k\sigma,\nu}\rangle\)} & {\(|e_{\mathbf k,l}\rangle\)} & {\(E^{\mathrm{SC},o}_{\mathbf k,\nu}-E^{\mathrm{SC},e}_{\mathbf k,l}\)} & {\(Z^{\mathrm{SC},+}_{\mathbf k,\nu l}=\left|u_{\mathbf k,\nu}^{*}x_{\mathbf k,l}+v_{\mathbf k,\nu}^{*}y_{\mathbf k,l}/\sqrt2\right|^2\)} \\
{\(|e_{\mathbf k,l}\rangle\)} & {\(|o_{\pm\mathbf k\sigma,\nu}\rangle\)} & {\(E^{\mathrm{SC},e}_{\mathbf k,l}-E^{\mathrm{SC},o}_{\mathbf k,\nu}\)} & {\(Z^{\mathrm{SC},-}_{\mathbf k,\nu l}=\left|u_{\mathbf k,\nu}^{*}y_{\mathbf k,l}/\sqrt2+v_{\mathbf k,\nu}^{*}z_{\mathbf k,l}\right|^2\)} \\
{\(|o_{\pm\mathbf k\sigma,\nu}\rangle\)} & {\(\begin{gathered}|2_{\mathbf k,t_{\pm1}}\rangle,\ |2_{\pm\mathbf k}\rangle\\|2_{\mathbf k,t_0}\rangle\end{gathered}\)} & {\(E^{\mathrm{SC},o}_{\mathbf k,\nu}-2\xi_{\mathbf k}\)} & {\(\begin{gathered}|v_{\mathbf k,\nu}|^2\\|v_{\mathbf k,\nu}|^2/2\end{gathered}\)} \\
{\(\begin{gathered}|2_{\mathbf k,t_{\pm1}}\rangle,\ |2_{\pm\mathbf k}\rangle\\|2_{\mathbf k,t_0}\rangle\end{gathered}\)} & {\(|o_{\pm\mathbf k\sigma,\nu}\rangle\)} & {\(2\xi_{\mathbf k}-E^{\mathrm{SC},o}_{\mathbf k,\nu}\)} & {\(\begin{gathered}|u_{\mathbf k,\nu}|^2\\|u_{\mathbf k,\nu}|^2/2\end{gathered}\)} \\
\bottomrule\end{tabular}}\par\endgroup
\smallskip
 
\hypertarget{sm-sc-spectral-zeeman-content}{}\hypertarget{sm-sc-spectral-zeeman}{}
\subsection*{L.2. One-electron ARPES under magnetic field}

\hypertarget{sm-table-l2-sc-zeeman}{}\par\medskip

\noindent{\small Table L2 \textbar{} Block eigenstates and eigenenergies of the correlated SC state under a magnetic field.\par}\nopagebreak[4]
\begingroup\small\setlength{\tabcolsep}{3.5pt}\renewcommand{\arraystretch}{1.30}
\noindent\FittedTable{\begin{tabular}{@{}ccccc@{}}\toprule
{\shortstack[c]{Block fermion\\ parity \(\pi_{\mathbf k,\lambda}\)}} & {Block eigenstate} & {\shortstack[c]{Block spin\\ \(\mathbb S^z_{\mathbf k}\)}} & {Block eigenenergy} & {Degeneracy} \\
\midrule
{\(+1\)} & {\(|e_{\mathbf k,l}\rangle\)} & {\(0\)} & {\(E^{\mathrm{SC},e}_{\mathbf k,l}\)} & {1 for each \(l\)} \\
{\(+1\)} & {\(|2_{\mathbf k,t_{+1}}\rangle\)} & {\(+1\)} & {\(2\xi_{\mathbf k}-B_z\)} & {1} \\
{\(+1\)} & {\(|2_{\mathbf k,t_0}\rangle,\ |2_{\pm\mathbf k}\rangle\)} & {\(0\)} & {\(2\xi_{\mathbf k}\)} & {3} \\
{\(+1\)} & {\(|2_{\mathbf k,t_{-1}}\rangle\)} & {\(-1\)} & {\(2\xi_{\mathbf k}+B_z\)} & {1} \\
{\(-1\)} & {\(|o_{\pm\mathbf k\sigma,\nu}\rangle\)} & {\(\sigma/2\)} & {\(E^{\mathrm{SC},o}_{\mathbf k,\nu}-\sigma B_z/2\)} & {2 for each \(\nu,\sigma\)} \\
\bottomrule\end{tabular}}\par\endgroup
\smallskip
 
A uniform magnetic field \(B\) is applied to the homogeneous correlated spin-singlet SC state, and the field-dependent SC block Hamiltonian is

\hypertarget{sm-eq-sc-zeeman-hamiltonian}{}%
\FittedEquation{L6}{%
\hat h_{\mathbf k}^{\mathrm{SC}}(B_z)
=\xi_{\mathbf k}\hat N_{\mathbf k}
-2\mathcal J_{\mathbf k}\hat\Delta_{\mathbf k}^{\dagger}\hat\Delta_{\mathbf k}
-2\left(\Delta_{\mathbf k,d}\hat\Delta_{\mathbf k}^{\dagger}
+\Delta_{\mathbf k,d}^{*}\hat\Delta_{\mathbf k}\right)
-B_z\hat{\mathbb S}^{z}_{\mathbf k},
}
 
where \(B_z=g_e\mu_{\mathrm B}B\) is the Zeeman energy and \(\hat{\mathbb S}^{z}_{\mathbf k}=\hat S^z_{\mathbf k}+\hat S^z_{-\mathbf k}\) is the block spin operator along the magnetic field axis. The SC block eigenstates and eigenenergies under a magnetic field are listed in \hyperlink{sm-table-l2-sc-zeeman}{Table L2} (cf. \hyperlink{sm-sc-block-eigenstates}{Table 2}).

Since \([\hat\pi_{\mathbf k},\hat{\mathbb S}^{z}_{\mathbf k}]=0\), the magnetic field preserves block fermion parity while reducing spin rotation symmetry from \(SU(2)\) to \(U(1)\) rotations about the field axis. The three even-fermion-parity states \(|e_{\mathbf k,l}\rangle\) are spin singlets, whereas each odd-fermion-parity state \(|o_{\pm\mathbf k\sigma,\nu}\rangle\) carries spin projection \(\sigma/2\). The matrices in \hyperlink{sm-eq-sc-even-matrix}{Eq. (J3)} and \hyperlink{sm-eq-sc-odd-matrix}{Eq. (J11)} become

\hypertarget{sm-eq-sc-zeeman-block-matrices}{}%
\FittedEquation{L7}{%
h^{\mathrm{SC},e}_{\mathbf k}(B_z)=h^{\mathrm{SC},e}_{\mathbf k},
}
 
\hypertarget{sm-eq-sc-zeeman-odd-matrix}{}%
\FittedEquation{L8}{%
h^{\mathrm{SC},o}_{\mathbf k,\sigma}(B_z)=h^{\mathrm{SC},o}_{\mathbf k}-\frac{\sigma B_z}{2}\mathbf I_2.
}
 
The spin-triplet two-electron states \(|2_{\mathbf k,t_{\pm1}}\rangle\) and \(|2_{\mathbf k,t_0}\rangle\) split under a magnetic field, whereas the same-momentum spin-singlet two-electron states \(|2_{\pm\mathbf k}\rangle\) remain unshifted. Summing all sixteen SC block eigenstates in \hyperlink{sm-table-l2-sc-zeeman}{Table L2} gives the field-dependent SC block partition function

\hypertarget{sm-eq-sc-zeeman-partition}{}%
\FittedEquation{L9}{%
\begin{aligned}
Z^{\mathrm{SC}}_{\mathbf k}(B_z)
&=\operatorname{Tr}e^{-\beta\hat h_{\mathbf k}^{\mathrm{SC}}(B_z)}\\
&=\sum_{l=0}^{2}e^{-\beta E^{\mathrm{SC},e}_{\mathbf k,l}}
+4\cosh\!\left(\frac{\beta B_z}{2}\right)\sum_{\nu=\pm1}e^{-\beta E^{\mathrm{SC},o}_{\mathbf k,\nu}}
+[3+2\cosh(\beta B_z)]e^{-2\beta\xi_{\mathbf k}}.
\end{aligned}
}
 
A field-dependent SC block eigenstate is still denoted by \(|\pi_{\mathbf k,\lambda}\rangle\), with block fermion parity, spin projection and eigenenergy \(\pi_{\mathbf k,\lambda}\), \(m_{\mathbf k,\lambda}\) and \(E^{\mathrm{SC}}_{\mathbf k,\lambda}(B_z)=E^{\mathrm{SC}}_{\mathbf k,\lambda}-B_zm_{\mathbf k,\lambda}\), respectively. The field-dependent thermal occupation probability for each state in \hyperlink{sm-table-l2-sc-zeeman}{Table L2} is \(p^{\mathrm{SC}}_{\mathbf k,\lambda}(B_z)=e^{-\beta E^{\mathrm{SC}}_{\mathbf k,\lambda}(B_z)}/Z^{\mathrm{SC}}_{\mathbf k}(B_z)\), with

\hypertarget{sm-eq-sc-zeeman-probabilities}{}\hypertarget{sm-eq-sc-zeeman-odd-probability}{}\hypertarget{sm-eq-sc-zeeman-inactive-probability}{}%
\FittedEquation{L10}{%
p^{\mathrm{SC},e}_{\mathbf k,l}(B_z)=\frac{e^{-\beta E^{\mathrm{SC},e}_{\mathbf k,l}}}{Z^{\mathrm{SC}}_{\mathbf k}(B_z)},\quad p^{\mathrm{SC},o}_{\mathbf k,\nu\sigma}(B_z)=\frac{e^{-\beta(E^{\mathrm{SC},o}_{\mathbf k,\nu}-\sigma B_z/2)}}{Z^{\mathrm{SC}}_{\mathbf k}(B_z)},\quad p^{\mathrm{SC}}_{\mathbf k,2,m}(B_z)=\frac{e^{-\beta(2\xi_{\mathbf k}-mB_z)}}{Z^{\mathrm{SC}}_{\mathbf k}(B_z)},
}
 
where \(m=\pm1,0\), and \(p^{\mathrm{SC}}_{\mathbf k,2,0}(B_z)\) applies to \(|2_{\mathbf k,t_0}\rangle\) and \(|2_{\pm\mathbf k}\rangle\).

The one-particle lesser Green's function of the correlated SC state under a magnetic field is

\hypertarget{sm-eq-sc-zeeman-lesser}{}%
\FittedEquation{L11}{%
\begin{aligned}
G^<_{\mathrm{SC}}(\mathbf k\sigma,\mathbf k'\sigma';t,t';B_z)
&\equiv i\langle\hat c_{\mathbf k'\sigma'}^{\dagger}(t')\hat c_{\mathbf k\sigma}(t)\rangle_{\mathrm{SC},B_z}=\delta_{\mathbf k,\mathbf k'}\delta_{\sigma,\sigma'}G^<_{\mathrm{SC},\sigma}(\mathbf k;t,t';B_z),
\end{aligned}
}
 
\hypertarget{sm-eq-sc-zeeman-lesser-block-lehmann}{}%
\FittedEquation{L12}{%
G^<_{\mathrm{SC},\sigma}(\mathbf k;t,t';B_z)=i\sum_{\lambda,\lambda'}^{\substack{\pi_{\mathbf k,\lambda'}=-\pi_{\mathbf k,\lambda}\\m_{\mathbf k,\lambda'}=m_{\mathbf k,\lambda}-\sigma/2}} p^{\mathrm{SC}}_{\mathbf k,\lambda}(B_z) |\langle\pi_{\mathbf k,\lambda'}|\hat c_{\mathbf k\sigma}|\pi_{\mathbf k,\lambda}\rangle|^2\,e^{-i[E^{\mathrm{SC}}_{\mathbf k,\lambda}(B_z)-E^{\mathrm{SC}}_{\mathbf k,\lambda'}(B_z)](t-t')}.
}
 
The spin rotation \(U(1)\) symmetry about the field axis keeps the normal Green's function diagonal in spin. Applying the long-duration limit in \hyperlink{sm-eq-sc-arpes-occupied-definition}{Eq. (L1)} gives the occupied spectrum

\hypertarget{sm-eq-sc-zeeman-occupied}{}%
\FittedEquation{L13}{%
\begin{aligned}
&f_{\mathrm F}(\omega)A_{\mathrm{SC},\sigma}(\mathbf k,\omega;B_z)=\frac{1}{2\pi i}\int_{-\infty}^{\infty}dt_r\,e^{i\omega t_r}G^<_{\mathrm{SC},\sigma}(\mathbf k;t_r,0;B_z)\\
&=\sum_{\lambda,\lambda'}^{\substack{\pi_{\mathbf k,\lambda'}=-\pi_{\mathbf k,\lambda}\\m_{\mathbf k,\lambda'}=m_{\mathbf k,\lambda}-\sigma/2}}p^{\mathrm{SC}}_{\mathbf k,\lambda}(B_z)|\langle\pi_{\mathbf k,\lambda'}|\hat c_{\mathbf k\sigma}|\pi_{\mathbf k,\lambda}\rangle|^2\delta\!\left[\omega-E^{\mathrm{SC}}_{\mathbf k,\lambda}(B_z)+E^{\mathrm{SC}}_{\mathbf k,\lambda'}(B_z)\right]\\
&=f_{\mathrm F}(\omega)\sum_{\lambda,\lambda'}^{\substack{\pi_{\mathbf k,\lambda'}=-\pi_{\mathbf k,\lambda}\\m_{\mathbf k,\lambda'}=m_{\mathbf k,\lambda}-\sigma/2}}[p^{\mathrm{SC}}_{\mathbf k,\lambda}(B_z)+p^{\mathrm{SC}}_{\mathbf k,\lambda'}(B_z)]|\langle\pi_{\mathbf k,\lambda'}|\hat c_{\mathbf k\sigma}|\pi_{\mathbf k,\lambda}\rangle|^2\delta\!\left[\omega-E^{\mathrm{SC}}_{\mathbf k,\lambda}(B_z)+E^{\mathrm{SC}}_{\mathbf k,\lambda'}(B_z)\right],
\end{aligned}
}
 
which yields the SC one-particle spectral function under a magnetic field

\hypertarget{sm-eq-sc-zeeman-spectral-explicit}{}\hypertarget{sm-eq-sc-zeeman-spectral-lehmann}{}%
\FittedEquation{L14}{%
\begin{aligned}
&A_{\mathrm{SC},\sigma}(\mathbf k,\omega;B_z)\\
&=\sum_{\lambda,\lambda'}^{\substack{\pi_{\mathbf k,\lambda'}=-\pi_{\mathbf k,\lambda}\\m_{\mathbf k,\lambda'}=m_{\mathbf k,\lambda}-\sigma/2}}[p^{\mathrm{SC}}_{\mathbf k,\lambda}(B_z)+p^{\mathrm{SC}}_{\mathbf k,\lambda'}(B_z)]|\langle\pi_{\mathbf k,\lambda'}|\hat c_{\mathbf k\sigma}|\pi_{\mathbf k,\lambda}\rangle|^2\delta\!\left[\omega-E^{\mathrm{SC}}_{\mathbf k,\lambda}(B_z)+E^{\mathrm{SC}}_{\mathbf k,\lambda'}(B_z)\right]\\
&=\sum_{l=0}^{2}\sum_{\nu=\pm1}
[p^{\mathrm{SC},e}_{\mathbf k,l}(B_z)+p^{\mathrm{SC},o}_{\mathbf k,\nu\sigma}(B_z)]Z^{\mathrm{SC},+}_{\mathbf k,\nu l}
\delta\!\left(\omega-E^{\mathrm{SC},o}_{\mathbf k,\nu}+E^{\mathrm{SC},e}_{\mathbf k,l}+\frac{\sigma B_z}{2}\right)\\
&+\sum_{l=0}^{2}\sum_{\nu=\pm1}
[p^{\mathrm{SC},e}_{\mathbf k,l}(B_z)+p^{\mathrm{SC},o}_{\mathbf k,\nu,-\sigma}(B_z)]Z^{\mathrm{SC},-}_{\mathbf k,\nu l}
\delta\!\left(\omega+E^{\mathrm{SC},o}_{\mathbf k,\nu}-E^{\mathrm{SC},e}_{\mathbf k,l}+\frac{\sigma B_z}{2}\right)\\
&+\sum_{\nu=\pm1}|v_{\mathbf k,\nu}|^2
\left\{p^{\mathrm{SC},o}_{\mathbf k,\nu,-\sigma}(B_z)+p^{\mathrm{SC}}_{\mathbf k,2,-\sigma}(B_z)
+\left(1+1/2\right)[p^{\mathrm{SC},o}_{\mathbf k,\nu\sigma}(B_z)+p^{\mathrm{SC}}_{\mathbf k,2,0}(B_z)]\right\}\\
&\qquad\times\delta\!\left(\omega-E^{\mathrm{SC},o}_{\mathbf k,\nu}+2\xi_{\mathbf k}+\frac{\sigma B_z}{2}\right)\\
&+\sum_{\nu=\pm1}|u_{\mathbf k,\nu}|^2
\left\{p^{\mathrm{SC},o}_{\mathbf k,\nu\sigma}(B_z)+p^{\mathrm{SC}}_{\mathbf k,2,\sigma}(B_z)
+\left(1+1/2\right)[p^{\mathrm{SC},o}_{\mathbf k,\nu,-\sigma}(B_z)+p^{\mathrm{SC}}_{\mathbf k,2,0}(B_z)]\right\}\\
&\qquad\times\delta\!\left(\omega+E^{\mathrm{SC},o}_{\mathbf k,\nu}-2\xi_{\mathbf k}+\frac{\sigma B_z}{2}\right).
\end{aligned}
}
 
The four classes of allowed one-electron removal processes between SC block eigenstates under a magnetic field are listed in \hyperlink{sm-table-l3-sc-zeeman-transitions}{Table L3}. In general, \(A_{\uparrow}(\mathbf k,\omega;B_z)\ne A_{\downarrow}(\mathbf k,\omega;B_z)\) for a nonzero magnetic field, whereas \(A_{\bar\sigma}(\mathbf k,\omega;-B_z)=A_{\sigma}(\mathbf k,\omega;B_z)\).

\hypertarget{sm-table-l3-sc-zeeman-transitions}{}\par\medskip

\noindent{\small Table L3 \textbar{} One-electron removal processes between SC block eigenstates under a magnetic field.\par}\nopagebreak[4]
\begingroup\small\setlength{\tabcolsep}{3.5pt}\renewcommand{\arraystretch}{1.30}
\noindent\FittedTable{\begin{tabular}{@{}cccc@{}}\toprule
{\shortstack[c]{Initial state\\ \(|\pi_{\mathbf k,\lambda}\rangle\)}} & {\shortstack[c]{Final state\\ \(|\pi_{\mathbf k,\lambda'}\rangle\)}} & {\shortstack[c]{Removal energy\\ \(E^{\mathrm{SC}}_{\mathbf k,\lambda}(B_z)-E^{\mathrm{SC}}_{\mathbf k,\lambda'}(B_z)\)}} & {\shortstack[c]{Squared matrix elements\\ \(|\langle\pi_{\mathbf k,\lambda'}|\hat c_{\mathbf k\sigma}|\pi_{\mathbf k,\lambda}\rangle|^2\)}} \\
\midrule
{\(|o_{\mathbf k\sigma,\nu}\rangle\)} & {\(|e_{\mathbf k,l}\rangle\)} & {\(E^{\mathrm{SC},o}_{\mathbf k,\nu}-E^{\mathrm{SC},e}_{\mathbf k,l}-\sigma B_z/2\)} & {\(Z^{\mathrm{SC},+}_{\mathbf k,\nu l}\)} \\
{\(|e_{\mathbf k,l}\rangle\)} & {\(|o_{-\mathbf k,-\sigma,\nu}\rangle\)} & {\(E^{\mathrm{SC},e}_{\mathbf k,l}-E^{\mathrm{SC},o}_{\mathbf k,\nu}-\sigma B_z/2\)} & {\(Z^{\mathrm{SC},-}_{\mathbf k,\nu l}\)} \\
{\(|o_{\mathbf k,-\sigma,\nu}\rangle\)} & {\(|2_{\mathbf k,t_{-\sigma}}\rangle\)} & {\multirow{3}{*}{\(E^{\mathrm{SC},o}_{\mathbf k,\nu}-2\xi_{\mathbf k}-\sigma B_z/2\)}} & {\(|v_{\mathbf k,\nu}|^2\)} \\
{\(|o_{-\mathbf k\sigma,\nu}\rangle\)} & {\(|2_{-\mathbf k}\rangle\)} &  & {\(|v_{\mathbf k,\nu}|^2\)} \\
{\(|o_{\mathbf k\sigma,\nu}\rangle\)} & {\(|2_{\mathbf k,t_0}\rangle\)} &  & {\(|v_{\mathbf k,\nu}|^2/2\)} \\
{\(|2_{\mathbf k,t_{\sigma}}\rangle\)} & {\(|o_{-\mathbf k\sigma,\nu}\rangle\)} & {\multirow{3}{*}{\(2\xi_{\mathbf k}-E^{\mathrm{SC},o}_{\mathbf k,\nu}-\sigma B_z/2\)}} & {\(|u_{\mathbf k,\nu}|^2\)} \\
{\(|2_{\mathbf k}\rangle\)} & {\(|o_{\mathbf k,-\sigma,\nu}\rangle\)} &  & {\(|u_{\mathbf k,\nu}|^2\)} \\
{\(|2_{\mathbf k,t_0}\rangle\)} & {\(|o_{-\mathbf k,-\sigma,\nu}\rangle\)} &  & {\(|u_{\mathbf k,\nu}|^2/2\)} \\
\bottomrule\end{tabular}}\par\endgroup
\smallskip
 
\hypertarget{sm-appendix-sc-two-particle}{}\hypertarget{sm-appendix-sc-two-particle-title}{}\hypertarget{sm-sc-two-photon}{}
\section*{Appendix M. cARPES of the correlated SC state}

\hypertarget{sm-sc-sequential-removal}{}
\subsection*{M.1. Two-photon cARPES}

The two-particle removal spectral function of the correlated SC state measured by two-photon cARPES is

\hypertarget{sm-eq-sc-carpes-fourier-definition}{}%
\FittedEquation{M1}{%
\begin{aligned}
&A^{(2)<}_{\mathrm{SC},\sigma_1\sigma_2}(\mathbf k_1,\mathbf k_2;\omega_1,\omega_2)
=\lim_{\mathcal T\to\infty}^{\mathcal D'}\frac{1}{(2\pi i)^2(\mathcal T)^2}
\iiiint_{-\frac{\mathcal T}{2}}^{\frac{\mathcal T}{2}}dt_1\,dt_2\,dt'_1\,dt'_2\\
&\quad\times e^{i[\omega_1(t_1-t'_1)+\omega_2(t_2-t'_2)]}
G^{(2)<}_{\mathrm{SC},\sigma_1\sigma_2}(\mathbf k_1,\mathbf k_2;t_1,t_2,t'_1,t'_2).
\end{aligned}
}
 
The two-particle lesser Green's function is defined as

\hypertarget{sm-eq-sc-carpes-lesser-definition}{}%
\FittedEquation{M2}{%
G^{(2)<}_{\mathrm{SC},\sigma_1\sigma_2}(\mathbf k_1,\mathbf k_2;t_1,t_2,t'_1,t'_2)
\equiv i^2\Bigl\langle
T_{t'}[\hat c^\dagger_{\mathbf k_1\sigma_1}(t'_1)\hat c^\dagger_{\mathbf k_2\sigma_2}(t'_2)]
T_t[\hat c_{\mathbf k_2\sigma_2}(t_2)\hat c_{\mathbf k_1\sigma_1}(t_1)]\Bigr\rangle_{\mathrm{SC}},
}
 
where \(T_t\) and \(T_{t'}\) denote fermionic time ordering and anti-time ordering, respectively. For generic back-to-back momenta \(\mathbf k_1=-\mathbf k_2=\mathbf k\), the two-particle lesser Green's function is evaluated from the SC block eigenstates. Define the time-ordered two-electron removal amplitude

\hypertarget{sm-eq-sc-carpes-time-amplitude}{}%
\FittedEquation{M3}{%
\varphi^{(2)}_{\mathrm{SC},\lambda\lambda'';\sigma_1\sigma_2}(\mathbf k,-\mathbf k;t_1,t_2)
\equiv\langle\pi_{\mathbf k,\lambda''}|T_t[\hat c_{-\mathbf k\sigma_2}(t_2)\hat c_{\mathbf k\sigma_1}(t_1)]|\pi_{\mathbf k,\lambda}\rangle,
}
 
and express the two-particle lesser Green's function as

\hypertarget{sm-eq-sc-carpes-amplitude-lehmann}{}%
\FittedEquation{M4}{%
\begin{aligned}
&G^{(2)<}_{\mathrm{SC},\sigma_1\sigma_2}(\mathbf k,-\mathbf k;t_1,t_2,t'_1,t'_2)\\
&=i^2\sum_{\lambda,\lambda''}^{\pi_{\mathbf k,\lambda''}=\pi_{\mathbf k,\lambda}}
p^{\mathrm{SC}}_{\mathbf k,\lambda}\,
\varphi^{(2)}_{\mathrm{SC},\lambda\lambda'';\sigma_1\sigma_2}(\mathbf k,-\mathbf k;t_1,t_2)
\bigl[\varphi^{(2)}_{\mathrm{SC},\lambda\lambda'';\sigma_1\sigma_2}(\mathbf k,-\mathbf k;t'_1,t'_2)\bigr]^*.
\end{aligned}
}
 
Inserting the intermediate SC block eigenstates gives the time-ordered removal amplitude

\hypertarget{sm-eq-sc-carpes-ordered-amplitude}{}%
\FittedEquation{M5}{%
\begin{aligned}
&\varphi^{(2)}_{\mathrm{SC},\lambda\lambda'';\sigma_1\sigma_2}(\mathbf k,-\mathbf k;t_1,t_2)=\Theta(t_2-t_1)
\sum_{\lambda'}^{\pi_{\mathbf k,\lambda'}=-\pi_{\mathbf k,\lambda}}
\langle\pi_{\mathbf k,\lambda''}|\hat c_{-\mathbf k\sigma_2}|\pi_{\mathbf k,\lambda'}\rangle
\langle\pi_{\mathbf k,\lambda'}|\hat c_{\mathbf k\sigma_1}|\pi_{\mathbf k,\lambda}\rangle\\
&\quad\times e^{i(E^{\mathrm{SC}}_{\mathbf k,\lambda''}-E^{\mathrm{SC}}_{\mathbf k,\lambda'})t_2}
e^{i(E^{\mathrm{SC}}_{\mathbf k,\lambda'}-E^{\mathrm{SC}}_{\mathbf k,\lambda})t_1}\\
&\quad-\Theta(t_1-t_2)
\sum_{\lambda'}^{\pi_{\mathbf k,\lambda'}=-\pi_{\mathbf k,\lambda}}
\langle\pi_{\mathbf k,\lambda''}|\hat c_{\mathbf k\sigma_1}|\pi_{\mathbf k,\lambda'}\rangle
\langle\pi_{\mathbf k,\lambda'}|\hat c_{-\mathbf k\sigma_2}|\pi_{\mathbf k,\lambda}\rangle\\
&\qquad\times e^{i(E^{\mathrm{SC}}_{\mathbf k,\lambda''}-E^{\mathrm{SC}}_{\mathbf k,\lambda'})t_1}
e^{i(E^{\mathrm{SC}}_{\mathbf k,\lambda'}-E^{\mathrm{SC}}_{\mathbf k,\lambda})t_2}.
\end{aligned}
}
 
Intermediate states \(|\pi_{\mathbf k,\lambda'}\rangle\) with the same energy \(E^{\mathrm{SC}}_{\mathbf k,\lambda'}=E^{\Xi_{\mathrm{SC}}}_{\mathbf k}\) are grouped into a degenerate subspace \(\Xi_{\mathrm{SC}}\) satisfying \(\pi_{\mathbf k,\lambda'}=-\pi_{\mathbf k,\lambda}\). The coherent sums of matrix-element products for the two removal orderings are

\hypertarget{sm-eq-sc-sequential-amplitudes}{}%
\FittedEquation{M6}{%
M^{\sigma_2\sigma_1}_{\lambda,\lambda'';\Xi_{\mathrm{SC}}}(\mathbf k)
=\sum_{\lambda'\in\Xi_{\mathrm{SC}}}
\langle\pi_{\mathbf k,\lambda''}|\hat c_{-\mathbf k\sigma_2}|\pi_{\mathbf k,\lambda'}\rangle
\langle\pi_{\mathbf k,\lambda'}|\hat c_{\mathbf k\sigma_1}|\pi_{\mathbf k,\lambda}\rangle.
}
 
\hypertarget{sm-eq-sc-sequential-amplitudes-reverse}{}%
\FittedEquation{M7}{%
M^{\sigma_1\sigma_2}_{\lambda,\lambda'';\Xi_{\mathrm{SC}}}(-\mathbf k)
=\sum_{\lambda'\in\Xi_{\mathrm{SC}}}
\langle\pi_{\mathbf k,\lambda''}|\hat c_{\mathbf k\sigma_1}|\pi_{\mathbf k,\lambda'}\rangle
\langle\pi_{\mathbf k,\lambda'}|\hat c_{-\mathbf k\sigma_2}|\pi_{\mathbf k,\lambda}\rangle.
}
 
The intermediate-state energy offset is

\hypertarget{sm-eq-sc-carpes-intermediate-offset}{}%
\FittedEquation{M8}{%
\omega^{\Xi_{\mathrm{SC}}}_{\mathbf k,\lambda\lambda''}
=\frac{E^{\mathrm{SC}}_{\mathbf k,\lambda}+E^{\mathrm{SC}}_{\mathbf k,\lambda''}}2
-E^{\Xi_{\mathrm{SC}}}_{\mathbf k}.
}
 
The allowed two-electron removal processes between SC block eigenstates are listed in \hyperlink{sm-table-l4-sc-equal-spin-transitions}{Table M1} for the equal-spin channel and \hyperlink{sm-table-l5-sc-opposite-spin-transitions}{Table M2} for the opposite-spin channel.

\hypertarget{sm-sc-carpes-equal-spin-table-block}{}\hypertarget{sm-table-l4-sc-equal-spin-transitions}{}\par\medskip

\noindent{\small Table M1 \textbar{} Two-electron removal between SC block eigenstates from the fixed equal-spin modes \((\mathbf k,\uparrow)\) and \((-\mathbf k,\uparrow)\).\par}\nopagebreak[4]
\begingroup\small\setlength{\tabcolsep}{3.5pt}\renewcommand{\arraystretch}{1.30}
\noindent\FittedTable{\begin{tabular}{@{}ccccc@{}}\toprule
{\(|\pi_{\mathbf k,\lambda}\rangle\to|\pi_{\mathbf k,\lambda''}\rangle\)} & {\(E^{\Xi_{\mathrm{SC}}}_{\mathbf k}\)} & {\(\omega^{\Xi_{\mathrm{SC}}}_{\mathbf k,\lambda\lambda''}\)} & {\(M^{\sigma_2\sigma_1}_{\lambda,\lambda'';\Xi_{\mathrm{SC}}}(\mathbf k)\)} & {\(M^{\sigma_1\sigma_2}_{\lambda,\lambda'';\Xi_{\mathrm{SC}}}(-\mathbf k)\)} \\
\midrule
{\(|e_{\mathbf k,l}\rangle\to|2_{\mathbf k,t_{-1}}\rangle\)} & {\(E^{\mathrm{SC},o}_{\mathbf k,\nu}\)} & {\(E^{\mathrm{SC},e}_{\mathbf k,l}/2+\xi_{\mathbf k}-E^{\mathrm{SC},o}_{\mathbf k,\nu}\)} & {\(\begin{gathered}v_{\mathbf k,\nu}\\{}\times \Bigl(u_{\mathbf k,\nu}^{*}y_{\mathbf k,l}/\sqrt2+v_{\mathbf k,\nu}^{*}z_{\mathbf k,l}\Bigr)\end{gathered}\)} & {\(\begin{gathered}-v_{\mathbf k,\nu}\\{}\times \Bigl(u_{\mathbf k,\nu}^{*}y_{\mathbf k,l}/\sqrt2+v_{\mathbf k,\nu}^{*}z_{\mathbf k,l}\Bigr)\end{gathered}\)} \\
{\(|2_{\mathbf k,t_{+1}}\rangle\to|e_{\mathbf k,l'}\rangle\)} & {\(E^{\mathrm{SC},o}_{\mathbf k,\nu}\)} & {\(E^{\mathrm{SC},e}_{\mathbf k,l'}/2+\xi_{\mathbf k}-E^{\mathrm{SC},o}_{\mathbf k,\nu}\)} & {\(\begin{gathered}\Bigl(x_{\mathbf k,l'}^{*}u_{\mathbf k,\nu}+y_{\mathbf k,l'}^{*}v_{\mathbf k,\nu}/\sqrt2\Bigr)\\{}\times u_{\mathbf k,\nu}^{*}\end{gathered}\)} & {\(\begin{gathered}-\Bigl(x_{\mathbf k,l'}^{*}u_{\mathbf k,\nu}+y_{\mathbf k,l'}^{*}v_{\mathbf k,\nu}/\sqrt2\Bigr)\\{}\times u_{\mathbf k,\nu}^{*}\end{gathered}\)} \\
{\(|2_{\mathbf k,t_{+1}}\rangle\to|2_{\mathbf k,t_0}\rangle\)} & {\(E^{\mathrm{SC},o}_{\mathbf k,\nu}\)} & {\(2\xi_{\mathbf k}-E^{\mathrm{SC},o}_{\mathbf k,\nu}\)} & {\(-(1/\sqrt2)u_{\mathbf k,\nu}^{*}v_{\mathbf k,\nu}\)} & {\(-(1/\sqrt2)u_{\mathbf k,\nu}^{*}v_{\mathbf k,\nu}\)} \\
{\(|2_{\mathbf k,t_0}\rangle\to|2_{\mathbf k,t_{-1}}\rangle\)} & {\(E^{\mathrm{SC},o}_{\mathbf k,\nu}\)} & {\(2\xi_{\mathbf k}-E^{\mathrm{SC},o}_{\mathbf k,\nu}\)} & {\(-(1/\sqrt2)u_{\mathbf k,\nu}^{*}v_{\mathbf k,\nu}\)} & {\(-(1/\sqrt2)u_{\mathbf k,\nu}^{*}v_{\mathbf k,\nu}\)} \\
{\(|o_{\mathbf k\uparrow,\nu}\rangle\to|o_{\mathbf k\downarrow,\nu'}\rangle\)} & {\(E^{\mathrm{SC},e}_{\mathbf k,l}\)} & {\((E^{\mathrm{SC},o}_{\mathbf k,\nu}+E^{\mathrm{SC},o}_{\mathbf k,\nu'})/2-E^{\mathrm{SC},e}_{\mathbf k,l}\)} & {\(\begin{gathered}-\Bigl(u_{\mathbf k,\nu'}^{*}y_{\mathbf k,l}/\sqrt2+v_{\mathbf k,\nu'}^{*}z_{\mathbf k,l}\Bigr)\\{}\times \Bigl(x_{\mathbf k,l}^{*}u_{\mathbf k,\nu}+y_{\mathbf k,l}^{*}v_{\mathbf k,\nu}/\sqrt2\Bigr)\end{gathered}\)} & {\(0\)} \\
{\(|o_{\mathbf k\uparrow,\nu}\rangle\to|o_{\mathbf k\downarrow,\nu'}\rangle\)} & {\(2\xi_{\mathbf k}\)} & {\((E^{\mathrm{SC},o}_{\mathbf k,\nu}+E^{\mathrm{SC},o}_{\mathbf k,\nu'})/2-2\xi_{\mathbf k}\)} & {\(-(1/2)u_{\mathbf k,\nu'}^{*}v_{\mathbf k,\nu}\)} & {\(u_{\mathbf k,\nu'}^{*}v_{\mathbf k,\nu}\)} \\
{\(|o_{-\mathbf k\uparrow,\nu}\rangle\to|o_{-\mathbf k\downarrow,\nu'}\rangle\)} & {\(E^{\mathrm{SC},e}_{\mathbf k,l}\)} & {\((E^{\mathrm{SC},o}_{\mathbf k,\nu}+E^{\mathrm{SC},o}_{\mathbf k,\nu'})/2-E^{\mathrm{SC},e}_{\mathbf k,l}\)} & {\(0\)} & {\(\begin{gathered}-\Bigl(u_{\mathbf k,\nu'}^{*}y_{\mathbf k,l}/\sqrt2+v_{\mathbf k,\nu'}^{*}z_{\mathbf k,l}\Bigr)\\{}\times \Bigl(x_{\mathbf k,l}^{*}u_{\mathbf k,\nu}+y_{\mathbf k,l}^{*}v_{\mathbf k,\nu}/\sqrt2\Bigr)\end{gathered}\)} \\
{\(|o_{-\mathbf k\uparrow,\nu}\rangle\to|o_{-\mathbf k\downarrow,\nu'}\rangle\)} & {\(2\xi_{\mathbf k}\)} & {\((E^{\mathrm{SC},o}_{\mathbf k,\nu}+E^{\mathrm{SC},o}_{\mathbf k,\nu'})/2-2\xi_{\mathbf k}\)} & {\(u_{\mathbf k,\nu'}^{*}v_{\mathbf k,\nu}\)} & {\(-(1/2)u_{\mathbf k,\nu'}^{*}v_{\mathbf k,\nu}\)} \\
\bottomrule\end{tabular}}\par\endgroup
\smallskip

{\small Each row specifies an initial state \(|\pi_{\mathbf k,\lambda}\rangle\), a final state \(|\pi_{\mathbf k,\lambda''}\rangle\), and an intermediate energy \(E^{\Xi_{\mathrm{SC}}}_{\mathbf k}\). With spin rotation symmetry, the spectral functions for \((\downarrow,\downarrow)\) are identical to those for \((\uparrow,\uparrow)\).\par}
 
For the equal-spin channel \(\sigma_1=\sigma_2=\sigma\),

 \hypertarget{sm-eq-sc-carpes-time-kernel-v}{}\hypertarget{sm-eq-sc-carpes-time-kernel-u}{}\hypertarget{sm-eq-sc-carpes-time-kernel-d}{}\hypertarget{sm-eq-sc-carpes-time-kernel-c}{}\hypertarget{sm-eq-sc-carpes-time-kernel-q}{}\hypertarget{sm-eq-sc-carpes-time-kernel-r}{}
\hypertarget{sm-eq-sc-carpes-equal-spin-time}{}%
\FittedEquation{M9}{%
\begin{aligned}
&\frac{G^{(2)<}_{\mathrm{SC},\sigma\sigma}(\mathbf k,-\mathbf k;t_1,t_2,t'_1,t'_2)}{i^2}\\
&=\sum_{l=0}^{2}p^{\mathrm{SC},e}_{\mathbf k,l}e^{-\frac i2(E^{\mathrm{SC},e}_{\mathbf k,l}-2\xi_{\mathbf k})(t_1+t_2-t'_1-t'_2)}\,e^{\frac i2(E^{\mathrm{SC},e}_{\mathbf k,l}+2\xi_{\mathbf k})(|t_2-t_1|-|t'_2-t'_1|)}\\
&\quad\times\sum_{\nu,\nu'=\pm1}v_{\mathbf k,\nu}v_{\mathbf k,\nu'}^*\Bigl(\frac{u_{\mathbf k,\nu}^*y_{\mathbf k,l}}{\sqrt2}+v_{\mathbf k,\nu}^*z_{\mathbf k,l}\Bigr)\Bigl(\frac{u_{\mathbf k,\nu'}^*y_{\mathbf k,l}}{\sqrt2}+v_{\mathbf k,\nu'}^*z_{\mathbf k,l}\Bigr)^*\,e^{-iE^{\mathrm{SC},o}_{\mathbf k,\nu}|t_2-t_1|+iE^{\mathrm{SC},o}_{\mathbf k,\nu'}|t'_2-t'_1|}\\
&+p^{\mathrm{SC}}_{\mathbf k,2}\sum_{l=0}^{2}e^{-\frac i2(2\xi_{\mathbf k}-E^{\mathrm{SC},e}_{\mathbf k,l})(t_1+t_2-t'_1-t'_2)}\,e^{\frac i2(2\xi_{\mathbf k}+E^{\mathrm{SC},e}_{\mathbf k,l})(|t_2-t_1|-|t'_2-t'_1|)}\\
&\quad\times\sum_{\nu,\nu'=\pm1}u_{\mathbf k,\nu}^*u_{\mathbf k,\nu'}\Bigl(x_{\mathbf k,l}^*u_{\mathbf k,\nu}+\frac{y_{\mathbf k,l}^*v_{\mathbf k,\nu}}{\sqrt2}\Bigr)\Bigl(x_{\mathbf k,l}^*u_{\mathbf k,\nu'}+\frac{y_{\mathbf k,l}^*v_{\mathbf k,\nu'}}{\sqrt2}\Bigr)^*\,e^{-iE^{\mathrm{SC},o}_{\mathbf k,\nu}|t_2-t_1|+iE^{\mathrm{SC},o}_{\mathbf k,\nu'}|t'_2-t'_1|}\\
&+p^{\mathrm{SC}}_{\mathbf k,2}\bigl[\Theta(t_2-t_1)-\Theta(t_1-t_2)\bigr]\bigl[\Theta(t'_2-t'_1)-\Theta(t'_1-t'_2)\bigr]\,e^{2i\xi_{\mathbf k}(|t_2-t_1|-|t'_2-t'_1|)}\\
&\quad\times\sum_{\nu,\nu'=\pm1}u_{\mathbf k,\nu}^*v_{\mathbf k,\nu}u_{\mathbf k,\nu'}v_{\mathbf k,\nu'}^*e^{-iE^{\mathrm{SC},o}_{\mathbf k,\nu}|t_2-t_1|+iE^{\mathrm{SC},o}_{\mathbf k,\nu'}|t'_2-t'_1|}\\
&+\sum_{\nu,\nu'=\pm1}p^{\mathrm{SC},o}_{\mathbf k,\nu}e^{-\frac i2(E^{\mathrm{SC},o}_{\mathbf k,\nu}-E^{\mathrm{SC},o}_{\mathbf k,\nu'})(t_1+t_2-t'_1-t'_2)}\,e^{\frac i2(E^{\mathrm{SC},o}_{\mathbf k,\nu}+E^{\mathrm{SC},o}_{\mathbf k,\nu'})(|t_2-t_1|-|t'_2-t'_1|)}\\
&\quad\times\Biggl\{\bigl[\Theta(t_2-t_1)\Theta(t'_2-t'_1)+\Theta(t_1-t_2)\Theta(t'_1-t'_2)\bigr]\sum_{l,l'=0}^{2}\\
&\qquad\quad\times\Bigl(\frac{u_{\mathbf k,\nu'}^*y_{\mathbf k,l}}{\sqrt2}+v_{\mathbf k,\nu'}^*z_{\mathbf k,l}\Bigr)\Bigl(x_{\mathbf k,l}^*u_{\mathbf k,\nu}+\frac{y_{\mathbf k,l}^*v_{\mathbf k,\nu}}{\sqrt2}\Bigr)\\
&\qquad\quad\times\left[\Bigl(\frac{u_{\mathbf k,\nu'}^*y_{\mathbf k,l'}}{\sqrt2}+v_{\mathbf k,\nu'}^*z_{\mathbf k,l'}\Bigr)\Bigl(x_{\mathbf k,l'}^*u_{\mathbf k,\nu}+\frac{y_{\mathbf k,l'}^*v_{\mathbf k,\nu}}{\sqrt2}\Bigr)\right]^*\,e^{-iE^{\mathrm{SC},e}_{\mathbf k,l}|t_2-t_1|+iE^{\mathrm{SC},e}_{\mathbf k,l'}|t'_2-t'_1|}\\
&\qquad+\left(1-\frac12\bigl[\Theta(t_2-t_1)\Theta(t'_2-t'_1)+\Theta(t_1-t_2)\Theta(t'_1-t'_2)\bigr]\right)\sum_{l=0}^{2}\Biggl[\\
&\qquad\quad \Bigl(\frac{u_{\mathbf k,\nu'}^*y_{\mathbf k,l}}{\sqrt2}+v_{\mathbf k,\nu'}^*z_{\mathbf k,l}\Bigr)\Bigl(x_{\mathbf k,l}^*u_{\mathbf k,\nu}+\frac{y_{\mathbf k,l}^*v_{\mathbf k,\nu}}{\sqrt2}\Bigr)u_{\mathbf k,\nu'}v_{\mathbf k,\nu}^*\,e^{-iE^{\mathrm{SC},e}_{\mathbf k,l}|t_2-t_1|+i2\xi_{\mathbf k}|t'_2-t'_1|}\\
&\qquad\quad+u_{\mathbf k,\nu'}^*v_{\mathbf k,\nu}\left[\Bigl(\frac{u_{\mathbf k,\nu'}^*y_{\mathbf k,l}}{\sqrt2}+v_{\mathbf k,\nu'}^*z_{\mathbf k,l}\Bigr)\Bigl(x_{\mathbf k,l}^*u_{\mathbf k,\nu}+\frac{y_{\mathbf k,l}^*v_{\mathbf k,\nu}}{\sqrt2}\Bigr)\right]^*\,e^{-i2\xi_{\mathbf k}|t_2-t_1|+iE^{\mathrm{SC},e}_{\mathbf k,l}|t'_2-t'_1|}\Biggr]\\
&\qquad+\left(1+\frac14\bigl[\Theta(t_2-t_1)\Theta(t'_2-t'_1)+\Theta(t_1-t_2)\Theta(t'_1-t'_2)\bigr]\right)\\
&\qquad\quad\times|u_{\mathbf k,\nu'}|^2|v_{\mathbf k,\nu}|^2e^{-2i\xi_{\mathbf k}(|t_2-t_1|-|t'_2-t'_1|)}\Biggr\}.
\end{aligned}
}
 
\hypertarget{sm-sc-carpes-opposite-spin-table-block}{}\hypertarget{sm-table-l5-sc-opposite-spin-transitions}{}\par\medskip

\noindent{\small Table M2 \textbar{} Two-electron removal between SC block eigenstates from the fixed opposite-spin modes \((\mathbf k,\uparrow)\) and \((-\mathbf k,\downarrow)\).\par}\nopagebreak[4]
\begingroup\small\setlength{\tabcolsep}{3.5pt}\renewcommand{\arraystretch}{1.30}
\noindent\FittedTable{\begin{tabular}{@{}ccccc@{}}\toprule
{\(|\pi_{\mathbf k,\lambda}\rangle\to|\pi_{\mathbf k,\lambda''}\rangle\)} & {\(E^{\Xi_{\mathrm{SC}}}_{\mathbf k}\)} & {\(\omega^{\Xi_{\mathrm{SC}}}_{\mathbf k,\lambda\lambda''}\)} & {\(M^{\sigma_2\sigma_1}_{\lambda,\lambda'';\Xi_{\mathrm{SC}}}(\mathbf k)\)} & {\(M^{\sigma_1\sigma_2}_{\lambda,\lambda'';\Xi_{\mathrm{SC}}}(-\mathbf k)\)} \\
\midrule
{\(|e_{\mathbf k,l}\rangle\to|e_{\mathbf k,l'}\rangle\)} & {\(E^{\mathrm{SC},o}_{\mathbf k,\nu}\)} & {\((E^{\mathrm{SC},e}_{\mathbf k,l}+E^{\mathrm{SC},e}_{\mathbf k,l'})/2-E^{\mathrm{SC},o}_{\mathbf k,\nu}\)} & {\(\begin{gathered}-\Bigl(x_{\mathbf k,l'}^{*}u_{\mathbf k,\nu}+y_{\mathbf k,l'}^{*}v_{\mathbf k,\nu}/\sqrt2\Bigr)\\{}\times \Bigl(u_{\mathbf k,\nu}^{*}y_{\mathbf k,l}/\sqrt2+v_{\mathbf k,\nu}^{*}z_{\mathbf k,l}\Bigr)\end{gathered}\)} & {\(\begin{gathered}\Bigl(x_{\mathbf k,l'}^{*}u_{\mathbf k,\nu}+y_{\mathbf k,l'}^{*}v_{\mathbf k,\nu}/\sqrt2\Bigr)\\{}\times \Bigl(u_{\mathbf k,\nu}^{*}y_{\mathbf k,l}/\sqrt2+v_{\mathbf k,\nu}^{*}z_{\mathbf k,l}\Bigr)\end{gathered}\)} \\
{\(|e_{\mathbf k,l}\rangle\to|2_{\mathbf k,t_0}\rangle\)} & {\(E^{\mathrm{SC},o}_{\mathbf k,\nu}\)} & {\(E^{\mathrm{SC},e}_{\mathbf k,l}/2+\xi_{\mathbf k}-E^{\mathrm{SC},o}_{\mathbf k,\nu}\)} & {\(\begin{gathered}-(1/\sqrt2)v_{\mathbf k,\nu}\\{}\times \Bigl(u_{\mathbf k,\nu}^{*}y_{\mathbf k,l}/\sqrt2+v_{\mathbf k,\nu}^{*}z_{\mathbf k,l}\Bigr)\end{gathered}\)} & {\(\begin{gathered}(1/\sqrt2)v_{\mathbf k,\nu}\\{}\times \Bigl(u_{\mathbf k,\nu}^{*}y_{\mathbf k,l}/\sqrt2+v_{\mathbf k,\nu}^{*}z_{\mathbf k,l}\Bigr)\end{gathered}\)} \\
{\(|2_{\mathbf k,t_0}\rangle\to|e_{\mathbf k,l'}\rangle\)} & {\(E^{\mathrm{SC},o}_{\mathbf k,\nu}\)} & {\(E^{\mathrm{SC},e}_{\mathbf k,l'}/2+\xi_{\mathbf k}-E^{\mathrm{SC},o}_{\mathbf k,\nu}\)} & {\(\begin{gathered}\Bigl(x_{\mathbf k,l'}^{*}u_{\mathbf k,\nu}+y_{\mathbf k,l'}^{*}v_{\mathbf k,\nu}/\sqrt2\Bigr)\\{}\times u_{\mathbf k,\nu}^{*}/\sqrt2\end{gathered}\)} & {\(\begin{gathered}-\Bigl(x_{\mathbf k,l'}^{*}u_{\mathbf k,\nu}+y_{\mathbf k,l'}^{*}v_{\mathbf k,\nu}/\sqrt2\Bigr)\\{}\times u_{\mathbf k,\nu}^{*}/\sqrt2\end{gathered}\)} \\
{\(|2_{\mathbf k,t_0}\rangle\to|2_{\mathbf k,t_0}\rangle\)} & {\(E^{\mathrm{SC},o}_{\mathbf k,\nu}\)} & {\(2\xi_{\mathbf k}-E^{\mathrm{SC},o}_{\mathbf k,\nu}\)} & {\((1/2)u_{\mathbf k,\nu}^{*}v_{\mathbf k,\nu}\)} & {\(-(1/2)u_{\mathbf k,\nu}^{*}v_{\mathbf k,\nu}\)} \\
{\(|2_{\mathbf k,t_{+1}}\rangle\to|2_{\mathbf k,t_{+1}}\rangle\)} & {\(E^{\mathrm{SC},o}_{\mathbf k,\nu}\)} & {\(2\xi_{\mathbf k}-E^{\mathrm{SC},o}_{\mathbf k,\nu}\)} & {\(u_{\mathbf k,\nu}^{*}v_{\mathbf k,\nu}\)} & {\(0\)} \\
{\(|2_{\mathbf k}\rangle\to|2_{\mathbf k}\rangle\)} & {\(E^{\mathrm{SC},o}_{\mathbf k,\nu}\)} & {\(2\xi_{\mathbf k}-E^{\mathrm{SC},o}_{\mathbf k,\nu}\)} & {\(u_{\mathbf k,\nu}^{*}v_{\mathbf k,\nu}\)} & {\(0\)} \\
{\(|2_{\mathbf k,t_{-1}}\rangle\to|2_{\mathbf k,t_{-1}}\rangle\)} & {\(E^{\mathrm{SC},o}_{\mathbf k,\nu}\)} & {\(2\xi_{\mathbf k}-E^{\mathrm{SC},o}_{\mathbf k,\nu}\)} & {\(0\)} & {\(-u_{\mathbf k,\nu}^{*}v_{\mathbf k,\nu}\)} \\
{\(|2_{-\mathbf k}\rangle\to|2_{-\mathbf k}\rangle\)} & {\(E^{\mathrm{SC},o}_{\mathbf k,\nu}\)} & {\(2\xi_{\mathbf k}-E^{\mathrm{SC},o}_{\mathbf k,\nu}\)} & {\(0\)} & {\(-u_{\mathbf k,\nu}^{*}v_{\mathbf k,\nu}\)} \\
{\(|o_{\mathbf k\uparrow,\nu}\rangle\to|o_{\mathbf k\uparrow,\nu'}\rangle\)} & {\(E^{\mathrm{SC},e}_{\mathbf k,l}\)} & {\((E^{\mathrm{SC},o}_{\mathbf k,\nu}+E^{\mathrm{SC},o}_{\mathbf k,\nu'})/2-E^{\mathrm{SC},e}_{\mathbf k,l}\)} & {\(\begin{gathered}\Bigl(u_{\mathbf k,\nu'}^{*}y_{\mathbf k,l}/\sqrt2+v_{\mathbf k,\nu'}^{*}z_{\mathbf k,l}\Bigr)\\{}\times \Bigl(x_{\mathbf k,l}^{*}u_{\mathbf k,\nu}+y_{\mathbf k,l}^{*}v_{\mathbf k,\nu}/\sqrt2\Bigr)\end{gathered}\)} & {\(0\)} \\
{\(|o_{\mathbf k\uparrow,\nu}\rangle\to|o_{\mathbf k\uparrow,\nu'}\rangle\)} & {\(2\xi_{\mathbf k}\)} & {\((E^{\mathrm{SC},o}_{\mathbf k,\nu}+E^{\mathrm{SC},o}_{\mathbf k,\nu'})/2-2\xi_{\mathbf k}\)} & {\(-(1/2)u_{\mathbf k,\nu'}^{*}v_{\mathbf k,\nu}\)} & {\(0\)} \\
{\(|o_{-\mathbf k\downarrow,\nu}\rangle\to|o_{-\mathbf k\downarrow,\nu'}\rangle\)} & {\(E^{\mathrm{SC},e}_{\mathbf k,l}\)} & {\((E^{\mathrm{SC},o}_{\mathbf k,\nu}+E^{\mathrm{SC},o}_{\mathbf k,\nu'})/2-E^{\mathrm{SC},e}_{\mathbf k,l}\)} & {\(0\)} & {\(\begin{gathered}-\Bigl(u_{\mathbf k,\nu'}^{*}y_{\mathbf k,l}/\sqrt2+v_{\mathbf k,\nu'}^{*}z_{\mathbf k,l}\Bigr)\\{}\times \Bigl(x_{\mathbf k,l}^{*}u_{\mathbf k,\nu}+y_{\mathbf k,l}^{*}v_{\mathbf k,\nu}/\sqrt2\Bigr)\end{gathered}\)} \\
{\(|o_{-\mathbf k\downarrow,\nu}\rangle\to|o_{-\mathbf k\downarrow,\nu'}\rangle\)} & {\(2\xi_{\mathbf k}\)} & {\((E^{\mathrm{SC},o}_{\mathbf k,\nu}+E^{\mathrm{SC},o}_{\mathbf k,\nu'})/2-2\xi_{\mathbf k}\)} & {\(0\)} & {\((1/2)u_{\mathbf k,\nu'}^{*}v_{\mathbf k,\nu}\)} \\
{\(|o_{\mathbf k\downarrow,\nu}\rangle\to|o_{\mathbf k\downarrow,\nu'}\rangle\)} & {\(2\xi_{\mathbf k}\)} & {\((E^{\mathrm{SC},o}_{\mathbf k,\nu}+E^{\mathrm{SC},o}_{\mathbf k,\nu'})/2-2\xi_{\mathbf k}\)} & {\(-u_{\mathbf k,\nu'}^{*}v_{\mathbf k,\nu}\)} & {\(u_{\mathbf k,\nu'}^{*}v_{\mathbf k,\nu}\)} \\
{\(|o_{-\mathbf k\uparrow,\nu}\rangle\to|o_{-\mathbf k\uparrow,\nu'}\rangle\)} & {\(2\xi_{\mathbf k}\)} & {\((E^{\mathrm{SC},o}_{\mathbf k,\nu}+E^{\mathrm{SC},o}_{\mathbf k,\nu'})/2-2\xi_{\mathbf k}\)} & {\(-u_{\mathbf k,\nu'}^{*}v_{\mathbf k,\nu}\)} & {\(u_{\mathbf k,\nu'}^{*}v_{\mathbf k,\nu}\)} \\
\bottomrule\end{tabular}}\par\endgroup
\smallskip

{\small Each row specifies an initial state \(|\pi_{\mathbf k,\lambda}\rangle\), a final state \(|\pi_{\mathbf k,\lambda''}\rangle\), and an intermediate energy \(E^{\Xi_{\mathrm{SC}}}_{\mathbf k}\). With spin rotation symmetry, the spectral functions for \((\downarrow,\uparrow)\) are identical to those for \((\uparrow,\downarrow)\).\par}
 
For the opposite-spin channel \(\sigma_1=\bar\sigma_2=\sigma\),

\hypertarget{sm-eq-sc-carpes-opposite-spin-time}{}%
\FittedEquation{M10}{%
\begin{aligned}
&\frac{G^{(2)<}_{\mathrm{SC},\sigma\bar\sigma}(\mathbf k,-\mathbf k;t_1,t_2,t'_1,t'_2)}{i^2}\\
&=\sum_{l,l'=0}^{2}p^{\mathrm{SC},e}_{\mathbf k,l}e^{-\frac i2(E^{\mathrm{SC},e}_{\mathbf k,l}-E^{\mathrm{SC},e}_{\mathbf k,l'})(t_1+t_2-t'_1-t'_2)}\,e^{\frac i2(E^{\mathrm{SC},e}_{\mathbf k,l}+E^{\mathrm{SC},e}_{\mathbf k,l'})(|t_2-t_1|-|t'_2-t'_1|)}\sum_{\nu,\nu'=\pm1}\\
&\qquad\times\Bigl(x_{\mathbf k,l'}^*u_{\mathbf k,\nu}+\frac{y_{\mathbf k,l'}^*v_{\mathbf k,\nu}}{\sqrt2}\Bigr)\Bigl(\frac{u_{\mathbf k,\nu}^*y_{\mathbf k,l}}{\sqrt2}+v_{\mathbf k,\nu}^*z_{\mathbf k,l}\Bigr)\\
&\qquad\times\left[\Bigl(x_{\mathbf k,l'}^*u_{\mathbf k,\nu'}+\frac{y_{\mathbf k,l'}^*v_{\mathbf k,\nu'}}{\sqrt2}\Bigr)\Bigl(\frac{u_{\mathbf k,\nu'}^*y_{\mathbf k,l}}{\sqrt2}+v_{\mathbf k,\nu'}^*z_{\mathbf k,l}\Bigr)\right]^*\,e^{-iE^{\mathrm{SC},o}_{\mathbf k,\nu}|t_2-t_1|+iE^{\mathrm{SC},o}_{\mathbf k,\nu'}|t'_2-t'_1|}\\
&+\frac12\sum_{l=0}^{2}p^{\mathrm{SC},e}_{\mathbf k,l}e^{-\frac i2(E^{\mathrm{SC},e}_{\mathbf k,l}-2\xi_{\mathbf k})(t_1+t_2-t'_1-t'_2)}\,e^{\frac i2(E^{\mathrm{SC},e}_{\mathbf k,l}+2\xi_{\mathbf k})(|t_2-t_1|-|t'_2-t'_1|)}\\
&\quad\times\sum_{\nu,\nu'=\pm1}v_{\mathbf k,\nu}v_{\mathbf k,\nu'}^*\Bigl(\frac{u_{\mathbf k,\nu}^*y_{\mathbf k,l}}{\sqrt2}+v_{\mathbf k,\nu}^*z_{\mathbf k,l}\Bigr)\Bigl(\frac{u_{\mathbf k,\nu'}^*y_{\mathbf k,l}}{\sqrt2}+v_{\mathbf k,\nu'}^*z_{\mathbf k,l}\Bigr)^*\,e^{-iE^{\mathrm{SC},o}_{\mathbf k,\nu}|t_2-t_1|+iE^{\mathrm{SC},o}_{\mathbf k,\nu'}|t'_2-t'_1|}\\
&+\frac12p^{\mathrm{SC}}_{\mathbf k,2}\sum_{l=0}^{2}e^{-\frac i2(2\xi_{\mathbf k}-E^{\mathrm{SC},e}_{\mathbf k,l})(t_1+t_2-t'_1-t'_2)}\,e^{\frac i2(2\xi_{\mathbf k}+E^{\mathrm{SC},e}_{\mathbf k,l})(|t_2-t_1|-|t'_2-t'_1|)}\\
&\quad\times\sum_{\nu,\nu'=\pm1}u_{\mathbf k,\nu}^*u_{\mathbf k,\nu'}\Bigl(x_{\mathbf k,l}^*u_{\mathbf k,\nu}+\frac{y_{\mathbf k,l}^*v_{\mathbf k,\nu}}{\sqrt2}\Bigr)\Bigl(x_{\mathbf k,l}^*u_{\mathbf k,\nu'}+\frac{y_{\mathbf k,l}^*v_{\mathbf k,\nu'}}{\sqrt2}\Bigr)^*\,e^{-iE^{\mathrm{SC},o}_{\mathbf k,\nu}|t_2-t_1|+iE^{\mathrm{SC},o}_{\mathbf k,\nu'}|t'_2-t'_1|}\\
&+p^{\mathrm{SC}}_{\mathbf k,2}\left(\frac14+2\bigl[\Theta(t_2-t_1)\Theta(t'_2-t'_1)+\Theta(t_1-t_2)\Theta(t'_1-t'_2)\bigr]\right)\,e^{2i\xi_{\mathbf k}(|t_2-t_1|-|t'_2-t'_1|)}\\
&\quad\times\sum_{\nu,\nu'=\pm1}u_{\mathbf k,\nu}^*v_{\mathbf k,\nu}u_{\mathbf k,\nu'}v_{\mathbf k,\nu'}^*e^{-iE^{\mathrm{SC},o}_{\mathbf k,\nu}|t_2-t_1|+iE^{\mathrm{SC},o}_{\mathbf k,\nu'}|t'_2-t'_1|}\\
&+\sum_{\nu,\nu'=\pm1}p^{\mathrm{SC},o}_{\mathbf k,\nu}e^{-\frac i2(E^{\mathrm{SC},o}_{\mathbf k,\nu}-E^{\mathrm{SC},o}_{\mathbf k,\nu'})(t_1+t_2-t'_1-t'_2)}\,e^{\frac i2(E^{\mathrm{SC},o}_{\mathbf k,\nu}+E^{\mathrm{SC},o}_{\mathbf k,\nu'})(|t_2-t_1|-|t'_2-t'_1|)}\\
&\quad\times\Biggl\{\bigl[\Theta(t_2-t_1)\Theta(t'_2-t'_1)+\Theta(t_1-t_2)\Theta(t'_1-t'_2)\bigr]\sum_{l,l'=0}^{2}\\
&\qquad\quad\times\Bigl(\frac{u_{\mathbf k,\nu'}^*y_{\mathbf k,l}}{\sqrt2}+v_{\mathbf k,\nu'}^*z_{\mathbf k,l}\Bigr)\Bigl(x_{\mathbf k,l}^*u_{\mathbf k,\nu}+\frac{y_{\mathbf k,l}^*v_{\mathbf k,\nu}}{\sqrt2}\Bigr)\\
&\qquad\quad\times\left[\Bigl(\frac{u_{\mathbf k,\nu'}^*y_{\mathbf k,l'}}{\sqrt2}+v_{\mathbf k,\nu'}^*z_{\mathbf k,l'}\Bigr)\Bigl(x_{\mathbf k,l'}^*u_{\mathbf k,\nu}+\frac{y_{\mathbf k,l'}^*v_{\mathbf k,\nu}}{\sqrt2}\Bigr)\right]^*\,e^{-iE^{\mathrm{SC},e}_{\mathbf k,l}|t_2-t_1|+iE^{\mathrm{SC},e}_{\mathbf k,l'}|t'_2-t'_1|}\\
&\qquad-\frac12\bigl[\Theta(t_2-t_1)\Theta(t'_2-t'_1)+\Theta(t_1-t_2)\Theta(t'_1-t'_2)\bigr]\sum_{l=0}^{2}\Biggl[\\
&\qquad\quad \Bigl(\frac{u_{\mathbf k,\nu'}^*y_{\mathbf k,l}}{\sqrt2}+v_{\mathbf k,\nu'}^*z_{\mathbf k,l}\Bigr)\Bigl(x_{\mathbf k,l}^*u_{\mathbf k,\nu}+\frac{y_{\mathbf k,l}^*v_{\mathbf k,\nu}}{\sqrt2}\Bigr)u_{\mathbf k,\nu'}v_{\mathbf k,\nu}^*\,e^{-iE^{\mathrm{SC},e}_{\mathbf k,l}|t_2-t_1|+i2\xi_{\mathbf k}|t'_2-t'_1|}\\
&\qquad\quad+u_{\mathbf k,\nu'}^*v_{\mathbf k,\nu}\left[\Bigl(\frac{u_{\mathbf k,\nu'}^*y_{\mathbf k,l}}{\sqrt2}+v_{\mathbf k,\nu'}^*z_{\mathbf k,l}\Bigr)\Bigl(x_{\mathbf k,l}^*u_{\mathbf k,\nu}+\frac{y_{\mathbf k,l}^*v_{\mathbf k,\nu}}{\sqrt2}\Bigr)\right]^*\,e^{-i2\xi_{\mathbf k}|t_2-t_1|+iE^{\mathrm{SC},e}_{\mathbf k,l}|t'_2-t'_1|}\Biggr]\\
&\qquad+\left(2+\frac14\bigl[\Theta(t_2-t_1)\Theta(t'_2-t'_1)+\Theta(t_1-t_2)\Theta(t'_1-t'_2)\bigr]\right)\\
&\qquad\quad\times|u_{\mathbf k,\nu'}|^2|v_{\mathbf k,\nu}|^2e^{-2i\xi_{\mathbf k}(|t_2-t_1|-|t'_2-t'_1|)}\Biggr\}.
\end{aligned}
}
 \hypertarget{sm-eq-sc-carpes-center-times}{}\hypertarget{sm-eq-sc-carpes-relative-times}{}\hypertarget{sm-eq-sc-carpes-time-domain-bounds}{} 
Introducing the center and relative times as in \hyperlink{sm-eq-c22}{Eq. (D23)}, the center-time integral at fixed relative times gives

\hypertarget{sm-eq-sc-carpes-center-average}{}%
\FittedEquation{M11}{%
\begin{aligned}
&\lim_{\mathcal T\to\infty}\frac{1}{\mathcal T^2}
\prod_{j=1}^{2}\left[\int_{-(\mathcal T-|t_{rj}|)/2}^{(\mathcal T-|t_{rj}|)/2}dt_{cj}\right]
G^{(2)<}_{\mathrm{SC},\sigma_1\sigma_2}(\mathbf k,-\mathbf k;t_1,t_2,t'_1,t'_2)\\
&=\frac{i^2}{2}\sum_{\lambda,\lambda'',\Xi_{\mathrm{SC}}}^{\pi_{\mathbf k,\lambda''}=\pi_{\mathbf k,\lambda}}
p^{\mathrm{SC}}_{\mathbf k,\lambda}\,
\Bigl[|M^{\sigma_2\sigma_1}_{\lambda,\lambda'';\Xi_{\mathrm{SC}}}(\mathbf k)|^2
e^{-i(E^{\mathrm{SC}}_{\mathbf k,\lambda}-E^{\Xi_{\mathrm{SC}}}_{\mathbf k})t_{r1}}
e^{-i(E^{\Xi_{\mathrm{SC}}}_{\mathbf k}-E^{\mathrm{SC}}_{\mathbf k,\lambda''})t_{r2}}\\
&\qquad+|M^{\sigma_1\sigma_2}_{\lambda,\lambda'';\Xi_{\mathrm{SC}}}(-\mathbf k)|^2
e^{-i(E^{\Xi_{\mathrm{SC}}}_{\mathbf k}-E^{\mathrm{SC}}_{\mathbf k,\lambda''})t_{r1}}
e^{-i(E^{\mathrm{SC}}_{\mathbf k,\lambda}-E^{\Xi_{\mathrm{SC}}}_{\mathbf k})t_{r2}}\Bigr].
\end{aligned}
}
 
The center-time integral eliminates terms with phase factors \(e^{\pm i(E^{\Xi_{\mathrm{SC}}}_{\mathbf k}-E^{\Xi'_{\mathrm{SC}}}_{\mathbf k})(t_{c1}-t_{c2})}\) arising from distinct degenerate subspaces \(\Xi_{\mathrm{SC}}\neq\Xi_{\mathrm{SC}}'\) with different intermediate energies, \(E^{\Xi_{\mathrm{SC}}}_{\mathbf k}\neq E^{\Xi'_{\mathrm{SC}}}_{\mathbf k}\). Intermediate states within the same degenerate SC eigenspace \(\Xi_{\mathrm{SC}}\) remain coherent. Introducing the total and relative energies as in \hyperlink{sm-eq-c24}{Eq. (D25)}, the remaining Fourier transforms over the relative times yield

 \hypertarget{sm-eq-sc-carpes-total-energy}{}\hypertarget{sm-eq-sc-carpes-relative-energy}{} 
\hypertarget{sm-eq-sc-spectral-sequential}{}%
\FittedEquation{M12}{%
\begin{aligned}
&A^{(2)<}_{\mathrm{SC},\sigma_1\sigma_2}(\mathbf k,-\mathbf k;\Omega,\omega_{\mathrm r})
=\frac12\sum_{\lambda,\lambda'',\Xi_{\mathrm{SC}}}^{\pi_{\mathbf k,\lambda''}=\pi_{\mathbf k,\lambda}}
p^{\mathrm{SC}}_{\mathbf k,\lambda}\,
\delta(\Omega-E^{\mathrm{SC}}_{\mathbf k,\lambda}+E^{\mathrm{SC}}_{\mathbf k,\lambda''})\\
&\quad\times\Bigl[|M^{\sigma_2\sigma_1}_{\lambda,\lambda'';\Xi_{\mathrm{SC}}}(\mathbf k)|^2
\delta\!\left(\omega_{\mathrm r}-\omega^{\Xi_{\mathrm{SC}}}_{\mathbf k,\lambda\lambda''}\right)
+|M^{\sigma_1\sigma_2}_{\lambda,\lambda'';\Xi_{\mathrm{SC}}}(-\mathbf k)|^2
\delta\!\left(\omega_{\mathrm r}+\omega^{\Xi_{\mathrm{SC}}}_{\mathbf k,\lambda\lambda''}\right)\Bigr].
\end{aligned}
}
 
\hypertarget{sm-sc-carpes-explicit-spin-spectra}{}For generic back-to-back momenta with \(\mathcal J_{\mathbf k}>0\), a nonzero local pair field \(\Delta_d\ne0\) and no additional intermediate-energy degeneracies, the two-particle removal spectral function at finite temperature for the equal-spin channel \(\sigma_1=\sigma_2=\sigma\) is

\hypertarget{sm-eq-sc-carpes-equal-spin-spectrum}{}%
\FittedEquation{M13}{%
\begin{aligned}
&A^{(2)<}_{\mathrm{SC},\sigma\sigma}(\mathbf k,-\mathbf k;\Omega,\omega_{\mathrm r})=\frac12\sum_{s=\pm1}\Biggl\{\sum_{l=0}^{2}\sum_{\nu=\pm1}\Biggl[p^{\mathrm{SC},e}_{\mathbf k,l}|v_{\mathbf k,\nu}|^2Z^{\mathrm{SC},-}_{\mathbf k,\nu l}\delta(\Omega-E^{\mathrm{SC},e}_{\mathbf k,l}+2\xi_{\mathbf k})\\
&\quad+p^{\mathrm{SC}}_{\mathbf k,2}|u_{\mathbf k,\nu}|^2Z^{\mathrm{SC},+}_{\mathbf k,\nu l}\delta(\Omega-2\xi_{\mathbf k}+E^{\mathrm{SC},e}_{\mathbf k,l})\Biggr]\delta\!\left[\omega_{\mathrm r}-s\left(\frac{E^{\mathrm{SC},e}_{\mathbf k,l}}{2}+\xi_{\mathbf k}-E^{\mathrm{SC},o}_{\mathbf k,\nu}\right)\right]\\
&\qquad+p^{\mathrm{SC}}_{\mathbf k,2}\sum_{\nu=\pm1}\left|u_{\mathbf k,\nu}v_{\mathbf k,\nu}\right|^2\delta(\Omega)\delta\!\left[\omega_{\mathrm r}-s\left(2\xi_{\mathbf k}-E^{\mathrm{SC},o}_{\mathbf k,\nu}\right)\right]\\
&\qquad+\sum_{\nu,\nu'=\pm1}p^{\mathrm{SC},o}_{\mathbf k,\nu}\Biggl[\sum_{l=0}^{2}Z^{\mathrm{SC},-}_{\mathbf k,\nu' l}Z^{\mathrm{SC},+}_{\mathbf k,\nu l}\delta\!\left[\omega_{\mathrm r}-s\left(\frac{E^{\mathrm{SC},o}_{\mathbf k,\nu}+E^{\mathrm{SC},o}_{\mathbf k,\nu'}}{2}-E^{\mathrm{SC},e}_{\mathbf k,l}\right)\right]\\
&\quad+\frac54|u_{\mathbf k,\nu'}|^2|v_{\mathbf k,\nu}|^2\delta\!\left[\omega_{\mathrm r}-s\left(\frac{E^{\mathrm{SC},o}_{\mathbf k,\nu}+E^{\mathrm{SC},o}_{\mathbf k,\nu'}}{2}-2\xi_{\mathbf k}\right)\right]\Biggr]\delta(\Omega-E^{\mathrm{SC},o}_{\mathbf k,\nu}+E^{\mathrm{SC},o}_{\mathbf k,\nu'})\Biggr\},
\end{aligned}
}
 
and for the opposite-spin channel \(\sigma_1=\bar\sigma_2=\sigma\) is

\hypertarget{sm-eq-sc-carpes-opposite-spin-spectrum}{}%
\FittedEquation{M14}{%
\begin{aligned}
&A^{(2)<}_{\mathrm{SC},\sigma\bar\sigma}(\mathbf k,-\mathbf k;\Omega,\omega_{\mathrm r})=\frac12\sum_{s=\pm1}\Biggl\{\sum_{l,l'=0}^{2}\sum_{\nu=\pm1}p^{\mathrm{SC},e}_{\mathbf k,l}Z^{\mathrm{SC},+}_{\mathbf k,\nu l'}Z^{\mathrm{SC},-}_{\mathbf k,\nu l}\\
&\quad\times\delta(\Omega-E^{\mathrm{SC},e}_{\mathbf k,l}+E^{\mathrm{SC},e}_{\mathbf k,l'})\delta\!\left[\omega_{\mathrm r}-s\left(\frac{E^{\mathrm{SC},e}_{\mathbf k,l}+E^{\mathrm{SC},e}_{\mathbf k,l'}}{2}-E^{\mathrm{SC},o}_{\mathbf k,\nu}\right)\right]\\
&\qquad+\frac12\sum_{l=0}^{2}\sum_{\nu=\pm1}\Biggl[p^{\mathrm{SC},e}_{\mathbf k,l}|v_{\mathbf k,\nu}|^2Z^{\mathrm{SC},-}_{\mathbf k,\nu l}\delta(\Omega-E^{\mathrm{SC},e}_{\mathbf k,l}+2\xi_{\mathbf k})\\
&\quad+p^{\mathrm{SC}}_{\mathbf k,2}|u_{\mathbf k,\nu}|^2Z^{\mathrm{SC},+}_{\mathbf k,\nu l}\delta(\Omega-2\xi_{\mathbf k}+E^{\mathrm{SC},e}_{\mathbf k,l})\Biggr]\delta\!\left[\omega_{\mathrm r}-s\left(\frac{E^{\mathrm{SC},e}_{\mathbf k,l}}{2}+\xi_{\mathbf k}-E^{\mathrm{SC},o}_{\mathbf k,\nu}\right)\right]\\
&\qquad+\frac94p^{\mathrm{SC}}_{\mathbf k,2}\sum_{\nu=\pm1}\left|u_{\mathbf k,\nu}v_{\mathbf k,\nu}\right|^2\delta(\Omega)\delta\!\left[\omega_{\mathrm r}-s\left(2\xi_{\mathbf k}-E^{\mathrm{SC},o}_{\mathbf k,\nu}\right)\right]\\
&\qquad+\sum_{\nu,\nu'=\pm1}p^{\mathrm{SC},o}_{\mathbf k,\nu}\Biggl[\sum_{l=0}^{2}Z^{\mathrm{SC},-}_{\mathbf k,\nu' l}Z^{\mathrm{SC},+}_{\mathbf k,\nu l}\delta\!\left[\omega_{\mathrm r}-s\left(\frac{E^{\mathrm{SC},o}_{\mathbf k,\nu}+E^{\mathrm{SC},o}_{\mathbf k,\nu'}}{2}-E^{\mathrm{SC},e}_{\mathbf k,l}\right)\right]\\
&\quad+\frac94|u_{\mathbf k,\nu'}|^2|v_{\mathbf k,\nu}|^2\delta\!\left[\omega_{\mathrm r}-s\left(\frac{E^{\mathrm{SC},o}_{\mathbf k,\nu}+E^{\mathrm{SC},o}_{\mathbf k,\nu'}}{2}-2\xi_{\mathbf k}\right)\right]\Biggr]\delta(\Omega-E^{\mathrm{SC},o}_{\mathbf k,\nu}+E^{\mathrm{SC},o}_{\mathbf k,\nu'})\Biggr\}.
\end{aligned}
}
 
Here \(Z^{\mathrm{SC},\pm}_{\mathbf k,\nu l}\) are the one-electron transition weights defined in \hyperlink{sm-eq-sc-spectral-transition-weights}{Eqs. (69)--(70)}. The sum over \(s=\pm1\) includes the two removal orderings.

For \(\mathcal J_{\mathbf k}=0\) and a nonzero pair field, the spectral function reduces to the conventional BCS limit in \hyperlink{sm-eq-b26}{Eq. (C27)}, which cannot be obtained by directly taking \(\mathcal J_{\mathbf k}\to0\) in \hyperlink{sm-eq-sc-carpes-equal-spin-spectrum}{Eqs. (M13)--(M14)}. The reason is that additional energy degeneracies among the intermediate states require their amplitudes to be summed coherently before taking the modulus squared.
 
\hypertarget{sm-sc-singlet-carpes}{}
\subsection*{M.2. Spin-singlet two-particle and pair-removal spectral functions}

\hypertarget{sm-sc-simultaneous-removal}{}For generic back-to-back momenta, the spin-singlet pair-removal correlation of the correlated SC state has the Lehmann representation

\hypertarget{sm-eq-sc-pair-time-lehmann}{}%
\FittedEquation{M15}{%
\begin{aligned}
S^<_{\mathrm{SC},\Delta}(\mathbf k;t,t')
&\equiv\langle\hat\Delta_{\mathbf k}^{\dagger}(t')\hat\Delta_{\mathbf k}(t)\rangle_{\mathrm{SC}}\\
&=\sum_{\lambda,\lambda''}^{\pi_{\mathbf k,\lambda''}=\pi_{\mathbf k,\lambda}}
p^{\mathrm{SC}}_{\mathbf k,\lambda}
|\langle\pi_{\mathbf k,\lambda''}|\hat\Delta_{\mathbf k}|\pi_{\mathbf k,\lambda}\rangle|^2
e^{-i(E^{\mathrm{SC}}_{\mathbf k,\lambda}-E^{\mathrm{SC}}_{\mathbf k,\lambda''})(t-t')}.
\end{aligned}
}
 
The pair operator preserves spin and block fermion parity. Using \(\langle e_{\mathbf k,l'}|\hat\Delta_{\mathbf k}|e_{\mathbf k,l}\rangle=x_{\mathbf k,l'}^*y_{\mathbf k,l}+y_{\mathbf k,l'}^*z_{\mathbf k,l}\) and \(\langle o_{\pm\mathbf k\sigma,\nu'}|\hat\Delta_{\mathbf k}|o_{\pm\mathbf k\sigma,\nu}\rangle=u_{\mathbf k,\nu'}^*v_{\mathbf k,\nu}/\sqrt2\) gives

\hypertarget{sm-eq-sc-pair-time-explicit}{}%
\FittedEquation{M16}{%
\begin{aligned}
S^<_{\mathrm{SC},\Delta}(\mathbf k;t,0)
&=\sum_{l,l'=0}^{2}p^{\mathrm{SC},e}_{\mathbf k,l}
|x_{\mathbf k,l'}^*y_{\mathbf k,l}+y_{\mathbf k,l'}^*z_{\mathbf k,l}|^2
e^{-i(E^{\mathrm{SC},e}_{\mathbf k,l}-E^{\mathrm{SC},e}_{\mathbf k,l'})t}\\
&\quad+4\times\frac12\sum_{\nu,\nu'=\pm1}p^{\mathrm{SC},o}_{\mathbf k,\nu}
|u_{\mathbf k,\nu'}^*v_{\mathbf k,\nu}|^2
e^{-i(E^{\mathrm{SC},o}_{\mathbf k,\nu}-E^{\mathrm{SC},o}_{\mathbf k,\nu'})t}.
\end{aligned}
}
 
Fourier transformation gives the nonnegative pair-removal spectral function

\hypertarget{sm-eq-sc-pair-lehmann}{}%
\FittedEquation{M17}{%
\begin{aligned}
S^<_{\mathrm{SC},\Delta}(\mathbf k,\Omega)
&\equiv\int_{-\infty}^{\infty}\frac{dt}{2\pi}\,e^{i\Omega t}S^<_{\mathrm{SC},\Delta}(\mathbf k;t,0)\\
&=\sum_{\lambda,\lambda''}^{\pi_{\mathbf k,\lambda''}=\pi_{\mathbf k,\lambda}}
p^{\mathrm{SC}}_{\mathbf k,\lambda}
|\langle\pi_{\mathbf k,\lambda''}|\hat\Delta_{\mathbf k}|\pi_{\mathbf k,\lambda}\rangle|^2
\delta(\Omega-E^{\mathrm{SC}}_{\mathbf k,\lambda}+E^{\mathrm{SC}}_{\mathbf k,\lambda''})\\
&=\sum_{l,l'=0}^{2}p^{\mathrm{SC},e}_{\mathbf k,l}
|x_{\mathbf k,l'}^*y_{\mathbf k,l}+y_{\mathbf k,l'}^*z_{\mathbf k,l}|^2
\delta(\Omega-E^{\mathrm{SC},e}_{\mathbf k,l}+E^{\mathrm{SC},e}_{\mathbf k,l'})\\
&\quad+2\sum_{\nu,\nu'=\pm1}p^{\mathrm{SC},o}_{\mathbf k,\nu}
|u_{\mathbf k,\nu'}^*v_{\mathbf k,\nu}|^2
\delta(\Omega-E^{\mathrm{SC},o}_{\mathbf k,\nu}+E^{\mathrm{SC},o}_{\mathbf k,\nu'}).
\end{aligned}
}
 
\hypertarget{sm-eq-sc-pair-sumrules}{}which satisfies the equal-time sum rule \(\int d\Omega\,S^<_{\mathrm{SC},\Delta}(\mathbf k,\Omega) =\langle\hat\Delta_{\mathbf k}^{\dagger}\hat\Delta_{\mathbf k}\rangle_{\mathrm{SC}}\).

The correlated SC state still has spin rotation symmetry, and the spin-singlet component of the two-particle lesser Green's function is (cf. \hyperlink{sm-eq-c34}{Eq. (D39)})

\hypertarget{sm-eq-sc-singlet-lesser-projection}{}%
\FittedEquation{M18}{%
G^{(2)<}_{\mathrm{SC},\Delta}(\mathbf k;t_1,t_2,t'_1,t'_2) =2G^{(2)<}_{\mathrm{SC},\uparrow\downarrow}(\mathbf k,-\mathbf k;t_1,t_2,t'_1,t'_2) -G^{(2)<}_{\mathrm{SC},\uparrow\uparrow}(\mathbf k,-\mathbf k;t_1,t_2,t'_1,t'_2).
}
 
\hypertarget{sm-eq-sc-singlet-equal-time}{}The equal-time relation is \(G^{(2)<}_{\mathrm{SC},\Delta}(\mathbf k;t,t,t',t')=i^2S^<_{\mathrm{SC},\Delta}(\mathbf k;t,t')\).

Applying the four-time Fourier transform and long-duration normalization gives the nonnegative spin-singlet two-particle removal spectral function

\hypertarget{sm-eq-sc-sequential-singlet}{}%
\FittedEquation{M19}{%
A^{(2)<}_{\mathrm{SC},\Delta}(\mathbf k,-\mathbf k;\Omega,\omega_{\mathrm r})=2A^{(2)<}_{\mathrm{SC},\uparrow\downarrow}(\mathbf k,-\mathbf k;\Omega,\omega_{\mathrm r})-A^{(2)<}_{\mathrm{SC},\uparrow\uparrow}(\mathbf k,-\mathbf k;\Omega,\omega_{\mathrm r}).
}
 
For generic back-to-back momenta with \(\mathcal J_{\mathbf k}>0\), a nonzero local pair field \(\Delta_d\ne0\) and no additional intermediate-energy degeneracies, substituting \hyperlink{sm-eq-sc-carpes-equal-spin-spectrum}{Eqs. (M13)--(M14)} into \hyperlink{sm-eq-sc-sequential-singlet}{Eq. (M19)} gives

\hypertarget{sm-eq-sc-carpes-singlet-explicit}{}%
\FittedEquation{M20}{%
\begin{aligned}
&A^{(2)<}_{\mathrm{SC},\Delta}(\mathbf k,-\mathbf k;\Omega,\omega_{\mathrm r})=\frac12\sum_{s=\pm1}\Biggl\{2\sum_{l,l'=0}^{2}\sum_{\nu=\pm1}p^{\mathrm{SC},e}_{\mathbf k,l}
Z^{\mathrm{SC},+}_{\mathbf k,\nu l'}
Z^{\mathrm{SC},-}_{\mathbf k,\nu l}\\
&\qquad\times\delta(\Omega-E^{\mathrm{SC},e}_{\mathbf k,l}+E^{\mathrm{SC},e}_{\mathbf k,l'})
\delta\!\left[\omega_{\mathrm r}-s\left(\frac{E^{\mathrm{SC},e}_{\mathbf k,l}+E^{\mathrm{SC},e}_{\mathbf k,l'}}2-E^{\mathrm{SC},o}_{\mathbf k,\nu}\right)\right]\\
&\quad+\frac72p^{\mathrm{SC}}_{\mathbf k,2}\sum_{\nu=\pm1}|u_{\mathbf k,\nu}v_{\mathbf k,\nu}|^2
\delta(\Omega)\delta\!\left[\omega_{\mathrm r}-s(2\xi_{\mathbf k}-E^{\mathrm{SC},o}_{\mathbf k,\nu})\right]\\
&\quad+\sum_{\nu,\nu'=\pm1}p^{\mathrm{SC},o}_{\mathbf k,\nu}
\Biggl[\sum_{l=0}^{2}Z^{\mathrm{SC},-}_{\mathbf k,\nu' l}
Z^{\mathrm{SC},+}_{\mathbf k,\nu l}\delta\!\left[\omega_{\mathrm r}-s\left(\frac{E^{\mathrm{SC},o}_{\mathbf k,\nu}+E^{\mathrm{SC},o}_{\mathbf k,\nu'}}2-E^{\mathrm{SC},e}_{\mathbf k,l}\right)\right]\\
&\qquad+\frac{13}{4}|u_{\mathbf k,\nu'}^*v_{\mathbf k,\nu}|^2
\delta\!\left[\omega_{\mathrm r}-s\left(\frac{E^{\mathrm{SC},o}_{\mathbf k,\nu}+E^{\mathrm{SC},o}_{\mathbf k,\nu'}}2-2\xi_{\mathbf k}\right)\right]\Biggr]\delta(\Omega-E^{\mathrm{SC},o}_{\mathbf k,\nu}+E^{\mathrm{SC},o}_{\mathbf k,\nu'})\Biggr\}.
\end{aligned}
}
 
The three contributions describe transitions within the even-fermion-parity subspace \(\{|e_{\mathbf k,l}\rangle\}\), within the even-fermion-parity subspace \(\{|2_{\mathbf k,t_{\pm1}}\rangle,|2_{\mathbf k,t_0}\rangle,|2_{\pm\mathbf k}\rangle\}\), and within the odd-fermion-parity subspace \(\{|o_{\pm\mathbf k\sigma,\nu}\rangle\}\). The coefficients are \(7/2=2(9/4)-1\) and \(13/4=2(9/4)-5/4\). Integrating over relative energy yields

\hypertarget{sm-eq-sc-carpes-singlet-integrated}{}%
\FittedEquation{M21}{%
\begin{aligned}
\overline A^{(2)<}_{\mathrm{SC},\Delta}(\mathbf k,\Omega)
&\equiv\int d\omega_{\mathrm r}\,A^{(2)<}_{\mathrm{SC},\Delta}(\mathbf k,-\mathbf k;\Omega,\omega_{\mathrm r})\\
&=2\sum_{l,l'=0}^{2}\sum_{\nu=\pm1}p^{\mathrm{SC},e}_{\mathbf k,l}
Z^{\mathrm{SC},+}_{\mathbf k,\nu l'}
Z^{\mathrm{SC},-}_{\mathbf k,\nu l}\delta(\Omega-E^{\mathrm{SC},e}_{\mathbf k,l}+E^{\mathrm{SC},e}_{\mathbf k,l'})\\
&\quad+\frac72p^{\mathrm{SC}}_{\mathbf k,2}\sum_{\nu=\pm1}|u_{\mathbf k,\nu}v_{\mathbf k,\nu}|^2\delta(\Omega)\\
&+\sum_{\nu,\nu'=\pm1}p^{\mathrm{SC},o}_{\mathbf k,\nu}\Biggl[\sum_{l=0}^{2}
Z^{\mathrm{SC},-}_{\mathbf k,\nu' l}Z^{\mathrm{SC},+}_{\mathbf k,\nu l}+\frac{13}{4}|u_{\mathbf k,\nu'}^*v_{\mathbf k,\nu}|^2\Biggr]
\delta(\Omega-E^{\mathrm{SC},o}_{\mathbf k,\nu}+E^{\mathrm{SC},o}_{\mathbf k,\nu'}).
\end{aligned}
}
 
In the correlated SC state, the relation between the spin-singlet two-particle removal and the pair-removal spectral functions is (cf. \hyperlink{sm-eq-c39}{Eq. (D43)})

\hypertarget{sm-eq-sc-singlet-pair-relation}{}%
\FittedEquation{M22}{%
\overline A^{(2)<}_{\mathrm{SC},\Delta}(\mathbf k,\Omega)
=S^<_{\mathrm{SC},\Delta}(\mathbf k,\Omega)
+\mathcal I_{\mathrm{SC},\Delta}(\mathbf k,\Omega),
}
 
where the interference correction is

\hypertarget{sm-eq-sc-singlet-interference-correction}{}%
\FittedEquation{M23}{%
\begin{aligned}
\mathcal I_{\mathrm{SC},\Delta}(\mathbf k,\Omega)
&=-\frac12\sum_{\lambda,\lambda''}^{\pi_{\mathbf k,\lambda''}=\pi_{\mathbf k,\lambda}}
p^{\mathrm{SC}}_{\mathbf k,\lambda}
\delta(\Omega-E^{\mathrm{SC}}_{\mathbf k,\lambda}+E^{\mathrm{SC}}_{\mathbf k,\lambda''})\\
&\quad\times\operatorname{Re}\sum_{s=\pm1}\sum_{\Xi_{\mathrm{SC}}\ne\Xi'_{\mathrm{SC}}}
\mathcal M^{s}_{\mathbf k,\lambda\lambda'';\Xi_{\mathrm{SC}}}
\mathcal M^{s*}_{\mathbf k,\lambda\lambda'';\Xi'_{\mathrm{SC}}}.
\end{aligned}
}
 
Here \(\Xi_{\mathrm{SC}}\) and \(\Xi'_{\mathrm{SC}}\) label distinct intermediate-state subspaces. The amplitudes within each degenerate energy subspace are summed coherently before taking the modulus squared, and the two ordering amplitudes are

\hypertarget{sm-eq-sc-singlet-ordering-amplitude}{}%
\FittedEquation{M24}{%
\mathcal M^{s}_{\mathbf k,\lambda\lambda'';\Xi_{\mathrm{SC}}}
=\frac{s}{\sqrt2}\Bigl[
M^{\downarrow\uparrow}_{\lambda,\lambda'';\Xi_{\mathrm{SC}}}(s\mathbf k)
-M^{\uparrow\downarrow}_{\lambda,\lambda'';\Xi_{\mathrm{SC}}}(s\mathbf k)\Bigr].
}
 
The signs \(s=+1\) and \(-1\) label the removal orderings \(21\) and \(12\), respectively. Completeness gives \(\sum_{\Xi_{\mathrm{SC}}}\mathcal M^{s}_{\mathbf k,\lambda\lambda'';\Xi_{\mathrm{SC}}}=-s\langle\pi_{\mathbf k,\lambda''}|\hat\Delta_{\mathbf k}|\pi_{\mathbf k,\lambda}\rangle\). For fixed initial and final states and a given removal ordering, long-duration sequential removal gives \(\sum_{\Xi_{\mathrm{SC}}}|\mathcal M^{s}_{\mathbf k,\lambda\lambda'';\Xi_{\mathrm{SC}}}|^2\), whereas simultaneous pair removal gives \(\left|\sum_{\Xi_{\mathrm{SC}}}\mathcal M^{s}_{\mathbf k,\lambda\lambda'';\Xi_{\mathrm{SC}}}\right|^2\). Their difference, arising from interference between paths through distinct intermediate-state subspaces of different intermediate energies, gives the interference correction.

The interference correction also reflects the order of the time integrations. Integrating over \(\omega_{\mathrm r}\) enforces \(t_1-t'_1=t_2-t'_2\), whereas simultaneous pair removal imposes \(t_1=t_2\) and \(t'_1=t'_2\). The long-duration center-time average removes interference between different intermediate energies before the relative-energy integral is taken. Consequently, that integral does not generally recover the simultaneous pair-removal spectrum.

For \(\mathcal J_{\mathbf k}=0\) and a nonzero pair field, the interference correction reduces to the conventional BCS limit

\hypertarget{sm-eq-sc-singlet-bcs-interference}{}%
\FittedEquation{M25}{%
\mathcal I_{\mathrm{BCS},\Delta}(\mathbf k,\Omega)=4|u_{\mathbf k}v_{\mathbf k}|^2f_{\mathrm F}(E^{\mathrm{BCS}}_{\mathbf k})\bigl[1-f_{\mathrm F}(E^{\mathrm{BCS}}_{\mathbf k})\bigr]\delta(\Omega),
}
 
whose weight is exponentially small at low temperatures \(k_{\mathrm B}T\ll E^{\mathrm{BCS}}_{\mathbf k}\) and vanishes at zero temperature.
 
\hypertarget{sm-appendix-sc-nodal-dos}{}\hypertarget{sm-sc-nodal-density-of-states}{}
\section*{Appendix N. Nodal density of states of the correlated SC state}

\hypertarget{sm-sc-dos-exact}{}
\subsection*{N.1. Exact one-particle density of states}

The exact finite-temperature one-particle density of states (DOS) per spin in the correlated SC state is obtained from \hyperlink{sm-eq-sc-spectral-one}{Eq. (76)}:

\hypertarget{sm-eq-sc-dos-definition}{}%
\FittedEquation{N1}{%
\begin{aligned}
\rho_{\mathrm{SC}}(\omega,T)&\equiv\int_{\mathrm{BZ}}\frac{d^2\mathbf k}{(2\pi)^2}\,A_{\mathrm{SC}}(\mathbf k,\omega;T)\\
&=\int_{\mathrm{BZ}}\frac{d^2\mathbf k}{(2\pi)^2}\sum_{\nu=\pm1}
\Biggl\{\sum_{l=0}^{2}\bigl(p^{\mathrm{SC},e}_{\mathbf k,l}+p^{\mathrm{SC},o}_{\mathbf k,\nu}\bigr)
\Bigl[Z^{\mathrm{SC},+}_{\mathbf k,\nu l}\delta(\omega-E^{\mathrm{SC},o}_{\mathbf k,\nu}+E^{\mathrm{SC},e}_{\mathbf k,l})\\
&\qquad\qquad+Z^{\mathrm{SC},-}_{\mathbf k,\nu l}\delta(\omega+E^{\mathrm{SC},o}_{\mathbf k,\nu}-E^{\mathrm{SC},e}_{\mathbf k,l})\Bigr]\\
&\quad+\frac52\bigl(p^{\mathrm{SC}}_{\mathbf k,2}+p^{\mathrm{SC},o}_{\mathbf k,\nu}\bigr)
\Bigl[|v_{\mathbf k,\nu}|^2\delta(\omega{-}E^{\mathrm{SC},o}_{\mathbf k,\nu}{+}2\xi_{\mathbf k})+|u_{\mathbf k,\nu}|^2\delta(\omega{+}E^{\mathrm{SC},o}_{\mathbf k,\nu}{-}2\xi_{\mathbf k})\Bigr]\Biggr\}.
\end{aligned}
}
 
\hypertarget{sm-sc-dos-local-scaling}{}
\subsection*{N.2. Local expansion near the nodes}

Performing the local expansion near the node \(\mathbf k_n=(k_n,k_n)\) in the moving tangent--normal frame of \hyperlink{sm-dos-i-coordinates}{Appendix E.2} gives

\hypertarget{sm-eq-sc-dos-local-expansion}{}%
\FittedEquation{N2}{%
\begin{aligned}
\xi_{\mathbf k}&=v_n k_\perp+O(k_\perp k_\parallel^2,k_\perp^2),
\end{aligned}
}
 
\hypertarget{sm-eq-sc-dos-local-rml-expansion}{}%
\FittedEquation{N3}{%
\begin{aligned}
\mathcal J_{\mathbf k}&=u_n k_\parallel^2+O(k_\perp k_\parallel^2,k_\parallel^4),
\end{aligned}
}
 
\hypertarget{sm-eq-sc-dos-local-pairing-expansion}{}%
\FittedEquation{N4}{%
\begin{aligned}
\sqrt2\Delta_{\mathbf k,d}&=v_\Delta k_\parallel+O(k_\perp k_\parallel,k_\parallel^3),
\end{aligned}
}
 
where the coordinates \(k_\parallel\) and \(k_\perp\) are the signed arc length from the node and the normal displacement from the Fermi-surface contour, respectively. Here \(v_n=|\boldsymbol\nabla_{\mathbf k}\xi_{\mathbf k}|_{\mathbf k_n}\), \(u_n=\mathcal J_d\sin^2 k_n/2\), and \(v_\Delta=\sqrt2\,\left.\partial_{k_\parallel}\Delta_{\mathbf k,d}\right|_{\mathbf k_n}=|\Delta_d\sin k_n|\), with the phase of \(\Delta_d\) and the tangential orientation chosen so that \(v_\Delta>0\). Comparing the linear SC pairing scale \(v_\Delta|k_\parallel|\) with the quadratic RML interaction scale \(u_n k_\parallel^2\) defines the crossover energy scale

\hypertarget{sm-eq-sc-dos-crossover-energy}{}%
\FittedEquation{N5}{%
E_*\equiv\frac{v_\Delta^2}{u_n}=\frac{2|\Delta_d|^2}{\mathcal J_d}.
}
 
The one-particle excitation gap \(\Delta_{\mathrm{1p}}^{\mathrm{SC}}(\mathbf k)\) reduces to the Dirac dispersion

\hypertarget{sm-eq-sc-dos-dirac-energy}{}%
\FittedEquation{N6}{%
E=\sqrt{v_n^2k_\perp^2+v_\Delta^2k_\parallel^2},
}
 
and

\hypertarget{sm-eq-sc-dos-rml-relative-scale}{}%
\FittedEquation{N7}{%
\frac{\mathcal J_{\mathbf k}}{E}\simeq\frac{E}{E_*}\frac{v_\Delta^2k_\parallel^2}{E^2}\lesssim\frac{E}{E_*}.
}
 
For \(T,|\omega|\ll E_*\) and fixed nonzero \(\Delta_d\), the RML interaction is therefore subleading. To zeroth order in the RML interaction near the node, the retarded Green's function reduces to the BCS form of \hyperlink{sm-eq-sc-zero-temperature-green-bcs}{Eq. (74)}:

\hypertarget{sm-eq-sc-dos-dirac-retarded}{}%
\FittedEquation{N8}{%
\begin{aligned}
G_{\mathrm{SC}}^{R,(0)}(\mathbf k,\omega)
&=\frac{\omega+i0^++\xi_{\mathbf k}}{(\omega+i0^+)^2-\xi_{\mathbf k}^2-2|\Delta_{\mathbf k,d}|^2}\\
&\simeq\frac{\omega+i0^++v_nk_\perp}{(\omega+i0^+-E)(\omega+i0^++E)}\\
&=\frac12\left(1+\frac{v_nk_\perp}{E}\right)\frac{1}{\omega+i0^+-E}
+\frac12\left(1-\frac{v_nk_\perp}{E}\right)\frac{1}{\omega+i0^++E}.
\end{aligned}
}
 
Then the zeroth-order one-particle spectral function is

\hypertarget{sm-eq-sc-dos-dirac-spectral}{}%
\FittedEquation{N9}{%
A_{\mathrm{SC}}^{(0)}(\mathbf k,\omega)
=\frac12\left(1+\frac{v_nk_\perp}{E}\right)\delta(\omega-E)
+\frac12\left(1-\frac{v_nk_\perp}{E}\right)\delta(\omega+E).
}
 
Let \(v_nk_\perp=E\cos\vartheta\) and \(v_\Delta k_\parallel=E\sin\vartheta\), so \(dk_\perp\,dk_\parallel=E\,dE\,d\vartheta/(v_nv_\Delta)\). Performing the energy and angular integrals yields

\hypertarget{sm-eq-sc-dos-dirac-integral}{}%
\FittedEquation{N10}{%
\begin{aligned}
\rho_{\mathrm{SC}}^{(0)}(\omega,T)
&=4\times\frac1{(2\pi)^2v_nv_\Delta}
\int_0^\infty E\,dE\int_0^{2\pi}d\vartheta\,
\frac{1+\operatorname{sgn}(\omega)\cos\vartheta}{2}\delta(|\omega|-E)\\
&=\frac{|\omega|}{\pi v_nv_\Delta}.
\end{aligned}
}
 
\hypertarget{sm-sc-dos-scaling-function}{}
\subsection*{N.3. SC scaling function}

Introduce the dimensionless variables \(x=v_nk_\perp/T\), \(y=v_\Delta k_\parallel/T\), \(z=\omega/T\), and \(g_*=T/E_*=u_nT/v_\Delta^2\). The dimensionless SC block Hamiltonian near the nodes is

\hypertarget{sm-eq-sc-dos-scaled-hamiltonian}{}%
\FittedEquation{N11}{%
\frac{\hat h^{\mathrm{SC}}_{\mathbf k}}T
\simeq x\hat N_{\mathbf k}-2g_* y^2\hat\Delta_{\mathbf k}^{\dagger}\hat\Delta_{\mathbf k}
-\sqrt2y(\hat\Delta_{\mathbf k}^{\dagger}+\hat\Delta_{\mathbf k}).
}
 
Denote the corresponding dimensionless spectral function by \(a_{\mathrm{SC}}(x,y,z;g_*)\equiv T A_{\mathrm{SC}}(\xi_{\mathbf k},\mathcal J_{\mathbf k},\Delta_{\mathbf k,d},\omega;T)\), where the factor \(T\) comes from rescaling the spectral delta functions in \hyperlink{sm-eq-sc-spectral-one}{Eq. (76)}. With \(dk_\parallel\,dk_\perp=T^2\,dx\,dy/(v_nv_\Delta)\), the two-variable SC scaling function that retains the RML interaction in the local nodal expansion is defined by

\hypertarget{sm-eq-sc-dos-scaling}{}%
\FittedEquation{N12}{%
\mathcal R_{\mathrm{SC}}(z,g_*)\equiv\int_{-\infty}^{\infty}dx\int_{-\infty}^{\infty}dy\,a_{\mathrm{SC}}(x,y,z;g_*),
}
 
where the infinite integration ranges correspond to the leading-order nodal approximation. Summing the four symmetry-related nodes then gives

\hypertarget{sm-eq-sc-dos-scaling-density}{}%
\FittedEquation{N13}{%
\rho_{\mathrm{SC}}(\omega,T)\simeq\frac{T}{\pi^2v_nv_\Delta}\mathcal R_{\mathrm{SC}}(z,g_*).
}
 
In the limit \(T\to0\) at fixed \(z\) and fixed nonzero \(\Delta_d(0)\), \(g_*\to0\), and

\hypertarget{sm-eq-sc-dos-leading-scaling}{}%
\FittedEquation{N14}{%
\mathcal R_{\mathrm{SC}}(z,0)=\pi|z|.
}
 
The SC scaling function is even in \(z\), \(\mathcal R_{\mathrm{SC}}(z,g_*)=\mathcal R_{\mathrm{SC}}(-z,g_*)\), but has no constant term at \(z=g_*=0\): \(\mathcal R_{\mathrm{SC}}(0,0)=0\) (cf. \(\mathcal R_{\mathrm N}(z)\) in \hyperlink{sm-eq-dos-i-small-series}{Eq. (E22)}).

\hypertarget{sm-sc-dos-frequency-limits}{}
\subsection*{N.4. Small- and large-\(|\omega|/T\) expansions}

For \(T,|\omega|\ll E_*\), the leading RML interaction correction to the zeroth-order one-particle DOS per spin in \hyperlink{sm-eq-sc-dos-dirac-integral}{Eq. (N10)} follows by expanding the dimensionless spectral function \(a_{\mathrm{SC}}\) in \hyperlink{sm-eq-sc-dos-scaling}{Eq. (N12)} to first order in \(g_*\) at fixed nonzero \(z\). The zeroth-order normal and anomalous contractions are

\hypertarget{sm-eq-sc-dos-zeroth-order-occupation}{}%
\FittedEquation{N15}{%
n^{(0)}=\left.\left\langle\hat c_{\mathbf k\sigma}^{\dagger}\hat c_{\mathbf k\sigma}\right\rangle\right|_{g_*=0}
=\frac12\left[1-\frac{x}{\sqrt{x^2+y^2}}\tanh\!\left(\frac{\sqrt{x^2+y^2}}2\right)\right],
}
 
\hypertarget{sm-eq-sc-dos-zeroth-order-anomalous}{}%
\FittedEquation{N16}{%
F^{(0)}=\left.\left\langle\hat c_{\mathbf k\uparrow}\hat c_{-\mathbf k\downarrow}\right\rangle\right|_{g_*=0}
=\frac{y}{2\sqrt{x^2+y^2}}\tanh\!\left(\frac{\sqrt{x^2+y^2}}2\right).
}
 
Apply Wick's theorem to the zeroth-order BCS state. The RML interaction \(-2g_*y^2\hat\Delta_{\mathbf k}^{\dagger}\hat\Delta_{\mathbf k}\) shifts the normal and anomalous coefficients by \(\delta x=-g_*y^2 n^{(0)}\) and \(\delta y=2g_*y^2 F^{(0)}\), respectively, giving

\hypertarget{sm-eq-sc-dos-first-order-coefficients}{}%
\FittedEquation{N17}{%
X=x+\delta x=x-\frac{g_*y^2}{2}\left[1-\frac{x}{\sqrt{x^2+y^2}}\tanh\!\left(\frac{\sqrt{x^2+y^2}}{2}\right)\right],
}
 
\hypertarget{sm-eq-sc-dos-first-order-pairing-coefficient}{}%
\FittedEquation{N18}{%
Y=y+\delta y=y+\frac{g_*y^3}{\sqrt{x^2+y^2}}\tanh\!\left(\frac{\sqrt{x^2+y^2}}{2}\right).
}
 
Then the dimensionless spectral function becomes

\hypertarget{sm-eq-sc-dos-first-order-spectral}{}%
\FittedEquation{N19}{%
\begin{aligned}
a_{\mathrm{SC}}(x,y,z;g_*)
&=\frac12\left(1+\frac{X}{\sqrt{X^2+Y^2}}\right)\delta\!\left(z-\sqrt{X^2+Y^2}\right)\\
&\quad+\frac12\left(1-\frac{X}{\sqrt{X^2+Y^2}}\right)\delta\!\left(z+\sqrt{X^2+Y^2}\right)+O(g_*^2).
\end{aligned}
}
 
Changing variables from \((x,y)\) to \((X,Y)\) to first order in \(g_*\) gives

\hypertarget{sm-eq-sc-dos-first-order-measure}{}%
\FittedEquation{N20}{%
dx\,dy=\left[1-\partial_x\delta x-\partial_y\delta y+O(g_*^2)\right]\,dX\,dY,
}
 
where

\hypertarget{sm-eq-sc-dos-first-order-jacobian}{}%
\FittedEquation{N21}{%
\begin{aligned}
\frac{\partial_x\delta x+\partial_y\delta y}{g_*}
&=\frac{y^2(6x^2+5y^2)}{2(x^2+y^2)^{3/2}}\tanh\!\left(\frac{\sqrt{x^2+y^2}}{2}\right)\\
&\quad+\frac{y^2(x^2+2y^2)}{4(x^2+y^2)}\operatorname{sech}^2\!\left(\frac{\sqrt{x^2+y^2}}{2}\right).
\end{aligned}
}
 
Performing the integrals over \(X\) and \(Y\) gives

\hypertarget{sm-eq-sc-dos-first-correction}{}%
\FittedEquation{N22}{%
\mathcal R_{\mathrm{SC}}(z,g_*)
=\pi|z|\left\{1-\frac{7g_*}{16}\left[3|z|\tanh\!\left(\frac{|z|}{2}\right)+\frac{z^2}{2}\operatorname{sech}^2\!\left(\frac{|z|}{2}\right)\right]\right\}+O(g_*^2).
}
 
For \(|\omega|\ll T\ll E_*\), expanding the coefficient of \(g_*\) at small \(|z|\) gives (cf. \hyperlink{sm-eq-dos-i-small-frequency-result}{Eq. (E23)})

\hypertarget{sm-eq-sc-dos-small-z}{}%
\FittedEquation{N23}{%
\mathcal R_{\mathrm{SC}}(z,g_*)
=\pi|z|\left[1-\frac{7g_*}{8}z^2+\frac{7g_*}{64}z^4+O(g_* z^6)\right]+O(g_*^2).
}
 
For \(T\ll|\omega|\ll E_*\), expanding the coefficient of \(g_*\) at large \(|z|\) gives (cf. \hyperlink{sm-eq-dos-i-zero-temperature-result}{Eq. (E27)})

\hypertarget{sm-eq-sc-dos-large-z}{}%
\FittedEquation{N24}{%
\mathcal R_{\mathrm{SC}}(z,g_*)
=\pi|z|\left[1-\frac{21g_*|z|}{16}+\frac{7g_*}{8}(3|z|-z^2)e^{-|z|}
+O\!\left(g_*z^2e^{-2|z|}\right)\right]+O(g_*^2).
}
 
The leading relative correction is \(-21g_*|z|/16=-21|\omega|/(16E_*)\), whereas the thermal correction is exponentially small in \(|\omega|/T\).
 
\hypertarget{sm-appendix-sc-thermodynamics}{}\hypertarget{sm-sc-thermodynamics}{}
\section*{Appendix O. Thermodynamics of the correlated SC state}

\hypertarget{sm-sc-thermo-entropy}{}
\subsection*{O.1. Entropy, free energy, internal energy, and thermodynamic critical field}

The SC Gibbs entropy per site at fixed electron density \(n_e\) is

\hypertarget{sm-eq-sc-entropy}{}%
\FittedEquation{O1}{%
\begin{aligned}
s_{\mathrm{SC}}(T,n_e)&=-\frac1{N_s}\sum_{\{\mathbf k,-\mathbf k\}}\sum_{\lambda}
p^{\mathrm{SC}}_{\mathbf k,\lambda}\ln p^{\mathrm{SC}}_{\mathbf k,\lambda}\\
&=-\frac1{N_s}\sum_{\{\mathbf k,-\mathbf k\}}\left[
\sum_{l=0}^{2}p^{\mathrm{SC},e}_{\mathbf k,l}\ln p^{\mathrm{SC},e}_{\mathbf k,l}
+5p^{\mathrm{SC}}_{\mathbf k,2}\ln p^{\mathrm{SC}}_{\mathbf k,2}
+4\sum_{\nu=\pm1}p^{\mathrm{SC},o}_{\mathbf k,\nu}\ln p^{\mathrm{SC},o}_{\mathbf k,\nu}\right].
\end{aligned}
}
 
The SC Helmholtz free energy per site at fixed \(n_e\) [\hyperlink{sm-eq-sc-free-energy}{Eq. (J17)}] is

\hypertarget{sm-eq-sc-thermo-block-free-energy}{}%
\FittedEquation{O2}{%
f_{\mathrm{SC}}(T,n_e)=\frac{|\Delta_d|^2}{2J}-\frac{T}{N_s}\sum_{\{\mathbf k,-\mathbf k\}}\ln Z^{\mathrm{SC}}_{\mathbf k}+\mu_{\mathrm{SC}}n_e.
}
 
\hypertarget{sm-sc-thermo-energy}{}The SC internal energy per site at fixed \(n_e\) is

\hypertarget{sm-eq-sc-thermo-internal-energy}{}%
\FittedEquation{O3}{%
\begin{aligned}
u_{\mathrm{SC}}(T,n_e)&=f_{\mathrm{SC}}(T,n_e)+Ts_{\mathrm{SC}}(T,n_e)\\
&=\frac{|\Delta_d|^2}{2J}-\frac{T}{N_s}\sum_{\{\mathbf k,-\mathbf k\}}\ln Z^{\mathrm{SC}}_{\mathbf k}+\mu_{\mathrm{SC}}n_e
-\frac{T}{N_s}\sum_{\{\mathbf k,-\mathbf k\}}\sum_\lambda p^{\mathrm{SC}}_{\mathbf k,\lambda}\ln p^{\mathrm{SC}}_{\mathbf k,\lambda}\\
&=\frac{|\Delta_d|^2}{2J}-\frac{T}{N_s}\sum_{\{\mathbf k,-\mathbf k\}}\ln Z^{\mathrm{SC}}_{\mathbf k}+\mu_{\mathrm{SC}}n_e\\
&\quad+\frac{1}{N_s}\sum_{\{\mathbf k,-\mathbf k\}}\sum_\lambda p^{\mathrm{SC}}_{\mathbf k,\lambda}E^{\mathrm{SC}}_{\mathbf k,\lambda}
+\frac{T}{N_s}\sum_{\{\mathbf k,-\mathbf k\}}\ln Z^{\mathrm{SC}}_{\mathbf k}\\
&=\frac{|\Delta_d|^2}{2J}+\frac{1}{N_s}\sum_{\{\mathbf k,-\mathbf k\}}\sum_\lambda p^{\mathrm{SC}}_{\mathbf k,\lambda}E^{\mathrm{SC}}_{\mathbf k,\lambda}+\mu_{\mathrm{SC}}n_e,
\end{aligned}
}
 
where \(p^{\mathrm{SC}}_{\mathbf k,\lambda}=e^{-E^{\mathrm{SC}}_{\mathbf k,\lambda}/T}/Z^{\mathrm{SC}}_{\mathbf k}\). An equivalent expression is

\hypertarget{sm-eq-sc-thermo-interaction-energy}{}%
\FittedEquation{O4}{%
\begin{aligned}
u_{\mathrm{SC}}(T,n_e)&=\frac1{N_s}\sum_{\{\mathbf k,-\mathbf k\}}
\left\langle\varepsilon_{\mathbf k}\hat N_{\mathbf k}-2\mathcal J_{\mathbf k}\hat\Delta_{\mathbf k}^{\dagger}\hat\Delta_{\mathbf k}\right\rangle_{\mathrm{SC}}
-2J\langle\hat\Lambda_d^\dagger\rangle_{\mathrm{SC}}\langle\hat\Lambda_d\rangle_{\mathrm{SC}}\\
&=\frac1{N_s}\sum_{\{\mathbf k,-\mathbf k\}}
\left\langle\varepsilon_{\mathbf k}\hat N_{\mathbf k}-2\mathcal J_{\mathbf k}\hat\Delta_{\mathbf k}^{\dagger}\hat\Delta_{\mathbf k}\right\rangle_{\mathrm{SC}}
-\frac{|\Delta_d|^2}{2J}.
\end{aligned}
}
 
Setting \(\Delta_d=0\) and replacing the label \(\mathrm{SC}\) with \(\mathrm N\) gives the corresponding thermodynamic quantities of the nodal Bose metal at the same electron density. For the continuous SC transition,

\hypertarget{sm-eq-sc-thermo-continuity-at-transition}{}%
\FittedEquation{O5}{%
s_{\mathrm{SC}}(T_{c,d})-s_{\mathrm N}(T_{c,d})=f_{\mathrm{SC}}(T_{c,d})-f_{\mathrm N}(T_{c,d})=u_{\mathrm{SC}}(T_{c,d})-u_{\mathrm N}(T_{c,d})=0.
}
 
\hypertarget{sm-sc-thermo-field}{}\hypertarget{sm-sc-thermo-free-condensation}{}The zero-temperature condensation energy per site is

\hypertarget{sm-eq-sc-thermo-condensation-energy-definition}{}%
\FittedEquation{O6}{%
\begin{aligned}
\epsilon_{\mathrm{conden}}&\equiv f_{\mathrm N}(0)-f_{\mathrm{SC}}(0)=u_{\mathrm N}(0)-u_{\mathrm{SC}}(0)\\
&=\int_0^{T_{c,d}}\!\left[s_{\mathrm N}(T')-s_{\mathrm{SC}}(T')\right]\,dT'
=\int_0^{T_{c,d}}\!\left[c_{\mathrm{SC}}(T')-c_{\mathrm N}(T')\right]\,dT'>0,
\end{aligned}
}
 
where \(s=-\left.\partial f/\partial T\right|_{n_e}\), \(c=T\left.\partial s/\partial T\right|_{n_e}=\left.\partial u/\partial T\right|_{n_e}\), and \(s(0)=0\).

The thermodynamic critical field is determined by the condensation free energy. In Gaussian units,

\hypertarget{sm-eq-sc-thermo-critical-field}{}%
\FittedEquation{O7}{%
\frac{H_c^2(T)}{8\pi}=\frac{f_{\mathrm N}(T)-f_{\mathrm{SC}}(T)}{v_0},
}
 
where \(v_0\) is the physical volume per site. An equivalent expression is

\hypertarget{sm-eq-sc-thermo-critical-field-ratio}{}%
\FittedEquation{O8}{%
\frac{H_c(T)}{H_c(0)}=\sqrt{\frac{f_{\mathrm N}(T)-f_{\mathrm{SC}}(T)}{\epsilon_{\mathrm{conden}}}}.
}
 
\hypertarget{sm-sc-thermo-specific-heat}{}
\subsection*{O.2. Specific heat jump}

The SC specific heat per site at fixed electron density \(n_e\) is

\hypertarget{sm-eq-sc-specific-heat-definition}{}%
\FittedEquation{O9}{%
c_{\mathrm{SC}}=T\left.\frac{\partial s_{\mathrm{SC}}}{\partial T}\right|_{n_e}=-\frac{T}{N_s}\sum_{\{\mathbf k,-\mathbf k\}}\sum_{\lambda}\left(1+\ln p^{\mathrm{SC}}_{\mathbf k,\lambda}\right)\left.\frac{d p^{\mathrm{SC}}_{\mathbf k,\lambda}}{dT}\right|_{n_e}.
}
 
Taking the logarithm of the SC thermal occupation probability gives

\hypertarget{sm-eq-sc-log-block-occupation}{}%
\FittedEquation{O10}{%
\ln p^{\mathrm{SC}}_{\mathbf k,\lambda}=-\frac{E^{\mathrm{SC}}_{\mathbf k,\lambda}}{T}-\ln Z^{\mathrm{SC}}_{\mathbf k}.
}
 
Differentiating \hyperlink{sm-eq-sc-log-block-occupation}{Eq. (O10)} at fixed electron density gives

\hypertarget{sm-eq-sc-log-block-occupation-derivative}{}%
\FittedEquation{O11}{%
\frac{1}{p^{\mathrm{SC}}_{\mathbf k,\lambda}}\left.\frac{d p^{\mathrm{SC}}_{\mathbf k,\lambda}}{dT}\right|_{n_e}
=\frac{E^{\mathrm{SC}}_{\mathbf k,\lambda}}{T^2}-\frac{1}{T}\left.\frac{d E^{\mathrm{SC}}_{\mathbf k,\lambda}}{dT}\right|_{n_e}
-\left.\frac{d\ln Z^{\mathrm{SC}}_{\mathbf k}}{dT}\right|_{n_e}.
}
 
At fixed \(n_e\), the SC block eigenenergies \(E^{\mathrm{SC}}_{\mathbf k,\lambda}\) can themselves vary with \(T\) because both the order parameter \(\Delta_d(T)\) and the chemical potential \(\mu_{\mathrm{SC}}(T)\) can change with temperature.

From the SC block partition function \(Z^{\mathrm{SC}}_{\mathbf k}=\sum_{\lambda'}e^{-E^{\mathrm{SC}}_{\mathbf k,\lambda'}/T}\), the logarithmic derivative is

\hypertarget{sm-eq-sc-log-block-partition-derivative}{}%
\FittedEquation{O12}{%
\left.\frac{d\ln Z^{\mathrm{SC}}_{\mathbf k}}{dT}\right|_{n_e}
=\sum_{\lambda'}p^{\mathrm{SC}}_{\mathbf k,\lambda'}
\left[\frac{E^{\mathrm{SC}}_{\mathbf k,\lambda'}}{T^2}-\frac{1}{T}\left.\frac{d E^{\mathrm{SC}}_{\mathbf k,\lambda'}}{dT}\right|_{n_e}\right].
}
 
Substituting \hyperlink{sm-eq-sc-log-block-partition-derivative}{Eq. (O12)} into \hyperlink{sm-eq-sc-log-block-occupation-derivative}{Eq. (O11)} and using \(\sum_{\lambda'}p^{\mathrm{SC}}_{\mathbf k,\lambda'}=1\) gives

\hypertarget{sm-eq-sc-probability-temperature-derivative}{}%
\FittedEquation{O13}{%
\left.\frac{d p^{\mathrm{SC}}_{\mathbf k,\lambda}}{dT}\right|_{n_e}=\frac{p^{\mathrm{SC}}_{\mathbf k,\lambda}}{T^2}\sum_{\lambda'}p^{\mathrm{SC}}_{\mathbf k,\lambda'}\Biggl[E^{\mathrm{SC}}_{\mathbf k,\lambda}-E^{\mathrm{SC}}_{\mathbf k,\lambda'}-T\left.\frac{d(E^{\mathrm{SC}}_{\mathbf k,\lambda}-E^{\mathrm{SC}}_{\mathbf k,\lambda'})}{dT}\right|_{n_e}\Biggr].
}
 
Using \(\sum_{\lambda}\left.d p^{\mathrm{SC}}_{\mathbf k,\lambda}/dT\right|_{n_e}=0\), substituting \hyperlink{sm-eq-sc-probability-temperature-derivative}{Eq. (O13)} into \hyperlink{sm-eq-sc-specific-heat-definition}{Eq. (O9)} and symmetrizing the sum over \(\lambda\) and \(\lambda'\) gives

\hypertarget{sm-eq-sc-heat-capacity}{}%
\FittedEquation{O14}{%
\begin{aligned}
c_{\mathrm{SC}}
&=\frac{1}{N_s}\sum_{\{\mathbf k,-\mathbf k\},\lambda}
 E^{\mathrm{SC}}_{\mathbf k,\lambda}\left.\frac{d p^{\mathrm{SC}}_{\mathbf k,\lambda}}{dT}\right|_{n_e}\\
&=\frac{1}{2N_s T^2}\sum_{\{\mathbf k,-\mathbf k\}}\sum_{\lambda,\lambda'}
p^{\mathrm{SC}}_{\mathbf k,\lambda}p^{\mathrm{SC}}_{\mathbf k,\lambda'}
\Biggl[\left(E^{\mathrm{SC}}_{\mathbf k,\lambda}-E^{\mathrm{SC}}_{\mathbf k,\lambda'}\right)^2-T\left(E^{\mathrm{SC}}_{\mathbf k,\lambda}-E^{\mathrm{SC}}_{\mathbf k,\lambda'}\right)
\left.\frac{d(E^{\mathrm{SC}}_{\mathbf k,\lambda}-E^{\mathrm{SC}}_{\mathbf k,\lambda'})}{dT}\right|_{n_e}\Biggr].
\end{aligned}
}
 
The Hellmann--Feynman theorem gives

\hypertarget{sm-eq-sc-level-temperature-derivative}{}%
\FittedEquation{O15}{%
\left.\frac{dE_{\mathbf k,\lambda}^{\mathrm{SC}}}{dT}\right|_{n_e}
=-\left.\frac{d\mu_{\mathrm{SC}}}{dT}\right|_{n_e}\langle\mathbf k,\lambda|\hat N_{\mathbf k}|\mathbf k,\lambda\rangle
-2\phi_{\mathbf k,d}\left.\frac{d\Delta_d}{dT}\right|_{n_e}\langle\mathbf k,\lambda|\hat\Delta_{\mathbf k}^{\dagger}+\hat\Delta_{\mathbf k}|\mathbf k,\lambda\rangle.
}
 
The gap and density equations in \hyperlink{sm-eq-sc-appendix-gap-equation}{Eqs. (J18)} and \hyperlink{sm-eq-sc-appendix-density-equation}{(J19)} are obtained by imposing the stationarity conditions:

\hypertarget{sm-eq-sc-thermo-fixed-density-stationarity}{}%
\FittedEquation{O16}{%
\frac{\partial f_{\mathrm{SC}}}{\partial\Delta_d}=\frac{\partial f_{\mathrm{SC}}}{\partial\mu_{\mathrm{SC}}}=0,
}
 
where the global phase is chosen so that \(\Delta_d\) is real. At fixed \(n_e\), differentiating \hyperlink{sm-eq-sc-thermo-fixed-density-stationarity}{Eq. (O16)} with respect to \(T\) gives

\hypertarget{sm-eq-sc-thermo-gap-total-derivative}{}%
\FittedEquation{O17}{%
\left.\frac{d}{dT}\!\left(\frac{\partial f_{\mathrm{SC}}}{\partial\Delta_d}\right)\right|_{n_e}
=\frac{\partial^2 f_{\mathrm{SC}}}{\partial T\,\partial\Delta_d}
+\frac{\partial^2 f_{\mathrm{SC}}}{\partial\Delta_d^2}\left.\frac{d\Delta_d}{dT}\right|_{n_e}
+\frac{\partial^2 f_{\mathrm{SC}}}{\partial\mu_{\mathrm{SC}}\,\partial\Delta_d}\left.\frac{d\mu_{\mathrm{SC}}}{dT}\right|_{n_e}=0,
}
 
\hypertarget{sm-eq-sc-thermo-density-total-derivative}{}%
\FittedEquation{O18}{%
\left.\frac{d}{dT}\!\left(\frac{\partial f_{\mathrm{SC}}}{\partial\mu_{\mathrm{SC}}}\right)\right|_{n_e}
=\frac{\partial^2 f_{\mathrm{SC}}}{\partial T\,\partial\mu_{\mathrm{SC}}}
+\frac{\partial^2 f_{\mathrm{SC}}}{\partial\Delta_d\,\partial\mu_{\mathrm{SC}}}\left.\frac{d\Delta_d}{dT}\right|_{n_e}
+\frac{\partial^2 f_{\mathrm{SC}}}{\partial\mu_{\mathrm{SC}}^2}\left.\frac{d\mu_{\mathrm{SC}}}{dT}\right|_{n_e}=0.
}
 
The derivatives \(\left.d\Delta_d/dT\right|_{n_e}\) and \(\left.d\mu_{\mathrm{SC}}/dT\right|_{n_e}\) are obtained by solving

\hypertarget{sm-eq-sc-thermo-implicit-derivatives}{}%
\FittedEquation{O19}{%
\begin{pmatrix}
\dfrac{\partial^2 f_{\mathrm{SC}}}{\partial\Delta_d^2}
&\dfrac{\partial^2 f_{\mathrm{SC}}}{\partial\mu_{\mathrm{SC}}\,\partial\Delta_d}\\[5pt]
\dfrac{\partial^2 f_{\mathrm{SC}}}{\partial\Delta_d\,\partial\mu_{\mathrm{SC}}}
&\dfrac{\partial^2 f_{\mathrm{SC}}}{\partial\mu_{\mathrm{SC}}^2}
\end{pmatrix}
\begin{pmatrix}
\left.\dfrac{d\Delta_d}{dT}\right|_{n_e}\\[5pt]
\left.\dfrac{d\mu_{\mathrm{SC}}}{dT}\right|_{n_e}
\end{pmatrix}
=-\begin{pmatrix}
\dfrac{\partial^2 f_{\mathrm{SC}}}{\partial T\,\partial\Delta_d}\\[5pt]
\dfrac{\partial^2 f_{\mathrm{SC}}}{\partial T\,\partial\mu_{\mathrm{SC}}}
\end{pmatrix}.
}
 
For \(T>T_{c,d}\), \(\Delta_d=0\) and \hyperlink{sm-eq-sc-heat-capacity}{Eq. (O14)} reduces to \hyperlink{sm-eq-thermo-f-heat-exact}{Eq. (F50)} in the nodal Bose metal. The specific-heat jump at the continuous SC transition at fixed \(n_e\) is

\hypertarget{sm-eq-sc-heat-capacity-jump}{}%
\FittedEquation{O20}{%
\Delta c=c_{\mathrm{SC}}(T_{c,d}^{-})-c_{\mathrm N}(T_{c,d}^{+}).
}
 
\hypertarget{sm-sc-thermo-nodal-limit}{}
\subsection*{O.3. Low-temperature behavior}

At fixed \(n_e\) and \(T\ll\min\{T_{c,d},E_*(0)\}\), the leading thermal contributions come from regions near the four gapless Dirac nodes [\hyperlink{sm-eq-sc-dos-dirac-energy}{Eq. (N6)}], whereas the RML interaction is subleading at excitation energies \(\omega\sim T\) [\hyperlink{sm-eq-sc-dos-rml-relative-scale}{Eq. (N7)}]. The thermodynamic quantities therefore have the same leading low-temperature behavior as those of a conventional nodal BCS state. For \(\omega>0\), the positive- and negative-energy poles of the zeroth-order one-particle spectral function [\hyperlink{sm-eq-sc-dos-dirac-spectral}{Eq. (N9)}] contribute to the Bogoliubov quasiparticle DOS per spin as follows:

\hypertarget{sm-eq-sc-thermo-nodal-electron-dos}{}%
\FittedEquation{O21}{%
\rho_{\mathrm{SC}}^{(0)}(\omega,0)=\frac{1}{N_s}\sum_{\mathbf k}u_{\mathbf k}^{2}\,\delta(\omega-E_{\mathbf k}),
}
 
\hypertarget{sm-eq-sc-thermo-nodal-negative-energy-dos}{}%
\FittedEquation{O22}{%
\rho_{\mathrm{SC}}^{(0)}(-\omega,0)=\frac{1}{N_s}\sum_{\mathbf k}v_{\mathbf k}^{2}\,\delta(-\omega+E_{\mathbf k}).
}
 
Here \(E_{\mathbf k}>0\) is the Bogoliubov quasiparticle dispersion. The electron and hole spectral weights satisfy \(u_{\mathbf k}^2+v_{\mathbf k}^2=(1+v_n k_\perp/E_{\mathbf k})/2+(1-v_n k_\perp/E_{\mathbf k})/2=1\). Adding the two spectral weights thus counts each Bogoliubov quasiparticle once per spin. Including both spins gives the Bogoliubov quasiparticle DOS per site for \(\omega>0\),

\hypertarget{sm-eq-sc-thermo-nodal-quasiparticle-dos}{}%
\FittedEquation{O23}{%
\begin{aligned}
D_{\mathrm{qp}}(\omega)
&=2\big[\rho_{\mathrm{SC}}^{(0)}(\omega,0)+\rho_{\mathrm{SC}}^{(0)}(-\omega,0)\big]\\
&=\frac{2}{N_s}\sum_{\mathbf k}(u_{\mathbf k}^2+v_{\mathbf k}^2)\,\delta(\omega-E_{\mathbf k})
=\frac{1}{N_s}\sum_{\mathbf k,\sigma}\delta(\omega-E_{\mathbf k}).
\end{aligned}
}
 
In the leading-order nodal approximation, emergent particle--hole symmetry gives \(\rho_{\mathrm{SC}}^{(0)}(-\omega,0)=\rho_{\mathrm{SC}}^{(0)}(\omega,0)\). Hence \(D_{\mathrm{qp}}(\omega)=4\rho_{\mathrm{SC}}^{(0)}(\omega,0)=4\pi \omega/V(0)\) [\hyperlink{sm-eq-sc-dos-dirac-integral}{Eq. (N10)}], where \(V(0)=\pi^2v_n(0)v_\Delta(0)\) and the factor of four does not arise from counting the four gapless Dirac nodes.

Weighting the nodal Bogoliubov quasiparticles by their energy \(\omega\) and occupation \(1/(e^{\omega/T}+1)\) gives the leading thermal correction to the internal energy per site:

\hypertarget{sm-eq-sc-thermo-nodal-energy}{}%
\FittedEquation{O24}{%
\begin{aligned}
u_{\mathrm{SC}}(T,n_e)-u_{\mathrm{SC}}(0,n_e)
&\simeq\int_0^\infty\frac{D_{\mathrm{qp}}(\omega)\,\omega\,d\omega}{e^{\omega/T}+1}\\
&=\frac{4\pi}{V(0)}\int_0^\infty\frac{\omega^2\,d\omega}{e^{\omega/T}+1}
=\frac{6\pi\zeta(3)}{V(0)}T^3.
\end{aligned}
}
 
With \(s_{\mathrm{SC}}(0,n_e)=0\), integrating \(ds_{\mathrm{SC}}=du_{\mathrm{SC}}/T\) at fixed \(n_e\) gives the entropy per site:

\hypertarget{sm-eq-sc-thermo-nodal-entropy}{}%
\FittedEquation{O25}{%
s_{\mathrm{SC}}(T,n_e)=\int_0^T\frac{dT'}{T'}\left.\frac{\partial u_{\mathrm{SC}}(T',n_e)}{\partial T'}\right|_{n_e}
\simeq\frac{9\pi\zeta(3)}{V(0)}T^2.
}
 
The specific heat per site then is

\hypertarget{sm-eq-sc-thermo-nodal-heat-capacity}{}%
\FittedEquation{O26}{%
c_{\mathrm{SC}}(T,n_e)=T\left.\frac{\partial s_{\mathrm{SC}}}{\partial T}\right|_{n_e}
\simeq\frac{18\pi\zeta(3)}{V(0)}T^2.
}
 
The leading thermal correction to the Helmholtz free energy per site follows from \(f_{\mathrm{SC}}=u_{\mathrm{SC}}-Ts_{\mathrm{SC}}\):

\hypertarget{sm-eq-sc-thermo-nodal-free-energy}{}%
\FittedEquation{O27}{%
f_{\mathrm{SC}}(T,n_e)-f_{\mathrm{SC}}(0,n_e)\simeq-\frac{3\pi\zeta(3)}{V(0)}T^3.
}
 
With \(c_{\mathrm N}(T,n_e)=\gamma_0T+\gamma_1T^{3/2}+o(T^{3/2})\) from \hyperlink{sm-eq-mstj-heat-capacity-low-temperature}{Eq. (45)}, the leading low-temperature free-energy difference is

\hypertarget{sm-eq-sc-thermo-low-t-free-energy-difference}{}%
\FittedEquation{O28}{%
f_{\mathrm N}(T,n_e)-f_{\mathrm{SC}}(T,n_e)=\epsilon_{\mathrm{conden}}-\frac{\gamma_0}{2}T^2+o(T^2).
}
 
The thermodynamic critical field follows from \hyperlink{sm-eq-sc-thermo-critical-field-ratio}{Eq. (O8)}:

\hypertarget{sm-eq-sc-thermo-low-t-critical-field}{}%
\FittedEquation{O29}{%
\frac{H_c(T)}{H_c(0)}=\sqrt{\frac{\epsilon_{\mathrm{conden}}-\gamma_0 T^2/2+o(T^2)}{\epsilon_{\mathrm{conden}}}}\simeq 1-\frac{\gamma_0 T^2}{4\epsilon_{\mathrm{conden}}}.
}
 
\hypertarget{sm-appendix-experimental-arpes-gap-data}{}\hypertarget{sm-appendix-experimental-arpes-gap-data-title}{}
\section*{Appendix P. ARPES gap data}

The appendix compiles published ARPES gap data along the Fermi surfaces of the \(n\)-layer Bi cuprates \(\mathrm{Bi}22(n-1)n\), with \(n=1,2,3\): Bi2201 in \hyperlink{sm-fig-p-gap-scaling-bi2201-reviewed}{Fig. P1} from Refs. [\hyperlink{sm-app-p-ref-wei-2008}{P1}--\hyperlink{sm-app-p-ref-kondo-2007}{P7}], Bi2212 in \hyperlink{sm-fig-p-gap-scaling-bi2212-reviewed}{Fig. P2} from Refs. [\hyperlink{sm-app-p-ref-ding-1996}{P8}--\hyperlink{sm-app-p-ref-vishik-2009}{P16}], and Bi2223 in \hyperlink{sm-fig-p-gap-scaling-bi2223-reviewed}{Fig. P3} from Refs. [\hyperlink{sm-app-p-ref-luo-2023}{P17}--\hyperlink{sm-app-p-ref-kunisada-2017}{P22}]. The compilation uses published numerical source data where available. For the remaining datasets, the reported gap values and visible error bars are extracted from the original figures.

Extensive experimental evidence, particularly from phase-sensitive measurements [\hyperlink{sm-app-p-ref-wollman-1993}{P23}--\hyperlink{sm-app-p-ref-tsuei-kirtley-2000}{P26}], establishes predominantly \(d_{x^2-y^2}\) pairing symmetry in cuprate superconductors. In theoretical studies, the pairing field is often taken to be linear in the \(d\)-wave form factor, \(\Delta_{\mathbf k}^{\mathrm{SC}}=\Delta_{\mathrm{SC}}\phi_{\mathbf k,d}\), with \(\phi_{\mathbf k,d}=(\cos k_x-\cos k_y)/2\). The appendix uses the compiled ARPES gap data to examine the phenomenological basis for the quadratic form \(\mathcal J_{\mathbf k}^{\mathrm N}=\mathcal J_{\mathrm N}\phi_{\mathbf k,d}^2\) of the RML interaction.

Each dataset retains its original horizontal coordinate, either the \(d\)-wave form factor or the Fermi-surface angle. The two types of datasets are analyzed separately, using \(u=|\phi_{\mathbf k,d}|\) or \(u=|\cos2\theta|\), respectively, where \(\theta=0\) at the antinode and \(\theta=45^\circ\) at the node. For each data group, equal-weight least squares in the original energy units, with the intercept fixed at zero, determines \(\Delta_{\mathrm{SC}}\) from \(E=\Delta_{\mathrm{SC}} u\) at \(T<T_c\), or \(\mathcal J_{\mathrm N}\) from \(E=\mathcal J_{\mathrm N} u^2\) at \(T_c<T<T^*\). All retained points, including gapless points, enter the fits. Different samples, temperatures, and resolved planes or bands retain separate fitted amplitudes. The plotted gaps and reported vertical errors are divided by the corresponding amplitude, giving \(E/\Delta_{\mathrm{SC}}\) or \(E/\mathcal J_{\mathrm N}\); fit-amplitude uncertainty is not propagated into the displayed error bars. The amplitudes \(\Delta_{\mathrm{SC}}\) and \(\mathcal J_{\mathrm N}\) provide a common normalization for comparing gap shapes. The original gap data are also fitted to a power law, \(E\propto u^\alpha\), allowing the exponent to vary without imposing an a priori normalization scheme. The power-law fits use the same equal-weight least-squares criterion in the original energy units.

\par\medskip
\hypertarget{sm-fig-p-gap-scaling-bi2201-reviewed}{}
\hypertarget{sm-fig-p-gap-scaling-bi2201-reviewed-sc-form-factor}{}
\hypertarget{sm-fig-p-gap-scaling-bi2201-reviewed-pg-form-factor}{}
\hypertarget{sm-fig-p-gap-scaling-bi2201-reviewed-sc-angle}{}
\hypertarget{sm-fig-p-gap-scaling-bi2201-reviewed-pg-angle}{}
\hypertarget{sm-fig-p-gap-scaling-bi2201-reviewed-sc-exponent}{}
\hypertarget{sm-fig-p-gap-scaling-bi2201-reviewed-pg-exponent}{}
\PrintAsset{bi2201-reviewed-gap-triptych}
{\small\noindent \textbf{Figure P1. Bi2201 normalized gaps and fitted power-law exponents.} (a,b) Gaps versus the \(d\)-wave form factor \(\phi_{\mathbf k,d}=(\cos k_x-\cos k_y)/2\) and (c,d) gaps versus the Fermi-surface angle \(\theta\), below and above \(T_c\), respectively. Open symbols and bars show the normalized gaps and reported errors; legends identify the sample, temperature, and fitted amplitude in meV. In sample labels, UD, OP (or OPT), and OD denote underdoped, optimally doped, and overdoped samples, respectively; the following number gives \(T_c\) in kelvin. Dark navy long-dashed and deep magenta short-dashed curves show the linear and quadratic guides. (e,f) Best-fit exponents below and above \(T_c\), respectively. Filled circles and triangles denote form-factor and angular fits, respectively; bars show nominal 95\% bootstrap intervals. The solid line and gray band indicate the median and interquartile range of the best-fit exponents, respectively. Intervals extending beyond \(\alpha=10\) are clipped.\par}
\medskip

\par\medskip
\hypertarget{sm-fig-p-gap-scaling-bi2212-reviewed}{}
\hypertarget{sm-fig-p-gap-scaling-bi2212-sc-form-factor}{}
\hypertarget{sm-fig-p-gap-scaling-bi2212-pg-form-factor}{}
\hypertarget{sm-fig-p-gap-scaling-reviewed-sc}{}
\hypertarget{sm-fig-p-gap-scaling-bi2212-reviewed-pg-angle}{}
\hypertarget{sm-fig-p-gap-scaling-bi2212-reviewed-sc-exponent}{}
\hypertarget{sm-fig-p-gap-scaling-bi2212-reviewed-pg-exponent}{}
\PrintAsset{bi2212-reviewed-gap-quad}
{\small\noindent \textbf{Figure P2. Bi2212 normalized gaps and fitted power-law exponents.} (a,b) Gaps versus the \(d\)-wave form factor \(\phi_{\mathbf k,d}=(\cos k_x-\cos k_y)/2\) and (c,d) gaps versus the Fermi-surface angle \(\theta\), below and above \(T_c\), respectively. (e,f) Best-fit exponents below and above \(T_c\), respectively. BB and AB denote bonding and antibonding bands. Symbols, guide curves, and uncertainty conventions follow \hyperlink{sm-fig-p-gap-scaling-bi2201-reviewed}{Fig. P1}.\par}
\medskip

\par\medskip
\hypertarget{sm-fig-p-gap-scaling-bi2223-reviewed}{}
\hypertarget{sm-fig-p-gap-scaling-bi2223-reviewed-sc-form-factor}{}
\hypertarget{sm-fig-p-gap-scaling-bi2223-reviewed-pg-form-factor}{}
\hypertarget{sm-fig-p-gap-scaling-bi2223-reviewed-sc-angle}{}
\hypertarget{sm-fig-p-gap-scaling-bi2223-reviewed-sc-exponent}{}
\hypertarget{sm-fig-p-gap-scaling-bi2223-reviewed-pg-exponent}{}
\PrintAsset{bi2223-reviewed-gap-triptych}
{\small\noindent \textbf{Figure P3. Bi2223 normalized gaps and fitted power-law exponents.} (a,b) Gaps versus the \(d\)-wave form factor \(\phi_{\mathbf k,d}=(\cos k_x-\cos k_y)/2\) below and above \(T_c\), respectively; (c) gaps versus the Fermi-surface angle \(\theta\) below \(T_c\). Data for the outer plane (OP) and inner plane (IP), or for resolved bands, are shown separately. (d,e) Best-fit exponents below and above \(T_c\), respectively. Symbols, guide curves, and uncertainty conventions follow \hyperlink{sm-fig-p-gap-scaling-bi2201-reviewed}{Fig. P1}.\par}
\medskip

Qualitatively, the gaps above \(T_c\) decrease more rapidly on approaching the node than the gaps below \(T_c\). The power-law fits reflect this superlinear behavior: 53 of the 55 data groups above \(T_c\) have best-fit \(\alpha>1\). Below \(T_c\), the data lie closer to the linear guide curve, with median exponents \(1.52\), \(1.35\), and \(1.21\) for Bi2201, Bi2212, and Bi2223, respectively. Thus the median approaches unity across the single-, double-, and triple-layer compounds. Above \(T_c\), the corresponding medians are \(4.09\), \(3.46\), and \(1.69\), approaching the quadratic guide curve most closely in Bi2223.

The deviations from the guide curves are asymmetric near the node: below \(T_c\), more data lie below the linear guide curve than above it; above \(T_c\), most lie below the quadratic guide curve. This tendency occurs in all three materials. Across Bi2201, Bi2212, and Bi2223, the median root-mean-square deviation from the respective guide curves across data groups falls from \(0.109\) to \(0.056\) to \(0.045\) below \(T_c\), and from \(0.154\) to \(0.118\) to \(0.027\) above \(T_c\), with both medians decreasing as \(n\) increases in \(\mathrm{Bi}22(n-1)n\) cuprates.

Spatial phase fluctuations play an important role in cuprates [\hyperlink{sm-app-p-ref-emery-kivelson-1995}{P27}] and can provide a possible interpretation of the near-nodal suppression. With a finite local pairing amplitude, slowly varying SC phase fluctuations can Doppler broaden the spectrum and fill the gap [\hyperlink{sm-app-p-ref-franz-millis-1998}{P28}]. Comparable Doppler broadening has a larger relative effect on smaller near-nodal gaps.

The closer agreement with the guide curves accompanies the well-known increase in the maximum transition temperature \(T_{c,\max}\) from Bi2201 to Bi2223 [\hyperlink{sm-app-p-ref-wang-2023-charge-transfer}{P29}]. One possible interpretation is reduced thermal phase disorder in coupled layers: when their relative phases are locked, the layer stiffnesses add for the common phase mode, while interlayer Josephson coupling penalizes relative-phase fluctuations. Although the compiled ARPES gap data are insufficient to establish a universal quadratic law or uniquely determine the RML interaction, \(\mathcal J_{\mathbf k}^{\mathrm N}=\mathcal J_{\mathrm N}\phi_{\mathbf k,d}^2\) remains a plausible phenomenological gap scale for modeling the pseudogap regime.

 \hypertarget{sm-appendix-p-references}{}
\begingroup
\renewcommand{\refname}{References for Appendix P}

\endgroup

\end{document}